\documentclass[headsepline,footsepline,letterpaper]{scrreprt}
\usepackage{varioref}
\usepackage[latin1]{inputenc}
\usepackage[T1]{fontenc}
\usepackage{color}
\usepackage{capt-of}
\usepackage[font=small,labelfont=bf]{caption}[2003/12/20]
\usepackage{mathpazo}
\usepackage[scaled=.95]{helvet}
\usepackage{courier}
\usepackage{epsfig}
\usepackage{afterpage}
\usepackage{graphics}
\usepackage{amsmath}
\usepackage{amsfonts}
\usepackage{hyperref}
\usepackage{upgreek}
\usepackage{amssymb}
\usepackage{textcomp}
\usepackage[numbers,square,sort&compress,super]{natbib}

\def\st{\item[$\bullet$]}

\def\ra{$\rightarrow$ }
\def\be{\begin{equation}}
\def\ee{\end{equation}}
\def\bi{\begin{itemize}}
\def\ei{\end{itemize}}
\def\bmx{\begin{matrix}}
\def\emx{\end{matrix}}
\newcommand{\sub}[1]{\ensuremath{_{\text{{#1}}}}}
\begin{document}
\clubpenalty = 10000 \widowpenalty = 10000 \displaywidowpenalty = 10000

\pagenumbering{roman}

\begin{titlepage}
\thispagestyle{empty}

\begin{center}
\mbox{}

\vspace{0.5cm}
\huge \textsf{\textbf{Search for Cosmic-Ray Antiparticles\\ with Balloon-borne and Space-borne Experiments}}

\vspace{2.5cm}
\normalsize Von der Fakultät für Mathematik, Informatik und Naturwissenschaften der RWTH Aachen University zur Erlangung des akademischen Grades eines Doktors der Naturwissenschaften\\ genehmigte Dissertation\\ 

\vspace{2.5cm}
vorgelegt  von

\vspace{1cm}
Diplom-Physiker\\ Philip von Doetinchem

\vspace{1cm}
aus Mülheim an der Ruhr

\vspace{4cm}
\begin{tabular}{ll}
Berichter: 	& Universitätsprofessor Dr. Stefan Schael\\
		& Universitätsprofessor Dr. Klaus Lübelsmeyer
\end{tabular}

\vspace{2cm}
Tag der mündlichen Prüfung: 14. Mai 2009

\vspace{2.5cm}
Diese Dissertation ist auf den Internetseiten der Hochschulbibliothek online verfügbar.
\end{center}

\end{titlepage}

\thispagestyle{empty}

\newpage
\thispagestyle{empty}
\mbox{}
\newpage
\thispagestyle{empty}

\begin{center}
\mbox{}
\vspace{4cm}

\centerline{\epsfig{file=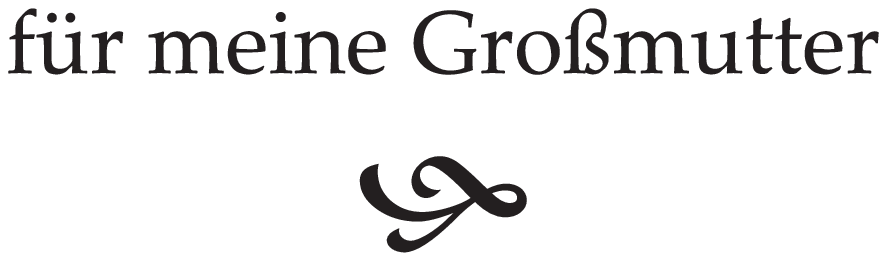,width=6cm}}
\end{center}

\newpage
\thispagestyle{empty}
\mbox{}
\newpage
\thispagestyle{empty}

\thispagestyle{empty}
\chapter*{Abstract}
\thispagestyle{empty}
This thesis discusses two different approaches for the measurement of cosmic-ray antiparticles in the GeV to TeV energy range. 

The first part studies the prospects of antiparticle flux  measurements with the proposed PEBS detector. The project allots long duration balloon flights at one of Earth's poles at an altitude of 40\,km. The detector consists of a transition radiation detector, a time of flight system, a combined silicon and scintillating fiber tracker and an electromagnetic calorimeter. All detectors are placed in a superconducting magnet which creates a field of 0.8\,T. Several flights are discussed starting from 2012 to get a total of 100\, days measurement time with an acceptance of 0.4\,m$^2$sr. This work calculates the systematic effects due to interactions of cosmic rays in the atmosphere. GEANT4 simulations were carried out which determine the atmospheric background and attenuation especially for antiparticles. Projected results taking the effects of the atmosphere and the detector properties into account are discussed.

The second part covers the AMS-02 experiment which will be installed in 2010 on the International Space Station at an altitude of about 400\,km for about three years to measure cosmic rays without the influence of Earth's atmosphere. The detector consists of several subdetectors for the determination of particle properties, namely a transition radiation detector, a time of flight system, a cylindrical silicon tracker with eight layers surrounded by an anticoincidence counter in a superconducting magnet with a field of about 0.8\,T strength, a ring image \v{C}erenkov detector and an electromagnetic calorimeter. The total acceptance with and without the electromagnetic calorimeter is 0.095\,m$^2$sr and 0.45\,m$^2$sr, respectively. The present work focuses on the anticoincidence counter system (ACC). The ACC is needed to reduce the trigger rate during periods of high fluxes and to reject events with external particles crossing the tracker from the side or with particles resulting from interactions within the detector which have possibly disturbed charge and momentum measurements. The last point is especially important for the measurement of anti\-nuclei and antiparticles. The ACC has a modular design of 16 singular plastic scintillator panels which form a cylinder around the tracker with a a diameter of 1100\,mm, a height of 830\,mm and a thickness of 8\,mm. The scintillator light is guided by wavelength shifting and clear fibers to photomultiplier tubes. The detector has to cope with several challenges: Fast response and stable operation in a high magnetic field are required. The detector has to withstand a launch with a Space Shuttle into space. The power consumption is only 800\,mW and the total weight is 54\,kg. The qualification and performance tests of the panels, fibers, photomultiplier tubes and flight electronics are described. The ACC detection efficiency for charged particles is extracted from testbeam and atmospheric muon measurements and simulations and enters as input to the antimatter measurement performance. In addition, the AMS-02 possibilities for the detection of positrons and antiprotons are studied.

\thispagestyle{empty}
\newpage
\thispagestyle{empty}
\mbox{}
\newpage

\thispagestyle{empty}
\chapter*{Zusammenfassung}
\thispagestyle{empty}
Diese Arbeit diskutiert zwei verschiedene Ansätze zur Messung von Antiteilchen in der kos\-mischen Strahlung im GeV bis TeV Energiebereich.

Im ersten Teil wird die Fähigkeit zur Messung des Antiteilchenflusses mit dem vorgeschlagenen PEBS Detektor diskutiert. Das Projekt sieht Langzeitballonflüge an einem der Erdpole in 40\,km Höhe vor. Der Detektor besteht aus einem Übergangsstrahlungsdetektor, einem Flugzeitzähler, einer Spurkammer aus Silizium und szintillierenden Fasern und einem elektromagnetischen Kalorimeter. Die Detektoren befinden sich innerhalb eines supraleitenden Magneten, der ein Feld von 0.8\,T erzeugt. Es sollen beginnend von 2012 mehrere Flüge stattfinden, um insgesamt 100\,Tage Messzeit mit einer Akzeptanz von 0.4\,m$^2$sr zu erreichen. Hier werden die systematischen Effekte von Wechselwirkungen der kosmischen Strahlung in der Atmosphäre berechnet. Dafür wurden GEANT4 Simulationen durchgeführt, die den atmosphärischen Untergrund und die atmosphärische Abschwächung insbesondere für Antiteilchen bestimmen. Simulier\-te Daten werden unter Berücksichtigung atmosphärischer Effekte und Detektoreigenschaften diskutiert.

Der zweite Teil behandelt das AMS-02 Experiment, das 2010 auf der Internationalen Raumstation in ca. 400\,km Höhe für drei Jahre installiert werden wird, um kosmische Strahlung ohne den Einfluss der Erdatmophäre zu messen. Der Detektor besteht ebenfalls aus mehreren Subdetektoren, um die Teilcheneigenschaften zu bestimmen. Es handelt sich um einen Über\-gangs\-\-strahlungs\-detektor, einen Flugzeitzähler, eine zylindrische Siliziumspurkammer mit acht Lagen umgeben von einem Antikoinzidenzzähler innerhalb eines supraleitenden Magneten mit 0.8\,T Feldstärke, einem Ringbild \v{C}erenkov Detektor und einem elektromagnetischen Kalorimeter. Die Gesamt\-ak\-zep\-tanz beträgt 0.095\,m$^2$sr mit elektromagnetischen Kalorimeter und 0.45\,m$^2$sr ohne. Dieser Teil der Arbeit kon\-zen\-triert sich auf den Antikoinzidenzzähler (ACC). Der ACC wird benötigt, um die Triggerrate in Phasen hohen Flusses zu reduzieren und um Ereignisse mit externen Teilchen, die den Detektor von der Seite durchfliegen, oder mit Teilchen aus Wechsel\-wirk\-ungen im Detektor zu verwerfen, wo eine saubere Ladungs- und Impulsbestimmung möglicherweise gestört ist. Der letzte Punkt ist besonders wichtig für die Messung von Antikernen und Antiteilchen. Der ACC ist modular aus 16 einzelnen Kunststoffszintillationszählern aufgebaut, die einen Zylinder um die Spurkammer mit einem Durchmesser von 1100\,mm, einer Höhe von 830\,mm und einer Dicke von 8\,mm formen. Das Szintillationslicht wird mit wellenlängenschiebenden und klaren Fasern an Photovervielfacherröhren weitergeleitet. Der Detektor hat einige Herausforderungen zu bewältigen: Schnelles Ansprechen und stabiler Betrieb in einem hohen Magnetfeld, der Detektor muss nach dem Start mit einem Space Shuttle im Weltall funktionieren und die Gesamtleistung darf nur 800\,mW und das Gewicht 54\,kg betragen. Es werden die Qualifikations- und Verhaltenstests der Szintillationszähler, Fasern, Photo\-ver\-viel\-facher\-röhren und der Flugelektronik beschrieben und die ACC-Detektionseffizienz für geladene Teilchen durch einen Strahltest, Messungen mit atmos\-phä\-rischen Myonen und Simulationen bestimmt. Diese Effizienz geht in die Bestimmung der Fähigkeit zur Antimateriemessung ein. Zusätzlich werden die Mög\-lich\-kei\-ten für Positron- und Antiprotondetektion studiert.

\newpage
\thispagestyle{empty}
\mbox{}
\thispagestyle{empty}
\textheight40cm
\topmargin-2cm
\newpage

\thispagestyle{empty}
\tableofcontents
\thispagestyle{empty}

\textheight24.7cm

\newpage
\thispagestyle{empty}
\mbox{}
\topmargin-1cm
\newpage
\thispagestyle{empty}

\pagenumbering{arabic}

\chapter{Introduction}

The observation of the sky and the research in the structure of matter has a long tradition in mankind. The modern fields of particle physics and astrophysics are able to describe very well a lot of the properties of matter and galactic and extragalactic objects but when looking deeper new and so far unanswered questions arise. What we know nowadays is that the Universe has formed from the Big Bang where space and time were created. The expansion of the Universe, primordial nucleosynthesis and the cosmic microwave background support this theory strongly. The discipline of particle physics tries to reveal the nature of matter and its constituents by observing reactions at energies which prevailed very close to the Big Bang. The standard model of particle physics describes especially the electroweak interaction successfully. Astrophysics works on scales up to the size of the Universe and is attempting to explain the functionality of objects like planets and stars up to the structure of galaxy clusters. In the last decades both disciplines started working together and sharing analysis methods, observation techniques and data in order to constrain existing theories.

The asymmetry between matter and antimatter and the nature of dark matter are only two of the big mysteries in nature. The explanation of the observed asymmetry between matter and antimatter is the explanation for the existence of our matter dominated Universe. Matter and antimatter annihilate when they meet such so the Universe would be completely made out of photons in the case of a perfect symmetry. It is maybe possible that matter and antimatter have been separated at the time of the Big Bang. Otherwise, new so far unknown processes must be introduced to describe the excess of matter above antimatter dynamically. Currently approved theories do not provide mechanisms capable of generating the correct excess of matter over antimatter. In addition, various measurements show a deviation from the expected behavior of visible matter which can be explained by an additional form of matter composing about 23\,\% of the total mass budget of the Universe. This matter have not been measured yet directly by any experiment and is called dark matter. Dark matter is also needed to explain the formation of structure in the Universe. The challenge is now to find a viable dark matter candidate in particle physics without violating constraints from astrophysical observations. 

Cosmic rays arising from astrophysical objects are accelerated and transported in the cosmos and can be used as messengers of important properties of our Universe. This is also the case for dark matter and antimatter theories which can be constrained by measurements of cosmic rays (especially antiparticles). Particles are much more abundant than antiparticles. Therefore experiments with long measurement times, large angular acceptance and large discrimination power are needed for a high precision determination of the particle and antiparticle fluxes in cosmic rays. The Alpha Magnetic Spectrometer (AMS-02) experiment will measure cosmic rays for about three years on the International Space Station. The Positron Electron Balloon Spectrometer (PEBS) detector, similar in concept, is proposed to measure cosmic rays at one of Earth's poles during several balloon flights in the atmosphere. Chapter \ref{c-ap} will summarize the key points of current theories in particle physics and astrophysics which influence the missions concepts.

The PEBS experiment described in chapter \ref{c-pebs} focuses on antiproton and positron measurements during balloon flights in Earth's atmosphere at an altitude of 40\,km. The effects of the atmosphere and the geomagnetic field are important for the analysis of such experiments. Particles interact with the atmosphere leading to primary flux changes. This effect must be taken into account  because cosmic rays cross about 30\,\% of the total amount of matter on their way from the astrophysical source to the detector for a flight at an altitude of 40\,km in the atmosphere. To this purpose a simulation with the GEANT4 based PLANETOCOSMICS package was done. The details of the simulation and the atmospheric and magnetic models used are described and the results compared with earlier measurements. Taking the properties of PEBS into account the results are calculated at the South Pole.

Chapter \ref{c-ams} will give an overview of the space-based AMS-02 mission and discuss the design, qualification and performance of the anticoincidence counter (ACC) system in detail. The ACC is a plastic scintillator detector surrounding the silicon tracker which is used to measure particle momenta in the AMS-02 experiment. The light of the scintillator is guided to photodetectors via plastic optical fibers. Tests of the individual components were carried out in Aachen and electronics tests and a complete detector test were performed in Geneva.  The ACC will contribute to the detection of antimatter or the determination of upper bounds for the existence of antimatter by assuring very clean and undisturbed measurements. To improve current bounds the design of the ACC requires a detection efficiency of 99.99\,\% for all charged particle species. All tests focus on reaching the needed efficiency by maximizing the signal output of the detector and at the same time reducing the noise of the electronics to get a high resolution even for small signals.

\chapter{Astroparticle Physics\label{c-ap}}

The discipline of astroparticle physics is developing quite fast in the last years and a lot of interesting results have been obtained. Methods of particle physics are employed in the fields of astrophysics and astronomy while new kinds of particles may exist in the Universe which have not yet been detected on Earth. The following sections will summarize important theories and current measurements in the area of interplay of these subjects.

\section{Basics in Particle Physics}

The standard model is the current theoretical framework of particle physics. It is based on gauge symmetries with
matter fields for quarks and leptons, fields for gauge bosons and one scalar Higgs field. So far all interactions of elementary particle physics up to energies of $\cal{O}$(200\,GeV) can be described within the standard model \citep{Weinberg:1967tq,salam,Glashow:1970gm,berger}. Four forces are known: the electromagnetic, the weak, the strong and the gravitational force, the latter not being part of the standard model. Elementary particles are the constituents of ordinary matter without any substructure. They are grouped in quarks, leptons and gauge bosons. Quarks and leptons exist in three families. Each of these particles is exactly described by its quantum numbers. Quarks cannot be observed freely because they are confined. Baryons (e.g. protons, neutrons) are made up of three quarks and mesons (e.g. pions, kaons) out of one quark and one antiquark. Antibaryons (e.g. antiprotons) are made of three antiquarks.

Quarks and leptons feel the electromagnetic and the weak forces which are unified in the electroweak theory. The interaction is mediated by massless and chargeless photons and by neutral $Z^0$ bosons and charged $W^\pm$ bosons with mass. In addition, quarks carry the color charge which is the source of the strong force. Three different colors are exchanged by eight gluons. An overview on the particle content of the standard model is given in Tab.~\ref{t-weak}. The symmetries of the elementary particles are described by unitary groups:
\be SU(3)\sub C\otimes SU(2)\sub L\otimes U(1)\sub Y.\ee
$SU(3)\sub C$ is the group for the strong force and $SU(2)\sub L\otimes U(1)\sub Y$ is the group of the electroweak interaction. The strength of a force is defined by the coupling constant which depends on the energy scale of the interaction. So far only the electromagnetic and the weak forces could be unified in one theory. The particle masses result probably from a Yukawa interaction with the scalar higgs field but this higgs particle has not been detected yet.

An important fact is the existence of antiparticles because of symmetry reasons. With respect to the particles, antiparticles have opposite additive quantum numbers e.g. charge, baryon or lepton number but they have e.g. the same spin, mass and lifetime. Particles and antiparticles can annihilate in pairs producing photons. It is known that the particle antiparticle symmetry is not completely perfect for electroweak processes and that $CP$ symmetry is violated. $C$ stands for the charge conjugation which transforms a particle into its antiparticle, and $P$ for parity which creates the mirror image of a physical system. $CP$ violation is needed to create matter-antimatter imbalance \cite{sak,pdgbook}.

\begin{table}
\captionof{table}[Elementary particles in the standard model]{\label{t-weak}Quantum numbers of elementary particles in the standard model. Fermions are grouped in left-handed doublets and right-handed singlets. The left-handed primed down type quarks are not the physical mass eigenstates. The observed mixing is described by the Cabbibo-Kobayashi-Maskawa matrix. $Q$ is the charge, $Y_W$ is the weak hyper charge and $T_3$ is the third component of the weak isospin.}
\begin{center}
\begin{tabular}{l|c|c|c|c|c|c}
\hline
\hline
\multicolumn{7}{c}{Fermions (Spin 1/2)}\\
\hline
\hline
	&	\multicolumn{3}{c|}{Families} 	&	\multicolumn{3}{c}{Quantum numbers}\\
\hline
	&	1.	& 2.	& 3.	& $Q$	& $T_3$	&	$Y_W$\\
\hline
	&		&		&		&		&		&		\\
 	& $\left(\begin{matrix}u \\ d^\prime\end{matrix}\right)\sub L$ & $\left(\begin{matrix}c \\ s^\prime\end{matrix}\right)\sub L$ & 
$\left(\begin{matrix}t \\ b^\prime\end{matrix}\right)\sub L$& $\begin{matrix}2/3 \\ -1/3\end{matrix}$ & $\begin{matrix}1/2 \\ -1/2\end{matrix}$ & 
$\begin{matrix}1/3 \\ 1/3\end{matrix}$\\
Quarks	&		&		&		&		&		&		\\
	& $\begin{matrix}u\sub R \\ d\sub R^\prime\end{matrix}$ & $\begin{matrix}c\sub R \\ s\sub R^\prime\end{matrix}$ & $\begin{matrix}t\sub R \\ b\sub 
R^\prime\end{matrix}$ & $\begin{matrix}2/3 \\ -1/3\end{matrix}$ & $\begin{matrix}0 \\ 0\end{matrix}$ & $\begin{matrix}4/3 \\ -2/3\end{matrix}$\\
	&		&		&		&		&		&		\\
\hline
	&		&		&		&		&		&		\\
& $\left(\begin{matrix}\nu_e \\ e\end{matrix}\right)\sub L$ & $\left(\begin{matrix}\nu_\mu \\ \mu\end{matrix}\right)\sub L$ & 
$\left(\begin{matrix}\nu_\tau\\ \tau\end{matrix}\right)\sub L$& $\begin{matrix}0 \\ -1\end{matrix}$ & $\begin{matrix}1/2 \\ -1/2\end{matrix}$ & 
$\begin{matrix}-1 \\ -1\end{matrix}$\\
Leptons 	&		&		&		&		&		&		\\
	& $e\sub R$ & $\mu\sub R$ & $\tau\sub R$ & -1 & 0 & 2\\
	&		&		&		&		&		&		\\
\hline
\hline
\multicolumn{7}{c}{Bosons (Spin 1)}\\
\hline
\hline
\multicolumn{2}{l|}{Interaction} & \multicolumn{2}{c|}{Mediator} & $Q$	& $T_3$	&	$Y_W$\\
\hline
\multicolumn{2}{c|}{} & \multicolumn{2}{c|}{} &		&		&		\\
\multicolumn{2}{l|}{electromagnetic} & \multicolumn{2}{c|}{$\gamma$} & 0	& 0	&	0\\
\multicolumn{2}{l|}{weak} & \multicolumn{2}{c|}{$\begin{matrix}Z^0 \\ W^\pm\end{matrix}$} & $\begin{matrix}0 \\ \pm1\end{matrix}$	& 
$\begin{matrix}0 \\ -1\end{matrix}$	&	$\begin{matrix}0 \\ \pm1\end{matrix}$ \\
\multicolumn{2}{l|}{strong} & \multicolumn{2}{c|}{$g_{1\dots8}$} & 0	& 0	&	0\\
\multicolumn{2}{c|}{} & \multicolumn{2}{c|}{} &		&		&		\\
\hline
\end{tabular}
\end{center}
\end{table}

\section{Basics in Cosmology and Astrophysics}

Cosmological models try to understand the evolution and structure formation of the Universe \cite{rubcos,goenner,olive-1994,bertone-2005-405}. Because of the large distances between astrophysical objects gravitation is the most important interaction while the microscopic processes are described by particle physics. So far no quantum theory of gravitation could be successfully developed. A good description is Einstein's field equation:
\be R_{\mu\nu}-\frac12 Rg_{\mu\nu}=8\pi G T_{\mu\nu}+\Lambda g_{\mu\nu}\label{e-art}\ee
where $R_{\mu\nu}$ is the Ricci tensor which gives the deviation from a flat Minkowski space. The distribution of matter is described by the energy momentum tensor $T_{\mu\nu}$. The metric tensor $g_{\mu\nu}$ describes the space time geometry and $R$ is the Ricci scalar. The cosmological constant $\Lambda$ is related to the vacuum energy and $G$ is the gravitational constant. Experiments show a large homogeneity and isotropy of the Universe for large scales which leads to the following line element (Robertson Walker metric):
\be \text ds^2 = \text dt^2-R^2(t)\left[\frac{\text dr^2}{1-kr^2}+r^2\text d\theta^2+r^2\sin^2\theta \text d\phi^2\right].\ee
$(t,r,\phi,\theta)$ are the spacetime coordinates and $k$ describes the curvature. $k$ is the crucial parameter to understand the further evolution of the Universe. The Universe would be closed for $k=1$ and collapse because of gravitation. For $k=-1$ the Universe would expand and cool down indefinitely. For $k=0$ the expansion will slow down indefinitely and cease asymptotically. The expansion is given by Hubble's law:
\be H=\frac{\dot R(t)}{R(t)}\ee
with the expansion velocity $H$ at time $t$. Another important parameter is the energy density of the Universe $\rho$ which measures the energy and matter content in the Universe. The density parameter $\Omega$ is the ratio of the energy density $\rho$ to the critical ratio $\rho_c$. At the critical energy density $\rho_c$ the Universe is flat. $\rho$ is smaller than the critical density for an open, indefinitely expanding Universe and larger for a closed, collapsing one.

By making the assumption that an extrapolation of particle physics and general relativity back in time is reasonable, the Big Bang theory \cite{lem-1927} explains the evolution of the Universe with an expansion of a very hot singularity. The earliest times that can be discussed are $10^{-43}$\,s after the Big Bang. Prior to this time a complete theory of quantum gravity would be needed. The fact that our Universe is large and homogeneous can only be explained if the increase of the radius did not happen at a constant velocity but with an exponential expansion in the very first stage. This is called inflation \cite{PhysRevD.23.347}. Theories predict a new scalar inflaton field driving this process. Vacuum fluctuations of the inflaton field can lead to primordial fluctuations needed in the order of
\be\frac{\delta\rho}{\rho}\approx10^{-5}\ee
where $\rho$ is the density \cite{rubcos}. The primordial fluctuations were transformed into sound waves and built up structures from small to large objects. These fluctuations of the inflationary stage stay constant until the decoupling of photons from equilibrium. Particles have to interact sufficiently to stay in equilibrium while the temperature drops during the expansion. If the interaction rate decreases it is possible that a particle species freezes out. The present relic density of a specific particle type can be calculated by using the appropriate interactions and their properties \cite{1988NuPhB.310..693S}. 

\begin{table}
\begin{center}
\captionof{table}{\label{t-unihistory}Short history of the Universe for the time after the Big Bang.}
\begin{tabular}{c|c|l}
\hline
\hline
Time [s]	&Temperature [eV]	& Action\\
\hline
0		& $\infty$ 	& Big Bang\\
$10^{-43}$	& $10^{27}$	& electroweak, strong and gravitational force unified\\
$10^{-35}$	& $10^{26}$	& unified gauge group of SM breaks down\\
		& $10^{23}$ 	& possible baryon-number-violating processes\\
$10^{-10}$	& $10^{11}$	& electroweak symmetry breaking \ra origin of baryogenesis\\
		& $10^9-10^{11}$ & possible weakly interacting dark matter candidates decouple\\
$10^{-5}$	& $3\cdot10^8$	& QCD transition \ra confinement of quarks and gluons\\
		& $10^6$	& neutrons decouple\\
600		& $10^5$	& nucleosynthesis: formation of light elements\\
$2\cdot10^{12}$	& $1$		& matter domination \ra structure formation starts\\
$9\cdot10^{12}$	& $0.4$		& photons decouple \ra cosmic microwave background\\
$10^{17}$	& $10^{-4}$	& temperature of cosmic microwave background today\\
\hline
\end{tabular}
\end{center}
\end{table}

Tab.~\ref{t-unihistory} gives a short overview of the important epochs. Most important for this work are the origins of baryogenesis and the decoupling of dark matter which will be discussed later.

\begin{figure}
\begin{center}
\centerline{\epsfig{file=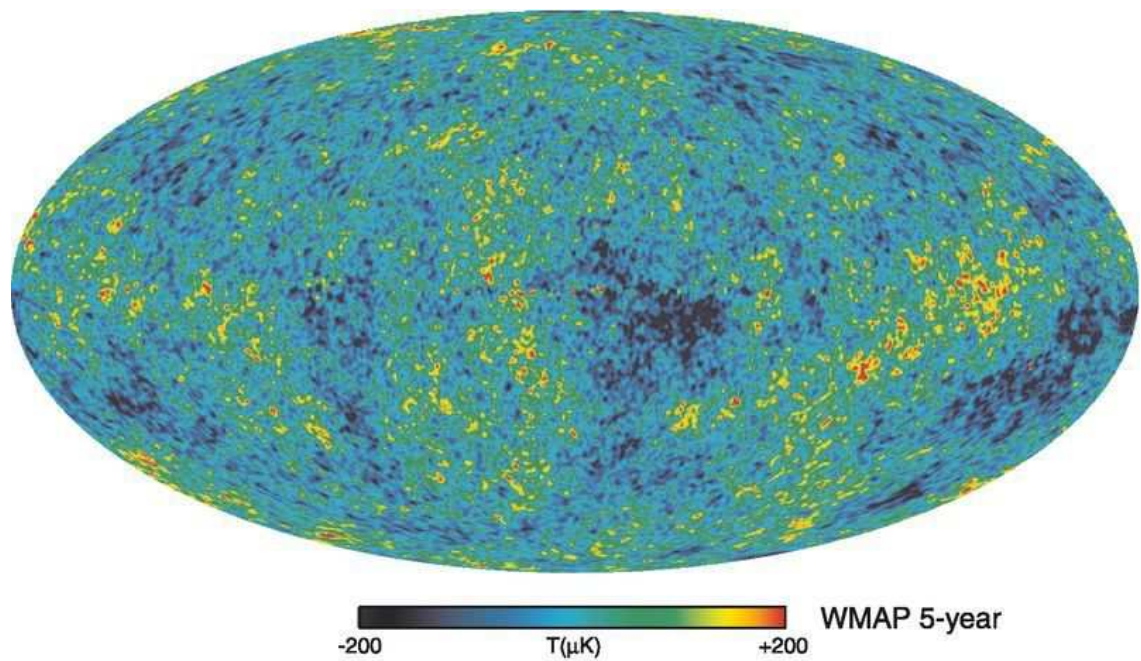,width=15cm}}
\captionof{figure}{\label{f-wmap_map_5yr}The foreground-reduced Internal Linear Combination (ILC) map based on the 5 year WMAP data\cite{wmap_basic}. Credit: NASA/WMAP Science Team.}
\end{center}
\end{figure}

\begin{figure}
\begin{center}
\captionof{table}{\label{t-wmap_basic}Composition of the density parameter\cite{wmap_basic}.}

\bigskip

\begin{minipage}[c]{0.3\linewidth}
\epsfig{file=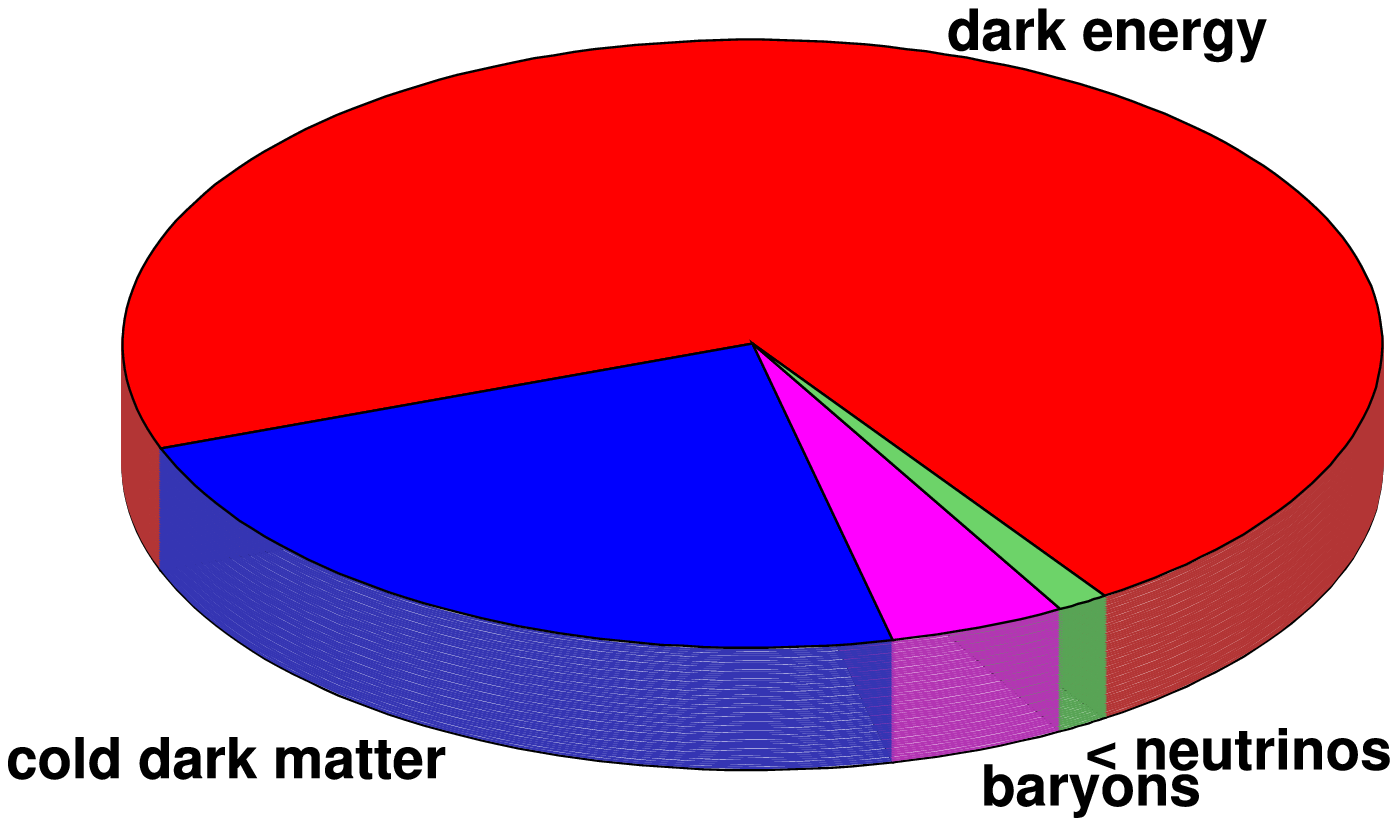,width=1.1\linewidth}
\end{minipage}
\hspace{.1\linewidth}
\begin{minipage}[c]{0.5\linewidth}
\begin{tabular}{l||c|c}
\hline
\hline
Type		& Sign		& Density\\
\hline
dark energy	& $\Omega_\Lambda$ & $0.721\pm0.015$\\
cold dark matter& $\Omega\sub{cdm}$	& $0.233\pm0.0013$\\
baryons		& $\Omega_b$	& $0.0462\pm0.0015$\\
neutrinos	& $\Omega_\nu$	& $<0.0133$ (95\,\% C.L.)\\
\hline
\textbf{total}	& \textbf{$\Omega$} & \textbf{$1.0052\pm0.0064$}\\
\hline
\end{tabular}
\end{minipage}
\end{center}
\end{figure}

The cosmic microwave background (CMB) is sensitive to a lot of the properties of the Universe e.g. primordial density perturbations, matter content and spatial geometry. It provides a picture of the young Universe (300000\,yr) where photons froze out from thermal equilibrium at the time of the recombination of protons and electrons. The CMB anisotropy is due to density perturbations during the recombination which are sound waves at different wavelengths. Physical models for these first years of the Universe are well understood and cosmological parameters can be extracted. The five year data of the WMAP satellite are shown in Fig.~\ref{f-wmap_map_5yr}. In addition to other parameters, the composition of the energy density $\Omega$ was determined by carrying out a combined analysis of WMAP with the Baryonic Acoustic Oscillations (BAO) \cite{perbao} and the supernova (SN)  data from Supernova "Gold Sample" \cite{riesn}, Supernova Legacy Survey \cite{astsup} and SNEssence (Tab.~\ref{t-wmap_basic}). BAO measured from galaxy surveys can be used to measure the geometry of the Universe through the distance-redshift relation and Supernovae are used as standard candles to measure the acceleration of the Universe. The combined analysis of these different observations allows a precise determination of cosmological parameters. The known baryonic matter and neutrinos can only add up to about 6\,\% of the total density parameter. The rest is composed of non-relativistic cold dark matter ($\approx 23$\,\%) and unclustered dark energy ($\approx 72$\,\%). The latter can be interpreted as the repulsive force to power the expansion of the Universe. Even if there is a lot of evidence from different observations for the existence of dark matter its nature is so far unknown. Some popular ideas will be discussed in Sec.~\ref{ss-darkmatter}. So far only some ideas for the nature of dark energy exist.

\section{Beyond the Standard Model\label{s-beyond}}

Although the models of particle physics and cosmology work well for a lot of phenomena, there are important unexplained features, e.g.:

\bi
\st Where does the observed asymmetry between matter and antimatter in the Universe come from?
\st What is the nature of dark matter and dark energy?
\st Is an unification of all forces including gravitation possible?
\ei

Popular theories addressing the issues listed above will be briefly discussed in the following sections.

\subsection{Baryogenesis}

One of the most puzzling questions is the existence of baryonic matter in the Universe today. Within the standard models of particle physics and cosmology, equal amounts of matter and antimatter were produced in the Big Bang. This symmetry was obviously not perfect. The baryon asymmetry is expressed through the measurement of the number densities of baryonic particles $n_B$ and photons $n_\gamma$ in the current Universe \cite{pdgbook}:
\be4.7\cdot10^{-10}<\eta=\frac{n_b-n_{\bar b}}{n_\gamma}=\frac{n_B}{n_\gamma}<6.5\cdot10^{-10}.\ee
where $n_{b/\bar b}$ is the number density of baryons and antibaryons. This tiny excess explains all baryonic matter in the Universe. Explanations for this observation must satisfy the three Sakharov conditions \cite{sak}:

\bi
\st Baryon number is not conserved.
\st Interaction rates for baryons and antibaryons are different. Therefore, charge conjugation ($C$) or charge and parity conjugation ($CP$) must be violated.
\st The Universe cannot be in thermal equilibrium.
\ei

Processes creating the baryon asymmetry are called baryogenesis. Many mechanisms following different approaches have been proposed. It is nowadays believed that the baryogenesis was inhomogeneous which means baryogenesis started from a baryon asymmetry with small amounts of antimatter contained in matter dominated regions \cite{kurkisuonio-2000-62}. Matter and antimatter domains were separated such that pair annihilation of baryons and antibaryons could be avoided. The measurement of antimatter or upper bounds on the existence of antimatter will put constraints on these theories. The calculation of annihilation reactions between matter and antimatter as well as possible antimatter sources must be taken into account.

Mechanisms for the inhomogeneous behavior can be e.g. due to grand unified theories (GUT) \cite{nanopoulos-1978} at an energy scale of $10^{15}$ - $10^{16}$\,GeV. There it could be possible that ultra-heavy particles decay while violating $CP$ and baryon number $B$. Later on electroweak anomalies violate baryon number to explain the observed fraction $\eta$. This kind of baryogenesis could have happened in the reheating phase of the Universe at the end of the inflationary phase. It is also possible that baryogenesis is completely based on electroweak anomalies. These kind of processes will be switched off at $\cal{O}$(100\,GeV) \cite{rubcos} but there must be enough $CP$ violation before and during the phase transition to generate the asymmetry. Supersymmetric theories may deliver these additional sources for $CP$ violating processes \cite{kraml-2007}.

\subsection{Dark Matter\label{ss-darkmatter}}

Besides the fraction of dark matter in the total energy density determined from SN, BAO and CMB as already mentioned, direct evidence for the existence of dark matter was found in clusters of galaxies. The gravitating mass can be determined by different techniques, e.g. X-ray measurements \cite{bradac-2006-652}, rotation curves \cite{sofue-2001-39}, velocity of galaxies in clusters  and gravitational lensing \cite{kneib-2003-598}. These measurements give an estimate for $\Omega\sub{cdm}$ similar to the value in Tab.~\ref{t-wmap_basic} by applying the mass to light ratio in clusters of galaxies to the whole Universe \cite{rubcos}. It is also seen from the measurements that the distribution of dark matter does not follow the visible matter. Dark matter is clustered in galaxies and in the gas clouds around galaxies. The form of dark matter substructure plays an important role in several scenarios \cite{bertone-2005-405}. Shapes from isothermal to cuspy \cite{burkert-1995-447,1996ApJ...462..563N} or even ring-like structures \cite{deboer-2005-444} of dark matter in galaxies are discussed and influence the relic density calculations and the detection probability. Therefore it is important to understand the density distribution of dark matter.

The nature of dark matter is crucial for the understanding of structure formation in the Universe. Current observations and theories exclude baryonic matter as a dark matter candidate. Also neutrinos would wash out the observed small scale structures because of their high velocities. The favored dark matter type is of non-baryonic and non-relativistic nature with a mass large enough to explain the structure in the Universe and is called cold dark matter (CDM). At the moment only hypothetical candidates for dark matter particles exist.

The field of candidates arising from different theories is huge. Most extensions of the standard model of particle physics, also provide a candidate for dark matter. All these candidates must be heavy enough to explain the observed structure and relic density of our Universe and interact only very weakly because they have not been detected yet by any experiment.

\subsubsection{Supersymmetry}

Supersymmetry (SUSY) introduces a symmetry between fermions and bosons \cite{weinberg-2000,drees-1996}. Each fer\-mion (boson) of the standard model has a supersymmetric bosonic (fermionic) partner. In contrast to the standard model, two complex scalar higgs fields are needed to avoid anomalies. They are distinguished by the multiplicative $R$ parity quantum number which is 1 for the standard model particles and -1 for their superpartners. Supersymmetric particles have not been observed so far, so the symmetry must be broken and the superpartners must have larger masses. The breaking can be done by a variety of mechanisms, the particular choice having a crucial influence on further properties of the corresponding model. The theory is able to unify the electroweak and strong coupling constants at the scale of grand unified theories (GUT) at about $10^{16}$\,GeV and introduces large $CP$ violating phases. Another very important feature of supersymmetry is the cancellation of quadratic divergences of the standard model, e.g. in the higgs sector. The physical supersymmetric particles are linear combinations of the appropriate mass eigenstates. The neutral particle called neutralino can be the lightest supersymmetric particle and is a popular candidate for dark matter in the case of $R$ parity conserving models \cite{jungman-1996-267,bergstrom-2000-63}. It consists out of the bino $\widetilde B$, wino $\widetilde W^3$ and the higgsinos $\widetilde H^0_{1/2}$, the fermionic super partners of the corresponding standard model bosons:
\be\chi_i^0=N_{i,1}\widetilde B+N_{i,2}\widetilde W^3+N_{i,3}\widetilde H^0_1+N_{i,4}\widetilde H^0_2\ee
where $N_{i,j}$ are complex numbers. The charged supersymmetric fermions are made up of the charged fermionic mass eigenstates (charg\-inos). Further superpartners are called sleptons, squarks and gluinos.

A popular version widely studied is the minimal supersymmetric model (MSSM). A common approach is to break SUSY by coupling to a yet unknown supergravity theory (mSUGRA) which reduces the number of new parameters down to five. They are the masses of the gauginos $m_0$ and scalars $m_{1/2}$ at the GUT scale, $\tan\beta$ the ratio of vacuum expectation values for the neutral components of the higgsinos, $\text{sign}(\mu)$ the sign of the higgsino mass parameter and the trilinear coupling $A_0$ which describes the coupling of the higgs fields among each other.

\subsubsection{Kaluza-Klein Extra Dimensions}

Kaluza and Klein (KK) originally introduced a theory to unify the electroweak and the gravitational force, but did not succeed \cite{klein-1926}. Nowadays Kaluza-Klein theories can be used to provide a viable dark matter candidate in the model of universal extra dimensions \cite{cheng-2002-89,bringmann-2005}. This allows all standard model fields to propagate in extra dimensions. The extra dimensions cannot be detected and the extra momentum will be observed as additional mass. The Klein-Gordon equation is expressed in the case of a field $\Phi(x^\mu,y)$ of mass $m$ with the common four dimensional space $x^\mu$ and one extra dimension $y$ by:
\be\left(\square-\frac{\displaystyle\partial^2}{\displaystyle\partial y^2}+m^2\right)\Phi(x^\mu,y)=0.\ee
The extra dimensions are compactified on the compactification radius $R$. The Fourier decomposition of the field makes the idea of a stack of excitations with extra mass compared to the standard model particle visible:
\be\Phi(x^\mu,y)=\sum_n\Phi^{(n)}(x^\mu)\exp\left(-i\frac{\displaystyle 2\pi n}{\displaystyle R}\right)\;\wedge\;m_n^2=m^2+\left(\frac{\displaystyle n}{\displaystyle R}\right)^2\ee
with the excitation stage $n$, the mass of the excitated stage $m_n$ and the standard model particle mass $m$. $n=0$ indicates the standard model particles. A good candidate for the lightest Kaluza-Klein dark matter candidate (LKP) is the first excitation of the hypercharge boson of the electroweak theory $B^{(1)}$. This particle is stable due to the $KK$ parity arising from the fact that $(-1)^n$ is conserved in the theory such that it cannot decay to a standard model particle with $n=0$.

Unlike the neutralino, the SUSY dark matter candidate, $B^{(1)}$ is a boson. This influences the detection probability and the extracted dark matter density shape (Sec.~\ref{s-gev}) .

\subsubsection{More Candidates}

Some more dark matter candidates which are not detectable by the experiments described in this thesis are explained briefly in the following.

Axions are believed to solve the strong $CP$ problem, i.e. the absence of $CP$ violation in strong interactions \cite{kim-2000}. This is done dynamically by introducing the new scalar axion field. The axions are believed to have a mass between $10^{-6}\,\text{eV}<m_a<10^{-3}\,\text{eV}$. They can be considered as non-relativistic cold dark matter because they formed from a Bose condensate and have never been in thermal equilibrium \cite{pdgbook}. So far no axions have been observed \cite{RevModPhys.75.777}.

The technicolor theory \cite{hill-2003} was introduced to explain the electroweak symmetry breaking dynamically by introducing again a new symmetry. This turned out to be very problematic but recent developments in the minimal walking technicolor framework with a very slowly running ("walking") coupling constant over a large energy range are looking more promising \cite{gudnason-2006-73,gudnason-2006-74}. These new technipartners of the standard model particles can be dark matter candidates even if they cannot account for all of the dark matter energy density. It depends on the particular theory whether a technibaryon or a technineutrino is the lightest techniparticle and thus the favored dark matter candidate.

Even more exotic are decaying ultra heavy dark matter candidates which seem to be ruled out by current measurements \cite{aloisio-2007,semikoz-2007} and Q-balls which are localized stable field configurations that could have formed in the early Universe \cite{enqvist-2002-526}.

\subsection{Ideas for Dark Energy}

There are only a few ideas for the nature of dark energy on the market\cite{rubcos,peebles-2003-75}. From the Einstein equation one knows that dark energy works as a kind of repulsive force delivering the energy for the expansion of the Universe. A huge positive amount of dark energy would have triggered the expansion earlier and a large negative one would have caused an early recollapse. In both cases there would not have been enough time for structure formation as we know it today. One idea could be a light scalar field called quintessence which could have driven the expansion \cite{ziaeepour-2004-69}. In this case the density of the dark energy would not be constant. Another idea argues with deviations from general relativity on distance scales in the order of the Universe \cite{rubcos}. Both ideas are quite unnatural and it is not very likely to reveal the nature of dark energy in the near future.

\section{Probing Particle Physics and Astrophysics with cosmic Rays\label{s-cr}}

\begin{figure}
\begin{center}
\centerline{\epsfig{file=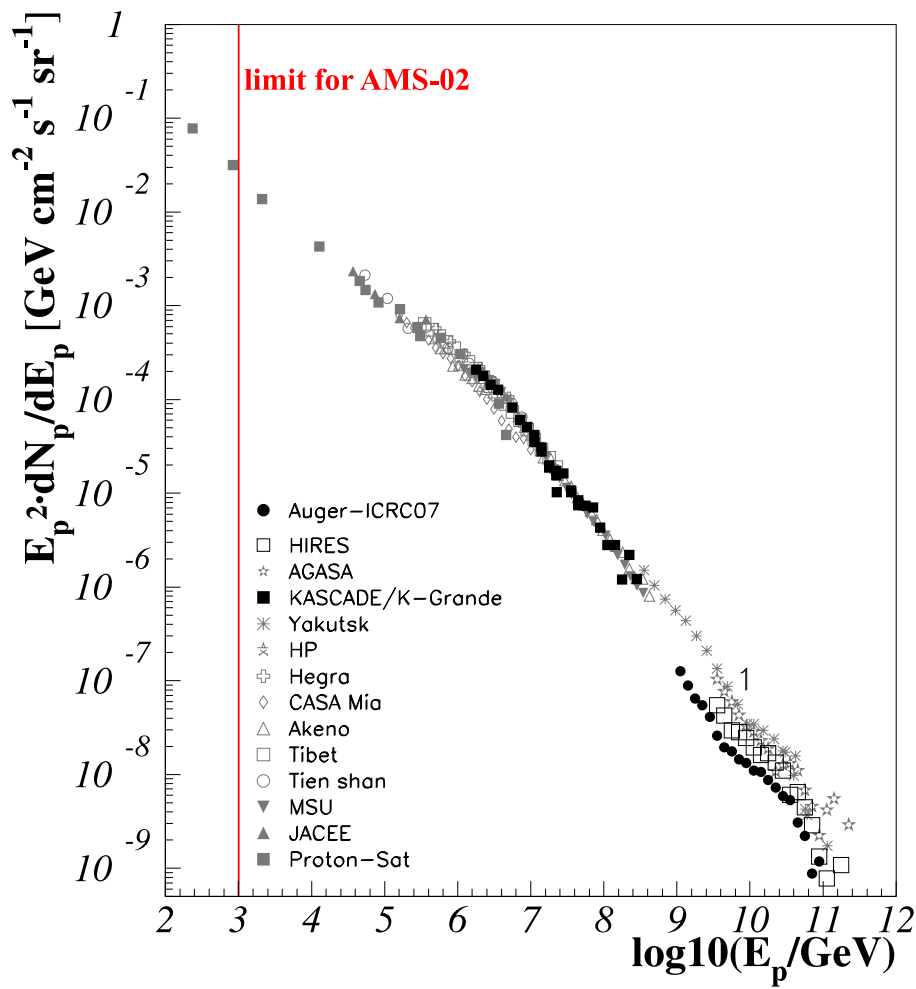,width=12cm}}
\captionof{figure}{\label{f-sciam1}Differential cosmic-ray flux. Reprinted figure with permission from Elsevier \cite{becker-2007}.}
\end{center}
\end{figure}

Recent measurements of cosmic rays over a large energy range have given access to different aspects of particle physics, astrophysics and cosmology \cite{parizot} and will be briefly discussed in this section. Cosmic protons are measured over an energy range of 20 orders of magnitude (Fig.~\ref{f-sciam1}). The challenge is to explain in a consistent theory the formation of cosmic rays in astrophysical objects, their acceleration and propagation in the intergalactic and interstellar medium. Three different properties of cosmic rays are accessible to measurement: the angular distribution, the composition and the differential energy flux.

\subsection{Sources, Acceleration and Propagation\label{ss-source}}

\begin{figure}
\begin{center}
\centerline{\epsfig{file=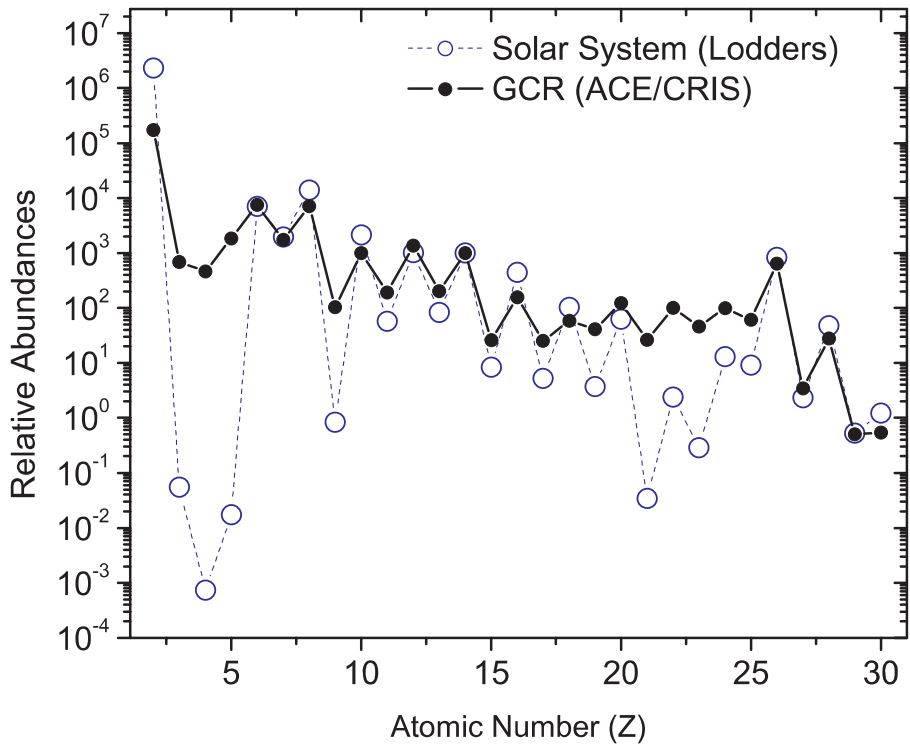,width=12cm}}
\captionof{figure}{\label{f-sol_cr_comp}Composition of cosmic rays in comparison to solar system abundances. GCR denotes galactic cosmic rays. Reprinted figure with permission from Elsevier \cite{2005NuPhA.758..201I}.}
\end{center}
\end{figure}

For energies below $10^{15}$\,eV galactic cosmic rays can be explained as originating from nucleosynthesis in the stellar atmospheres \cite{parizot}. Within our galaxy the acceleration processes are e.g. gravitation (accretion discs, neutron stars, pulsars, black holes) but the most efficient acceleration occurs in magnetic shock waves caused by supernovae (Fermi acceleration). The charged particles are very often scattered head on in the moving magnetic shock fronts of supernovae such that they gain energy in a collisionless way. About $25$ - $30$\,\% of the energy of a supernova powers cosmic rays. Their maximum energy depends on the acceleration processes mentioned above and their confinement time in the accelerator. The sources of extragalactic cosmic rays with energies above $10^{15}$\,eV are not understood and must have been accelerated via so far unknown extragalactic processes.

The angular distribution of the directions of charged cosmic rays is isotropic due to the interstellar magnetic fields which deflect the charged particles and randomize them over the complete sky. Other processes which influence the energy distribution and the abundance are: 
\bi
\st decay of unstable particles
\st particle escape due to diffusion processes
\st interactions with the interstellar medium: spallation, hadronic interactions, friction, inverse Compton scattering, pair production, photo disintegration, radioactive decay
\st reacceleration
\ei

Protons, helium, heavy nuclei and electrons are considered to be primary cosmic rays while positrons, photons and antiprotons arise from secondary interactions of the primary cosmic rays with the interstellar medium. Fig.~\ref{f-sol_cr_comp} shows the abundance of elements in cosmic rays at an energy of $\cal O$(0.1\,GeV) in comparison to the solar system. All stable elements occur in cosmic rays and the most elements are nearly as abundant as in the solar system. This implies that cosmic rays are accelerated from a mixed interstellar matter compatible with the solar system \cite{2005NuPhA.758..201I}. Nuclei with odd atomic number $Z$ are more weakly bound and more frequent products in nuclear reactions. The abundance minimum of lithium, beryllium and boron for the solar system is not visible for cosmic rays because light elements can be produced in spallation processes of cosmic rays with the interstellar medium. 

The propagation time of cosmic rays and the halo size of the galaxy can be determined from the ratios of stable and unstable isotopes of the same element. The age of cosmic rays is about $\cal{O}$$(2\cdot 10^7\text{\,yr})$ and the halo size is 3 - 7\,kpc. The grammage of matter cosmic rays have gone through can be deduced from the secondary to primary ratios and is on average 6 - 10\,g/cm$^2$.  These measurements also give the average density of the propagation medium to be about 0.2\,H\,atoms/cm$^{3}$. This is much less than the mean density of the galactic disk (1\,H\,atom/cm$^{3}$). Therefore, cosmic rays spend a lot of their propagation time in low-density regions like the galactic halo.

\subsection{Cosmic Rays in the GeV to TeV Range\label{s-gev}}

\begin{figure}
\begin{center}
\begin{minipage}[b]{.4\linewidth}
\centerline{\epsfig{file=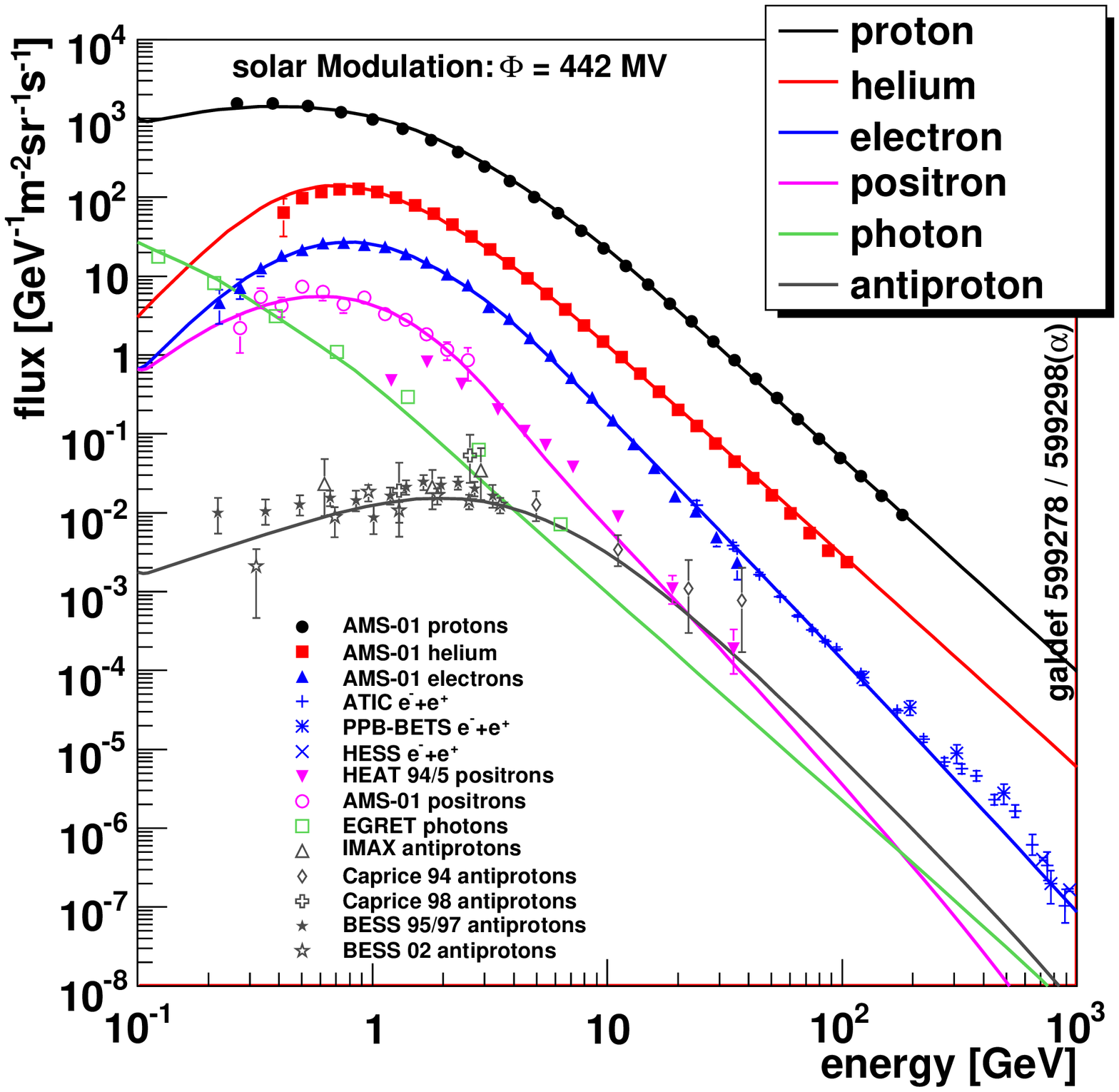,width=8cm}}
\captionof{figure}{\label{f-fluxes_geo_mod_galprop_conv}Current status of cosmic-ray flux measurements in the GeV range \cite{ams01, barwick-1998-498,egret,orito-2000-84,PhysRevLett.76.3057,atic,torii-2008,hess_electron,collaboration-2001-561,olzem-2007}.}
\end{minipage}
\hspace{.1\linewidth}
\begin{minipage}[b]{.4\linewidth}
\centerline{\epsfig{file=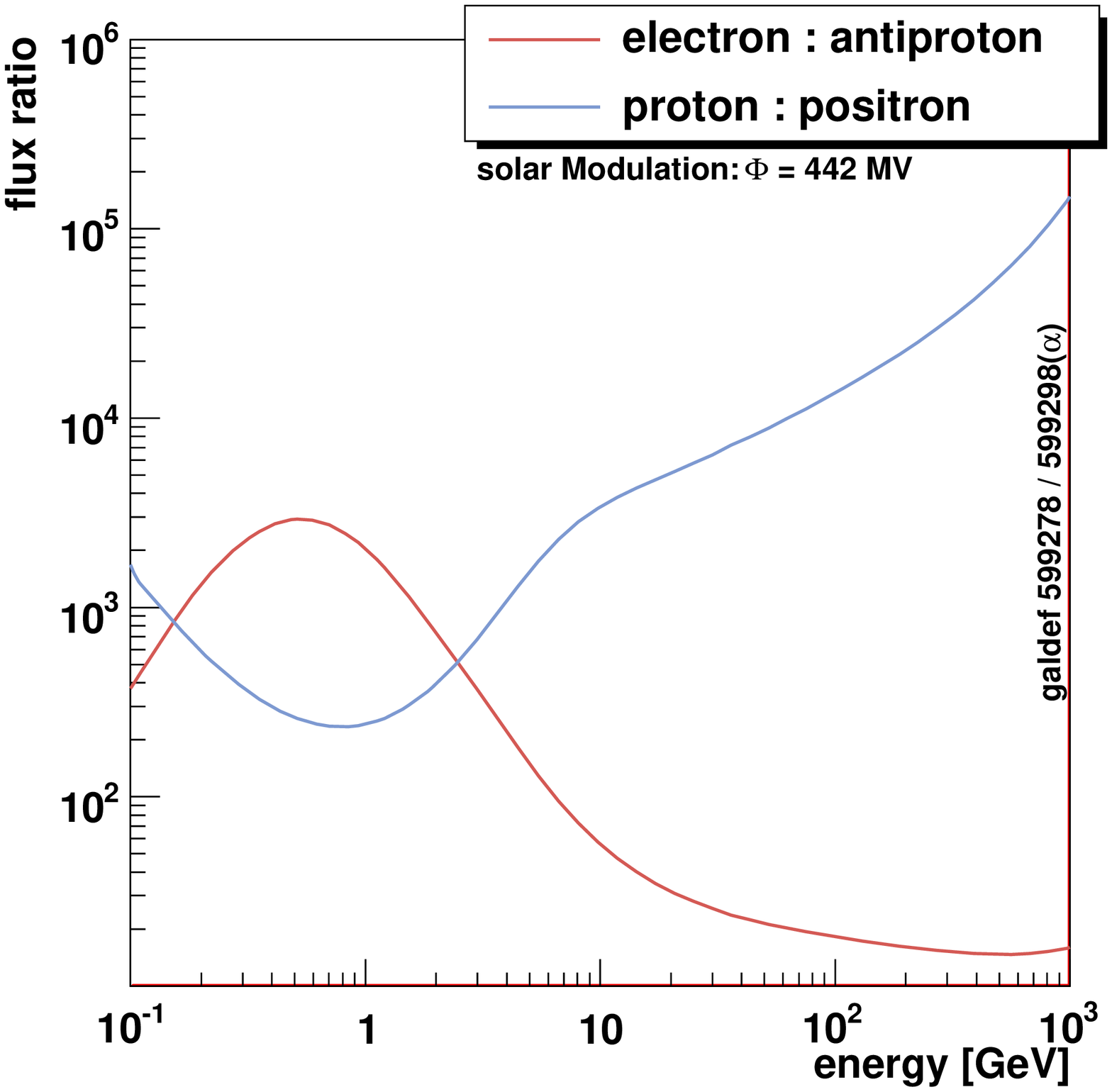,width=8cm}}
\captionof{figure}{\label{f-ratio_mod}Ratio of particle to antiparticle fluxes.}
\end{minipage}
\end{center}
\end{figure}

The measurement of cosmic rays in the range up to $10^3$\,GeV is accessible to direct observations (Fig.~\ref{f-fluxes_geo_mod_galprop_conv}). This is an advantage for the determination of particle properties like mass, charge, energy and direction. The detection of antiparticles is challenging because of the large background from the corresponding particles (Fig.~\ref{f-ratio_mod}). There are no known primary sources of antiparticles in the Universe and the antiparticles arise from secondary interactions of primary cosmic rays with the interstellar medium and are therefore much less abundant. The main background component for positrons are protons and for antiproton measurements electrons. The flux ratios at the desired energy determine the requirements on the detector for a reliable discrimination. The conventional reacceleration GALPROP cosmic-ray propagation model \cite{strong-2001-27,strong-1998-509} describes the measured differential flux spectra well, but unexplained features do exist for positrons, photons and even for electrons.

\begin{figure}
\begin{center}
\centerline{\epsfig{file=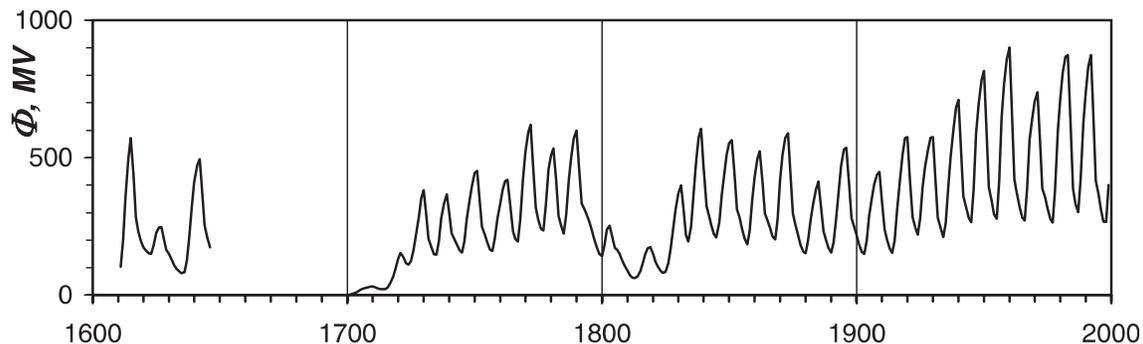,width=15cm}}
\captionof{figure}{\label{f-solMod}Solar modulation depending on time. Reprinted figure with permission from American Geophysical Union \cite{Usoskin-2002}.}
\end{center}
\end{figure}

\subsubsection{Solar Modulation of Flux}

An important effect for measurements in the GeV energy range is the modulation by the magnetic field of the sun which depends on the solar cycle. The effect can be roughly described with a force-field approximation \cite{gleeson-1967}:
\be F\sub{mod}(E=E\sub{LIS}-|Z|e\Phi)=F(r,E\sub{LIS}-|Z|e\Phi)=F(\infty,E\sub{LIS})\cdot\frac{(E\sub{LIS}-|Z|e\Phi)^2-m_0^2}{E\sub{LIS}^2-m_0^2}\ee
where $F(\infty,E\sub{LIS})$ is the flux in the local interstellar medium, $E\sub{LIS}$ is the energy of the particle with mass $m_0$ and charge $|Z|e$ and the effective solar modulation parameter $\Phi$ for all particle species. This can be correlated with the solar cycle (Fig.~\ref{f-solMod}). $\Phi$ has been calculated from the observation of sun spots. 

\begin{table}
\begin{center}
\captionof{table}{\label{t-geo}Fit parameters for the low energy range of cosmic-ray spectra.}
\begin{tabular}{l||c|c}
\hline
\hline
Type 			&C [GeV]	&a\\
\hline
electron\cite{gast-2008}		&0.720		&-1.874\\
positron\cite{gast-2008}		&1.587		&-1.021\\
helium			& 0.5	&-2.7\\
\hline
\end{tabular}
\end{center}
\end{table}

In addition, the GALPROP models for electrons, positrons and helium nuclei are adapted to the low energy range <1\,GeV. Using the fit parameters $C$ and $a$, the final flux $F\sub{geo}(E)$ with $E=E\sub{LIS}-|Z|e\Phi$ is given by:
\be F\sub{geo}(E)=F\sub{mod}(E)\cdot\frac{\displaystyle1}{\displaystyle1+\left(\frac{\displaystyle E}{\displaystyle C}\right)^a}.\ee
The fitted values for the parameters are shown in Tab.~\ref{t-geo}. 

\subsubsection{Influence of the geomagnetic Field}

\begin{figure}
\begin{center}
\centerline{\epsfig{file=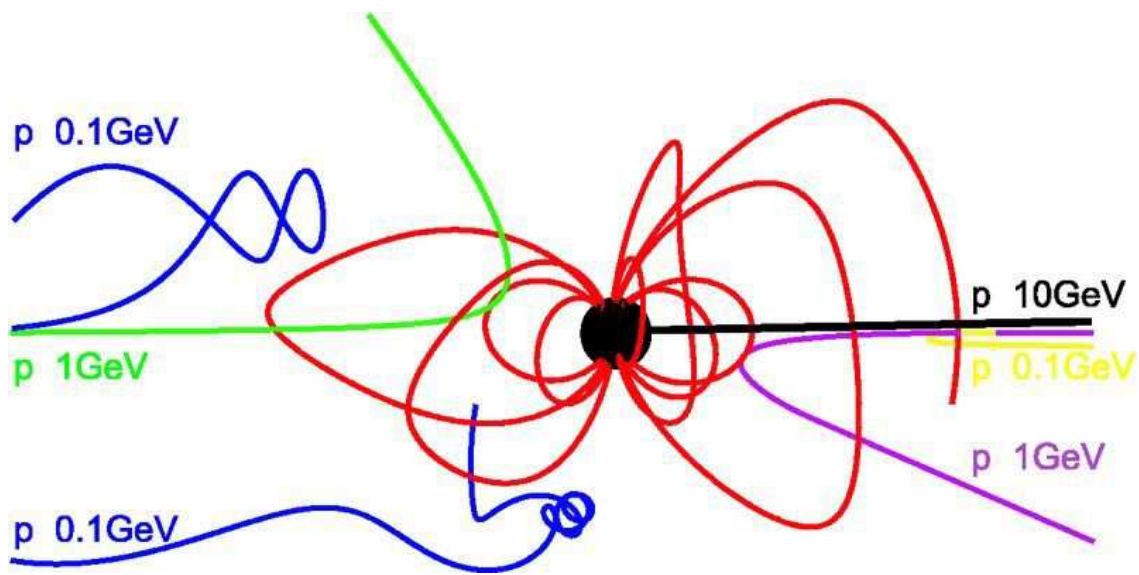,width=15cm}}
\captionof{figure}{\label{f-particle_magnetic_field}Behavior of charged particles in Earth's magnetic field.}
\end{center}
\end{figure}

\begin{figure}
\begin{center}
\centerline{\epsfig{file=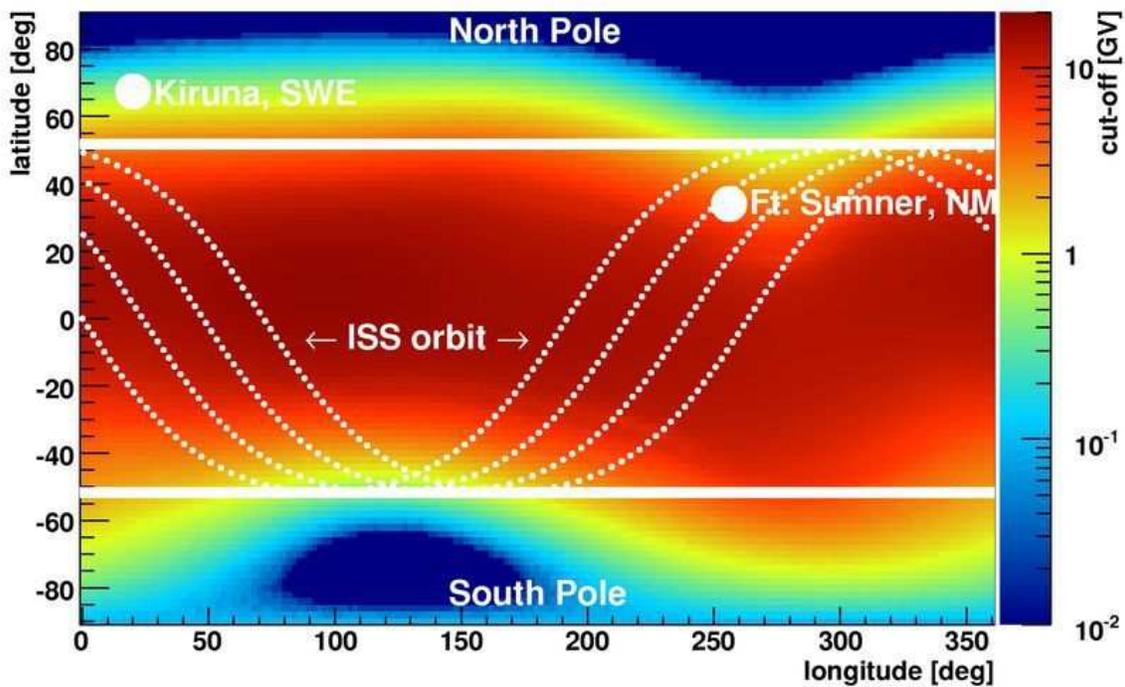,width=15cm}}
\captionof{figure}{\label{f-cutoff}Geomagnetic cut-off at 40\,km altitude for tracks with 0° zenith angle calculated with PLANETOCOSMICS \cite{laurent-2005}. The ISS orbit is at an altitude of about 390\,km.}
\end{center}
\end{figure}

\begin{figure}
\begin{center}
\centerline{\epsfig{file=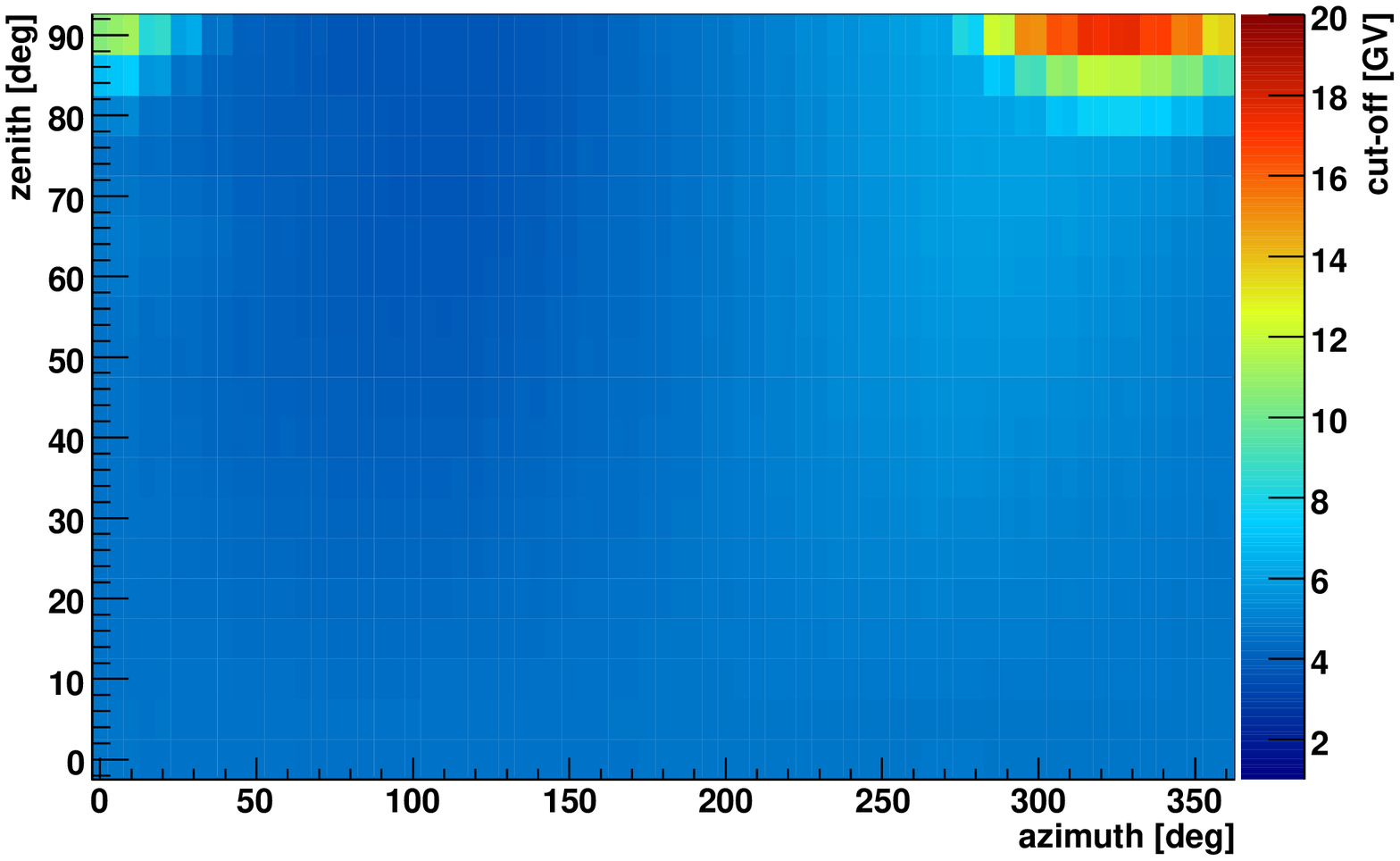,width=15cm}}
\captionof{figure}{\label{f-cutoff_az_zen_34_256}Geomagnetic cut-off at Ft. Sumner with respect to the azimuth and zenith angle of incident at 500\,km altitude.}
\end{center}
\end{figure}

Closer to the Earth, the geomagnetic field becomes important and defines cut-off rigidities, the ratio of the momentum to the particle charge, depending on position and incident angle of the particle. The geomagnetic field is approximately a dipole field tilted with respect to the geographic poles. The field lines are approximately perpendicular to Earth's surface at the poles and parallel at the equator. The magnetic North Pole is at about 82°N, 113°W (coordinate system used in the following: lat.: 82°, long.: 247°) and the South Pole at about 64°S, 138°E (lat.: -64°, long.: 138°) \cite{pole_geo}. The field is described by the International Geomagnetic Reference Field (IGRF) derived from satellite magnetic missions and is a mathematical representation of the main field and its rate of changes. 

In general, deflection of particles is strong at the equator and weak at the poles but this depends additionally on the rigidity. Fig.~\ref{f-particle_magnetic_field} shows the simulated trajectories of different particles started at the same altitude and latitude (equator) with 0° zenith angle. Particles with $\cal O$\,(0.1\,GV) momentum can be trapped in the Van-Allen radiation belts of the Earth. 1\,GeV particles are deflected while for 10\,GeV particles the magnetic effect can be neglected. Therefore, the geomagnetic field modulates the differential flux and the detection rate depends on the geographical position and it is especially for the low energetic part of the spectra important to correct for the geomagnetic cut-off effect. The cut-offs are calculated by tracing particles back to the outside of Earth's magnetosphere. Particles of a specific rigidity making several complex loops before reaching a desired position are 'forbidden' and cannot escape the magnetosphere while trajectories of particles escaping Earth's magnetosphere are 'allowed' trajectories. The rigidity corresponding to the last allowed trajectory for a certain geographic position and direction is called cut-off rigidity. The cut-offs from a PLANETOCOSMICS calculation (further explanations in Chap.~\ref{c-pebs}) during December 2005  are shown in Fig.~\ref{f-cutoff} at 40\,km altitude and 0° zenith angle. The dependence on the geographic position is obvious and supports the planning of balloon flights at the poles because of the very small cut-offs. It is also important to take the cut-offs as a function of the direction into account (Fig.~\ref{f-cutoff_az_zen_34_256}). The structure in Fig.~\ref{f-cutoff_az_zen_34_256} can be explained by the local magnetic field, e.g. particles with large zenith and azimuth angles are strongly deflected up to rigidities of 20\,GV.

\begin{figure}
\begin{center}
\begin{minipage}[b]{.4\linewidth}
\centerline{\epsfig{file=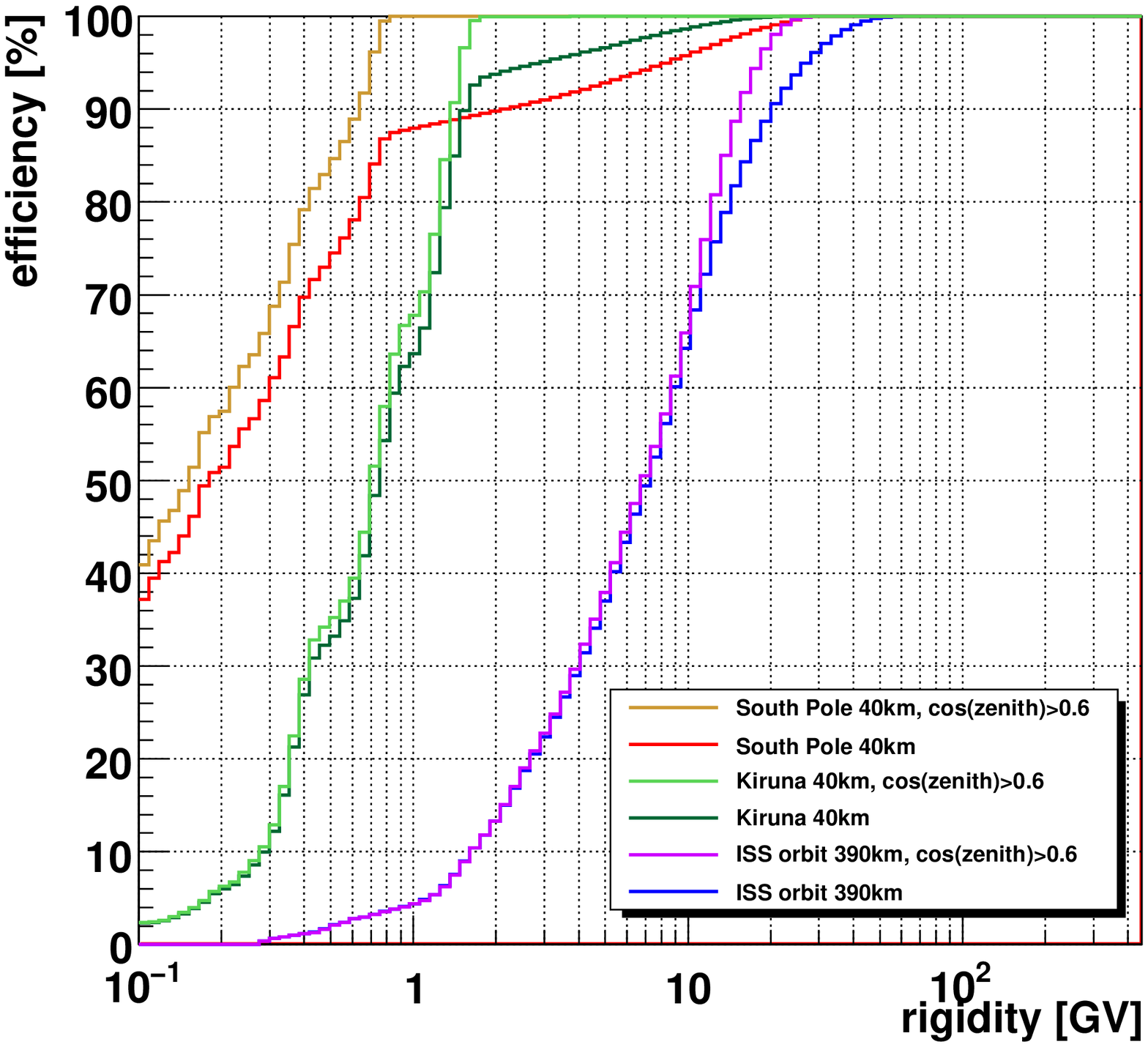,width=8cm}}\captionof{figure}{\label{f-cutoff_issorbit}Comparison between the efficiencies for geomagnetic cut-offs at South Pole, Kiruna and at ISS orbit for the full zenith angle range and $\cos(\text{zenith})>0.6$ (detector acceptance).}
\end{minipage}
\hspace{.1\linewidth}
\begin{minipage}[b]{.4\linewidth}
\centerline{\epsfig{file=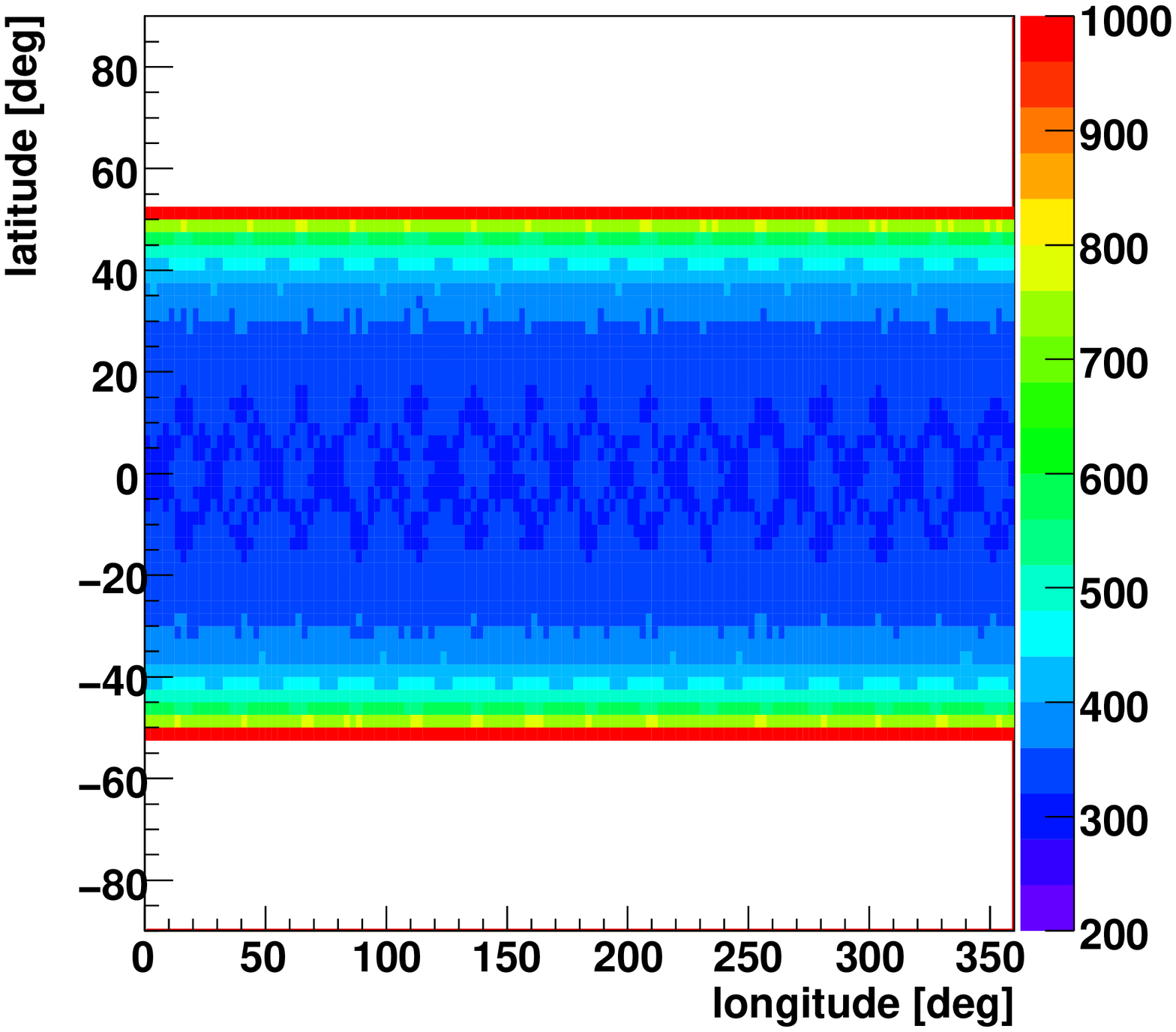,width=8cm}}\captionof{figure}{\label{f-h_longitude_latitude}Occupancy of geographic positions used for the ISS orbit cut-off calculation. The color code on the right shows the number of entries.}
\end{minipage}
\end{center}
\end{figure}

The following chapters~\ref{c-pebs} and \ref{c-ams} discuss the PEBS experiment planned to fly at Earth's poles and the AMS-02 experiment confirmed to be installed on the International Space Station (ISS), respectively. Therefore, the probability for a certain cut-off as a function of detector trajectory at the poles and at ISS orbit are calculated. A scan of geomagnetic cut-offs for the South Pole for all geographic positions at 40\,km altitude and for all incident angles with reasonable step sizes was performed (position angles: 10°, direction angles: 5°). A random position following an uniform distribution in longitude and $\sin(\text{latitude})$ (South Pole: interval [-90°,-75°]) was chosen together with a random pair of isotropic zenith and azimuth direction angles. The cut-offs for random positions and particle directions are calculated by interpolation. The results are shown in Fig.~\ref{f-cutoff_issorbit} for the full zenith angle range and for $\cos(\text{zenith})>0.6$ which corresponds roughly to the angular acceptance of the PEBS detector. The efficiency distribution for the cut-offs in the full zenith angle range shows a much steeper increase from small cut-offs to about 1\,GV than from 1\,GV to about 10\,GV. Here, nearly only particles with large zenith angles are cut away as the comparison with the distribution using the zenith angle constraint shows. This is due to the shape of the geomagnetic field which is nearly perpendicular to Earth's surface at the poles. Respecting the isotropic particle incidence, the cut-off is negligible at the South Pole for the full zenith angle range starting from about 10\,GV. Taking also the angular acceptance of PEBS into account the cut-off is negligible starting from 0.8\,GV and at 0.15\,GV about 50\,\% of the particles are cut away. A similar study was also carried out for typical particle trajectories for a launch from Kiruna, Sweden near the North Pole (latitude interval [60°,75°], longitude interval [25°, 240°]) \cite{blast}. Here, respecting PEBS's acceptance a cut-off efficiency of 50\,\% is expected at 0.75\,GV and the cut-off is negligible starting from about 1.8\,GV. Particles with large zenith angles near the North Pole show a similar behavior as at the South Pole. The effects at the North and South Pole are small in comparison to the AMS-02 detector at ISS orbit which will be discussed in the following. 

The cut-offs relevant for AMS-02 are calculated for an ISS altitude of 390\,km. The ISS has an inclination $\alpha=51.7$°\cite{iss} and the angular velocity around Earth $\omega\sub{ISS}$ is calculated by:\be\omega\sub{ISS}(h)=\sqrt{\frac{G m_\oplus}{(R_\oplus+h)^3}}.\ee $m_\oplus$ is the mass and $R_\oplus$ the radius of the Earth, $G$ is the gravitational constant and $h$ the altitude of the ISS ($\omega\sub{ISS}(390\,\text{km})=1.1\cdot10^{-3}$\,Hz). The geographic position at a given time $t$ of the ISS is calculated by:

\be \left(\begin{matrix}\cos\theta\cos\phi \\ \cos\theta\sin\phi\\\sin \theta\end{matrix}\right) = 
\left(\begin{matrix} 
\cos(\omega_\oplus t) 	& -\sin(\omega_\oplus t) 	& 0 \\ 
\sin(\omega_\oplus t) 	& \cos(\omega_\oplus t)		& 0 \\ 
0 			& 0				& 1 
\end{matrix}\right)\cdot
\left(\begin{matrix} 
1 	& 0 		& 0 \\ 
0 	& \cos\alpha	& -\sin\alpha \\ 
0 	& \sin\alpha	& \cos\alpha 
\end{matrix}\right)\cdot
\left(\begin{matrix}\cos(\omega\sub{ISS}t) \\ \sin(\omega\sub{ISS}t) \\0\end{matrix}\right) 
\label{e-issorbit}
\ee
where $\phi$ and $\theta$ are the geographic longitude and the latitude, respectively, and $\omega_\oplus=7.3\cdot10^{-5}$\,Hz is the angular velocity around the axis of the Earth. The simulation followed the ISS orbit for 300 days which corresponds to about 4500 orbits. A pair of zenith and azimuth direction angles was again randomly chosen following an isotropic distribution for each position every 10\,s. The occupancy of geographic positions is shown in Fig.~\ref{f-h_longitude_latitude}. The cut-off efficiencies averaged over all positions and direction angles at ISS orbit is also shown in Fig.~\ref{f-cutoff_issorbit}. The probability to measure particles below 0.25\,GV is nearly 0, 50\,\% of the particles are cut away at about 6.5\,GV and the geomagnetic effect is insignificant starting from about 25\,GV. 

\subsubsection{Antimatter and Antiparticles}

\begin{figure}
\begin{center}
\begin{minipage}[b]{.4\linewidth}
\centerline{\epsfig{file=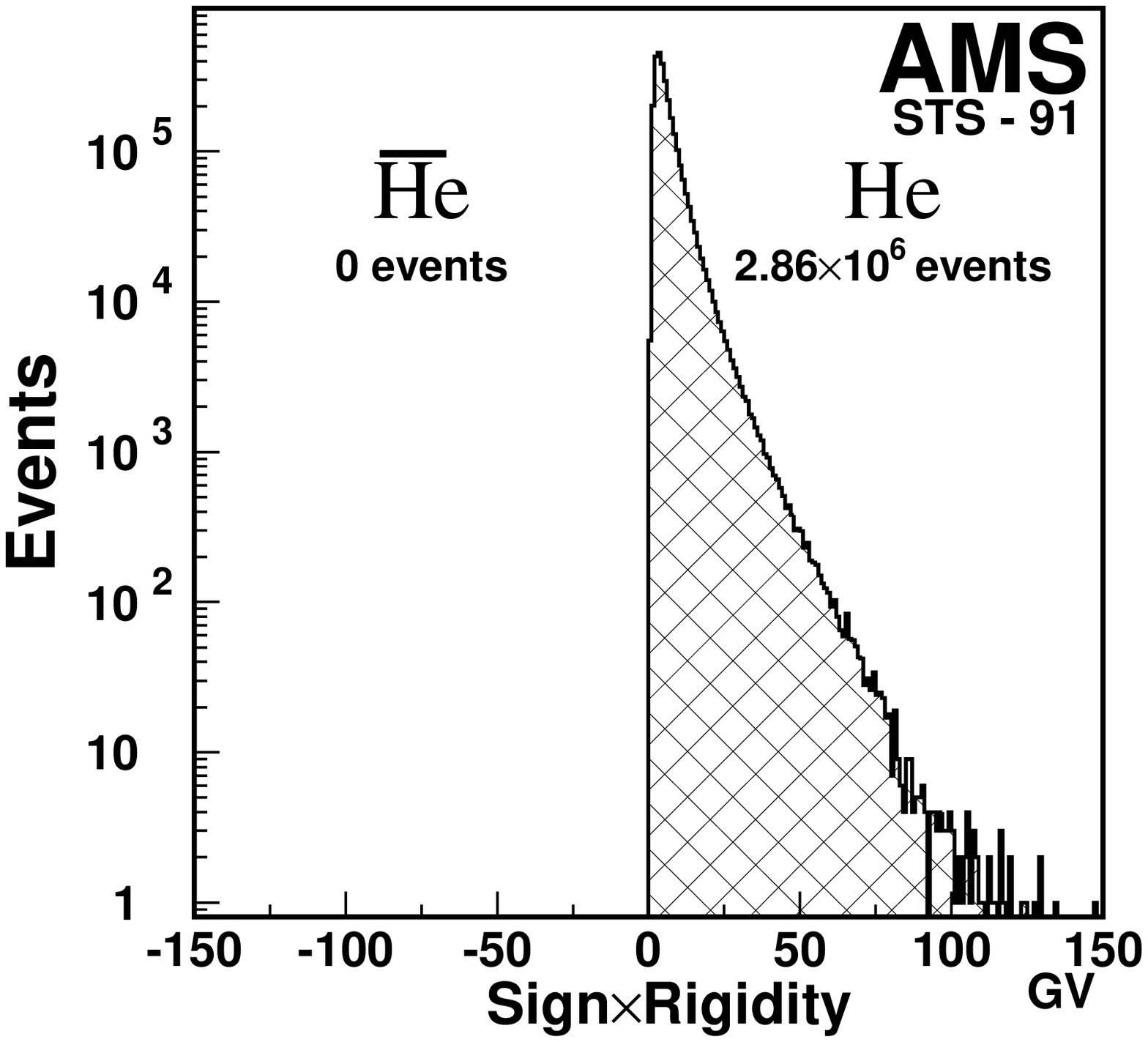,width=8cm}}
\captionof{figure}{\label{f-antihelium_ams1}Measured helium and antihelium spectra in cosmic rays by AMS-01. Reprinted figure with permission from Elsevier \cite{ams01}.}
\end{minipage}
\hspace{.1\linewidth}
\begin{minipage}[b]{.4\linewidth}
\centerline{\epsfig{file=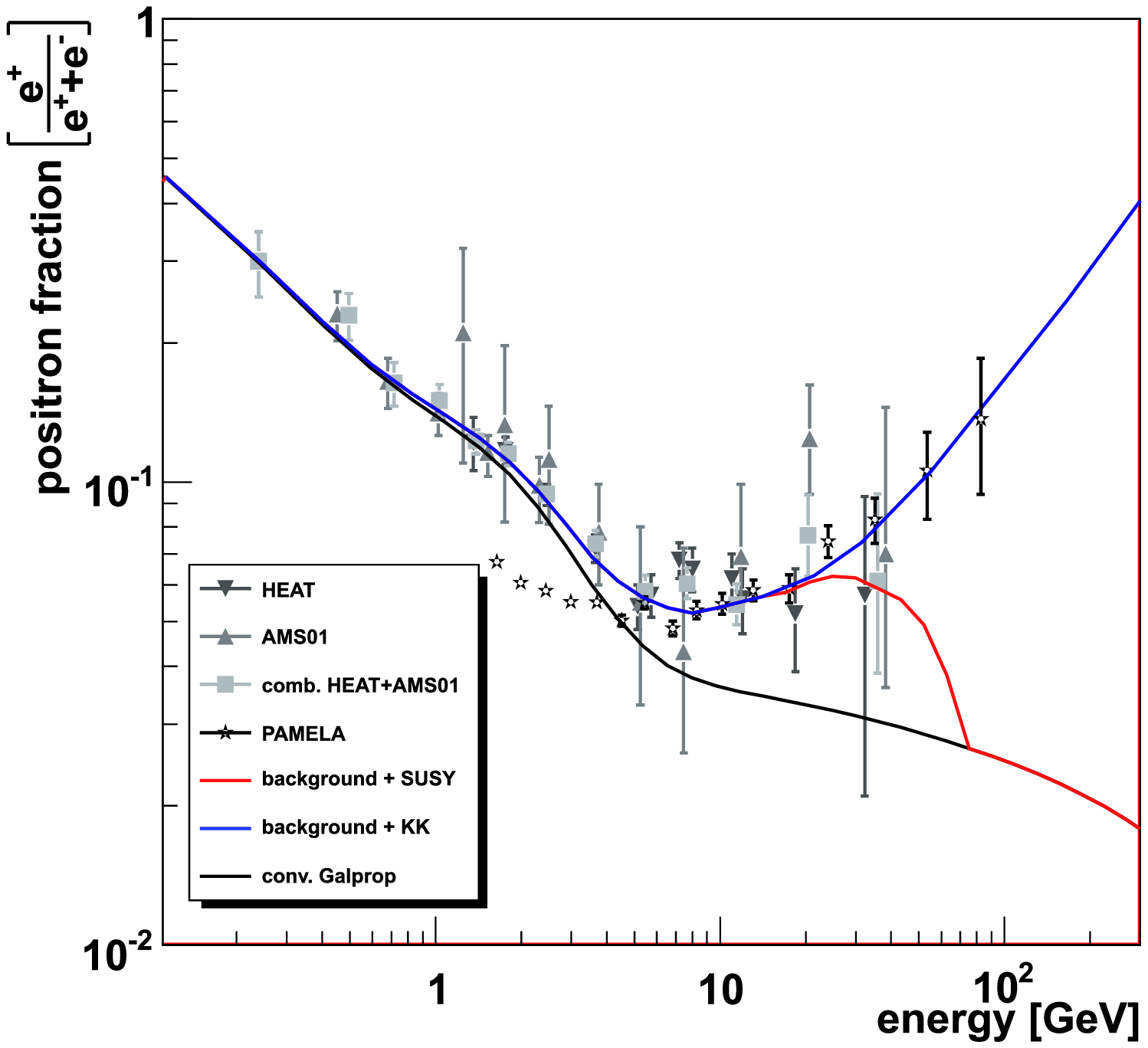,width=8cm}\vspace{0.2cm}}
\captionof{figure}{\label{f-positron_fraction}Current status of positron fraction measurements with models for supersymmetric and Kaluza-Klein dark matter candidates\cite{Barwick-1997,ams01,olzem-2007,pamela_ep,hooper-2004,gast-2008}.}
\end{minipage}
\end{center}
\end{figure}

Current theories concerning antimatter assume a matter-antimatter asymmetric Universe \cite{steigman-2008}. It consists predominantly of ordinary matter and contains only small amounts of antimatter. A production of antinuclei like antihelium in a matter environment is excluded on the time scale of our Universe and a measurement of antihelium would be a strong hint for the existence of primordial antihelium that was formed during the Big Bang. Carbon and heavier elements can only arise from stars. Assuming similar fusion reactions in stars made of antimatter (antistars) as in regular stars, the measurement of anticarbon nuclei constrains the existence of antigalaxies and antistars  \cite{swordy-1998}. For rigidities up to 15\,GV, the present upper bound for the existence of antihelium relative to helium is $6.8\cdot10^{-7}$ \cite{Sasaki-2002}. Fig.~\ref{f-antihelium_ams1} shows the AMS-01 helium measurements. Assuming a similar spectrum for antihelium as for helium, the upper limit on the existence of antihelium is $10^{-6}$ up to 150\,GV. Indirect evidence for the existence of antimatter would come from $\gamma$ rays produced during the annihilation of matter and antimatter but there is no indication for such mixed matter-antimatter regions on scales ranging from galaxies to groups and clusters of galaxies. Antimatter must be separated from matter by about $\cal O$(10\,Mpc) with a mass scale of $\cal O$$(10^{16}\,M_\odot)$ \cite{steigman-2008}. The observation of antinuclei like antihelium and heavier nuclei would have a large impact on the understanding of the Universe. 

\begin{figure}
\begin{center}
\begin{minipage}[b]{.4\linewidth}
\centerline{\epsfig{file=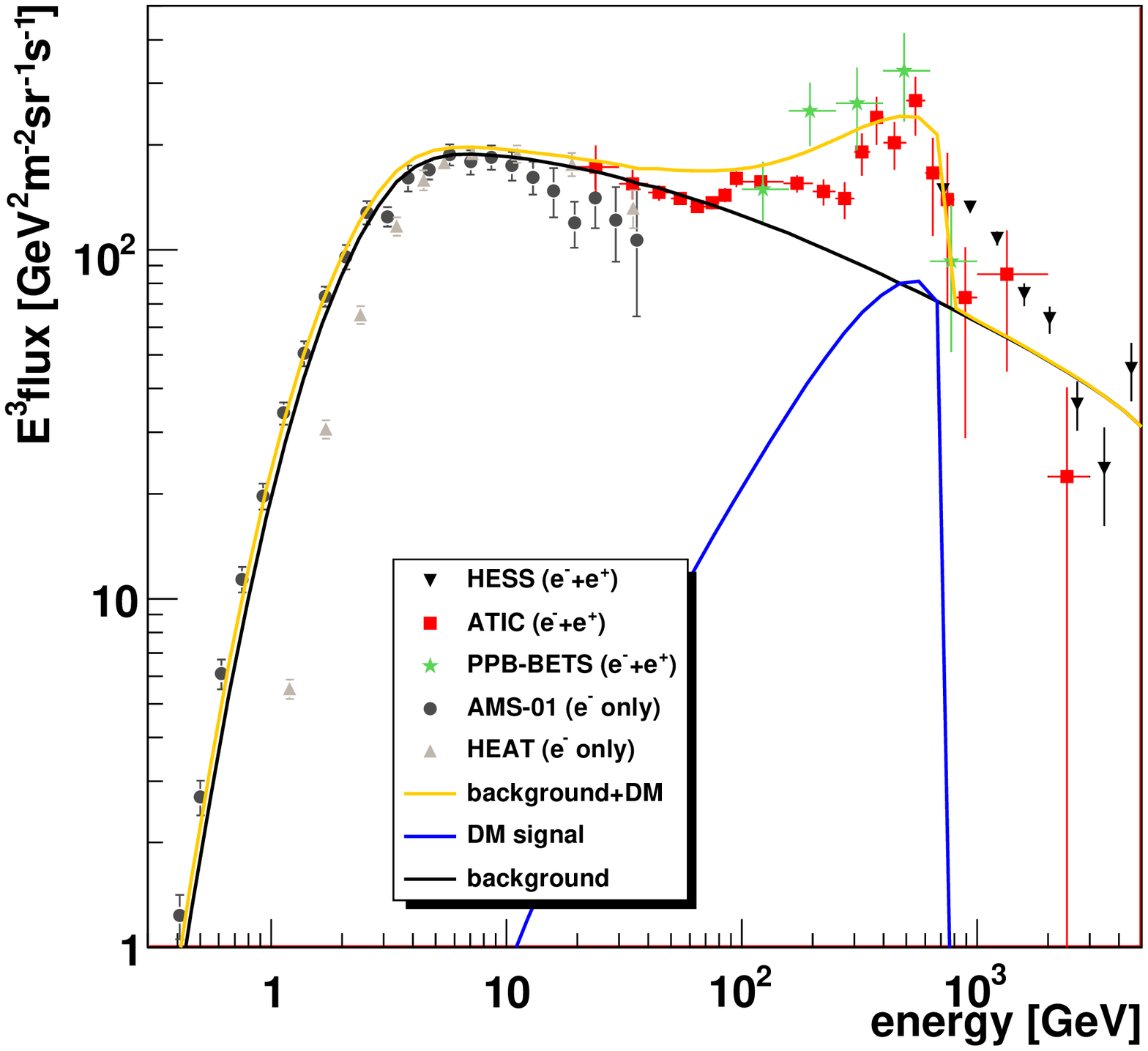,width=8cm}}\captionof{figure}{\label{f-cr_electron_spectrum}Electron (and positron) fluxes \cite{Barwick-1997,ams01,atic,torii-2008,hess_electron} with a model for a 1.4\,TeV dark matter candidate which decays directly to electrons and positrons \cite{chung-2009}.}
\end{minipage}
\hspace{.1\linewidth}
\begin{minipage}[b]{.4\linewidth}
\centerline{\epsfig{file=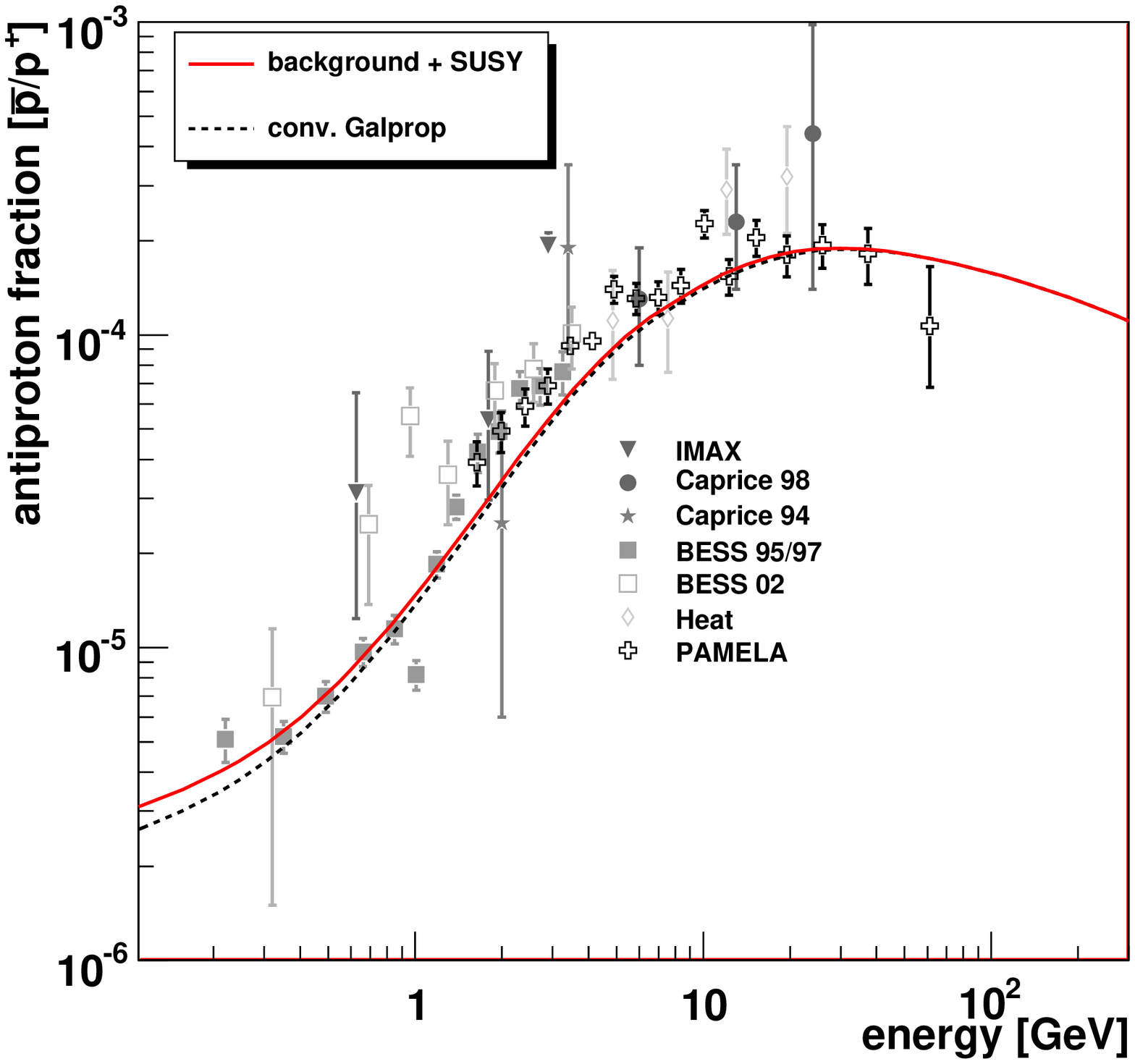,width=8cm}}\captionof{figure}{\label{f-antiproton_susy}Antiprotons in a supersymmetric theory \cite{PhysRevLett.76.3057,Barwick-1997,orito-2000-84,collaboration-2001-561,beach-2001,abe-2008,pamela_pbar,gast-2008}.}
\end{minipage}
\end{center}
\end{figure}

Since there are no known primary sources, the abundance of antiparticles in cosmic rays is sensitive to new effects. The observed antiparticle fluxes arise from secondary or tertiary reactions of the primary cosmic rays like protons, electrons, helium nuclei or light nuclei with the interstellar medium. Even small galactic or extragalactic effects which are not described by these reactions may be observed as a deviation in the antiparticle spectra. These effects would probably be symmetric for antiparticles and particles but as the particle fluxes are much larger than the antiparticle fluxes small effects would be better seen in the antiparticle spectra.

Positron and antiproton measurements can be used for indirect dark matter searches. Self annihilating dark matter candidates, e.g. from supersymmetric and Kaluza-Klein theories, can increase the antiparticle fluxes significantly. To cancel systematic effects it is advantageous to study the antiparticle to particle fraction. The positron fraction $f_{e^+}$ and antiproton fraction $f_{\bar p}$ are defined as:\be f_{e^+}=\frac{\displaystyle F_{e^+}}{\displaystyle F_{e^+}+F_{e^-}} \qquad\text{and}\qquad f_{\bar p}=\frac{\displaystyle F_{\bar p}}{\displaystyle F_{p}} \ee where $F_{x}$ is the corresponding differential particle flux. Recently published measurements of the satellite-borne PAMELA experiment \cite{pamela_ep} of the positron fraction (Fig.~\ref{f-positron_fraction}) and the feature in the electron spectrum measured by ATIC-2 \cite{atic} (Fig.~\ref{f-cr_electron_spectrum}) cannot be explained within conventional propagation models without introducing a fresh component. Fig.~\ref{f-positron_fraction} shows the positron fraction measurements and the background expectation together with the predictions of a SUSY model and a KK model for self-annihilating dark matter. The supersymmetric dark matter candidate (neutralino) is a fermion while the Kaluza-Klein candidate ($B^{(1)}$) is a boson. The direct annihilation of neutralinos is chirally suppressed and positrons could arise from decay chains via quark or vector boson pairs. This washes out a clear mass peak. A direct annihilation to $e^+e^-$ is allowed in the Kaluza-Klein case and will give a sharp rise and edge in the positron fraction. Also rapidly spinning and magnetized neutron stars (pulsars) could be responsible for the observed positron excess. The sum of all pulsars in the Milkyway could contribute significantly to the electron and positron fluxes and would also imply an anisotropy in the electron spectrum \cite{hooper-2008,pohl}. \v{C}erenkov telescopes may be able to discriminate between the Kaluza-Klein and the pulsar theory \cite{hess_electron,hall-2008}. Fig.~\ref{f-antiproton_susy} show a supersymmetric model for antiprotons. 
So far no significant discrepancy was found between measurements and theory up to 50\,GeV. Possible deviations at higher energies may be detectable in the future. As the recently published measurements still leave room for a lot of possible theories the following analyses and calculations of the PEBS and AMS-02 experiment capabilities will use exemplarily a supersymmetric model \cite{gast-2008}.

As noted above, it is known that dark matter does not follow the distribution of visible matter. Fluctuations in the local dark matter density could lead to enhancements in the local annihilation rate. The enhancement is expressed by the boost factor: \be BF=\frac{\displaystyle\int\rho\sub{DM}^2\text{d}V}{\left(\displaystyle\int\rho\sub{DM}\text{d}V\right)^2}\ee where $\rho\sub{DM}$ is the dark matter density and $V$ is the volume contributing to the flux. Antiprotons are much less attenuated than positrons in the interstellar medium and could thus arise from more distant sources. Photons are not deflected by magnetic fields and point back to their sources. Therefore, it may be possible to derive the shape of the dark matter density from flux measurements of the different (anti)particle species.

\subsubsection{Strangelets}

The standard model allows a new form of hadronic matter with a baryon number larger than 100. A state of $u$, $d$ and $s$ quarks in a hadronic bag (strangelet) is energetically favored \cite{amsyale}. If a neutron star is assumed to be a giant strangelet, fragments of such stars could be a part of cosmic rays exhibiting a large ratio of mass to charge.

\subsection{Ultra high Energy cosmic Rays}

\begin{figure}
\begin{center}
\begin{minipage}[b]{.4\linewidth}
\centerline{\epsfig{file=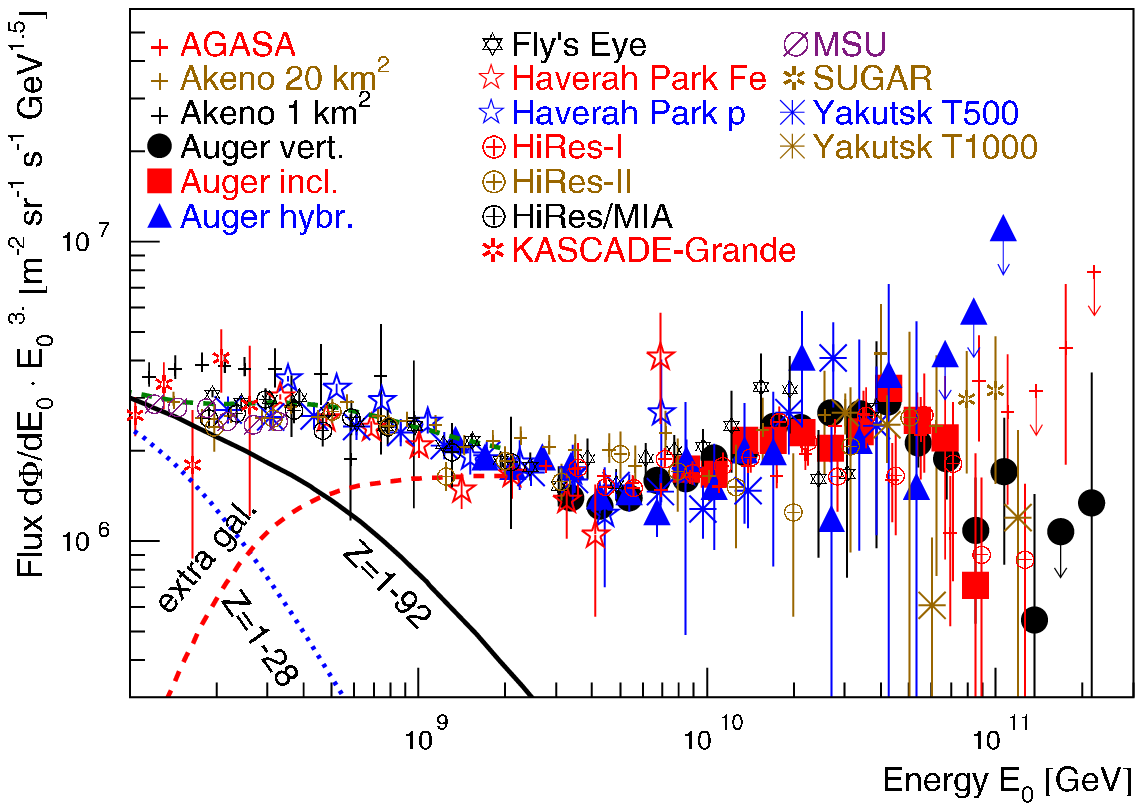,width=7.5cm}}
\captionof{figure}{\label{f-uhecr1}Differential cosmic-ray flux for ultra high energies. Reprinted figure with permission from the author \cite{hoerandel-2008}.}
\bigskip

\centerline{\epsfig{file=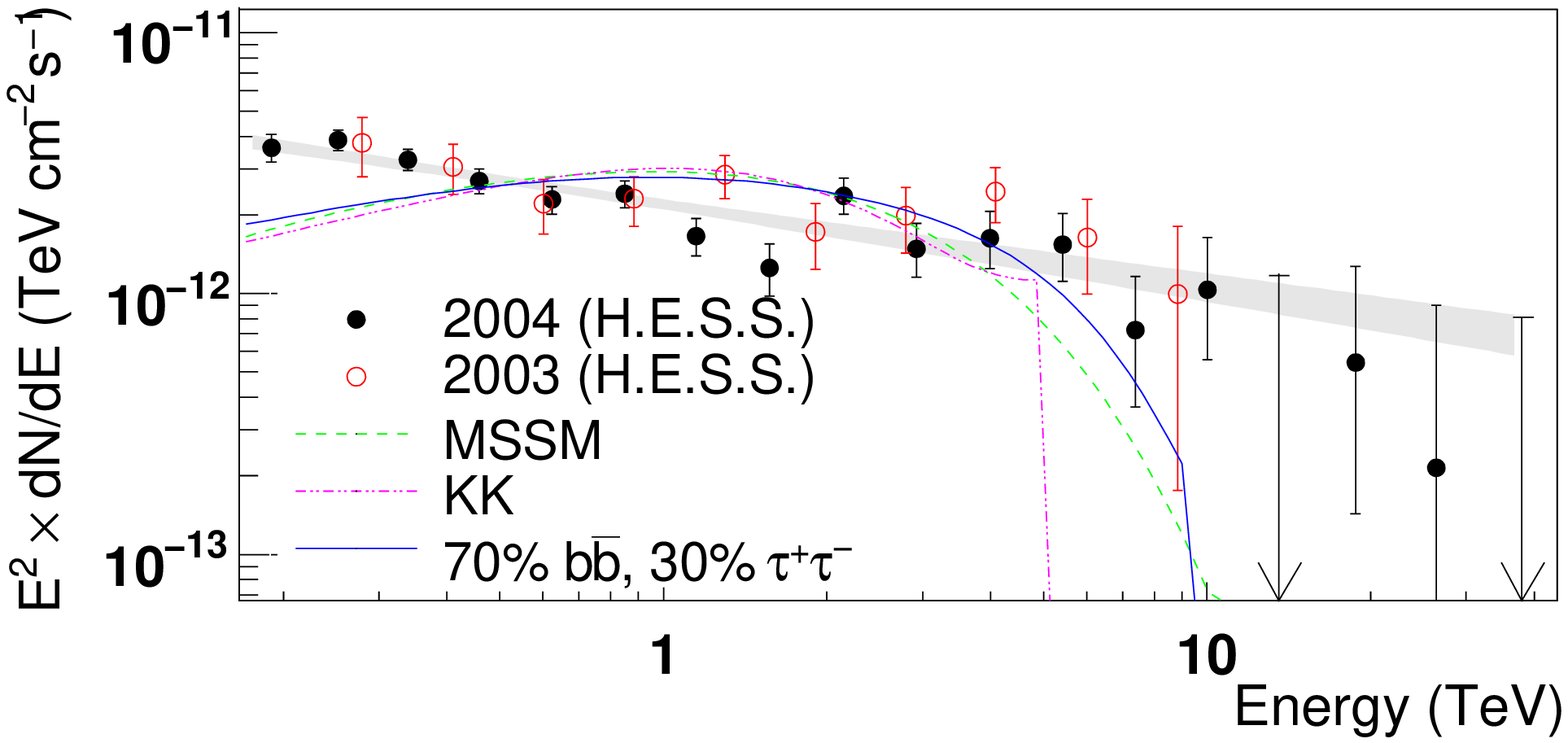,width=7.5cm}}
\captionof{figure}{\label{f-hess_dm}Differential $\gamma$-ray flux together with the predictions of dark matter models. Reprinted figure with permission from the American Physical Society \cite{collaboration-2007}.}
\end{minipage}
\hspace{.1\linewidth}
\begin{minipage}[b]{.4\linewidth}
\centerline{\epsfig{file=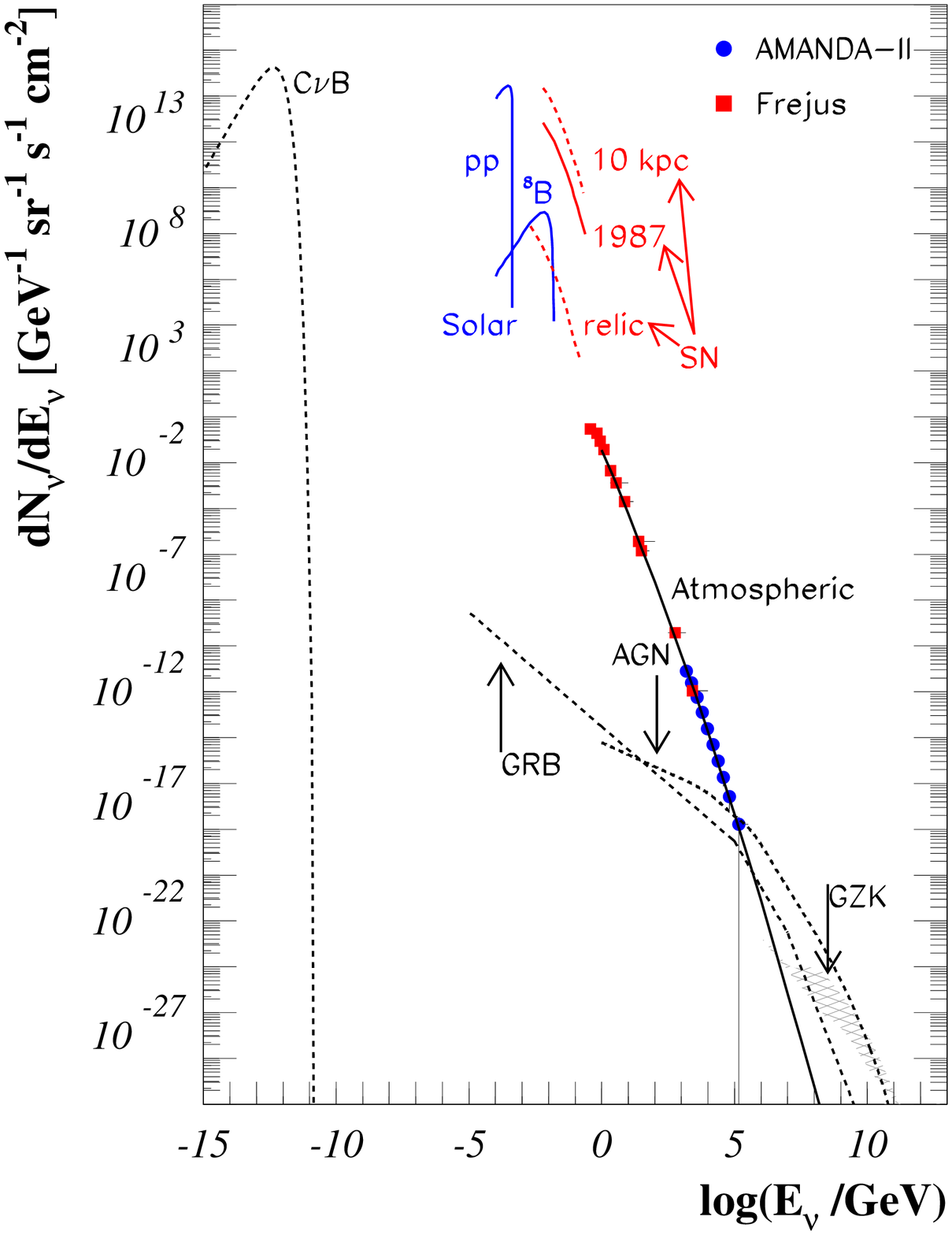,width=9cm}}
\captionof{figure}{\label{f-cr_nu}Differential cosmic neutrino ray flux showing the different contributions including the predictions for ultra high energies. Reprinted figure with permission from Elsevier \cite{becker-2007}.}
\end{minipage}
\end{center}
\end{figure}

Further interesting fields in astroparticle physics are the ground based observations of (ultra) high energy cosmic-ray particles (protons, iron, neutrinos and TeV $\gamma$-rays) \cite{parizot}. The fluxes are very low and need large area telescopes with a reasonable acceptance (Fig.~\ref{f-uhecr1} - \ref{f-cr_nu}). So far there are no known cosmic accelerators for protons with an energy of $10^{20}$\,eV. Even at these energies charged particles are deflected by magnetic fields, so it is also interesting to measure photons and neutrinos in addition. They could have formed in reactions of ultra high energy protons $p$ with the interstellar medium $n$ via pion $\pi$ production:
\be p+n\rightarrow \pi+X \rightarrow\begin{cases}\pi^0&\rightarrow \gamma +\gamma,\\\pi^\pm&\rightarrow \mu^\pm +\stackrel{_{\tiny{(-)\;\;}}}{\nu_\mu}.\end{cases}\ee
For energies up to a few  TeV the spectrum of $\gamma$-rays can still be explained by inverse Compton scattering and production in proton interactions. $\gamma$-rays are used to observe astronomical objects like gamma ray bursts or pulsar wind nebulae and are important for the understanding of cosmic-ray accelerators \cite{collaboration-2007}. It must be noted that the Universe is not transparent to photons with energies larger than about $10^{13}$\,eV because of absorption due to pair production with the cosmic microwave background. Therefore, the measurement of neutrinos would be very important. In addition, neutrinos can also be used to measure annihilation products of dark matter.

\subsection{Sky Coverage}

\begin{figure}
\begin{center}
\begin{minipage}[b]{.55\linewidth}
\centerline{\epsfig{file=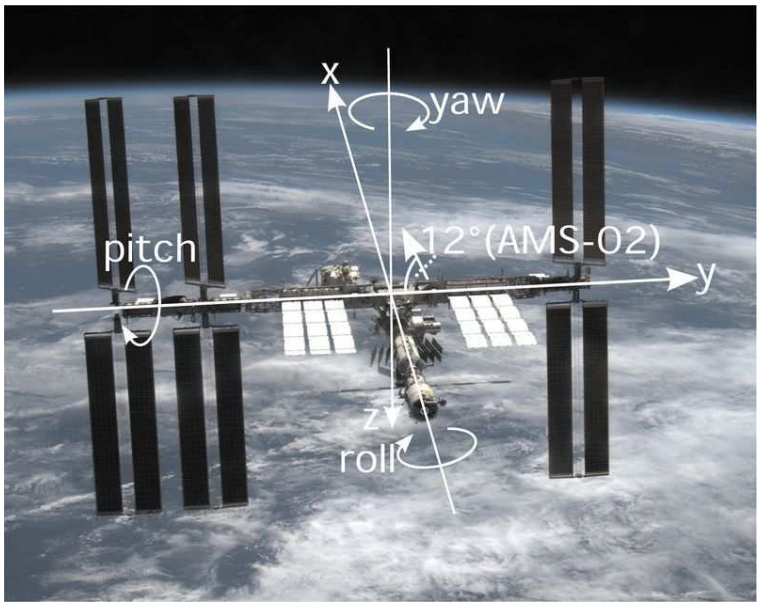,width=10cm}}
\vspace{0.8cm}
\captionof{figure}{\label{f-iss_coordinate}ISS coordinate system. Picture of the ISS taken during the NASA Space Shuttle Mission STS-126 in November 2008 \cite{sts-126}}
\end{minipage}
\hspace{.1\linewidth}
\begin{minipage}[b]{.25\linewidth}
\centerline{\begin{tabular}{c|c|c}
\hline
\hline
Roll [°] 			&Pitch[°]	&Yaw[°]\\
\hline
-4.7    &18.5   &-13\\
-0.8    &17.8   &6.1\\
-1.5    &17.7   &11.5\\
-1.1    &17.9   &167\\
-1.5    &18     &167\\
-0.9    &-3.3   &-5\\
-1.7    &20.54  &-11\\
-2.6    &19.99  &-11\\
-1.1    &21.92  &180\\
-2.8    &21.13  &176\\
-0.2    &0.01   &-4.4\\
-1.4    &21.25  &-13\\
-1.6    &22.31  &176\\
-1.5    &21.31  &176\\
-0.2    &0.31   &-4.4\\
-0.7    &19     &175\\
0.5     &-2     &-4\\
-0.5    &17.5   &-10\\
0.4     &-3     &-3\\
\hline
\end{tabular}}
\captionof{table}{\label{t-pry}Typical flight attitudes of the ISS\cite{clark-2009}.}
\end{minipage}
\end{center}
\end{figure}

It is interesting to map the galactic sky with charged particles in the GeV to TeV range because even if they are deflected in the galactic magnetic field it still might be possible to find anisotropies due to new so far unknown sources, e.g. nearby pulsars. This section studies the sky coverages of PEBS and AMS-02 which are different because of the different flight trajectories. The idea is to simulate particles from all possible positions and directions of the galactic sky and determine the exposure for each position. Here, magnetic fields are neglected and particles are assumed to follow straight lines. A simulation was carried out in a similar way as for the calculation of the geomagnetic cut-off efficiencies (Sec.~\ref{s-gev}). PEBS is planned to fly at Earth's poles and the detector positions are randomly generated following an uniform distribution in longitude and $\sin(\text{latitude})$ in the latitude interval [75°,90°] for the North Pole and in the interval [-90°,-75°] for the South Pole, respectively. The ISS orbit is calculated according to Eq.~\ref{e-issorbit}. The AMS-02 sky coverage calculation also has to take into account that firstly, AMS-02 is tilted by 12° towards the center of the ISS and secondly, the different flight attitudes of the ISS. The ISS coordinate system is shown in Fig.~\ref{f-iss_coordinate} and typical ISS flight attitudes for different operations, e.g. nominal conditions and orbiter rendezvous, are shown in Tab.~\ref{t-pry} \cite{clark-2009}. The ISS orientation is described with roll $\phi$, pitch $\theta$ and yaw $\psi$ angles and the orientation change is subdivided into three consecutive rotations. The corresponding rotation matrix $\mathcal R_{ypr}$ is calculated according to:

\be
\mathcal R_{ypr}=
\left(\begin{matrix} 
1 		& 0 		& 0 \\ 
0 		& \cos\phi 	& -\sin\phi \\ 
0	 	& \sin\phi 	& \cos\phi
\end{matrix}\right)
\cdot
\left(\begin{matrix} 
\cos\theta	& 0 		& \sin\theta \\ 
0 		& 1	 	& 0 \\ 
-\sin\theta 	& 0	 	& \cos\theta
\end{matrix}\right)
\cdot
\left(\begin{matrix} 
\cos\psi	& -\sin\psi	& 0 \\ 
\sin\psi 	& \cos\psi 	& 0 \\ 
0	 	& 0	 	& 1
\end{matrix}\right).
\ee

The simulation assumes an uniform distribution of the attitudes shown in table.~\ref{t-pry}. As for the cut-off calculation, a random pair of direction angles according to an isotropic distribution was chosen for each position and the exposure for a certain position and direction was weighted corresponding to the zenith angle acceptance of PEBS and AMS-02 \cite{gast-accep}. 

So far the trajectories were expressed in celestial coordinates which are now transformed to galactic coordinates. In astronomy, an epoch is a specific reference frame for which celestial coordinates are specified. The current epoch is called J2000.0 \cite{standish-1982}. Here, the celestial coordinates are converted from epoch J2000.0 ($x_{2000}$) to epoch B1950.0 ($x_{1950}$) with\cite{murray-1989}:

\be
\vec x_{1950}=
\left(\begin{matrix} 
0.9999257079523629	& 0.0111789381264276	& 0.0048590038414544\\
-0.0111789381377700	& 0.9999375133499888	& -0.0000271579262585\\
-0.0048590038153592	& -0.0000271625947142	& 0.9999881946023742
\end{matrix}\right)
\cdot\vec x_{2000}.
\ee

\begin{figure}
\begin{center}
\centerline{\epsfig{file=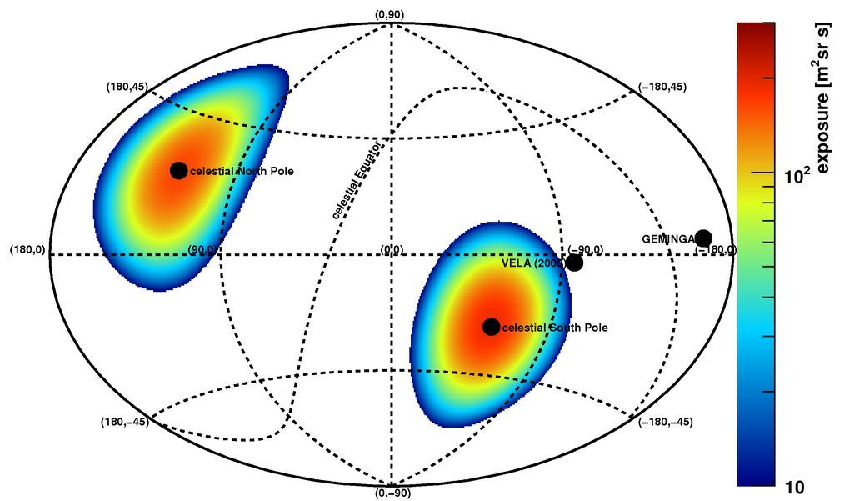,width=15cm}}
\captionof{figure}{\label{f-northsouthpole_longitude_latitude}Exposure of the sky in galactic coordinates (Aitoff projection) for the balloon-borne PEBS experiment with an acceptance of 0.4\,m$^2$sr at the North Pole and South Pole for a flight time of 50\,days each.}
\end{center}
\end{figure}

\begin{figure}
\begin{center}
\centerline{\epsfig{file=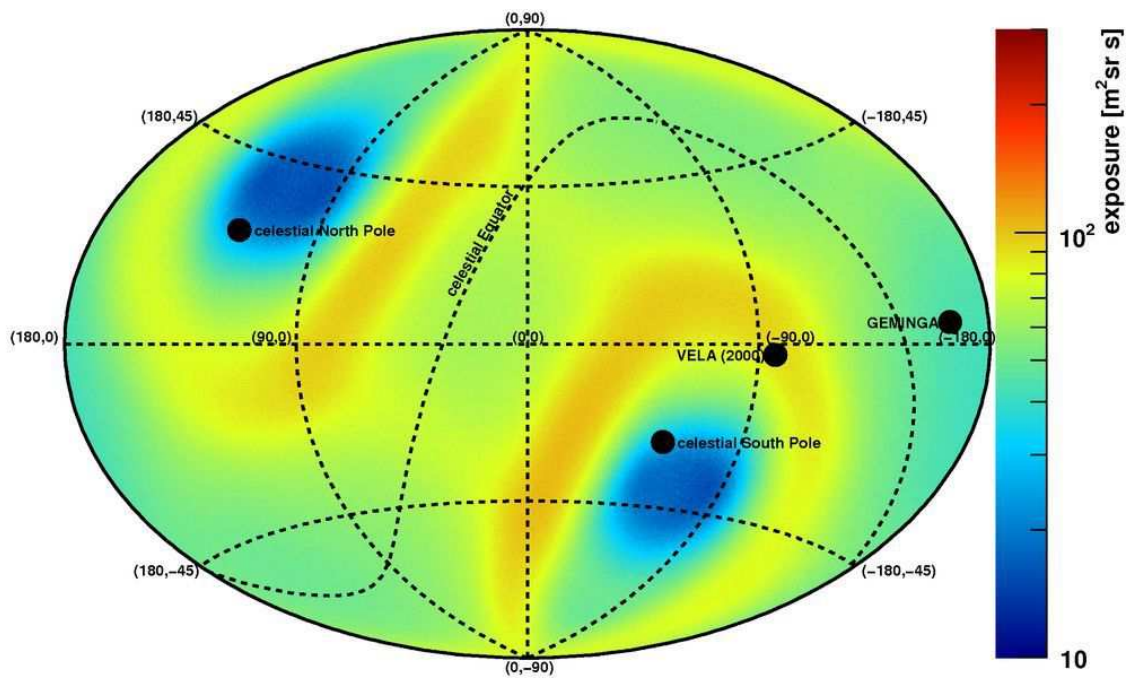,width=15cm}}
\captionof{figure}{\label{f-issorbit_longitude_latitude}Exposure of the sky in galactic coordinates (Aitoff projection) for the space-based AMS-02 experiment with an acceptance of 0.095\,m$^2$sr at ISS orbit for a flight time of 3\,years.}
\end{center}
\end{figure}

\begin{figure}
\begin{center}
\centerline{\epsfig{file=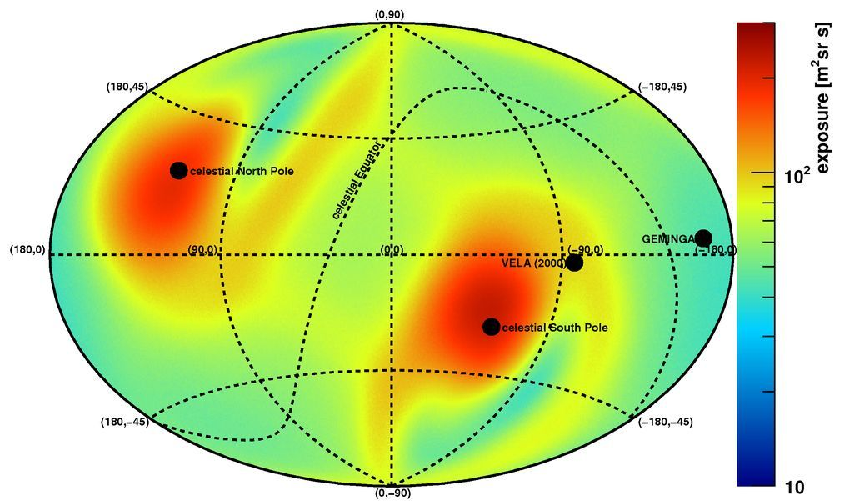,width=15cm}}
\captionof{figure}{\label{f-add_longitude_latitude}Added exposure of the sky in galactic coordinates (Aitoff projection) for the balloon-borne PEBS experiment with an acceptance of 0.4\,m$^2$sr at the North Pole and South Pole for a flight time of 50\,days each and for the space-based AMS-02 experiment with an acceptance of 0.095\,m$^2$sr at ISS orbit for a flight time of 3\,years.}
\end{center}
\end{figure}

The galactic coordinates are then calculated according to\cite{zeilik-1997}:

\be
\left(\begin{matrix} 
\cos(b)\cos(l-33^{\circ})\\
\cos(b)\sin(l-33^{\circ})\\
\sin(b)
\end{matrix}\right)=
\left(\begin{matrix} 
\cos(\delta)\cos(\alpha-282.25^{\circ})\\
\sin(\delta)\sin(62.6^{\circ})+\cos(\delta)\sin(\alpha-282.25^{\circ})\cos(62.6^{\circ})\\
\sin(\delta)\cos(62.6^{\circ})-\cos(\delta)\sin(\alpha-282.25^{\circ})\sin(62.6^{\circ})
\end{matrix}\right).
\ee

where $\delta$ is the declination and $\alpha$ is the right ascension (epoch B1950.0) and $b$ and $l$ are the galactic latitude and longitude, respectively. Aitoff projections \cite{Snyder-1997} of the exposures are shown in Fig.~\ref{f-northsouthpole_longitude_latitude} for PEBS with two flights (North Pole and South Pole, 50\,days each) and in Fig.~\ref{f-issorbit_longitude_latitude} for AMS-02 at ISS orbit for three years. The exposure is defined as the total number of entries normalized to the detector acceptance multiplied by the measurement time. The measurements with PEBS cover the regions around the celestial poles while AMS-02 covers nearly the complementary region. Only both experiments together are able to deliver a complete picture of the sky (Fig.~\ref{f-add_longitude_latitude}). These pictures can be interpreted as the expected diffuse charged particle background. A deeper study of possible new sources would have to take into account effects of the galactic magnetic field, e.g. as a function of distance to the source.
 
\newpage
\mbox{}
\newpage

\chapter{The Balloon-borne PEBS Mission\label{c-pebs}}

\section{Balloon-borne Cosmic-Ray Experiments}

Balloon-borne cosmic-ray experiments exist since the discovery of cosmic rays by Victor Hess in 1912. Recently several high-altitude balloon experiments at altitudes between 30 to 40\,km were carried out (Tab.~\ref{t-balloonflights}). They contributed to the cosmic-ray flux measurements in the GeV to TeV energy range. The challenge in such experiments is to maximize the flight time and flight altitude while using a large acceptance high-precision particle detector. The analysis must always take into account the interactions of cosmic rays in Earth's atmosphere. The particles of cosmic origin are attenuated and particles from secondary interactions within the atmosphere contribute to the total flux. The following will discuss some important balloon-borne experiments of the last years.

The Isotope Matter Antimatter Experiment (IMAX) had a successful flight from Lynn Lake, Manitoba, Canada in July 1992 with a flight time of 16\,hours. IMAX was able to measure protons, antiprotons, deuterium and helium nuclei in the energy range of 0.2 to 3.2\,GeV with a time of flight system, a superconducting magnet, multiwire proportional drift chambers and a streamer-tube brass-calorimeter \cite{PhysRevLett.76.3057}.

The TS93 experiment was based on IMAX but used a transition radiation detector and a silicon-tungsten calorimeter. It flew 25\,hours in September 1993 from Ft. Sumner, New Mexico, USA. The main purpose was to measure electrons and positrons from 4 to about 50\,GeV \cite{1996ApJ...457L.103G}.

The Cosmic AntiParticle Ring Imaging Cherenkov Experiment (CAPRICE) was again based on IMAX and TS93. It had in addition a ring image \v{C}erenkov counter (RICH) and an improved imaging silicon-tungsten calorimeter. The experiment had a flight time of 23\,hours and was started in July 1994 from Lynn Lake, Manitoba, Canada. It measured atmospheric muons, antiprotons, positrons and light isotopes. The successor experiment CAPRICE-2 was improved with driftchambers and a new gas RICH. The energy range was extended to 50\,GeV and two flights were carried out in May 1998 with a flight time of 22\,hours \cite{caprice-positron,PhysRevLett.82.4757,caprice-antiproton,boezio-2003-19}.

The High-Energy Antimatter Telescope (HEAT) was launched in May 1994 from Fort Sumner, New Mexico, USA and May 1995 from Lynn Lake, Manitoba in Canada and again in June 2000 from  Fort Sumner and had a time of flight system, a transition radiation detector, a drift tube hodoscope, an electromagnetic calorimeter and a superconducting magnet. It was able to measure the antiproton flux up to energies of 20\,GeV and the positron fraction up to 50\,GeV. A positron excess above the purely secondary production predicted by propagation models was measured with large error bars at the highest energies \cite{Barwick-1997, beatty-2004-93}.

The BESS program had in total nine successful flight since 1993 with different purposes and launch locations. The detector consisted of a large solenoidal superconducting magnet, a time of flight system of scintillation counter hodoscopes and drift chambers. BESS was able to measure protons, atmospheric muons, protons, antiprotons and light nuclei and to set limits on the existence of antihelium. Antiprotons were measured down to very low energies of 0.1\,GeV \cite{Sasaki-2002,abe-2007-645,abe-2008}.

Since 1968 also emulsion chambers, e.g. ECC, have been exposed via several balloon flights to measure the cosmic electron flux and atmospherically produced photons \cite{nishimura-1980,yoshida-2006}.

The Advanced Thin Ionization Calorimeter (ATIC) uses the principle of ionization calorimetry. A silicon matrix is used to determine the absolute value of the electrical charge. During 2000 and 2003 ATIC completed three successful flights \cite{atic}.

The PPB-BETS detector flew in 2004 for 13\,days in Antarctica and is an imaging calorimeter which consisted of scintillating fiber belts, plastic scintillators and lead plates. It was able to measure cosmic-ray electrons and atmospheric photons \cite{torii-2008}.

Balloon-borne experiments like TRACER \cite{boyle-2008} and CREAM \cite{Ahn:2008my} focus on the measurement of the elemental composition of cosmic rays. They use large transition radiation detectors for charge determination. The CREAM experiment had a record breaking ultra-long duration balloon flight in the Antarctica of 42\,days in December 2004 and January 2005.

\begin{table}
\begin{center}
\captionof{table}{\label{t-balloonflights}Balloon flights.}
\begin{tabular}{l||l|l|l}
\hline
\hline
Experiment	& Year	& Location	& Subdetectors\\
\hline
IMAX		& 1992		& Lynn Lake, Manitoba, Canada	& TOF, sc. magnet, drift chamber,\\ 
		&		&				& brass-cal.\\
TS93		& 1993		& Ft. Sumner, New Mexico, USA	& TOF, sc. magnet, drift chamber, \\
		&		&				& TRD, silicon-tungsten cal.\\
CAPRICE		& 1994, 1998	& Lynn Lake, Manitoba, Canada	& TOF, sc. magnet, drift chamber,  \\
		&		&				& TRD, RICH, silicon-tungsten \\
HEAT		& 1994, 2000	& Ft. Sumner, New Mexico, USA	& cal., TOF, sc. magnet, drift tube \\ 
		& 1995		& Lynn Lake, Manitoba, Canada	& hodoscope, ECAL\\
BESS		& 1993 - 2007	& different locations:		& TOF, sc. magnet, drift chamber\\
		&		& e.g. Ft. Sumner, South Pole	&\\
ECC		& since 1968	& different locations		& emulsion chamber\\
ATIC		& 2000 - 2003	& South Pole			& ionization calorimeter, silicon\\
		&		&				& matrix\\
PPB-BETS	& 2004		& South Pole			& imaging calorimeter\\
TRACER		& 1999		& Ft. Sumner, New Mexico, USA	& plastic scintillators, TRD \\
		& 2003		& South Pole			& \v{C}erenkov det.\\
CREAM		& 2004 - 2008	& South Pole			& timing charge det., \v{C}erenkov \\
		&		&				& det., TRD, silicon charge det., \\
		&		&				& scint. fiber, hodoscope, \\
		&		&				& tungsten-scint. cal. \\
\hline
\end{tabular}
\end{center}
\end{table}

\begin{figure}
\begin{center}
\centerline{\epsfig{file=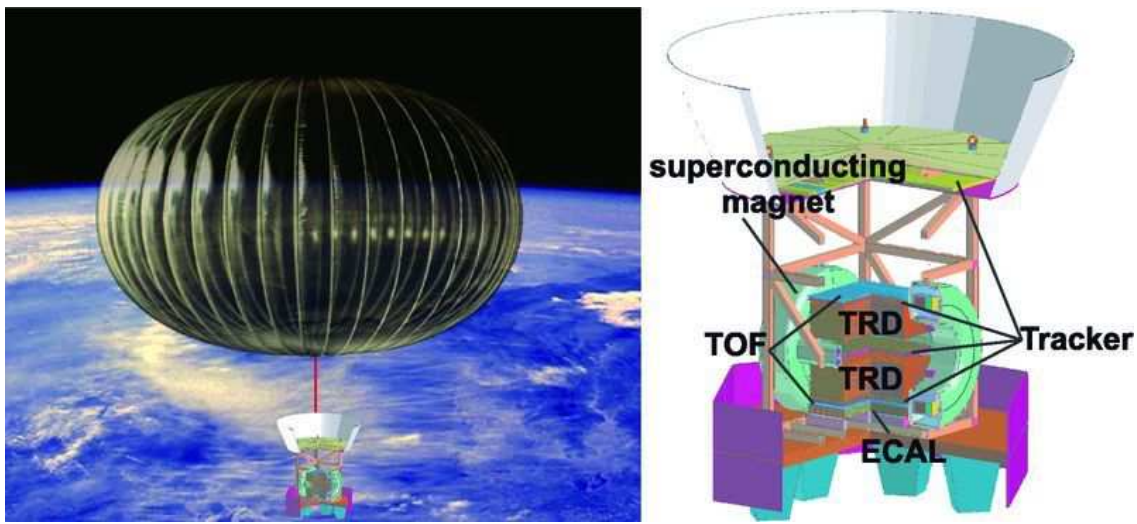,width=15cm}}
\captionof{figure}{\label{f-pebs}Schematic view of the PEBS experiment.}
\end{center}
\end{figure}

\section{The PEBS Experiment}

The Positron Electron Balloon Spectrometer (PEBS) is a proposal for a balloon-borne experiment to fly at an altitude of about 40\,km in Earth's atmosphere \cite{vondoetinchem-2007-581,gast-2008}. The long duration flights are planned for the North or South Pole. The poles have several advantages: the flight latitude stays quite stable during long duration flights, the landing position is predictable due to the stable circumpolar winds in the arctic or antarctic summer and the geomagnetic cut-off is small. The measurements must take place in summer to have good accessibility of the launching and landing sites and to generate the power with solar panels. The total measuring time of several flights is planned to add up to 100\,days.

PEBS (Fig.~\ref{f-pebs}) will consist of several subsystems. A time of flight system (TOF) with one lower and one upper plane is needed for triggering and for velocity measurements. The TOF is made out of plastic scintillators and is readout by silicon photomultiplier arrays. Two transition radiation detectors (TRD) have twelve layers each of proportional tubes filled with Xe/CO$_2$ (80\,\%/20\,\%) interleaved with fleece radiator. The TRD discriminates light and heavy particles on the basis of transition radiation which depends on the Lorentz factor $\gamma=E/m$. A combined silicon and scintillating fiber tracker is used for momentum measurement and a sandwich electromagnetic calorimeter (ECAL) with embedded fibers for the discrimination of particles using the shower shape. The detector is located inside a superconducting magnet with an average field of 0.8\,T. The overall weight should not exceed 2350\,kg including solar panels and readout electronics.

\section{Simulation of the Influence of the Atmosphere on Cosmic-Ray Measurements\label{s-atmo}}

It is important for balloon experiments to understand the interactions of cosmic-ray particles with Earth's atmosphere. Particles from secondary interactions contribute to the measured fluxes while the primary cosmic-ray particles are attenuated. This can falsify the interpretation of the data. To study these effects in detail the software package PLANETOCOSMICS \cite{laurent-2005} based on GEANT4 \cite{geant4-1,geant4-2} is used in the following. The electromagnetic physics are described with the standard physics list in Geant4 and hadronic physics with the quark-gluon string compound model using in addition the binary intranuclear cascade model and for elastic and inelastic scattering of neutrons with energies $< 20$\,MeV the HPNeutron model. The hadronic interaction of light ions with nuclei is described by an extension of the binary intranuclear cascade model.

\begin{figure}
\begin{center}
\centerline{\epsfig{file=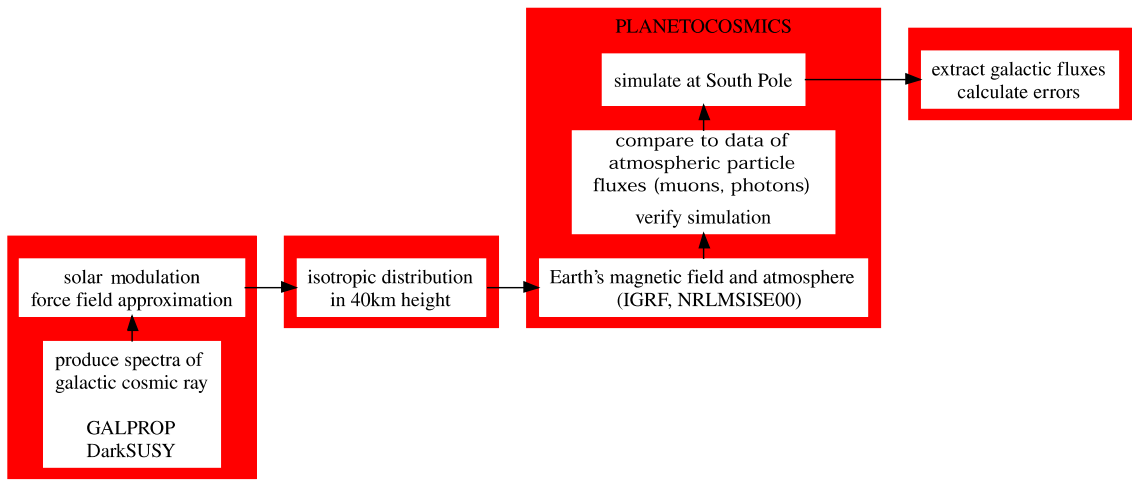,width=15cm}}
\captionof{figure}{\label{f-ana_scheme}Scheme of analysis.}
\end{center}
\end{figure}

Fig.~\ref{f-ana_scheme} shows the analysis scheme. In the first step the cosmic-ray spectra are calculated with GALPROP \cite{strong-2001-27,strong-1998-509}. Possible signals of supersymmetric neutralino annihilations are computed with DarkSUSY \cite{gondolo-2004-0407} (Fig.~\ref{f-fluxes_geo_mod_galprop_conv}) and serve as an example for a fresh cosmic-ray component. The supersymmetric model used here is favored by a study \cite{gast-2008} taking several constraints into account, namely the electroweak results from the Large Electron Positron collider, the relic density of the dark matter \cite{wmap_basic}, the top quark mass limits, the muon magnetic moment, the $b\rightarrow s\gamma$ branching ratio and the upper cross section bounds of direct dark matter detection experiments. The used top quark mass is 172.6\,GeV and the parameters of this supersymmetric model are:
\be m_{1/2}=260\,\text{GeV},\;m_{0}=1560\,\text{GeV},\;\tan\beta=40,\; \text{sign}(\mu) = +1,\;A_0 = 0.\label{eq-susypar}\ee 
The cosmic-ray fluxes must be solar modulated as described in Sec.~\ref{s-gev}. An isotropic distribution of these fluxes is then generated to be detected at different altitudes in Earth's atmosphere. The exact calculation will be explained later (Sec.~\ref{ss-atmosim}). The emphasis is on an altitude of 40\,km. This is the maximum altitude reachable at the poles during the summer for long duration balloon flights with a 2\,t payload.

For the simulation the current models for the atmospheric composition NRLMSISE00 \cite{picone-2002} and the magnetic field IGRF \cite{igrf} are used (Sec.~\ref{ss-atmoprop} and \ref{s-gev}).

A comparison with the atmospheric muon data from the BESS experiment \cite{abe-2007-645} taken at Ft. Sumner, New Mexico, USA at different altitudes in September 2001 is carried out to verify the simulations. In addition, a comparison with the photon data of the PPB-BETS experiment and the PPB-BETS model for atmospheric electron production is performed \cite{torii-2008}. The simulation is then carried out for PEBS at the South Pole and the cosmic-ray fluxes with errors are calculated by taking all atmospheric and detector properties with their corresponding errors into account.

\begin{figure}
\begin{center}
\centerline{\epsfig{file=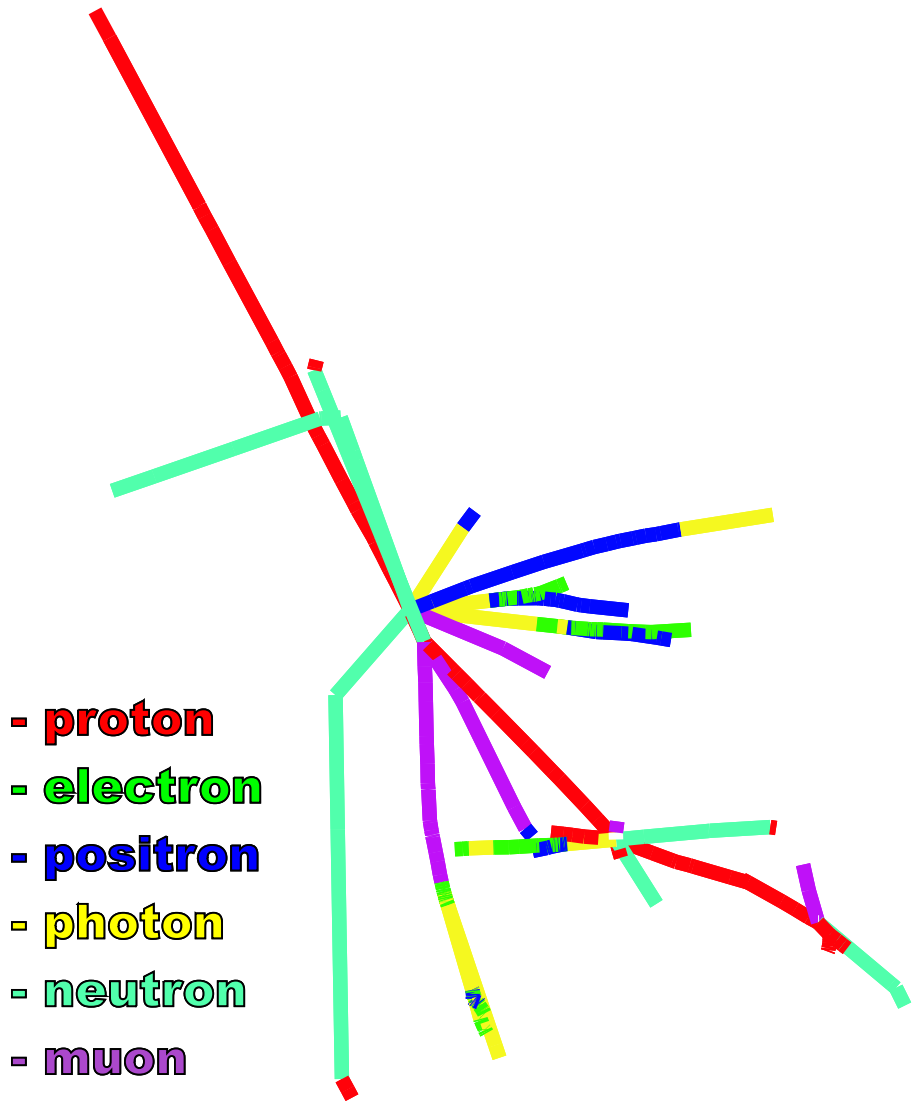,width=12cm}}
\captionof{figure}{\label{f-10_GeV_proton_ww_mag}Simulation of a 10\,GeV proton in Earth's atmosphere.}
\end{center}
\end{figure}

\subsection{Properties of Earth's Atmosphere\label{ss-atmoprop}}

Interactions of cosmic rays with Earth's atmosphere have to be studied to determine the cosmic flux modulation as a function of flight altitude for a detector in the atmosphere. The probabilities for interactions depend on density and composition of the atmosphere. Important physical processes in the atmosphere and properties of Earth's atmosphere will be discussed in the following.

Fig.~\ref{f-10_GeV_proton_ww_mag} shows a simulation of a 10\,GeV proton air shower in the atmosphere. A lot of particles arise from interactions mostly at lower altitudes. The primary proton is scattered and its kinetic energy decreases. Atmospherically produced particles are mostly decay products of pions $\pi^{\pm/0}$ which arise from interactions of primary cosmic rays like protons $p$, helium or heavier nuclei with atmospheric nuclei $n$:
\be p+n\rightarrow \pi+X \rightarrow\begin{cases}\pi^\pm&\rightarrow \mu^\pm +\stackrel{_{\tiny{(-)\;\;}}}{\nu_\mu}\rightarrow e^\pm+\stackrel{_{\tiny{(-)\;\;}}}{\nu_\mu}+\stackrel{_{\tiny{(-)\;\;}}}{\nu_e},\\\pi^0&\rightarrow \gamma +\gamma.\end{cases}\ee
The contributions to the atmospheric particle fluxes are small due to other mesons like $K$ and $\eta$ which are also produced in the primary interactions. Atmospheric antiprotons can form in proton-proton interactions like \cite{weber-1997}:
\be
p+p\rightarrow p+\bar p + p +p.
\ee
The kinematics of the production mechanisms do not favor low kinetic energies (<1\,GeV) for the antiprotons. In addition, losses of antiparticles due to annihilations and kinetic energy losses in the atmosphere must be respected. Heavy particles other than electrons lose energy mostly in nuclear interaction, e.g. ionization and atomic excitation while  bremsstrahlung losses dominate the energy loss for electrons and positrons \cite{pdgbook}.

The elemental composition and density of the atmosphere as a function of altitude enter as crucial parameters to the simulation of the cosmic-ray shower development. The PLANETOCOSMICS code makes use of the NRLMSISE00 model that has been developed on the basis of several sets of observations such as with satellites, rocket probes, incoherent scatter stations, molecular oxygen observations from sun occultations and the solar and magnetic activity. The model consists of parametrized analytical approximations to the observations.

\begin{figure}
\begin{center}
\begin{minipage}[b]{.4\linewidth}
\centerline{\epsfig{file=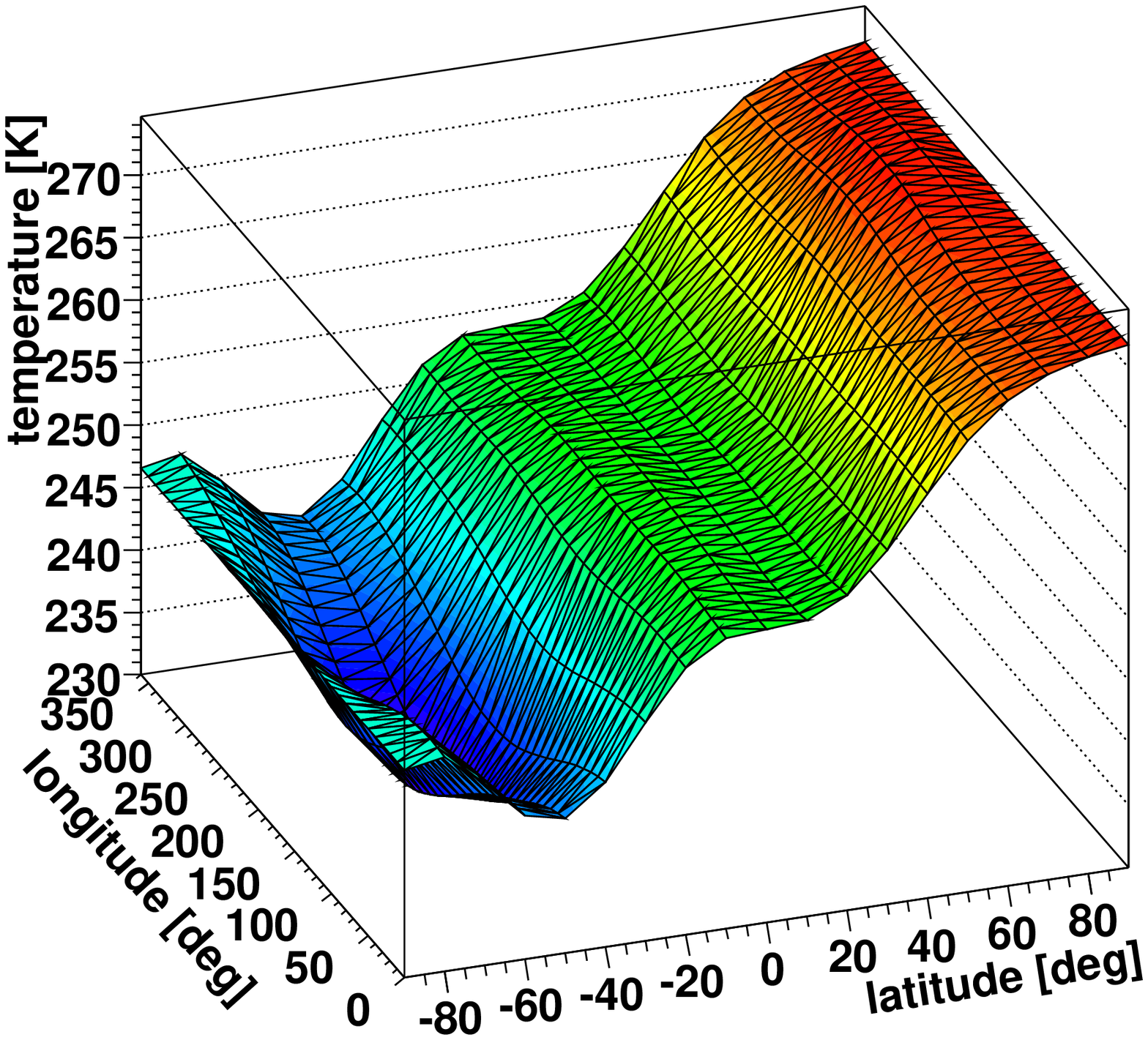,width=8cm}}
\captionof{figure}{\label{f-40km_4_178.4_187.4_0620_17_0_0_NRLMSISE00_temperature}Temperature profile at 40\,km altitude in Earth's atmosphere during June. The South Pole is at -90° lat. and the North Pole at 90° lat..}
\end{minipage}
\hspace{.1\linewidth}
\begin{minipage}[b]{.4\linewidth}
\centerline{\epsfig{file=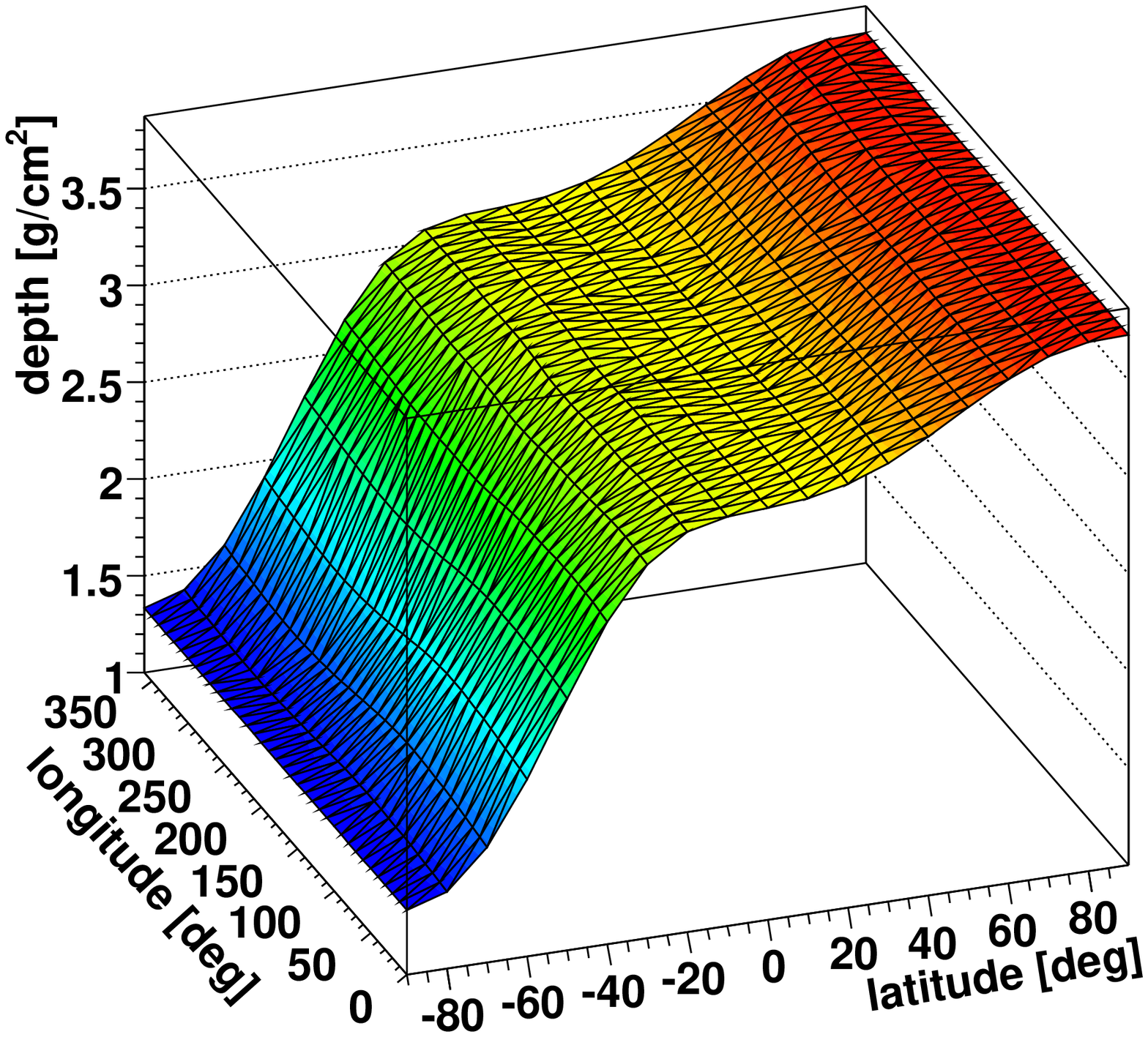,width=8cm}}
\captionof{figure}{\label{f-40km_4_178.4_187.4_0620_17_0_0_NRLMSISE00_depth}Atmospheric depth profile at 40\,km altitude in Earth's atmosphere during June. The South Pole is at -90° lat. and the North Pole at 90° lat..}
\end{minipage}
\end{center}
\end{figure}
\begin{figure}
\begin{center}
\centerline{\epsfig{file=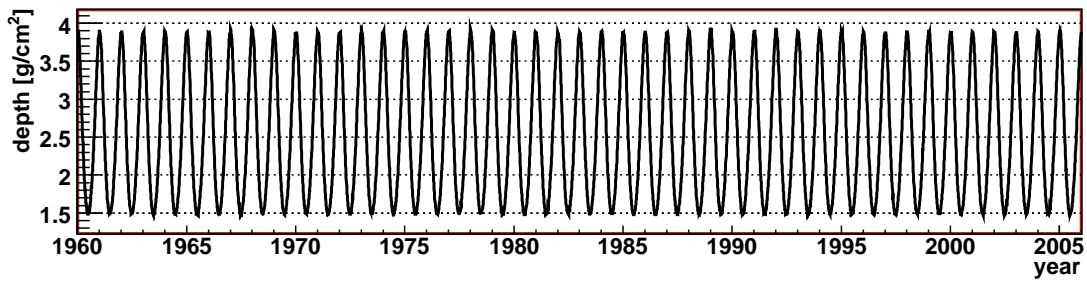,width=15cm}}
\captionof{figure}{\label{f-depth_time}Change of atmospheric depth with time at 40\,km altitude at the South Pole.}
\end{center}
\end{figure}

The temperature profile for Earth's atmosphere as a function of longitude and latitude at an altitude of 40\,km in June 2005 is shown in Fig.~\ref{f-40km_4_178.4_187.4_0620_17_0_0_NRLMSISE00_temperature}. The variation with longitude is small compared to the latitude variation. As expected, the northern hemisphere is warmer than the southern one during June. Even more interesting for the calculation of interactions is the amount of matter that the particles have to cross before detection. This quantity is called atmospheric depth $d$:\be d(H)=\int_H^\infty\rho(h)\text{d}h\ee where $\rho(h)$ is the altitude dependent atmospheric density and $H$ the detection altitude. Fig.~\ref{f-40km_4_178.4_187.4_0620_17_0_0_NRLMSISE00_depth} illustrates the depth variation for the same conditions as for the temperature profile above.  The depth during June at the North Pole is about three times larger than at the South Pole due to expansion of the atmosphere with increasing temperature. The behavior is the opposite in December. The depth calculation at an altitude of 40\,km at the South Pole with the appropriate magnetic and solar activities shows a regular behavior over the years. The maximum is always reached during the summer and the minimum during winter (Fig.~\ref{f-depth_time}). Here, the influence of the solar and magnetic activities is negligible.

\begin{figure}
\begin{center}
\begin{minipage}[b]{.4\linewidth}
\centerline{\epsfig{file=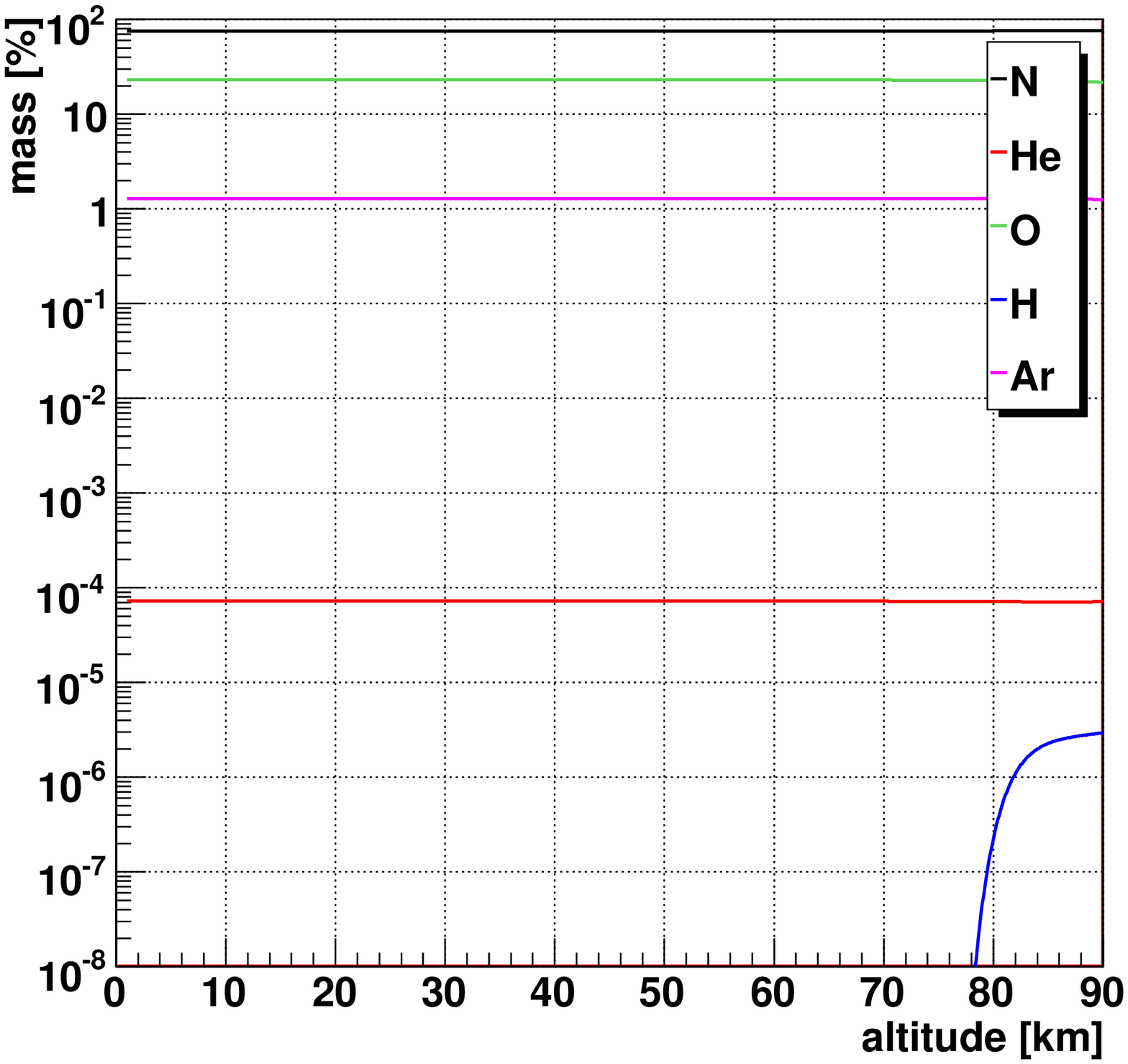,width=8cm}}
\captionof{figure}{\label{f-atmosphere_mass}Mass composition of the atmosphere during summer at the South Pole as a function of altitude.}
\end{minipage}
\hspace{.1\linewidth}
\begin{minipage}[b]{.4\linewidth}
\centerline{\epsfig{file=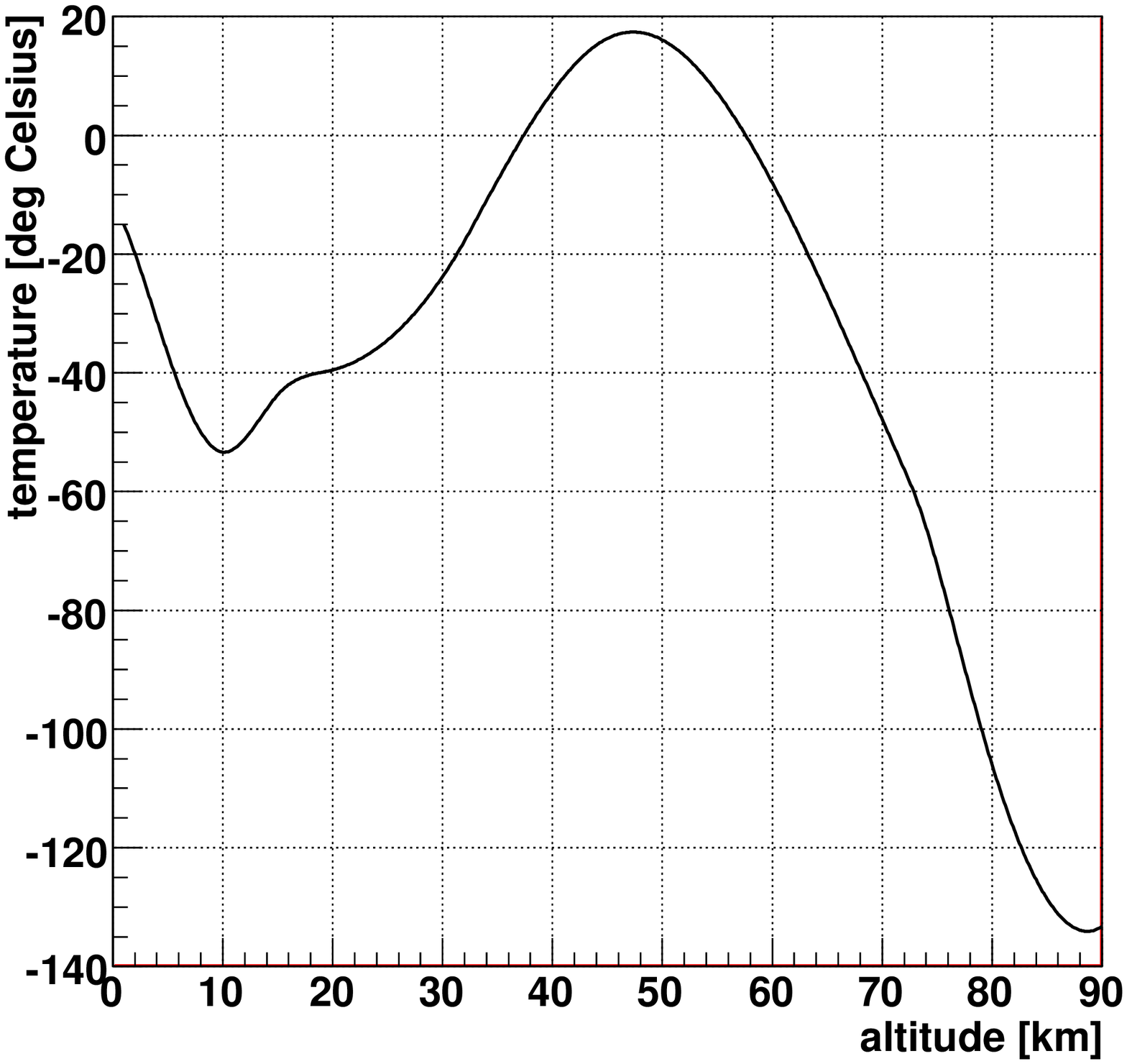,width=8cm}}
\captionof{figure}{\label{f-atmosphere_temp}Temperature in the atmosphere during summer at the South Pole as a function of altitude.}
\end{minipage}
\end{center}
\end{figure}
\begin{figure}
\begin{center}
\begin{minipage}[b]{.4\linewidth}
\centerline{\epsfig{file=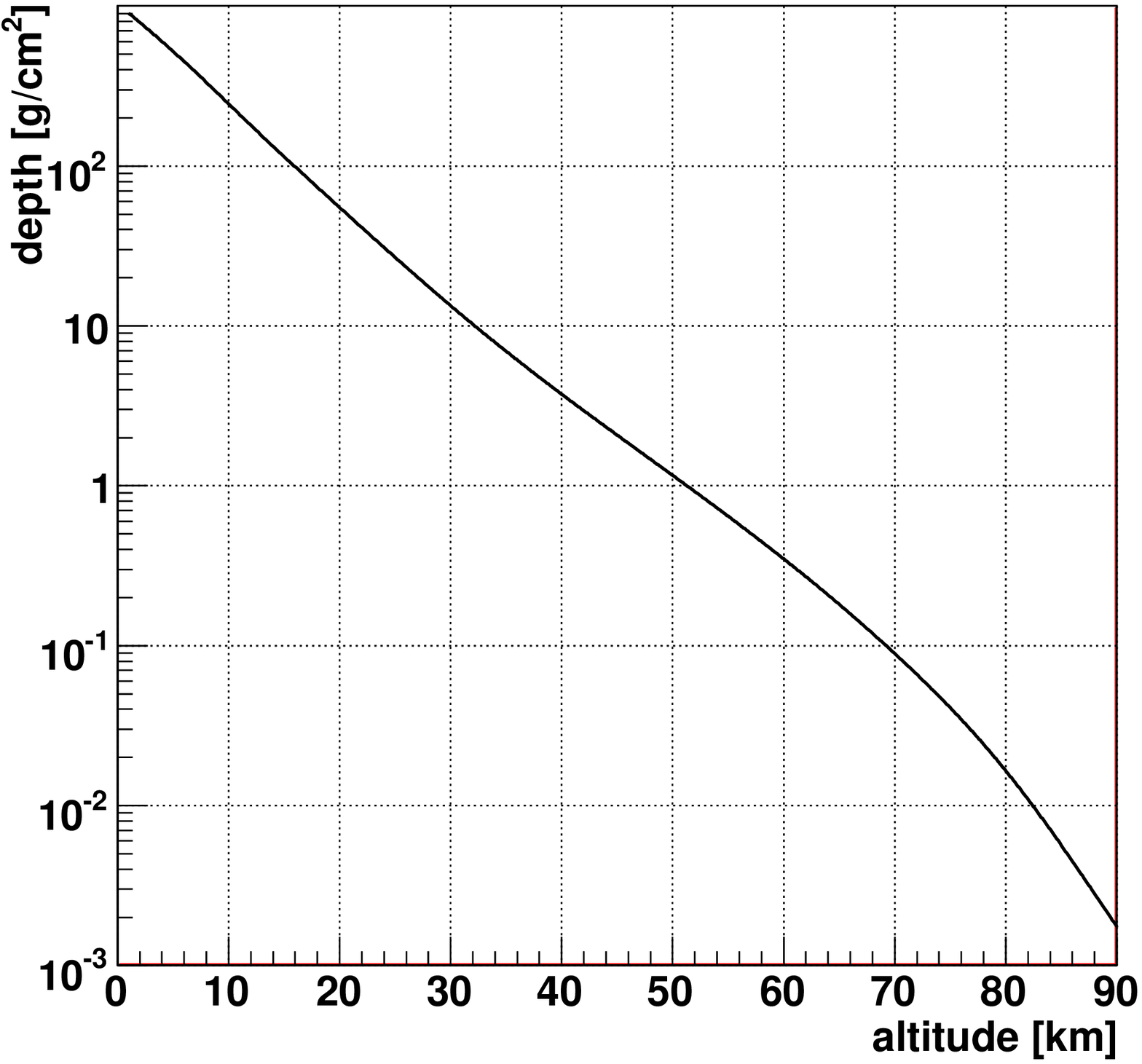,width=8cm}}
\captionof{figure}{\label{f-atmosphere_depth}Atmospheric depth in front of the detector during summer at the South Pole as a function of altitude for tracks with 0° zenith angle.}
\end{minipage}
\hspace{.1\linewidth}
\begin{minipage}[b]{.4\linewidth}
\centerline{\epsfig{file=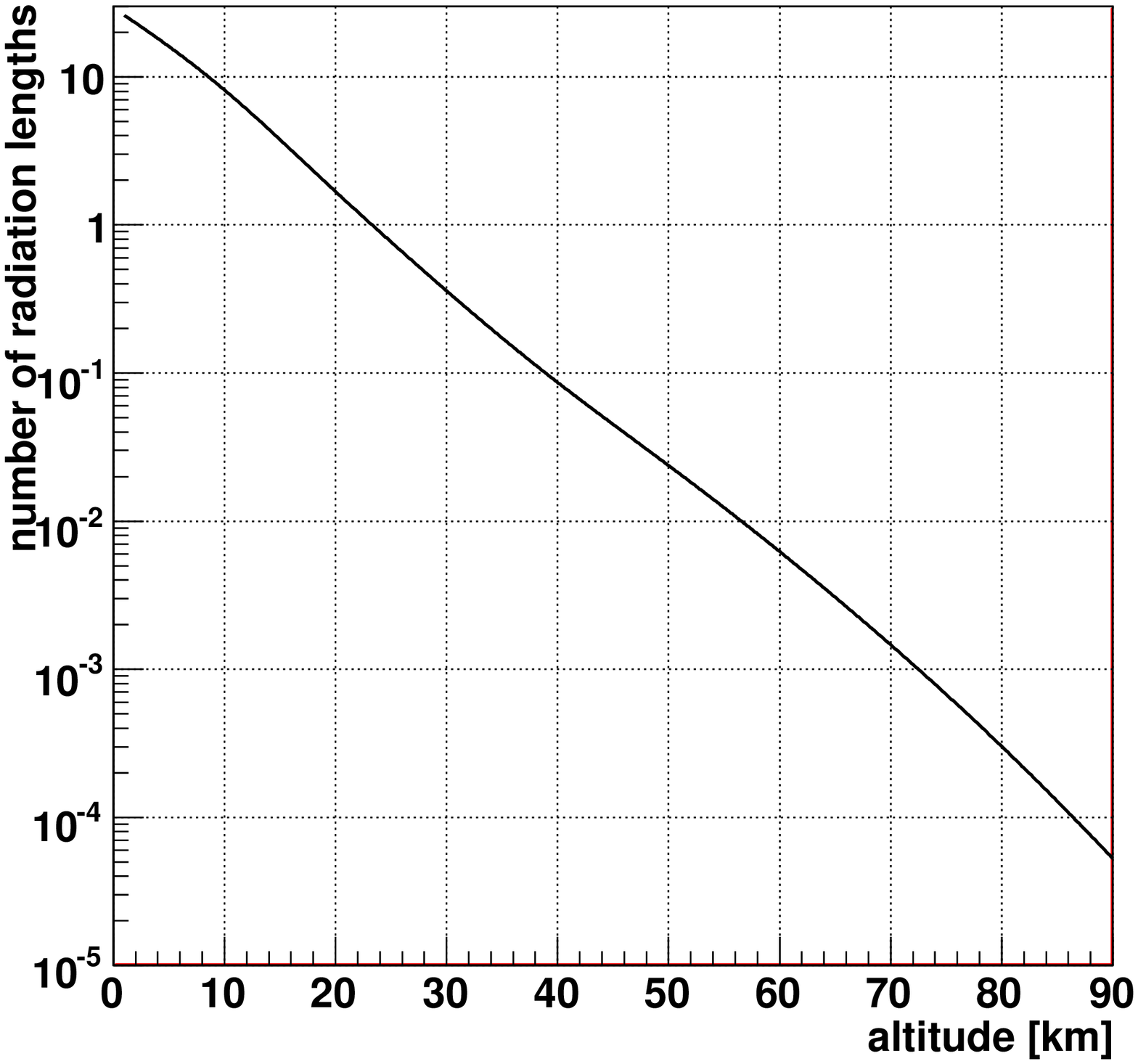,width=8cm}}
\captionof{figure}{\label{f-radiation_length_FtSumner_summer}Radiation length in front of the detector during summer at the South Pole as a function of altitude for tracks with 0° zenith angle.}
\end{minipage}
\end{center}
\end{figure}

The layer structure and variation with altitude in composition and densities of the atmosphere makes different kind of interactions possible\cite{barry-2003}. UV absorption effects start to be important at about 10\,km altitude in the so called tropopause and reheats the atmosphere. In addition to the UV light absorption, the ozone dissociation   at about 50\,km explains the maximum in temperature in the stratopause.  The change in mass composition and temperature with altitude during summer at the South Pole is shown in Fig.~\ref{f-atmosphere_mass} and \ref{f-atmosphere_temp}, respectively. 

The temperature at 40\,km is about 10°C which describes the kinetic energy of the atmospheric gas. This temperature has a negligible influence on the operation temperature of the detector as the atmosphere has a very small density of $\cal{O}$($5\cdot10^{-6}$\,g/cm$^3$) at 40\,km altitude. However the absorption of sun light by the detector is important and must be taken into account for stable operation.

\begin{figure}
\begin{center}
\begin{minipage}[b]{.4\linewidth}
\centerline{\epsfig{file=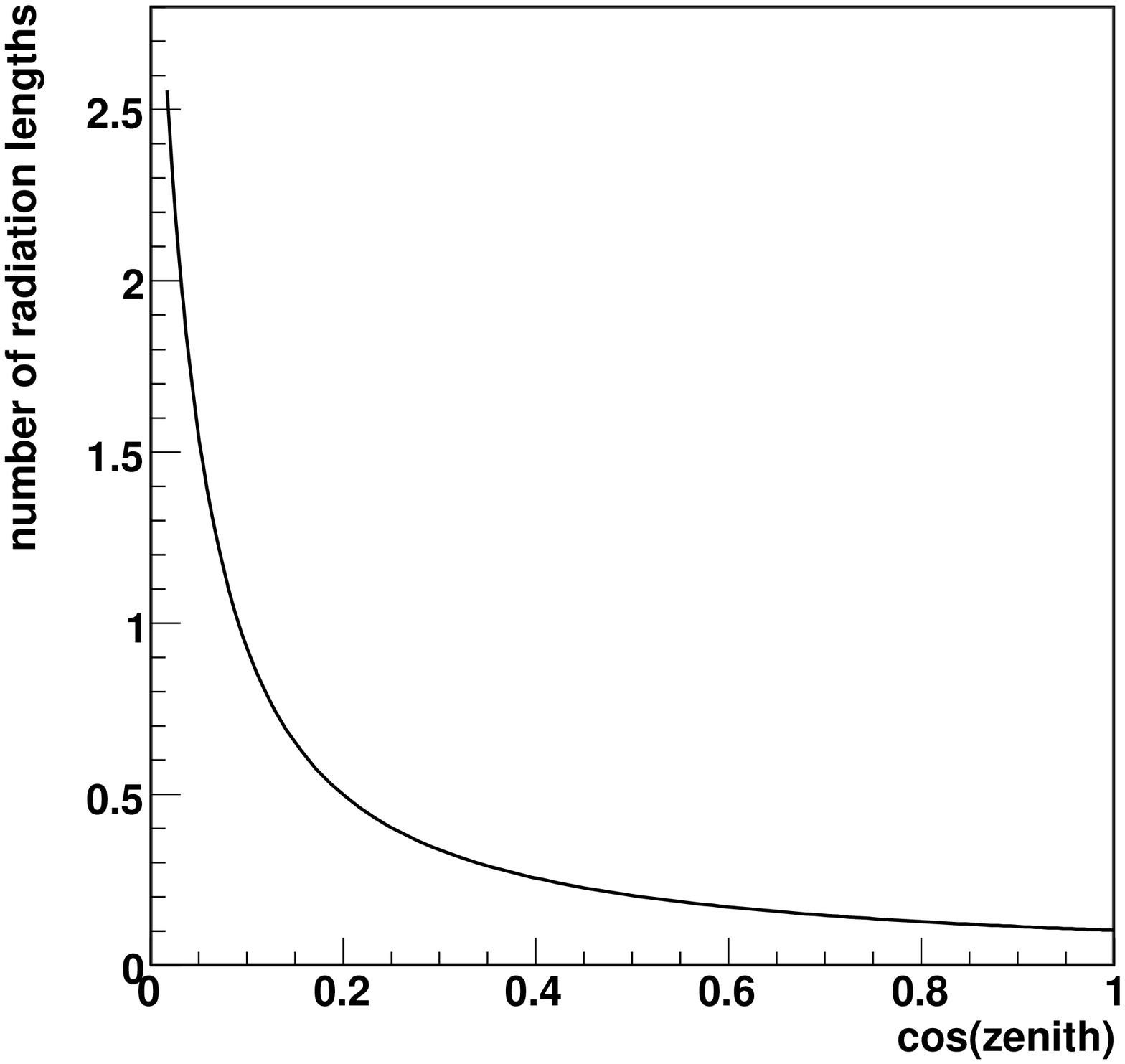,width=8cm}}
\captionof{figure}{\label{f-southpole_40_-89_0.01_0.01_zenithfixedPos.dat_cos_rad_length}Radiation lengths traversed before 40\,km altitude as a function of the zenith angle of the cosmic-ray particle trajectories.}
\end{minipage}
\hspace{.1\linewidth}
\begin{minipage}[b]{.4\linewidth}
\centerline{\epsfig{file=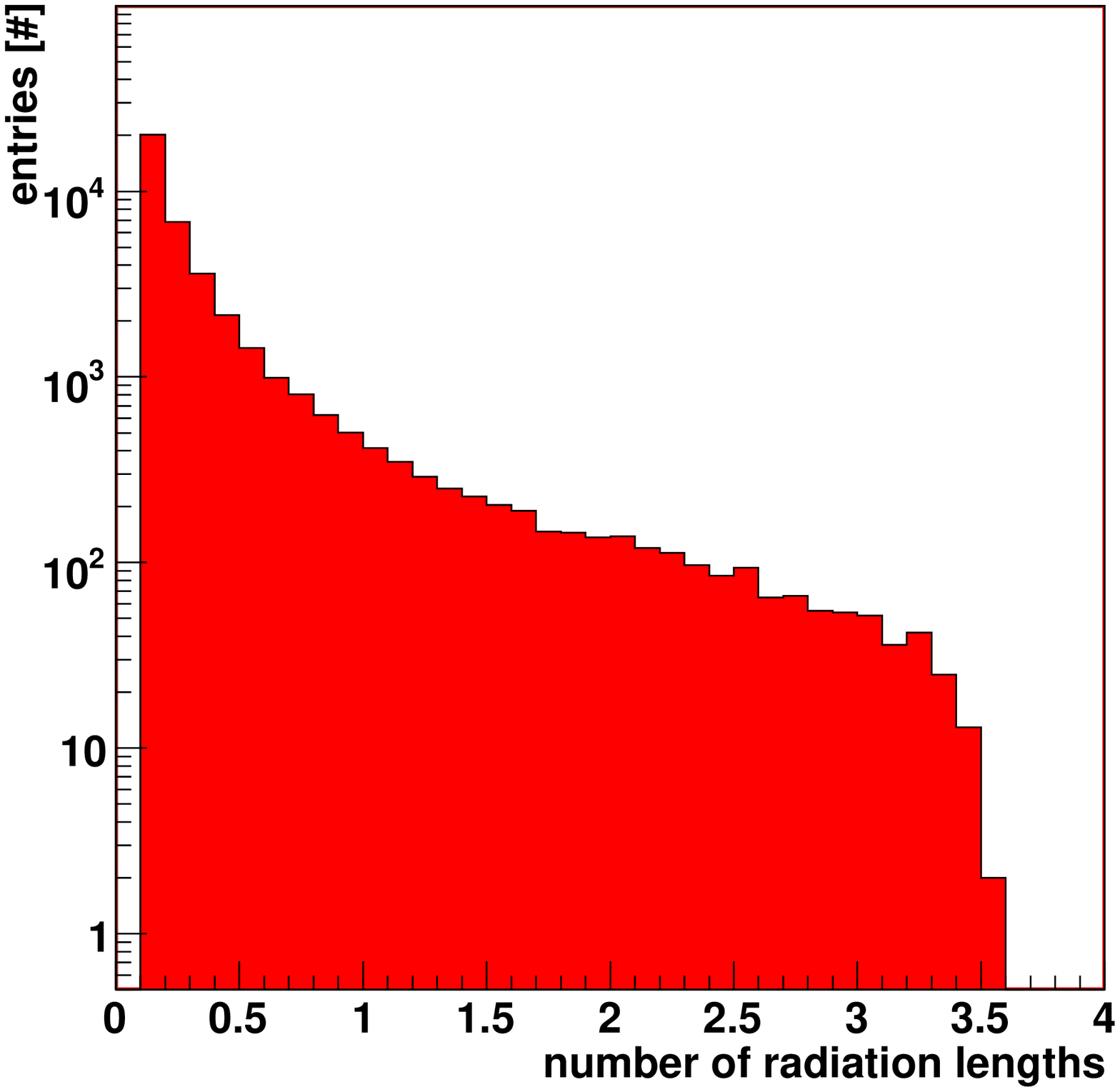,width=8cm}}
\captionof{figure}{\label{f-rad_length_southpole_40_-90_-75_0_360.dat_rad_length}Distribution of radiation lengths traversed before 40\,km altitude at the South Pole during the summer for an isotropic particle distribution.}
\end{minipage}
\end{center}
\end{figure}

The atmospheric depth a particle has to cross decreases nearly exponentially with altitude for perpendicular incident (0° zenith angle) such that the balloon should fly at the highest possible altitude (Fig.~\ref{f-atmosphere_depth}). At 40\,km the depth for perpendicular incident at the South Pole during summer is about 3.8\,g/cm$^2$. This is a non-negligible effect compared to the mean amount of matter traversed in the galaxy before entering the atmosphere (6 - 10\,g/cm$^2$). The radiation length $X_0$ can also be calculated from the composition and density of the atmosphere (Fig.~\ref{f-radiation_length_FtSumner_summer}) \cite{pdgbook}. Particles with 0° zenith angle traverse about 10\,\% of a radiation length before reaching 40\,km during the summer at the South Pole. It is obvious that the pathlength in the atmosphere and thus the radiation length increases at larger inclinations (Fig.~\ref{f-southpole_40_-89_0.01_0.01_zenithfixedPos.dat_cos_rad_length}). Several radiation lengths are possible for large zenith angles. The average number of radiation lengths can be calculated by simulating an isotropic particle distribution at 40\,km and tracing the particles back to space. Details of the calculation of the isotropic distribution can be found in Sec.~\ref{ss-atmosim}. A clear maximum can be found at small radiation lengths of about 10\,\%$X_0$ and the mean value is approximately 40\,\%$X_0$ (Fig.~\ref{f-rad_length_southpole_40_-90_-75_0_360.dat_rad_length}). Thus, electromagnetic showers cannot develop significantly in the atmosphere above an altitude of 40\,km. The particles resulting from atmospheric interactions can be interpreted to be most likely produced in the first interaction of the cosmic-ray particle with the atmosphere. This holds also for hadronic cascades as the corresponding nuclear mean free pathlength is about 5\,\% \cite{yoshida-2006}.

\subsection{Simulation of Cosmic-Ray Measurements in the Atmosphere\label{ss-atmosim}}

This section discusses features of the atmospheric simulation for New Mexico in September 2001 and South Pole in December 2005, respectively. The atmospheric depth is nearly the same in the arctic or antarctic summer at the corresponding pole. The geomagnetic cut-offs are very small at both poles. Thus the choice of pole does not affect the studies and further simulations have been carried out for the South Pole and would give very similar results at the North Pole. 

Starting positions and directions for particle trajectories at an altitude of 500\,km above Earth's surface are calculated such that an isotropic particle flux would be achieved in detection shells around the Earth without the atmospheric and the magnetic effects . Then the atmosphere and magnetic field is switched on for the simulation to study the effects. Protons, electrons, positrons, photons and antiprotons are simulated with energies between 0.1 and 450\,GeV for New Mexico and between 0.1 and 10,000\,GeV at the South Pole, respectively. The simulation of helium nuclei works only stably up to about 7.5\,GeV per nucleon due to the physics models implemented in GEANT4 for light ions. The cosmic-ray fluxes are taken from the reacceleration GALPROP model \cite{2006ApJ...642..902P} as shown in Fig.~\ref{f-fluxes_geo_mod_galprop_conv}. The signal from dark matter as calculated by DarkSUSY with the MSSM parameters (Eq.~\ref{eq-susypar}) is also included.

After comparisons between existing measurements to simulations of atmospheric secondaries particles, the simulation concentrates on cosmic-ray flux measurements with PEBS at the South Pole. Therefore, the detector properties are discussed because they have crucial impact on the quality of the measurement.

\subsubsection{Particle Positions and Directions for the Simulations}

\begin{figure}
\begin{center}
\centerline{\epsfig{file=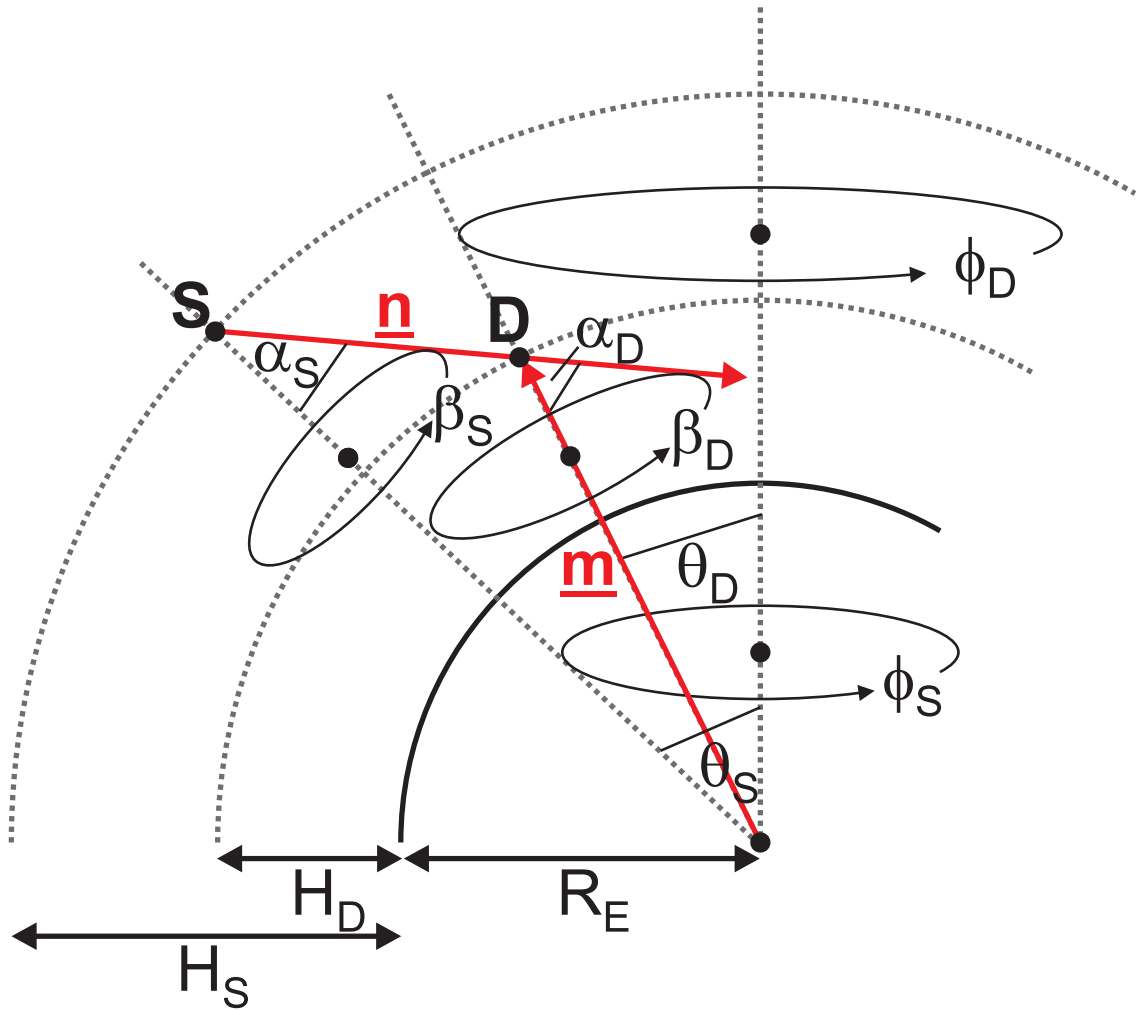,width=15cm}}
\captionof{figure}{\label{f-start_pos_calc}Definition of angles for the start position calculation.}
\end{center}
\end{figure}

The simulation described in the following assumes detection planes at several altitudes around the Earth. The particles will be started well above the atmosphere. The discussion of the properties of the atmosphere and the magnetic field (Sec.~\ref{s-gev}) showed that particle distributions play an important role in the atmospheric simulations. In order to keep the computation as fast as possible only particles crossing the desired detection planes in a certain range of latitude and longitude are simulated. The outline of the starting point calculation is described in the following. The goal is to derive the starting coordinate and direction of the particle trajectory at an altitude $H_S$ as a function of the detection coordinate and direction at altitude $H_D$. Fig.~\ref{f-start_pos_calc} gives an overview on the variables used for the calculation which are further described in Tab.~\ref{t-varposcalc}. 

\begin{figure}
\begin{center}
\captionof{table}{\label{t-varposcalc}Variables for the starting point calculation.}
\begin{tabular}{c|l}
\hline
\hline
Variable	& Description\\
\hline
$R_E$		& radius of the Earth\\
$H_D$		& detection altitude above the Earth surface\\
$H_S$		& starting altitude above the Earth surface\\
$\theta_S$	& geographical latitude of the starting position\\
$\phi_S$	& geographical longitude of the starting position\\
$\alpha_S$	& direction zenith angle at the starting position\\
$\beta_S$	& direction azimuth angle at the starting position\\
$\theta_D$	& geographical latitude of the detection position\\
$\phi_D$	& geographical longitude of the detection position\\
$\alpha_D$	& direction zenith angle at the detection position\\
$\beta_D$	& direction azimuth angle at the detection position\\
\hline
\end{tabular}
\end{center}
\end{figure}

The particle trajectory with the parameter $\tau$ crossing the detection altitude follows a straight line:
\be \vec x(H_D,\theta_D,\phi_D,\alpha_D,\beta_D) = (R_E+H_D)\cdot\vec m(\theta_D,\phi_D) +\tau\cdot\vec n_D(\theta_D,\phi_D,\alpha_D,\beta_D)\ee
where $\vec m(\theta_D,\phi_D)=[\cos\theta_D\cos\phi_D,\cos\theta_D\sin\phi_D,\sin\theta_D]$ is the unit vector at the position of the detector and $\vec n_D(\theta_D,\phi_D,\alpha_D,\beta_D)$ is the direction vector which can be derived with the following ansatz:
\begin{eqnarray} 
\vec n_D(\theta_D,\phi_D,\alpha_D,\beta_D) &=& \mathcal R(\theta_D,\phi_D, \beta_D)\cdot \vec z(\theta_D,\phi_D,\alpha_D)\\
\text{with }\;\vec z(\theta_D,\phi_D,\alpha_D)&=&\left(\begin{matrix}k_1(\theta_D,\alpha_D)\cos\phi_D \\ k_1(\theta_D,\alpha_D)\sin\phi_D\\k_2(\theta_D,\alpha_D)\end{matrix}\right).
\end{eqnarray}

The factors $k_1(\theta_D,\alpha_D)$ and $k_2(\theta_D,\alpha_D)$ can be calculated from the constraints:
\begin{eqnarray} 
k_1(\theta_D,\alpha_D)^2+k_2(\theta_D,\alpha_D)^2&=&1,\\
\vec m(\theta_D,\phi_D)\cdot\vec z(\theta_D,\phi_D,\alpha_D)&=&\cos\alpha_D. 
\end{eqnarray} 

Using the rotation matrix $\mathcal R(\theta_D,\phi_D, \beta_D)$, $\vec z(\theta_D,\phi_D,\alpha_D)$ is rotated in the next step by the azimuth angle $\beta_D$ around the geographical unit vector at the detection position $\vec m(\theta_D,\phi_D)=[m_1,m_2,m_3]$. The rotation matrix is:
\be \mathcal R(\theta_D,\phi_D, \beta_D) =
\left(\begin{matrix} 
c+m_1^2(1-c) 		& m_1 m_2(1-c)-m_3 s 	& m_1 m_3(1-c)+m_2 s \\ 
m_2 m_1(1-c)+m_3 s 	& c+m_2^2(1-c) 		& m_2 m_3(1-c)-m_1 s \\ 
m_3 m_1(1-c)-m_2 s 	& m_3 m_2(1-c)+m_1 s 	& c+m_3^2(1-c)
\end{matrix}\right)\ee

with $s=\sin\beta_D$ and $c=\cos\beta_D$. The calculation of the intersection between the particle trajectory and the starting plane determines the starting latitude $\theta_S$ and longitude $\phi_S$ for a certain set of zenith and azimuth angles $\alpha_D$ and $\beta_D$ and detection position $(\theta_D,\phi_D)$. 

The simulation requires that the direction vector of the trajectory must be transformed to the coordinate system having the starting position as origin. This is done with a rotation by $-\phi_S$ around the $z$ axis and a rotation by $\theta_S$ around the $y$ axis:

\be
\left(\begin{matrix}\cos\theta_S\cos\phi_S \\ \cos\theta_S\sin\phi_S\\\sin\theta_S\end{matrix}\right)
=\left(\begin{matrix} 
-\sin\theta_S 	& 0 		& \cos\theta_S \\ 
0 		& 1 		& 0 \\ 
-\cos\theta_S 	& 0 		& -\sin\theta_S
\end{matrix}\right)
\left(\begin{matrix} 
\cos(-\phi_S) 	& -\sin(-\phi_S)& 0 \\ 
\sin(-\phi_S)	& \cos(-\phi_S)	& 0 \\ 
0		& 0 		& 1
\end{matrix}\right)\cdot\vec n_D(\theta_D,\phi_D,\alpha_D,\beta_D).
\ee

\begin{figure}
\begin{center}
\begin{minipage}[b]{.4\linewidth}
\centerline{\epsfig{file=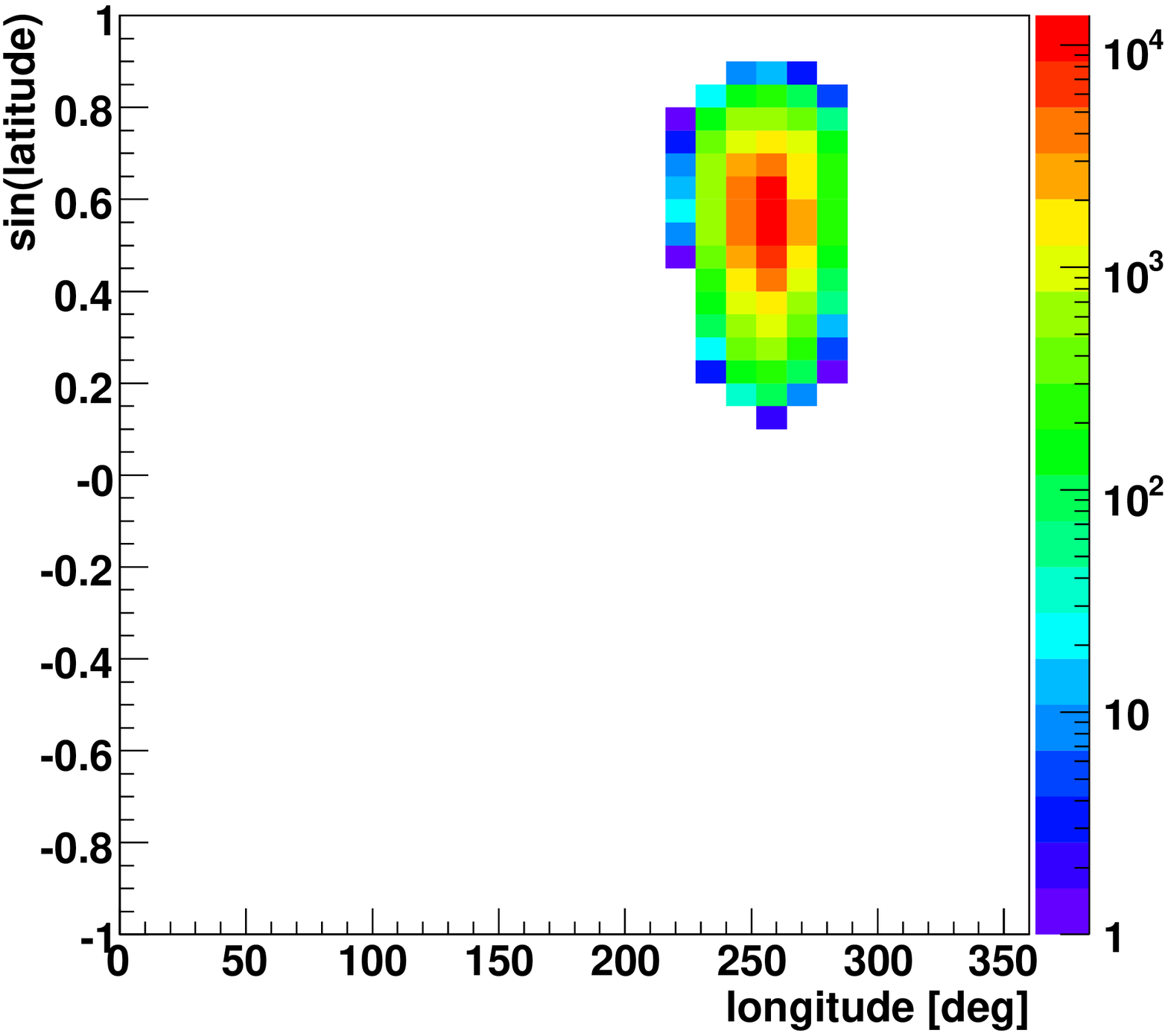,width=8cm}}
\captionof{figure}{\label{f-500_40_29_39_251_261.dat_sinlat_long}Starting positions at 500\,km altitude for simulations at Ft. Sumner, New Mexico. The color code on the right shows the number of simulated entries.}
\end{minipage}
\hspace{.1\linewidth}
\begin{minipage}[b]{.4\linewidth}
\centerline{\epsfig{file=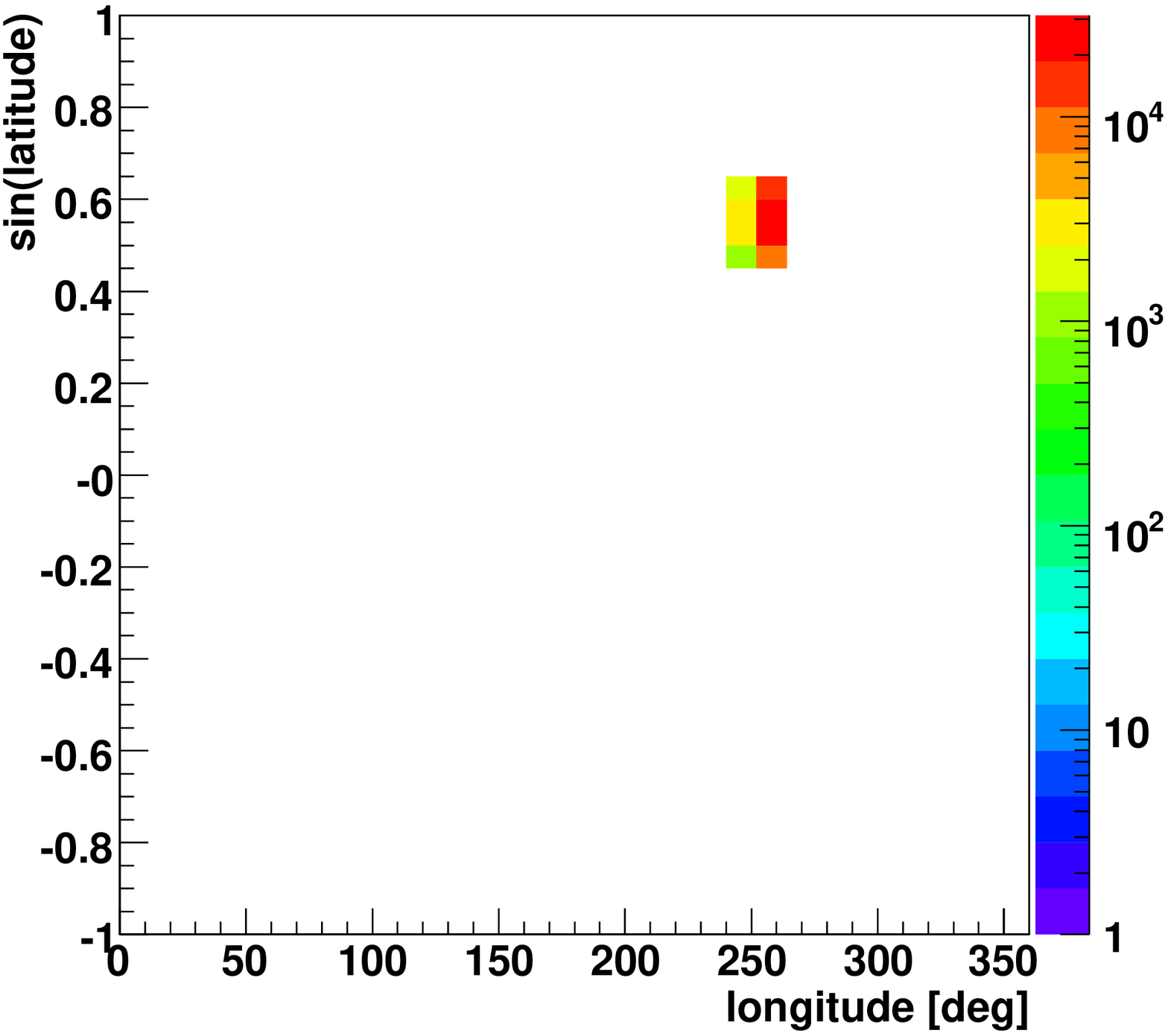,width=8cm}}
\captionof{figure}{\label{f-comp_40_500_40_29_39_251_261.dat_sinlat_long}Detection positions of tracks at 40\,km altitude for simulations at Ft. Sumner, New Mexico. The color code on the right shows the number of simulated entries.}
\end{minipage}
\end{center}
\end{figure}

\begin{figure}
\begin{center}
\begin{minipage}[b]{.4\linewidth}
\centerline{\epsfig{file=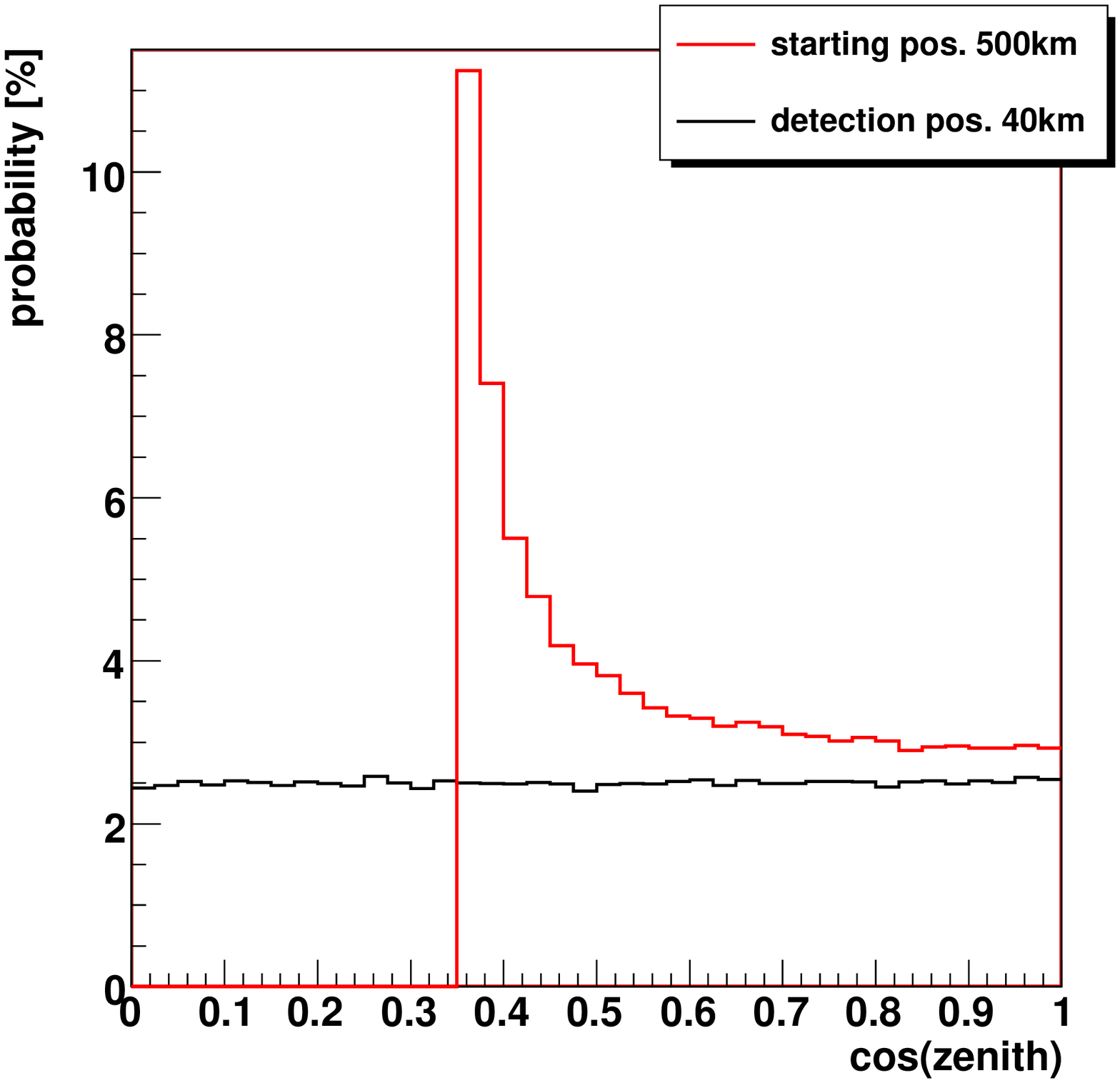,width=8cm}}
\captionof{figure}{\label{f-zenith_startpos}Probability distribution of zenith angles at the starting and detection positions.}
\end{minipage}
\hspace{.1\linewidth}
\begin{minipage}[b]{.4\linewidth}
\centerline{\epsfig{file=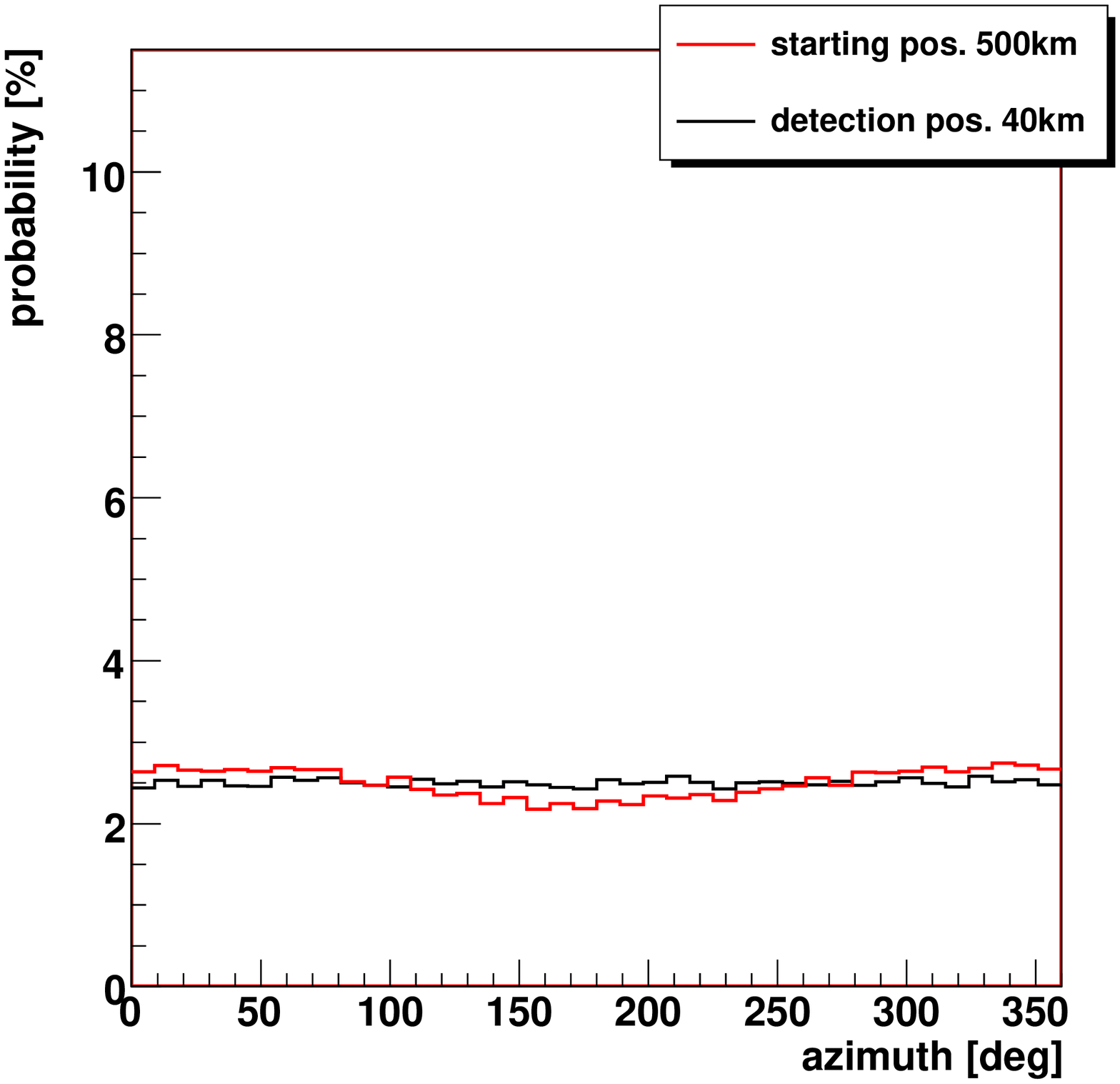,width=8cm}}
\captionof{figure}{\label{f-azimuth_startpos}Probability distribution of azimuth angles at the starting and detection positions.}
\end{minipage}
\end{center}
\end{figure}
\begin{figure}
\begin{center}
\centerline{\epsfig{file=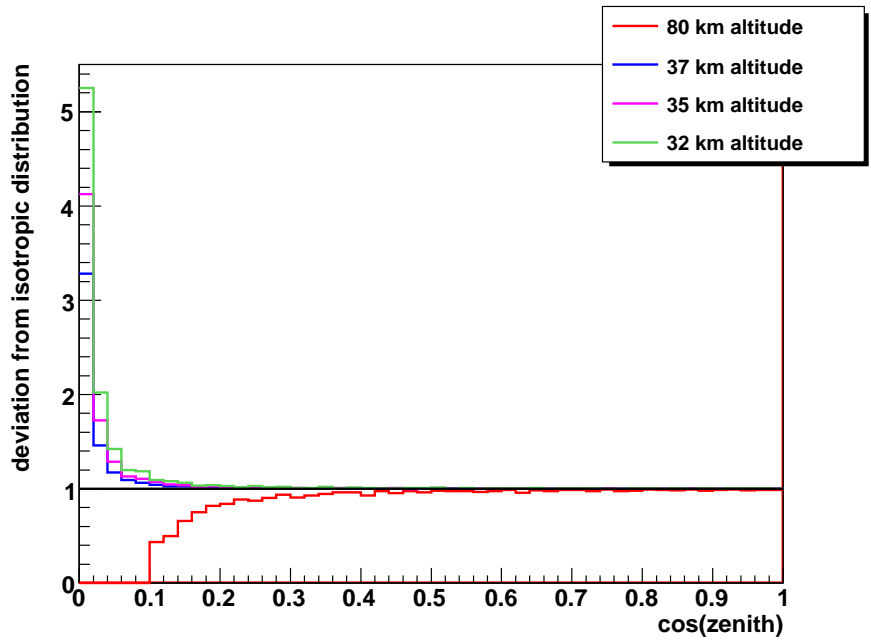,width=12cm}}
\captionof{figure}{\label{f-isotropic_scaling}Rescaling factors for an isotropic particle distribution at altitudes other than 40\,km.}
\end{center}
\end{figure}

For the simulation the coordinates at the detection altitude are isotropically distributed such that $\sin\theta_D$ and $\phi_D$ are uniformly distributed. For the starting altitude of 500\,km in New Mexico, Fig.~\ref{f-500_40_29_39_251_261.dat_sinlat_long} shows the distribution of the geographic starting positions which is wider in both latitude and longitude than for the detection altitude at 40\,km (Fig.~\ref{f-comp_40_500_40_29_39_251_261.dat_sinlat_long}).

The direction angles at the detection altitude are also isotropically distributed such that $\cos\theta_D$ and $\phi_D$ are uniformly distributed. The zenith angle direction distribution of the particles at the starting position at 500\,km altitude has a sharp peak at about $\cos\alpha_D=0.35$ ($\alpha_D\approx65$°) and is translated to the uniform isotropic distribution at the detection position at 40\,km altitude (Fig.~\ref{f-zenith_startpos}). Particles with zenith angles larger than 65° at the starting altitude cannot hit the detection altitude. Only a small modulation of the azimuth angle distribution in the starting plane compared to the uniform behavior in the detection plane is seen (Fig.~\ref{f-azimuth_startpos}). The range of the simulations is 29°-39° latitude and 251°-261° longitude for New Mexico and -90°-75° latitude and 0°-360° longitude for the South Pole, respectively.

The simulation is optimized for detection at 40\,km altitude such so the zenith angle distributions for other altitudes are not isotropic. At altitudes smaller than 40\,km not enough particles are generated at large zenith angles (small $cos\alpha_D$). At altitudes larger than 40\,km no particles are detected at large zenith angles down to altitude dependent cut-off zenith angles. Below the cut-off angle too many particles are detected. As noted above, the zenith angle has a large influence on the number of radiation lengths (Fig.~\ref{f-southpole_40_-89_0.01_0.01_zenithfixedPos.dat_cos_rad_length}) and to achieve isotropic distributions at all altitudes events must be rescaled corresponding to altitude and zenith angle (Fig.~\ref{f-isotropic_scaling}). Correction over the whole zenith angle range is possible for altitudes smaller than 40\,km but for larger altitudes only up to the cut-off zenith angle.

\subsubsection{Comparison between Simulations and Measurements}

\begin{figure}
\begin{center}
\begin{minipage}[b]{.4\linewidth}
\centerline{\epsfig{file=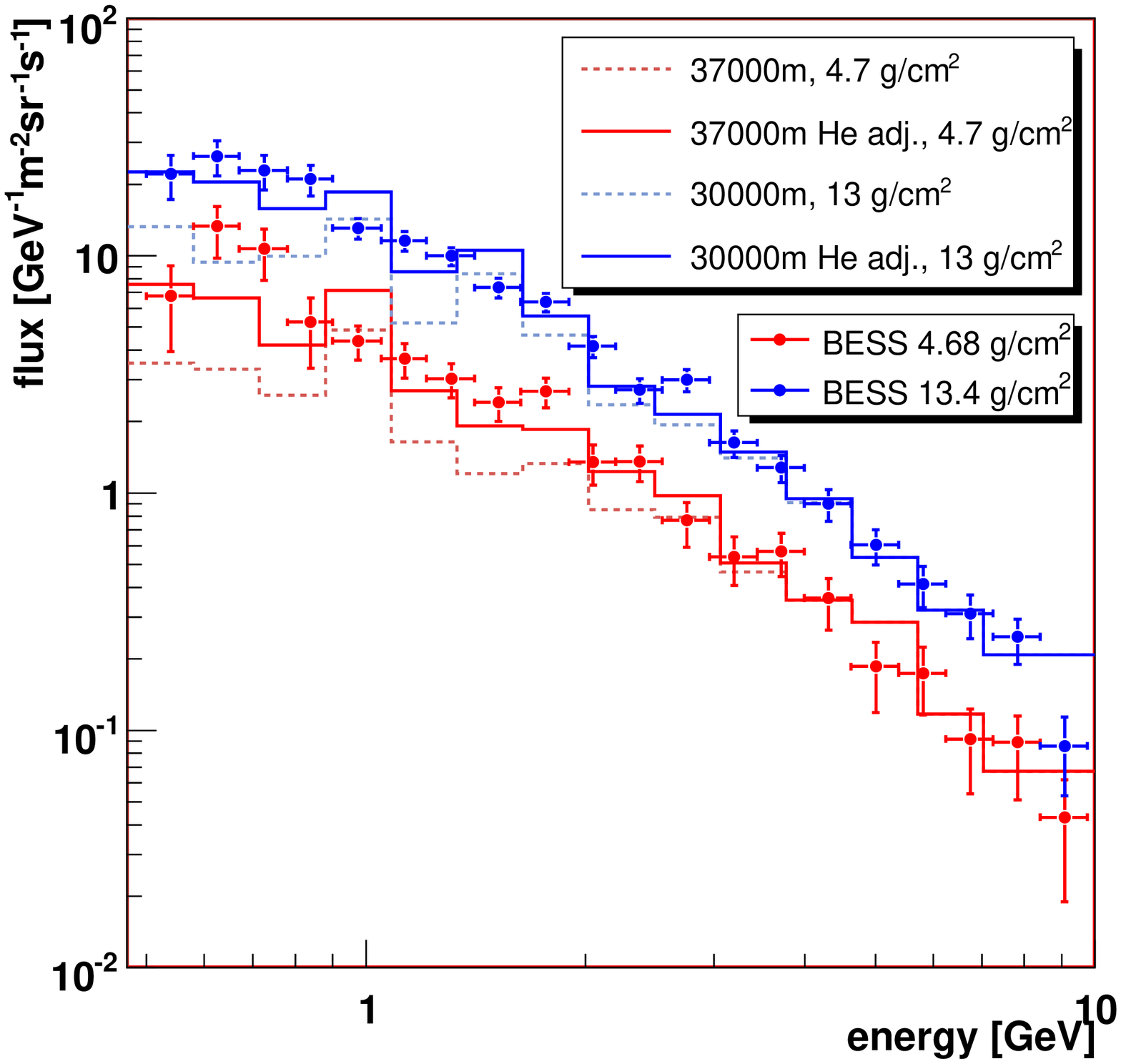,width=8cm}}
\captionof{figure}{\label{f-fluxes_compare_13_ftsumner}Comparison of simulated and measured muon fluxes.}
\end{minipage}
\hspace{.1\linewidth}
\begin{minipage}[b]{.4\linewidth}
\centerline{\epsfig{file=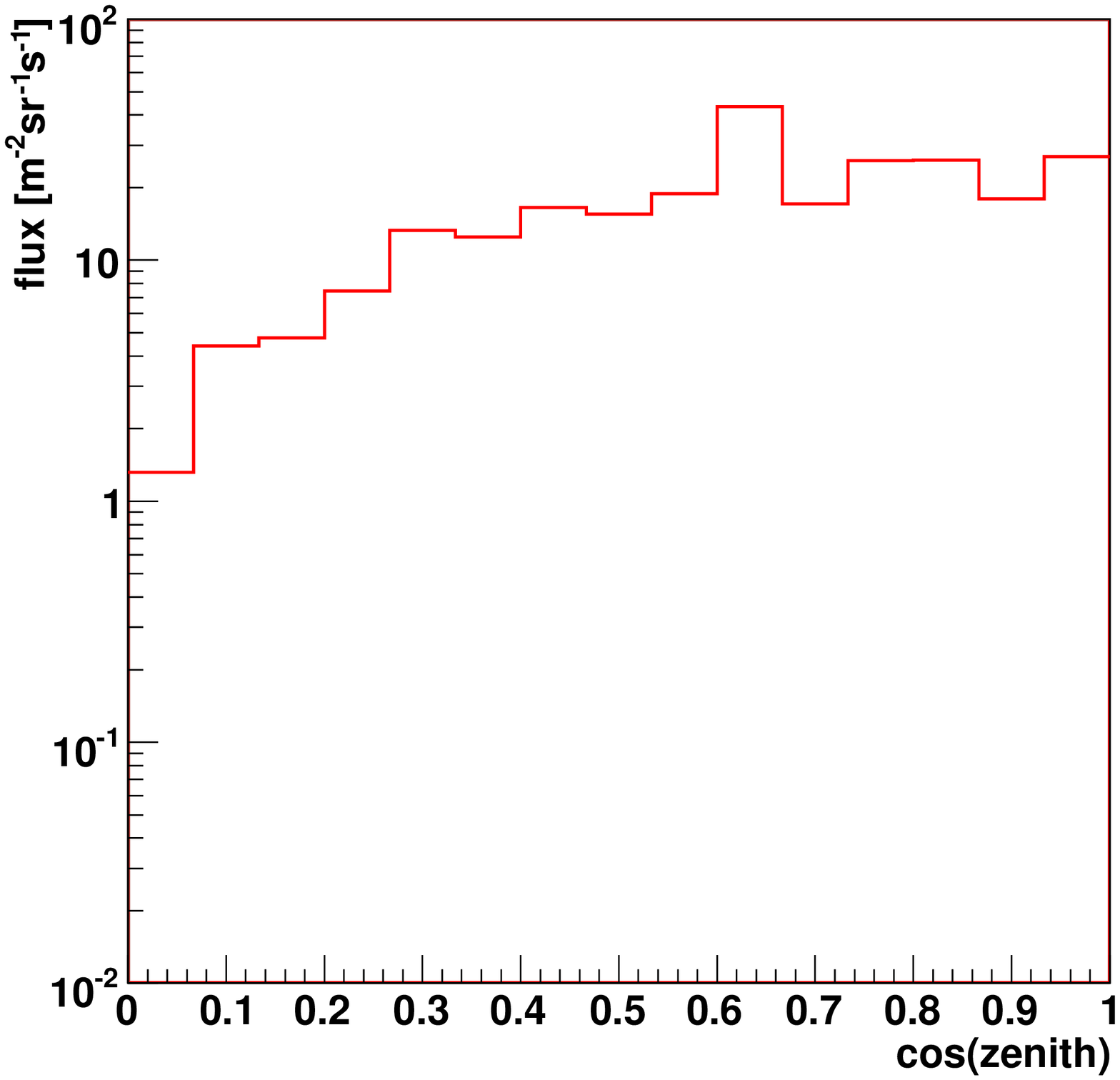,width=8cm}}
\captionof{figure}{\label{f-zenith_37000_m_13_ftsumner}Angular distribution of muon flux in 37000\,m altitude.}
\end{minipage}
\end{center}
\end{figure}

As aforementioned the particles were started at 500\,km altitude and only particles with rigidities larger than the geomagnetic cut-off were simulated (Fig.~\ref{f-cutoff_az_zen_34_256}). In addition, a solar modulation parameter of $\Phi=440$\,MV for New Mexico and $\Phi=550$\,MV for the South Pole was used. Several comparisons of the simulation to measurements and other models were carried out to validate the simulations.  

A comparison of the atmospheric muon flux measured by the BESS experiment \cite{abe-2007-645} for two different atmospheric depths with the fluxes simulated for New Mexico is shown in Fig.~\ref{f-fluxes_compare_13_ftsumner}. The dashed lines show the muon flux from the simulation without adjustment. For both depths the simulation lies below the data at lower energies. This is because GEANT4 works stably only up to a few GeV per nucleon for helium nuclei such so the secondary muon flux caused by helium is too small. A correction for this effect is estimated by comparing cosmic proton induced muon fluxes of different energy ranges. The cosmic-ray proton flux in the energy range of 0.1 - 450\,GeV give about a factor of 5 higher total atmospheric muon flux than protons with 0.1 - 30\,GeV. In the following all secondary particle fluxes induced by helium are multiplied by a factor 5. This effect is important for particle energies up to about 5\,GeV. The adjusted fluxes (solid lines) at 30\,km (13.0\,g/cm$^2$) and 37\,km (4.7\,g/cm$^2$) agree on average within about 15\,\% with the measurement. The simulated zenith angle distribution for muons at 37\,km altitude shows a nearly isotropic shape up to zenith angles of 75° ($\cos(\text{zenith})\approx0.3$) (Fig.~\ref{f-zenith_37000_m_13_ftsumner}). A measurement of the zenith angle distribution of muons would be an interesting test of atmospheric models and would be important for precise cosmic-ray background determinations.

\begin{figure}
\begin{center}
\begin{minipage}[b]{.4\linewidth}
\centerline{\epsfig{file=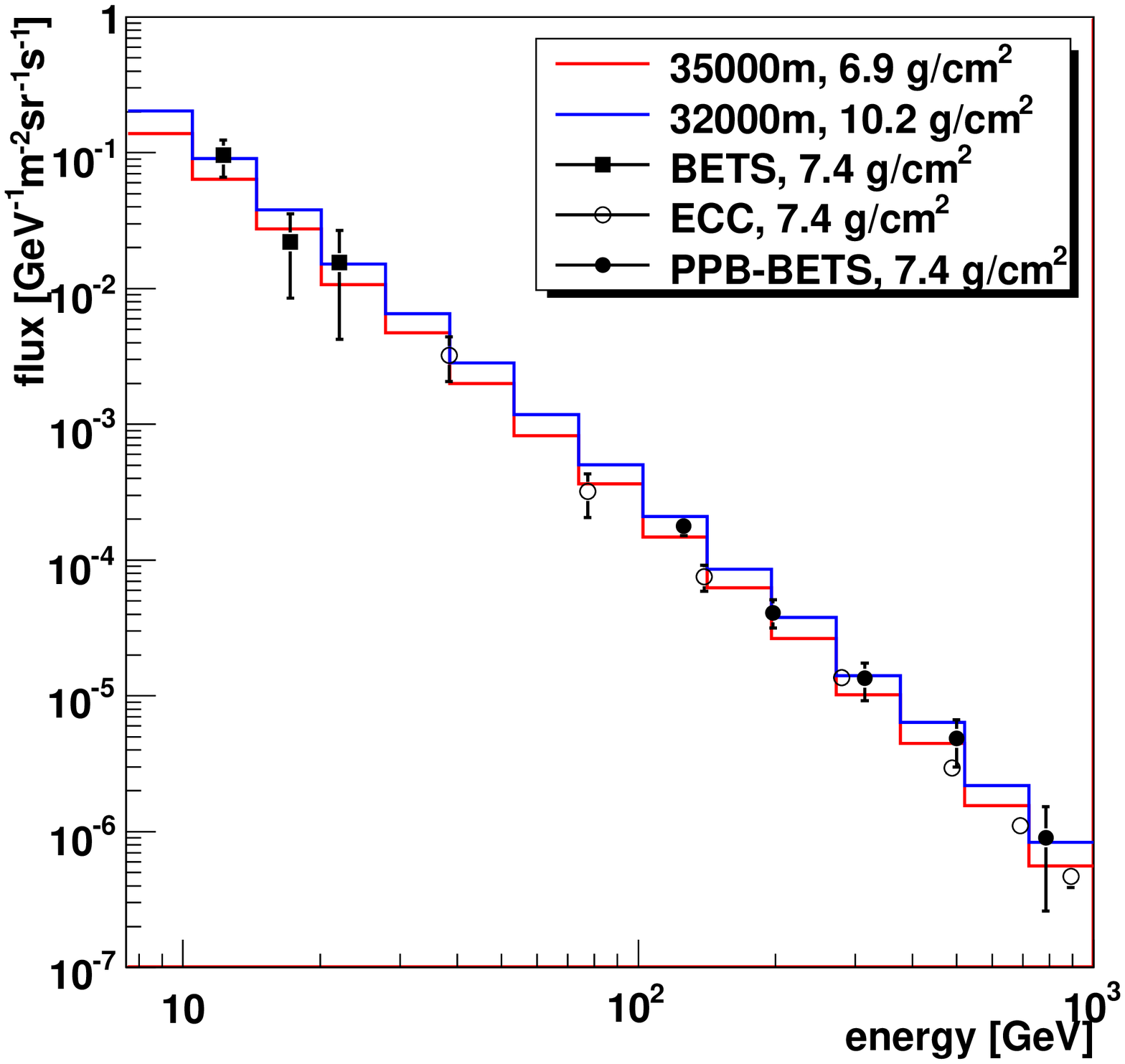,width=8cm}}
\captionof{figure}{\label{f-gamma_compare}Comparison of photon flux measurements and simulations.}
\end{minipage}
\hspace{.1\linewidth}
\begin{minipage}[b]{.4\linewidth}
\centerline{\epsfig{file=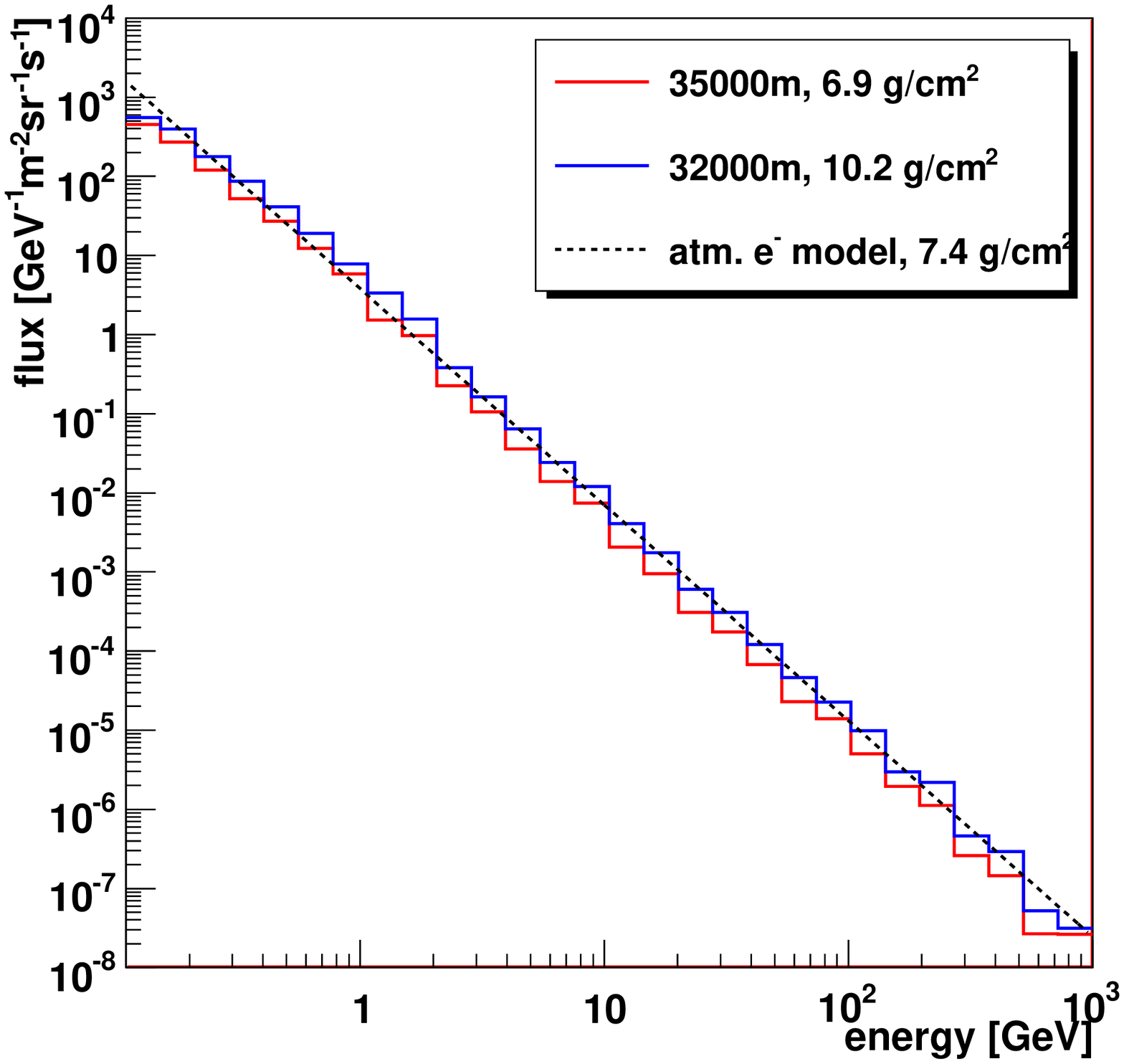,width=8cm}}
\captionof{figure}{\label{f-electron_compare}Comparison of secondary electron flux simulations and an analytical model.}
\end{minipage}
\end{center}
\end{figure}

In addition, the comparison of the observed atmospheric photon spectra at an atmospheric depth of 7.4\,g/cm$^2$ at the South Pole with the simulations at 32\,km (10.2\,g/cm$^2$) and 35\,km (6.9\,g/cm$^2$) at the South Pole shows good agreement within the errors (Fig.~\ref{f-gamma_compare}). In the next step, a model for the atmospherically induced electron flux used for the PPB-BETS experiment is compared to the simulated electron fluxes from atmospheric interactions \cite{nishimura-1980,yoshida-2006,torii-2008}:
\be F\sub{atm. $e^-$, PPB-BETS}=1.32\cdot10^{-5}\left(\frac E{100\text{\,GeV}}\right)^{-2.73}\,\text{GeV$^{-1}$m$^{-2}$sr$^{-1}$s$^{-1}$}\;\text{@ 7.4\,g/cm$^2$}.\ee
The agreement between the model and the simulation is again very well (Fig.~\ref{f-electron_compare}).

These comparisons are used for an estimation of the systematic uncertainty of the atmospheric simulations for the following analysis. The systematic error is assumed to be 15\,\% which is the same error that was assumed by the BESS collaboration \cite{bess-1995}.

\subsubsection{Particle Fluxes at the South Pole}

Following the good agreement with the atmospheric data, the simulation at the South Pole for the PEBS experiment were carried out using the parameters for the atmosphere and magnetic field for December 2005 and a solar modulation parameter of $\Phi=550$\,MV. Geomagnetic cut-offs are small at the poles and therefore neglected in the following.

\begin{figure}
\begin{center}
\begin{minipage}[b]{.4\linewidth}
\centerline{\epsfig{file=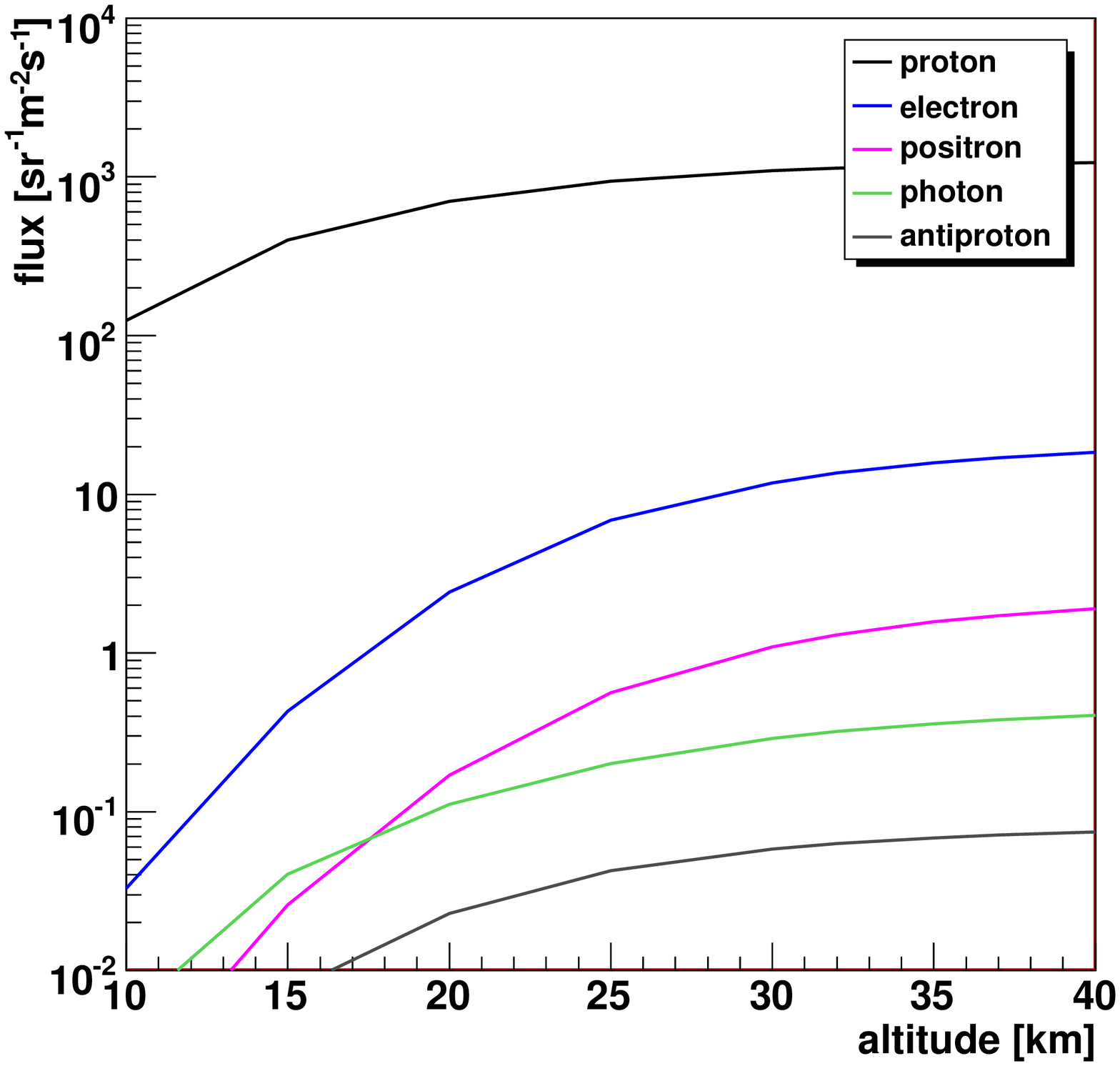,width=8cm}}
\captionof{figure}{\label{f-primary_flux_altitude_southpole}Altitude dependence of the attenuation of primary cosmic rays in Earth's atmosphere, integrated over the energy range 1 to 130\,GeV.}
\end{minipage}
\hspace{.1\linewidth}
\begin{minipage}[b]{.4\linewidth}
\centerline{\epsfig{file=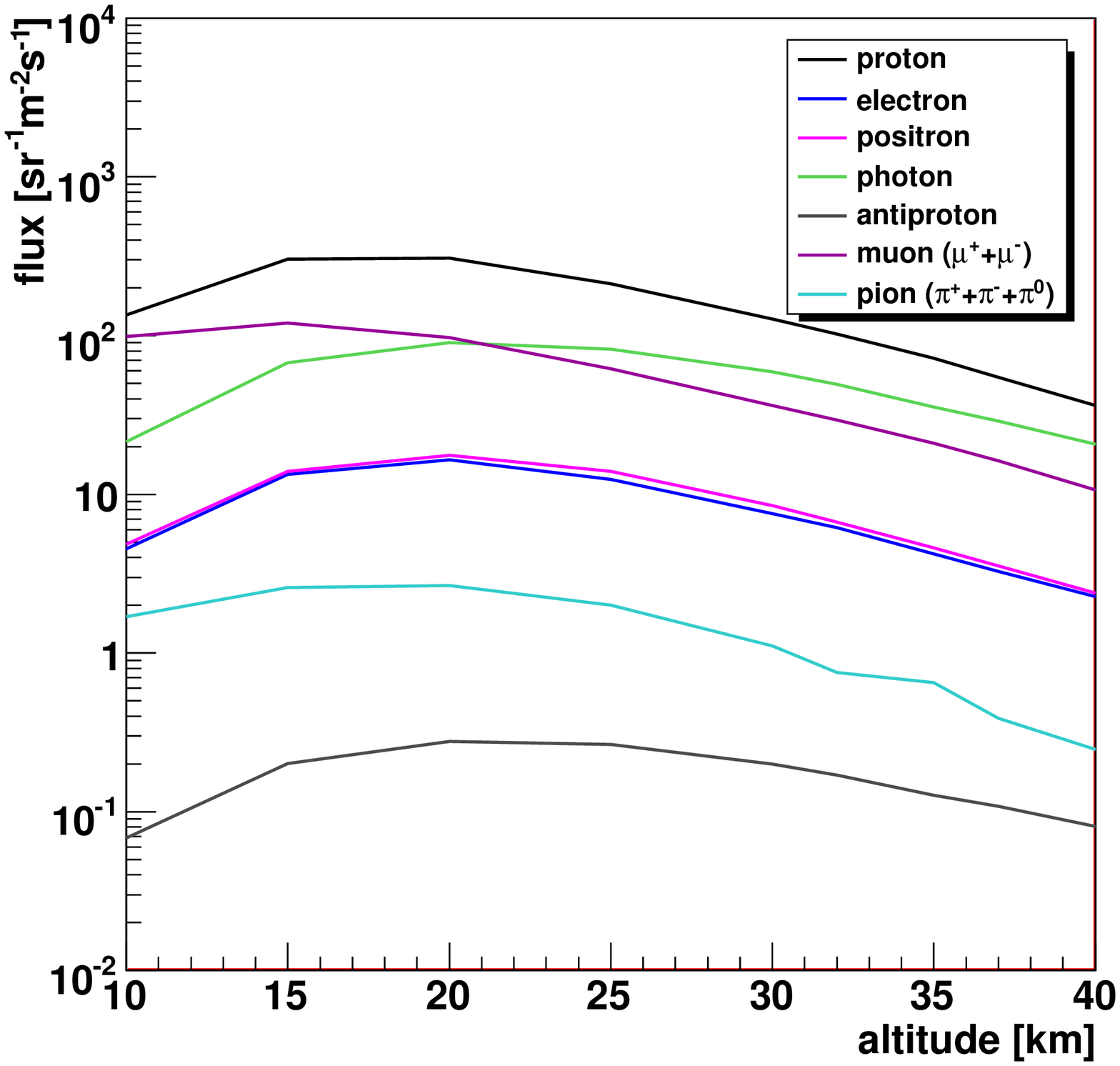,width=8cm}}
\captionof{figure}{\label{f-secondary_flux_altitude_southpole}Altitude dependence of secondary particle fluxes in Earth's atmosphere, integrated over the energy range 1 to 130\,GeV.}
\end{minipage}
\end{center}
\end{figure}

Cosmic-ray fluxes integrated over energy can be used for a first understanding of atmospheric influences on cosmic-ray antiparticle measurements in the atmosphere. The interactions with the atmosphere cause an attenuation of the primary\footnote{Here, the primary cosmic-ray fluxes refer to the fluxes at the top of the atmosphere and not to the fluxes at the source of some astrophysical object.} cosmic-ray particles (Fig.~\ref{f-primary_flux_altitude_southpole}). The atmospheric secondary fluxes increase down to an altitude of 15 to 20\,km. At that point the attenuation in the atmosphere starts to dominate the production of secondaries and the secondary fluxes start to decrease again (Fig.~\ref{f-secondary_flux_altitude_southpole}). The secondary fluxes show large contributions by muons and pions. For the extraction of the antiproton flux, muons and pions are an important source of background and their large abundances require good discrimination against them. In addition to atmospheric positron background, the positron measurement is contaminated by misidentified protons, as described later in the discussion of the detector performance. The separation of cosmic photons from the huge amount of atmospheric photons is not possible. Therefore, photon measurements are not further discussed.

\begin{figure}
\begin{center}
\begin{minipage}[b]{.4\linewidth}
\centerline{\epsfig{file=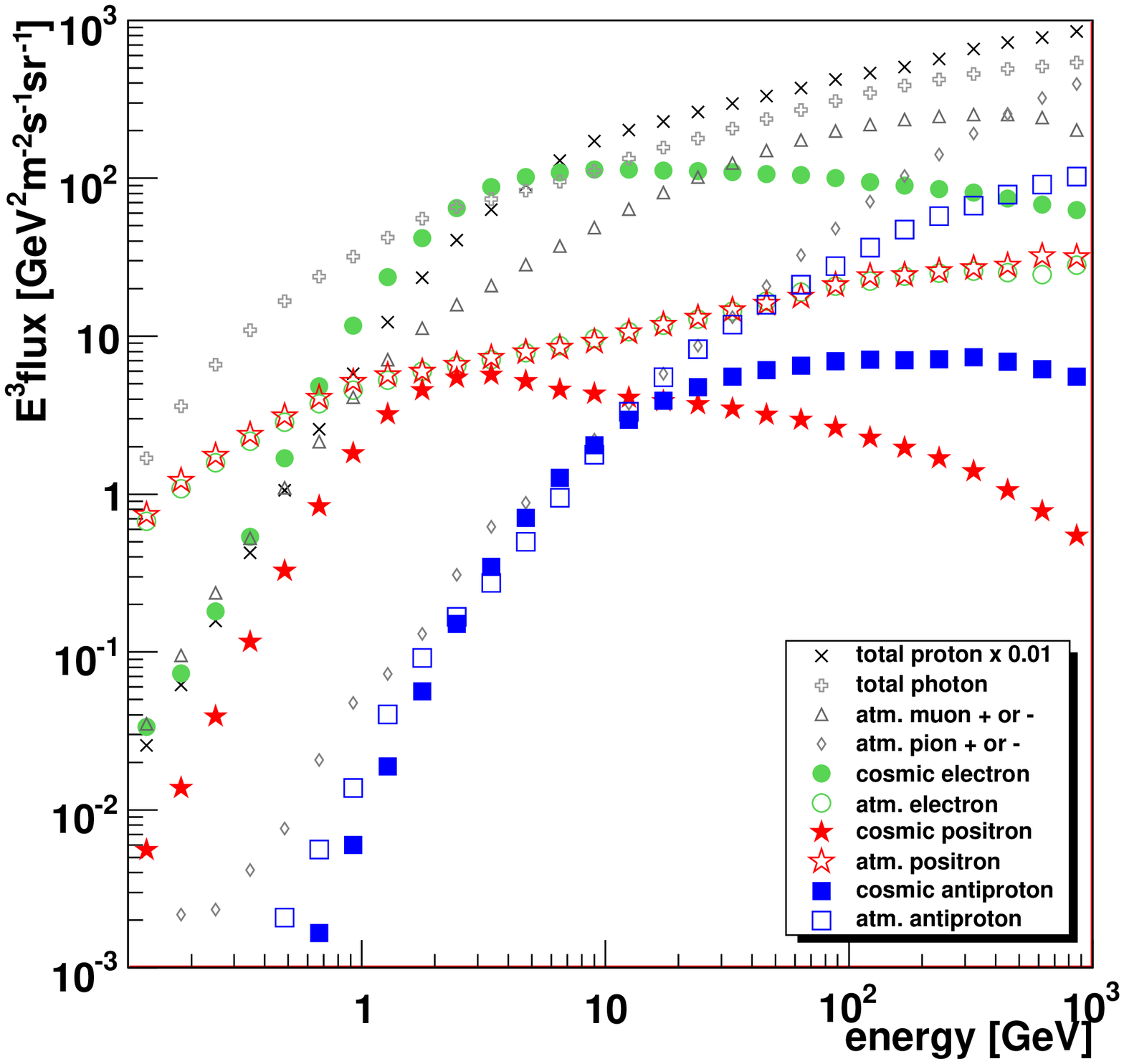,width=8cm}}
\captionof{figure}{\label{f-total_flux_40km_no_det}Fluxes (smoothed) at 40\,km altitude (3.8\,g/cm$^2$) at the South Pole.}
\end{minipage}
\hspace{.1\linewidth}
\begin{minipage}[b]{.4\linewidth}
\centerline{\epsfig{file=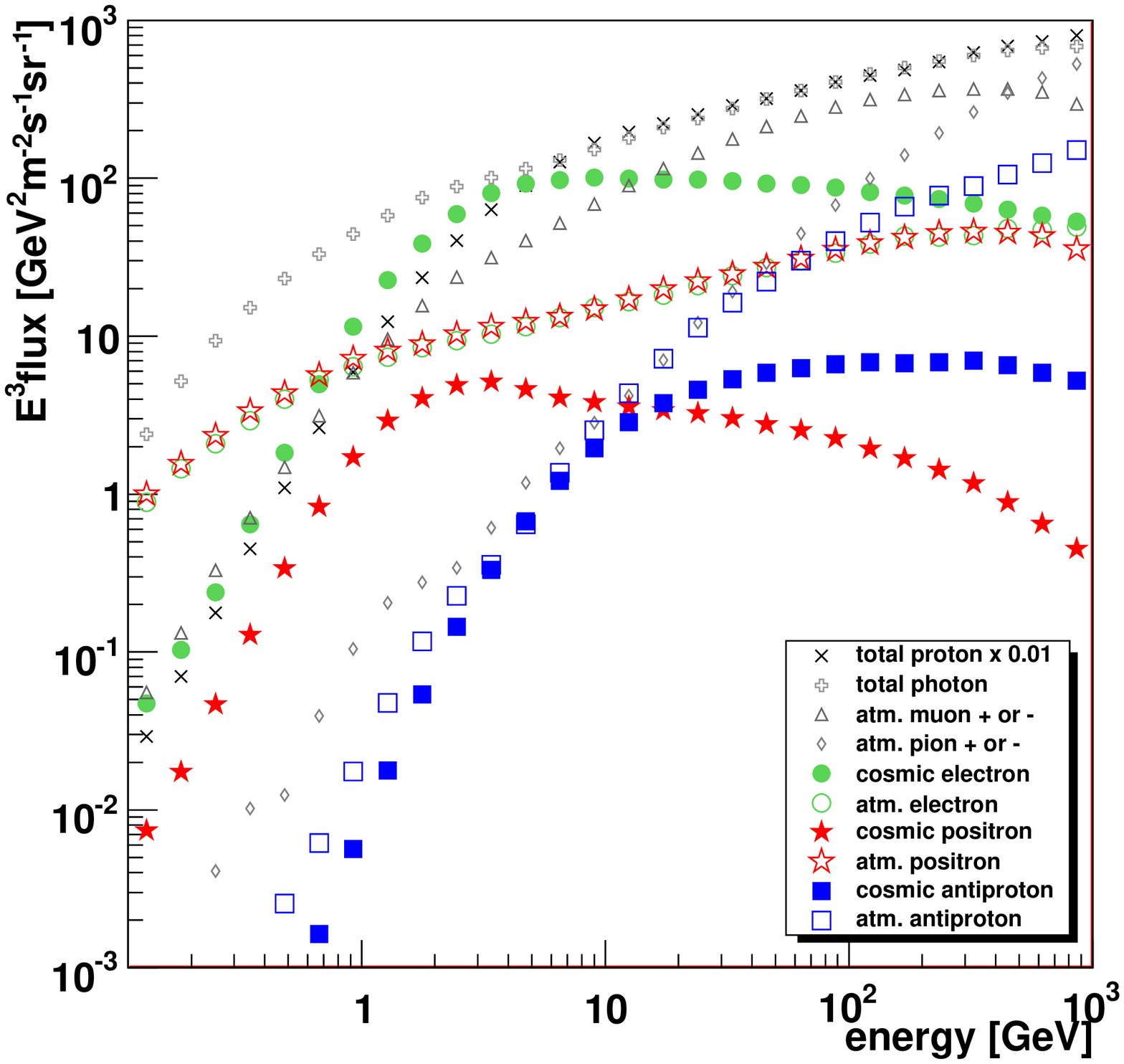,width=8cm}}
\captionof{figure}{\label{f-total_flux_37km_no_det}Fluxes (smoothed) at 37\,km altitude (5.4\,g/cm$^2$) at the South Pole.}
\end{minipage}
\end{center}
\end{figure}

Fig.~\ref{f-total_flux_40km_no_det} and \ref{f-total_flux_37km_no_det} show for electrons, positrons and antiprotons the cosmic and atmospheric energy differential fluxes separately and for the other particle types the sum of cosmic and atmospheric contributions at 40\,km and 37\,km altitude, respectively. As expected, the atmospherically induced fluxes increase from 40\,km to 37\,km because of the larger atmospheric depth. Cosmic antiprotons and positrons show large atmospheric backgrounds and the production of atmospheric electrons and positrons is approximately symmetric.

\begin{figure}
\begin{center}
\begin{minipage}[b]{.4\linewidth}
\centerline{\epsfig{file=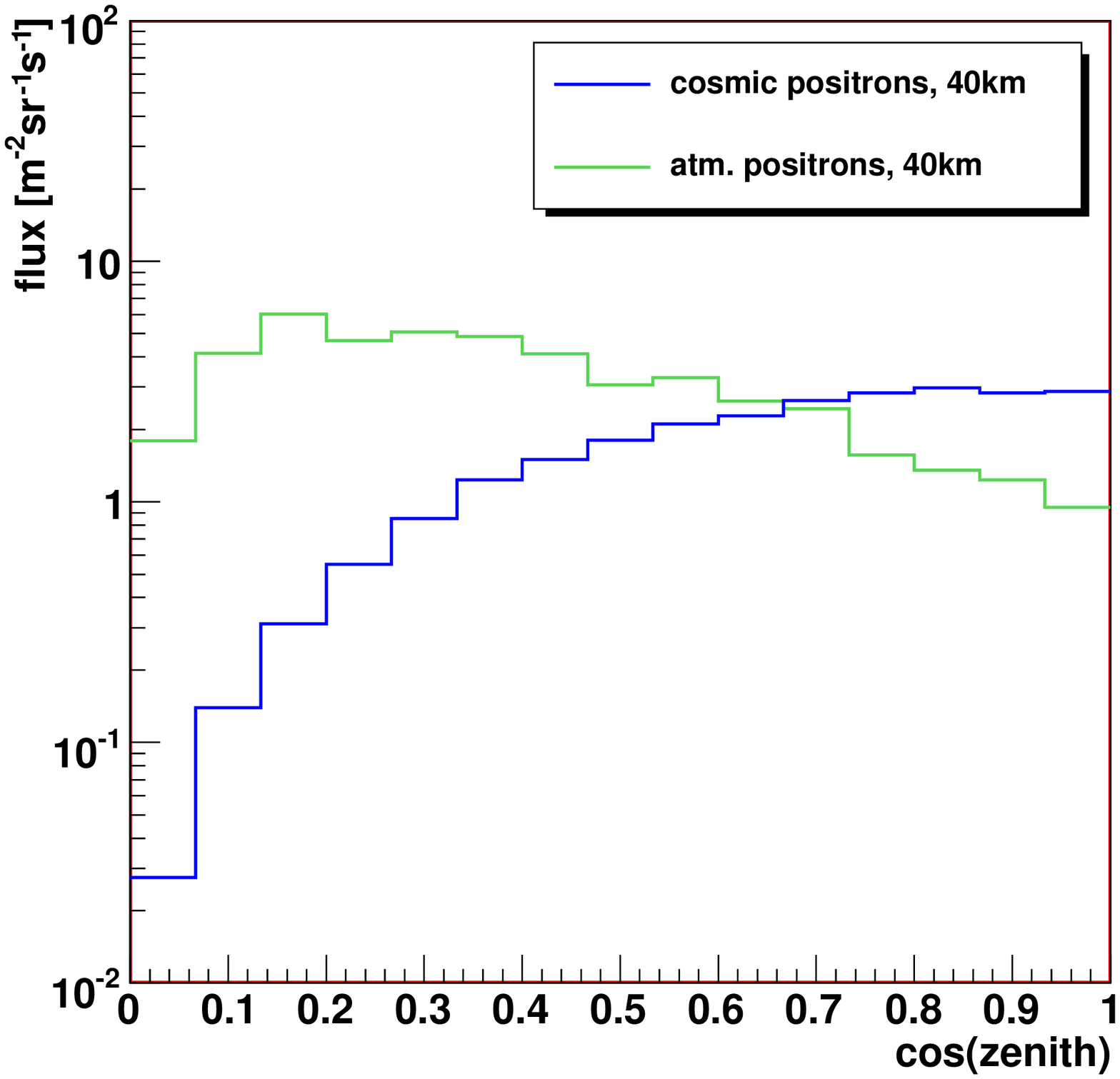,width=8cm}}
\captionof{figure}{\label{f-zenith_40000_m_-11_southpole}Primary and secondary positron fluxes at 40\,km altitude with respect to zenith angle, integrated over the energy range 1 to 130\,GeV.}
\end{minipage}
\hspace{.1\linewidth}
\begin{minipage}[b]{.4\linewidth}
\centerline{\epsfig{file=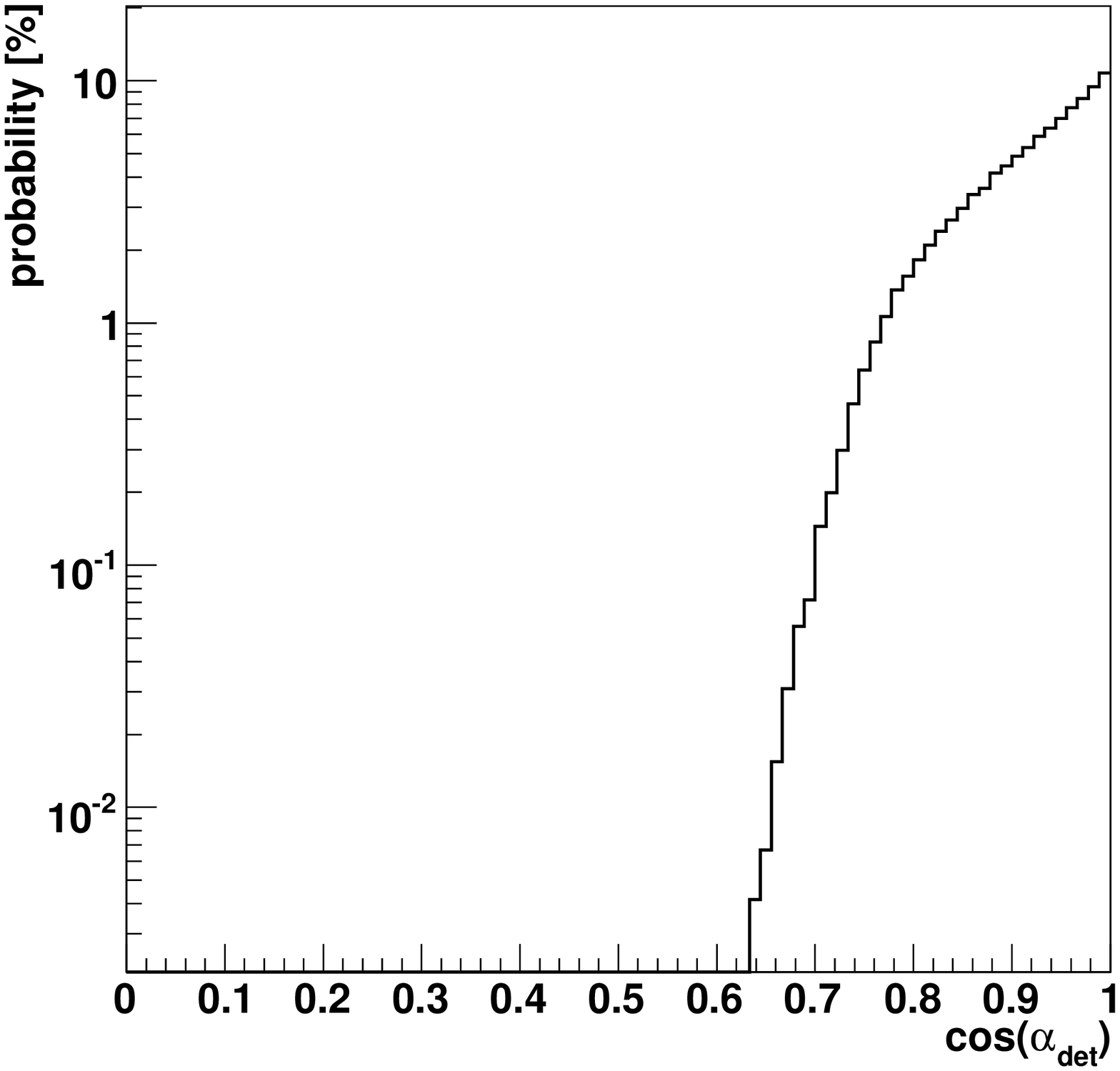,width=8cm}}
\captionof{figure}{\label{f-pebs_angular_res_cos_0804}Angular acceptance of PEBS.}
\end{minipage}
\end{center}
\end{figure}

\begin{figure}
\begin{center}
\begin{minipage}[b]{.4\linewidth}
\centerline{\epsfig{file=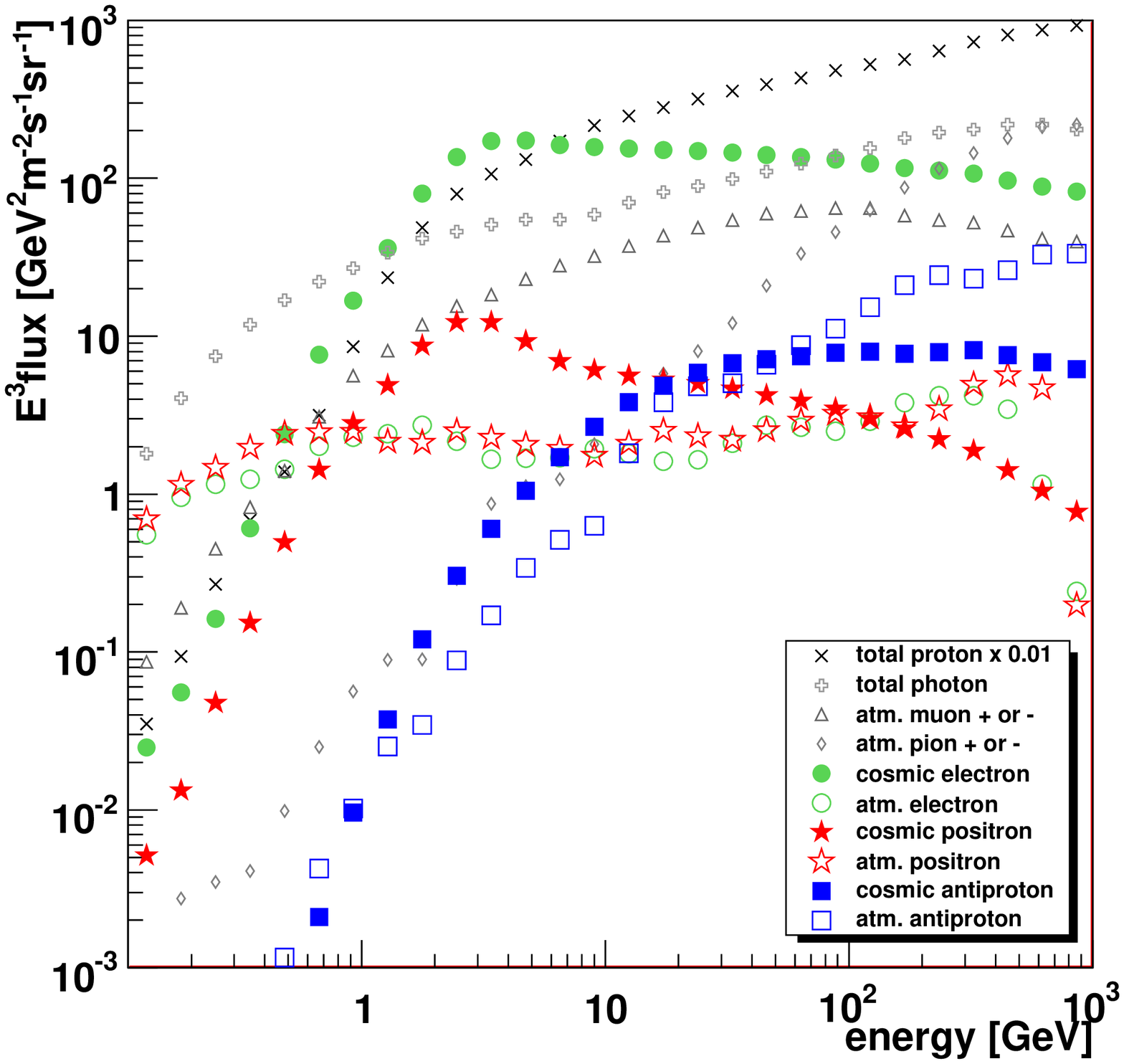,width=8cm}}
\captionof{figure}{\label{f-total_flux_40km}Fluxes (smoothed) at 40\,km altitude (3.8\,g/cm$^2$) at the South Pole respecting the detector acceptance.}
\end{minipage}
\hspace{.1\linewidth}
\begin{minipage}[b]{.4\linewidth}
\centerline{\epsfig{file=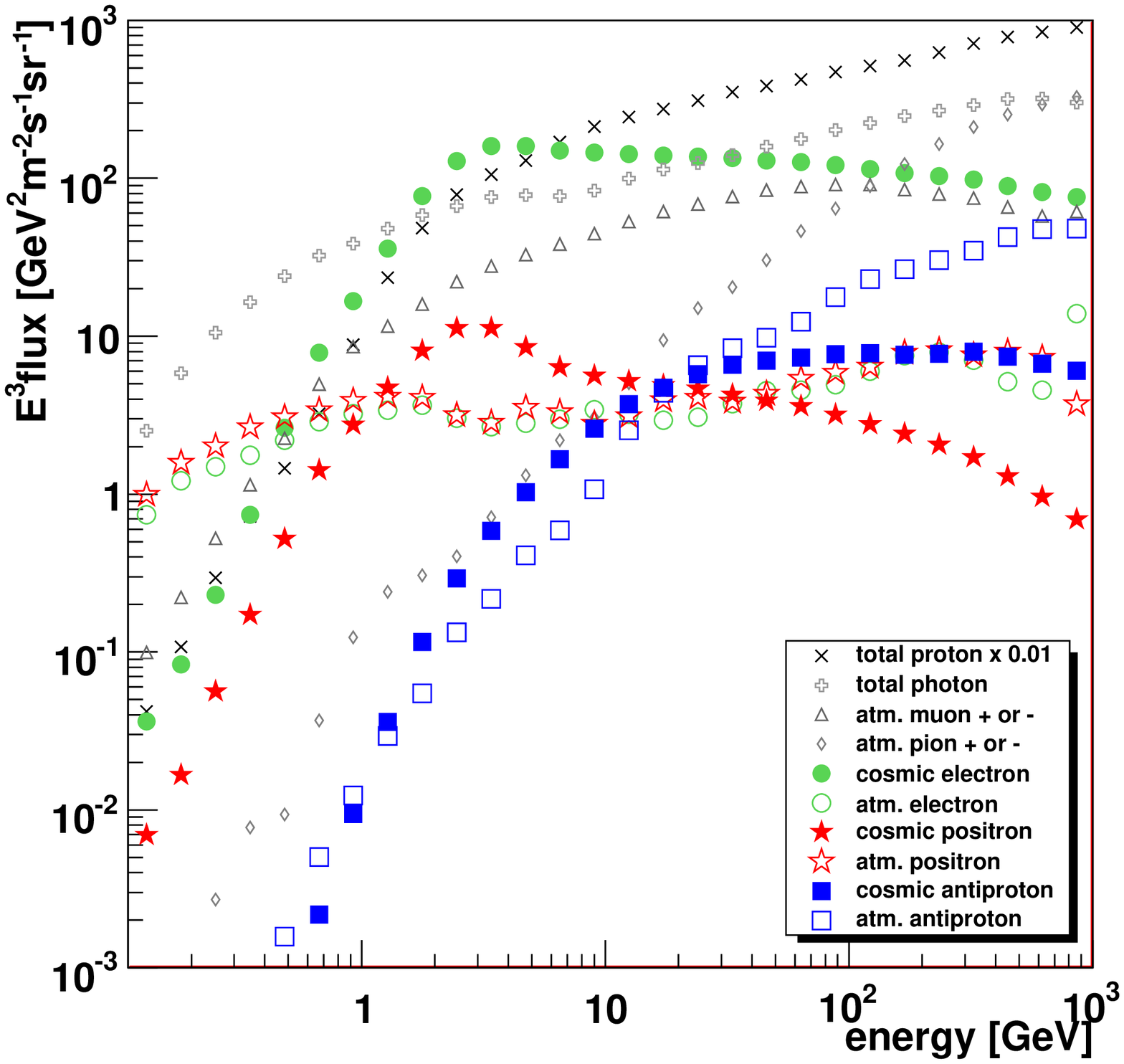,width=8cm}}
\captionof{figure}{\label{f-total_flux_37km}Fluxes (smoothed) at 37\,km altitude (5.4\,g/cm$^2$) at the South Pole respecting the detector acceptance.}
\end{minipage}
\end{center}
\end{figure}

Further interesting is the comparison of the zenith angle dependence of primary and secondary particle fluxes (Fig.~\ref{f-zenith_40000_m_-11_southpole}). As in the case of muons the secondary positron flux is nearly uniform, but primary positrons are strongly attenuated at large zenith angles. Assuming a detector perpendicular to the sky, so that the detector angle $\alpha\sub{det}$ is equal to the zenith angle, the angular acceptance of the PEBS experiment (Fig.~\ref{f-pebs_angular_res_cos_0804}) works as a filter on the secondary particles and reduces the background significantly by about an order of magnitude because atmospheric particles at large zenith angles cannot trigger the experiment. Fig.~\ref{f-total_flux_40km} and \ref{f-total_flux_37km} show the fluxes at 40\,km and 37\,km respecting the PEBS detector acceptance. The atmospheric backgrounds are significantly reduced. The following analysis respects always the detector acceptance but it remains still challenging to correct for atmospheric and detector effects to extract the cosmic fluxes.

\begin{figure}
\begin{center}
\begin{minipage}[b]{.4\linewidth}
\centerline{\epsfig{file=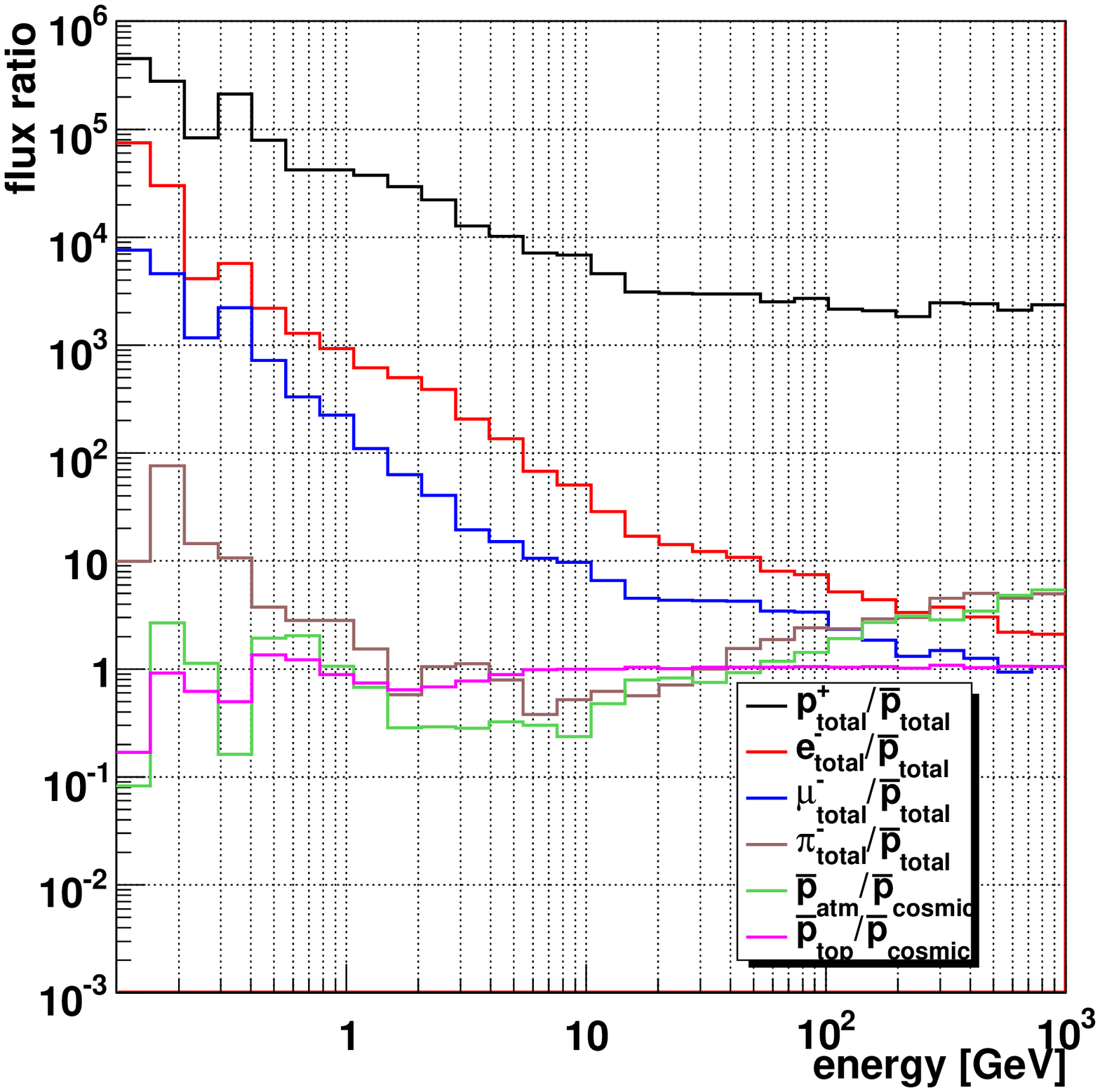,width=8cm}}
\end{minipage}
\hspace{.1\linewidth}
\begin{minipage}[b]{.4\linewidth}
\centerline{\epsfig{file=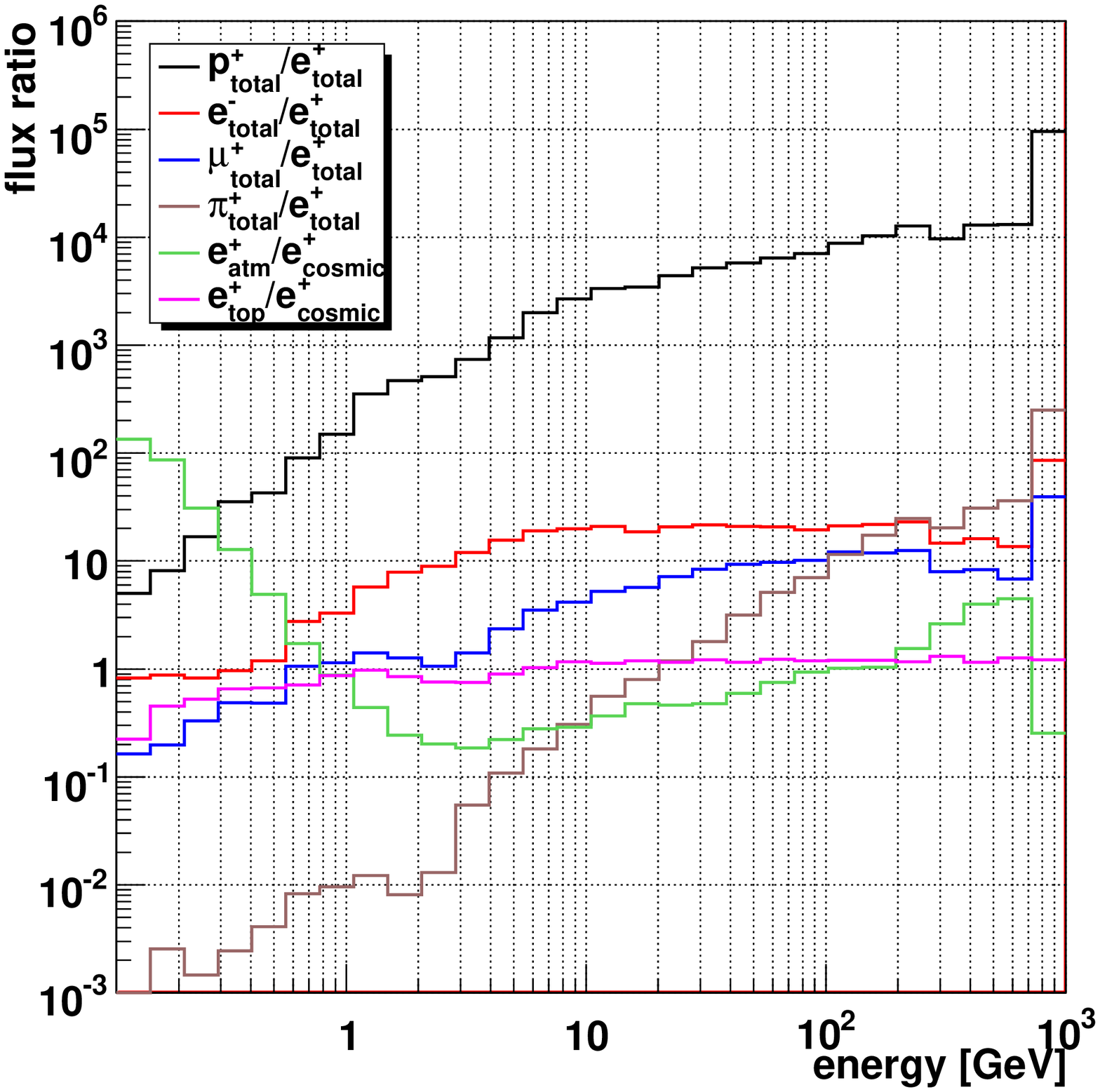,width=8cm}}
\end{minipage}
\captionof{figure}{\label{f-background_40km}Important flux ratios (smoothed) for the antiproton (\textbf{\textit{left}}) and positron (\textbf{\textit{right}}) measurements at 40\,km altitude (3.8\,g/cm$^2$). The label 'cosmic' depicts the attenuated flux of particles with cosmic origin at 40\,km, 'atm' depicts the atmospherically produced flux at 40\,km, 'total' depicts the sum of the cosmic and atmospheric flux at 40\,km and 'top' the flux before entering the atmosphere.}
\end{center}
\end{figure}

Fig.~\ref{f-background_40km} shows important flux ratios for the antiproton and positron measurements at 40\,km altitude. The subdetectors must be able to discriminate against protons, electrons, muons and pions. In addition, is is obvious that reliable predictions for the irreducible atmospheric effects are needed. A difference between antiprotons and positrons is clearly visible in the their attenuation in the atmosphere $\bar p\sub{top}/\bar p\sub{cosmic}$ ($e^+\sub{top}/e^+\sub{cosmic}$). The energy loss of positrons in the atmosphere is stronger than for antiprotons. The positron flux at 40\,km is at 0.1\,GeV about 5 times higher than at the top of the atmosphere and higher energies show decreased fluxes. This is because positrons lose energy strongly due to bremsstrahlung and are shifted to lower energies. In comparison, the energy loss of antiprotons mostly due to nuclear interactions is smaller in the atmosphere. The atmospheric background production of antiprotons (positrons) in the atmosphere is illustrated with the fraction $\bar p\sub{atm}/\bar p\sub{cosmic}$ ($e^+\sub{atm}/e^+\sub{cosmic}$). The atmospheric antiprotons below 1\,GeV exceed slightly the cosmic antiprotons. Between 1 and 60\,GeV the atmospheric antiprotons contribute about 30 - 100\,\%. At about 60\,GeV the contribution to the total antiproton flux by atmospheric antiprotons is equal to the cosmic contribution and at 1\,TeV the atmospheric flux is about 5 times as large as the cosmic flux. The secondary production of positrons below 1\,GeV is much stronger than for antiprotons and exceeds the cosmic positrons on top of the atmosphere below 1\,GeV by about $10^2$ and makes a reliable measurement of the cosmic positron flux very difficult. The atmospheric and cosmic contribution are equal at about 0.5\,GeV for electrons and positrons. The simulations predict a change of the slope at this point and it would be interesting to measure this shoulder precisely to constrain the atmospheric model. The atmospheric positron production up to energies of 100\,GeV is dominated by muon decay and the atmospheric to cosmic flux ratio at 40\,km altitude is about 10 - 20\,\%. Neutral pion decay to photons followed by electron-positron pair production becomes important from about 100\,GeV where the cosmic and atmospheric positron flux is nearly equal again.

\subsubsection{Flux Measurements with PEBS}

The total measured number of particles of a certain type at a certain altitude is influenced by interactions in the atmosphere and by detector misidentification. Misidentification becomes especially important if the background exceeds the signal as in the case of cosmic-ray particles and antiparticles. For example, the total number of particles classified as positrons can be calculated in the following way:

\be N_{e^+}^{\text{PEBS}} = N_{e^+}^{\text{prim}}\cdot \epsilon_{e^+}^{\text{atmo}}\cdot \epsilon_{e^+}^{\text{PEBS}}+N_{e^+}^{\text{sec}}\cdot \epsilon_{e^+}^{\text{PEBS}}+N_{e^-}^{\text{tot}}\cdot\epsilon_{e^-\rightarrow e^+}^{\text{PEBS}}+\frac{N_p^{\text{tot}}}{R_p}+\frac{N_{\mu^+}^{\text{tot}}}{R_{\mu^+}}+\frac{N_{\pi^+}^{\text{tot}}}{R_{\pi^+}}.\label{e-pebs_prim}\ee

In general, the detector properties depend on energy and the atmospheric particle numbers on energy and altitude. The number of particles $N_{x}$ are generated from the respective fluxes multiplied by the detector acceptance and the measurement time. The number of primary positrons $N_{e^+}^{\text{prim}}$ is modified by the attenuation of the atmosphere $\epsilon_{e^+}^{\text{atmo}}$ and by the positron identification efficiency $\epsilon_{e^+}^{\text{PEBS}}$. The contribution of secondarily produced positrons $N_{e^+}^{\text{sec}}$ must also be multiplied by this detection efficiency. Particles misidentified as positrons can be electrons, protons, muons and pions. In the case of electrons, misidentification arises from an incorrect charge reconstruction in the tracker with the probability $\epsilon_{e^-\rightarrow e^+}^{\text{PEBS}}$. Proton, muon and pion rejection against positrons ($R_p$, $R_{\mu^+}$, $R_{\pi^+}$) is done by the electromagnetic calorimeter, the transition radiation detector and the time of flight system. The detection efficiencies are taken to be the same for particles and antiparticles. Statistical errors are introduced by attributing the Poisson error $\sigma_{N_x}=\sqrt{N_x}$. The total number of particles classified as positrons $N_{e^+}^{\text{PEBS}}$ is smeared with a Gaussian distribution with the corresponding error as width for a more realistic simulation.  The number of primary cosmic-ray positrons before entering the atmosphere is now extracted by correcting for atmospheric and detector effects. Therefore eq. \ref{e-pebs_prim} is solved for $N_{e^+}^{\text{prim}}$. The atmospheric efficiency $\epsilon_{e^+}^{\text{atmo}}$ at a given energy is taken from the ratio of particles on top of the atmosphere to cosmic particles at a certain altitude (Fig.~\ref{f-background_40km}). The total fluxes of protons, electrons, muons and pions and atmospheric positrons are needed for the correction and are taken from the simulation as well as the fluxes of atmospheric positrons, muons and pions. Systematic uncertainties of 10\,\% for the detector effects are assumed while the systematic error for the atmospheric effects is set to 15\,\%, as noted above. The same calculation can be carried out for all other particle types.

\begin{figure}
\begin{center}
\begin{minipage}[b]{.4\linewidth}
\centerline{\epsfig{file=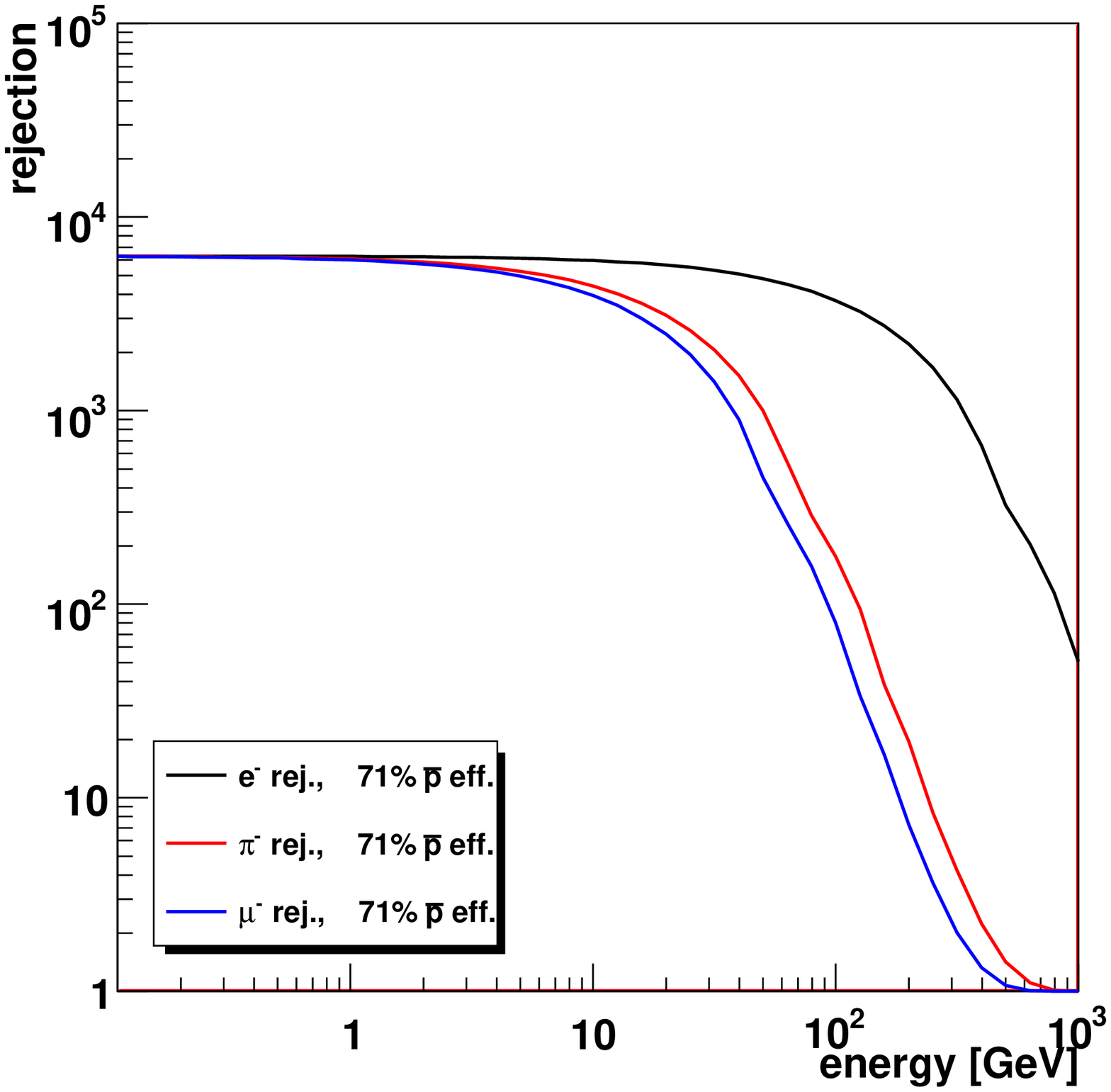,width=8cm}}
\captionof{figure}{\label{f-antiproton_rej_binomial}Electron (pion, muon) rejection against antiprotons vs. antiproton energy of the PEBS TRD.}
\end{minipage}
\hspace{.1\linewidth}
\begin{minipage}[b]{.4\linewidth}
\centerline{\epsfig{file=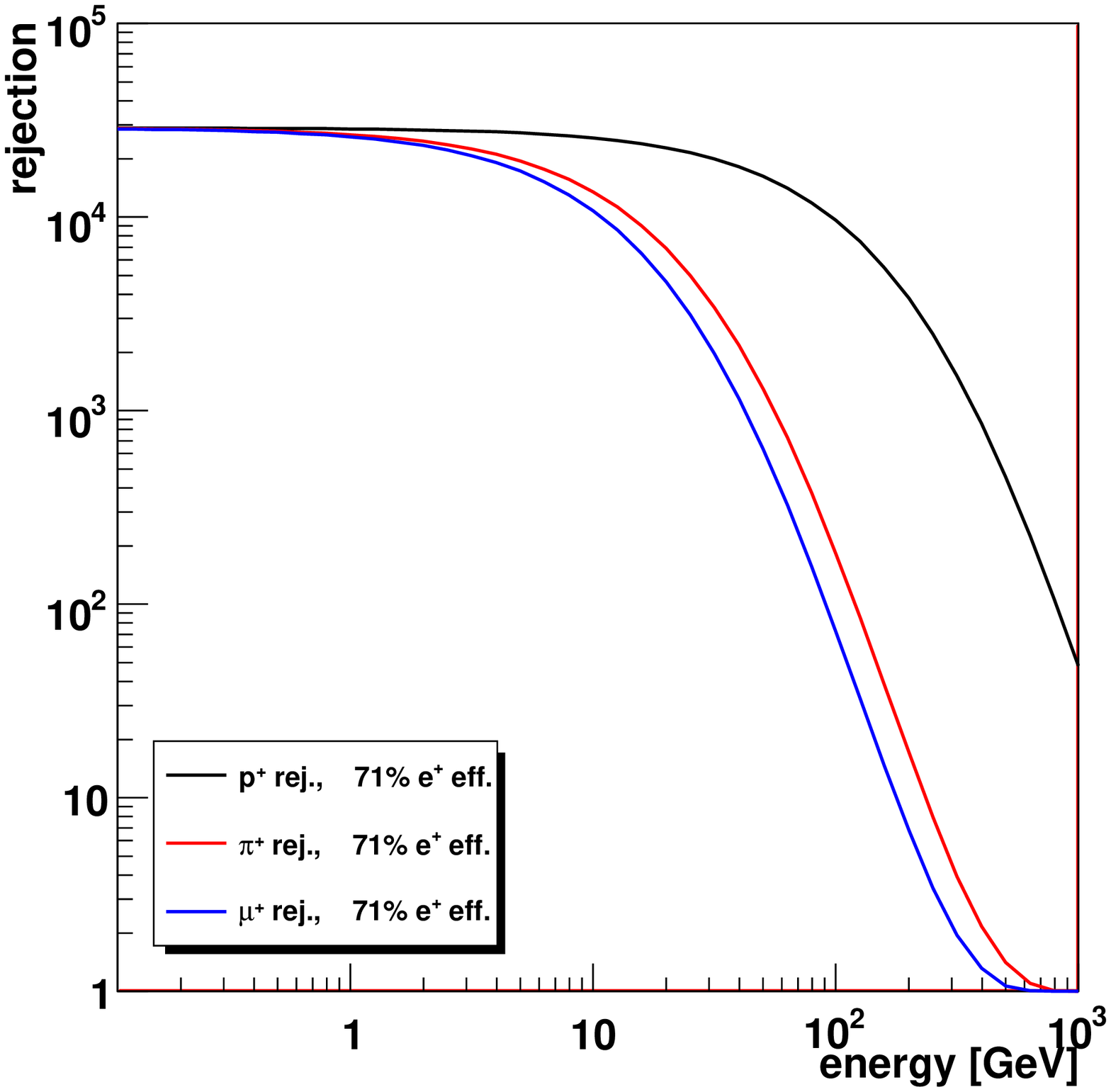,width=8cm}}
\captionof{figure}{\label{f-positron_eff_binomial}Proton (pion, muon) rejection against positrons vs. proton energy of the PEBS TRD.}
\end{minipage}
\end{center}
\end{figure}

\begin{figure}
\begin{center}
\begin{minipage}[b]{.4\linewidth}
\centerline{\epsfig{file=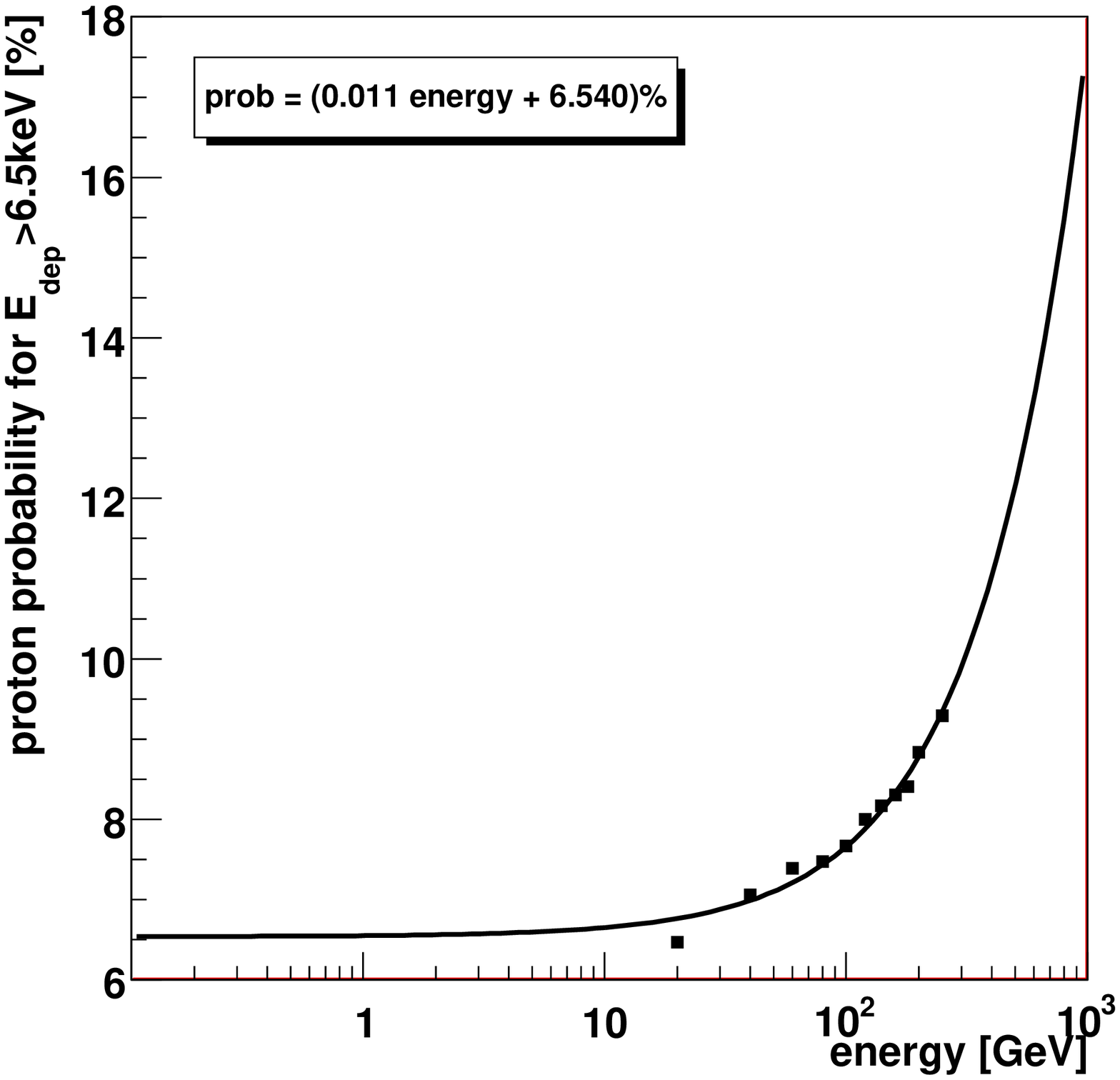,width=8cm}}
\captionof{figure}{\label{f-p_prob_6.5}Probability for protons to deposit more than 6.5\,keV in the proportional tubes of the TRD derived from AMS-02 TRD testbeam data \cite{doetinchem-2006-558}.}
\end{minipage}
\hspace{.1\linewidth}
\begin{minipage}[b]{.4\linewidth}
\centerline{\epsfig{file=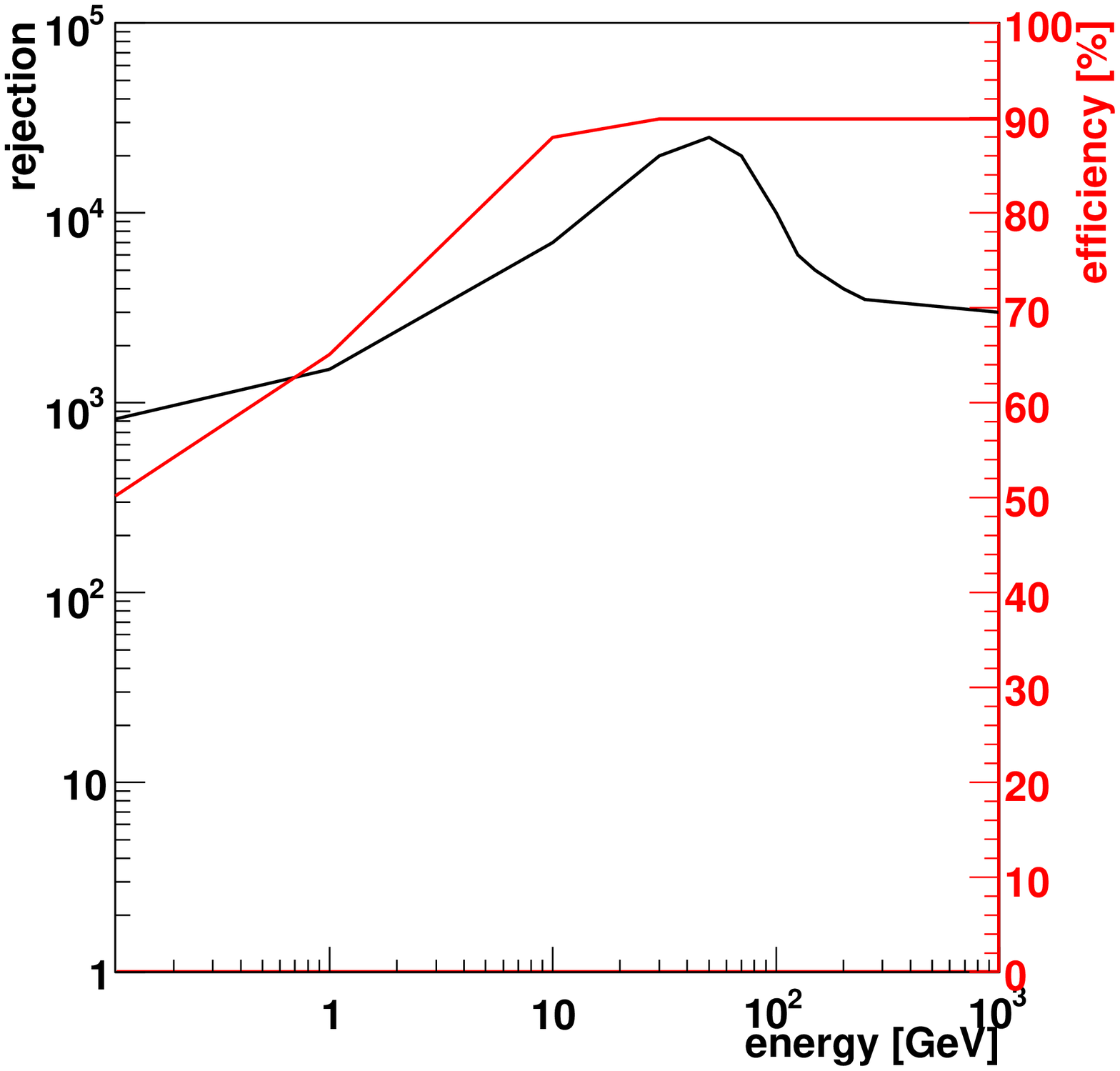,width=8cm}}
\captionof{figure}{\label{f-ecal_rej_eff} Electron (proton) rejection (black and antiproton (positron) efficiency (red) of the electromagnetic calorimeter vs. the reconstructed momentum \cite{gast-2008}.}
\end{minipage}
\end{center}
\end{figure}

The following discusses the detector properties. Fig.~\ref{f-antiproton_rej_binomial} and \ref{f-positron_eff_binomial} show the TRD electron (pion, muon) rejection factors against antiprotons and the proton (pion, muon) rejections against positrons as a function of energy at 71\,\% efficiency, respectively. They result from an analysis of testbeam data assuming a TRD consisting of two sets of twelve layers with a very similar design as the AMS-02~TRD (Sec.~\ref{ss-amsdetcomp})\cite{doetinchem-2006-558}. A cluster counting algorithm used the energy deposition in the proportional tubes to discriminate between light and heavy particles with the help of the transition radiation effect. It is assumed that the probability for electrons (positrons) to deposit energies above 6.5\,keV is 43.50\,\% for all kinetic energies. The probability $p$ to deposit energies above 6.5\,keV for protons (antiprotons) as a function of energy is shown in Fig.~\ref{f-p_prob_6.5} and fitted by:
\be
p(E)=((6.54\pm 0.09)+(1.12\pm 0.06)\cdot E)\,\%.
\ee
and was also used to extrapolate down to 0.1\,GeV and up to 1000\,GeV. 

The total probability $\mathcal P$ to measure $k$ energy depositions above 6.5\,keV in a TRD with $n$ layers is the cumulative binomial probability:
\be
\mathcal{P}=\sum_{j=k}^n \frac{n!}{j!(n-j)!}\cdot p^j \cdot (1-p)^{n-j}.
\ee
where the electron (proton) rejection is defined by $R_{e^-(p)}=1/\mathcal{P}_{e^-(p)}$ and the antiproton (posi\-tron) efficiency by $\epsilon_{\bar p(e^+)}^{\text{PEBS}} = \mathcal P_{\bar p(e^+)}$. It is assumed that the formation of transition radiation depends only on the Lorentz factor $\gamma=E/m$ such so the muon (pion) rejection can be calculated by scaling the probability $p_p$ to deposit energies above 6.5\,keV for protons with the mass ratio of protons to muons (pions) according to:
\be p_{\mu(\pi)}(E_{\mu(\pi)}) = p_{p}\left(E_{p}\cdot\frac{\displaystyle m_{p}}{\displaystyle m_{\mu(\pi)}}\right)\ee where $m_{\mu}$, $m_{\pi}$ and $m_{p}$ are the masses of the muon, pion and the proton. 

Fig.~\ref{f-ecal_rej_eff} shows the simulated proton rejection of the electromagnetic calorimeter and the corresponding positron efficiency as a function of the reconstructed particle momentum\cite{gast-2008}. Starting from 1\,GeV rejections between $10^3$ and $10^4$ can be achieved at detection efficiencies between 60\,\% and 90\,\%.

\begin{figure}
\begin{center}
\begin{minipage}[b]{.4\linewidth}
\centerline{\epsfig{file=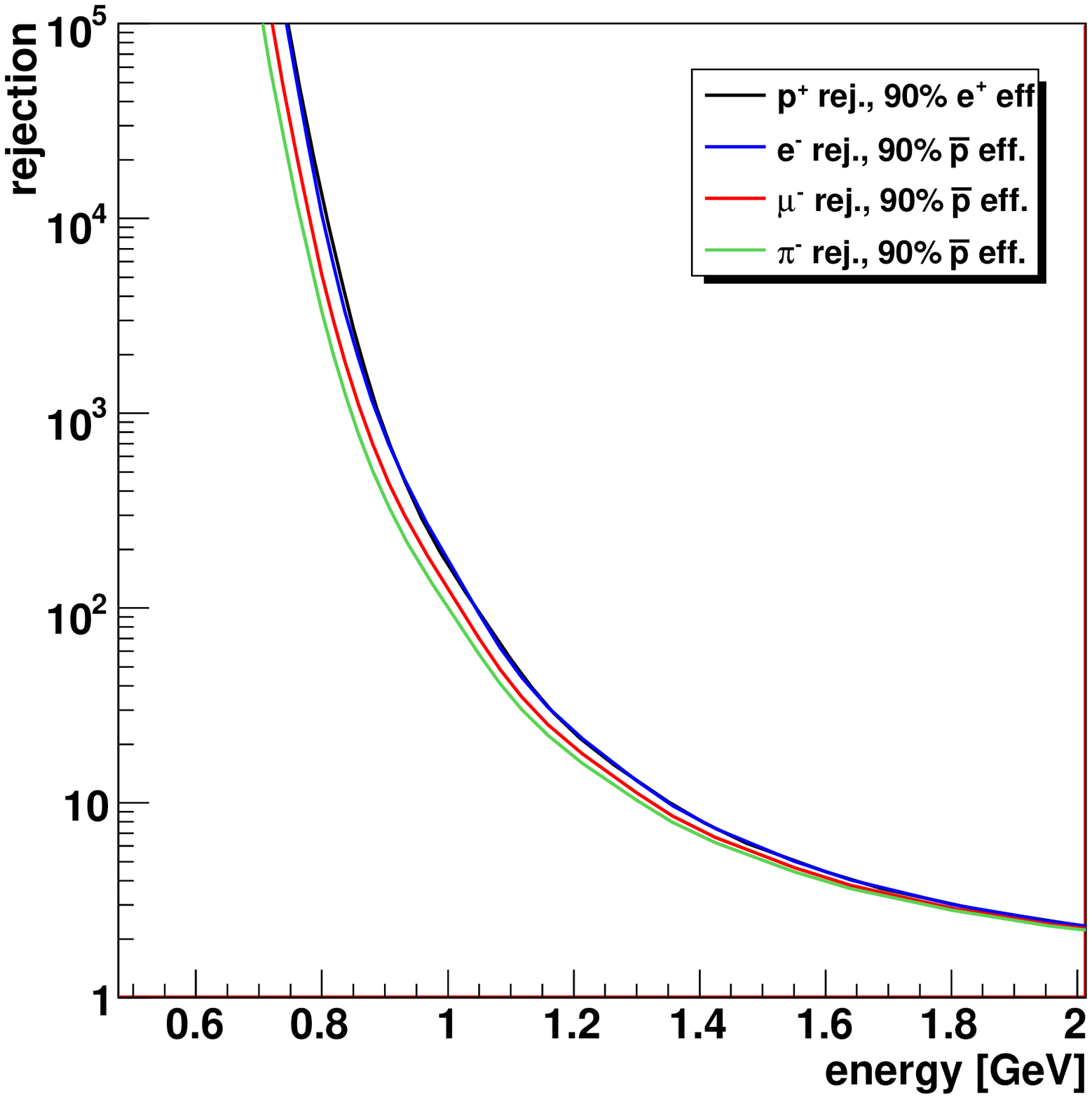,width=8cm}}
\captionof{figure}{\label{f-tof_pebs_rejection}Rejection of the PEBS TOF.}
\end{minipage}
\hspace{.1\linewidth}
\begin{minipage}[b]{.4\linewidth}
\centerline{\epsfig{file=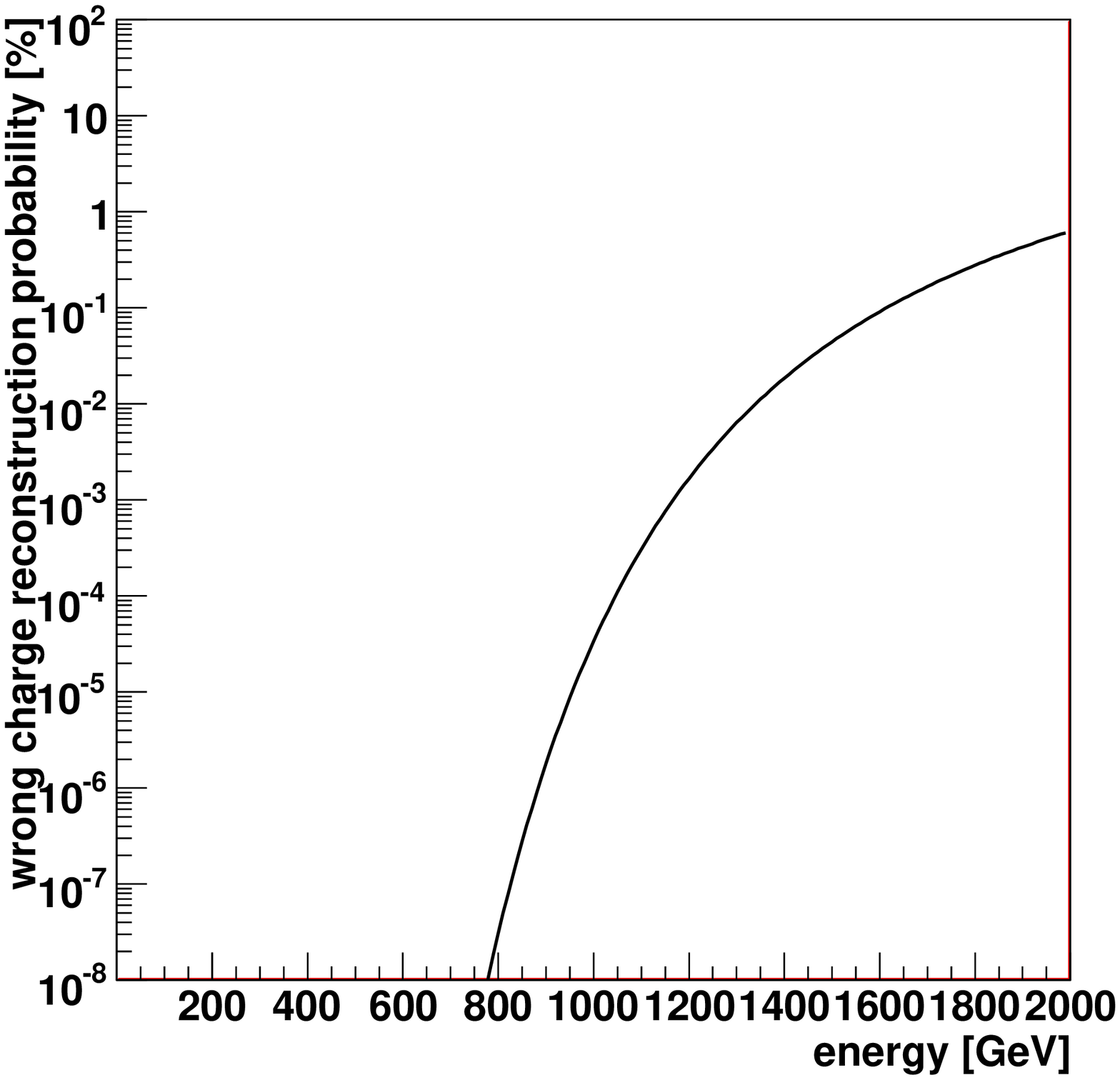,width=8cm}}
\captionof{figure}{\label{f-tracker_rej_0.02_2.3}Probability of wrong charge reconstruction in the PEBS tracker.}
\end{minipage}
\end{center}
\end{figure}

The TOF system can also be used to discriminate between heavy and light particles for momenta up to a few GeV (Fig.~\ref{f-tof_pebs_rejection}). The calculation of the rejection is based on the time resolution $\sigma_t$ of about 100\,ps and the distance $s=0.8$\,m between the TOF planes. The time distribution $T(t,E,m)$ is:
\be T(t,E,m) =\frac{\displaystyle 1}{\displaystyle\sigma_t\sqrt{2\pi}}\cdot\exp\left[-\frac{\displaystyle 1}{\displaystyle2}\left(\frac{\displaystyle t-\bar t}{\displaystyle\sigma_t}\right)^2\right].\label{e-tof1}\ee 
The mean time $\bar t$ is derived by:
\be
v=\frac{\displaystyle s}{\displaystyle \bar t}\quad\wedge\quad\gamma=\frac{\displaystyle 1}{\displaystyle \sqrt{1-\frac{\displaystyle v^2}{\displaystyle c^2}}}\quad\wedge\quad E=mc^2(\gamma-1)\quad \Longrightarrow\quad \bar t = \frac{\displaystyle s\left(E+mc^2\right)}{\displaystyle \sqrt{E\left(E+2mc^2\right)c}}
\ee
where $E$ is the kinetic energy of the particle and $c$ the speed of light. The time $t\sub{cut}$ for a given detection efficiency $\epsilon$ of positrons with mass $m_{e^+}$ and the fraction of protons with mass $m_{p}$ having $t>t\sub{cut}$ are calculated with:
\be\epsilon=\int_{-\infty}^{t\sub{cut}}T(t,E,m_{e^+})\text{d}t\quad\wedge\quad p\sub{above}=\int^{\infty}_{t\sub{cut}}T(t,E,m_p)\text{d}t.\ee
The rejection $R$ is defined as:
\be R=\frac{\displaystyle1}{\displaystyle 1-p\sub{above}}.\label{e-tof2}\ee 
The calculation of the discrimination power for other particle types follows the same principle.

\begin{figure}
\begin{center}
\begin{minipage}[b]{.4\linewidth}
\centerline{\epsfig{file=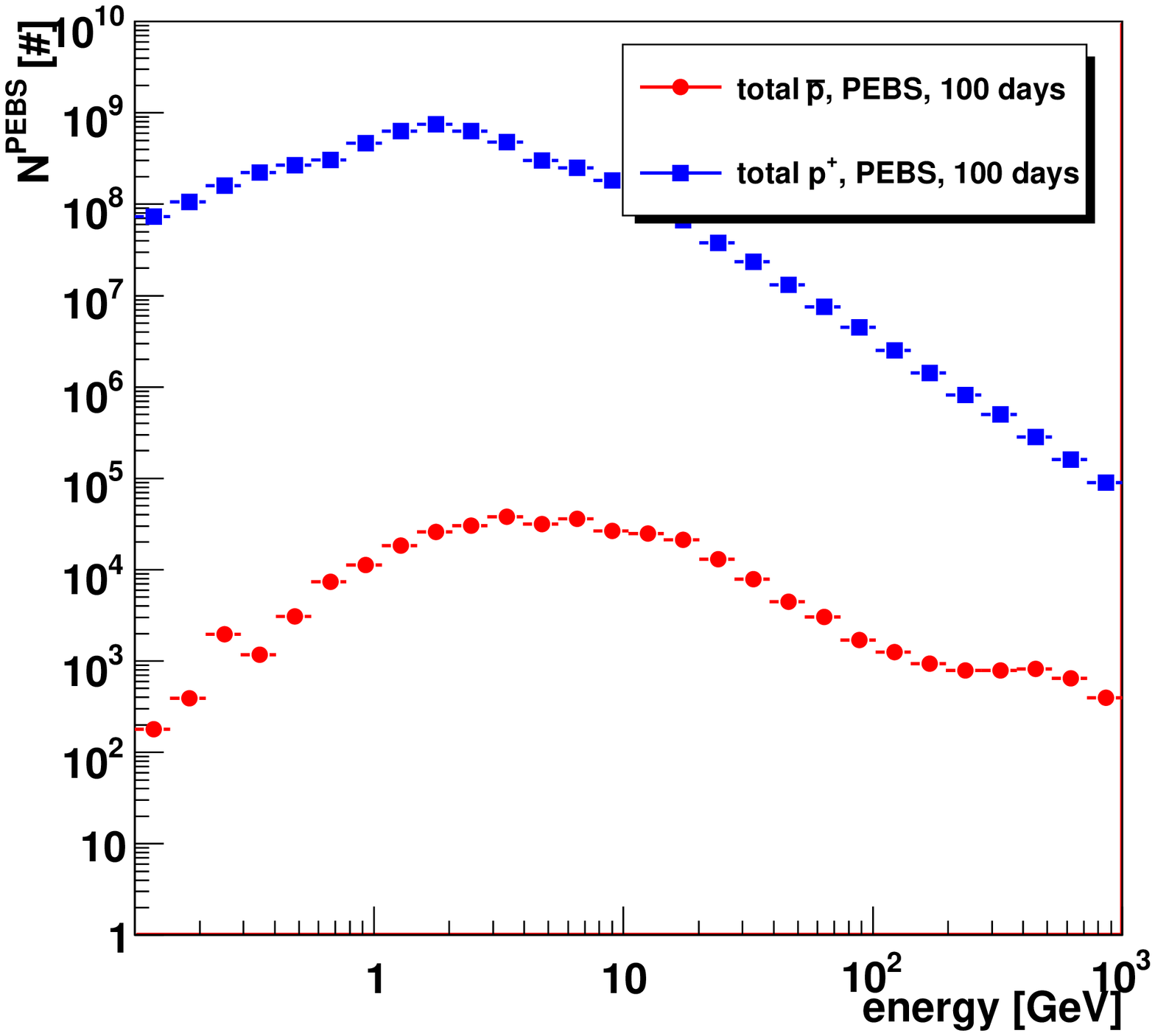,width=8cm}}
\captionof{figure}{\label{f-pbar_p_npebs_dm_hg_phi550_40000_m_southpole_100_days}Projected number of particles classified as protons and antiprotons with statistical errors.}
\end{minipage}
\hspace{.1\linewidth}
\begin{minipage}[b]{.4\linewidth}
\centerline{\epsfig{file=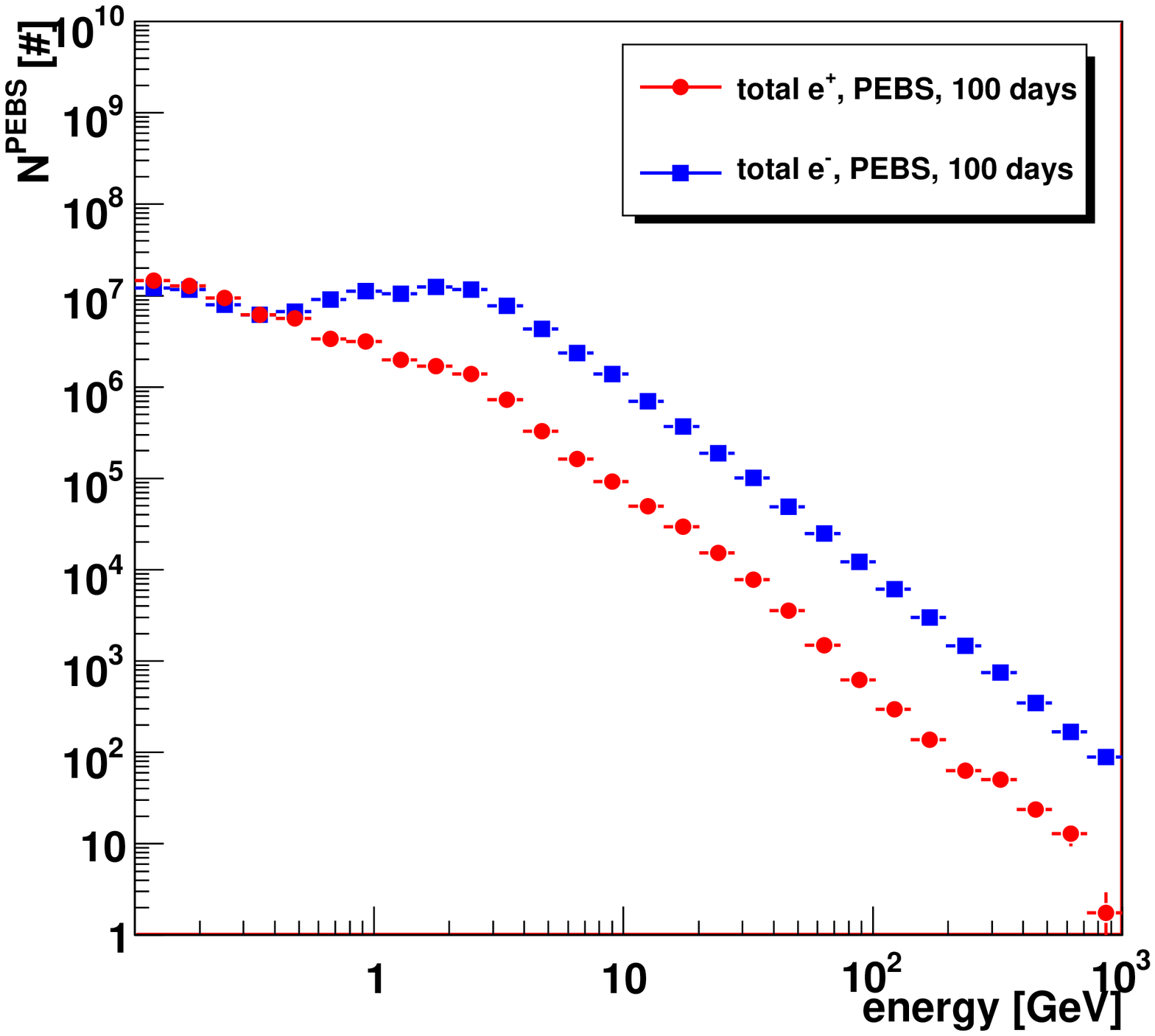,width=8cm}}
\captionof{figure}{\label{f-e+_e-_npebs_dm_hg_phi550_40000_m_southpole_100_days}Projected number of particles classified as electrons and positrons with statistical errors.}
\end{minipage}
\end{center}
\end{figure}

The probability for a wrong charge reconstruction (Fig.~\ref{f-tracker_rej_0.02_2.3}) is calculated from the momentum PEBS momentum resolution: 
\be\frac{\displaystyle\sigma_p}{\displaystyle p}=\frac{\displaystyle 0.02\,\%\cdot p}{\displaystyle\text{GeV}}\oplus2.3\,\%\ee 
by integration of the ratio $q$ of the measured to the real momentum with $\sigma_q=\sigma_p/p$: 
\be\epsilon_{e^-\rightarrow e^+}^{\text{PEBS}}(p)=\int_{-\infty}^0\frac{\displaystyle 1}{\displaystyle\frac{\displaystyle \sigma_p}{\displaystyle p}\sqrt{2\pi}}\cdot\exp\left[-\frac{\displaystyle 1}{\displaystyle2}\left(\frac{\displaystyle  q -1}{\displaystyle\frac{\displaystyle \sigma_p}{\displaystyle p}}\right)^2\right]\text{d} q.\ee
The probability for a wrong charge reconstruction is below 1\,\% in the energy range up to 2\,TeV and there will not be any significant contamination of the antiproton (positron) measurement by protons (electrons). In addition, the value of the momentum resolution quoted above also does not significantly affect the shape of the differential energy spectra and such an effect is not taken into account in the following.

The total measurement time of 100\,days from several flights with an acceptance of 0.4\,m$^2$sr\cite{gast-2008} allows the detection of about $10^9$ events in total. Protons are the main component. In addition, about $10^8$ electrons, $10^7$ positrons and $10^5$ antiprotons will be collected (Fig.~\ref{f-pbar_p_npebs_dm_hg_phi550_40000_m_southpole_100_days} and \ref{f-e+_e-_npebs_dm_hg_phi550_40000_m_southpole_100_days}). The atmospheric muon and pion background for the measurements can be reduced with the ECAL, the TRD and the TOF. As mentioned above, the ECAL delivers a rejection factor of about $10^3$ - $10^4$. In the energy range from 1 - 100\,GeV the pion to positron ratio is about 0.01 - 5 and the ratio between muon and positron fluxes is about 1.5 - 10. The combined TRD and ECAL rejection reduce the pion and muon background to less than 1\,\% up to 1000\,GeV such that muons and pions are neglected for the following positron analysis. It is very unlikely that muons and pions shower in the ECAL, so the discrimination of antiprotons against muons or pions can only be done with the TRD and the TOF and muon and pion background must be considered. 

In the energy range 1 - 60\,GeV the total antiproton flux has cosmic antiprotons as its main component followed by about 30\,\% atmospheric antiprotons and about 1\,\% misidentified muons. The muon and pion contribution are important starting from about 100\,GeV. The pion flux limits the reasonable antiproton measurement to about 200\,GeV. Contributions of electrons and protons due to detector inefficiencies are very small (Fig.~\ref{f-pbar_perc_dm_hg_phi550_40000_m_southpole_100_days}). Below 1\,GeV the positron measurement is dominated by atmospheric positrons. At higher energies about 80 - 90\,\% of the total positron flux are of cosmic origin. As noted above, the pion decay channel starts to contribute significantly to the atmospheric positron production from about 100\,GeV. The contribution of misidentified electrons is very small and wrong classified protons become important from about 700\,GeV (Fig.~\ref{f-e+_perc_dm_hg_phi550_40000_m_southpole_100_days}). Despite the atmospheric and detector effects the flux measurements of antiprotons and positrons are statistically limited. In the presented cosmic-ray theory with supersymmetric dark matter annihilations only $\approx3$ cosmic antiprotons and $\approx0.7$ cosmic positrons are expected in a measurement time of 100\,days in the energy interval around 600\,GeV. The extracted fluxes with statistical error bars are shown in Fig.~\ref{f-pbar_p_fluxes_dm_hg_phi550_40000_m_southpole_100_days} and \ref{f-e+_e-_fluxes_dm_hg_phi550_40000_m_southpole_100_days}.

Even more interesting are the fractions. Systematic effects are compensated by assuming the same rejection and atmospheric attenuation for both particles and antiparticles. The resulting corrected cosmic antiproton and positron fractions are shown in Fig.~\ref{f-pbar_p_fraction_dm_hg_phi550_40000_m_southpole_100_days} and \ref{f-e+_e-_fraction_dm_hg_phi550_40000_m_southpole_100_days} together with the systematic error bands composed of detector and atmospheric effects. The antiproton fraction should be well measured up to about 100\,GeV but a possible contribution by dark matter annihilations according to the model previously mentioned is likely to be too small to be resolved within the expected systematic errors. The systematic errors above 100\,GeV are dominated by the uncertainties of the pion and muon corrections and needs a careful detector calibration at high energies. The antiproton measurement can still be used to constrain galactic propagation models. On the other hand, the positron fraction will be measured precisely starting from 1\,GeV up to about 200 - 300\,GeV. The contribution from the supersymmetric model for dark matter annihilations would be well distinguishable from the background for an annihilation boost factor of 100 due to a clumpy dark matter distribution in the galaxy. In contrast to the antiproton fraction, the positron fraction shows at lower energies large systematic errors due to large atmospheric corrections resulting mainly from strong atmospheric positron production and from bremsstrahlung losses of the cosmic positrons in the atmosphere. The energy range below 1\,GeV must be treated very carefully and needs good models for the atmosphere

\pagebreak

\begin{center}
\begin{minipage}[b]{.4\linewidth}
\vspace{1cm}
\centerline{\epsfig{file=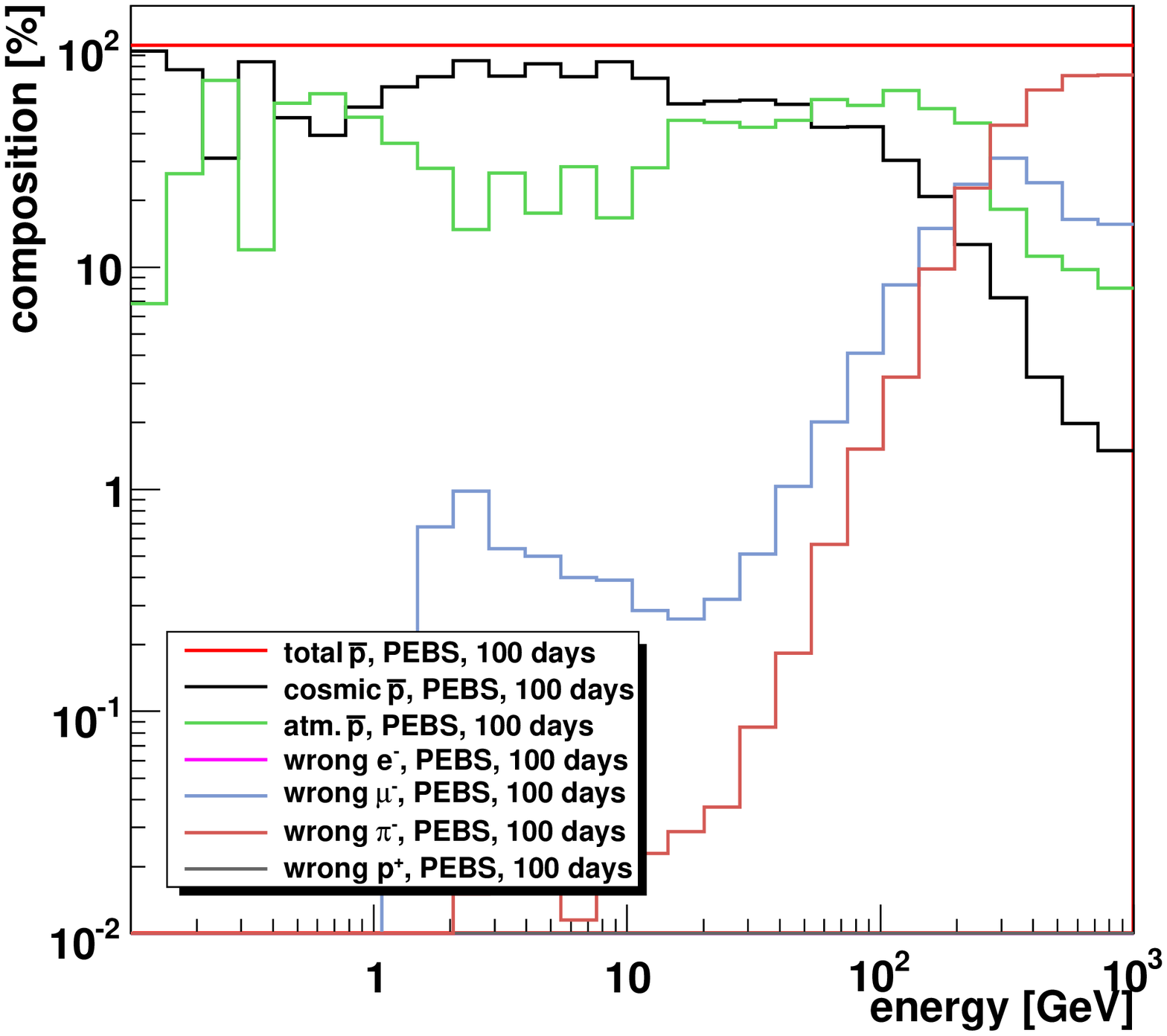,width=8cm}}
\captionof{figure}{\label{f-pbar_perc_dm_hg_phi550_40000_m_southpole_100_days}Composition of particles classified as antiprotons.}
\end{minipage}
\hspace{.1\linewidth}
\begin{minipage}[b]{.4\linewidth}
\vspace{1cm}
\centerline{\epsfig{file=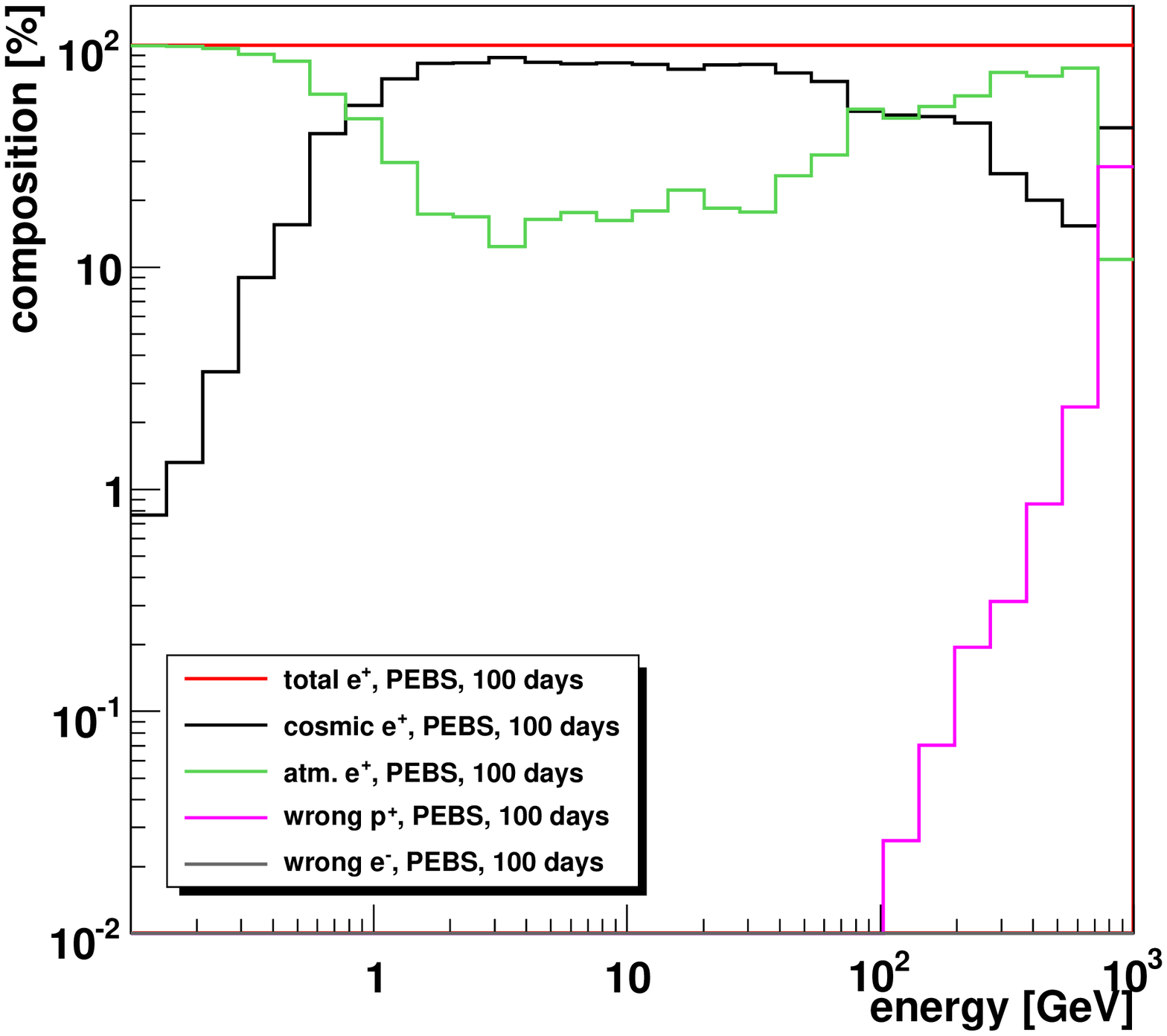,width=8cm}}
\captionof{figure}{\label{f-e+_perc_dm_hg_phi550_40000_m_southpole_100_days}Composition of particles classified as positrons.}
\end{minipage}
\end{center}

\vspace{2cm}

\begin{center}
\begin{minipage}[b]{.4\linewidth}
\centerline{\epsfig{file=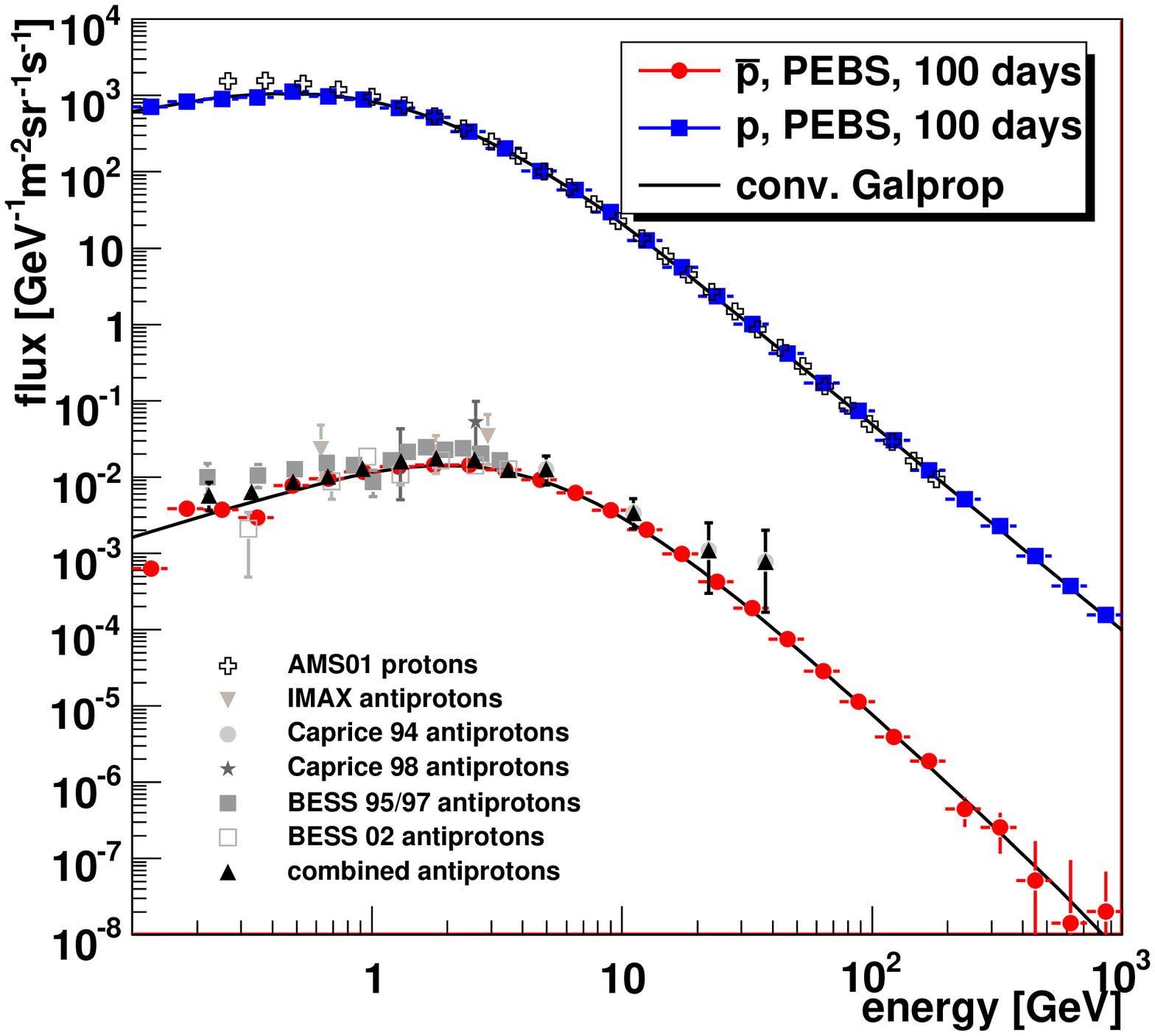,width=8cm}}
\captionof{figure}{\label{f-pbar_p_fluxes_dm_hg_phi550_40000_m_southpole_100_days}Projected proton and antiproton fluxes with statistical errors.}
\end{minipage}
\hspace{.1\linewidth}
\begin{minipage}[b]{.4\linewidth}
\centerline{\epsfig{file=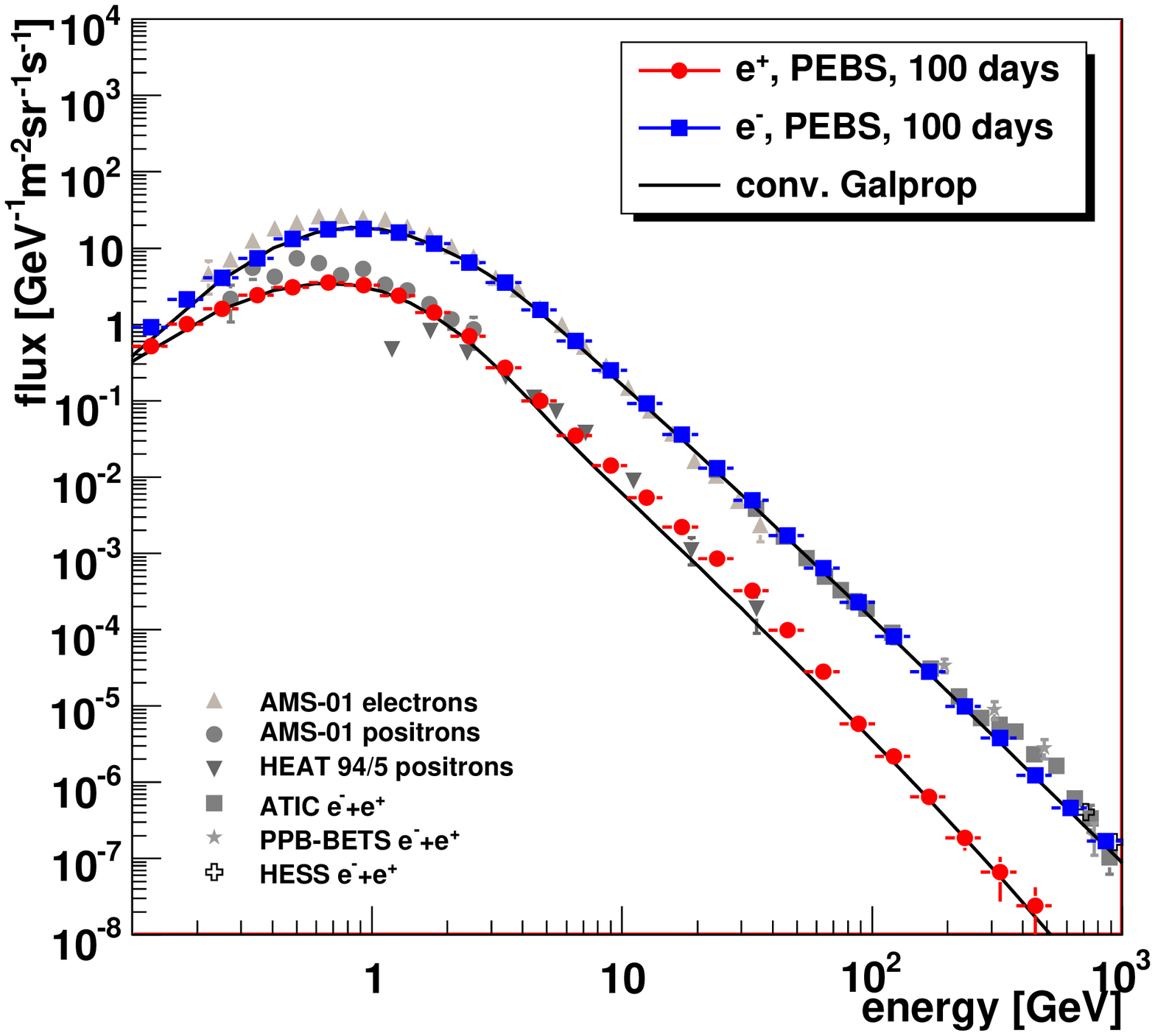,width=8cm}}
\captionof{figure}{\label{f-e+_e-_fluxes_dm_hg_phi550_40000_m_southpole_100_days}Projected electron and positron fluxes with statistical errors.}
\end{minipage}
\end{center}

\pagebreak

\begin{center}
\begin{minipage}[b]{.4\linewidth}
\centerline{\epsfig{file=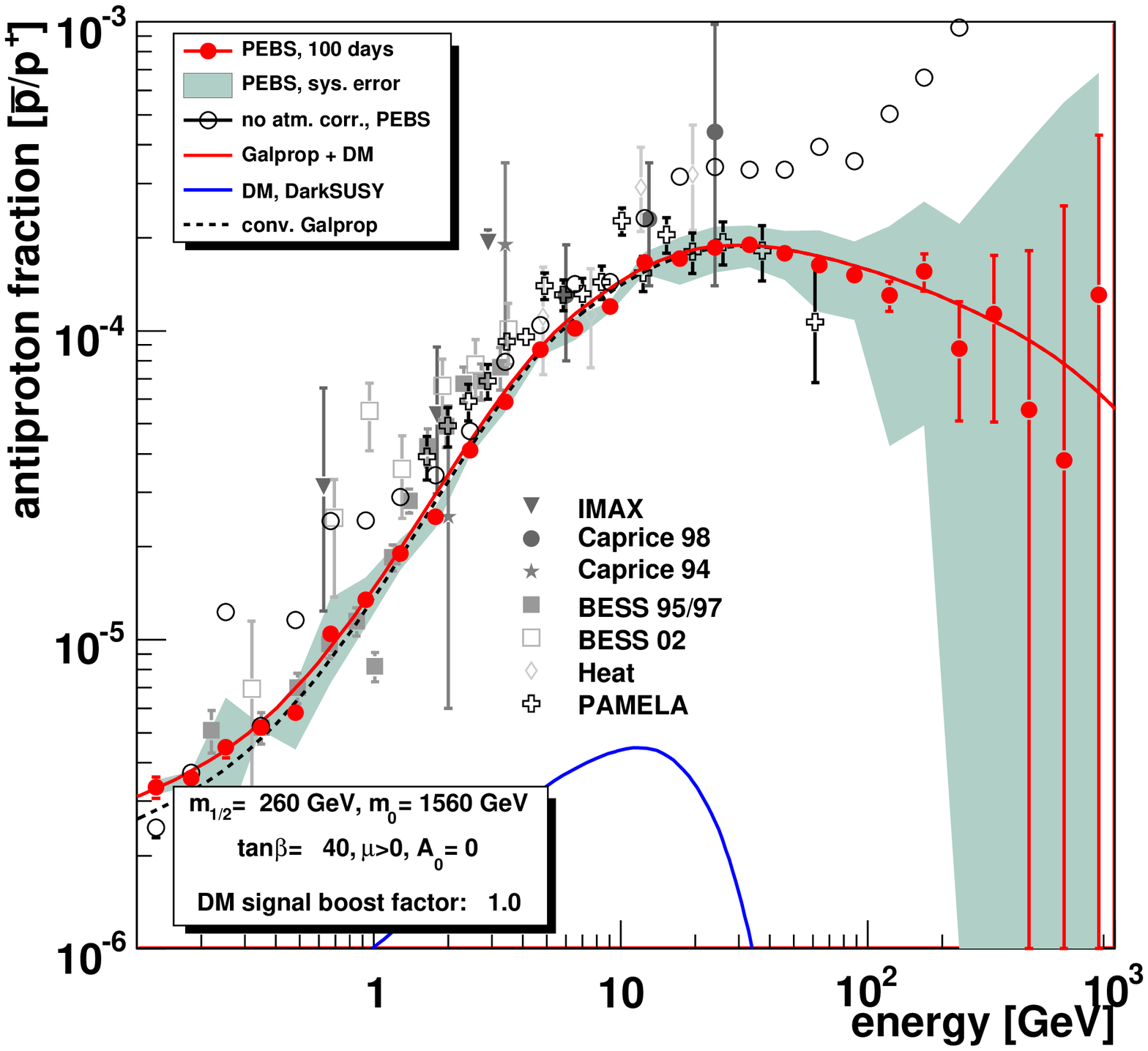,width=8cm}}
\captionof{figure}{\label{f-pbar_p_fraction_dm_hg_phi550_40000_m_southpole_100_days}Projected antiproton fraction with dark matter signal.}
\end{minipage}
\hspace{.1\linewidth}
\begin{minipage}[b]{.4\linewidth}
\centerline{\epsfig{file=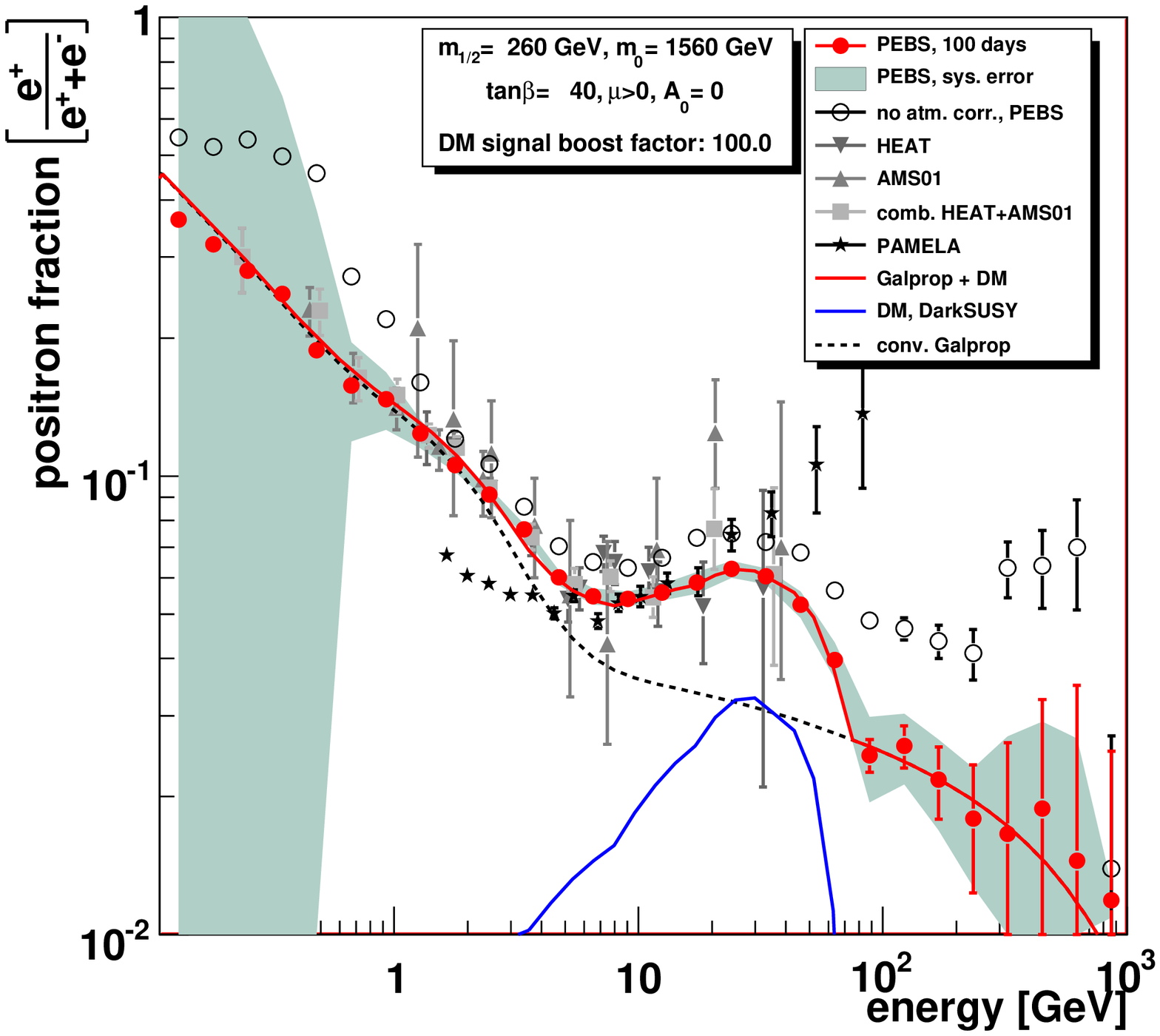,width=8cm}}
\captionof{figure}{\label{f-e+_e-_fraction_dm_hg_phi550_40000_m_southpole_100_days}Projected positron fraction with dark matter signal.}
\end{minipage}
\end{center}

and the solar modulation. The open circles in fig. ~\ref{f-pbar_p_fraction_dm_hg_phi550_40000_m_southpole_100_days} and \ref{f-e+_e-_fraction_dm_hg_phi550_40000_m_southpole_100_days} show the antiproton (positron) fraction without atmospheric corrections. It is clearly seen that the fractions are overestimated without corrections especially for high energies. Therefore, a good knowledge of atmospheric effects is indispensable for a reliable interpretation of the data.

The presented discussion of the PEBS measurement capabilities used a supersymmetric model which has large antiparticle flux contributions at $\cal O$(10\,GeV). Other theories predict large fluxes at much higher energies, e.g. Kaluza-Klein universal extra dimensions and pulsars. This influences the number of observable particles and therefore the statistical limitations of the experiment and emphasizes the need of a well calibrated detector up to high energies to discriminate between the different models.

\chapter{The Space-based AMS-02 Mission\label{c-ams}}

\begin{center}
\begin{minipage}[b]{0.55\linewidth}
\epsfig{file=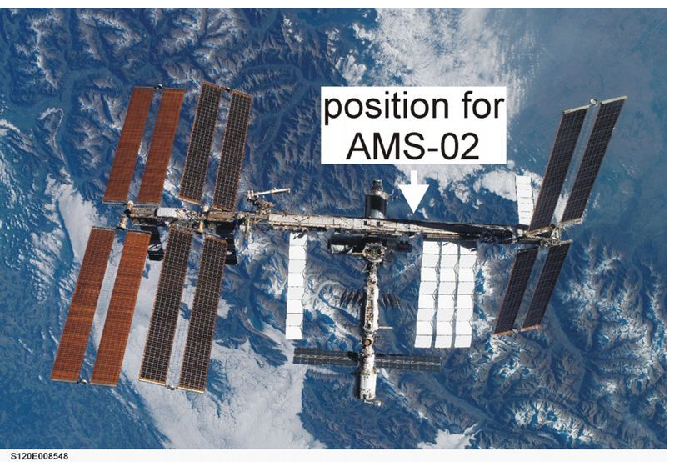,height=6.5cm}
\captionof{figure}{\label{f-ams_iss}Picture of the ISS taken during the NASA Space Shuttle Mission STS-120 in November 2007 \cite{sts-120}.}
\end{minipage}
\hspace{.1\linewidth}
\begin{minipage}[b]{0.28\linewidth}
\epsfig{file=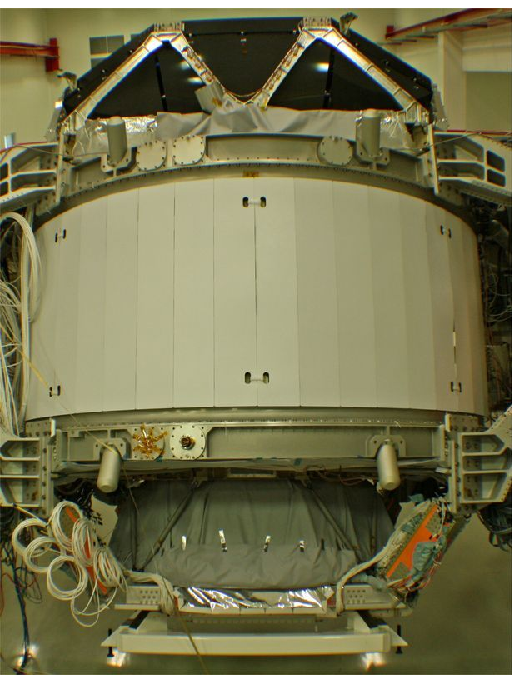,height=6.5cm}
\captionof{figure}{\label{f-pre-integration}AMS-02 during pre-integration.}
\end{minipage}
\end{center}

\section{Mission Overview}

The Alpha Magnetic Spectrometer (AMS-02) experiment is a space-based experiment which will be installed on the International Space Station (ISS) in 2010. The ISS is on an orbit at about 400\,km altitude above the Earth and can currently be reached by the American Space Shuttles, the Russian Soyuz spacecrafts and in the future by the European Automated Transfer Vehicle.

AMS-02 is designed to do precise spectroscopy of cosmic rays in the GeV - TeV energy range for several types of particle species. The launch and spacecraft flight to the ISS and operation in space provide a lot of challenges to cope with. The experiment must survive up to 9\,$g$ acceleration during the launch and must work properly in a temperature range of -180°C - 50°C in vacuum. All components have to fulfill National Aeronautics and Space Administration (NASA) restrictions on construction and materials used. For example there are restrictions on the outgassing rate of all materials used to assure unaltered operation for other experiments on the ISS and stress calculations on all bolts and screws. Other crucial parameters are the limits on weight (7000\,kg), height (3\,m), power consumption (2.5\,kW) and data link to Earth (2\,Mb/s). After installation on the ISS by the astronauts, AMS-02 will operate without further intervention for several years.

\begin{figure}
\begin{center}
\centerline{\epsfig{file=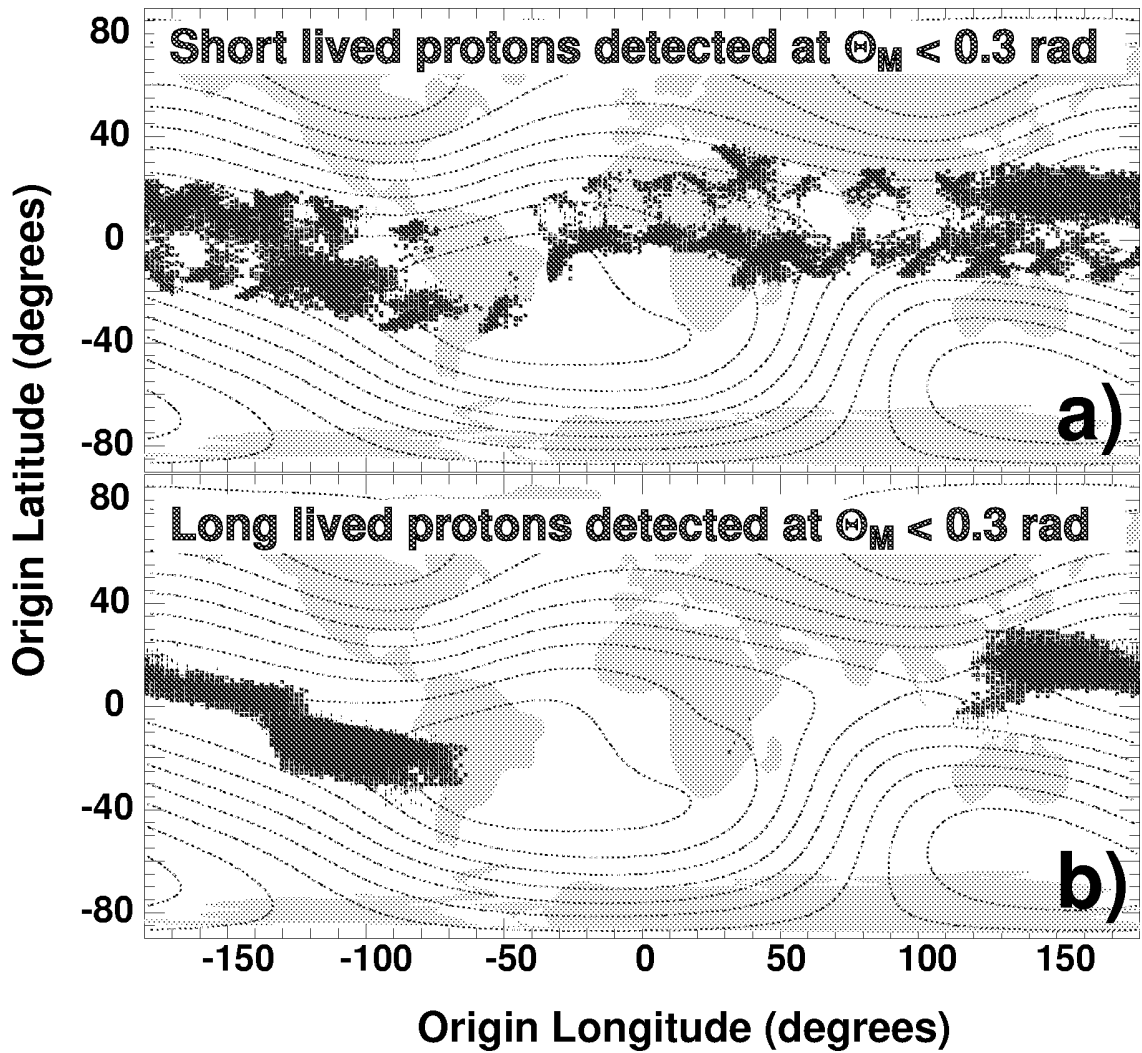,width=15cm}}
\captionof{figure}{\label{f-proton_belts_ams1}The geographical origin of a) short-lived ($<0.3$\,s between detection and production) and b) long-lived ($>0.3$\,s between detection and production) protons with $p < 3$\,GeV/c. The dashed lines indicate the geomagnetic field contours at 380\,km. Reprinted figure with permission from Elsevier \cite{ams01}.}
\end{center}
\end{figure}

As a test of the detector components, the precursor experiment AMS-01 was flown for 10\,days on a Space Shuttle in July 1998. Analysis of the data resulted in improved bounds on the existence of antimatter in the Universe, a discovery of a radiation belt of GeV protons around the Earth (Fig.~\ref{f-proton_belts_ams1}) and measurement of the energy spectra of protons, helium nuclei, electrons, positrons and antiprotons (Fig.~\ref{f-fluxes_geo_mod_galprop_conv}) \cite{ams01}.

\subsection{Detector Components \label{ss-amsdetcomp}}

\begin{figure}
\begin{center}
\centerline{\epsfig{file=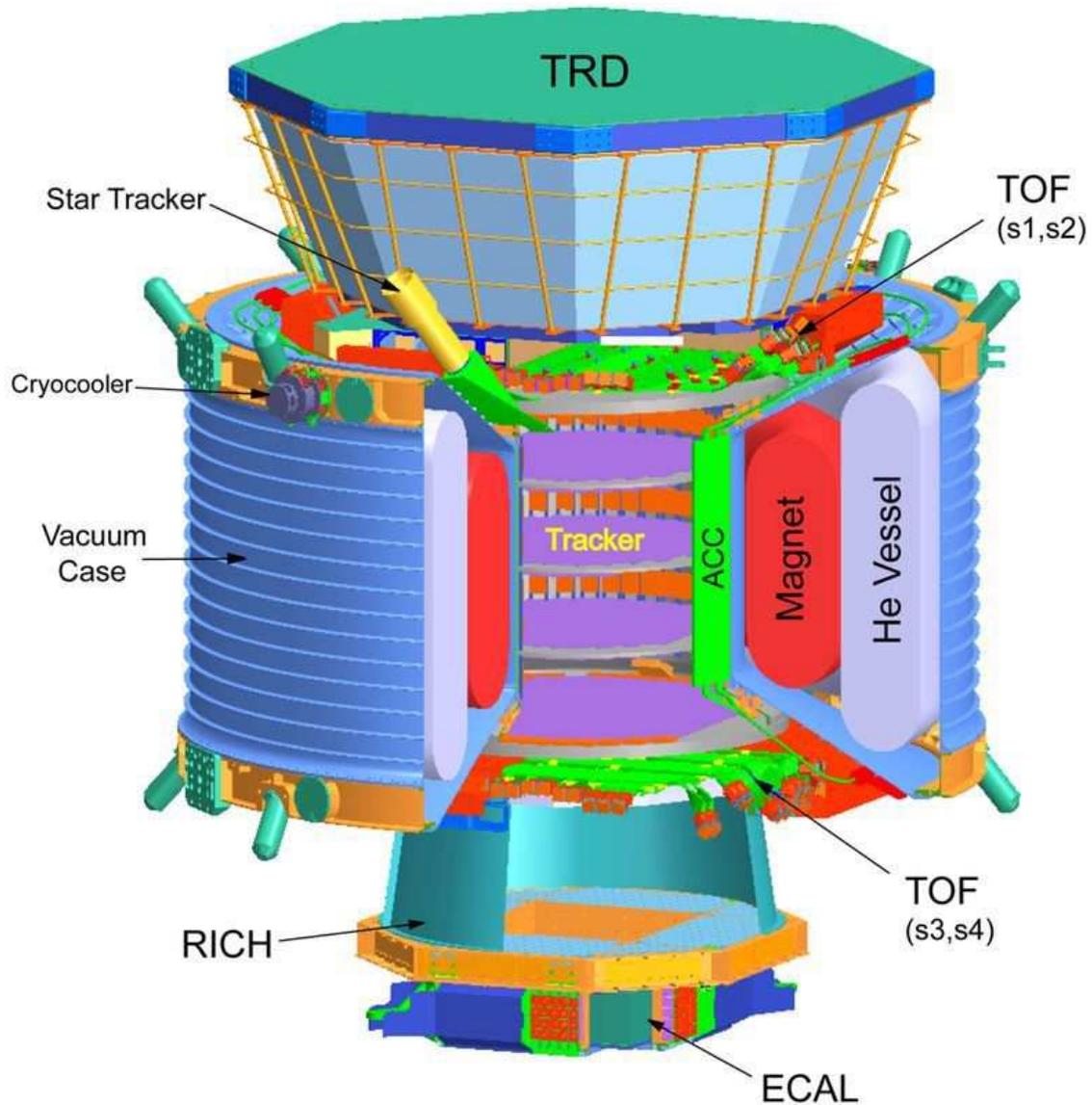,width=15cm}}
\captionof{figure}{\label{f-ams}The AMS-02 experiment\cite{ams}.}
\end{center}
\end{figure}

The AMS-02 detector (Fig.~\ref{f-ams}) has several components to determine the properties of traversing particles. The following sections will give a short overview of each subdetector. All parts have undergone space qualification tests to fulfill all NASA safety requirements and assure full functionality in a wide temperature range under vacuum. The different detectors have been built by institutes all over the world and have been integrated in a clean room at the European organization for nuclear research (CERN).

\subsubsection{Transition Radiation Detector (TRD)}

\begin{figure}
\begin{center}
\begin{minipage}[b]{.56\linewidth}
\centerline{\epsfig{file=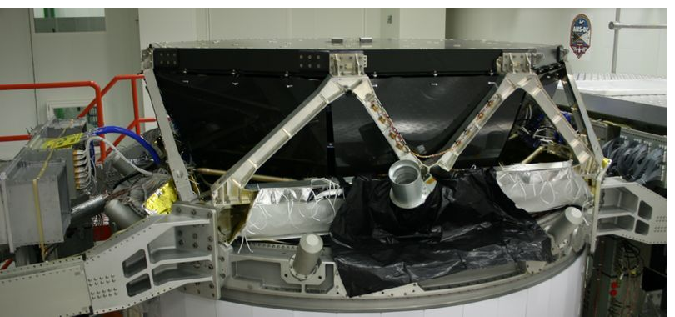,width=10cm}}\vspace{0.5cm}\captionof{figure}{\label{f-trd}TRD on top of AMS-02 during the pre-integration.}
\end{minipage}
\hspace{.1\linewidth}
\begin{minipage}[b]{.24\linewidth}
\centerline{\epsfig{file=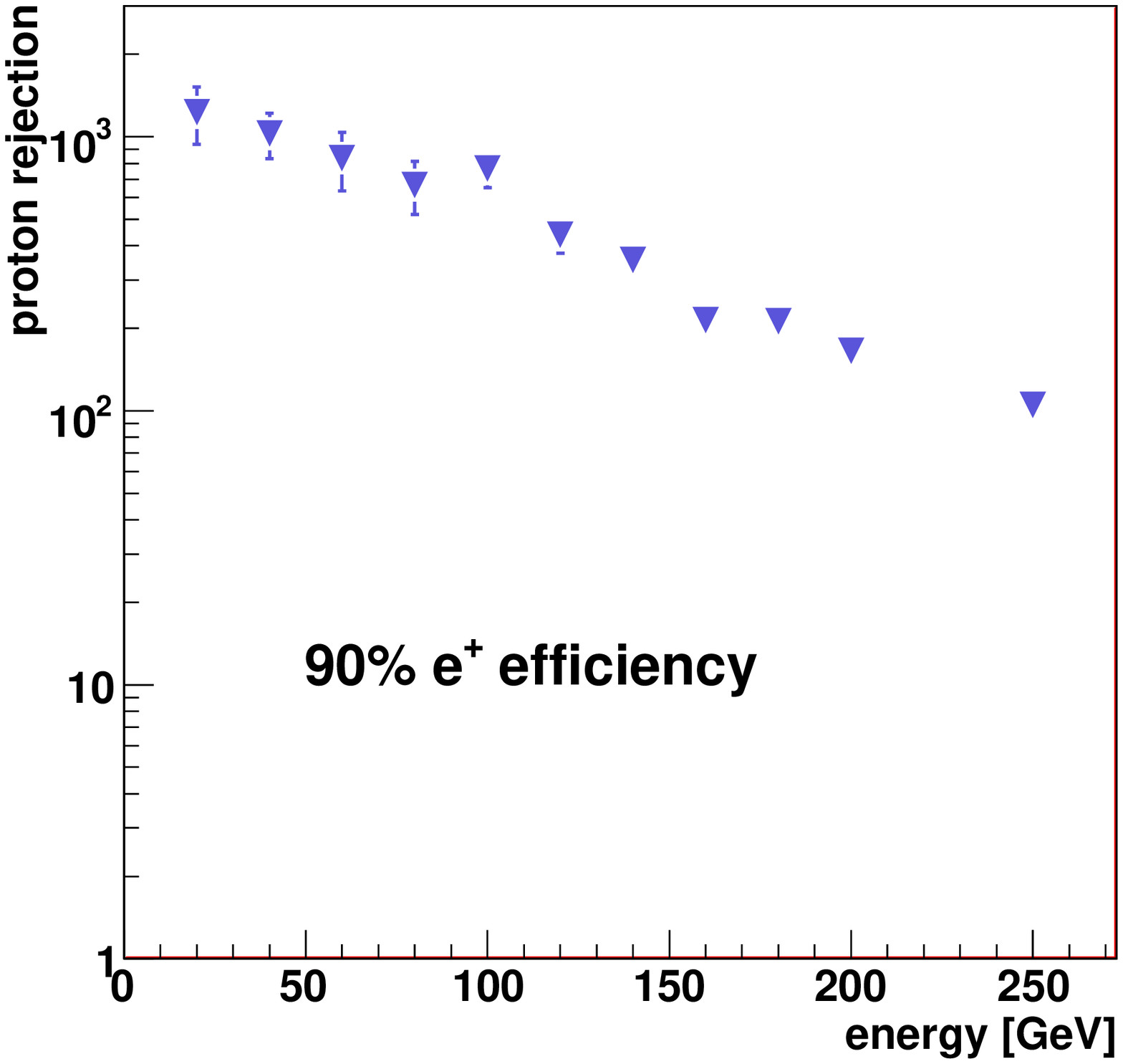,width=6cm}}\captionof{figure}{\label{f-combinedNNrej}Proton rejection at 90\,\% positron efficiency vs. proton energy \cite{doetinchem-2006-558}.}
\end{minipage}
\end{center}
\begin{center}
\centerline{\epsfig{file=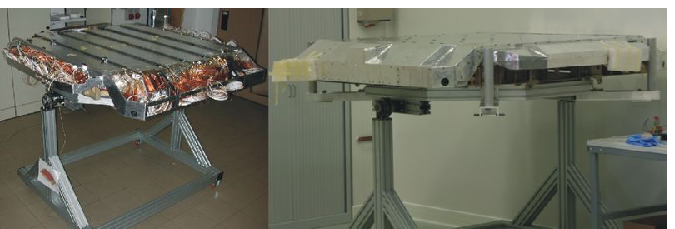,width=15cm}}
\captionof{figure}{\label{f-tof}Time of Flight detector: \textbf{\textit{Left)}} Lower TOF before packing. \textbf{\textit{Right)}} Upper TOF in the AMS-02 clean room.}
\end{center}
\begin{center}
\begin{minipage}[b]{.4\linewidth}
\centerline{\epsfig{file=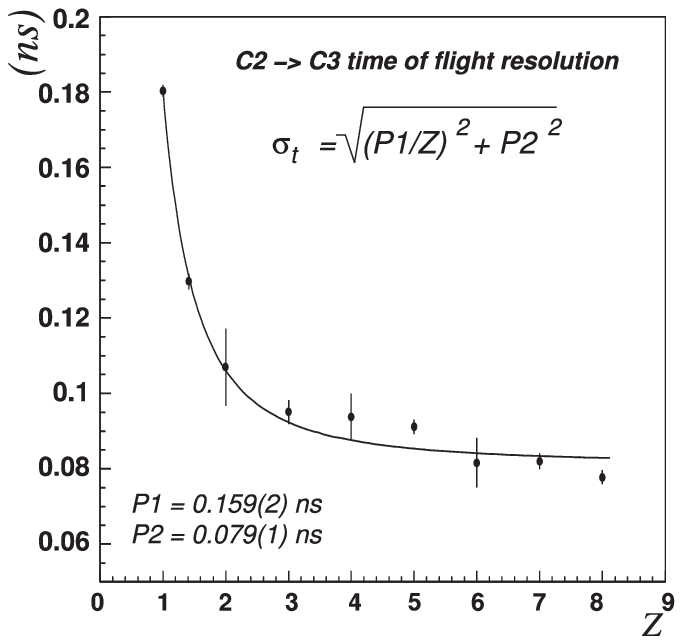,width=8cm}}\captionof{figure}{\label{f-tof_time}TOF: time resolution vs. charge \cite{bindi-2005}.}
\end{minipage}
\hspace{.1\linewidth}
\begin{minipage}[b]{.4\linewidth}
\centerline{\epsfig{file=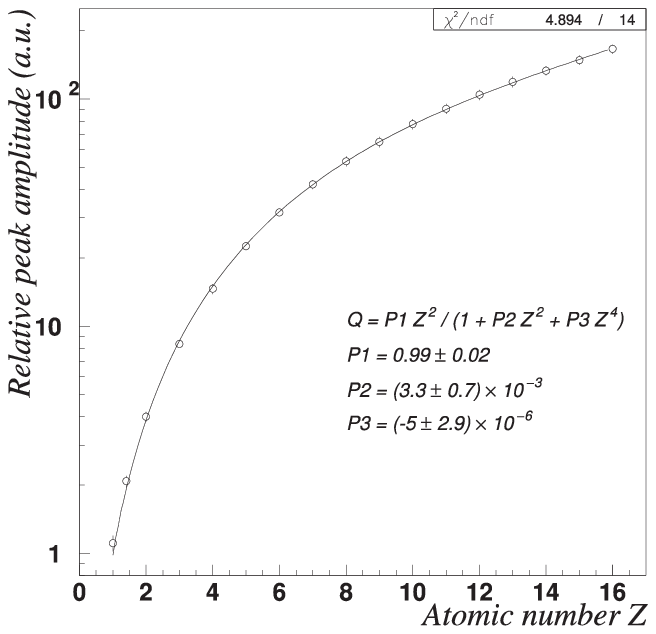,width=7.7cm}}\captionof{figure}{\label{f-tof_peak_z}TOF: amplitude vs. charge \cite{bindi-2005}.}
\end{minipage}
\end{center}
\end{figure}

The transition radiation detector (Fig.~\ref{f-trd}) is located at the very top of AMS-02. It discriminates between particles using the transition radiation effect \cite{ginzburg-1945}. The efficiency depends on the Lorentz factor $\gamma=E/m$. Therefore, this effect gives good discrimination between $e^+$ ($\bar p$) and $p$ ($e^-$) up to momenta of $\cal{O}$(100\,GeV). The TRD is needed together with the electromagnetic calorimeter to separate protons from positrons and electrons from antiprotons. The formation and detection probability of transition radiation is increased by using a detector of 20 layers radiator fleece interleaved with 20 layers of proportional tubes filled with Xe/CO$_2$ gas. The first and last four layers are rotated by 90° with respect to the twelve layers in the middle to gain three dimensional track information. The rejection is defined as the ratio of the total number of protons (electrons) to the misidentified number of protons (electrons) at a given detection efficiency for positrons (antiprotons). The rejection requirement on the detector is determined by the ratio of background to signal (Fig.~\ref{f-ratio_mod}). Fig.~\ref{f-combinedNNrej} shows the proton rejection at 90\,\% electron efficiency as measured in a testbeam. The requirement for a rejection in the order of $10^2$ - $10^3$ for particle energies in the range of 5 - 300\,GeV is fulfilled \cite{doetinchem-2006-558}. 

\subsubsection{Star Tracker}

Mounted on the TRD structure is a system of two star tracker cameras. They are needed for a precise AMS-02 position determination based on fix stars to reconstruct the coordinates in the sky of astronomical sources, e.g. of high energetic $\gamma$-rays.

\begin{figure}
\begin{center}
\centerline{\epsfig{file=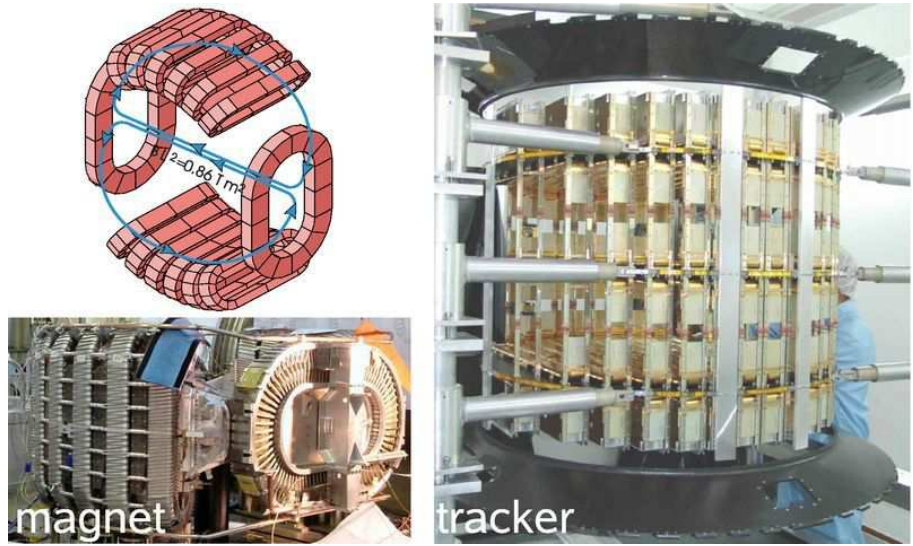,width=12cm}}
\captionof{figure}{\label{f-tr_magnet}Superconducting magnet and tracker: \textbf{\textit{Left)}} Design and manufactured magnet. \textbf{\textit{Right)}} Tracker \cite{ams}.}
\end{center}
\end{figure}

\subsubsection{Time of Flight (TOF)}

The main trigger of AMS-02 will be based on the time of flight system \cite{bindi-2005}. It consists of two sets of plastic scintillators at a distance of 1.3\,m. Each set is made out of two crossed planes to provide three dimensional information on the particle trajectory. The readout will be done with fine mesh photomultiplier tubes (PMT) which also work in the stray field of the superconducting magnet. The time resolution is shown in Fig.~\ref{f-tof_time}  as a function of the particle charge $Z$ and is of the order of 100\,ps. Using the energy deposition of a particle in matter which is proportional to the charge $Z^2$, the TOF is able to measure the absolute value of the charge. Fig.~\ref{f-tof_peak_z} shows the increase of the most probable (MOP) amplitudes of the PMTs with increasing charge.

\subsubsection{Magnet and Tracker}

The superconducting magnet produces a magnetic field of $B\approx0.8$\,T where aluminum enriched NbTi wires carry a current of 459\,A and will be cooled by evaporating superliquid helium. The tank will have enough helium to keep the magnet operational for 3 years (2500\,$\ell$) (Fig.~\ref{f-tr_magnet}, left).

The silicon microstrip tracker (Fig.~\ref{f-tr_magnet}, right) is located at the center of AMS-02 and is surrounded by the superconducting magnet \cite{ams}. The tracker is used for track reconstruction and momentum measurement in the magnetic field. It consists of eight thin layers of double-sided silicon microstrip detectors with a total area of 6.45\,m$^2$ and provides a resolution of 9\,\textmu m per layer in the bending plane of the magnet and 30\,\textmu m in the perpendicular plane \cite{Cecchi2004145}. Six layers are inside the magnetic field. The distance between the upper and lower inner layers is 0.8\,m and the distance between the upper and lower planes outside the magnetic field is 1\,m. The effective proton momentum resolution is:
\be\frac{\displaystyle\sigma_{p}}{\displaystyle p}=\frac{\displaystyle 0.04\,\%\cdot p}{\displaystyle\text{GeV}}\oplus1.5\,\%\label{e-tr}\ee
where $p$ is the proton momentum. Charge separation is possible up to iron nuclei ($Z=26$). Good knowledge of the ladder positions is very important for a high precision track reconstruction. Therefore, a laser alignment system is used which determines the relative position of the ladders to within 5\,\textmu m.

\subsubsection{Anticoincidence Counter (ACC)}

\begin{figure}
\begin{center}
\centerline{\epsfig{file=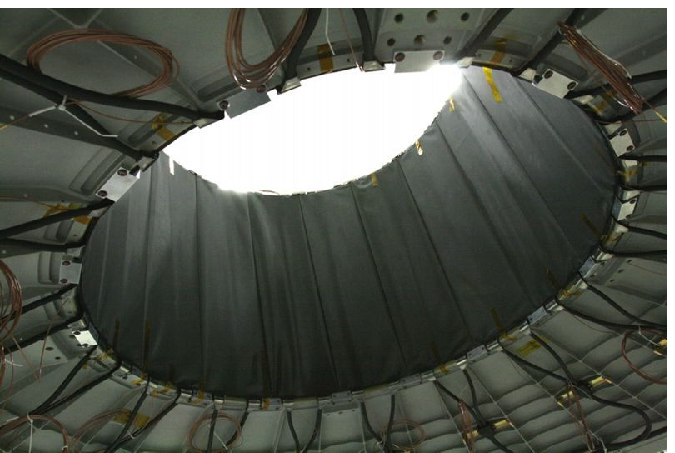,width=10cm}}\captionof{figure}{\label{f-acc}ACC during the pre-integration.}
\end{center}
\begin{center}
\centerline{\epsfig{file=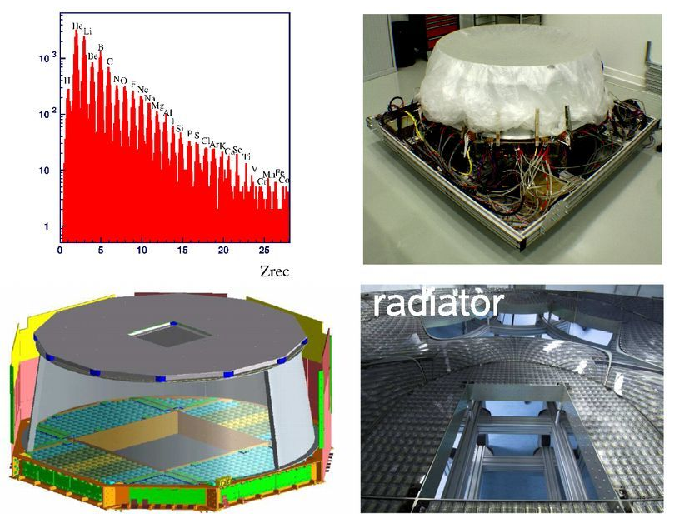,width=10cm}}\captionof{figure}{\label{f-rich_total}RICH detector: \textbf{\textit{Left)}} Charge resolution and design. \textbf{\textit{Right)}} Detector during the pre-integration and the radiator inside \cite{pereira-2008,ams}.}
\end{center}
\begin{center}
\begin{minipage}[b]{.375\linewidth}
\centerline{\epsfig{file=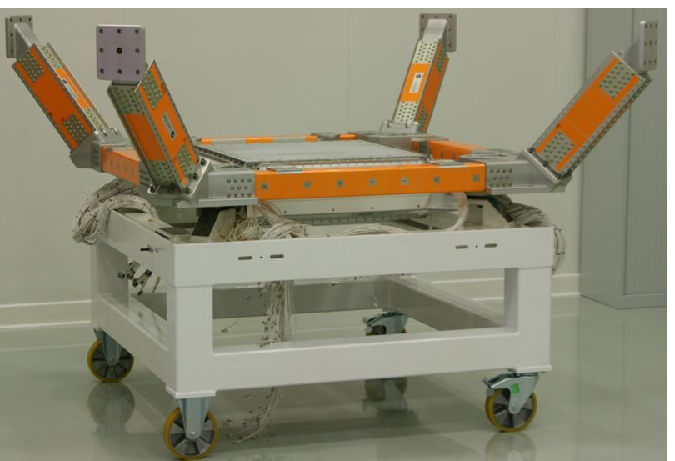,width=7.5cm}}\captionof{figure}{\label{f-ecal}Electromagnetic calorimeter in the AMS-02 clean room.}
\end{minipage}
\hspace{.1\linewidth}
\begin{minipage}[b]{.425\linewidth}
\centerline{\epsfig{file=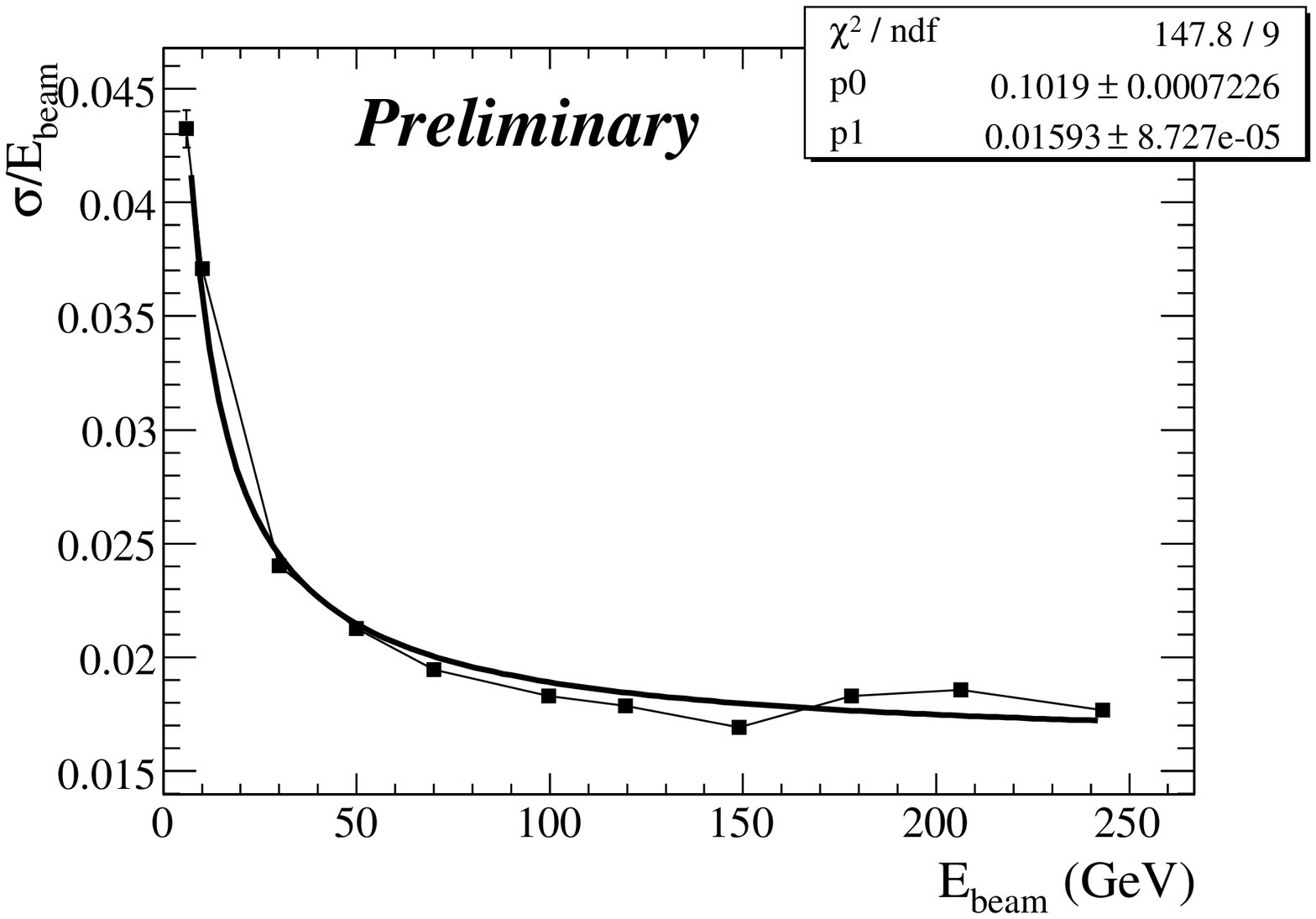,width=8.5cm}}\vspace{-0.4cm}\captionof{figure}{\label{f-ecal_res}Energy resolution of ECAL \cite{goy-2008}.}
\end{minipage}
\end{center}
\end{figure}

The anticoincidence counter (Fig.~\ref{f-acc}) is a cylinder made out of 16 plastic scintillator panels and surrounds the tracker in order to assure clean events. Events with particles crossing the detector from the side, backscattering from  the electromagnetic calorimeter or particles interacting within the tracker or elsewhere can produce hits in the ACC and can thus be rejected. The panels are read out by the same type of photomultiplier tubes used in the TOF. The ACC detector is treated in more detail in Sec.~\ref{s-acc}.

\subsubsection{Ring Imaging \v{C}erenkov Counter (RICH)}

The RICH (Fig.~\ref{f-rich_total}) is used to determine with high precision the velocities of particles with mass number $A<15-20$ in the momentum range $1\,\text{GeV}<p/A<12\,\text{GeV}$ \cite{pereira-2008}. While the small difference between ${}^9$Be and ${}^{10}$Be cannot be resolved by TOF and Tracker, the RICH is able to do so using the \v{C}erenkov effect, namely the emission of characteristic light by particles travelling faster than the speed of light in the propagation medium. The RICH measures the velocity with a precision of about 0.1\,\%. Together with the tracker momentum measurement the particle mass can be determined:
\be m= \frac p\beta\sqrt{1-\beta^2}.\ee

The characteristic \v{C}erenkov ring images form in the two radiator layers on the upper side of the RICH detector. One layer is made of Silica Aeorgel with a refractive index $n=1.05$ and the central blocks are made of NaF with $n=1.34$ which gives a wider \v{C}erenkov cone. This increases the photon detection efficiency because of the hole in the middle of the detection plane which is needed to assure undisturbed tracks for the electromagnetic calorimeter. The photons will be detected by photomultipliers. In addition, a conical mirror between radiator and detector layer increases the acceptance.

\subsubsection{Electromagnetic Calorimeter (ECAL)}

In combination with the transition radiation detector the electromagnetic calorimeter (Fig.~\ref{f-ecal}) is used to distinguish between light and heavy particles through the electromagnetic shower shape. The ECAL consists of nine super layers, each made of ten layers of lead with embedded scintillating fibers. To provide spatial resolution the superlayers are placed at 90° with respect to each other. The scintillation light is measured with photomultipliers placed at granularity of 0.5\,Molière radii in the $x$ and $y$ direction. The radiation length $X_0$ is about 10\,mm and the total ECAL length corresponds to 16.7\,$X_0$ giving an energy resolution of (Fig.~\ref{f-ecal_res}):
\be\frac\sigma{E}=\frac{10.2\,\%}{\sqrt{E/\text{GeV}}}\oplus 1.6\,\%.\ee
The different shower profiles for $\stackrel{_{\tiny{(-)}}}{p}$ and $e^\pm$ give a rejection of $10^2$ - $10^3$ for particle energies of $10$ - $500$\,GeV. A requirement on the $E/p$ ratio using the momentum measured by the tracker provides an additional factor of 10. The ECAL can also be used to provide a standalone trigger for photons.

\subsection{Mission Objectives}

\begin{figure}
\begin{center}
\begin{minipage}[b]{.4\linewidth}
\centerline{\epsfig{file=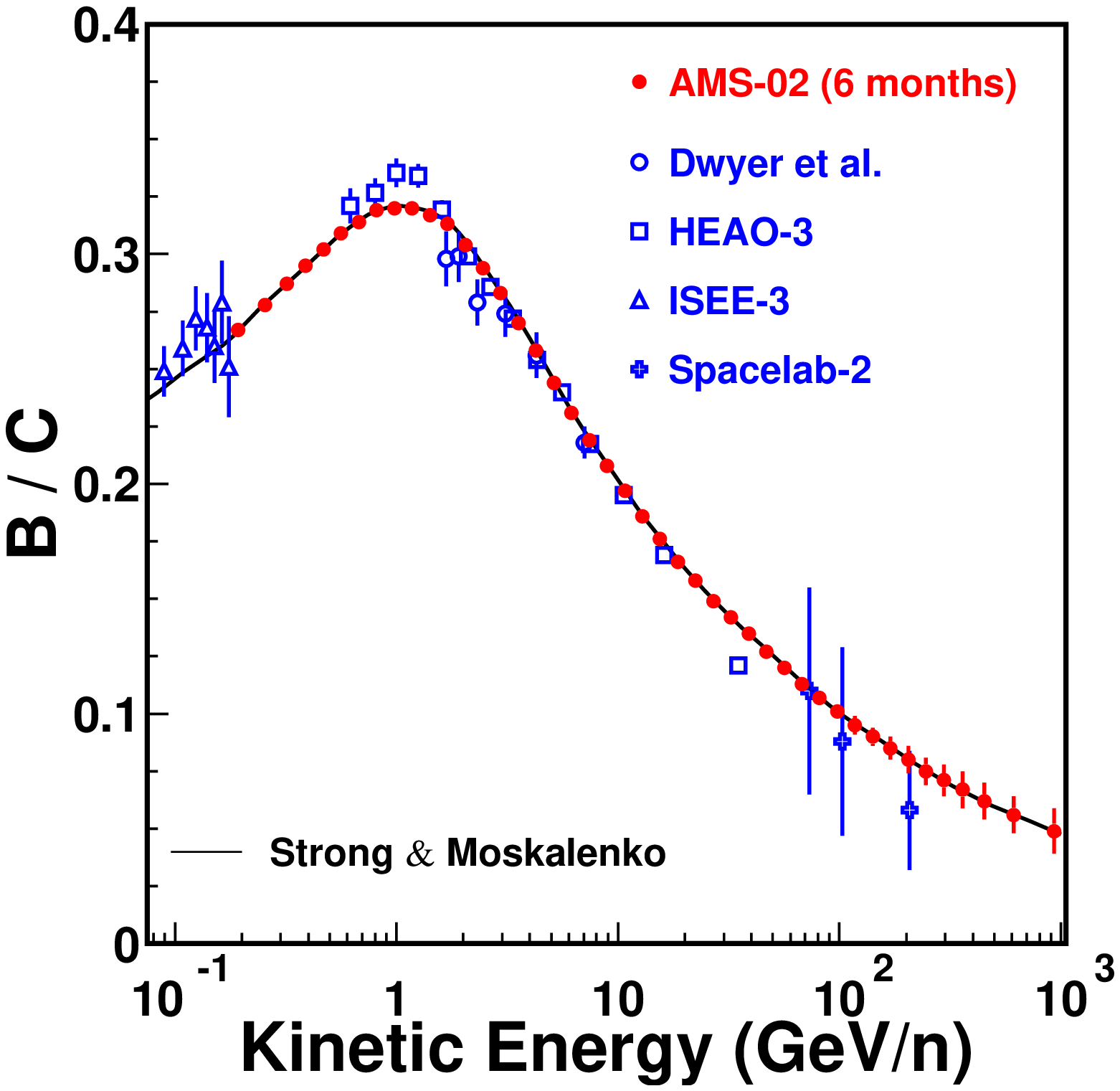,width=8cm}}\captionof{figure}{\label{f-b_c}Projected boron to carbon ratio of AMS-02 and current data \cite{ams}.}
\end{minipage}
\hspace{.1\linewidth}
\begin{minipage}[b]{.4\linewidth}
\centerline{\epsfig{file=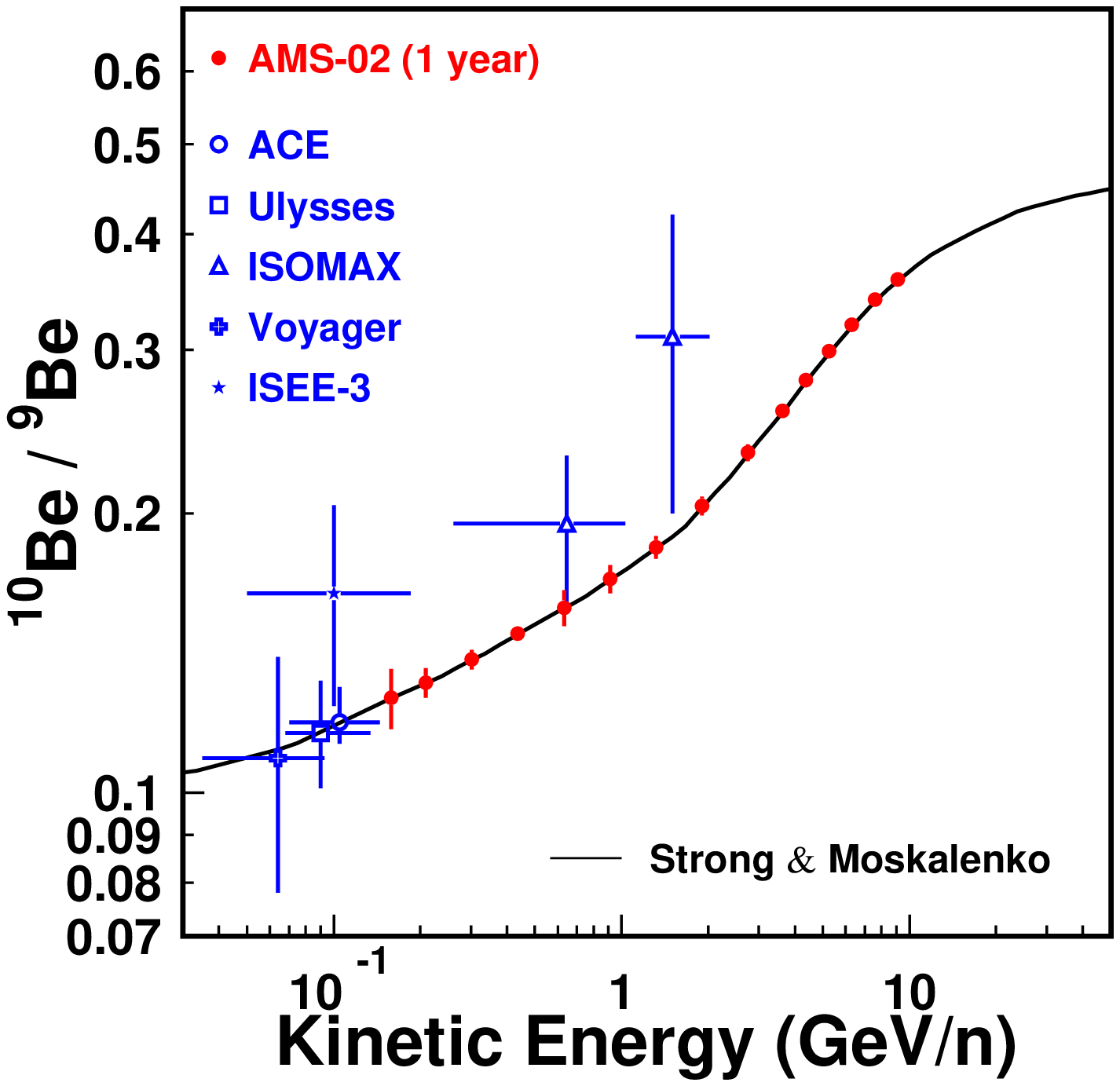,width=8cm}}\captionof{figure}{\label{f-10be_9be_AMS}Projected beryllium ${}^{10}$Be/${}^{9}$Be ratio and current data \cite{ams}.}
\end{minipage}
\end{center}
\end{figure}
\begin{figure}
\begin{center}
\begin{minipage}[b]{.4\linewidth}
\centerline{\epsfig{file=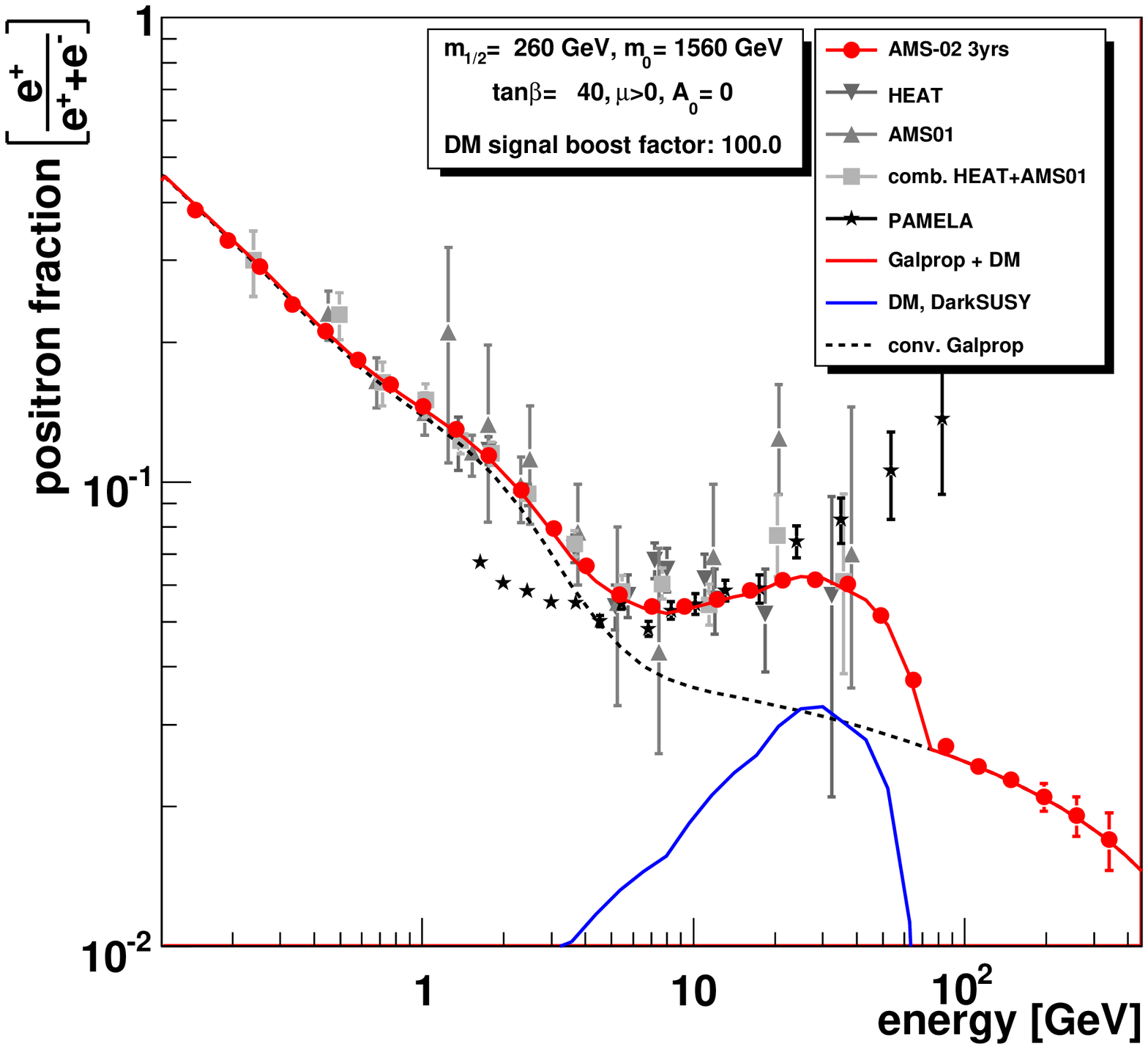,width=8cm}}
\captionof{figure}{\label{f-pf}Projected positron fraction of 3 year AMS-02 and current data including contribution of neutralino annihilations \cite{gast-2008}.}
\end{minipage}
\hspace{.1\linewidth}
\begin{minipage}[b]{.4\linewidth}
\centerline{\epsfig{file=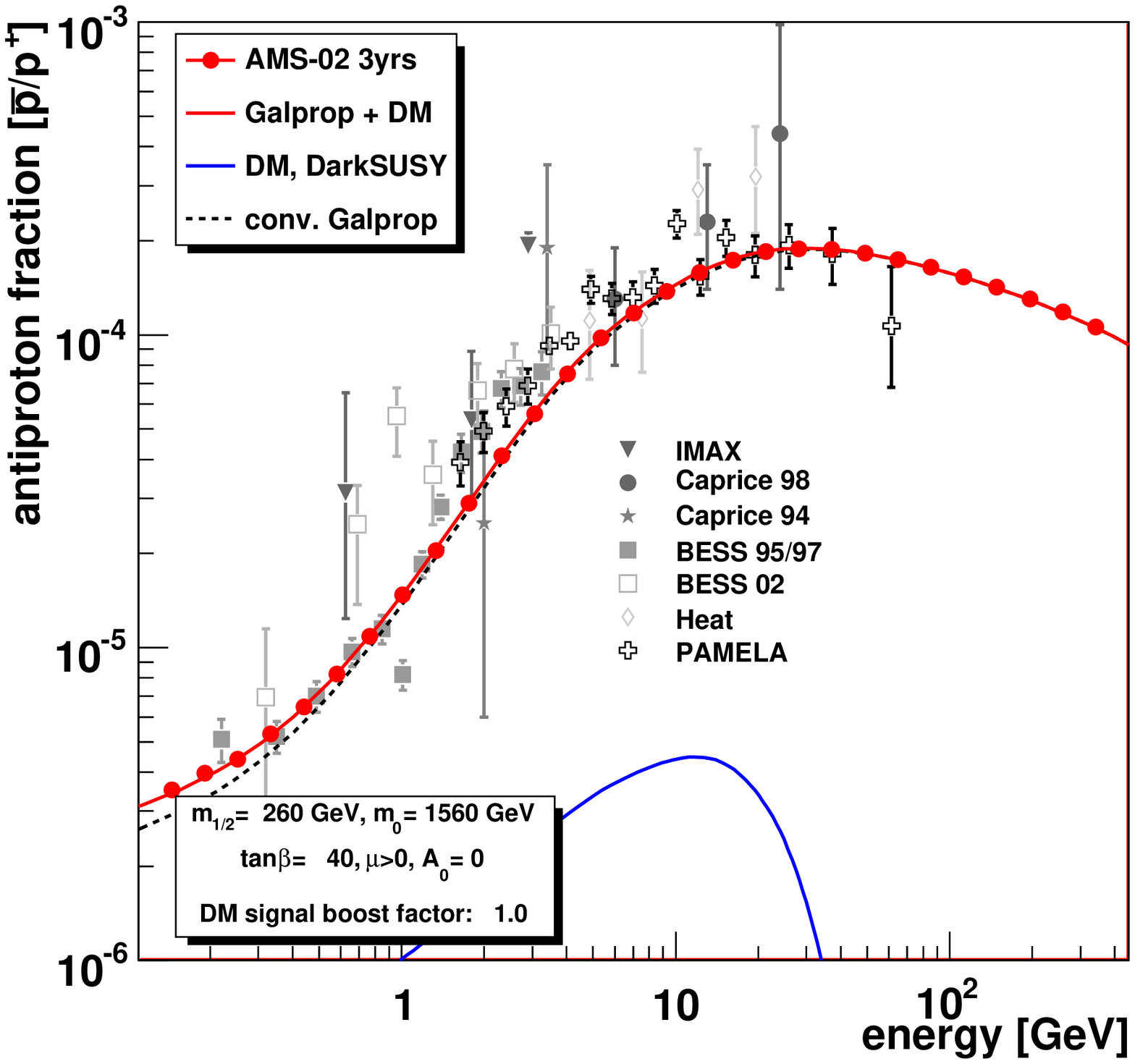,width=8cm}}
\captionof{figure}{\label{f-apf}Projected antiproton fraction of 3 year AMS-02 and current data including contribution of neutralino annihilations \cite{gast-2008}.}
\end{minipage}
\end{center}
\end{figure}
\begin{figure}
\begin{center}
\begin{minipage}[b]{.4\linewidth}
\centerline{\epsfig{file=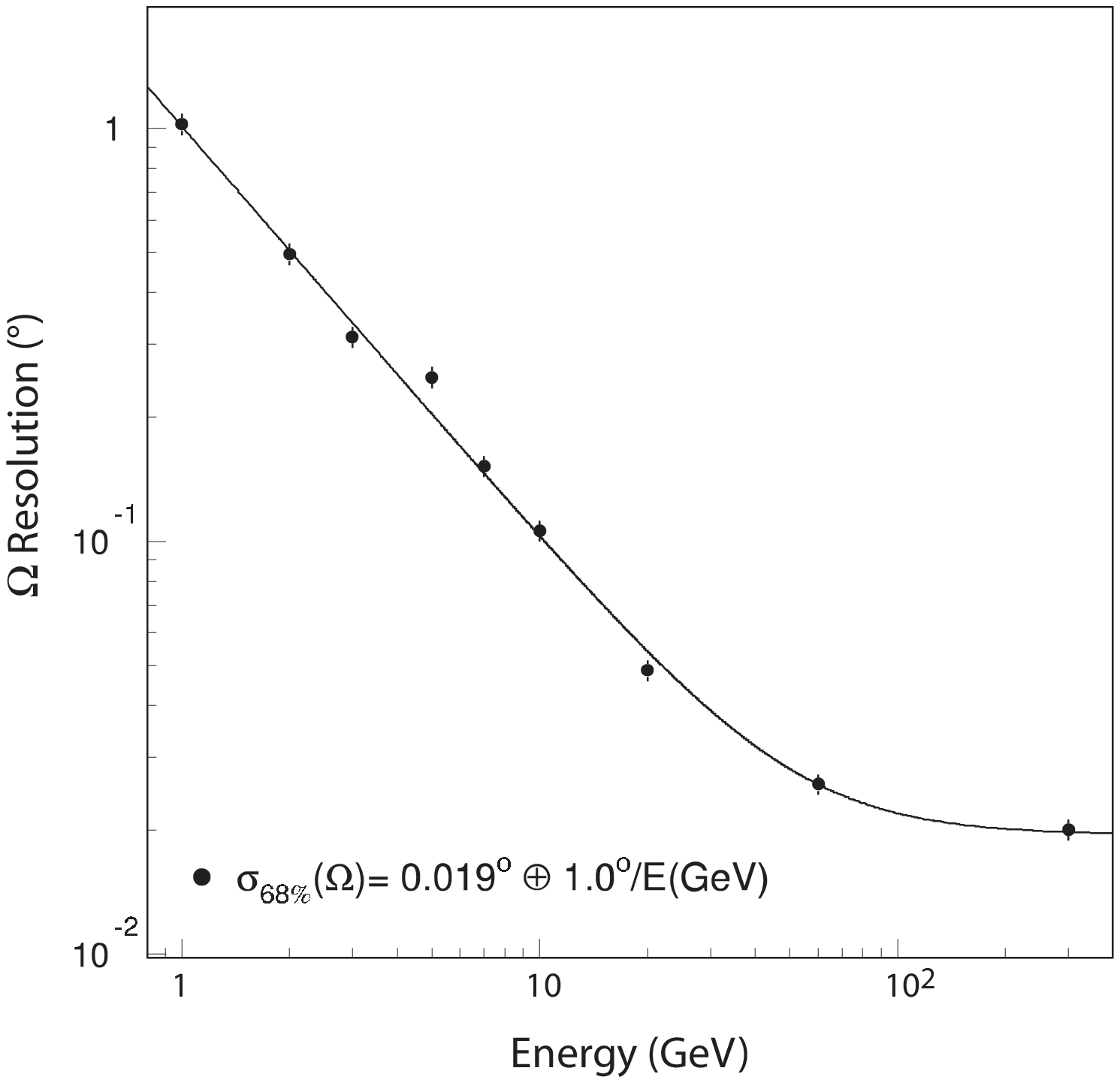,width=8cm}}
\captionof{figure}{\label{f-gamma_res_AMS}Angular resolution for photon detection\cite{ams}.}
\end{minipage}
\hspace{.1\linewidth}
\begin{minipage}[b]{.4\linewidth}
\centerline{\epsfig{file=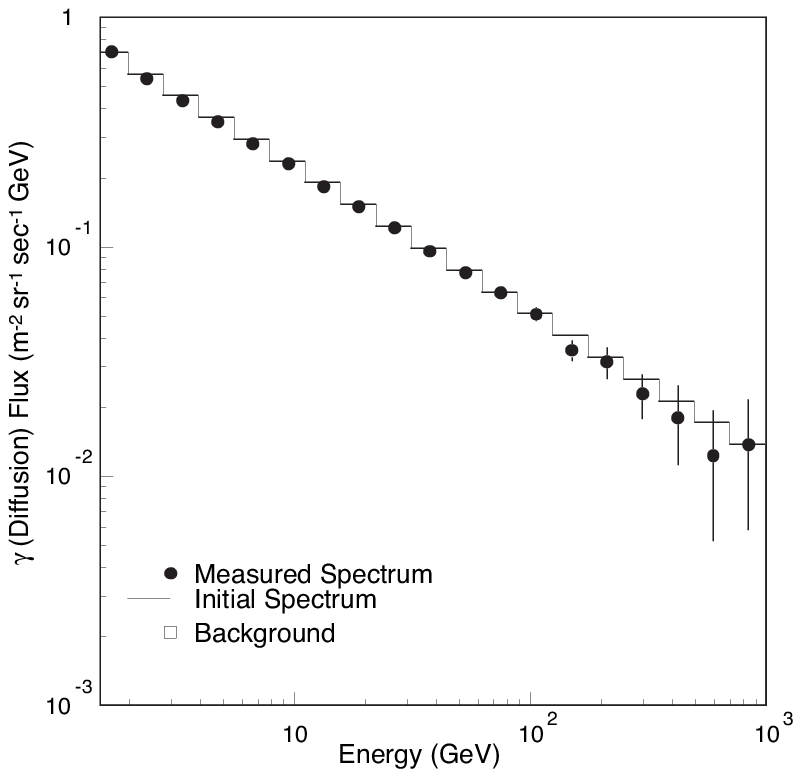,width=8cm}}
\vspace{0.1cm}
\captionof{figure}{\label{f-gamma_AMS}Projected diffuse galactic photon flux\cite{ams}.}
\end{minipage}
\end{center}
\end{figure}

The AMS-02 experiment has several objectives. It will measure over 3 years cosmic rays in the energy range up to 500\,GeV with very high precision. Given the large acceptance of 0.095\,m$^2$sr \cite{choutko2} approximately $10^{10}$ particles, nuclei and isotopes will be collected. These data will be used to test existing theories and to look for new phenomena.

One of the most important tasks is to fix parameters in the cosmic-ray propagation models, e.g. by looking at the boron to carbon ratio (Fig.~\ref{f-b_c}), the determination of the age of cosmic rays and the galactic halo size by analyzing the ratio of ${}^{10}$Be/${}^{9}$Be using the decay of ${}^{10}$Be with a half-life $\tau({}^{10}\text{Be})=3.6\cdot10^6$\,yr (Fig.~\ref{f-10be_9be_AMS}). The figures also show the projected AMS-02 results after a measurement time of 1\,year and 6\,months, respectively.

As for PEBS (Chap.~\ref{c-pebs}) another focus is the indirect search for dark matter through the measurement of the antiparticle and, in addition, photon fluxes. Fig.~\ref{f-pf} and \ref{f-apf} show the positron and antiproton fractions for a supersymmetric dark matter scenario \cite{gast-2008} based on the statistics expected from 3\,year of AMS-02 operation without detector efficiencies and geomagnetic modulation.

In addition, photons are important since they point back to their sources like active galactic nuclei (AGN) or gamma ray bursts (GRB). It will be possible to measure photons up to 1\,TeV. For low energies, $\gamma\rightarrow e^+e^-$ conversions before the tracker give the possibility to determine the initial photon direction with high precision. For high energies, the electromagnetic cascades of the photons in the ECAL are used for the analysis. The angular resolution is about 1° for low energies  and about $2\cdot10^{-2}$° at 300\,GeV (Fig.~\ref{f-gamma_res_AMS}). The projected diffuse galactic spectrum is shown in Fig.~\ref{f-gamma_AMS}.

The importance of antimatter searches has been explained above (Sec.~\ref{s-cr}). AMS-02 will be able to significantly improve the antihelium measurements. Within 3 years of measurement the limit for $\overline{\text{He}}/\text{He}$ will be $\cal{O}$($10^{-9}$) for rigidities up to about 150\,GV and $\cal{O}$($10^{-4}$) up to about 1\,TeV (Fig.~\ref{f-antihelium_AMS}).

Also interesting is the search for anomalously heavy nuclei like strangelets. They are believed to contain $s$ quarks besides $u$ and $d$ quarks. AMS-01 found one candidate with a charge to mass ratio of $Z/A=0.11$. If such particles really exist AMS-02 will find 200 events in 3 years of operation (Fig.~\ref{f-strangelet_AMS}).

In addition to the astrophysics results, AMS-02 will provide experience in the operation of superconducting magnets in space. This can be useful for field plasma rockets or for radiation protection in manned space flights.

\begin{center}
\begin{minipage}[b]{.4\linewidth}
\centerline{\epsfig{file=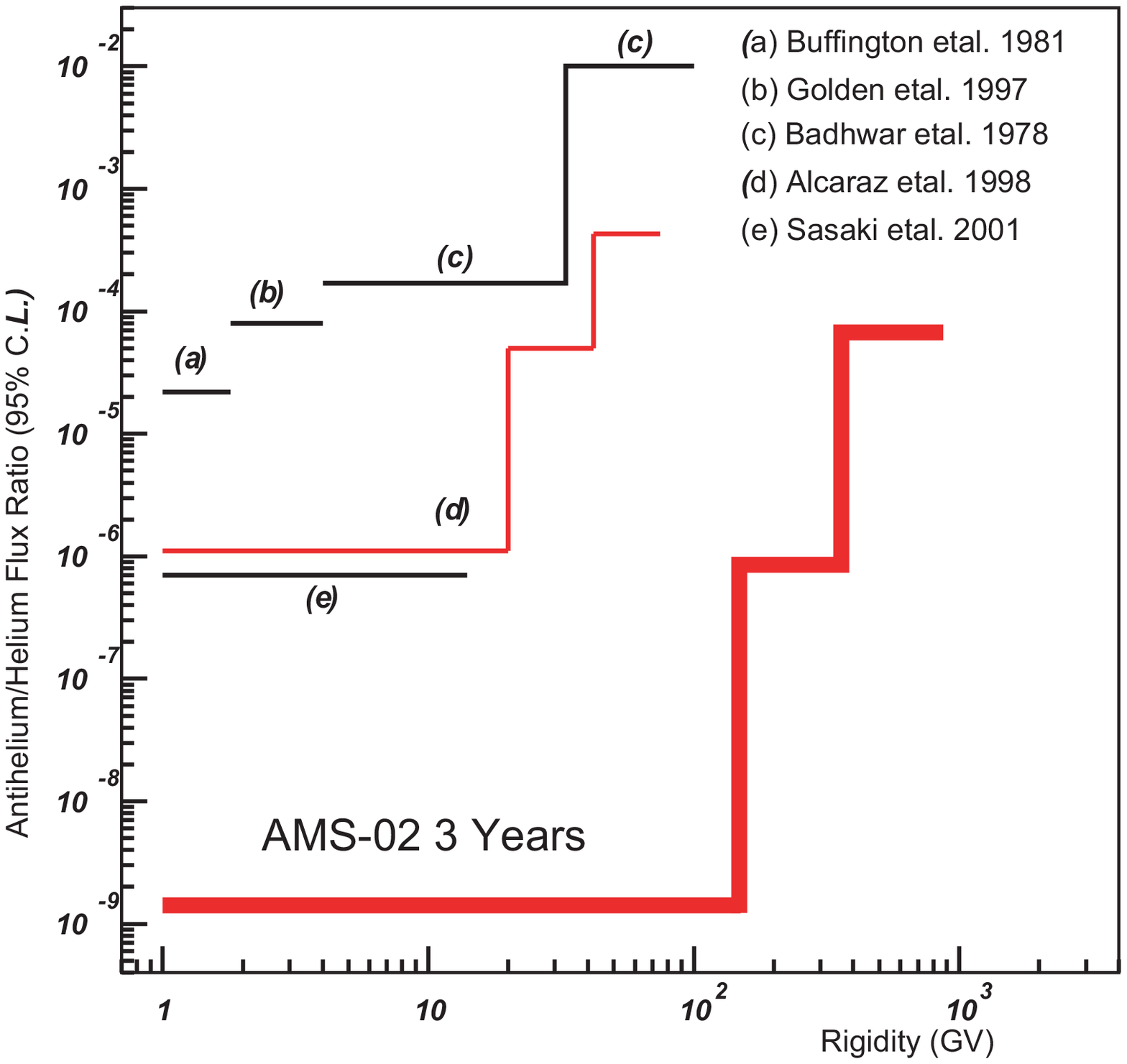,width=8cm}}\captionof{figure}{\label{f-antihelium_AMS}Antihelium search\cite{ams}.}
\end{minipage}
\hspace{.1\linewidth}
\begin{minipage}[b]{.4\linewidth}
\centerline{\epsfig{file=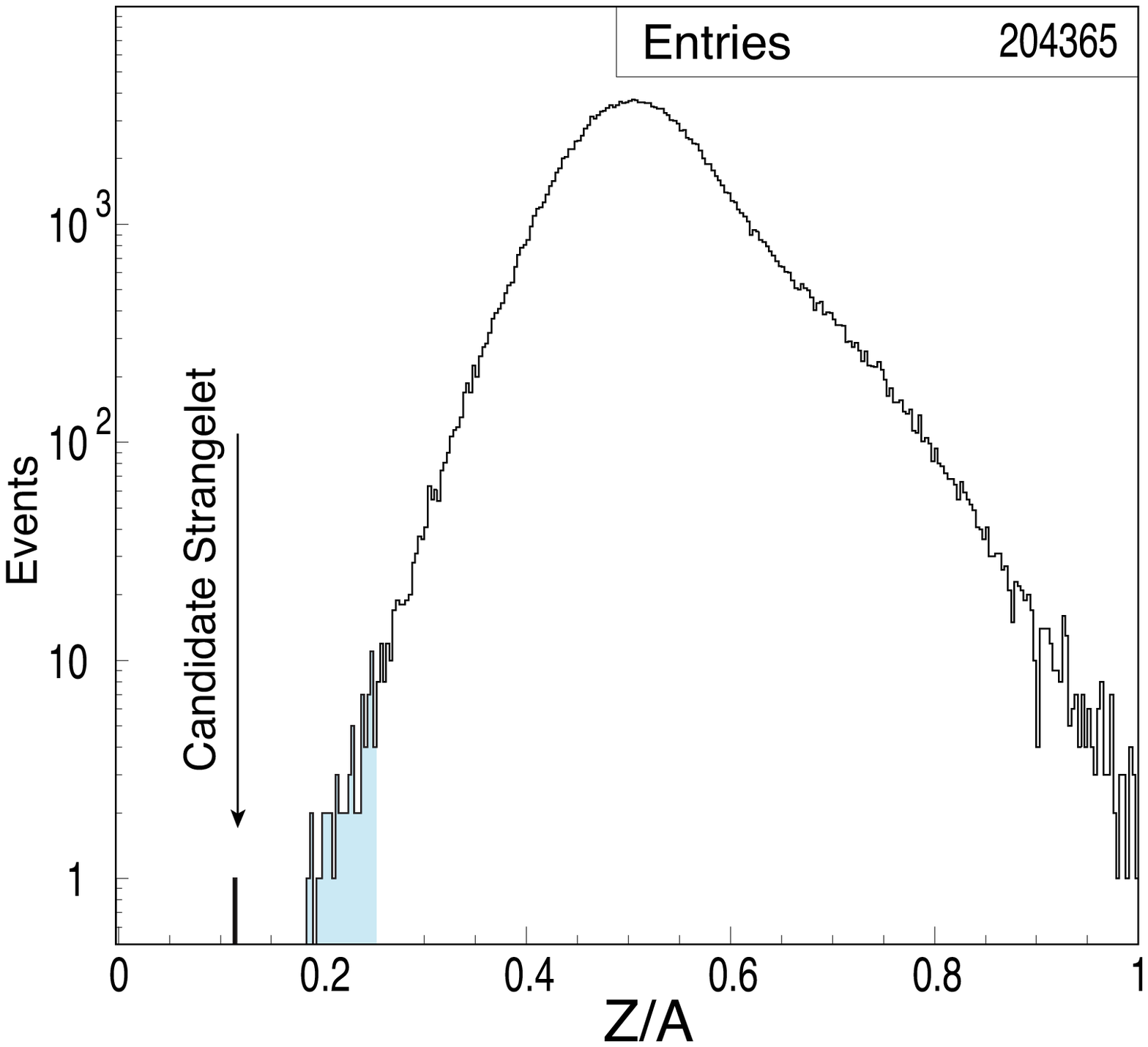,width=8cm}}\vspace{-0.1cm}\captionof{figure}{\label{f-strangelet_AMS}Strangelet search \cite{ams}.}
\end{minipage}
\end{center}

\afterpage{\clearpage}

\pagebreak

\centerline{\epsfig{file=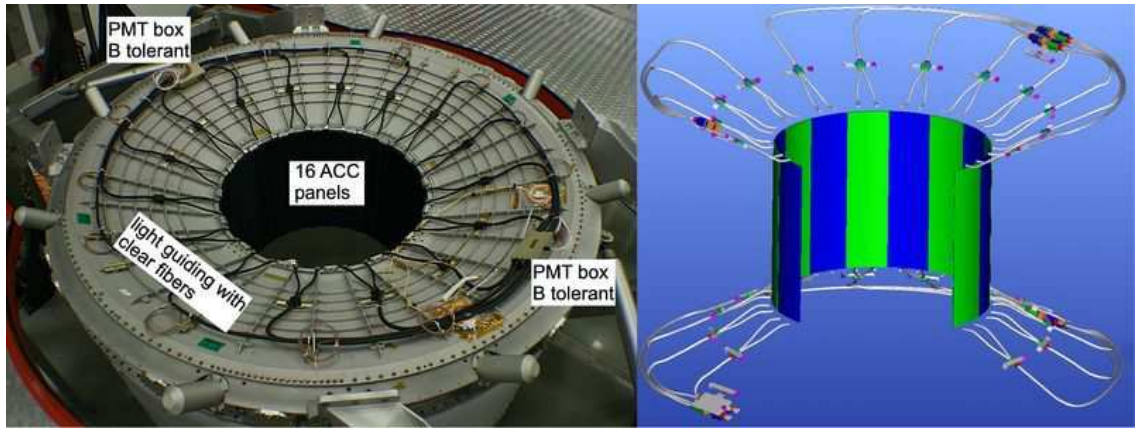,width=15cm}}
\captionof{figure}{\label{f-acc_overview}The AMS-02 anticoincidence counter system surrounds the silicon tracker.}

\section{The Anticoincidence Counter (ACC) \label{s-acc}}

\begin{figure}
\begin{center}
\begin{minipage}[b]{.4\linewidth}
\centerline{\epsfig{file=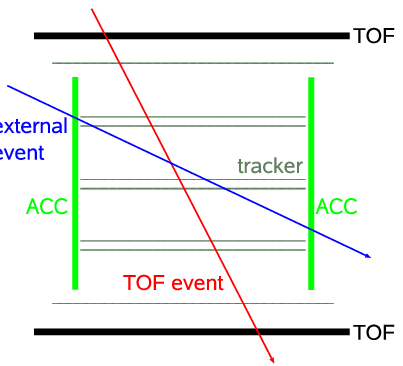,width=8cm}}
\captionof{figure}{\label{f-ext_event_acc}External event in the tracker.}
\end{minipage}
\hspace{.1\linewidth}
\begin{minipage}[b]{.4\linewidth}
\centerline{\epsfig{file=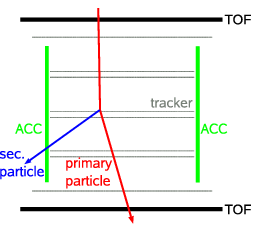,width=7.9cm}}
\vspace{0.1cm}
\captionof{figure}{\label{f-int_event_acc}Internal event in the tracker.}
\end{minipage}
\end{center}

\begin{center}
\centerline{\epsfig{file=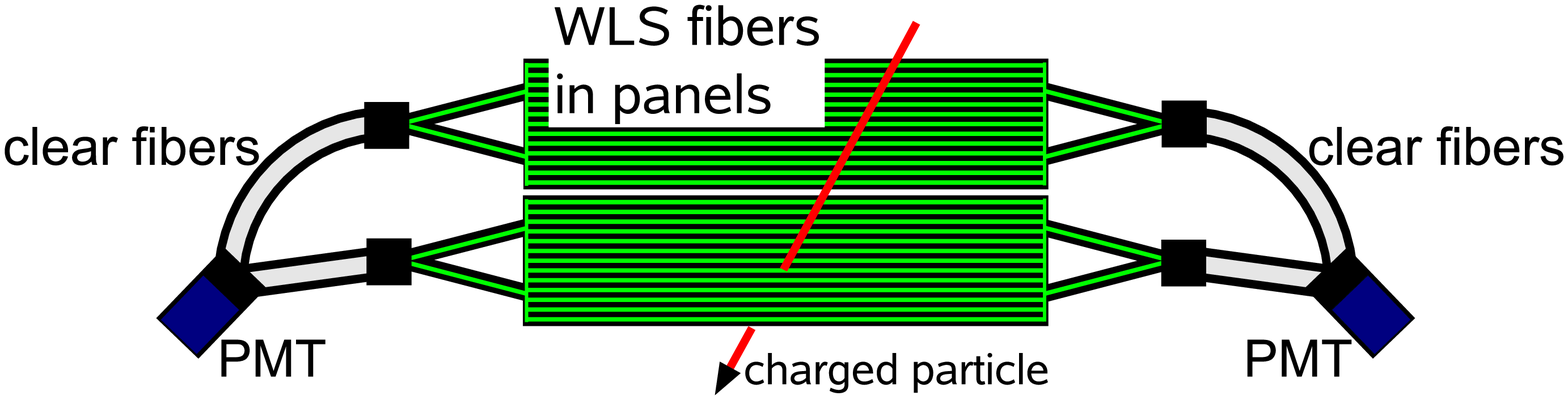,width=15cm}}
\captionof{figure}{\label{f-acc_principle}ACC principle.}
\end{center}
\end{figure}

The AMS-02 anticoincidence counter surrounds the silicon tracker with the purpose of vetoing to assure a clean track reconstruction (Fig.~\ref{f-acc_overview}). Particles entering the detector from the side or from interactions inside it could distort the charge measurement (Fig.~\ref{f-ext_event_acc} and \ref{f-int_event_acc}). This is especially essential for the antimatter measurement. To improve existing upper limits on antihelium an detection inefficiency smaller than $10^{-4}$ is needed. The inefficiency is the ratio of misses to the total number of particles crossing the ACC.

The second important task of the ACC is to reduce the trigger rate during periods of very large fluxes, e.g. in the South Atlantic Anomaly where the geomagnetic cut-off drops and the flux of low energy particles is drastically increased (Fig.~\ref{f-cutoff}). In this case, the ACC detector will be used as a veto for the trigger decision made by the TOF. For that purpose, it is important to use a detector with a fast response. 

The ACC system must meet crucial requirements: In addition to very high detection efficiency, the ACC must be able to withstand the liftoff conditions, operation in space and must tolerate a high magnetic field of 0.12\,T at the positions of the photomultipliers.

The ACC cylinder is made out of 16 scintillator panels with a diameter of 1.1\,m and a thickness of 8\,mm. The ultraviolet scintillation light through ionization losses of charged particles is absorbed by wavelength shifting (WLS) fibers, transformed to a different wavelength and coupled to clear fiber cables for the final transport to photomultiplier tubes (PMT). Light guides are needed because of the high magnetic field of the superconducting magnet. Although the field is self-compensating and $B$-field tolerant fine mesh PMTs have been chosen, photomultiplier operation closer to the panels would be too much distorted by the stray field. A set of two panels is read out by the same two photomultipliers, one on top and one on the bottom, via Y-shaped clear fiber cables in order to have redundancy and to save weight (Fig.~\ref{f-acc_principle}). The total weight of the ACC system is 53.7\,kg. The 16 scintillator panels with the WLS fibers have a weight of 29.2\,kg. Four PMT boxes each with four PMTs with signal, power cables and clear fiber cables contribute 14.4\,kg and additional supports and fixations have a weight of 10.1\,kg.

The ACC data processing and acquisition is done with the electronics inside the S-crate where the data of the TOF are also processed. There are four S-crates in total. One is mounted on bottom and top on each side (RAM ($-y$), WAKE ($+y$)). TOF and ACC share also the power distribution. In total four SHV-bricks are also mounted on bottom and top on each side. The AMS-02  readout trigger decision is taken in the J-crate. The logic has the option to veto the level\,1 trigger generation of the TOF (or ECAL) if the ACC shows a signal above an adjustable threshold.

The ACC signal can be estimated from the properties of its components which will be explained in detail in the next sections. The energy loss of a minimum ionizing particle in the scintillator is about 0.22\,MeV/mm and follows a Landau distribution \cite{pdgbook}. Tracks perpendicular to the panels have a pathlength of 8\,mm in the material which results in about 18,000 photons. The scintillation light is absorbed by the WLS fibers. The absorption efficiency of the fibers can be extracted from knowledge of the other components to be about 1.6\,\% as will be seen later. This includes the relation between active WLS fiber material and the passive scintillator and glue, the transmission probability of light to the core of the fiber and the angular acceptance for transport via total reflection. The WLS fibers are about 1.3\,m long such that the light is transported on average about 0.65\,m from the center of the panel to the PMT. This causes an attenuation by about 10\,\%. The coupling efficiency between the WLS and the clear fiber is on average 60\,\%. Then the photons strike the photocathode of the PMT and electrons arise from the photoelectric effect. The electrons are amplified and the signal at the PMT anode is still proportional to the initial number of photons. The PMTs have a quantum efficiency of about 10\,\% \cite{hamamatsu}. This results in a total expectation of about 16 photo-electrons at the photocathode for the most probable value.

\section{Fabrication, Test and Integration of the Anticoincidence Counter Components\label{s-fabacc}}

The following section describes the properties, fabrication and qualification tests of the ACC hardware components. It discusses also the interplay of all parts in a complete system test and the ACC pre-integration.

\subsection{Production and Test of the Scintillator Panels\label{ss-paneltest}}

\begin{figure}
\begin{center}
\centerline{\epsfig{file=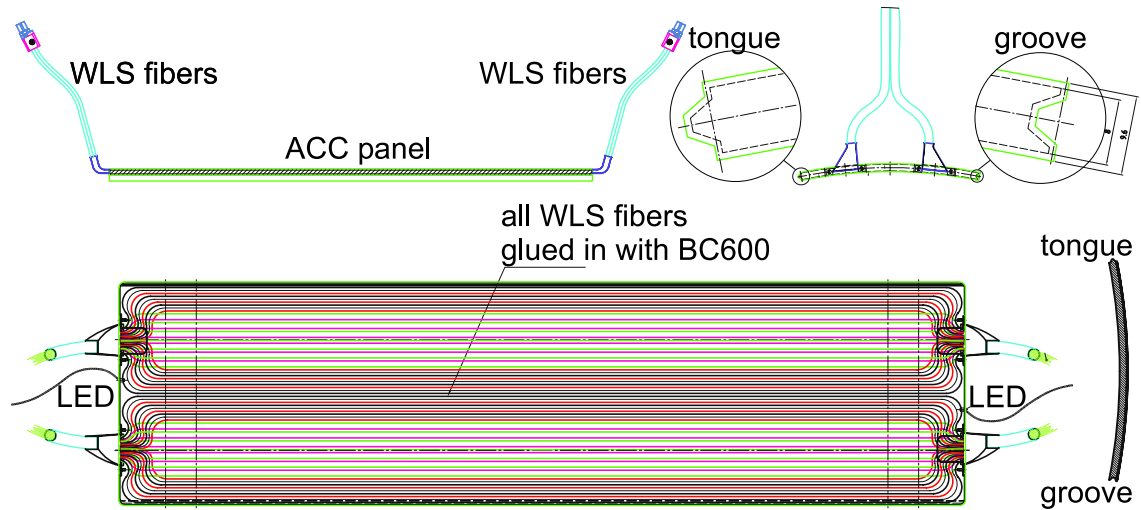,width=15cm}}
\captionof{figure}{\label{f-prod_panel}Design of the ACC panel.}
\end{center}
\end{figure}

\begin{figure}
\begin{center}
\centerline{\epsfig{file=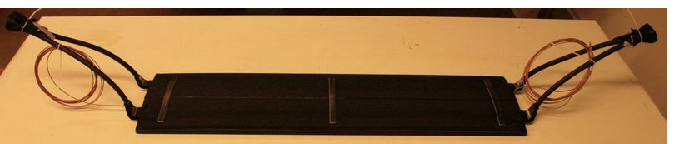,width=15cm}}
\captionof{figure}{\label{f-acc_panel}ACC panel.}
\end{center}
\end{figure}

\begin{figure}
\begin{center}
\centerline{\epsfig{file=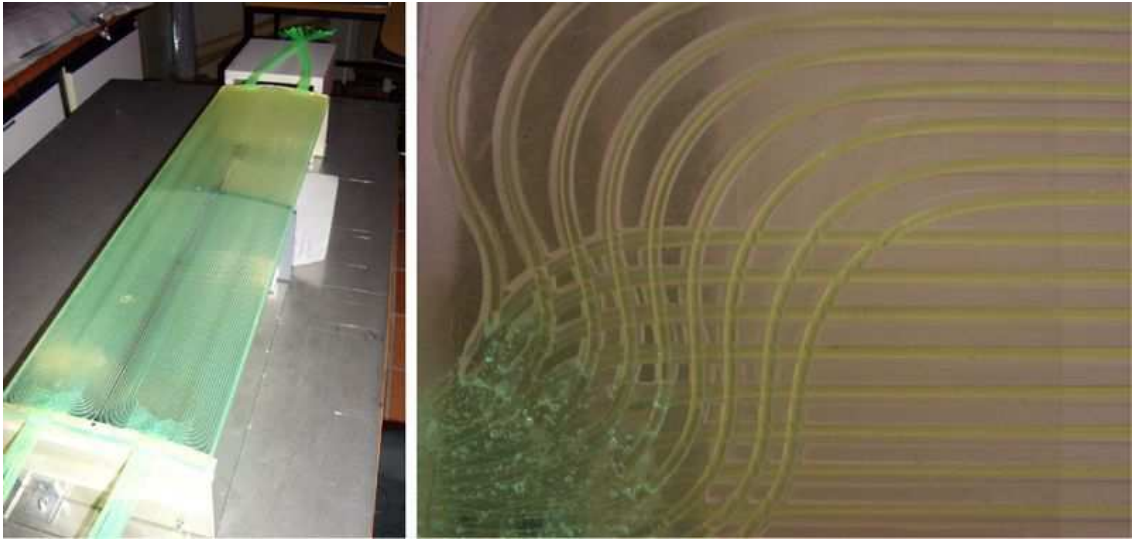,width=15cm}}
\captionof{figure}{\label{f-acc_fiber_glue}WLS fibers inside an ACC panel.}
\end{center}
\end{figure}

\begin{figure}
\begin{center}
\begin{minipage}[b]{.4\linewidth}
\centerline{\epsfig{file=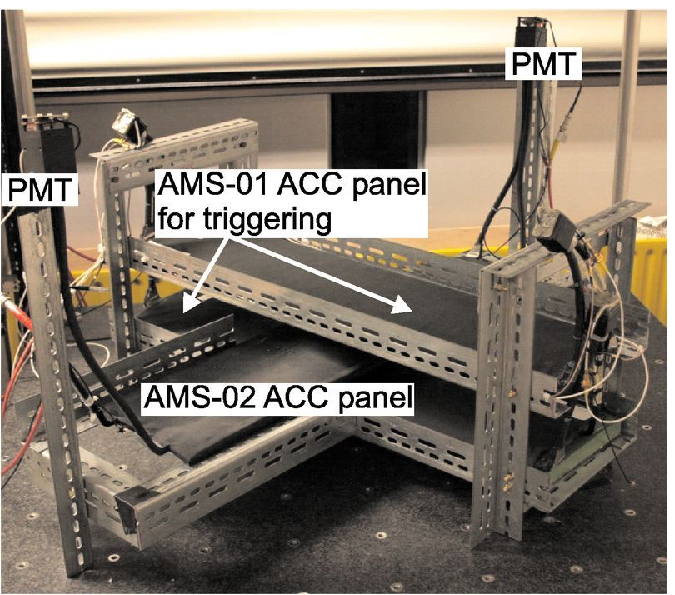,width=8cm}}\captionof{figure}{\label{f-pmt_panel_test}Test setup for the qualification of the scintillator panels and the photomultiplier tubes.}
\end{minipage}
\hspace{.1\linewidth}
\begin{minipage}[b]{.4\linewidth}
\centerline{\epsfig{file=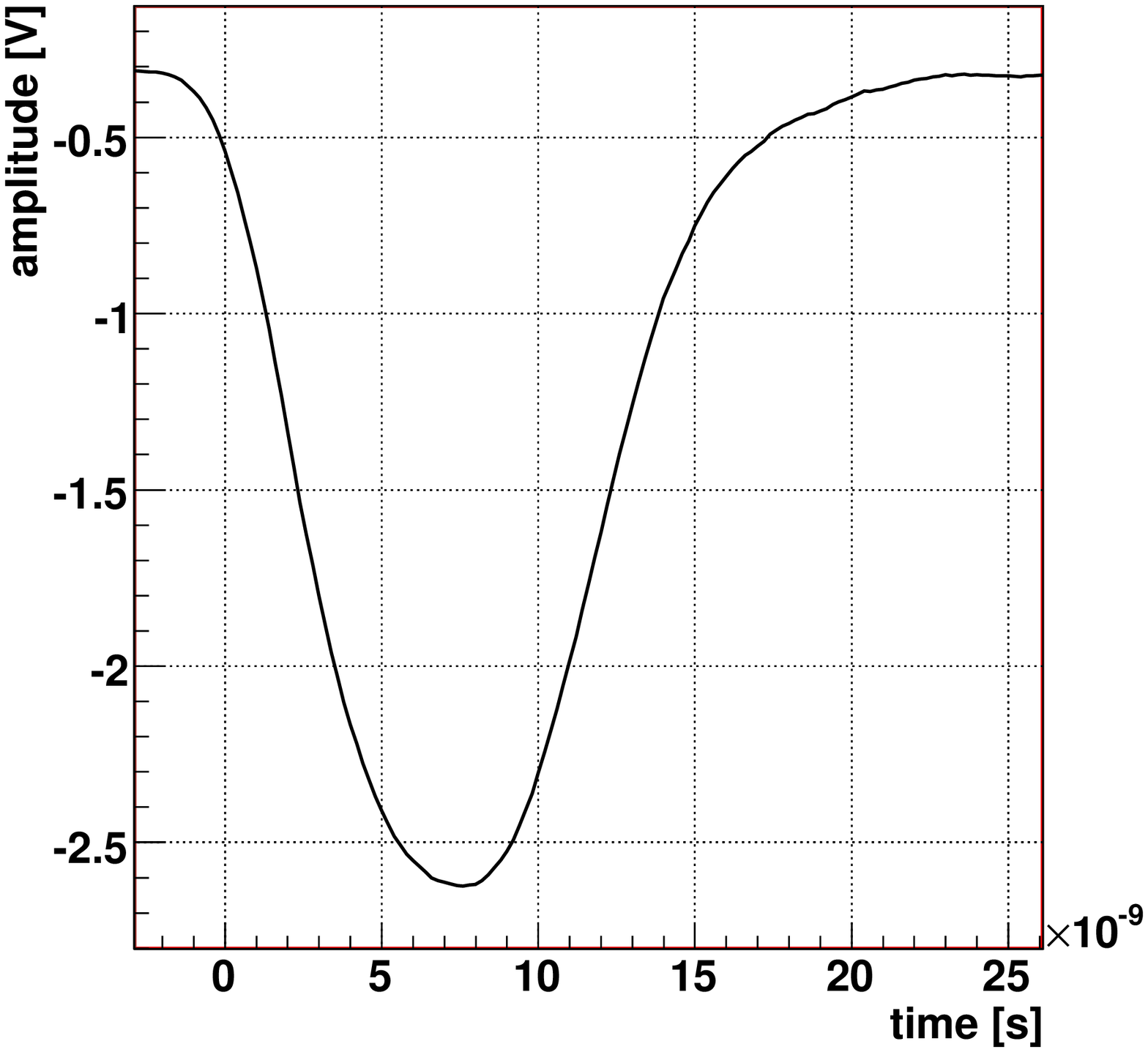,width=8cm}}\captionof{figure}{\label{f-070702_LED_noped_mean_pulse}Pulse for the excitation of the LED mounted inside a panel.}
\end{minipage}
\end{center}
\end{figure}

\subsubsection{Panel Production}

The ACC scintillator panels (Fig.~\ref{f-prod_panel} and \ref{f-acc_panel}) are made of 8\,mm thick, 830\,mm long and 220\,mm wide Bicron BC-414 material \cite{bc-414}. It is based on polyvinyltoluene (PVT) with a refractive index of 1.58 and a softening point at 70\,°C. This material can also be used in vacuum and has a fast signal rise time (0.7\,ns) with a short decay constant (1.8\,ns) with the emission maximum at 392\,nm. Due to the complex atomic structure of organic molecules the absorption of the energy loss of a charged particle generates an excitation and a deexcitation at a different wavelength (fluorescence). 

The panels have a bending radius of 0.55\,m and all 16 panels together form a cylinder. The upper right corner of Fig.~\ref{f-prod_panel} shows the tongue and groove of a panel. These are needed to connect the panels while maximizing the light output. The so called slot region between tongue and groove is crucial for the determination of the inefficiency because of less scintillator material and a smaller active to passive material ratio. For testing and calibration purposes UV LEDs are glued in on both ends of the panel. They emit light at 390\,nm wavelength with an opening angle of 10°. The LED cables will be cut after successful flight ACC integration in the AMS-02 detector.

The scintillation light is absorbed by Kuraray Y-11(200)M wavelength shifting fibers. In total 74 fibers are embedded in grooves milled into the panels. The fibers have a diameter of 1\,mm and are placed with a pitch of 2.9\,mm and are bended at both ends of the panel to exit at the same position. They have a polystyrene core ($n\sub{core} = 1.59$) and a multicladding made of two layers to provide larger light absorption probability and protection of the fiber surface (polymethylmethacrylate (PMMA) $n\sub{clad,1} = 1.49$, fluorinated polymer (FP) $n\sub{clad,2} = 1.42$). These properties lead to a large numerical aperture ($\text{\it NA}=0.72$) that characterizes the range of angles over which the system can accept or emit light.  The emitted scintillator light is shifted to a maximum of 476\,nm again due to the atomic structure of organic fiber material \cite{kuraray}. 

The fibers are glued using the optical cement Bicron BC-600 with a refractive index of 1.56 and a transmission of larger than 98\,\% for light wavelengths above 400\,nm \cite{bc-600}. Special care is taken to avoid inclusions of air bubbles in the glue which would reduce the light absorption probability of the fibers (Fig.~\ref{f-acc_fiber_glue}).

Two fiber bundles of 37 fibers each end up in a black polycarbonate (PC) connector on each side of the panel. The bundles are covered with space qualified Viton tubes. 

The complete panel is first wrapped with reflective mylar foil and then with black tissue to make it light tight. The mylar foil is needed to enhance the absorption efficiency because the light that would have left the scintillator material is reflected and the absorption probability in the WLS fibers is increased. All critical positions are additionally sealed with black silicone.

\subsubsection{Panel Test}

The produced panels were tested in the setup shown in Fig.~\ref{f-pmt_panel_test} using a NIM trigger logic and a CAMAC data acquisition system with an ADC converting the analog charge output of the PMT into a digital value. In this test the ACC panels were directly connected to the PMTs without the clear fiber cables. Two tests were carried out: In the first one the two integrated LEDs were used to inject light into the panels, in the second one the spectra of atmospheric muons were measured. For all panel measurements the same set of reference photomultipliers was used. The LED test was done before and after measuring pulseheight spectra of muons for crosscheck purposes.

\begin{figure}
\begin{center}
\begin{minipage}[b]{.4\linewidth}
\centerline{\epsfig{file=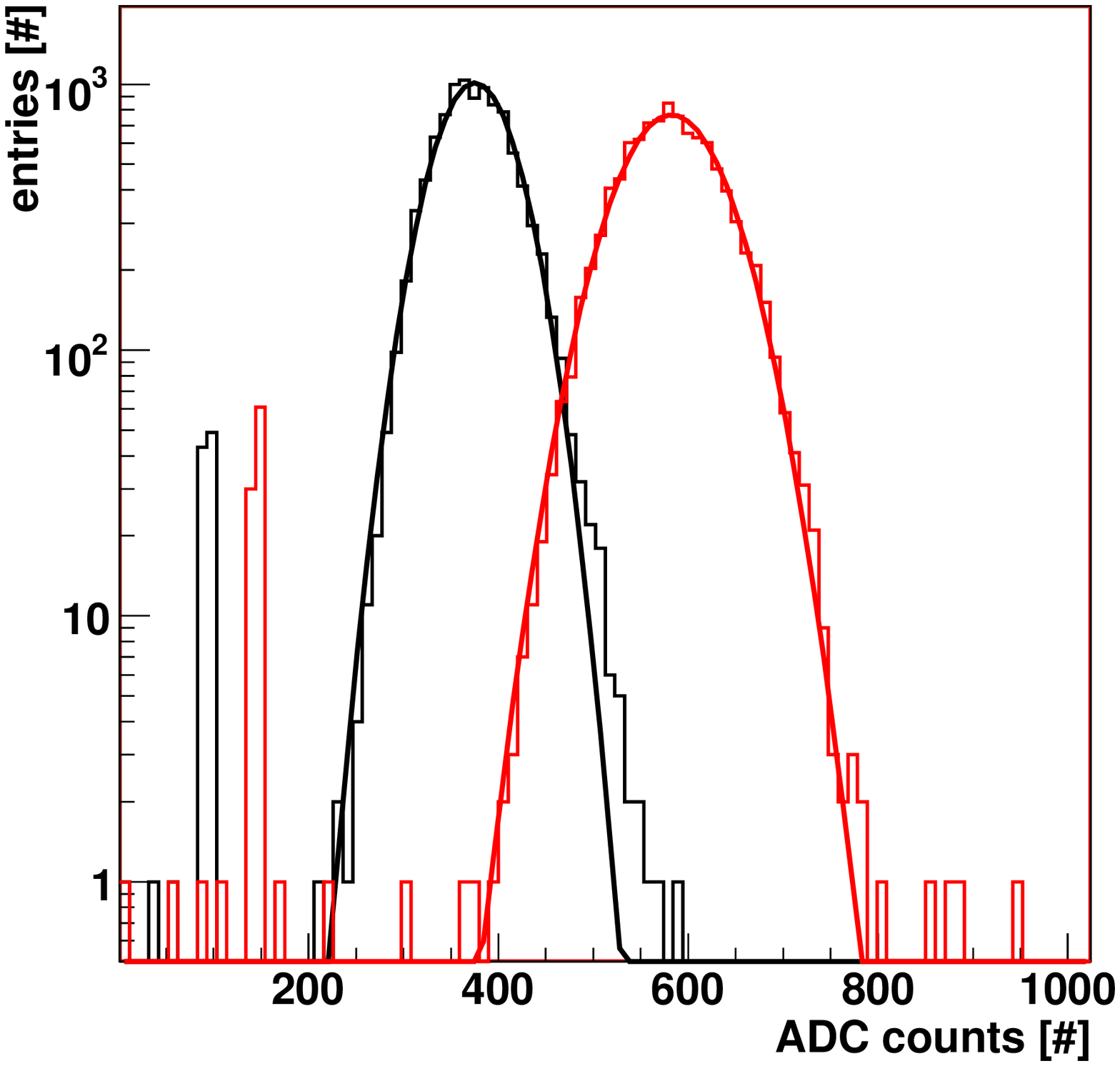,width=8cm}}\captionof{figure}{\label{f-P9C_R1.DAT_0_20000_P9L_R1.DAT_P9L_R2.DAT_led1}Collected LED spectra and pedestal positions of both PMTs for panel 9 without pedestal correction.}
\end{minipage}
\hspace{.1\linewidth}
\begin{minipage}[b]{.4\linewidth}
\centerline{\epsfig{file=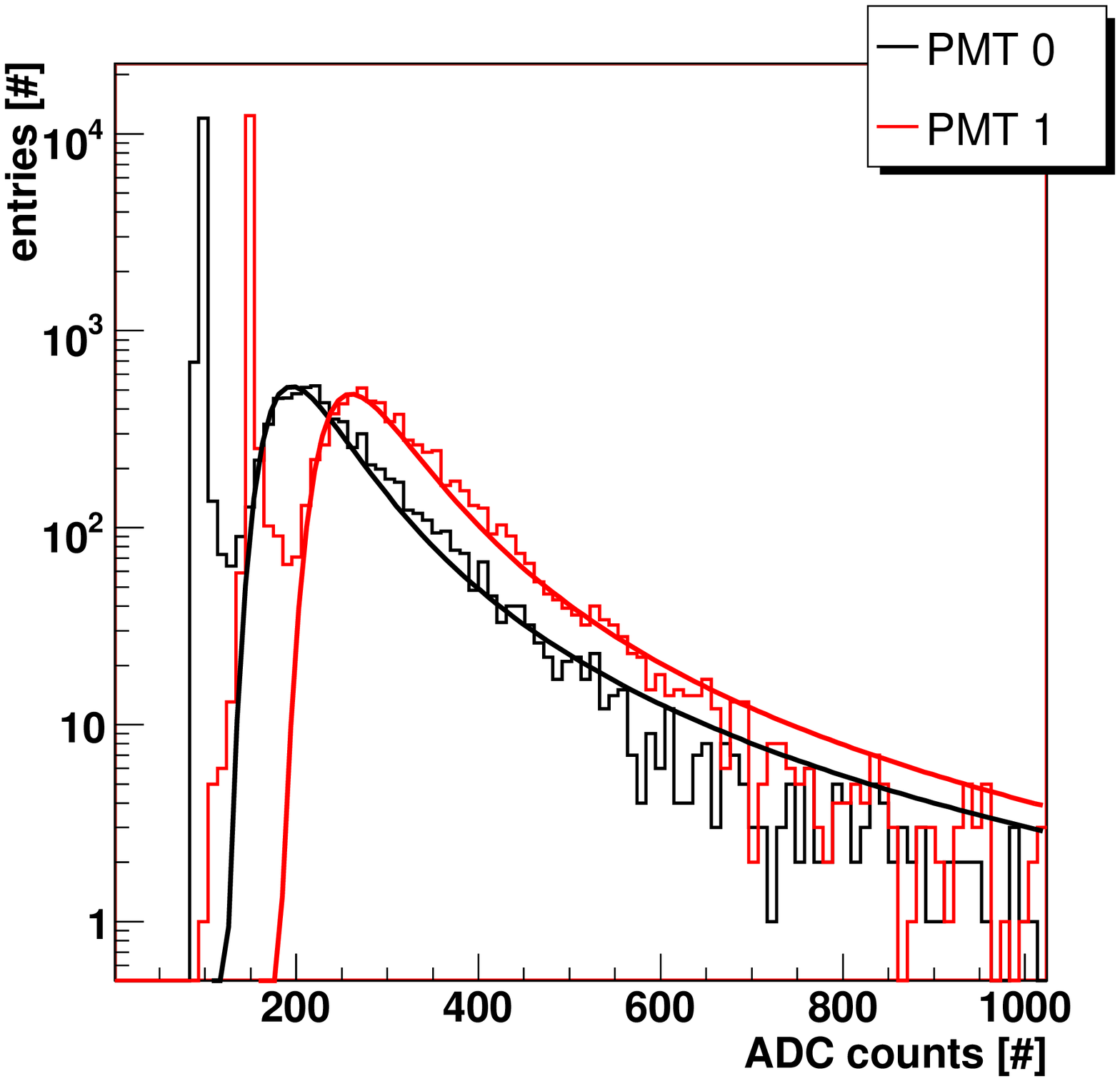,width=8cm}}\captionof{figure}{\label{f-P9C_R1.DAT_0_20000_P9L_R1.DAT_P9L_R2.DAT_acc}Collected atmospheric muons ADC spectra of both PMTs for panel 9 without pedestal correction.}
\end{minipage}
\end{center}
\end{figure}

For the LED test the readout trigger was given by the pulser used to excite the LED. Fig.~\ref{f-070702_LED_noped_mean_pulse} shows the excitation pulse which resulted in the ADC spectra of both PMTs shown in Fig.~\ref{f-P9C_R1.DAT_0_20000_P9L_R1.DAT_P9L_R2.DAT_led1}. Additionally shown are the pedestal positions generated with random triggers. For the muon test the trigger was created by two AMS-01 ACC panels with their four connected PMTs. These were crossed above and below the center of the AMS-02 ACC panel under test (Fig.~\ref{f-pmt_panel_test}). All four trigger PMTs must have fired for a trigger generation. In this way an equal distribution of light in both PMTs under test and the collection of pedestal entries defining the start of the dynamic range of the ADC is achieved at the same time. The result is shown in Fig.~\ref{f-P9C_R1.DAT_0_20000_P9L_R1.DAT_P9L_R2.DAT_acc}. The position of the pedestal and the Landau distribution of the energy deposition inside the AMS-02 panel are clearly seen.

\begin{figure}
\begin{center}
\begin{minipage}[b]{.4\linewidth}
\centerline{\epsfig{file=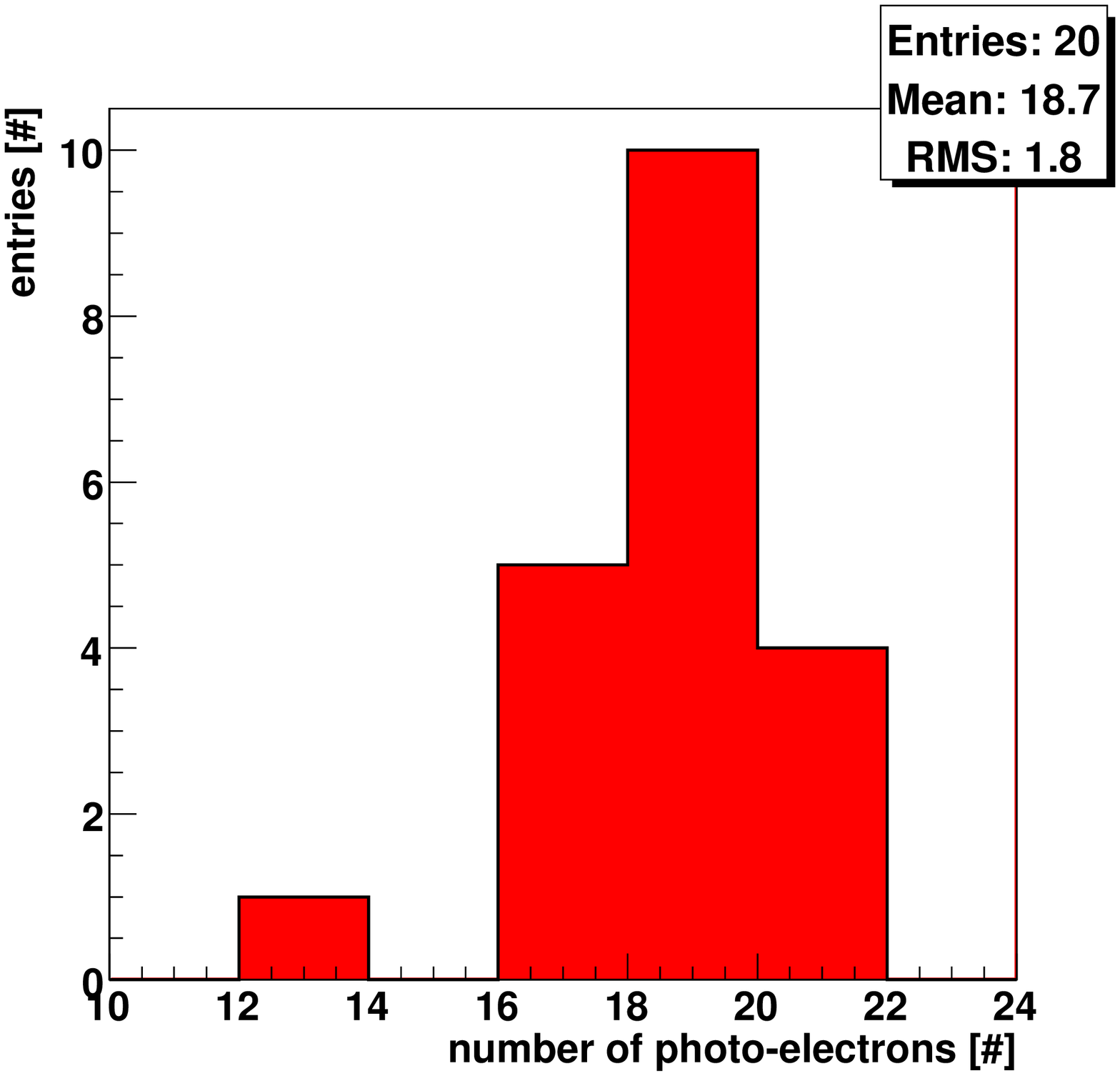,width=8cm}}\captionof{figure}{\label{f-paneldis}Distribution of the most probable number of photo-electrons for all panels tested with reference PMTs.}
\end{minipage}
\hspace{.1\linewidth}
\begin{minipage}[b]{.4\linewidth}
\centerline{\begin{tabular}{c|c||c|c}
\hline
\hline
Panel No.	&	p.e.	& Panel No. & p.e.\\
\hline
1	&13	&11	&19\\
2	&17	&12	&18\\
3	&17	&13	&17\\
4	&19	&14	&19\\
5	&19	&15	&20\\
6	&20	&16	&19\\
7	&19	&17	&20\\
8	&20	&18	&21\\
9	&19	&19	&18\\
10	&19	&20	&21\\
\hline
\end{tabular}}
\vspace{1cm}
\captionof{table}{\label{t-paneltest}Most probable number of photo-electrons obtained with reference PMTs. Panel 17 is smaller by 5\,mm. The integration order will be discussed in Sec.~\ref{ss-systemtest}.}
\end{minipage}
\end{center}
\end{figure}

The most probable (MOP) number of collected photo-electrons is calculated from the atmospheric muon spectra and is used to compare the panels to each other. The measured number of photo-electrons on the anode $N_A$ is proportional to the gain $G$ of the photomultiplier tubes and the number of photo-electrons $N\sub{pe}$ at the photocathode:
\be
N_A=N\sub{pe}\cdot G.
\ee
The number of photo-electrons created by the injection of the LED pulse follows a Poisson distribution. It is assumed that the error on the gain can be neglected in comparison to the error on the number of photo-electrons.
\be
N\sub{$A$,LED}=N\sub{pe,LED}\cdot G\qquad\wedge\qquad \sigma\sub{LED}=\sqrt{N\sub{pe,LED}}\cdot G
\ee
where $\sigma\sub{LED}$ is the width of the pulseheight distribution. From these two equations the gain is easily calculated to be:
\be
G=\frac{\sigma\sub{LED}^2}{N\sub{$A$,LED}}.
\ee
This gain is then used to extract the number of photo-electrons in the atmospheric muon test $N\sub{pe,$\mu$}$:
\be
N\sub{pe,$\mu$}=\frac{N\sub{$A$,LED}}{\sigma\sub{LED}^2}\cdot N_{A,\mu}.
\ee 
It is assumed that $N_A$ is proportional to the measured ADC values and the corresponding number of photo-electrons is calculated. The results for the most probable number of photo-electrons are shown in Fig.~\ref{f-paneldis} and Tab.~\ref{t-paneltest}. The lowest entry at 13 photo-electrons for panel 1 emphasizes the need of a good gluing technique for the wavelength shifting fibers to guarantee a large signal output. Air bubbles reduced the signal due to less scintillator material and due to a smaller total absorption probability in the fiber of scintillation light which depends on the refractive index of the adjacent medium. After the production of panel 1, the gluing technique was improved resulting directly in an increase of the light output. The panels with the 16 highest photo-electron numbers are used for the final ACC detector. Panel 17 is narrower by 5\,mm to balance mechanical tolerances in the integration of the ACC in complete AMS-02.

\subsection{Clear Fiber Cables and Coupling Optimization\label{ss-clear}}

\begin{figure}
\begin{center}
\centerline{\epsfig{file=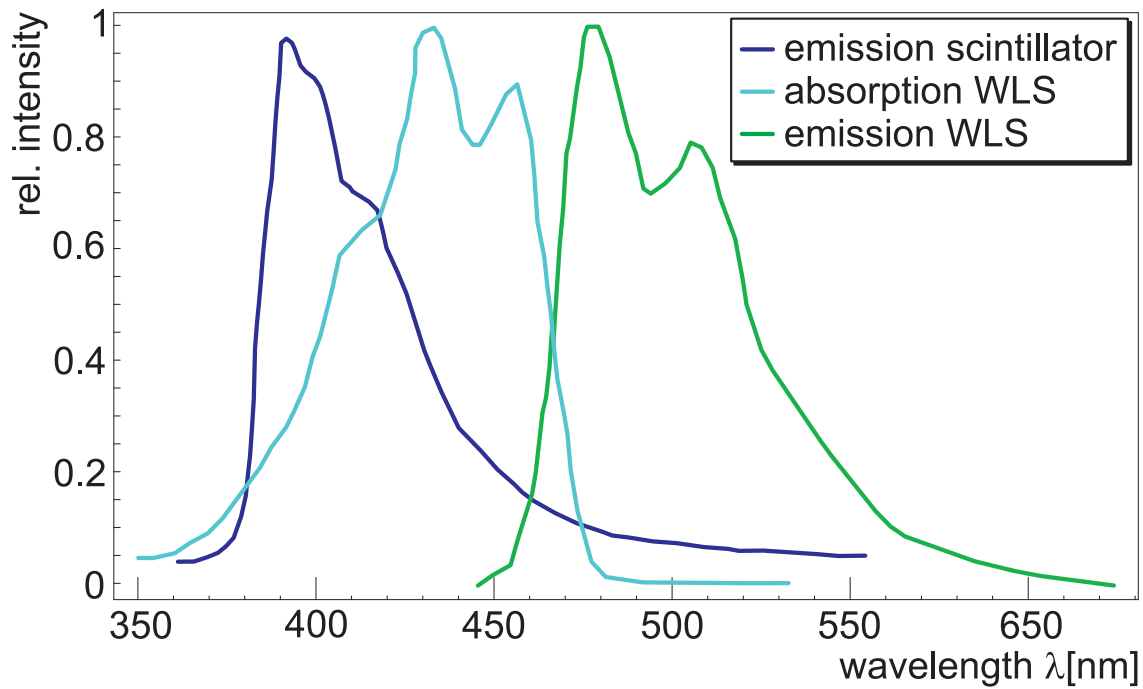,width=15cm}}\captionof{figure}{\label{f-wls_scint_spec}Emission and absorption spectra of the scintillator material and the wavelength shifting fibers \cite{bc-414,kuraray}.}
\end{center}
\end{figure}

The emission and absorption spectra of the scintillator material and the wavelength shifting fibers are shown in Fig.~\ref{f-wls_scint_spec}. Because light attenuation inside the WLS fibers in the wavelength range used is large (about 0.9\,dB/m) coupling to clear plastic optical fibers is employed to maximize the light output. The clear fibers have a much lower attenuation as discussed below. A good coupling between the two fiber types must be realized to keep the advantage of the clear fibers.  Therefore, the clear fibers must have good acceptance over the wavelength range and emission angles of the light emitted by the WLS fibers. 

\subsubsection{Fiber Optics}

\begin{figure}
\begin{center}
\centerline{\epsfig{file=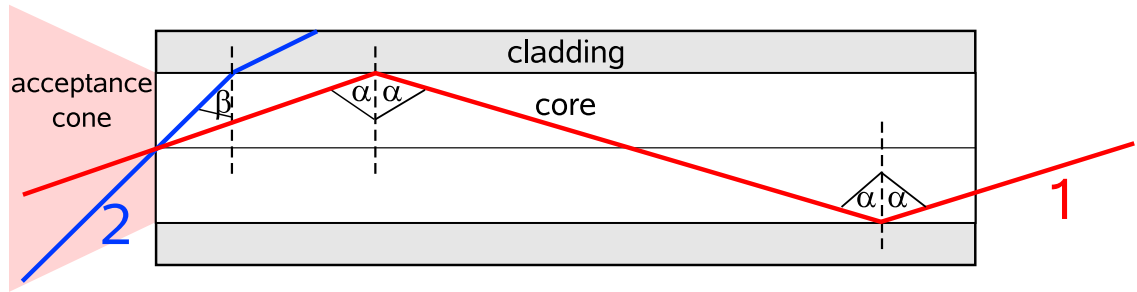,width=15cm}}\captionof{figure}{\label{f-fiber_ray}Light rays in fibers: Angle $\alpha$ of ray 1 is larger than the total reflection angle and angle $\beta$ is smaller than the total reflection angle.}
\end{center}
\end{figure}

The transport of two different kind of light rays in a fiber and the acceptance cone of the fiber is shown in Fig.~\ref{f-fiber_ray}. The acceptance cone is determined by the total reflection angle between the core and the cladding. Only rays with reflection angles between core and cladding larger or equal than the total reflection angle can be transported (ray 1). Rays with angles smaller than the total reflection angle penetrate the cladding (ray 2) and cannot be transported. The cladding has a smaller refractive index than the core and is especially needed to protect the core surface to assure clean reflections. The cladding material typically has large attenuation and light penetrating into the cladding is lost quickly. The fibers used here have a multicladding. Using geometrical optics \cite{kuzyk,daum,bludau} and Snell's law, the numerical aperture $\text{\it NA}$ with refractive indices of the adjacent medium $n$, the cladding $n\sub{clad}$ and the core $n\sub{core}$ is given by:
\be \text{\it \text{\it NA}} = n\sin(\theta\sub{max})=\sqrt{n\sub{core}^2-n\sub{clad}^2}\ee
where $\theta\sub{max}$ is the maximum acceptance angle of the fiber. This $\text{\it NA}$ value can be interpreted as the local numerical aperture of the fiber as explained in the following. A further crucial property is the number of modes $N$ in the fiber which accounts for the allowed electromagnetic field configurations. It can be approximated by \cite{daum}:
\be N\approx\frac12\cdot\left(2\pi\frac a\lambda\cdot \text{\it NA}\right)^2\ee
where $\lambda$ is the wavelength of the transported light and $a$ the core radius. For a plastic optical fiber with a typical $\text{\it NA}=0.5$, wavelength $\lambda=500$\,nm and a core radius $a=0.5$\,mm this results in $N\approx 80\cdot10^6$ modes. Each mode has a different path in the fiber which leads to different pathlengths in the core and in the cladding. Higher modes have more non-ideal reflections and deeper and longer paths inside the cladding with large attenuation. As a result the local $\text{\it NA}$ is not very useful for characterizing the fiber and to take  all effects into account one has to measure the effective $\text{\it NA}$ value over the total fiber length. Using these effective numerical apertures the damping $D$ can be calculated to be:
\be D=10\log\left(\frac{\text{\it NA}\sub{WLS fiber}}{\text{\it NA}\sub{clear fiber}}\right)^2\,\text{dB}.\ee

\begin{figure}
\begin{center}
\centerline{\epsfig{file=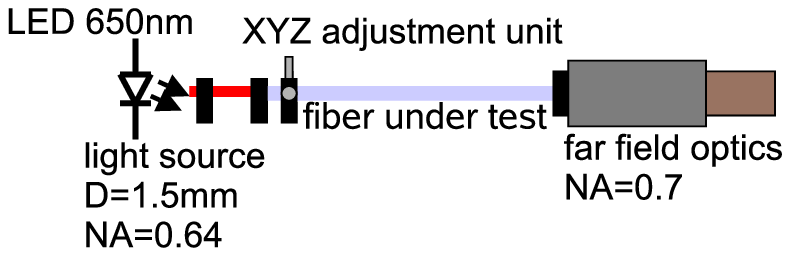,width=15cm}}
\captionof{figure}{\label{f-fiber_measure}Setup for fiber measurement.}
\end{center}
\begin{center}
\begin{minipage}[b]{.4\linewidth}
\centerline{\epsfig{file=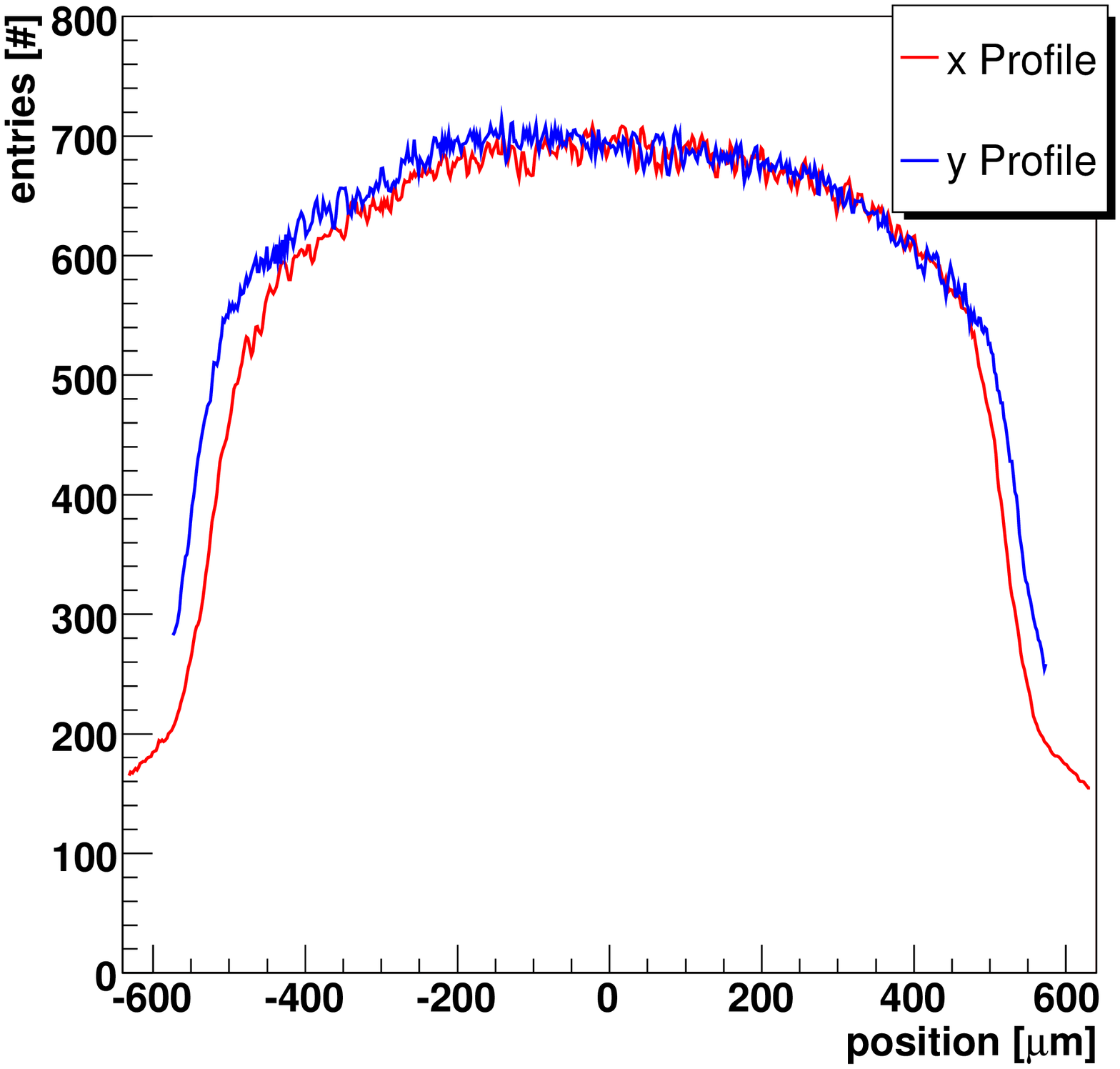,width=8cm}}\captionof{figure}{\label{f-WLS-Fiber_Nahfeld_mod.1D}Near field Kuraray WLS fiber.}
\end{minipage}
\hspace{.1\linewidth}
\begin{minipage}[b]{.4\linewidth}
\centerline{\epsfig{file=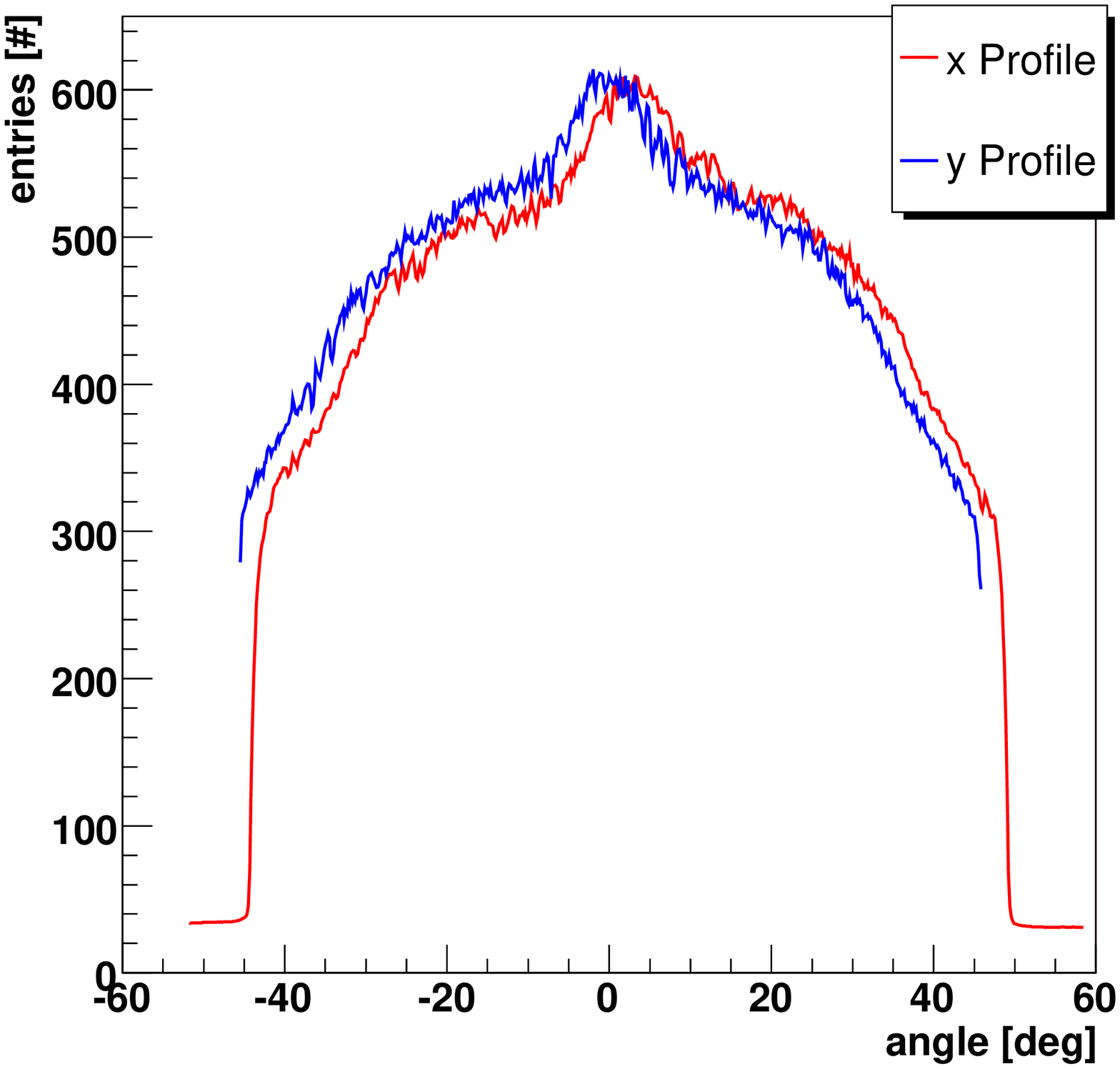,width=8cm}}\captionof{figure}{\label{f-WLS-Fiber_Fernfeld_mod.1D}Far field Kuraray WLS fiber.}
\end{minipage}
\end{center}
\begin{center}
\begin{minipage}[b]{.24\linewidth}
\centerline{\epsfig{file=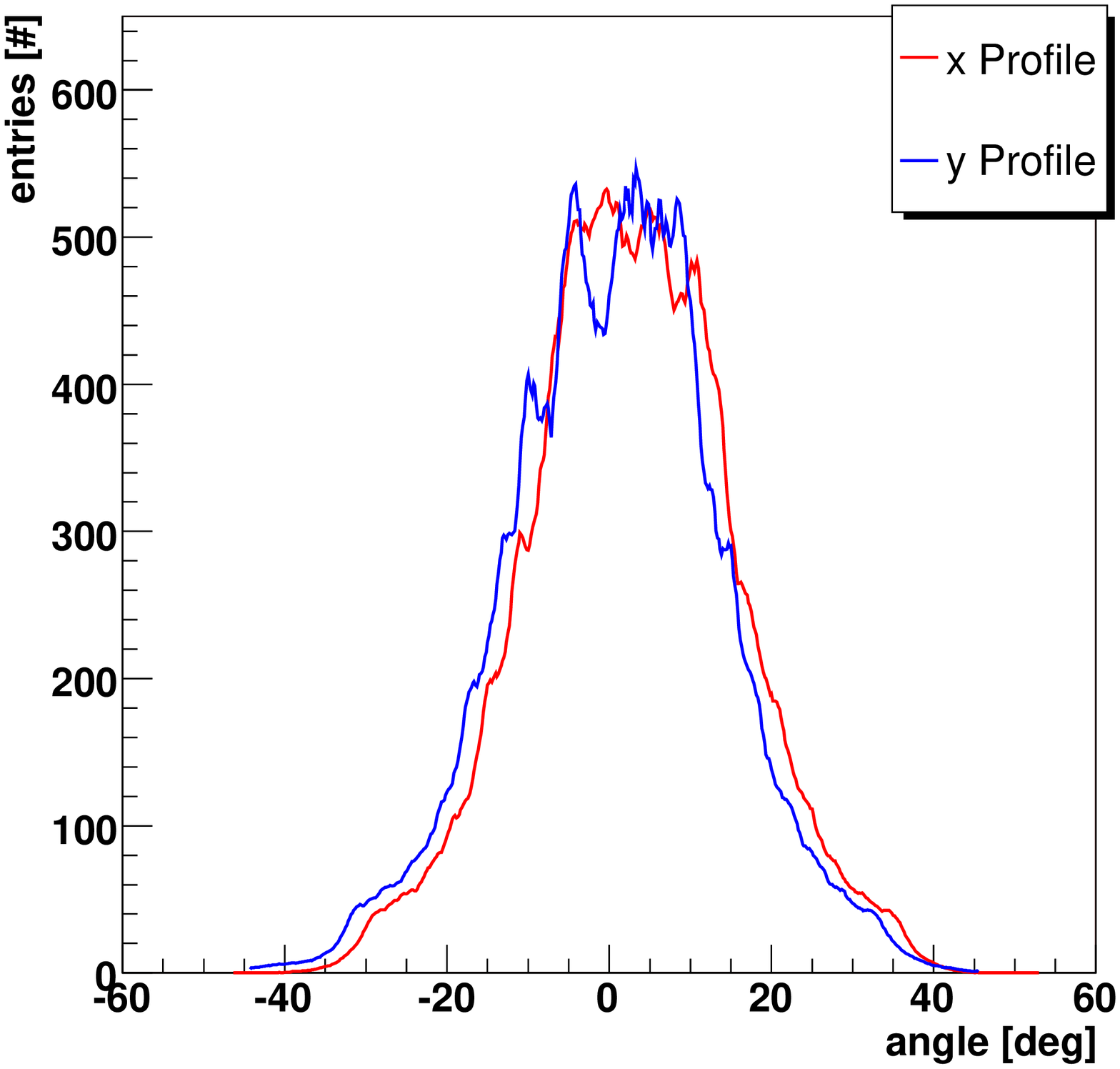,width=5.3cm}}\captionof{figure}{\label{f-Fernfeld_Klar1.5m_mod.1D}Far field Bicron clear fiber.}
\end{minipage}
\hspace{.1\linewidth}
\begin{minipage}[b]{.24\linewidth}
\centerline{\epsfig{file=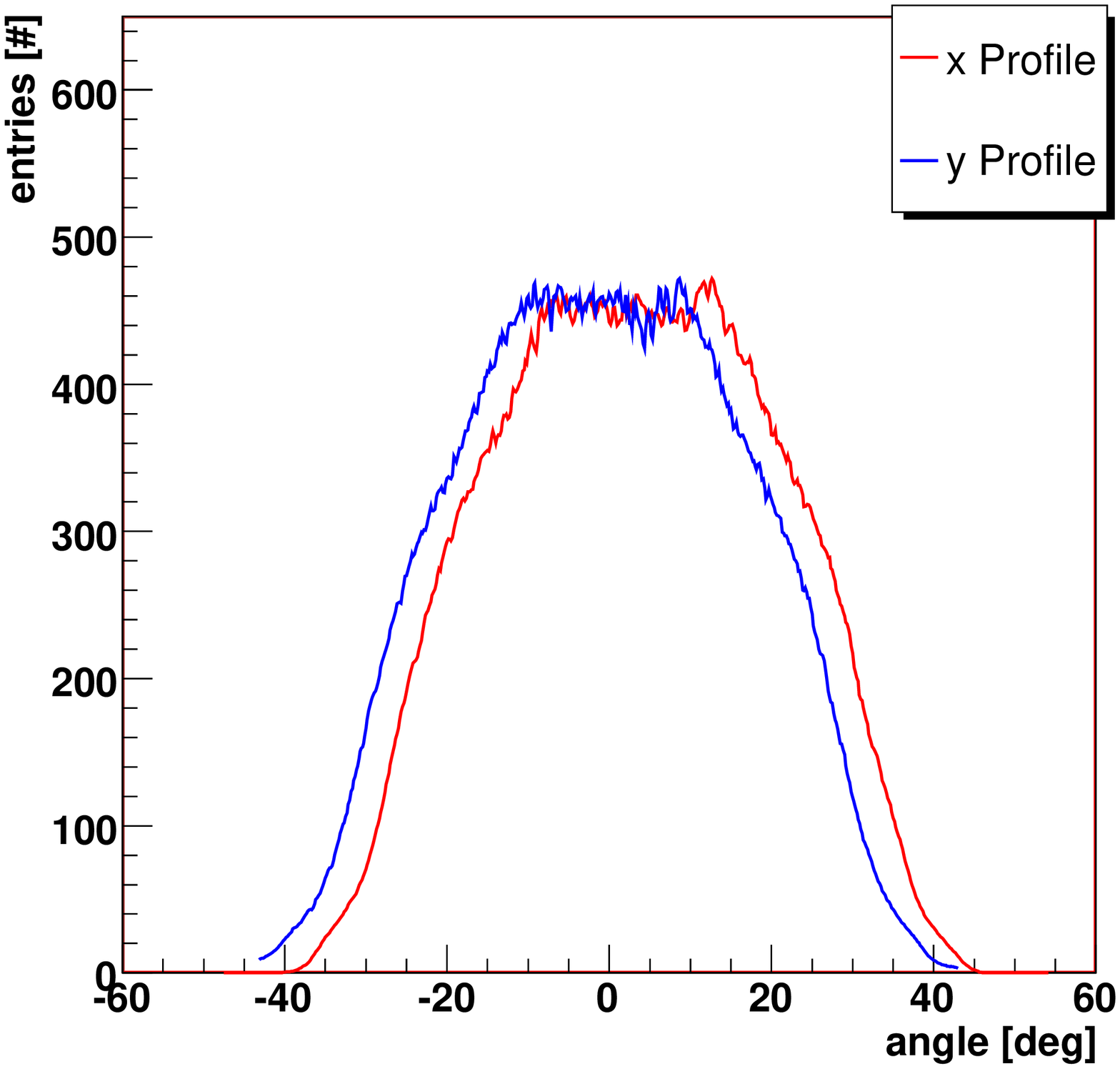,width=5.3cm}}\captionof{figure}{\label{f-Luminous_SHB-1000_mod.1D}Far field Luminous clear fiber.}
\end{minipage}
\hspace{.1\linewidth}
\begin{minipage}[b]{.24\linewidth}
\centerline{\epsfig{file=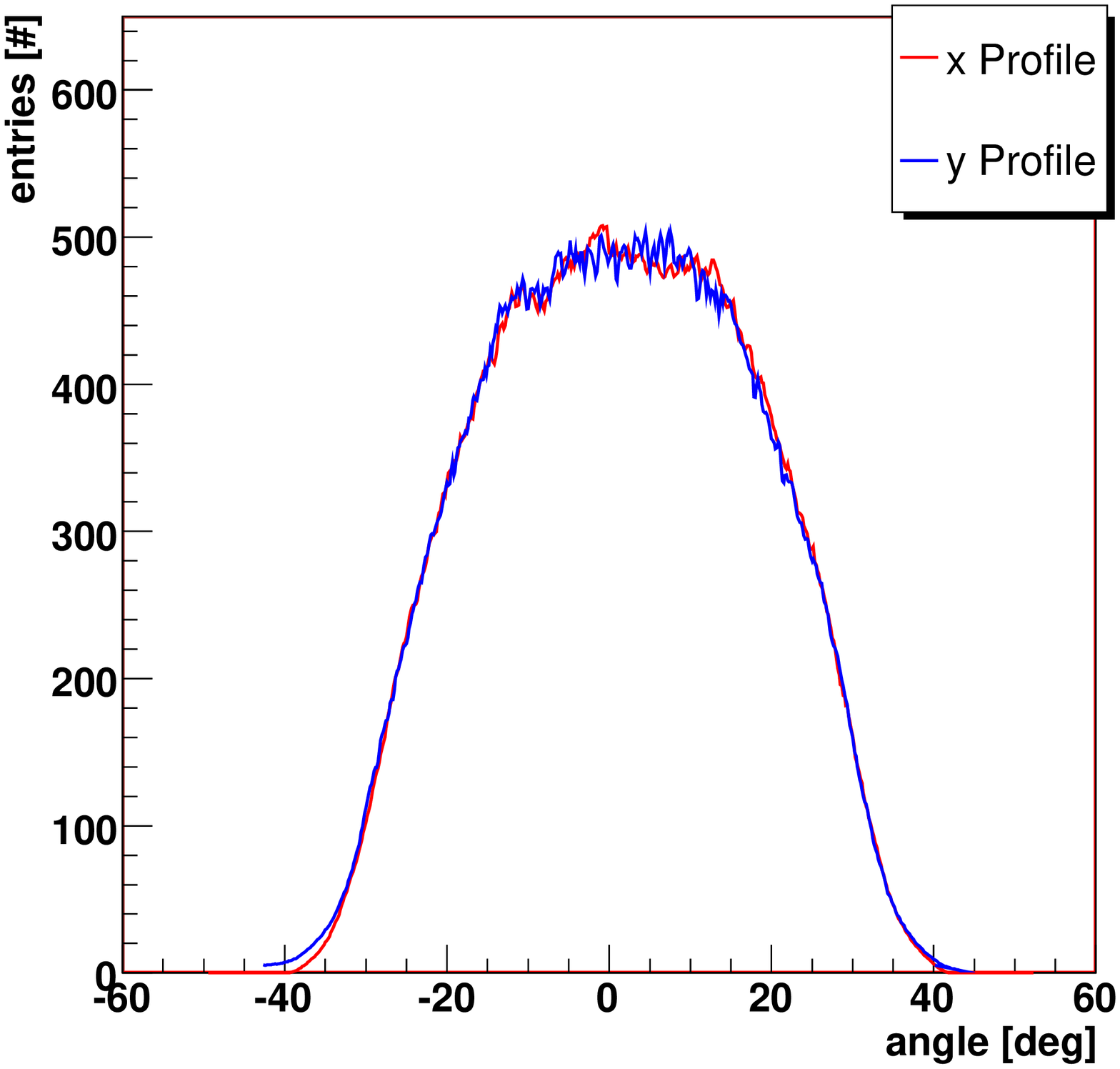,width=5.3cm}}\captionof{figure}{\label{f-Toray_PJU-FB_1000_mod.1D}Far field Toray clear fiber.}
\end{minipage}
\end{center}
\end{figure}

\subsubsection{Fiber Test}

The setup at the POFAC Nürnberg\footnote{polymer optical fiber working group in Nuremberg: www.pofac.de} (Fig.~\ref{f-fiber_measure}) was used to measure the angular far field distributions of different fiber types. Light was injected with an LED at 650\,nm with a  $\text{\it NA}=0.64$ at a diameter of 1.5\,mm. The far field optics are able to resolve numerical apertures up to 0.7. In addition, a scan of the near field was done with different optics in the same setup. The result of the near field measurement (Fig.~\ref{f-WLS-Fiber_Nahfeld_mod.1D}) shows the homogeneous light output of the WLS fiber on the surface and the far field measurement the angular distribution (Fig.~\ref{f-WLS-Fiber_Fernfeld_mod.1D}) of the light output. The angular distribution is quite wide because of non-ideal reflections and photon reabsorption in the fiber followed by isotropic fluorescence radiation at all angles. These effects are also responsible for the larger attenuation of the WLS fibers compared to the clear fibers. The requirements on the clear fiber are determined by the properties of the wavelength shifting fiber. Knowing the need for a wide angular acceptance one can choose the clear fiber type according to the far field distribution.  Fig.~\ref{f-Fernfeld_Klar1.5m_mod.1D} - \ref{f-Toray_PJU-FB_1000_mod.1D} show the results for three different fibers which are chosen because of high $\text{\it NA}$ according to the manufacturer. The Bicron fiber with a polyamide (PA) core shows a narrow distribution and therefore cannot accept much light from the WLS fiber although it is made from material with the same refractive indices as the WLS fiber. The fibers made of PMMA from Luminous and Toray show much wider distributions. This may be due to better production techniques because of the large number of applications, e.g. in the automotive industry.

\begin{figure}
\begin{center}
\centerline{\epsfig{file=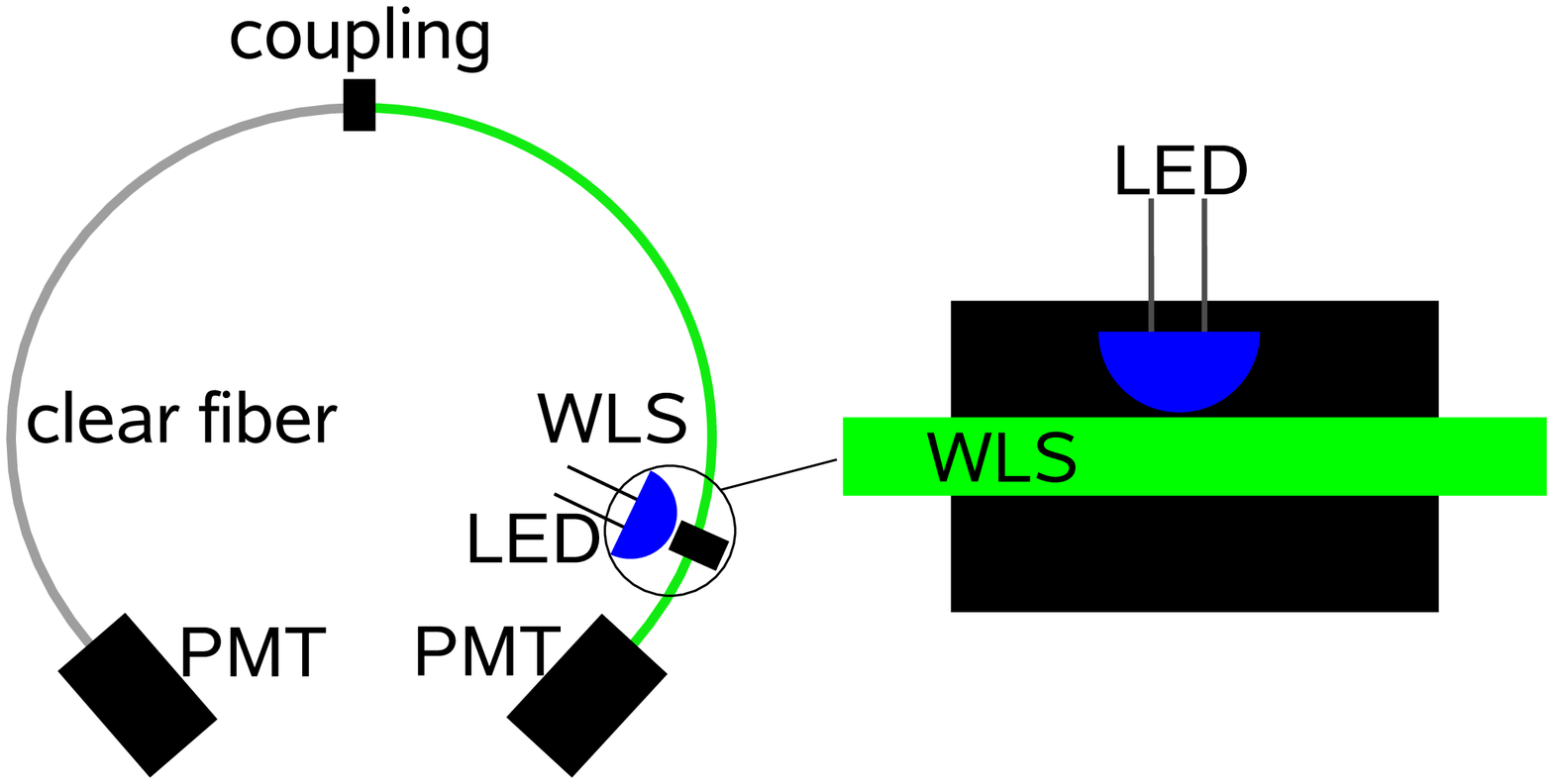,width=15cm}}
\captionof{figure}{\label{f-fiber_coupling}Setup for coupling measurement.}
\end{center}
\begin{center}
\captionof{table}{\label{t-fiber_coupling}Results of coupling measurements.}
\begin{tabular}{l||c|c}
\hline
\hline
Type	&Transmission efficiency [\%]	&Damping [dB]\\
\hline
Bicron			&42.5	&3.7\\
SHB Luminous		&59.9	&2.2\\
Toray PJU-FB1000	&71.2	&1.5\\
\hline
\end{tabular}
\end{center}
\end{figure}

\begin{figure}
\begin{center}
\begin{minipage}[b]{.4\linewidth}
\centerline{\epsfig{file=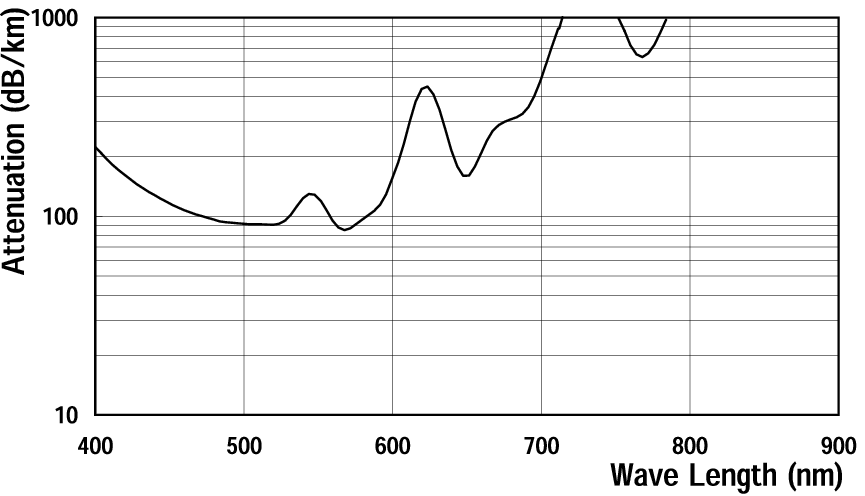,width=8cm}}\captionof{figure}{\label{f-toray_att}Attenuation length of Toray PJU-FB1000 clear fiber as a function of wavelength \cite{toray}.}
\end{minipage}
\hspace{.1\linewidth}
\begin{minipage}[b]{.4\linewidth}
\centerline{\epsfig{file=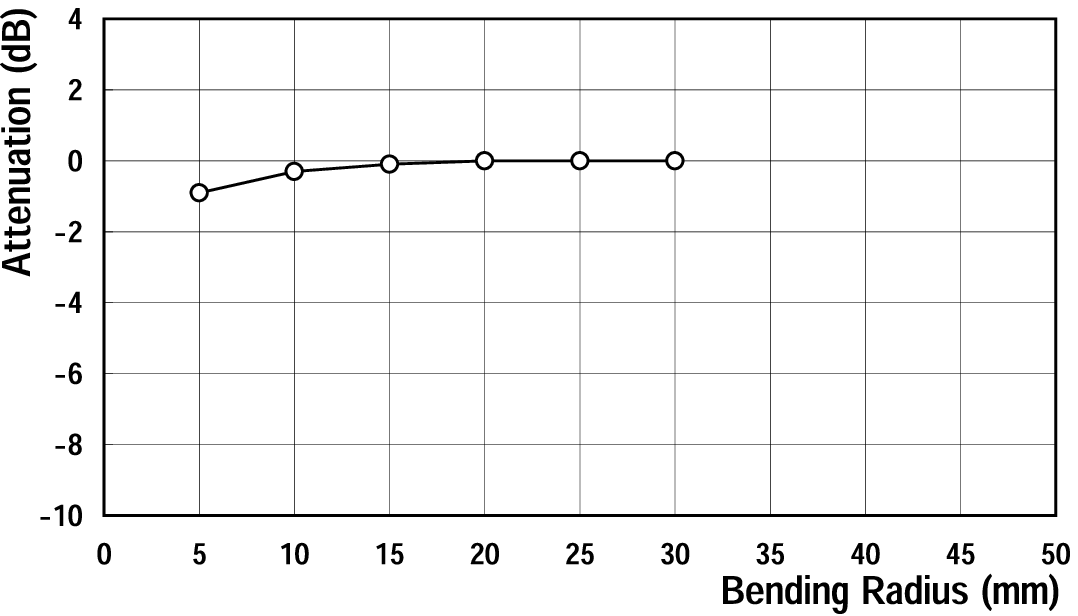,width=8cm}}\captionof{figure}{\label{f-toray_bending}Bending loss of Toray PJU-FB1000 clear fiber as a function of radius \cite{toray}.}
\end{minipage}
\end{center}

\begin{center}
\centerline{\epsfig{file=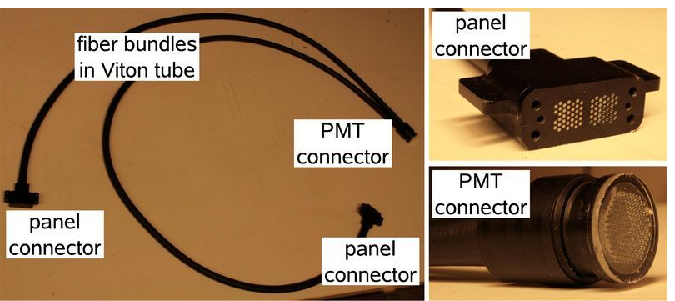,width=15cm}}
\captionof{figure}{\label{f-clear_cable}Clear fiber cable with connectors.}
\end{center}
\end{figure}

The final decision on the clear fiber was taken following the coupling measurements of the wavelength shifting fiber to the clear fiber with the setup outlined in Fig.~\ref{f-fiber_coupling}. Light was injected from a LED into three 1.5\,m long wavelength shifting fibers inside a black Viton tube. The output was measured at both ends with photomultipliers. Both ends of the WLS fibers are glued using optical BC-600 cement inside indentations in black polycarbonate endpieces. In the next step a black Viton tube with three clear fibers was coupled to the WLS fibers. One end of the clear fibers is glued in black polycarbonate endpieces while the other one is glued in clear polycarbonate endpieces because previous tests had shown that the light output at the PMT was increased by using transparent endpieces due to corona light around the fiber. The WLS fibers are connected to the black clear fiber endpieces with screws and shear pins for accurate positioning. From comparison between pulseheights with ($P\sub{wc}$) and without ($P\sub{w}$) clear fibers the transmission efficiencies $T$ were calculated:\be T=\frac{P\sub{wc}}{P\sub{w}}.\ee The results for the three clear fiber types are shown in Tab.~\ref{t-fiber_coupling} and favor the use of the Toray fiber. Even though the Luminous fiber angular acceptance is slightly wider than the Toray acceptance it shows larger damping because of larger attenuation inside the fiber. The attenuation as a function of wavelength for the Toray fiber is shown in Fig.~\ref{f-toray_att}. In addition, the fibers will have to be bent when mounted onto the helium vessel of the magnet cryostat. According to the manufacturer the Toray fiber can be bent without any significant increase in damping down to radii of $\approx15$\,mm (Fig.~\ref{f-toray_bending}), much smaller than the radii used. The Toray PJU-FB1000 also matches the NASA requirements for use on ISS \cite{chang-2007}.

\subsubsection{Clear Fiber Production}

\begin{figure}
\begin{center}
\centerline{\epsfig{file=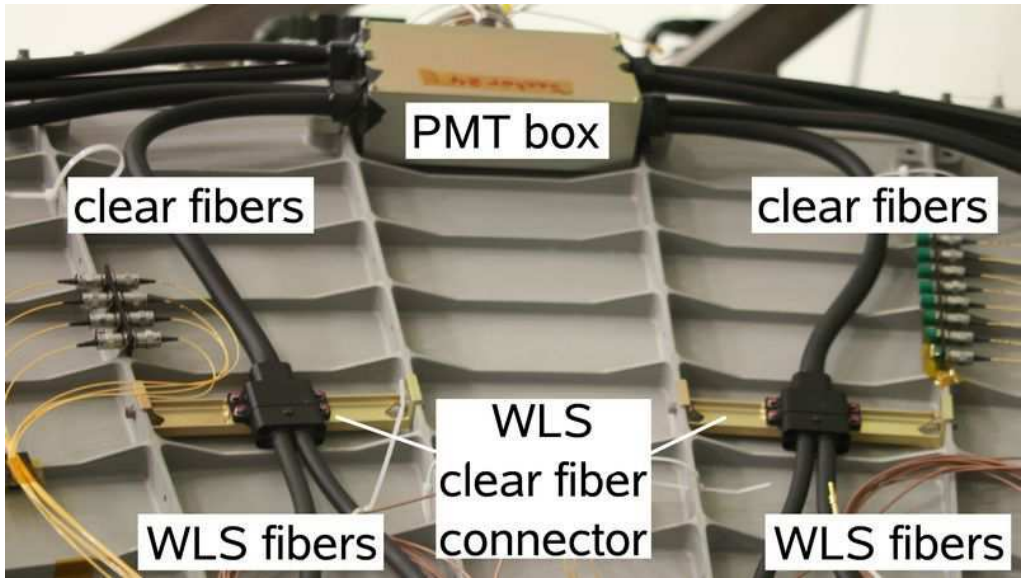,width=13.5cm}}
\captionof{figure}{\label{f-pmt_fiber}Connection between WLS fibers of the panels to the clear fiber cables and of clear fiber cables to PMT boxes.}
\end{center}
\end{figure}

Clear fiber cables with a Y-shape are needed for the connection of the ACC panels with the PMTs (Fig.~\ref{f-clear_cable}). Two bundles of 74 fibers each are glued into one connector which will be used to connect to one PMT. The other two ends are glued into two separate connectors for the connection to the panels. These are connectors of same type used for the WLS fibers on the panel. The fiber bundles are covered by a black space qualified Viton tube. The glue is again the optical cement BC-600. The surfaces of the connectors have been polished for uniform light output and better coupling. Fig.~\ref{f-pmt_fiber} shows the connection of the four PMTs inside a PMT box and the coupling  of WLS fibers of a panel to clear fiber cables on the helium tank. The connectors are precisely positioned with screws and shear pins. Eight long and eight short clear fiber cables are needed to reach all connection positions. The bundles of the long cables have lengths of 1355\,mm and 1825\,mm and the bundles of the short cables have lengths of 405\,mm and 911\,mm.

\begin{figure}
\begin{center}
\centerline{\epsfig{file=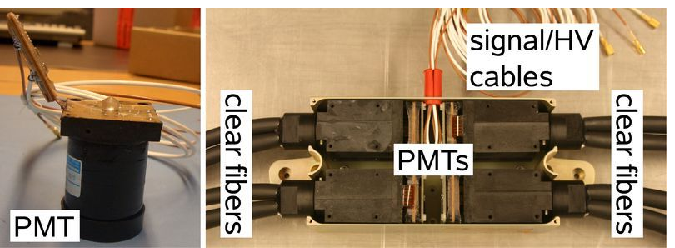,width=13.5cm}}
\captionof{figure}{\label{f-pmt_box_pic}\textbf{\textit{Left)}} Hamamatsu R5946 photomultiplier tube. \textbf{\textit{Right)}} PMT box with four PMTs with signal and high voltage cables and clear fiber cable connections.}
\end{center}
\begin{center}
\centerline{\epsfig{file=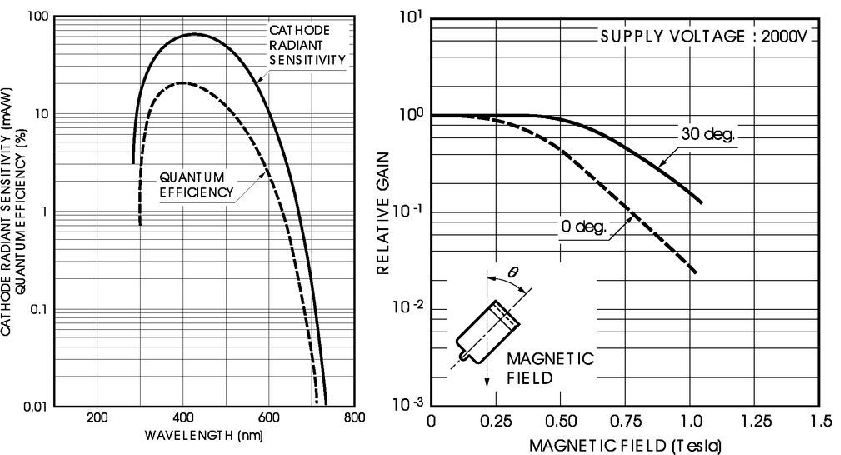,width=13.5cm}}
\captionof{figure}{\label{f-pmt_plots}\textbf{\textit{Left)}} PMT quantum efficiency vs. wavelength. \textbf{\textit{Right)}} Change of PMT gain in magnetic field \cite{hamamatsu}.}
\end{center}
\end{figure}

\subsection{Space Qualification of the Photomultiplier Tubes\label{ss-pmttest}}

Photomultiplier tubes (PMT) are used to amplify weak light signals. They have low noise and a quick response. A PMT consists of a photocathode, dynodes and an anode. The photons to be detected strike the photocathode and produce electrons as a consequence of the photoelectric effect. These electrons are amplified in several electrodes called dynodes. Each dynode is held at a higher voltage than the previous one and they are arranged so that the number of electrons increases after each dynode. The accumulated charge results in a sharp pulse at the anode. 

The ACC is instrumented with 16 Hamamatsu R5946 \cite{hamamatsu} photomultiplier tubes which are also used for the time of flight system (Fig.~\ref{f-pmt_box_pic}, left). The fine mesh dynodes of these PMTs allow operation in magnetic fields without a large distortion of the electron trajectories \cite{ams}. To even minimize this effect the PMTs are mounted parallel to the stray magnetic field of about 0.12\,T. Further properties of the PMTs are a bialkali photocathode and a borosilicate glass window and 16 bialkali dynodes. The operational voltage range is 1900\,V - 2300\,V and the power consumption is about 50\,mW. The quantum efficiency and performance in a magnetic field as given by Hamamatsu are shown in Fig.~\ref{f-pmt_plots} \cite{hamamatsu}. Four PMTs are placed inside each of four boxes. These PMT boxes are positioned close to the electronics crates on the upper and lower side of the vacuum tank (Fig.~\ref{f-pmt_fiber}). An inside view of one of these boxes is provided on the right side of Fig.~\ref{f-pmt_box_pic} where the signal and power cables are also visible. The PMTs are connected to the clear fiber cables inside the boxes which must be light tight. In addition, NASA requires the boxes to be glass particle tight in case a PMT breaks.

\subsubsection{Space Qualification}

\begin{figure}
\begin{center}
\centerline{\epsfig{file=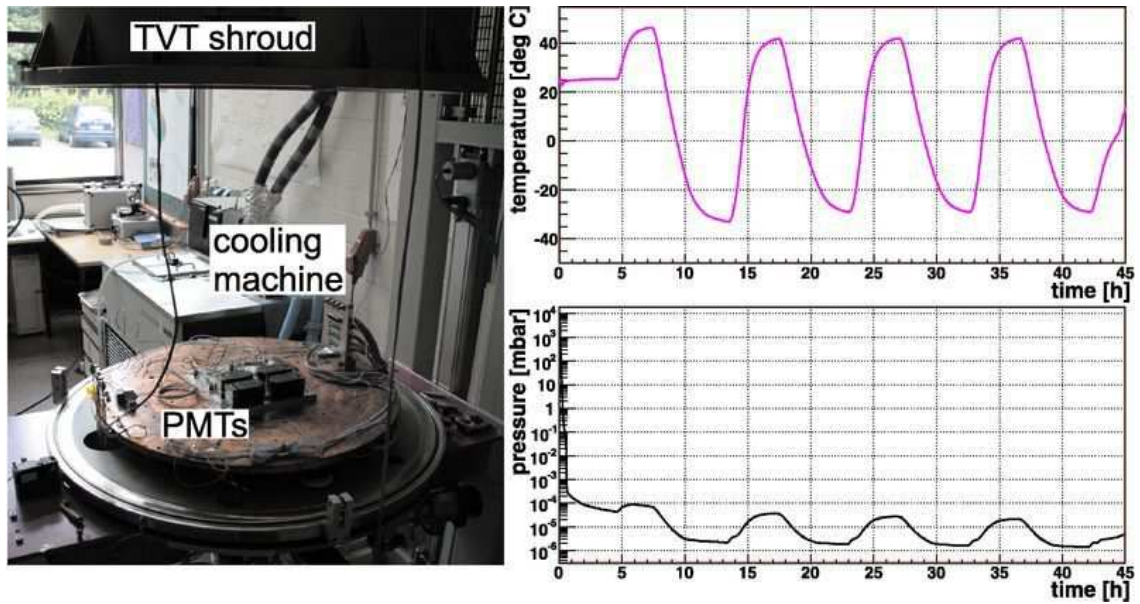,width=15cm}}
\captionof{figure}{\label{f-pmt_tvt_all}\textbf{\textit{Left)}} Photomultipliers mounted into the thermo vacuum test facility. \textbf{\textit{Right)}} Temperature cycles with the corresponding pressure inside the chamber as a function of time.}
\end{center}
\end{figure}
\begin{figure}
\begin{center}
\centerline{\epsfig{file=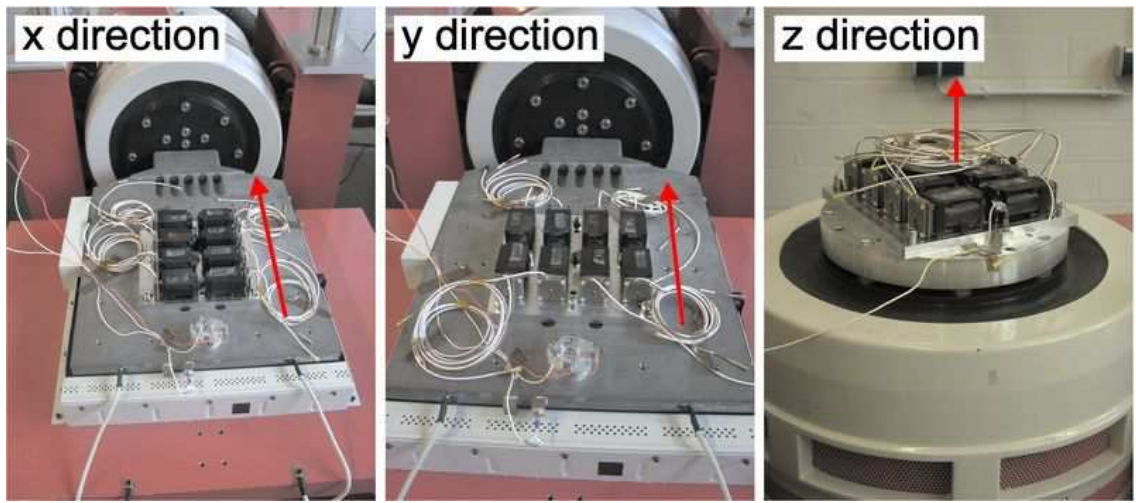,width=15cm}}
\captionof{figure}{\label{f-pmt_vib}Photomultipliers mounted on the vibration table.}
\end{center}
\end{figure}
\begin{figure}
\begin{center}
\captionof{table}{\label{t-pmtsq}Requirements for the space qualification of the photomultiplier tubes.}
\begin{tabular}{l||c|c}
\hline
\hline
Type		& Minimum [°C]	& Maximum [°C]\\
\hline
non-operational, HV off	& -35		& 50\\	
operational, HV on	& -30		& 45\\
\hline
\end{tabular}
\end{center}
\end{figure}

The PMTs had undergone space qualification tests to assure stable operation of the ACC in the space environment following the shuttle flight. The space qualification was done in two steps. First the PMTs were subjected to thermal cycles in the thermo-vacuum chamber shown in Fig.~\ref{f-pmt_tvt_all} (TVT test). The figure also shows the PMT temperature curve and the pressure curve inside the chamber. From thermal calculations for the whole experiment the temperature requirements for the PMTs are listed in Tab.~\ref{t-pmtsq}. On the first cycle the operational range was left and the PMTs are not switched on. The following three cycles are within the operational range and the high voltage for the PMTs was switched on to 2000\,V on the extremal positions.

\begin{figure}
\begin{center}
\begin{minipage}[b]{.4\linewidth}
\centerline{\epsfig{file=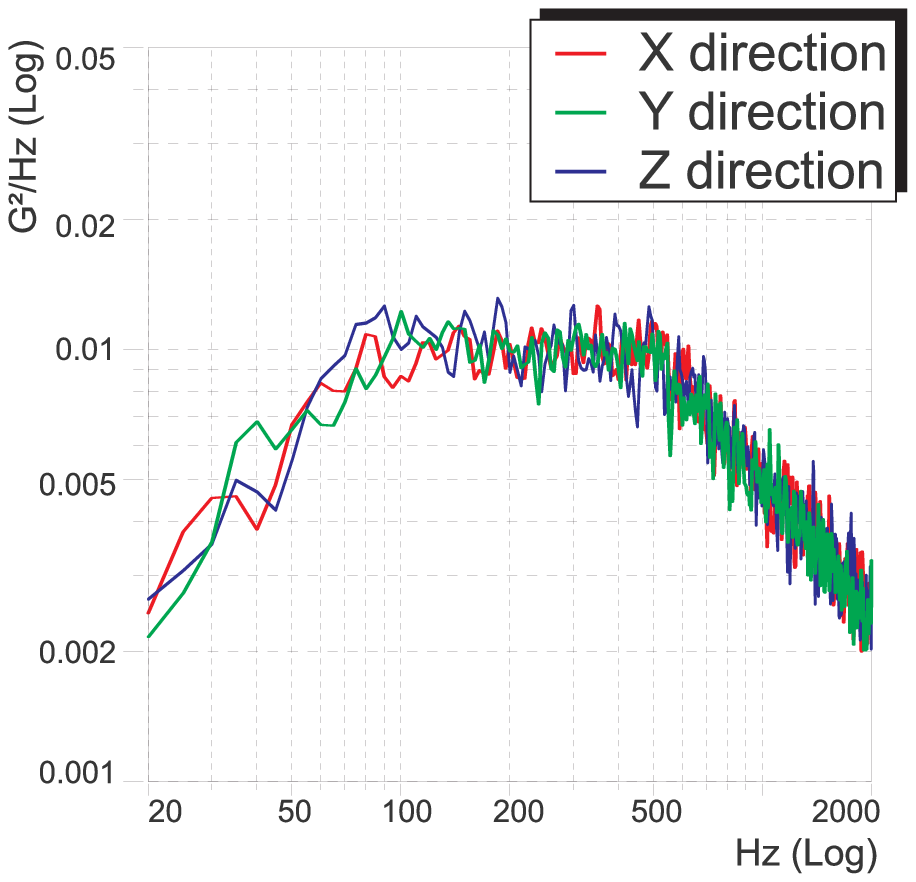,width=8cm}}\captionof{figure}{\label{f-pmt_vib_freq}Power spectrum during vibration. Average level in all directions: 3.4\,$g$.}
\end{minipage}
\hspace{.1\linewidth}
\begin{minipage}[b]{.4\linewidth}
\centerline{\epsfig{file=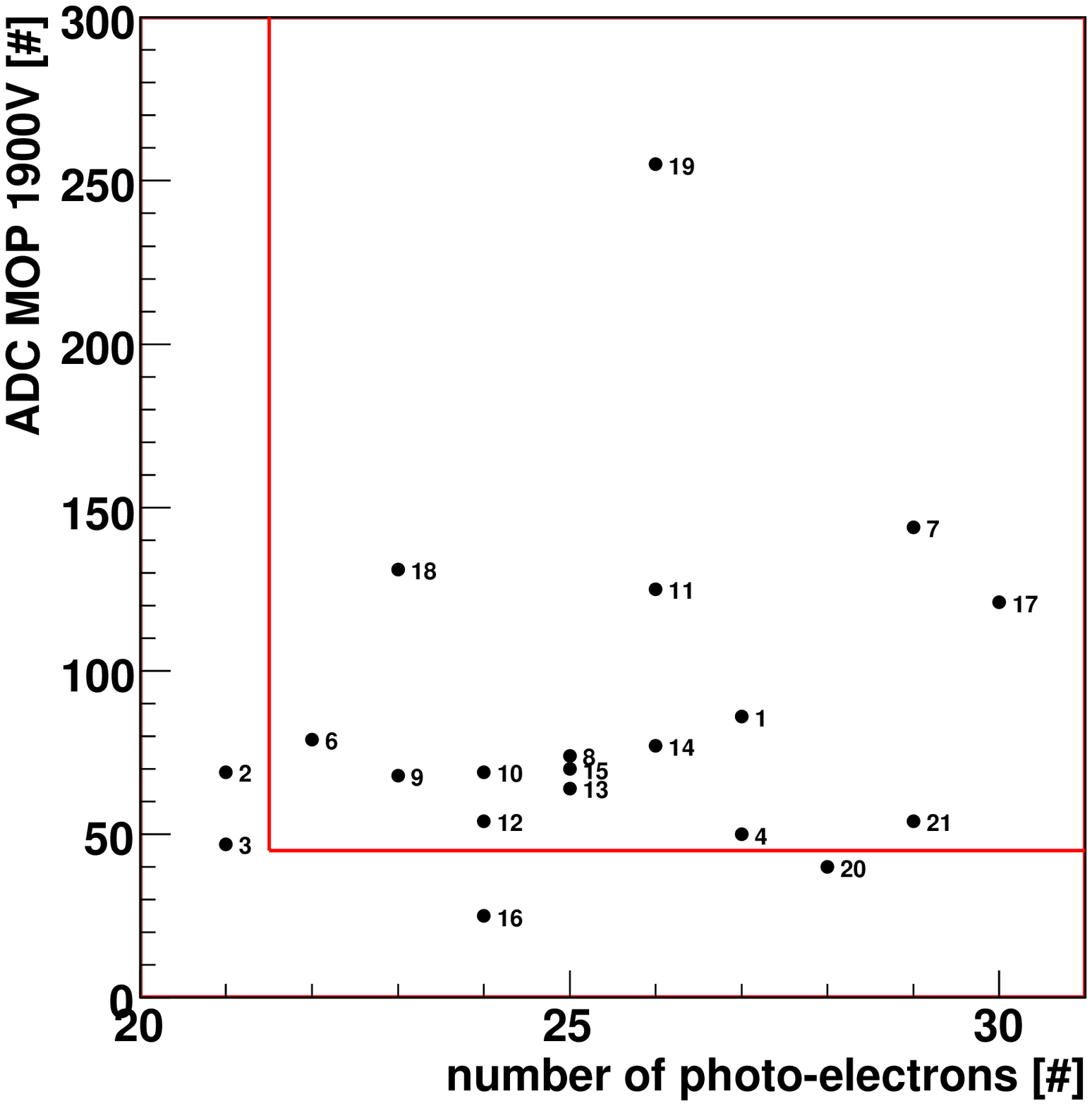,width=8cm}}\captionof{figure}{\label{f-pmt_pe_adc}PMT test: MOP of pulseheight spectra at 1900\,V vs. number of photo-electrons. The numbers next to the points indicate the PMT production numbers.}
\end{minipage}
\end{center}
\end{figure}

In the second step survival of the PMTs during space shuttle launch and landing was tested. To this purpose the PMTs were subjected to vibrations in all directions on a vibration table (Fig.~\ref{f-pmt_vib}). The requirement for the average acceleration is 3.4\,$g$. The random frequency spectrum on the PMTs in all directions is shown in Fig.~\ref{f-pmt_vib_freq}. The PMTs were tested after each step in the same way as the panels before (Sec.~\ref{ss-paneltest}). Therefore, they were mounted directly on the same reference panel (19) and then the LED and atmospheric muon pulseheight spectra were recorded. To obtain the gain as a function of the applied voltage the measurement was carried out for different voltages. The results for all PMTs are shown in Tab.~\ref{t-pmttest}. The variations in gain and in the most probable number of photo-electrons emitted by the photocathodes are due to the structure of the fine-mesh photomultiplier tubes. Imperfections in the meshes and different distances between the dynodes cause these variations. Within a range of about 5\,\% the PMTs do not show a variation before and after the space qualification test such that the space qualification was successful. PMT selection for integration in the final detector was based on the number of photo-electrons and the gain at 1900\,V. Fig.~\ref{f-pmt_pe_adc} shows the distribution and the selection criteria. PMTs not satisfying both requirements are kept as spare.

\begin{figure}
\begin{center}
\captionof{table}{\label{t-pmttest}Gain curves for PMTs measured with the same reference panel 19: most probable values of pulseheight spectra in ADC counts and corresponding number of photo-electrons.}
\begin{tabular}{c|c||c|c|c|c}
\hline
\hline
PMT production no.	&Serial No.	&MOP 1900\,V	&MOP 2100\,V	&MOP 2300\,V	&p.e.\\
\hline
1	&ZH5864	&86	&179	&359	&27\\
2	&ZH7116	&69	&147	&285	&21\\
3	&ZH5849	&47	&93	&171	&21\\
4	&ZH5769	&50	&97	&186	&27\\
6	&ZH5875	&79	&172	&331	&22\\
7	&ZH7110	&144	&150	&170	&29\\
8	&ZH5765	&74	&164	&321	&25\\
9	&ZH5877	&68	&147	&257	&23\\
10	&ZH5779	&69	&144	&276	&24\\
11	&ZH5773	&125	&267	&533	&26\\
12	&ZH5780	&54	&110	&206	&24\\
13	&ZH5854	&64	&137	&250	&25\\
14	&ZH5902	&77	&158	&298	&26\\
15	&ZH5770	&70	&139	&256	&25\\
16	&ZH5862	&25	&76	&119	&24\\
17	&ZH5959	&121	&246	&670	&30\\
18	&ZH5858	&131	&290	&620	&23\\
19	&ZH5879	&255	&-	&338	&26\\
20	&ZH7656	&40	&169	&130	&28\\
21	&ZH5776	&54	&112	&227	&29\\
\hline
\end{tabular}
\end{center}
\end{figure}

\subsection{Anticoincidence Counter System Test\label{ss-systemtest}}

\begin{figure}
\begin{center}
\centerline{\epsfig{file=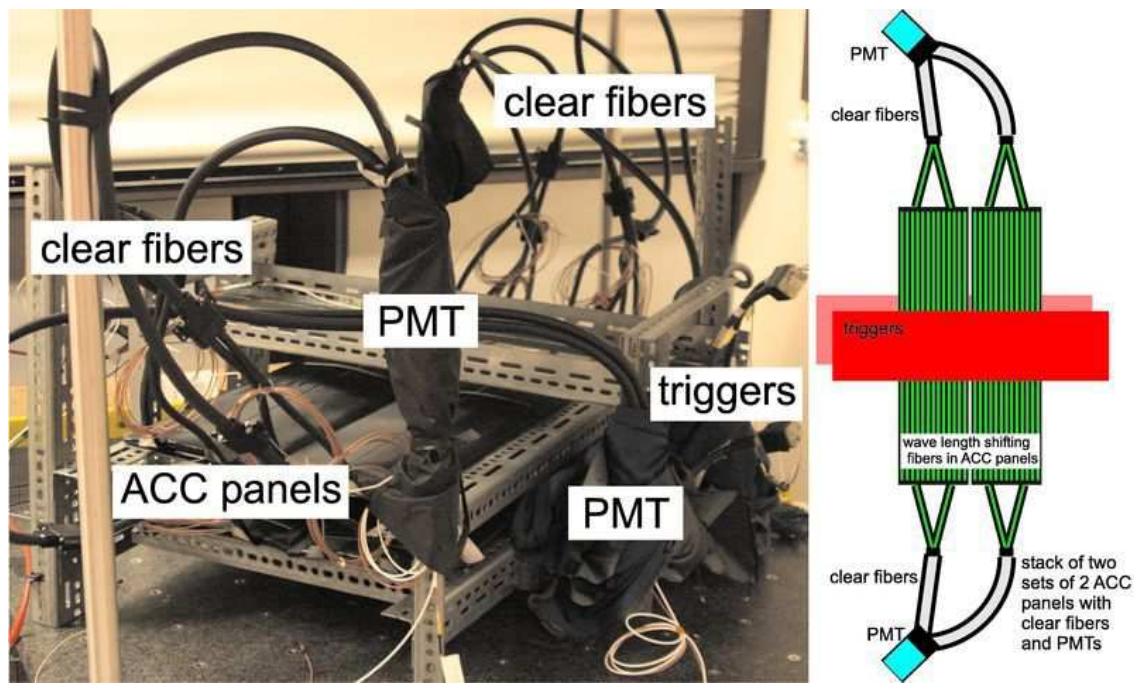,width=15cm}}
\captionof{figure}{\label{f-acc_systemtest}Setup for ACC system test.}
\end{center}
\end{figure}

\begin{figure}
\begin{center}
\begin{minipage}[b]{.4\linewidth}
\centerline{\epsfig{file=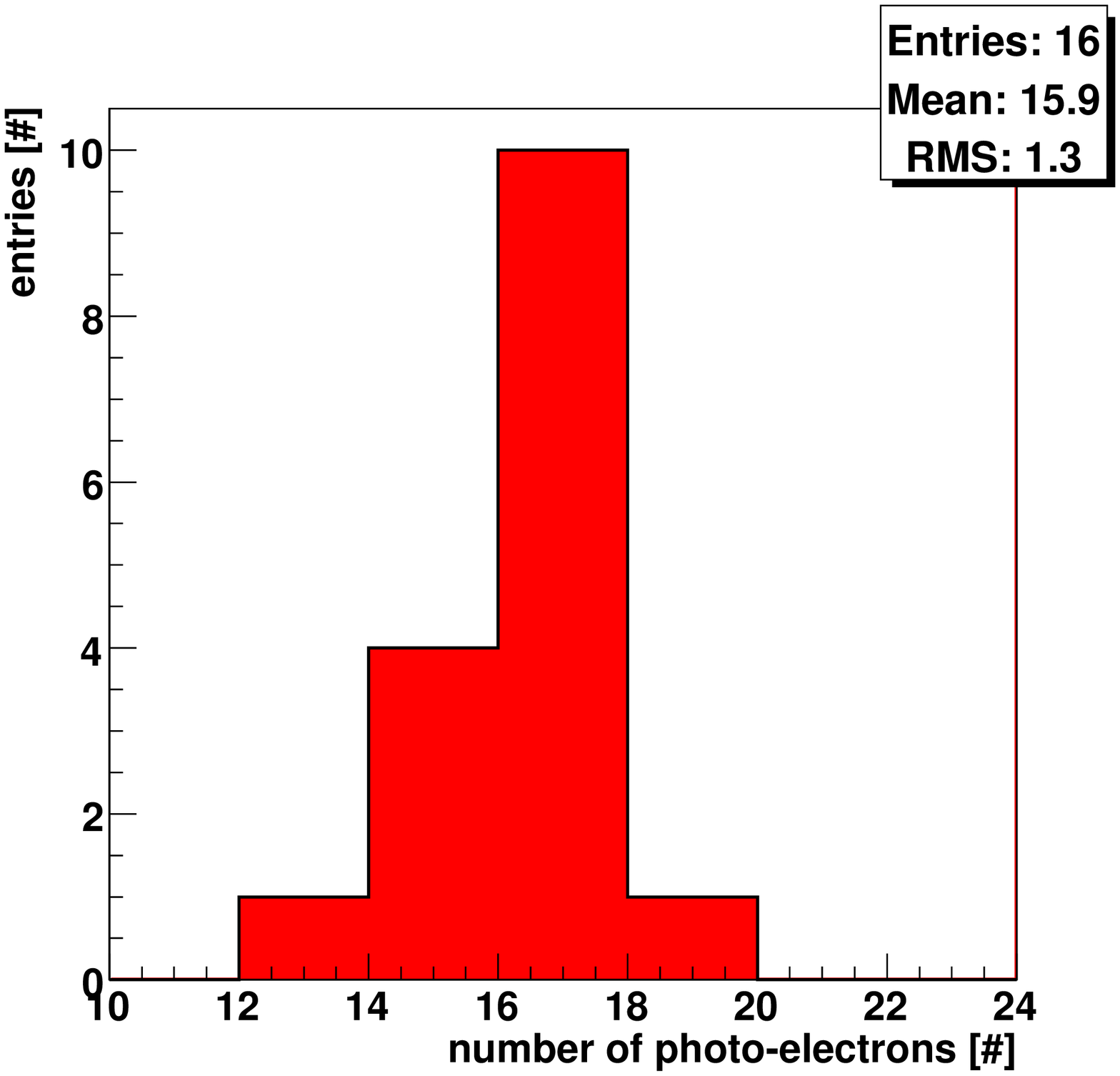,width=8cm}}\captionof{figure}{\label{f-paneldis_systemtest}Most probable number of photo-electrons for the 16 complete ACC panels.}
\end{minipage}
\hspace{.1\linewidth}
\begin{minipage}[b]{.4\linewidth}
\centerline{\epsfig{file=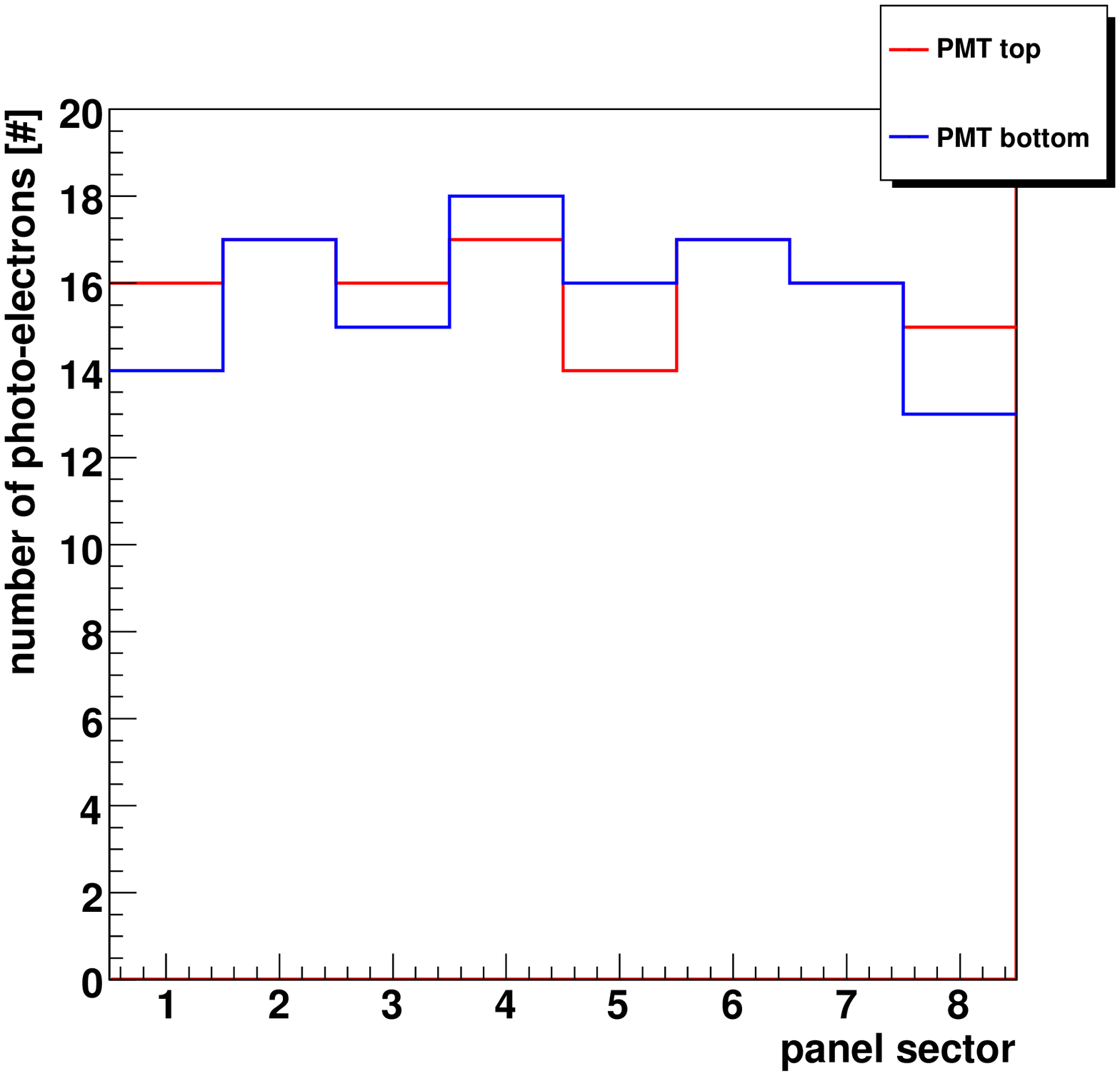,width=8cm}}\captionof{figure}{\label{f-acc_homo}Most probable number of photo-electrons as a function of panel sector.}
\end{minipage}
\end{center}
\end{figure}

As the next step, a system test with all components in flight configuration was performed. Two ACC scintillator panels were connected via two clear fiber cables to two photomultipliers and is called a set in the following (Fig.~\ref{f-acc_systemtest}). The tests were done in a similar way as for the panel and PMT classifications before (Sec.~\ref{ss-paneltest} and \ref{ss-pmttest}). Before and after the collection of atmospheric muons LED runs were done using all LEDs. Two complete sets could be tested at the same time. The trigger was made using the pulser for the LED runs or the AMS-01 scintillators with their PMTs for the muon runs. The number of photo-electrons for the AMS-02 ACC was again calculated. The distribution for all sets shown in Fig.~\ref{f-paneldis_systemtest} has an average of 15.9 with a root mean square (RMS) of 1.3 photo-electrons. Tab.~\ref{t-systemtest} lists the test results for the combination of panels, clear fiber cables and PMTs. The combination was chosen such that the PMTs with the highest gain were connected to the panels with the lowest photo-electron number (tables \ref{t-paneltest} and \ref{t-pmttest}) to assure as much as possible the homogeneity of the detector response assuming homogeneous quality of the clear fiber cables and negligible effects due to the different clear fiber lengths. Homogeneity is necessary because a combination of PMTs with lower gain and panels with lower light output would increase the average  ACC detection inefficiency drastically. The ACC inefficiency will be discussed in detail in Sec.~\ref{s-inefficiency}.

The individual transmission efficiencies $T$ of the WLS fibers to the clear fiber cables can now be extracted in the following way:

\be
T=\frac{\displaystyle N\sub{pe,sys}}{\displaystyle \frac{\displaystyle N\sub{pe,PMT}}{\displaystyle N\sub{pe,PMT$\sub{ ref}$}}\cdot\frac{\displaystyle 1}{\displaystyle 2}(N\sub{pe,panel$_1$}+N\sub{pe,panel$_2$})}
\ee

where $N\sub{pe,sys}$ is the measured number of photo-electrons in the system test, $N\sub{pe,PMT}$ is the number of measured photo-electrons on panel 19 for the flight PMTs, $N\sub{pe,PMT$\sub{ ref}$}$ is the number of measured photo-electrons on panel 19 with the reference PMTs in the panel test and $N\sub{pe,panel$_1$}$ and $N\sub{pe,panel$_2$}$ are the number of photo-electrons of the flight panels from the test with reference PMTs. The average transmission efficiency is 61\,\% with an RMS of 1\,\% which corresponds to a mean damping of 2.1\,dB with an RMS of 0.1\,dB.

As noted above, a homogeneous response of all PMTs is important for the operation of the ACC. The measured gain curves (Tab.~\ref{t-systemtest}) are used to calculate the voltage needed for each PMT by interpolating linearly between the points. Tab.~\ref{t-voltages} shows these voltages for different MOP values. Taking into account the operational range of the PMTs, an average MOP value of 100\,ADC counts is proposed for operation, corresponding to voltages ranging from 1865\,V to 2241\,V.

\begin{figure}
\begin{center}
\centerline{\epsfig{file=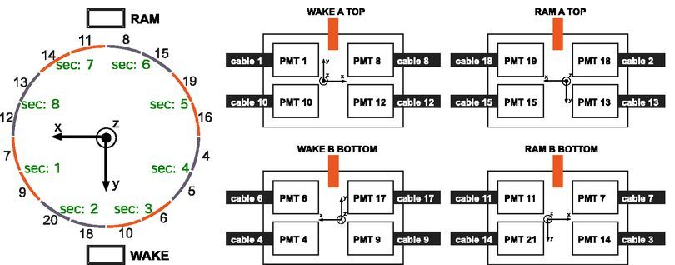,width=15cm}}\captionof{figure}{\label{f-acc_order}Order of ACC panels with corresponding sectors and PMT boxes.}
\end{center}
\begin{center}
\centerline{\epsfig{file=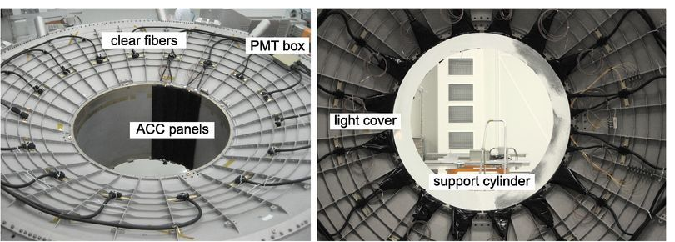,width=15cm}}\captionof{figure}{\label{f-acc_pre-integration}Pre-Integration: \textbf{\textit{Left)}} After mounting of two panels, the clear fiber cables and the PMT boxes. \textbf{\textit{Right)}} Completed pre-integration with light covers and support cylinder.}
\end{center}
\end{figure}

The homogeneous combination of ACC components effects the ordering of the panels as well as the assignment of the PMT boxes and is shown in Fig.~\ref{f-acc_homo} and \ref{f-acc_order}. The panel sector numbers in Fig.~\ref{f-acc_homo} correspond to the indicated sector numbers in Fig.~\ref{f-acc_order}.

\begin{table}
\begin{center}
\captionof{table}{\label{t-systemtest}Results of the complete system test in the final configuration with MOP values of the pulseheight spectra for different voltages and the number of photo-electrons (p.e.). P = panel production number, C = clear fiber cable production number where 's' denotes the short and 'l' the long side of the Y-shaped cable, PMT = photomultiplier production number, 1.9 - 2.3 are the used voltages in kV and the column gives the corresponding MOP value in ADC counts, T = transmission efficiency of the clear fiber cable. One row shows the combination of two panels with their clear fiber cables and PMTs and their performance.}
\begin{tabular}{c||c|c|c|c|c|c|c||c|c|c|c|c|c|c}
\hline
\hline
P			&C					&PMT	&1.9	&2.1	&2.3	&T	&p.e.	&C					&PMT	&1.9	&2.1	&2.3	&T	&p.e.\\
\hline
\multicolumn{15}{c}{long clear cable (s = 1355\,mm and l = 1825\,mm)}\\
\hline
$\bmx13\\12\emx$	&$\bmx18\text{ s}\\18\text{ l}\emx$	&19	&128	&290	&549	&0.61	&15	&$\bmx7\text{ s}\\7\text{ l}\emx$	&7	&70	&151	&291	&0.47	&13\\
\hline
$\bmx19\\16\emx$	&$\bmx2\text{ s}\\2\text{ l}\emx$	&18	&76	&170	&344	&0.63	&14	&$\bmx11\text{ s}\\11\text{ l}\emx$	&11	&60	&130	&277	&0.63	&16\\
\hline
$\bmx5\\4\emx$		&$\bmx1\text{ s}\\1\text{ l}\emx$	&1	&46	&97	&173	&0.63	&17	&$\bmx17\text{ s}\\17\text{ l}\emx$	&17	&69	&145	&224	&0.60	&18\\
\hline
$\bmx9\\7\emx$		&$\bmx8\text{ s}\\8\text{ l}\emx$	&8	&44	&89	&179	&0.64	&16	&$\bmx6\text{ s}\\6\text{ l}\emx$	&6	&53	&104	&194	&0.64	&14\\
\hline
\multicolumn{15}{c}{short clear cable (s = 405\,mm and l = 911\,mm)}\\
\hline
$\bmx11\\14\emx$	&$\bmx15\text{ s}\\15\text{ l}\emx$	&15	&44	&85	&162	&0.64	&16	&$\bmx3\text{ s}\\3\text{ l}\emx$	&14	&45	&96	&156	&0.62	&16\\
\hline
$\bmx10\\6\emx$		&$\bmx10\text{ s}\\10\text{ l}\emx$	&10	&44	&86	&167	&0.63	&16	&$\bmx9\text{ s}\\9\text{ l}\emx$	&9	&41	&83	&163	&0.62	&15\\
\hline
$\bmx8\\15\emx$		&$\bmx13\text{ s}\\13\text{ l}\emx$	&13	&43	&81	&151	&0.65	&17	&$\bmx14\text{ s}\\14\text{ l}\emx$	&21	&32	&64	&115	&0.56	&17\\
\hline
$\bmx18\\20\emx$	&$\bmx12\text{ s}\\12\text{ l}\emx$	&12	&37	&73	&135	&0.64	&17	&$\bmx4\text{ s}\\4\text{ l}\emx$	&4	&36	&71	&130	&0.57	&17\\
\hline
\multicolumn{15}{c}{spare parts: short (19)/long clear cable (21)}\\
\hline
$\bmx3\\17\emx$		&$\bmx19\text{ s}\\19\text{ l}\emx$	&2	&41	&87	&156	&0.68	&14	&$\bmx21\text{ s}\\21\text{ l}\emx$	&3	&40	&66	&117	&0.78	&16\\
\hline
\end{tabular}
\end{center}
\end{table}

\begin{table}
\begin{center}
\captionof{table}{\label{t-voltages}PMT voltage [V] for a homogeneous response (constant MOP values in ADC counts).}
\begin{tabular}{c||c|c|c|c|c}
\hline
\hline
PMT	&MOP=50	&MOP=75	&MOP=100	&MOP=125	&MOP=150\\
\hline
1	&1916	&2014	&2108	&2174	&2239\\
4	&1980	&2114	&2198	&2283	&2368\\
6	&1888	&1986	&2084	&2147	&2202\\
7	&1851	&1912	&1974	&2036	&2098\\
8	&1927	&2038	&2124	&2180	&2236\\
9	&1943	&2062	&2143	&2205	&2268\\
10	&1929	&2048	&2135	&2196	&2258\\
11	&1871	&1943	&2014	&2086	&2127\\
12	&1972	&2106	&2187	&2268	&2348\\
13	&1937	&2068	&2154	&2226	&2297\\
14	&1920	&2018	&2113	&2197	&2280\\
15	&1929	&2051	&2139	&2204	&2269\\
17	&1850	&1916	&1982	&2047	&2113\\
18	&1845	&1898	&1951	&2004	&2057\\
19	&1804	&1835	&1865	&1896	&1927\\
21	&2013	&2143	&2241	&2339	&2437\\
\hline
\end{tabular}
\end{center}
\end{table}

\subsection{Pre-Integration Process}

Since the superconducting magnet was delayed, a pre-integration of all AMS-02 subdetectors was carried out to investigate possible mechanical, electronic and detector performance issues. The pre-integration was performed in a clean room at CERN. The ACC was the first subdetector installed on the helium vessel (Fig.~\ref{f-acc_pre-integration}). In the first step the PMT boxes with the clear fiber cables were mounted with the vessel in the horizontal position. For the integration of the scintillator panels and connection of WLS fibers to the clear fiber cables the vessel was rotated to the vertical position. Panel 17 was produced with a smaller width than the other panels to take into account mechanical tolerances. The initially foreseen panel 11 was replaced by this smaller panel 17. Before inserting the carbon fiber support cylinder the laser alignment fibers were taped to panel 17. The first test results showed light leaks on the joint between the scintillator and the WLS fiber bundle tubes which had not occurred before and required additional covers. The light leaks were created during the integration on the helium vessel. For the pre-integration period black plastic bags were used to shade the light leaks. After the pre-integration phase all joints were additionally sealed with black tissue and black silicone.

\begin{figure}
\begin{center}
\centerline{\epsfig{file=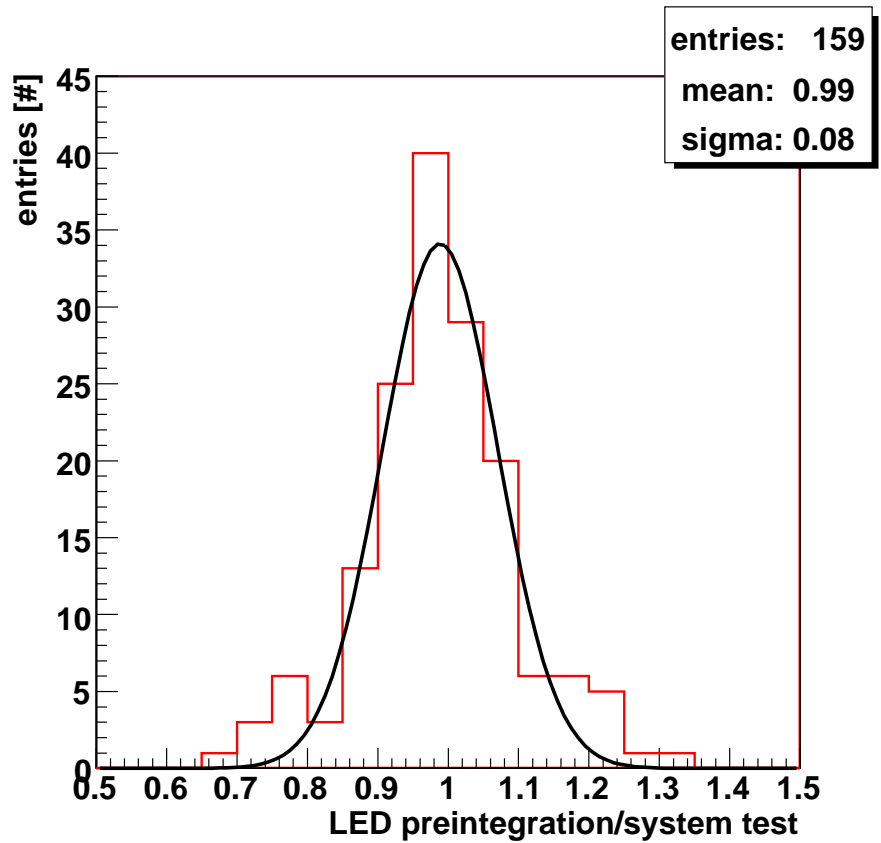,width=12cm}}\captionof{figure}{\label{f-rel_dev}Deviation between LED MOP values of LED spectra of the system test and the pre-integration test.}
\end{center}
\end{figure}

Pre-Integration was followed by the same tests carried out in the ACC system test (Sec.~\ref{ss-systemtest}). Muons were measured with the vessel in the horizontal position such that the distribution of the angle of incidence of atmospheric muons on the panels results in different pathlengths in the material and thus in shifted energy deposition spectra. Therefore, only the LED spectra can be used to compare the system test in Aachen and the pre-integration at CERN. The comparison of PMT mean values in the two tests for all LED runs at three different voltages is shown in Fig.~\ref{f-rel_dev}. Their ratio is on average 99\,\% with a standard deviation of 8\,\% such that the pre-integration was successful with the use of additional light covers. It can be concluded that dismounting, packing and transportation did not cause a change in the detector response.

\section{Performance and Inefficiency Studies with the Anticoincidence Counter \label{s-inefficiency}}

\begin{figure}
\begin{center}
\centerline{\rotatebox{270}{\epsfig{file=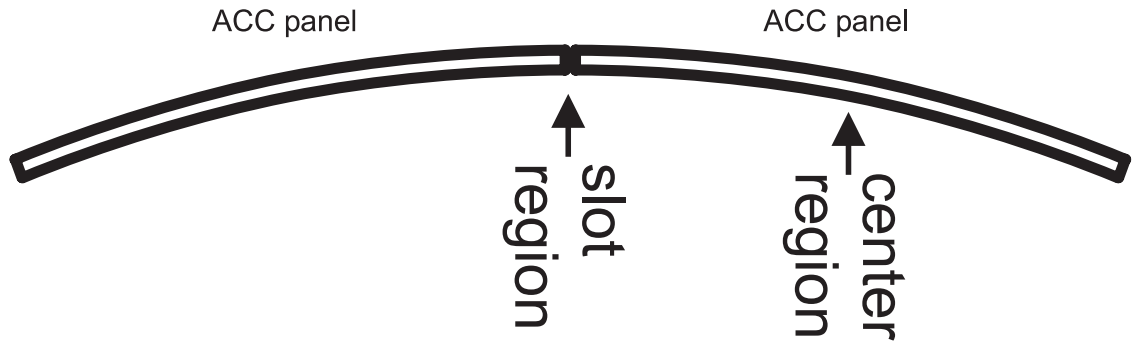,height=15cm}}}\captionof{figure}{\label{f-panel_side}Regions of interest for the inefficiency measurements.}
\end{center}
\end{figure}

The detection inefficiency of the anticoincidence counter is a crucial number for the determination of systematic uncertainties in antiparticle measurements with very small fluxes. The design of a modular detector consisting of 16 singular panels leads to the expectation that especially the slot regions between panels are very sensitive parts. The particles cross less material in these regions which results in a smaller energy deposition. In addition, the density of wavelength shifting fibers is smaller at the edges of the panels (Fig.~\ref{f-panel_side}). These facts necessitate a careful determination of the inefficiency as a function of the position.

This section studies effects of the final detection chain, electronics and data acquisition on the ACC detection efficiency using testbeam and atmospheric muon measurements. These results are then used to develop a model for the inefficiency determination.

\subsection{Testbeam Measurements\label{ss-testbeam}}

Testbeam measurements over the central region and the slot regions were performed in order to extract the mean inefficiency across the whole ACC panel.

\subsubsection{Setup and Trigger}

\begin{figure}
\begin{center}
\centerline{\epsfig{file=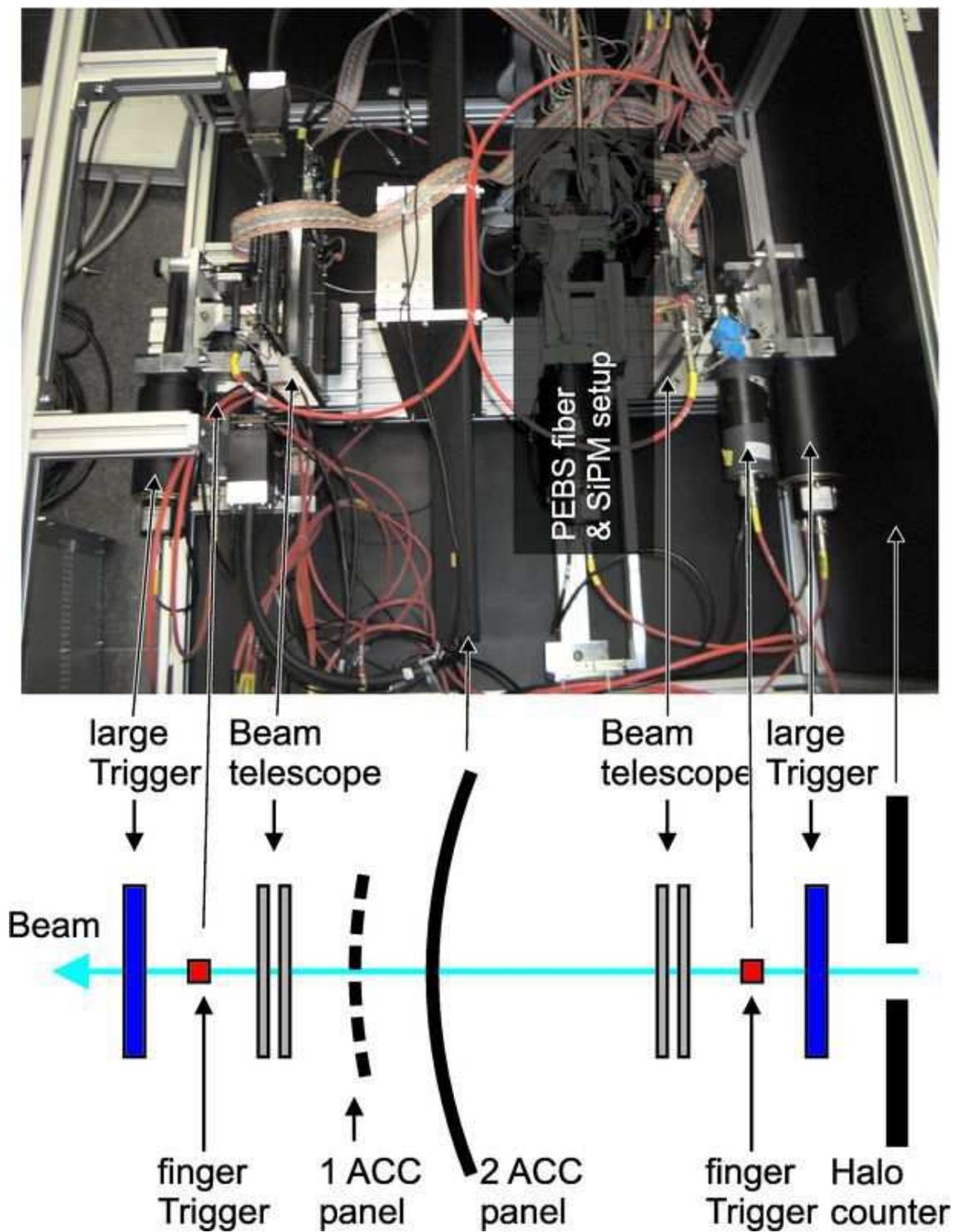,width=15cm}}
\captionof{figure}{\label{f-testbeam_overview}Testbeam setup.}
\end{center}
\end{figure}

The measurements were carried out at the PS T9 beam at CERN with 10\,GeV protons in October 2006 before the production of all panels, the clear fiber cables and the classification of PMTs was finished. Fig.~\ref{f-testbeam_overview} shows the setup, consisting of two large and two small finger trigger scintillation counters with photomultiplier readout, a beam telescope made out of four crossed CMS\footnote{The Compact Muon Solenoid is one of the experiments at the Large Hadron Collider at CERN} tracker endcap silicon microstrip detectors and a halo counter to veto particles which are not in the sensitive area of the beam telescope. This setup was used to test the central and the slot region of the ACC, as well as scintillating fibers with silicon photomultiplier readout for the PEBS experiment (Sec.~\ref{c-pebs}). The readout was done using NIM and CAMAC electronics and a LeCroy waverunner oscilloscope. Each ACC panel was connected directly to two photomultiplier tubes. The voltages for the PMTs were adjusted so that the spectra have similar MOP values.

\begin{figure}
\begin{center}
\centerline{\epsfig{file=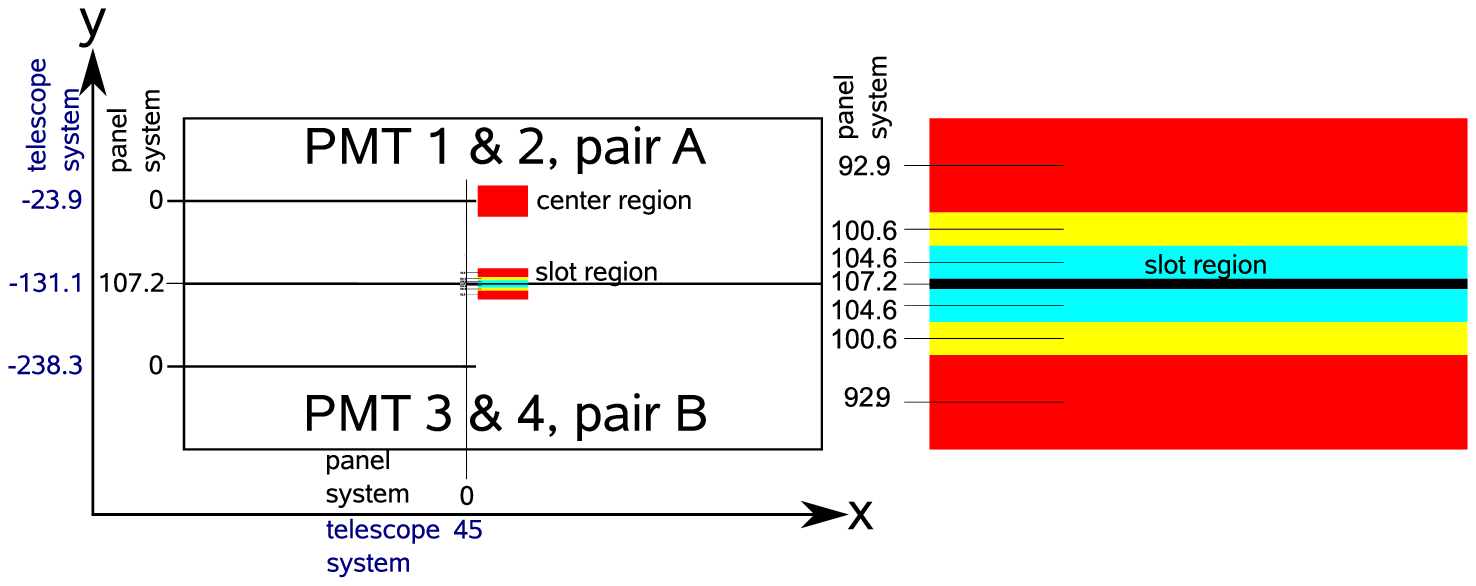,width=15cm}}\captionof{figure}{\label{f-coordinate_system_testbeam_overview_wide}Coordinate systems, all values are in mm: \textbf{\textit{Left) }}The two panel regions under investigation and the comparison of the beam telescope coordinate system with the panel coordinate system. \textbf{\textit{Right) }}Close up of the slot region and the subdivisions.}
\end{center}
\end{figure}

\begin{figure}
\begin{center}
\centerline{\epsfig{file=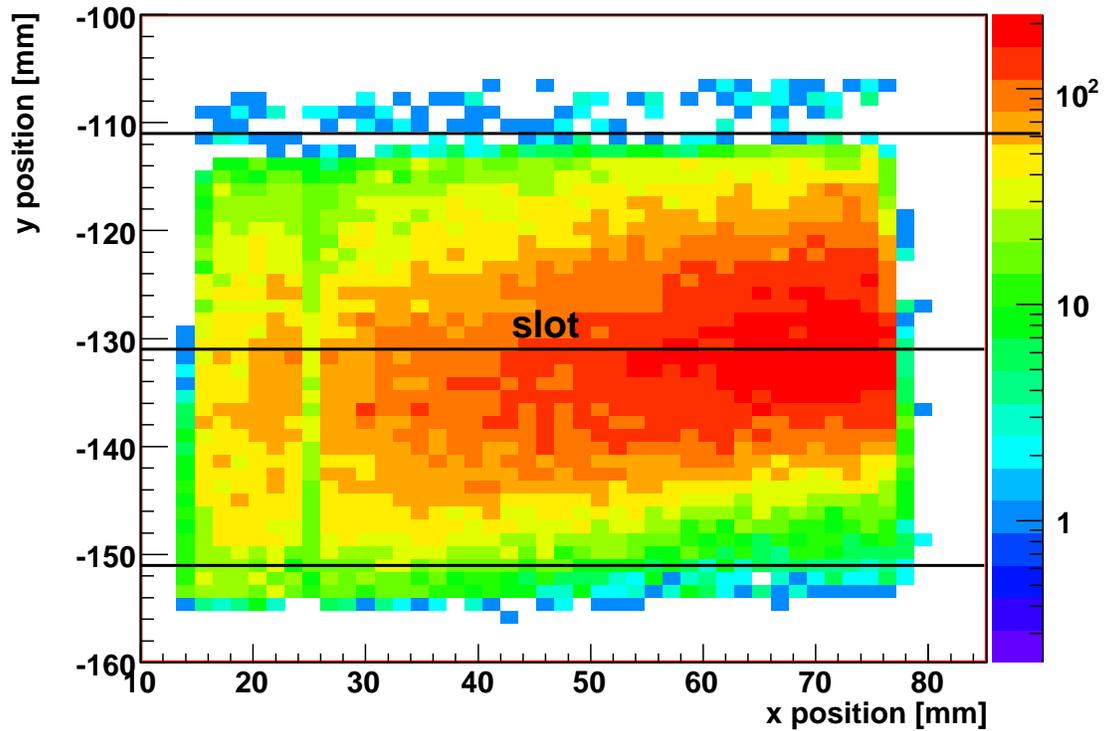,width=15cm}}\captionof{figure}{\label{f-beamprofile}Beam profile collected with the telescope and defined by the large trigger counters. The lines indicate the $y\sub{bt}$ range of hits used in the analysis. The color code on the right shows the number of entries.}
\end{center}
\end{figure}

The coordinate systems used for the ACC analysis are illustrated in Fig.~\ref{f-coordinate_system_testbeam_overview_wide} and \ref{f-beamprofile}. One is the panel system and one the beam telescope coordinate system. The coordinate system along the ACC panels is called the $x\sub{bt}$ direction. The beam spot is always in the center of the panel in $x\sub{bt}$ direction for the measurements described here. The $y\sub{bt}$ direction is perpendicular to the slot region between the panels and is varied throughout the testbeam. The $z\sub{bt}$ axis is aligned with the beam axis and therefore perpendicular to the beam telescope. In the following, the projection of the panel to the $(x\sub{bt},y\sub{bt})$ plane is used and the curvature of the panels is neglected because of a sensitive beam telescope area of only $40\times70$\,mm. The projected width of an ACC panel is 214.4\,mm. In addition, the center of the ACC panel is the origin of the panel coordinate system and the slot is defined to be at 107.2\,mm in $y\sub{p}$ direction in this system. The $x\sub{p}$ direction in the panel coordinate system is again along the panel. The panel coordinate system of the upper panel runs top-down and the panel coordinate system of the lower panel runs bottom-up such that both panels have the coordinate 107.2\,mm at the slot position in the corresponding panel system. The photomultipliers of the upper panel are called pair\,A with PMT\,1 and 2 while the lower panel photomultipliers are PMT\,3 and 4 and called pair\,B. This testbeam uses test PMTs and the numbers 1 - 4 have nothing to do with the production numbers introduced in Sec.~\ref{ss-pmttest}. The analysis of the slot region was subdivided into 7\,zones to check the inefficiency and signal behavior as a function of position.

The measurements were carried out in two trigger configurations. In the first one, only the large trigger counters were used. In the second one, the small finger trigger counters were put in coincidence with the large ones to study the ACC slot region. Only clean single track events in the beam telescope with large hits in the trigger counters are used for the analysis. Fig.~\ref{f-beamprofile} shows the beam profile obtained with the beam telescope when using only the large trigger counters.

\begin{figure}
\begin{center}
\begin{minipage}[b]{.4\linewidth}
\centerline{\epsfig{file=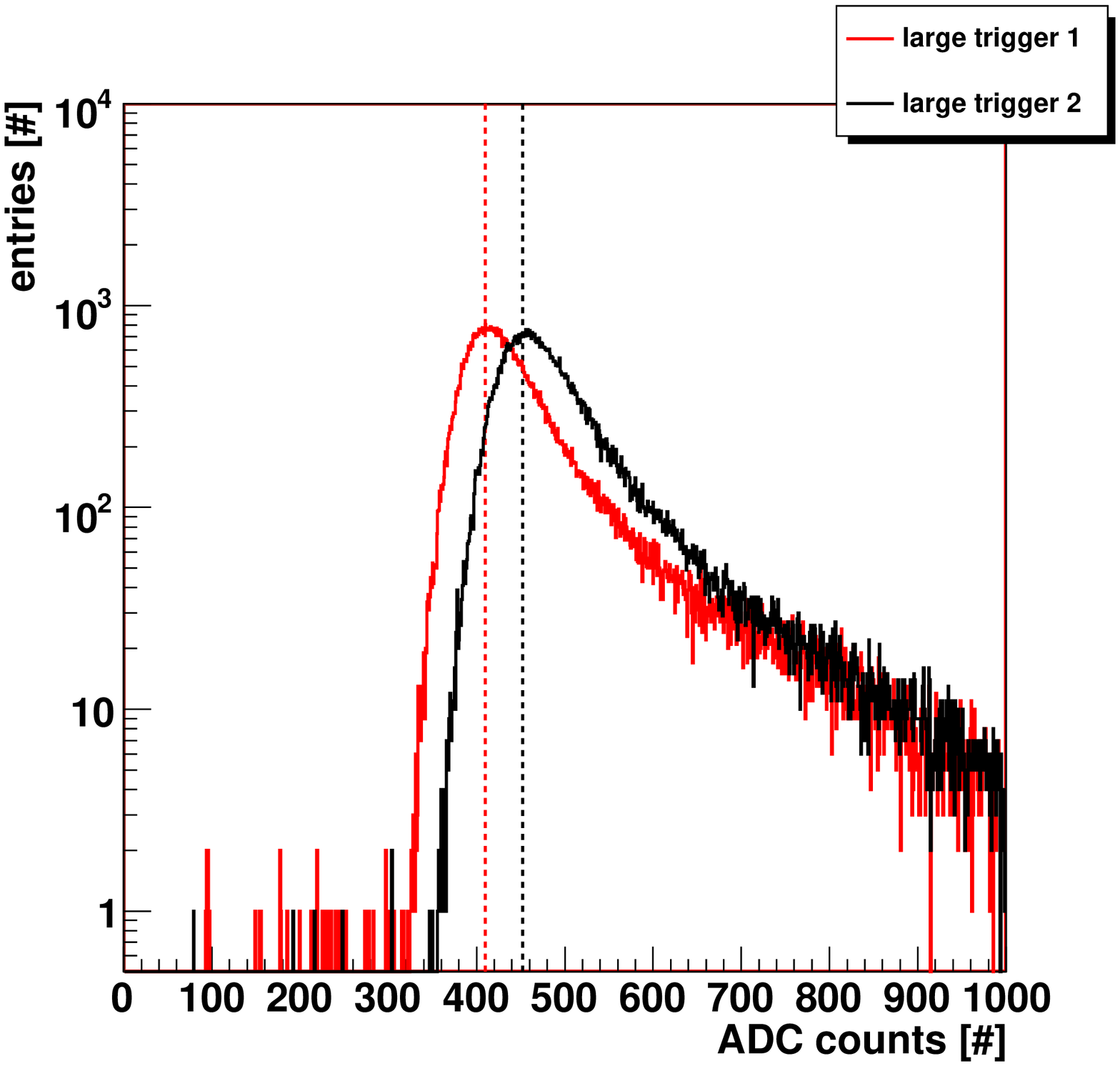,width=8cm}}\captionof{figure}{\label{f-trigger_2acc_ODER}Trigger spectra for large counters only. Cuts (dashed lines) are at 410 and 452 ADC counts.}
\end{minipage}
\hspace{.1\linewidth}
\begin{minipage}[b]{.4\linewidth}
\centerline{\epsfig{file=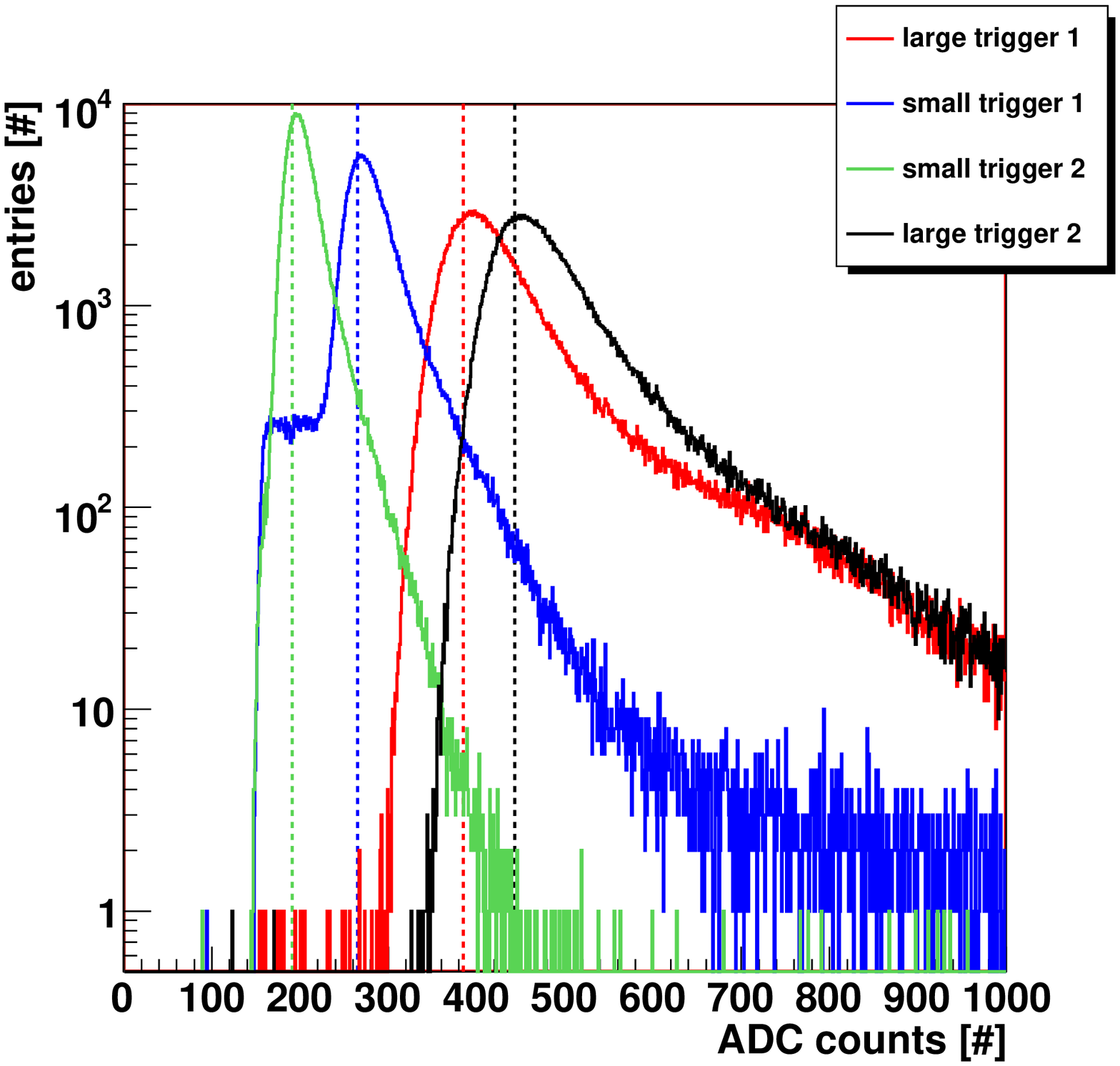,width=8cm}}\captionof{figure}{\label{f-trigger_2acc_UND}Trigger spectra for large and small counters. Cuts (dashed lines) are at 385, 265, 191 and 443 ADC counts.}
\end{minipage}
\end{center}
\end{figure}

\begin{figure}
\begin{center}
\begin{minipage}[b]{.4\linewidth}
\centerline{\epsfig{file=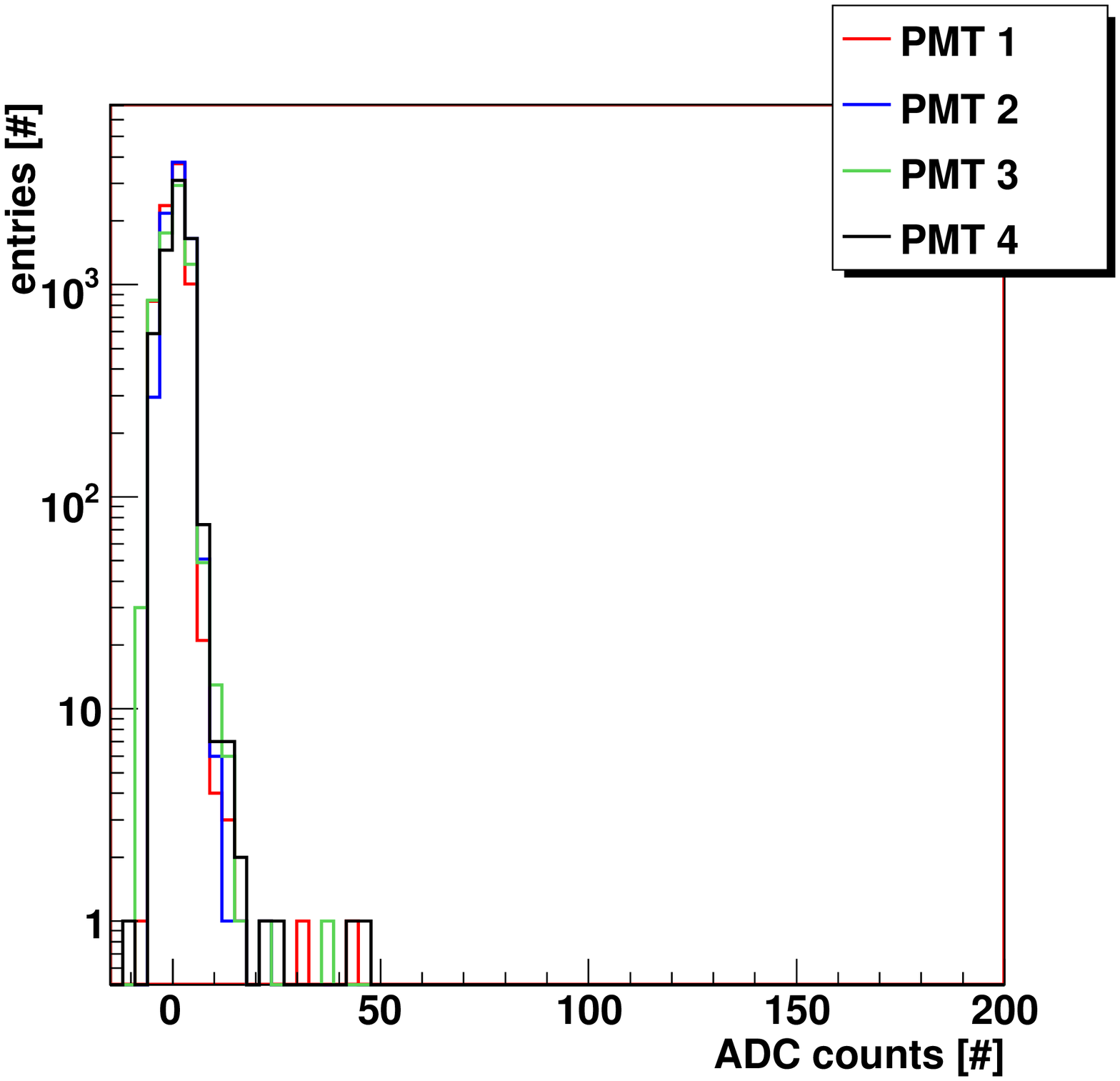,width=8cm}}\captionof{figure}{\label{f-0705_adc_pedestal_1acc_-1000_0}Pedestal measurements for all PMTs.}
\end{minipage}
\hspace{.1\linewidth}
\begin{minipage}[b]{.4\linewidth}
\centerline{\epsfig{file=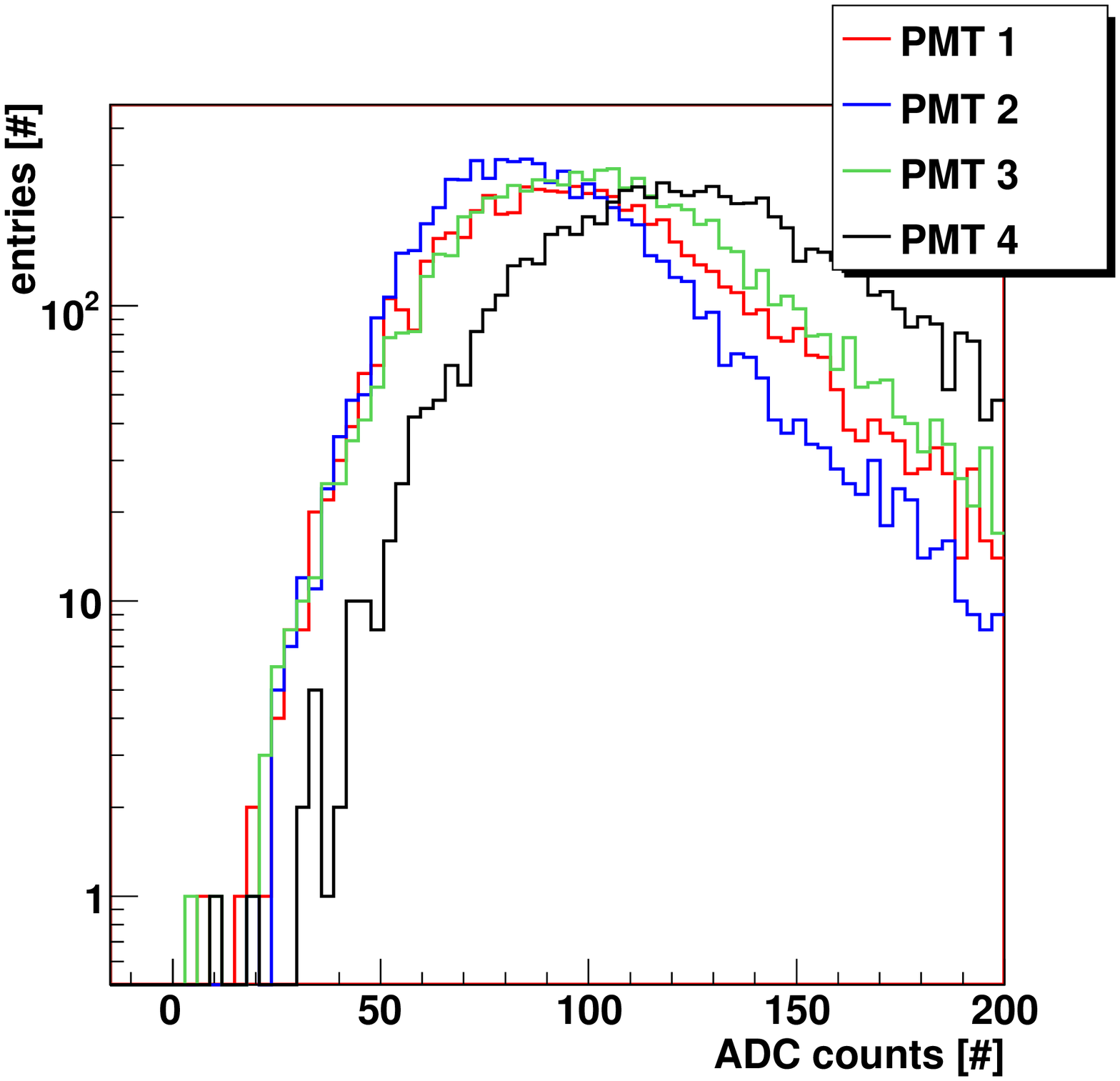,width=8cm}}\captionof{figure}{\label{f-0705_pedcorr_beam_1acc_ODER_-1000_0}Pedestal corrected ADC spectra for all PMTs at position $y\sub{p}=92.9$\,mm in panel coordinates.}
\end{minipage}
\end{center}
\end{figure}

The pulseheight spectra collected for the trigger counters for both trigger conditions are shown in Fig.~\ref{f-trigger_2acc_ODER} and \ref{f-trigger_2acc_UND} where the cuts applied are also indicated. Events are further analysed if for each counter the most probable value is exceeded. For the ACC panel photomultipliers Fig.~\ref{f-0705_adc_pedestal_1acc_-1000_0} and \ref{f-0705_pedcorr_beam_1acc_ODER_-1000_0} show the pedestal and signal spectra at an average position of $y\sub{p}=92.9$\,mm in panel coordinates. In the following, a good event in the ACC is defined by showing at least 1 pulseheight of any ACC photomultiplier above: \be p_i+3\cdot\sigma_i\ee where $p_i$ is the mean value and $\sigma_i$ the RMS of the pedestal distribution.  On average the cut for all pedestal corrected ADC spectra of the PMTs is at $3\sigma_i=7$\,ADC counts.

\subsubsection{Calibration of the CAMAC ADC}

\begin{figure}
\begin{center}
\centerline{\epsfig{file=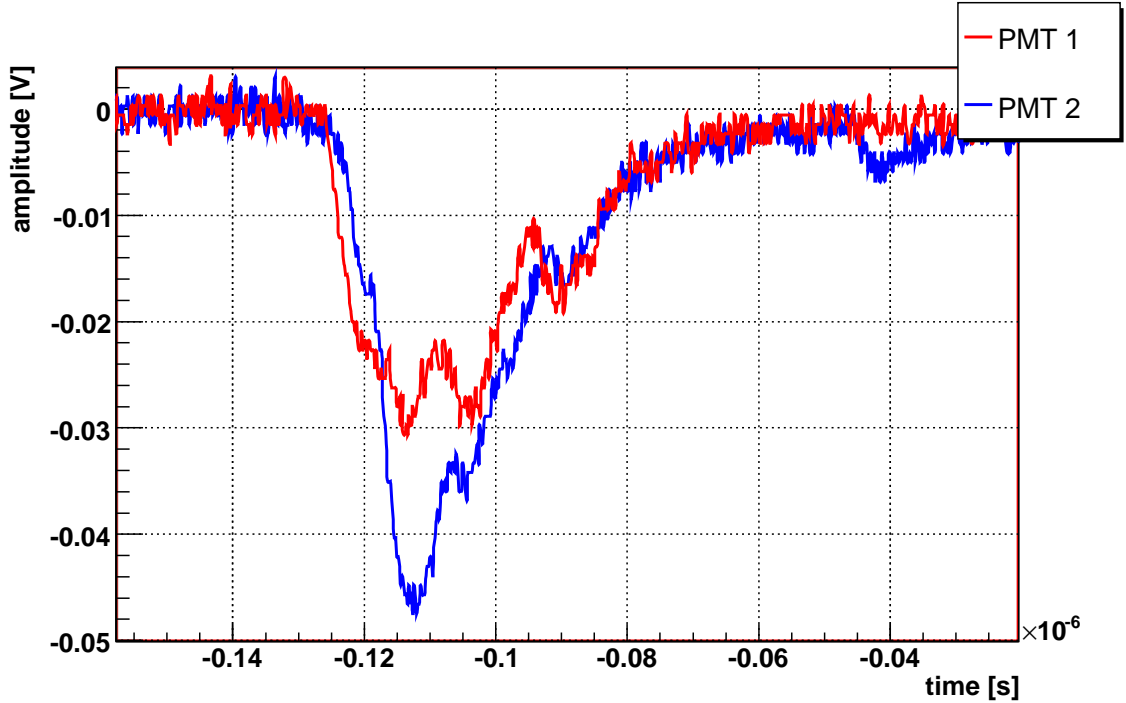,width=15cm}}
\captionof{figure}{\label{f-typical_event}Typical PMT signal shape recorded with the oscilloscope during the testbeam measurements.}
\end{center}
\end{figure}

\begin{figure}
\begin{center}
\begin{minipage}[b]{.4\linewidth}
\centerline{\epsfig{file=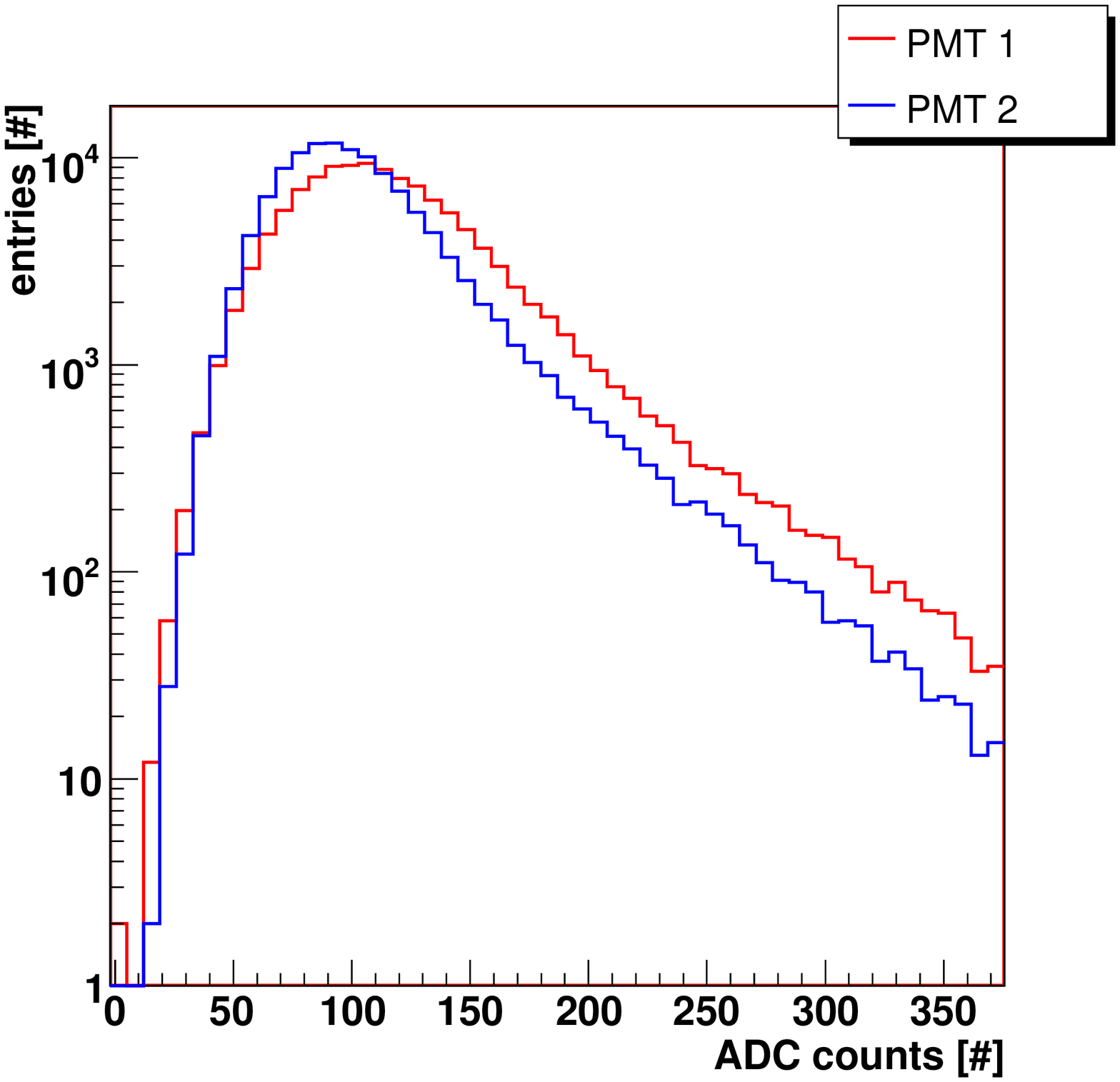,width=8cm}}\captionof{figure}{\label{f-0712_pedcorr_beam_1acc_ODER_-1000_0}Pedestal corrected ADC counts in the central region for both PMTs.}
\end{minipage}
\hspace{.1\linewidth}
\begin{minipage}[b]{.4\linewidth}
\centerline{\epsfig{file=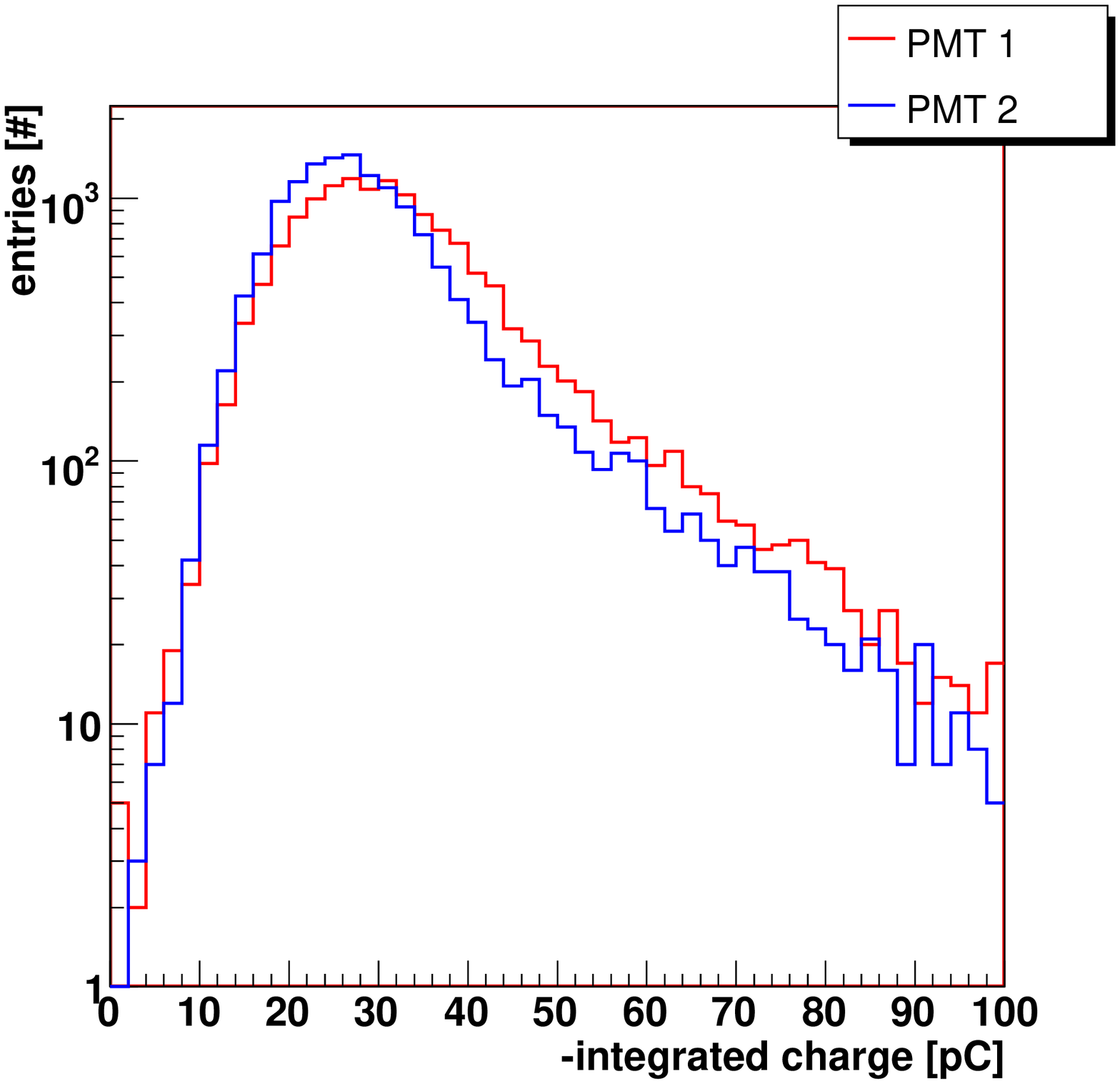,width=8cm}}\captionof{figure}{\label{f-integrated_charge_26_28}Integrated charges measured with an oscilloscope.}
\end{minipage}
\end{center}
\end{figure}

\begin{figure}
\begin{center}
\begin{minipage}[b]{.4\linewidth}
\centerline{\epsfig{file=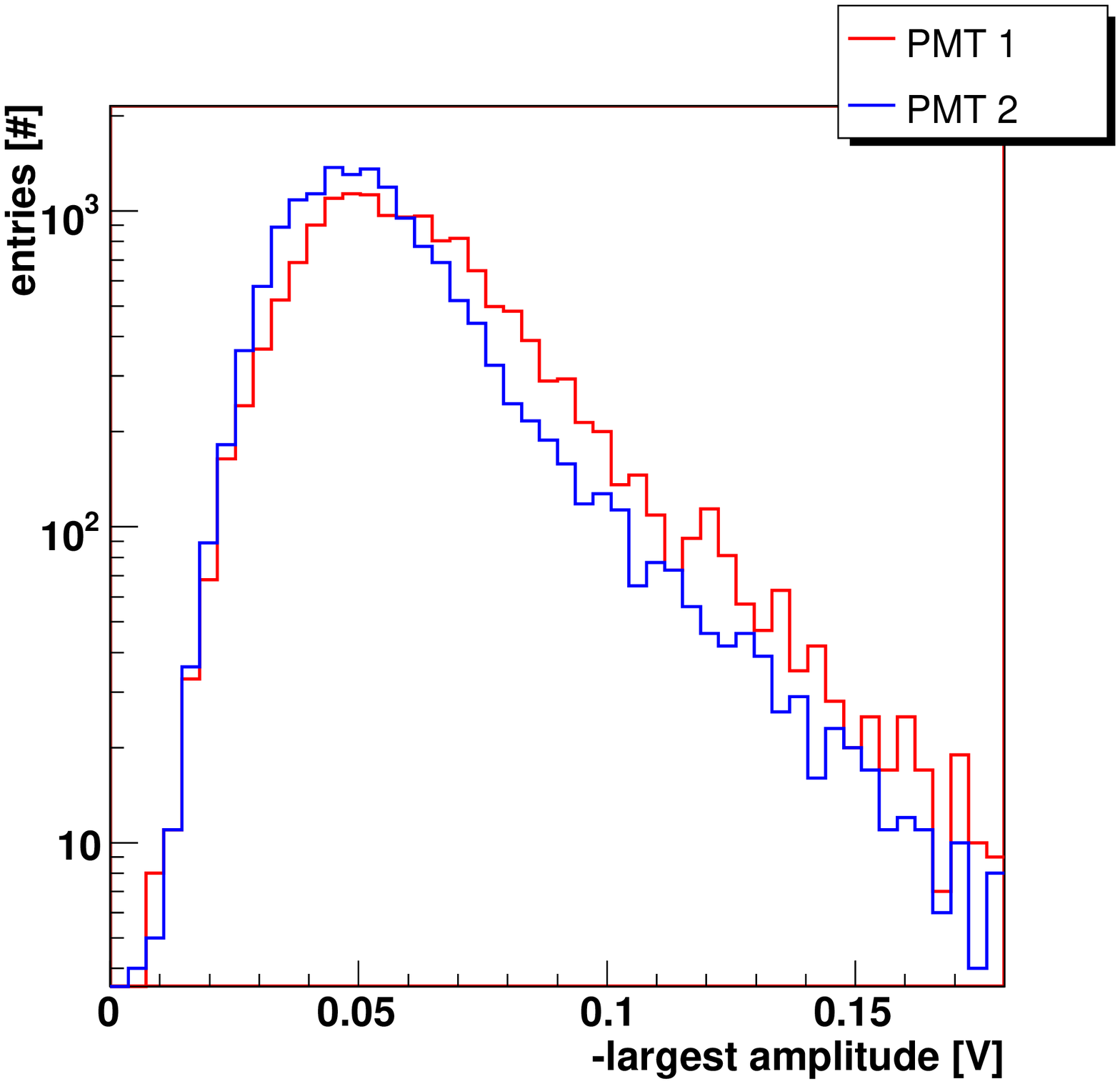,width=8cm}}\captionof{figure}{\label{f-largest_amplitude_26_28}Largest amplitude measured with an oscilloscope.}
\end{minipage}
\hspace{.1\linewidth}
\begin{minipage}[b]{.4\linewidth}
\centerline{\epsfig{file=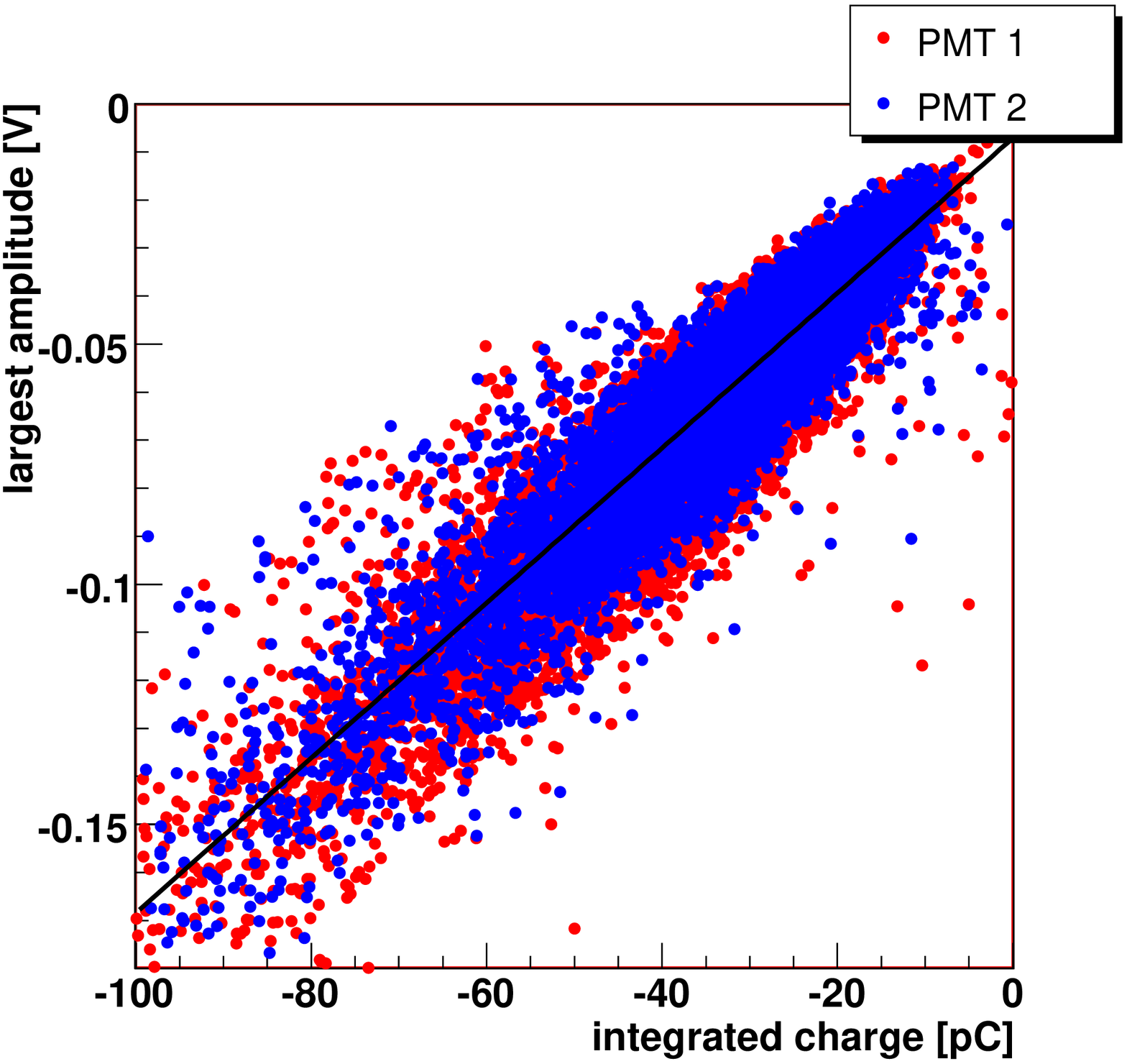,width=8cm}}\captionof{figure}{\label{f-la_ic_26_28}Correlation between largest amplitude and integrated charge.}
\end{minipage}
\hspace{.1\linewidth}
\end{center}
\end{figure}

Fig.~\ref{f-typical_event} shows a typical ACC pulse recorded with the oscilloscope during the testbeam measurements. Instead of a smooth behavior some spikes are observed. This is due to the small signals arriving at the PMTs. The distribution of pedestal corrected data of the oscilloscope and of the CAMAC ADC can be used to calibrate the latter. The integrated charge from the oscilloscope data was calculated according to \be Q= \frac {\Delta t} R \sum_i U_i\ee where $\Delta t$ is the time resolution, $U_i$ the voltage amplitude at time index $i$ and $R$ the termination resistance. Assuming linear behavior of the CAMAC ADC in the range used, the calibration  factor is determined by the ratio of the two most probable pedestal corrected values, ADC counts to integrated charge (Fig.~\ref{f-0712_pedcorr_beam_1acc_ODER_-1000_0} and \ref{f-integrated_charge_26_28}). This results in a charge calibration factor of $3.37\,\text{ADC}/\text{pC}$. In addition to an ADC charge measurement, the flight electronics will also apply a discriminator threshold on the pulse amplitude. It is therefore important to study the correlation between integrated charge and largest amplitude of a pulse. For the same data used in the calibration procedure, Fig.~\ref{f-largest_amplitude_26_28} shows the distribution of the largest pulse amplitudes. In general, the correlation between charge and largest amplitude is linear and can be reasonably fitted by (Fig.~\ref{f-la_ic_26_28}):
\be
U=(1.6\pm0.6)\frac{\displaystyle \text{mV}}{\displaystyle \text{pC}}\cdot Q-(7.1\pm20.0)\,\text{mV}.
\ee
It must be noted that many points scatter quite wide around this fit which is connected to the after-pulse issue that will be discussed in the Sec.~\ref{ss-accpretest}.

\subsubsection{Slot Determination}

\begin{figure}
\begin{center}
\begin{minipage}[b]{.4\linewidth}
\centerline{\epsfig{file=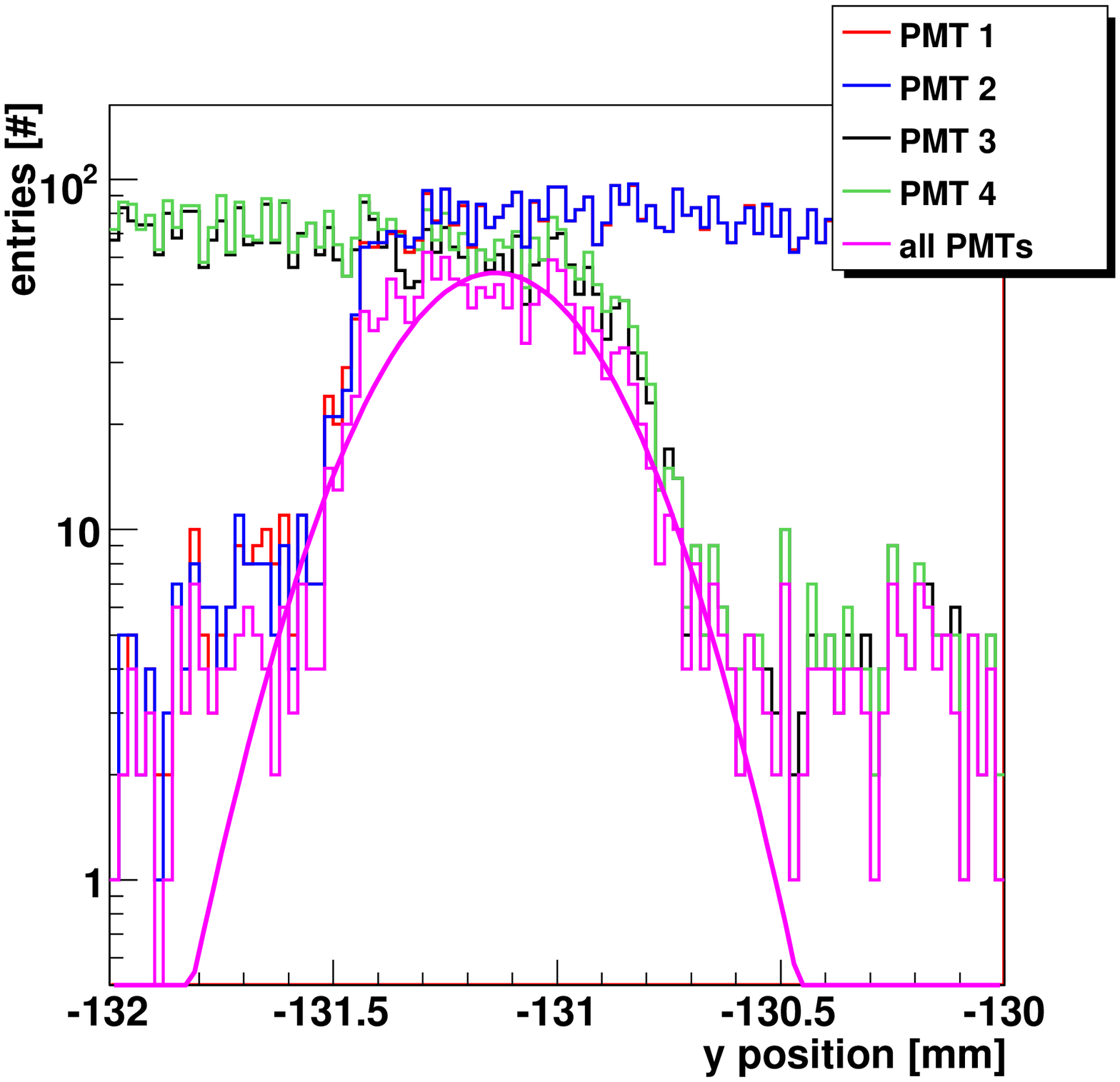,width=8cm}}\captionof{figure}{\label{f-0705_slit_beam_2acc_ODER_-151_-111_-151_-111}Number of events above $3\sigma_i$ for the PMTs with respect to the $y\sub{bt}$ position in beam telescope coordinates. Magenta distribution: events with all four PMTs above $3\sigma_i$.}
\end{minipage}
\hspace{.1\linewidth}
\begin{minipage}[b]{.4\linewidth}
\centerline{\epsfig{file=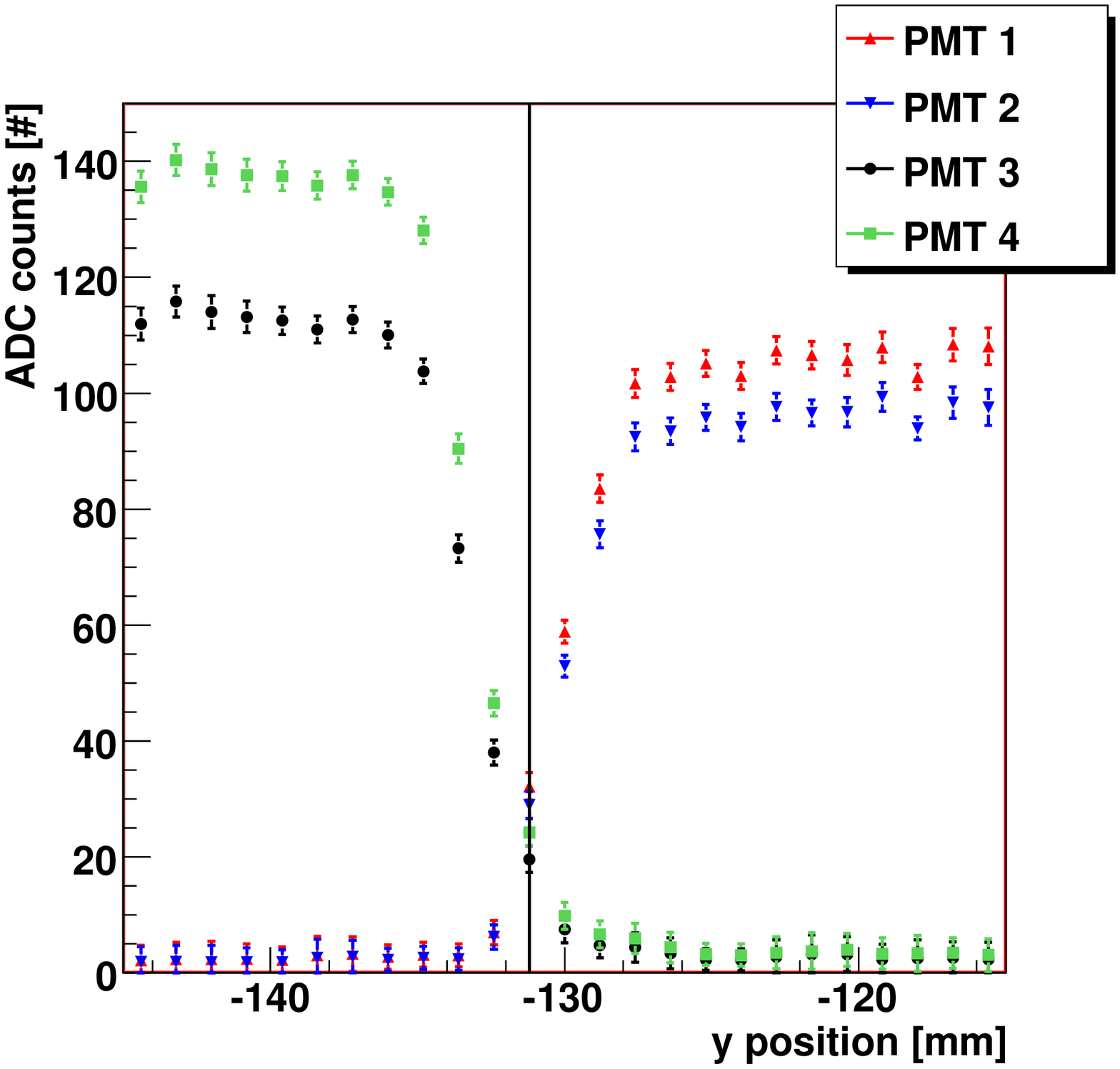,width=8cm}}\captionof{figure}{\label{f-0705_yadc24_beam_2acc_ODER_-151_-111_-151_-111}Mean pedestal corrected ADC counts as a function of the position in beam telescope coordinates across the two panels for all PMTs.}
\end{minipage}
\end{center}
\end{figure}

The slot position is determined by counting the number of events with pulseheights above $3\sigma_i$ for all photomultipliers as a function of the $y\sub{bt}$ coordinate in the beam telescope system (Fig.~\ref{f-0705_slit_beam_2acc_ODER_-151_-111_-151_-111}). A Gaussian fit to the distribution of events with good hits in all four PMTs determines the slot position to be at 131.1\,mm with a standard deviation of 0.2\,mm in beam telescope coordinates. This corresponds to $y\sub{p}=107.2$\,mm in the panel coordinate system. In addition, Fig.~\ref{f-0705_yadc24_beam_2acc_ODER_-151_-111_-151_-111} shows the mean ADC value of all PMTs across the slot in beam telescope coordinates. The mean ADC values of about 100 - 140\,ADC counts drop to nearly 0\,ADC counts for all PMTs as the slot is traversed.

\subsubsection{Quality Cuts}

\begin{figure}
\begin{center}
\begin{minipage}[b]{.4\linewidth}
\centerline{\epsfig{file=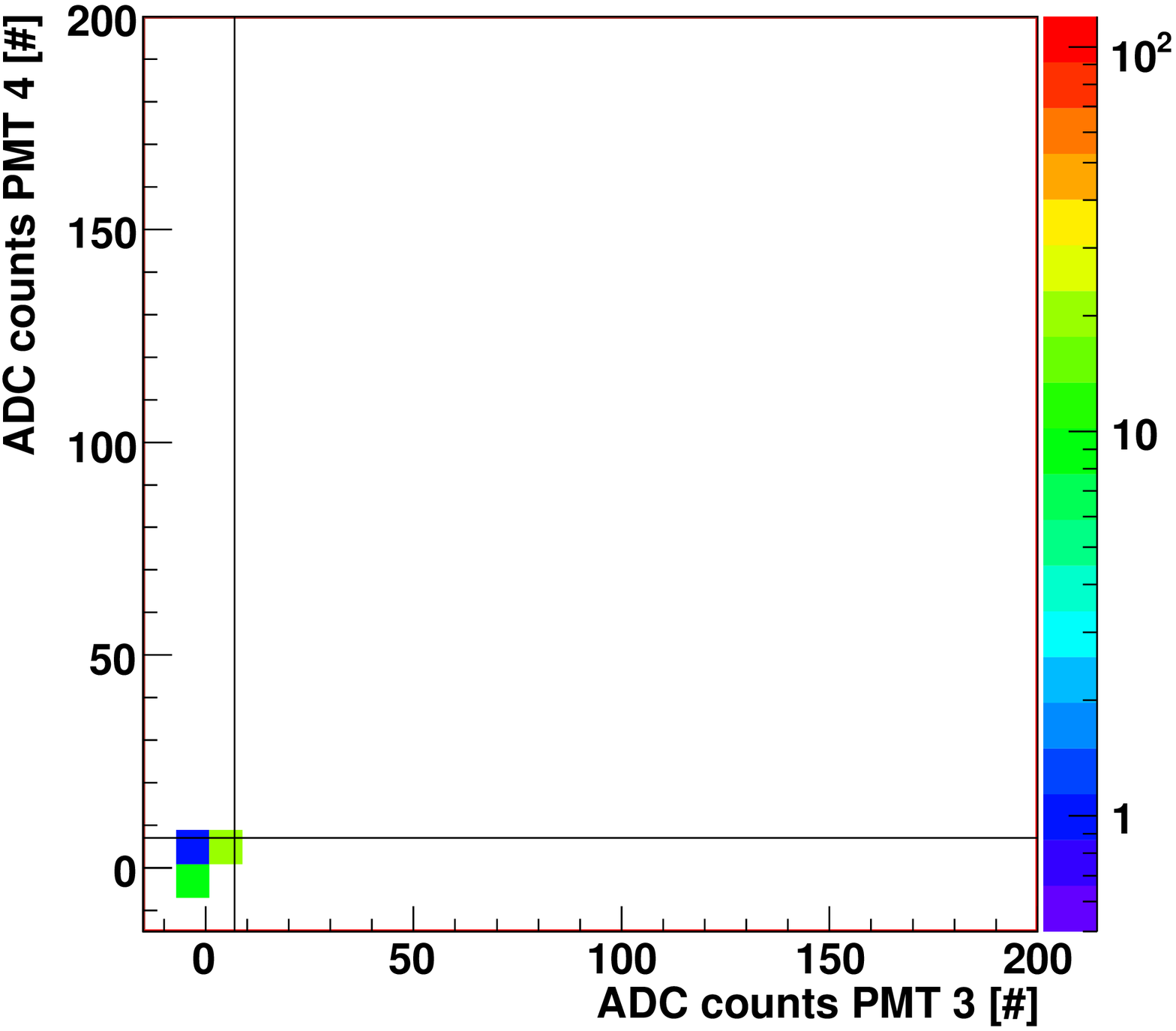,width=8cm}}\captionof{figure}{\label{f-0705_ADCscatter_under_Pair0_Cut1_beam_2acc_ODER_-139_-135_-139_-135}Pedestal corrected ADC counts of Pair\,B PMTs for PMT\,3 and PMT\,4 <$3\sigma_i$. Beam telescope track points to the interval [98.6\,mm,102.6\,mm] in panel coordinates. The color code on the right shows the number of entries. The black lines indicate the cuts for the definition of a good event.}
\end{minipage}
\hspace{.1\linewidth}
\begin{minipage}[b]{.4\linewidth}
\centerline{\epsfig{file=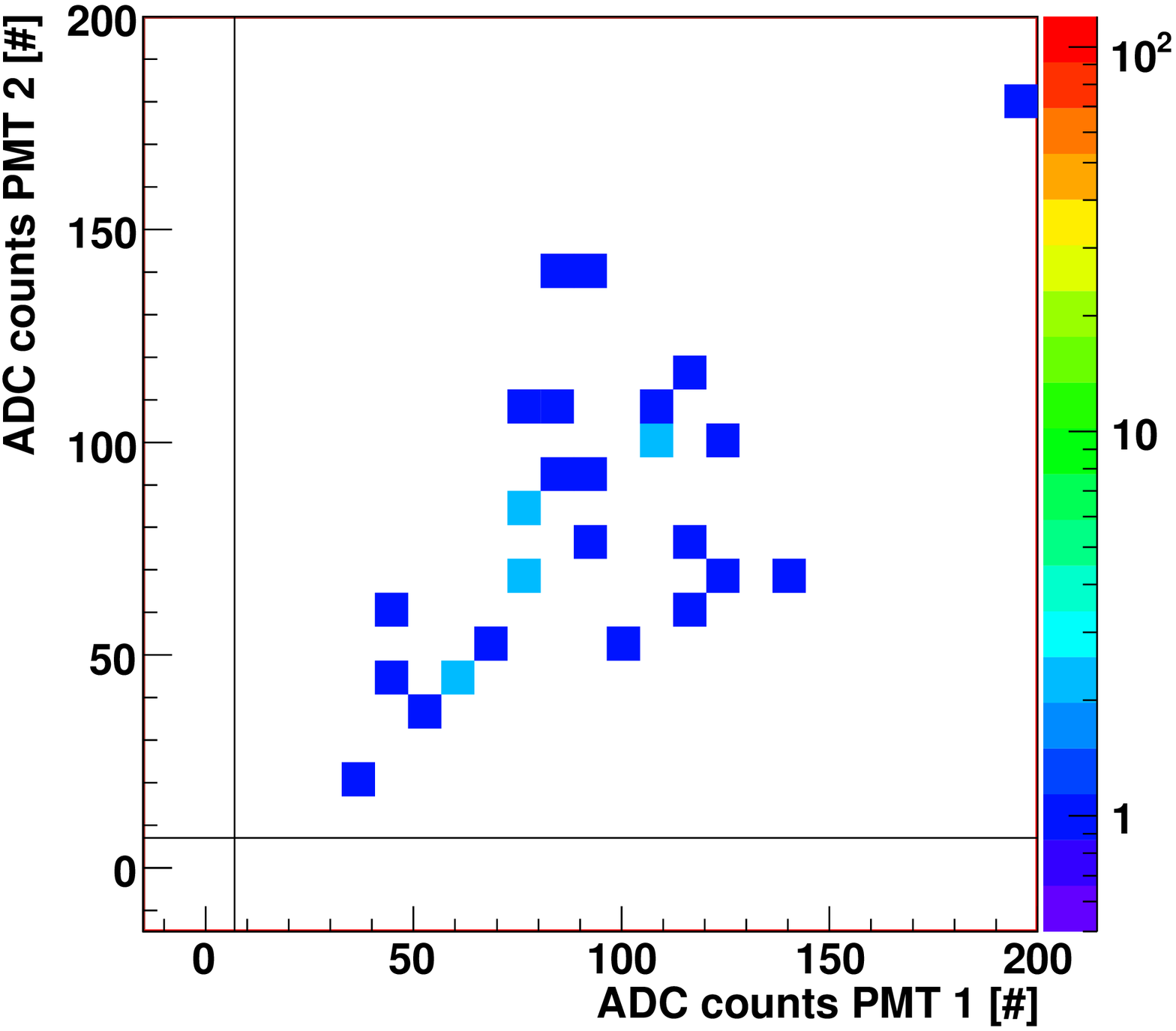,width=8cm}}\captionof{figure}{\label{f-0705_ADCscatter_over_Pair1_Cut1_beam_2acc_ODER_-139_-135_-139_-135}Pedestal corrected ADC counts of Pair\,A PMTs  for PMT\,3 and PMT\,4 <$3\sigma_i$. Beam telescope track points to the interval [98.6\,mm,102.6\,mm] in panel coordinates. The color code on the right shows the number of entries. The black lines indicate the cuts for the definition of a good event.}
\end{minipage}
\end{center}
\end{figure}

\begin{figure}
\begin{center}
\begin{minipage}[b]{.4\linewidth}
\centerline{\epsfig{file=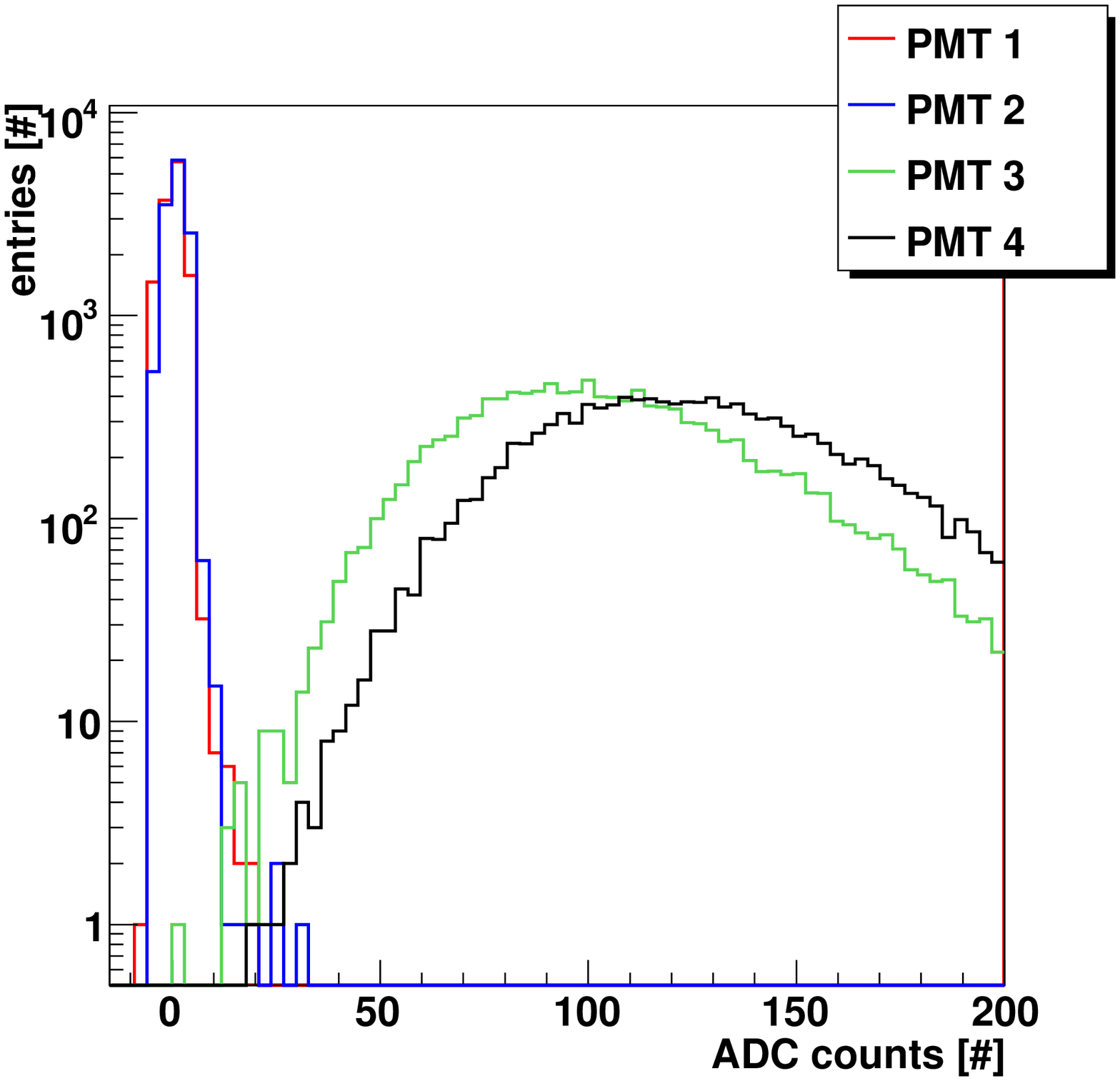,width=8cm}}\captionof{figure}{\label{f-adc_-139_-135_-139_-135}Pedestal corrected ADC counts for all ACC PMTs. Beam telescope track points to the interval [98.6\,mm,102.6\,mm] in panel coordinates.}
\end{minipage}
\hspace{.1\linewidth}
\begin{minipage}[b]{.4\linewidth}
\centerline{\epsfig{file=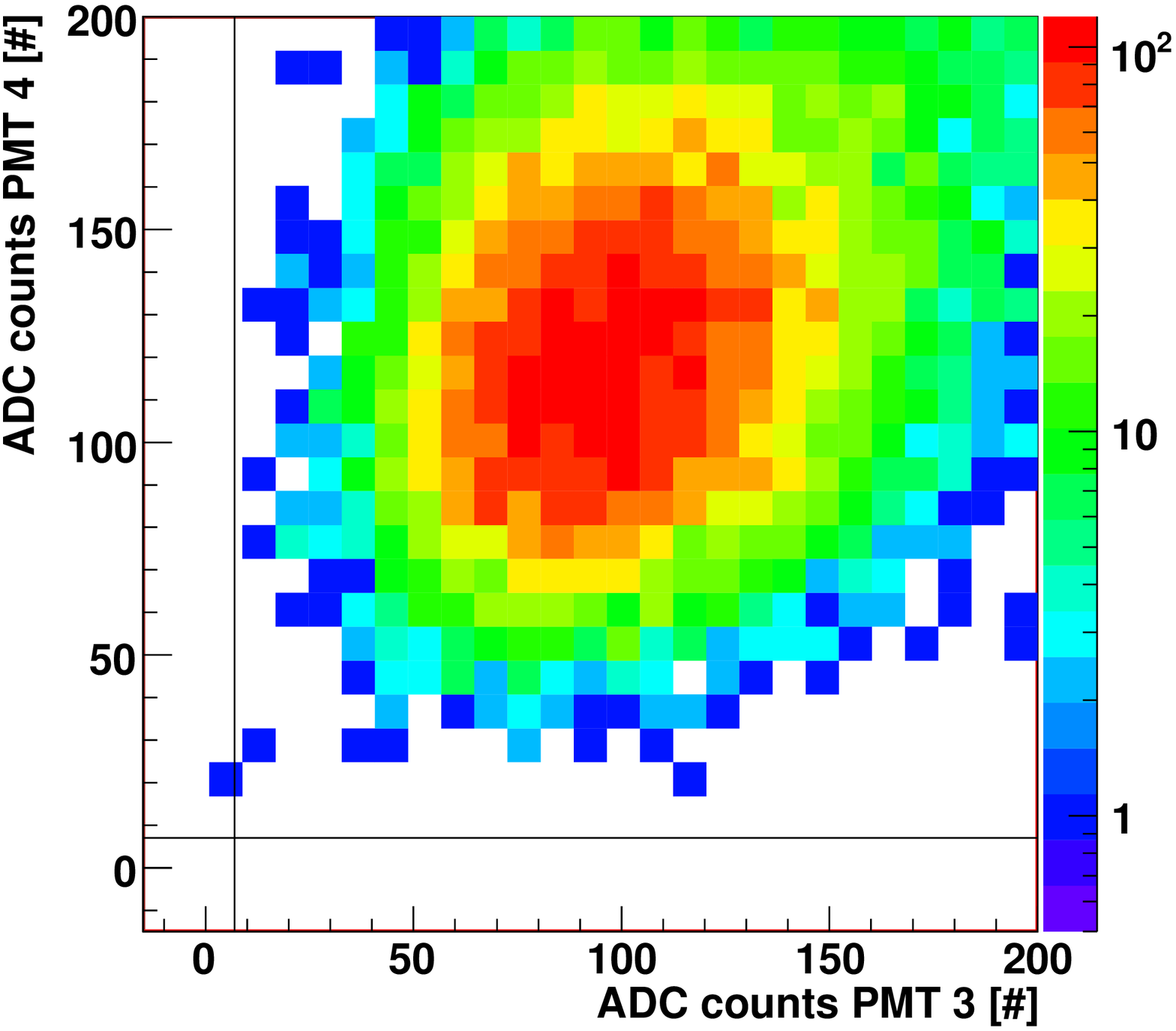,width=8cm}}\captionof{figure}{\label{f-adc_24_25_-139_-135_-139_-135}Pedestal corrected ADC counts of Pair\,B for clean single tracks. Beam telescope track points to the interval [98.6\,mm,102.6\,mm] in panel coordinates. The color code on the right shows the number of entries. The black lines indicate the cuts for the definition of a good event.}
\end{minipage}
\end{center}
\end{figure}

Additional event selection criteria are introduced to increase the data quality for the determination of the inefficiency. They are exemplarily explained for tracks believed to strike the ACC between $y\sub{p}=98.6$\,mm and $y\sub{p}=102.6$\,mm. In this case ACC hits are expected only for PMT pair\,B and not for PMT pair\,A. For events in which both pair B PMTs have pulseheights smaller than $3\sigma_i$ (Fig.~\ref{f-0705_ADCscatter_under_Pair0_Cut1_beam_2acc_ODER_-139_-135_-139_-135}), fig. \ref{f-0705_ADCscatter_over_Pair1_Cut1_beam_2acc_ODER_-139_-135_-139_-135} plots the pulseheights in pair A, showing clear ACC hits at about the MOP values. This leads to the assumption that the hit was in reality in pair\,A. This effect can be due to stray particles which have not been rejected by the halo counter or due to beam telescope inefficiencies. All events with unexpectedly large hits ($>3\sigma_i$) in both PMTs are rejected for the following analysis.  This requirement reduces the number of events by only $\approx 0.01$\,\%. For the remaining events Fig.~\ref{f-adc_-139_-135_-139_-135} shows the spectra of all four PMTs at the position under discussion here. Pair\,B (PMT 3 and 4) has the expected Landau shape while pair\,A (PMT 1 and 2) has only pedestal entries. The correlation of the pair\,B PMTs shows that the highest ADC value of the two pulseheights is always above $3\sigma_i$ (Fig.~\ref{f-adc_24_25_-139_-135_-139_-135}). Therefore, it should be possible to veto with the ACC detector on the basis of the highest ADC value for each event. 

\subsubsection{Energy Depositions across an ACC Panel\label{ss-depch}}

\begin{figure}
\begin{center}
\begin{minipage}[b]{.4\linewidth}
\centerline{\epsfig{file=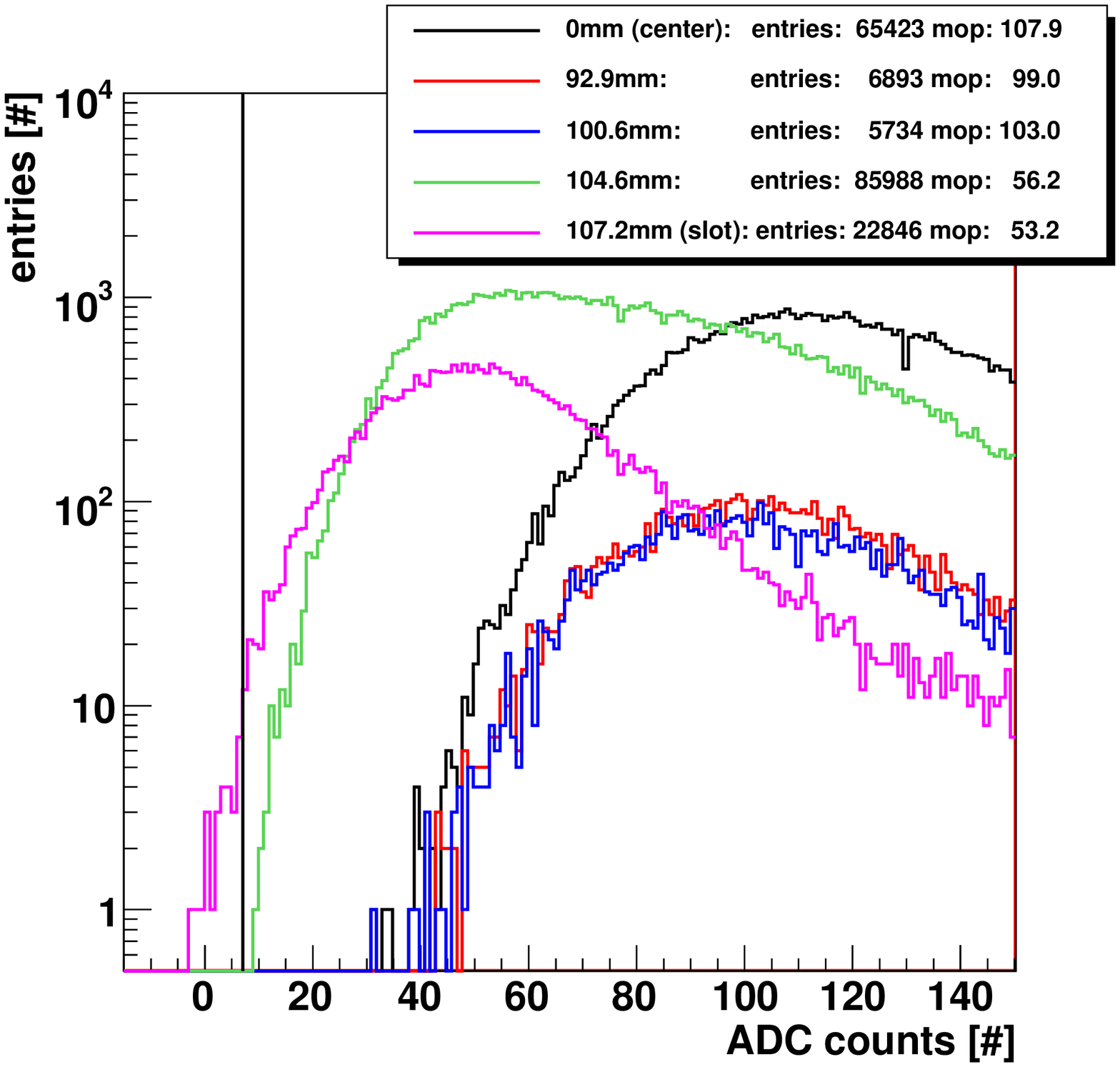,width=8cm}}\captionof{figure}{\label{f-adc_highest_testbeam_pair0}Highest pedestal corrected pulseheights for the pair\,B PMTs for various positions across the panel. The line indicates the cut at $3\sigma$ of the pedestal.}
\end{minipage}
\hspace{.1\linewidth}
\begin{minipage}[b]{.4\linewidth}
\centerline{\epsfig{file=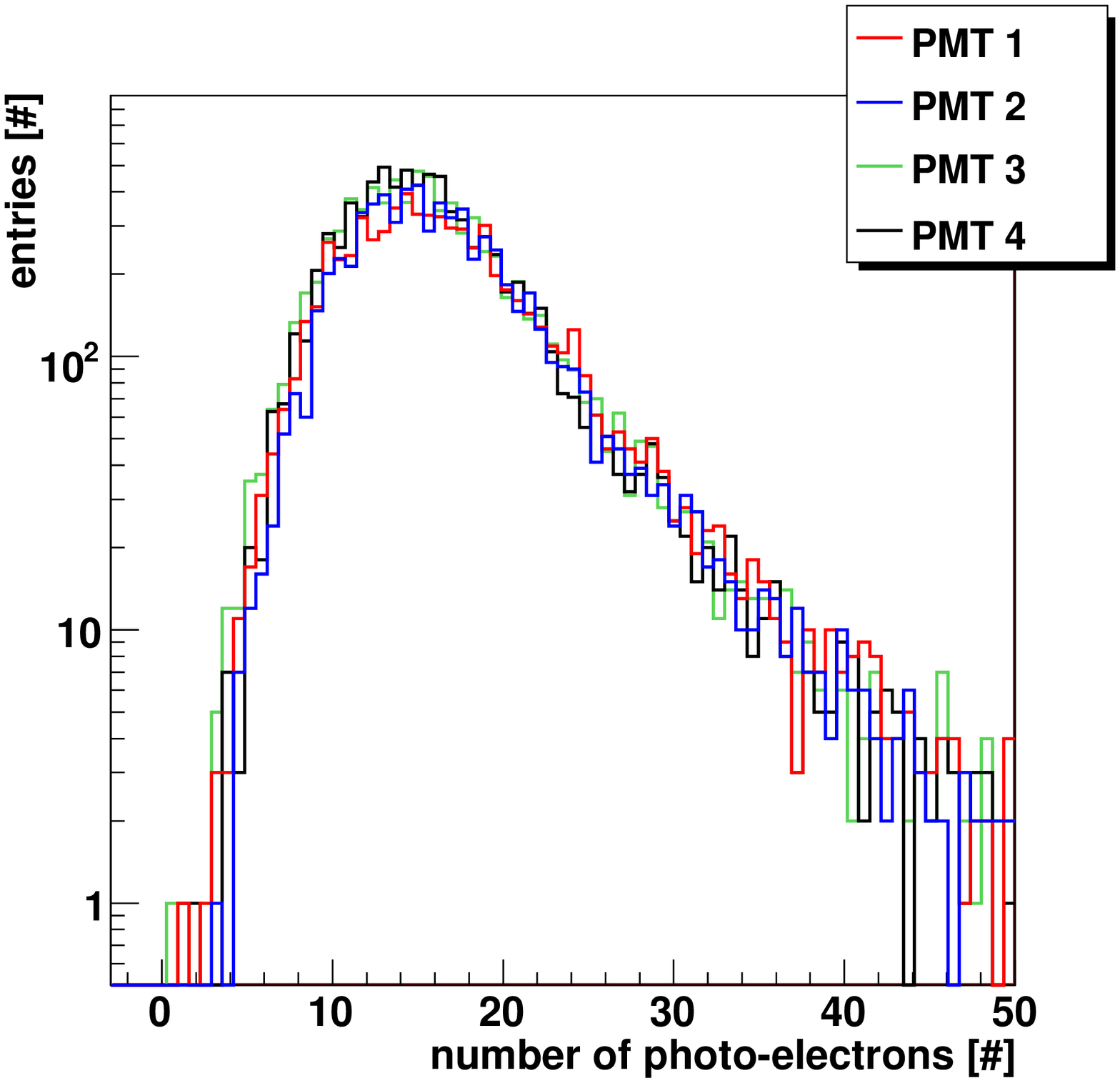,width=8cm}}\captionof{figure}{\label{f-pe_central}Number of photo-electrons for the positions 92.9\,mm away from the center of the panel.}
\end{minipage}
\hspace{.1\linewidth}
\end{center}
\end{figure}

For the panel crossed by the testbeam particle, Fig.~\ref{f-adc_highest_testbeam_pair0} shows the variation of the spectrum of the highest ADC value as the hit moves from the slot to the central region. The highest ADC value is defined as the highest ADC count value from the PMTs of the struck ACC panel. As expected, a decrease of the most probable ADC value is seen in going from the center of the panel ($y_p=0$\,mm) to the slot region ($y_p=107.2$\,mm). The drop by about 50\,\% in pulseheight at the slot is explained by less scintillator material and a smaller density than in the central region of WLS fibers relative to scintillator material. This drop is very sharp and sets in at a distance of about 2.6 - 6.6\,mm from the geometrical center of the slot.

\begin{figure}
\begin{center}
\begin{minipage}[b]{.4\linewidth}
\centerline{\epsfig{file=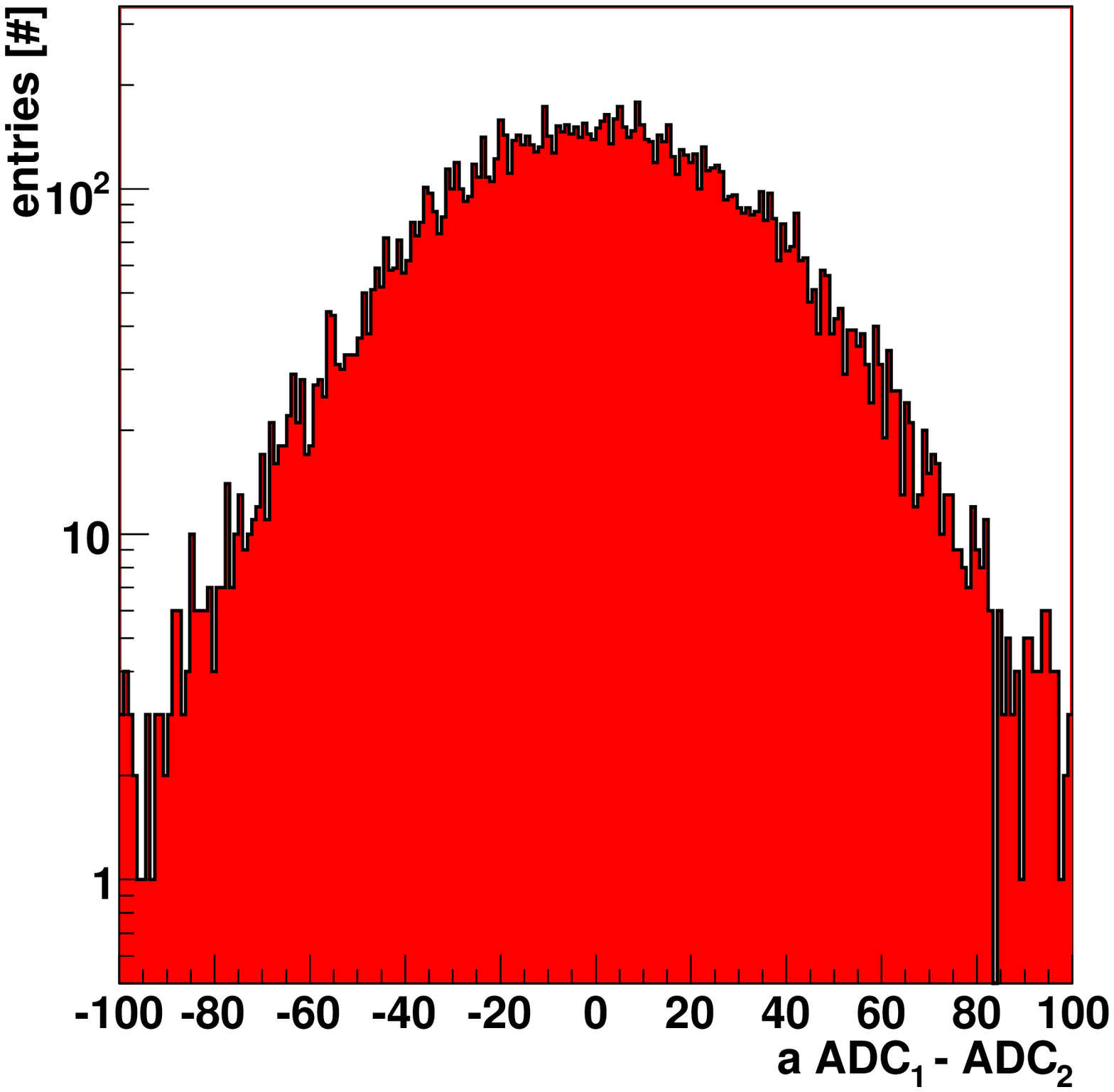,width=8cm}}\captionof{figure}{\label{f-ADCsubstract_Pair1_beam_2acc_ODER_-151_-135_-127_-111}Gain correction and width determination from the difference in signal of PMT pair A.}
\end{minipage}
\hspace{.1\linewidth}
\begin{minipage}[b]{.4\linewidth}
\centerline{\epsfig{file=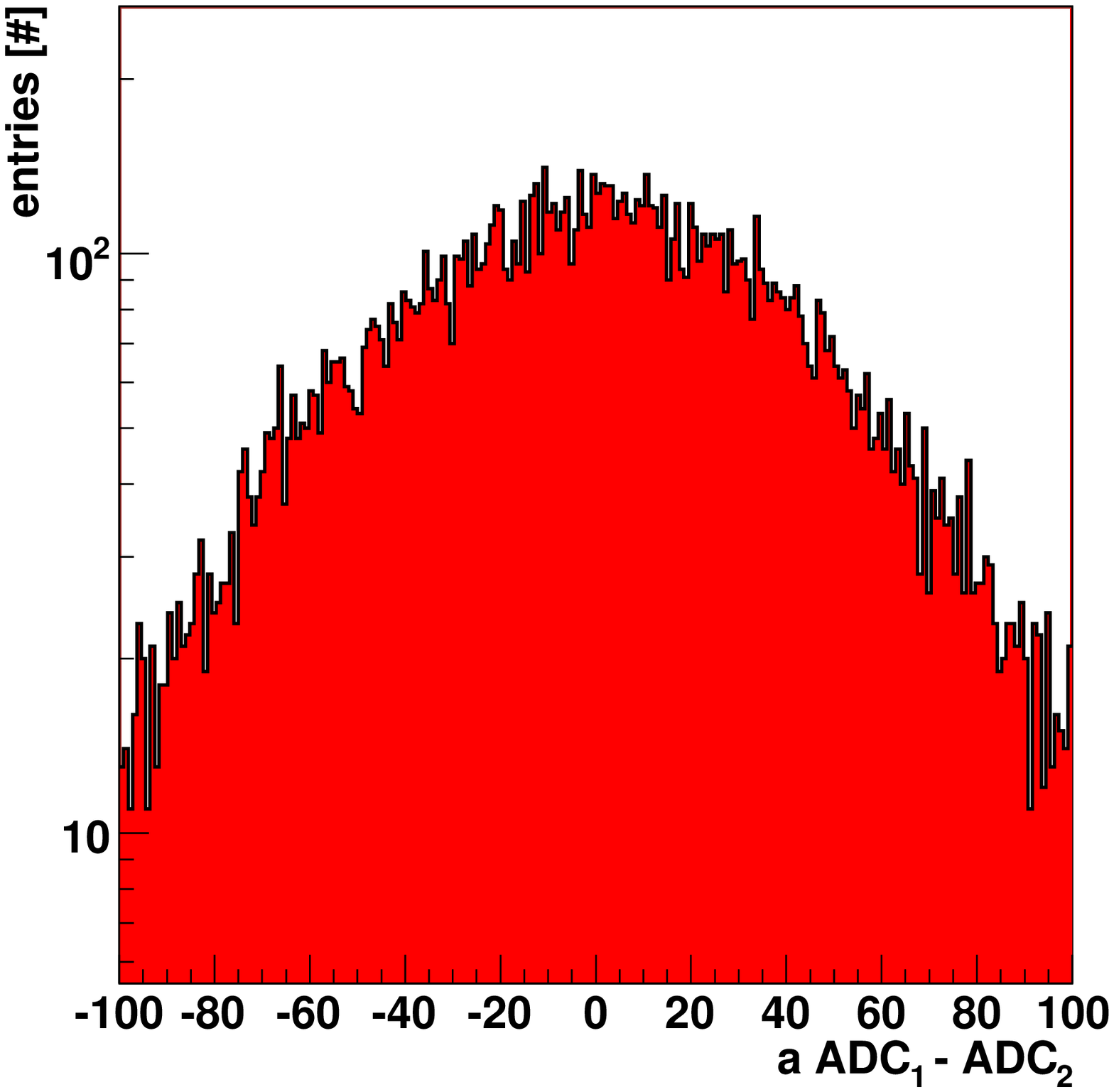,width=8cm}}\captionof{figure}{\label{f-ADCsubstract_Pair0_beam_2acc_ODER_-151_-135_-127_-111}Gain correction and width determination from the difference in signal of PMT pair B.}
\end{minipage}
\hspace{.1\linewidth}
\end{center}
\end{figure}

\subsubsection{Calculation of the Number of Photo-Electrons\label{ss-pecalc}}

\begin{figure}
\begin{center}
\captionof{table}{\label{t-pegauge}Parameters for photo-electron gauging.}
\begin{tabular}{c||c|c}
\hline
\hline
Pair &	$a$		&	G [ADC counts]\\
\hline
A	& 0.90 		&	5.56\\
B	& 1.23 		&	8.27\\
\hline
\end{tabular}
\end{center}
\end{figure}

To compare the testbeam measurements with the qualification tests of all panels and photomultipliers the number of photo-electrons has to be calculated and will be explained below. The distribution for all four photomultipliers at position $y\sub{p}=92.9$\,mm (panel coordinates) is shown in Fig.~\ref{f-pe_central}. On average the most probable number of photo-electrons is about 15. This is well compatible with the tests after the panel production (Sec.~\ref{ss-paneltest}).

The LED method explained before could not be used because there were no LEDs installed in the panels at the time of the testbeam measurements. However, the number of photo-electrons for each PMT can be calculated from the total signal $N_A$ at the anode from both PMTs connected to one panel and the assumption that the gains of both PMTs differ only by a factor $a$ \cite{bruch-2006}. Each PMT will collect on average half of the photo-electrons $N\sub{pe}$ at the central position:
\be
N_A=aN_{A,1}+ N_{A,2}=\frac12 N\sub{pe}\cdot \underbrace{aG}_{\displaystyle G_1} + \frac12 N\sub{pe}\cdot \underbrace{G}_{\displaystyle G_2}=\frac{\displaystyle 1+a}{\displaystyle2}N\sub{pe}G\label{e-npe}
\ee
with the signals $N_{A,i}$ and gains $G_i$ for the individual PMTs. It is assumed that the errors of the gain $G$ and the factor $a$ can be neglected. From error propagation with $\sigma_{N\sub{pe}}^2=N\sub{pe}$ one obtains:
\be
\sigma_{N_A}^2=\left(\frac{\displaystyle1+a}{\displaystyle2}\right)^2G^2N\sub{pe}\;\Longrightarrow\;G=\frac{\displaystyle 2}{\displaystyle 1+a}\frac{\displaystyle\sigma_{N_A}}{\displaystyle\sqrt{N\sub{pe}}}.
\ee
Putting this result into equation \ref{e-npe} leads to:
\begin{eqnarray}
N\sub{pe}=\left(\frac{\displaystyle N_A}{\displaystyle\sigma_{N_A}}\right)^2 &=& \left(\frac{\displaystyle aN_{A,1}+ N_{A,2}}{\displaystyle\sigma_{N_A}}\right)^2\\
&=&(aN_{A,1}+ N_{A,2})\cdot \frac{\displaystyle aN_{A,1}+ N_{A,2}}{\displaystyle \sigma_{N_A}^2} = (aN_{A,1}+ N_{A,2})\cdot\frac{\displaystyle 1}{\displaystyle G}.
\end{eqnarray}
The factor $a$ is chosen so that the distribution of $(aN_{A,1}-N_{A,2})$ is centered at zero (Fig.~\ref{f-ADCsubstract_Pair1_beam_2acc_ODER_-151_-135_-127_-111} and \ref{f-ADCsubstract_Pair0_beam_2acc_ODER_-151_-135_-127_-111}). The error $\sigma_{N_A}$ is the standard deviation of the distribution $(aN_{A,1}+N_{A,2})$ which is the same for the distribution  $(aN_{A,1}-N_{A,2})$. The gain is assumed to be the same over the complete signal range and is defined at the mean values $\bar N_{A,i}$:
\be G=\frac{\displaystyle \sigma_{N_A}^2}{\displaystyle a\bar N_{A,1}+ \bar N_{A,2}}.\ee
The ADC spectra of both PMTs can now be transformed to the number of photo-electrons with:
\be N\sub{pe,1}=\frac{\displaystyle aN_{A,1}}{\displaystyle G}\quad\wedge\quad N\sub{pe,2}=\frac{\displaystyle N_{A,2}}{\displaystyle G}.\ee
Averaging the results over all runs gives values of $a$ and $G$ needed to calculate the number of photo-electrons (Tab.~\ref{t-pegauge}).

\subsection{Tests with the Flight Detection Chain\label{ss-perfdet}}

\begin{figure}
\begin{center}
\centerline{\epsfig{file=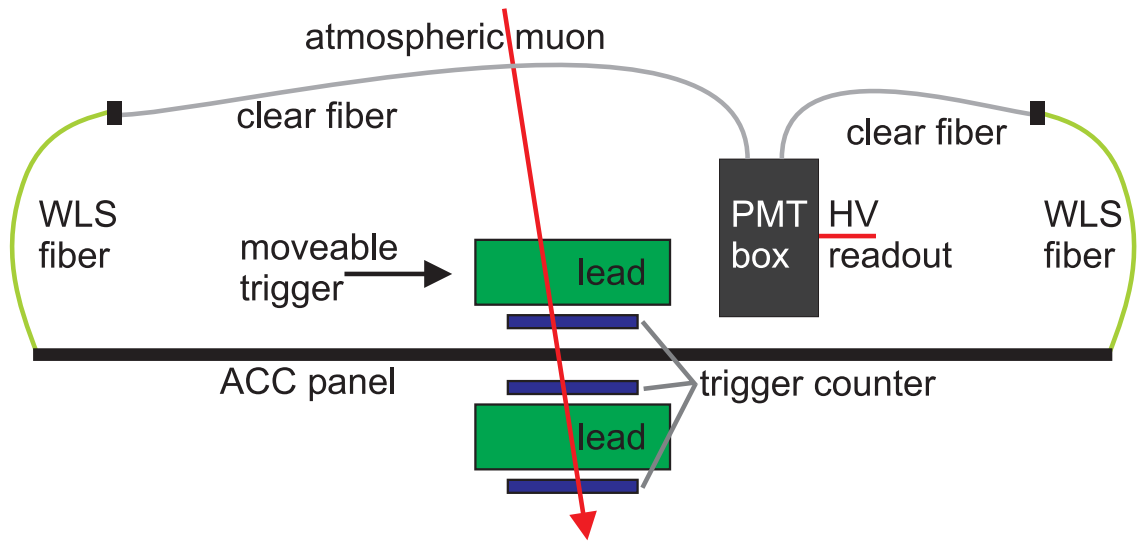,width=15cm}}\captionof{figure}{\label{f-electronic_ineff}Setup for ACC performance tests.}
\end{center}
\end{figure}

The testbeam runs described above were done during the design phase of the anticoincidence counter. After finishing all detector parts further measurements in the final design configuration were carried out (Fig.~\ref{f-electronic_ineff}). The goal of the measurements described in the following was to study the performance of the flight detection chain of the ACC system using flight spare parts. In particular the performance of the photomultipliers as a function of the applied voltage and the signal as a function of position along the panel and incidence angle on the panel was studied.

The setup consisted of flight spare parts (ACC panel: 11; PMTs: 2, 3; clear fiber cables: 19, 21). The studies described in this section were done using conventional CAMAC and NIM electronics. As for the flight configuration, the ACC panel was connected to the PMTs using clear fiber cables. The trigger for the measurements was generated by three plastic scintillator counters with photomultiplier readout, one placed above and two below the ACC panel. The trigger counters covered about 100\,cm$^2$ of the ACC panel and were moveable with respect to it. In addition, 5\,cm of lead absorber were placed above the upper trigger counter and also between the two lower trigger counters to filter low energy muons and to avoid events with secondary interactions. The CSDA range \footnote{The CSDA range is a very close approximation to the average pathlength travelled by a charged particle in material, calculated in the continuous-slowing-down approximation .} of 140\,MeV muons in lead is about 8.3\,cm and thus the kinetic energy cut-off of the setup is about 150\,MeV\cite{seltzer-1989}. Events suppressed by the lead could have charge exchange interactions generating neutral particles in the first trigger counter. Thus, there is no hit in the ACC panel. Behind the ACC panel the particle decays or converts again to low-energetic charged particles and produces signals in the lower trigger counters. The lead  between the lower trigger counters absorbs these low energy particles and prevents triggering. In addition to the coincidence of the three trigger counters an external random trigger is generated at 1\,Hz by a gate generator to measure pedestal values. The test results with the same setup but the flight electronics are described in Sec.~\ref{ss-flightelec}.

\begin{figure}
\begin{center}
\begin{minipage}[b]{.4\linewidth}
\centerline{\epsfig{file=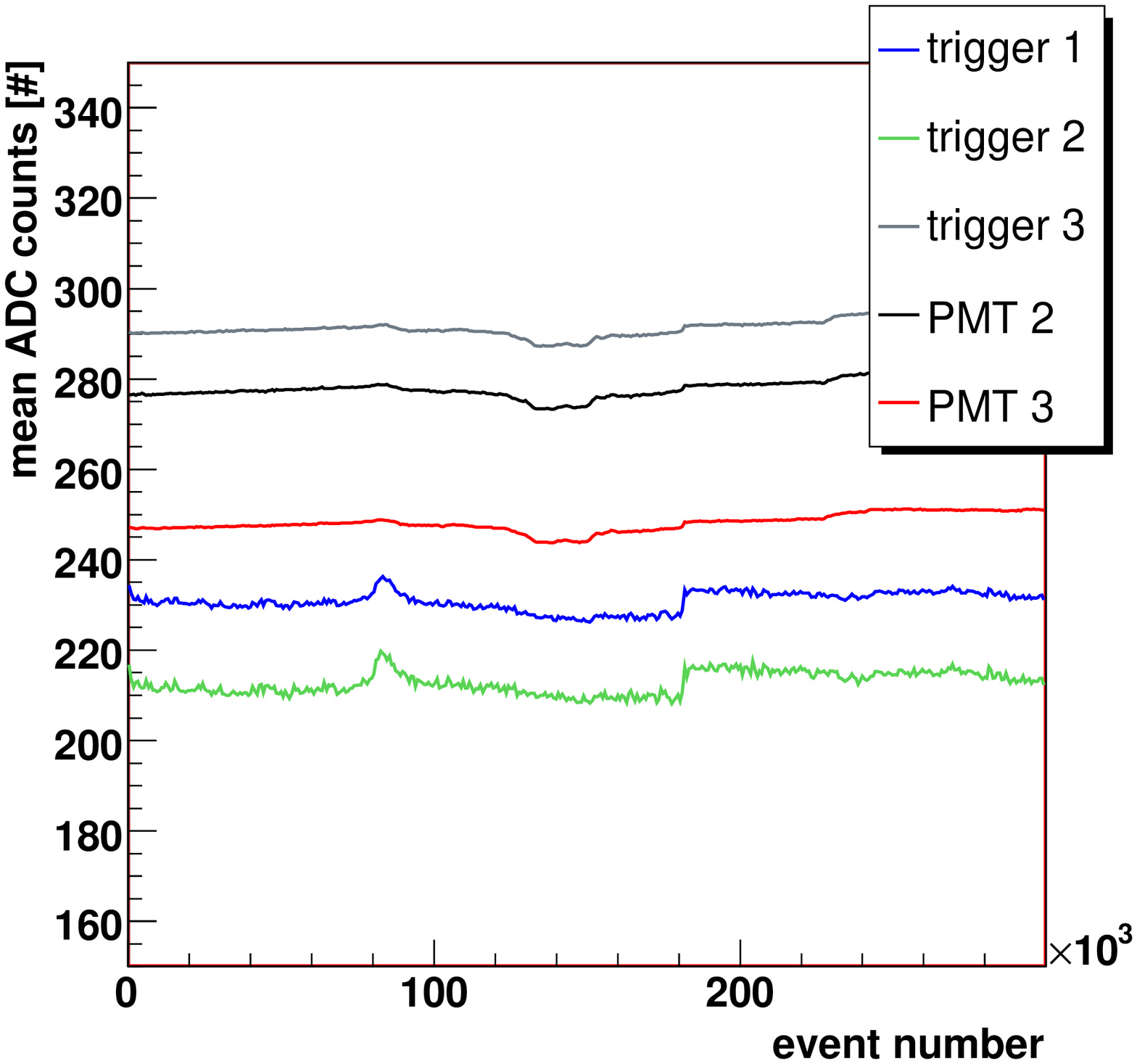,width=8cm}}\captionof{figure}{\label{f-eff_spe6_7_8m.dat_sample_ped}Pedestal measurement during run.}
\end{minipage}
\hspace{.1\linewidth}
\begin{minipage}[b]{.4\linewidth}
\centerline{\epsfig{file=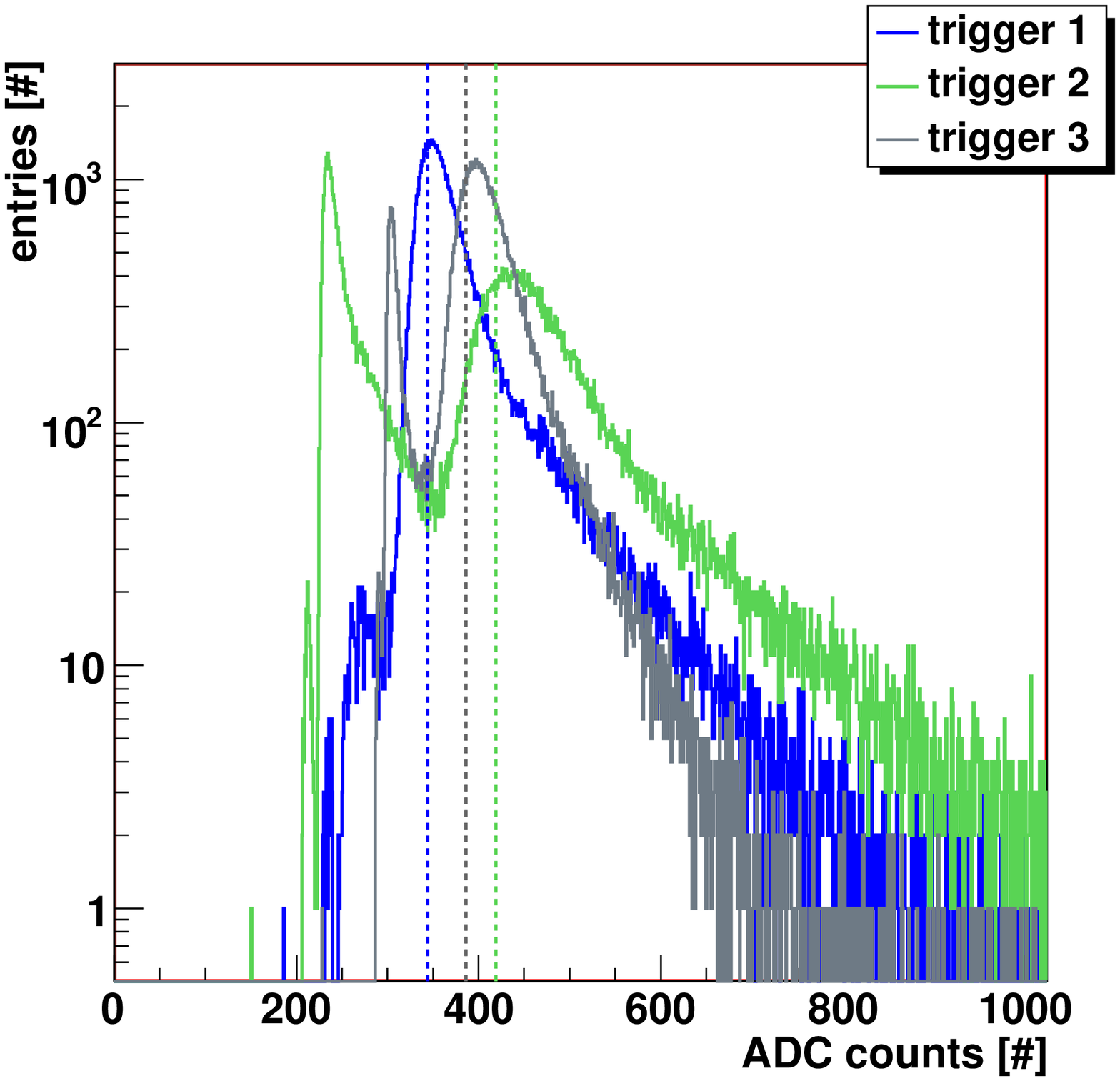,width=8cm}}\captionof{figure}{\label{f-eff_spe6_7_8m.dat_trig_spe}Spectra of trigger counters with cuts (dashed lines) used for the definition of a clean track.}
\end{minipage}
\end{center}
\end{figure}

Fig.~\ref{f-eff_spe6_7_8m.dat_sample_ped} shows the behavior of the pedestal position during the measurement and Fig.~\ref{f-eff_spe6_7_8m.dat_trig_spe} depicts the raw spectra of the trigger counters. The pedestal shift made it necessary to collect pedestal data during the measurements. The ADC spectra of all counters are corrected for the current pedestal position. In the following, an event with a clean track is defined by hits in all three trigger counters with pulseheights exceeding the corresponding MOP value.

\subsubsection{Inefficiency and Charge Resolution}

\begin{figure}
\begin{center}
\begin{minipage}[b]{.4\linewidth}
\centerline{\epsfig{file=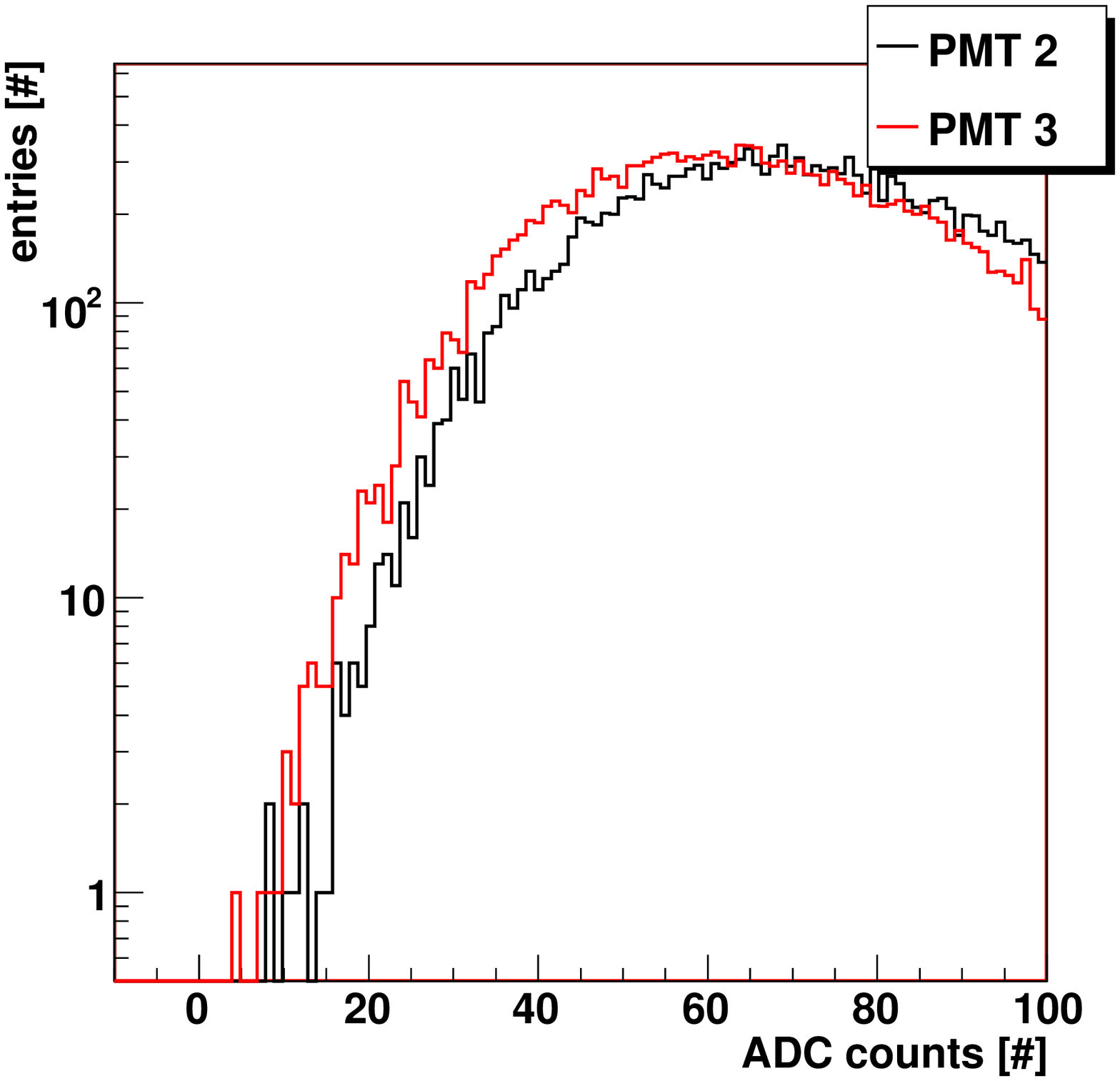,width=8cm}}\captionof{figure}{\label{f-eff_spe6_7_8m.dat_acc_subtract_spe}Pedestal corrected ADC spectra of the ACC PMTs for the central panel region.}
\end{minipage}
\hspace{.1\linewidth}
\begin{minipage}[b]{.4\linewidth}
\centerline{\epsfig{file=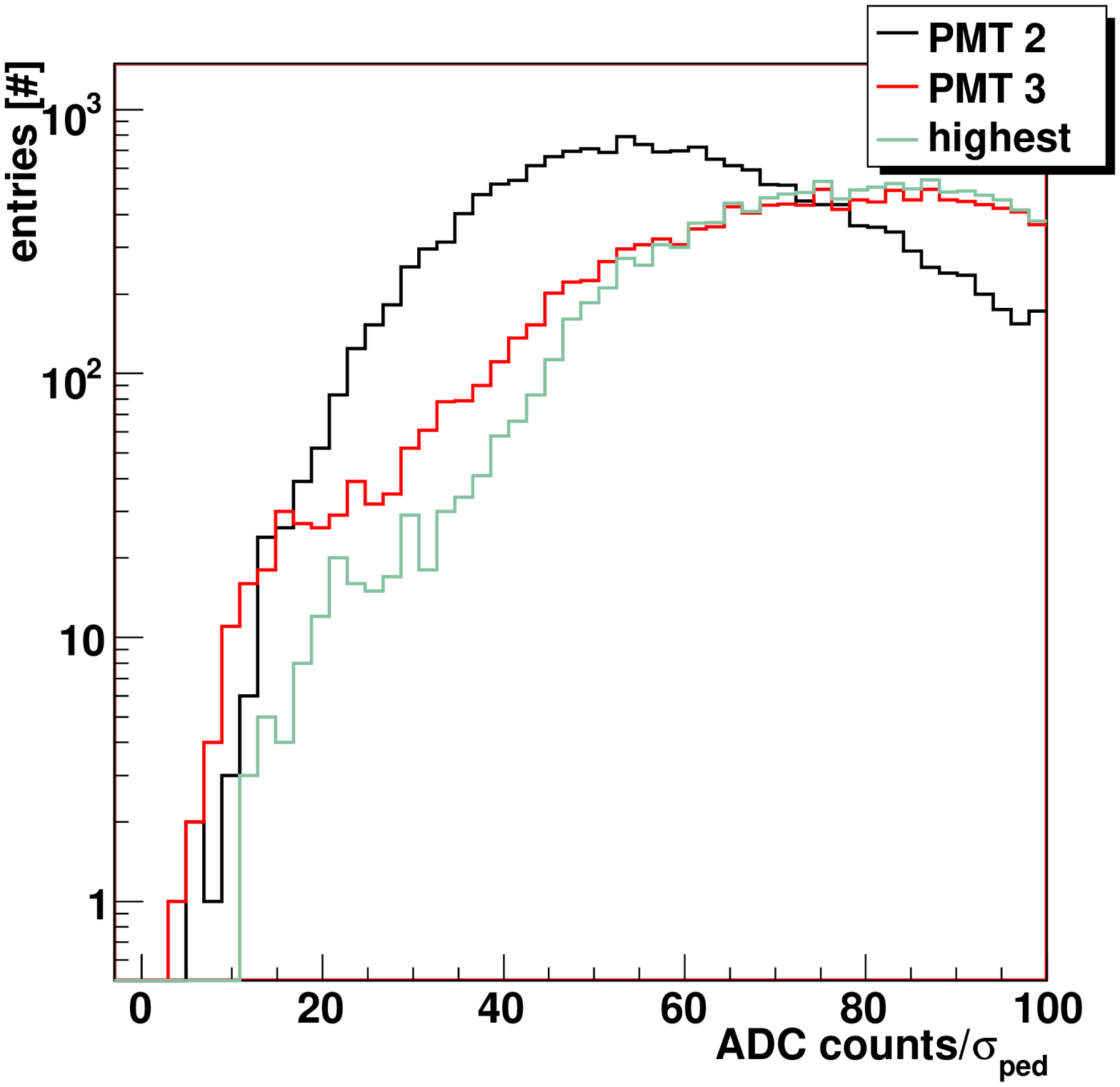,width=8cm}}\captionof{figure}{\label{f-eff_spe6_7_8m.dat_acc_sigma_spe}ADC spectra of the ACC PMTs normalized to the pedestal width. The spectrum labeled 'highest' depicts the highest of the two PMT values for each event.}
\end{minipage}
\end{center}
\end{figure}

The moveable trigger arrangement was first used to study the central region of the ACC panel. Fig.~\ref{f-eff_spe6_7_8m.dat_acc_subtract_spe} shows the pedestal subtracted spectra for both PMTs. The voltages were adjusted such that the MOP in ADC counts was nearly the same for both photomultipliers (PMT\,2: 2016\,V, PMT\,3: 2200\,V). A high charge resolution is especially important for the measurement of small charges. This resolution can be better understood by normalizing the ADC values to the pedestal width $\sigma\sub{ped}$. PMT\,3 shows a higher resolution than PMT\,2 (Fig.~\ref{f-eff_spe6_7_8m.dat_acc_sigma_spe}). Looking at the highest normalized ADC value of both PMTs, the distribution is dominated by PMT 3, having no entries below $11\sigma\sub{ped}$ for 20000 events. Thus in the central panel region the ACC detector design is seen to perform well, at least in a combination with conventional readout electronics (NIM, CAMAC).

\begin{figure}
\begin{center}
\begin{minipage}[b]{.4\linewidth}
\centerline{\epsfig{file=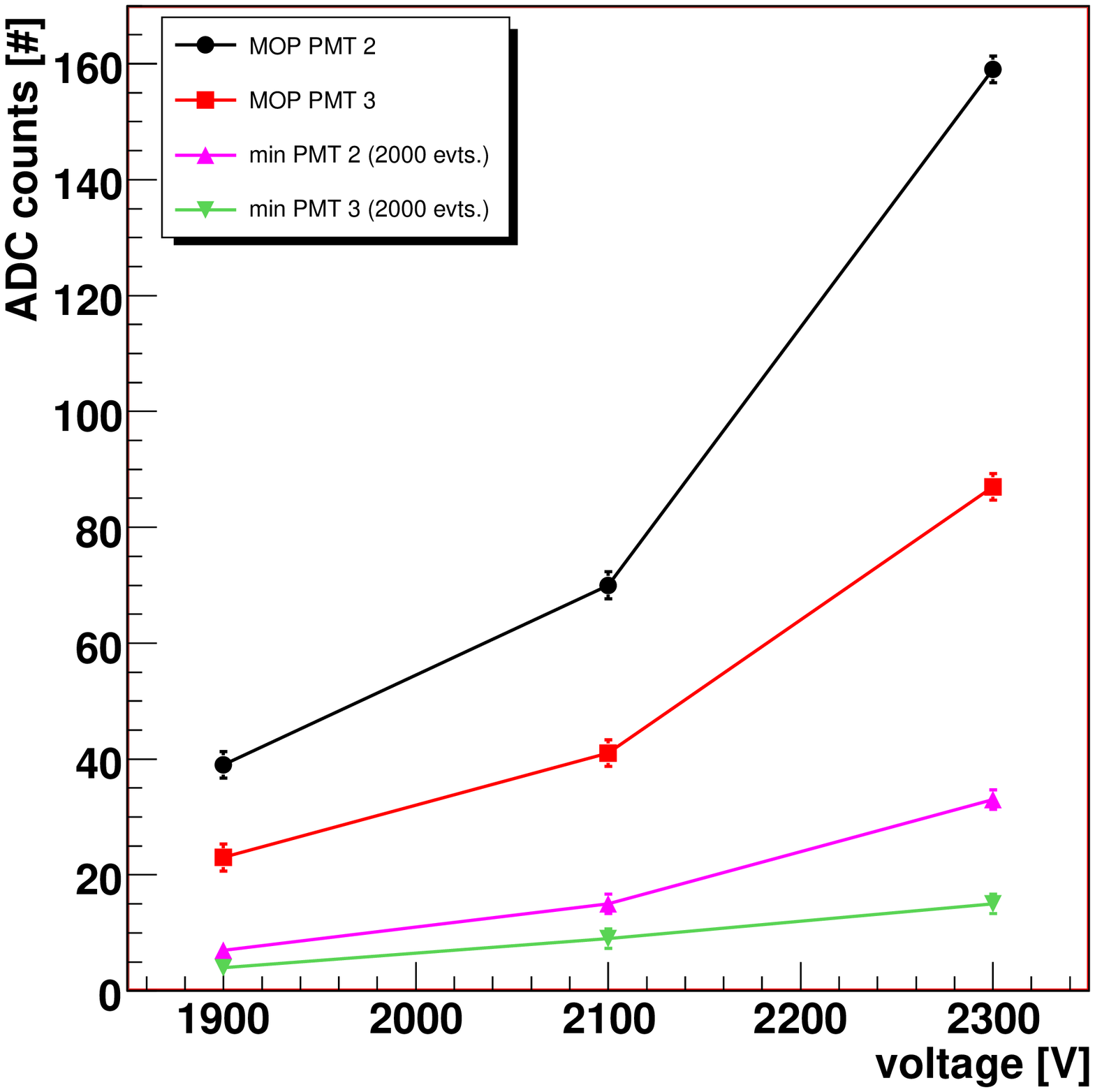,width=8cm}}
\captionof{figure}{\label{f-inefficiency_several_adc}MOP and minimum pedestal corrected ADC values at different voltages.}
\end{minipage}
\hspace{.1\linewidth}
\begin{minipage}[b]{.4\linewidth}
\centerline{\epsfig{file=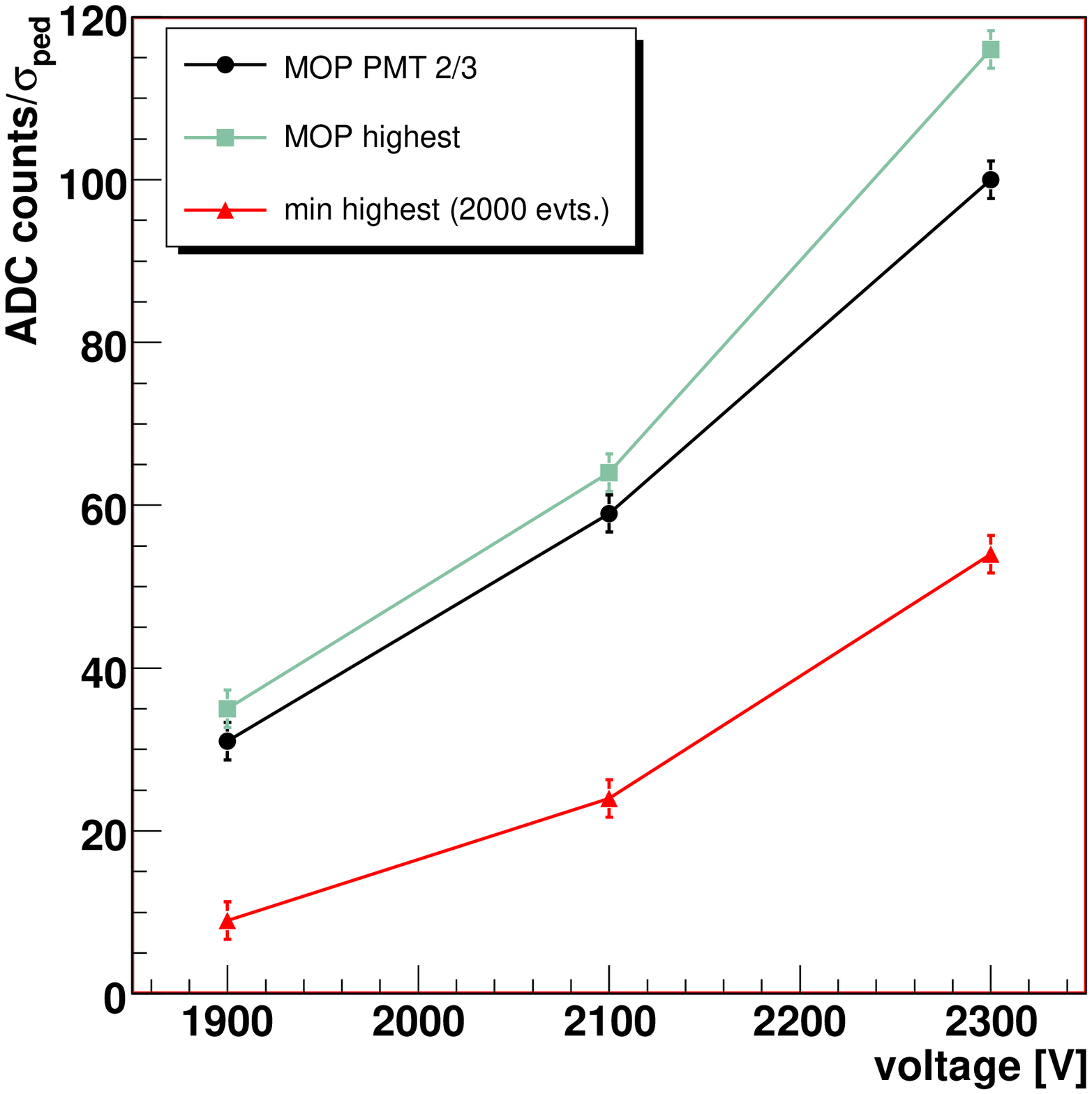,width=8cm}}
\captionof{figure}{\label{f-inefficiency_several_res}MOP and minimum normalized ADC values at different voltages. The normalized MOP values for PMT 2 and 3 are very close so that the mean value is shown. The curve labeled 'highest' depicts the highest of the two PMT values for each event.}
\end{minipage}
\end{center}
\end{figure}

As noted above, for minimal inefficiency of the ACC system it is important to operate with as high as possible charge resolution. Therefore, the behavior of the system at different voltages has been studied. The most probable values of the ADC spectra and the corresponding pedestal widths depend on the voltages applied to the photomultipliers (Fig.~\ref{f-inefficiency_several_adc}). The gains of the two PMTs used in this setup differ by a factor of about two at the same voltages. The minimum ADC value in 2000 events for each voltage increases slower than the MOP values. A large spread between the MOP and the minimum value implies that an increase in voltage improves the resolution for small energy depositions. The normalized ADC values $n\sub{ADC} = (\text{ADC}-p)/\sigma_p$ show the same behavior for the MOP and the minimum values of both PMTs for the allowed voltage range 1900\,V - 2300\,V (Fig.~\ref{f-inefficiency_several_res}). The highest of the two normalized MOP values is always about 15\,\% above the value for the distribution of PMT\,2 or 3. Again, the difference between the normalized MOP and the normalized minimum value increases when applying a higher voltage. The conclusion is that voltages as high as possible should be used for the best charge resolution. As measured before (Fig.~\ref{f-la_ic_26_28}), a voltage increase enhances not only the integrated charge but also the pulseheight which must exceed the discriminator threshold of the flight electronics. In the flight electronics it is not possible to set individual thresholds for each PMT so they should all be operated at the same gain to apply the same thresholds during flight. The gain must be adjusted such that no PMT exceeds the maximum allowed voltage of 2300\,V (Tab.~\ref{t-voltages}).

\subsubsection{Signal Behavior as a Function of Position and Incidence Angle}

\begin{figure}
\begin{center}
\begin{minipage}[b]{.4\linewidth}
\centerline{\epsfig{file=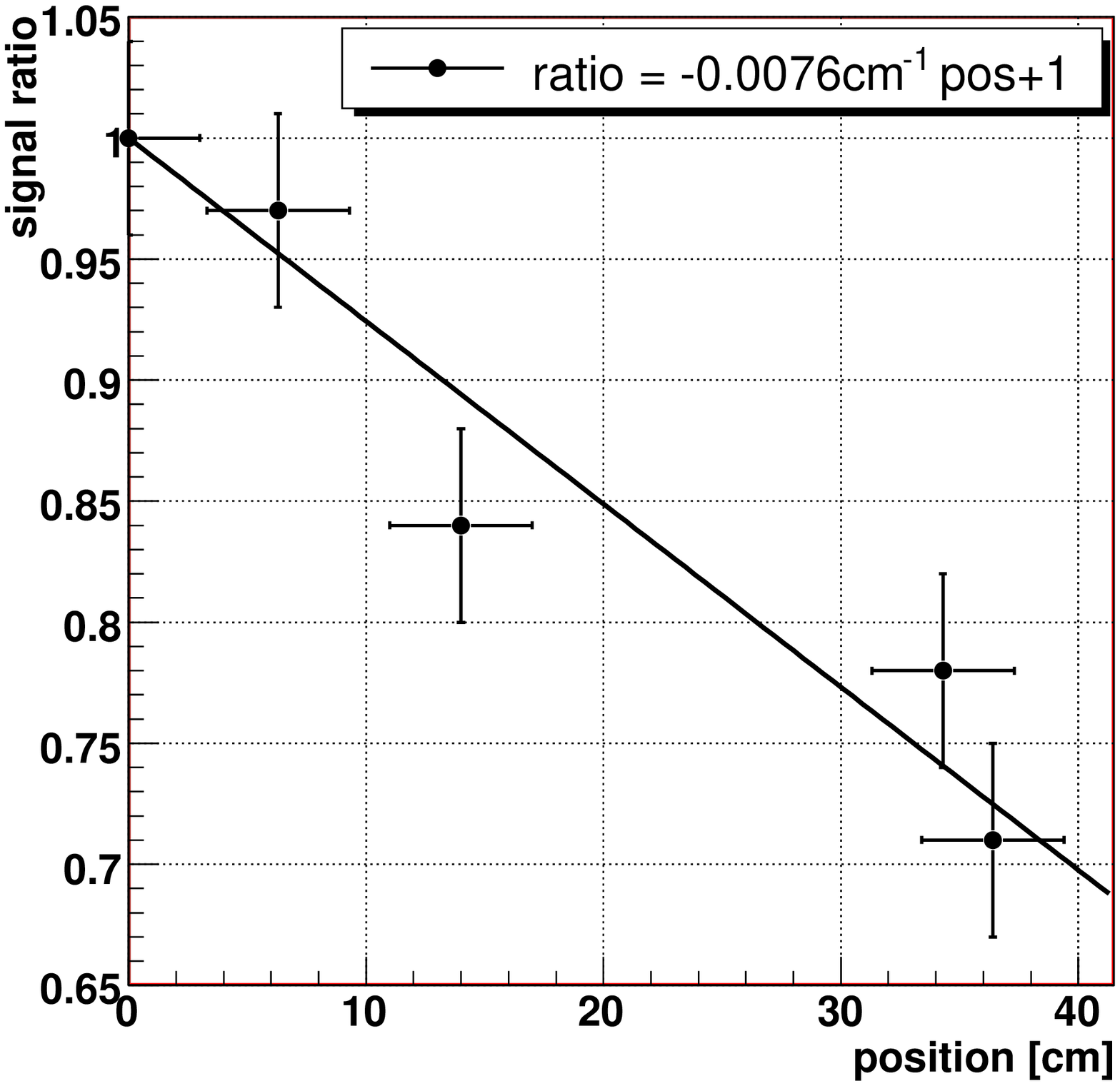,width=8cm}}
\captionof{figure}{\label{f-inefficiency_several_ratio}Ratio of the two PMT MOP values along a panel normalized to the center of the panel at 0\,cm.}
\end{minipage}
\hspace{.1\linewidth}
\begin{minipage}[b]{.4\linewidth}
\centerline{\epsfig{file=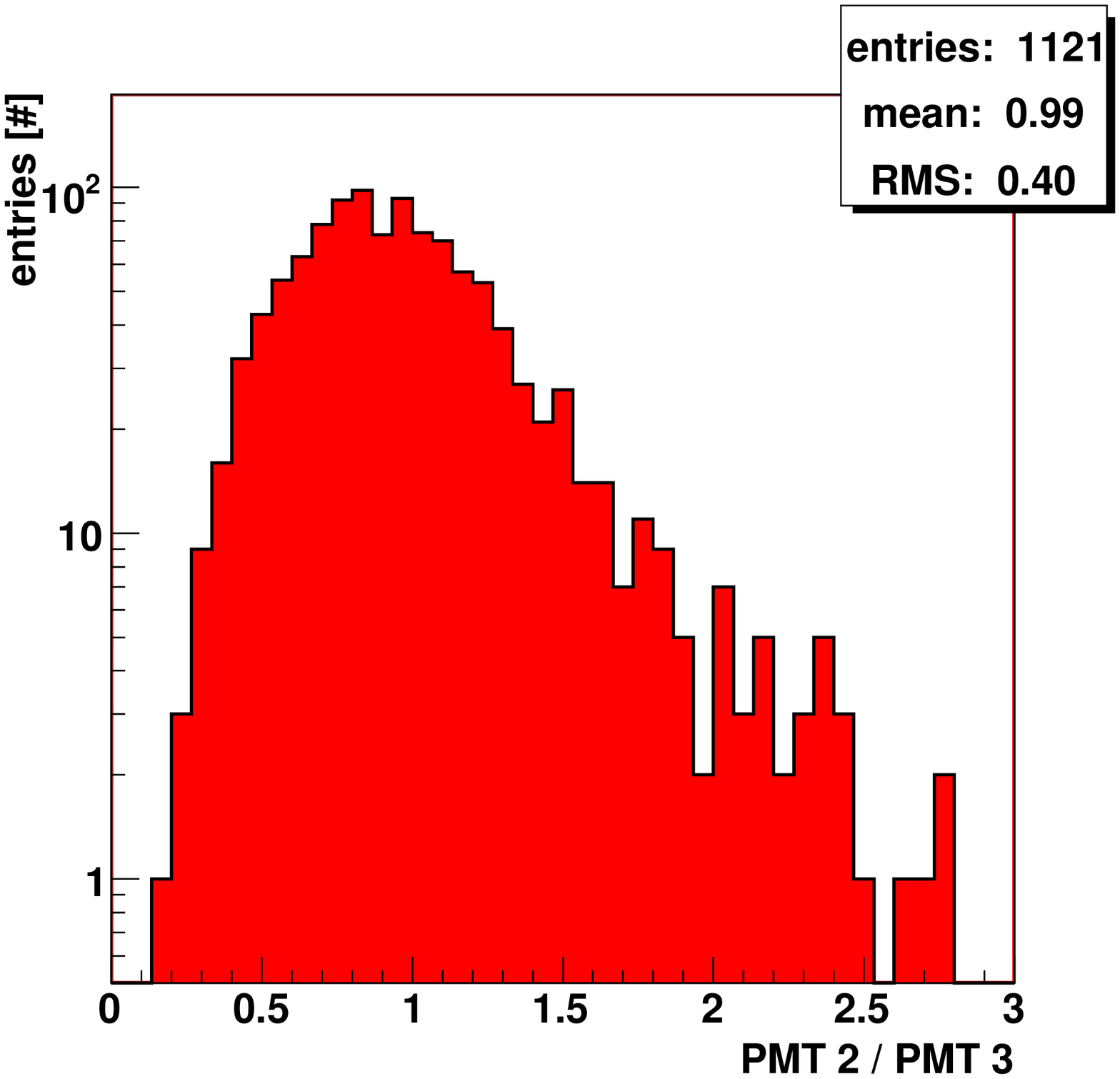,width=8cm}}
\captionof{figure}{\label{f-inefficiency_several_acc0_acc1_ratio}Pulseheight ratio of both PMTs at the central position.}
\end{minipage}
\end{center}
\end{figure}

\begin{figure}
\begin{center}
\begin{minipage}[b]{.4\linewidth}
\centerline{\epsfig{file=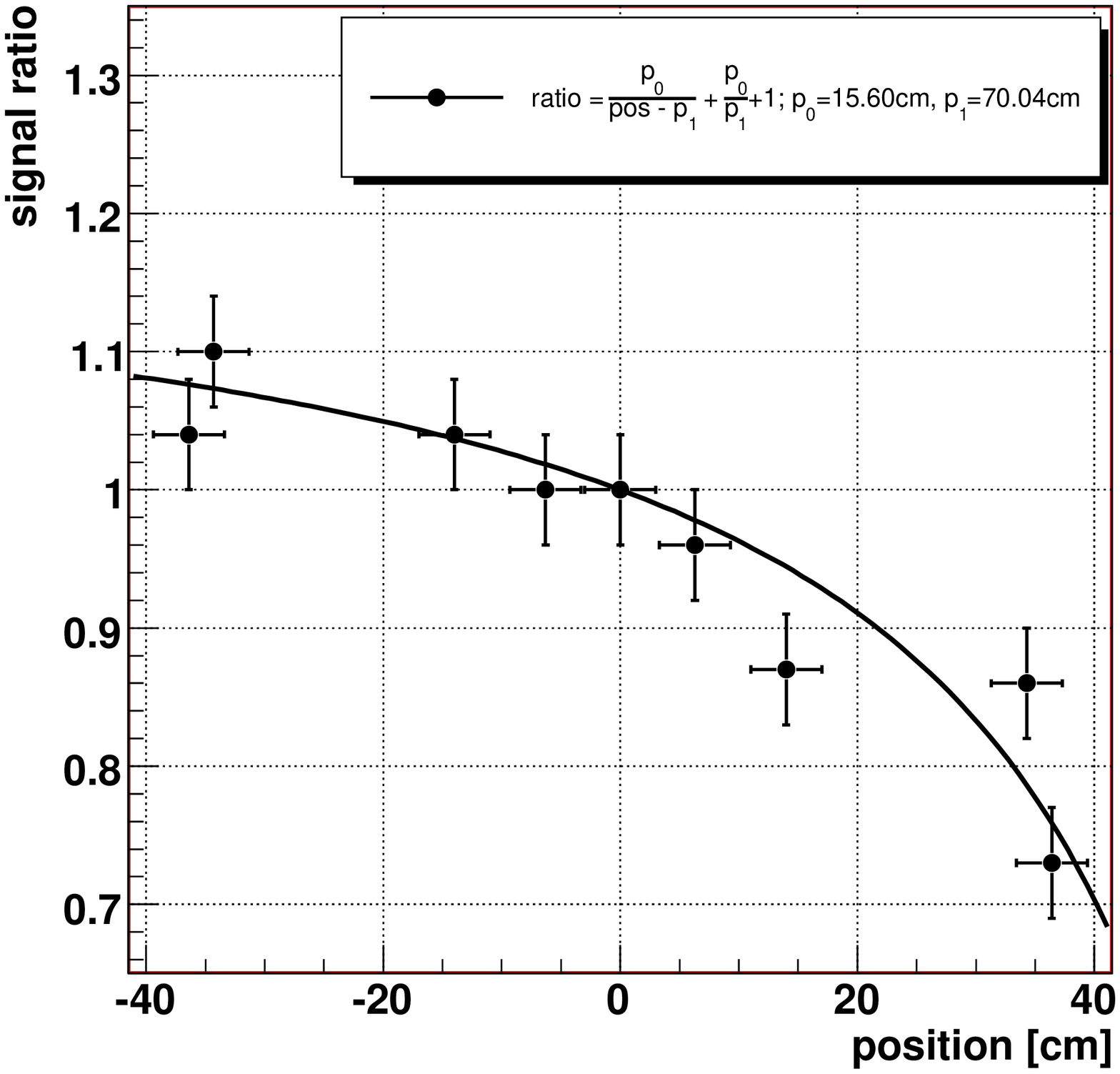,width=8cm}}
\captionof{figure}{\label{f-inefficiency_several_ratio_single}Behavior of the PMT MOP value along the panel normalized to the center of the panel at 0\,cm.}
\end{minipage}
\hspace{.1\linewidth}
\begin{minipage}[b]{.4\linewidth}
\centerline{\epsfig{file=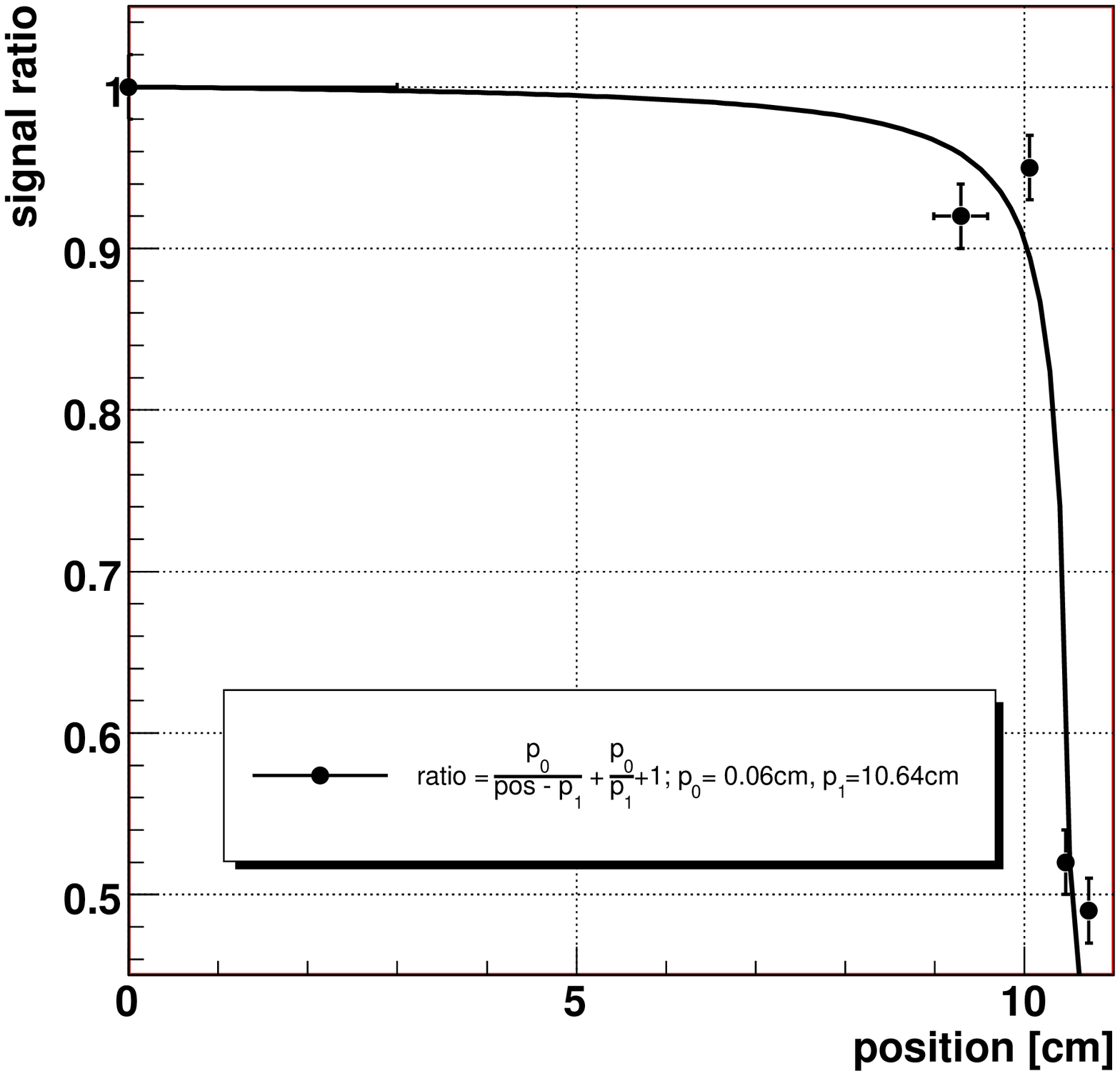,width=8cm}}
\captionof{figure}{\label{f-inefficiency_several_ratio_width}Behavior of one PMT MOP value across the panel from the testbeam measurements normalized to the center of the panel at 0\,cm. The slot position between two panels shows a signal drop by about 50\,\%.}
\end{minipage}
\end{center}
\end{figure}

The next measurements are used to study the ratio of the two PMT signals of one panel as a function of incident position and angle to develop a model for the variation of the signal. Such a model is important for the determination of the inefficiency of the complete ACC system (Sec.~\ref{ss-ineffdet}) and a realistic implementation of the ACC in the official AMS-02 Monte Carlo simulation code. The ratio $R$ of MOP values of the two PMTs were obtained by moving the three trigger counters along the panel while keeping all other conditions as for the measurement at the central position. The ratios for one half of the ACC panel are shown in Fig.~\ref{f-inefficiency_several_ratio} as a function of the position along the panel. They are normalized to 1 at the central position (0\,cm). The data can be used to extract the position $P$ of the hit along the panel. The behavior can be reasonably modeled by a straight line. A fit gives:
\be R = (-0.0076\pm0.0009)\,\text{cm}^{-1}\cdot P+1.\label{e-pmtratio}\ee
To calculate the precision of the position determination the distribution of the signal height ratios of both PMTs is needed (Fig.~\ref{f-inefficiency_several_acc0_acc1_ratio}). The RMS at the central position is about 40\,\%. This translates in a position determination accuracy. A good parametrization is:
\be P = (-131.6\pm15.6)\,\text{cm} \cdot ((1.00\pm0.40)\cdot R -1).\label{e-pmtratio_p}\ee 
Within a standard deviation the spread is larger than the panel length itself, so the signal ratio must be treated with great caution with regard to position determination. A similar study is shown in Fig.~\ref{f-inefficiency_several_ratio_single} but only for one PMT positioned at -40\,cm. The signal changes along the panel by about 30\,\% comparing the end of the panel with the center and can be parametrized by:
\be R = \frac{p_0}{P - p_1}+\frac{p_0}{p_1}+1\label{e-along}\ee
with the parameters $p_0 = (15.6\pm12.3)$\,cm and $p_1 = (70.0\pm20.2)$\,cm. The signal behaviour across the panel was studied in the testbeam (Sec.~\ref{ss-testbeam}) and is shown together with a fit of the same type (Eq.~\ref{e-along}) in Fig.~\ref{f-inefficiency_several_ratio_width}. The fitted parameters in this case are $p_0 = (0.06\pm0.02)$\,cm and $p_1 = (10.637\pm0.003)$\,cm.

\begin{figure}
\begin{center}
\begin{minipage}[b]{.4\linewidth}
\centerline{\epsfig{file=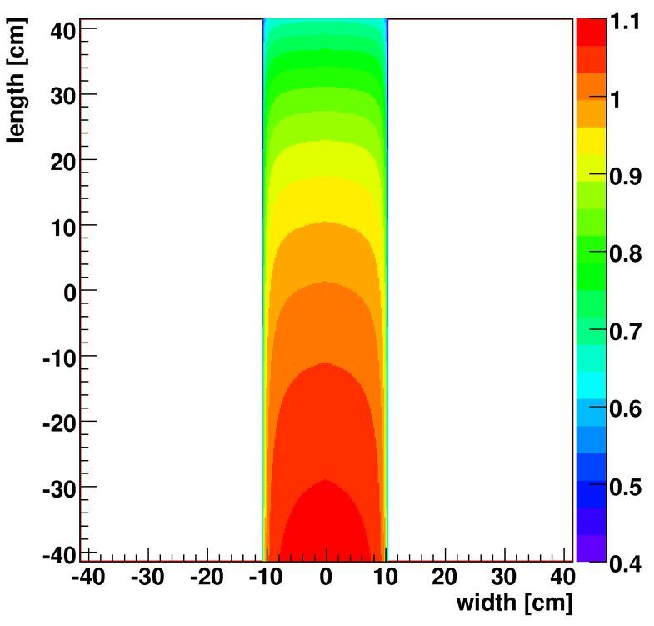,width=8cm}}
\captionof{figure}{\label{f-inefficiency_several_panel}Expected MOP value behavior for one PMT connected at the lower length position -40\,cm. The color code on the right indicates the relative signal change.}
\end{minipage}
\hspace{.1\linewidth}
\begin{minipage}[b]{.4\linewidth}
\centerline{\epsfig{file=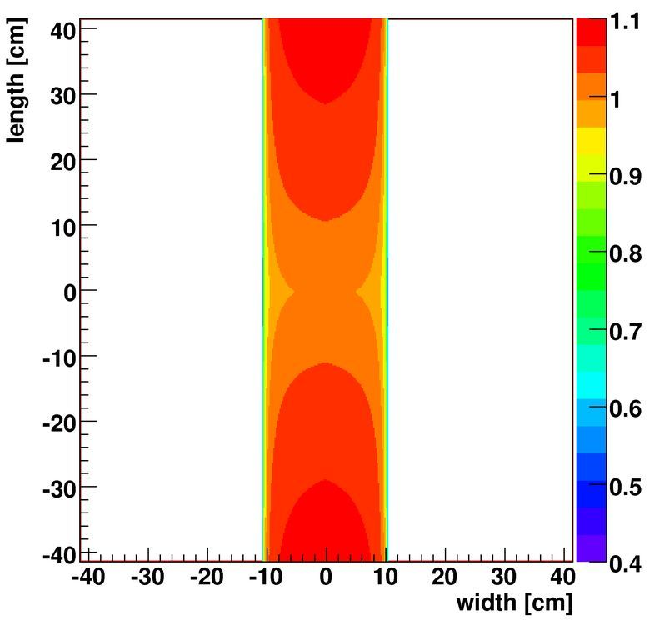,width=8cm}}
\captionof{figure}{\label{f-inefficiency_several_panel_both}Expected MOP value behavior for the highest ADC value with PMTs on both ends of the panel. The color code on the right indicates the relative signal change.}
\end{minipage}
\end{center}
\end{figure}

Putting the last two results together and using the fits to generate the expected average changes of the MOP value as a function of incident position on the ACC results in Fig.~\ref{f-inefficiency_several_panel}. The PMT under investigation is connected to the lower end of the panel and again the signal is normalized to 1 at the central position. The curvature of the panel shape is neglected in the figure. The signal drops to a minimum of about 50\,\% only in a very small region close to the slot. Taking both PMTs into account and always using the higher of the two values, the ratio drops below 1 only at the slot regions (Fig.~\ref{f-inefficiency_several_panel_both}).

\begin{figure}
\begin{center}
\begin{minipage}[b]{.4\linewidth}
\centerline{\epsfig{file=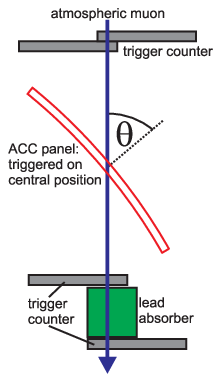,height=8cm}}
\captionof{figure}{\label{f-inefficiency_several_angle}Schematic setup for the variation of the incident angle.}
\end{minipage}
\hspace{.1\linewidth}
\begin{minipage}[b]{.4\linewidth}
\centerline{\epsfig{file=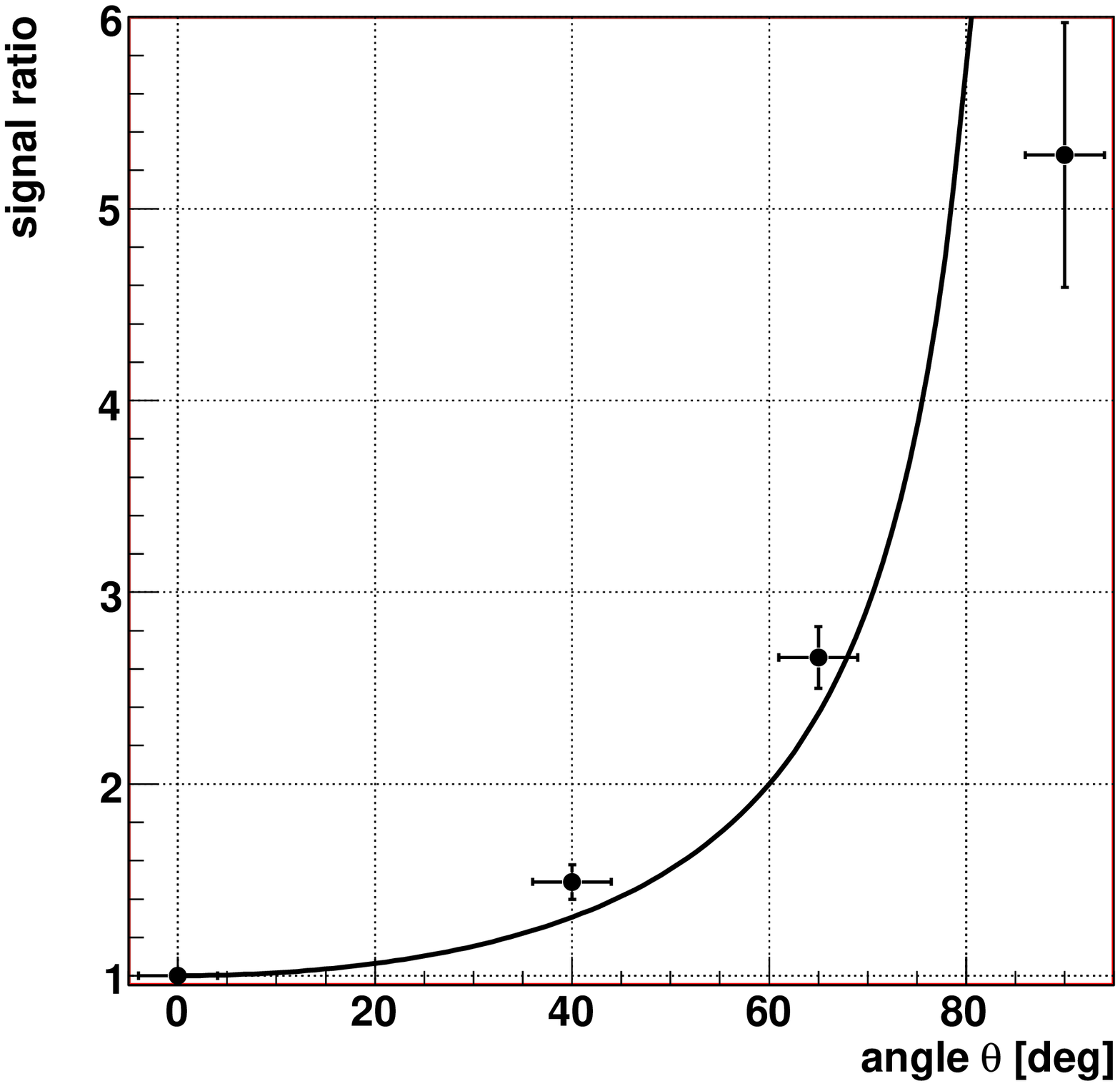,width=8cm}}
\captionof{figure}{\label{f-inefficiency_several_ratio_angle}Behavior of PMT MOP values as a function of the angle of incidence.}
\end{minipage}
\end{center}
\end{figure}

The angle of incidence of particles on the scintillator changes the pathlength travelled in the material. Fig.~\ref{f-inefficiency_several_angle} shows the test setup with the four trigger counters arranged to allow only nearly perpendicular muons to be measured while varying the angle of the ACC panel. This measurement is again carried out at the central panel position. It is expected that the relative MOP value changes linearly with the pathlength in the scintillator. As shown in Fig.~\ref{f-inefficiency_several_ratio_angle}, the signal height increases with the angle in agreement with the expected behavior $1/\cos\theta$. Measurements at large angles $\theta$ were difficult to perform in the test setup but for angles up to 65° the results are within 1 standard deviation of the theoretical expectations.

\subsection{Tests with the Flight Readout System for the ACC\label{ss-flightelec}}

This section describes the flight readout electronics of the ACC system and their calibration. Subsequently the tests with atmospheric muons performed with the setup of Sec.~\ref{ss-perfdet} are repeated with these final electronics.

\subsubsection{The ACC Readout System and Logic}

\begin{figure}
\begin{center}
\centerline{\epsfig{file=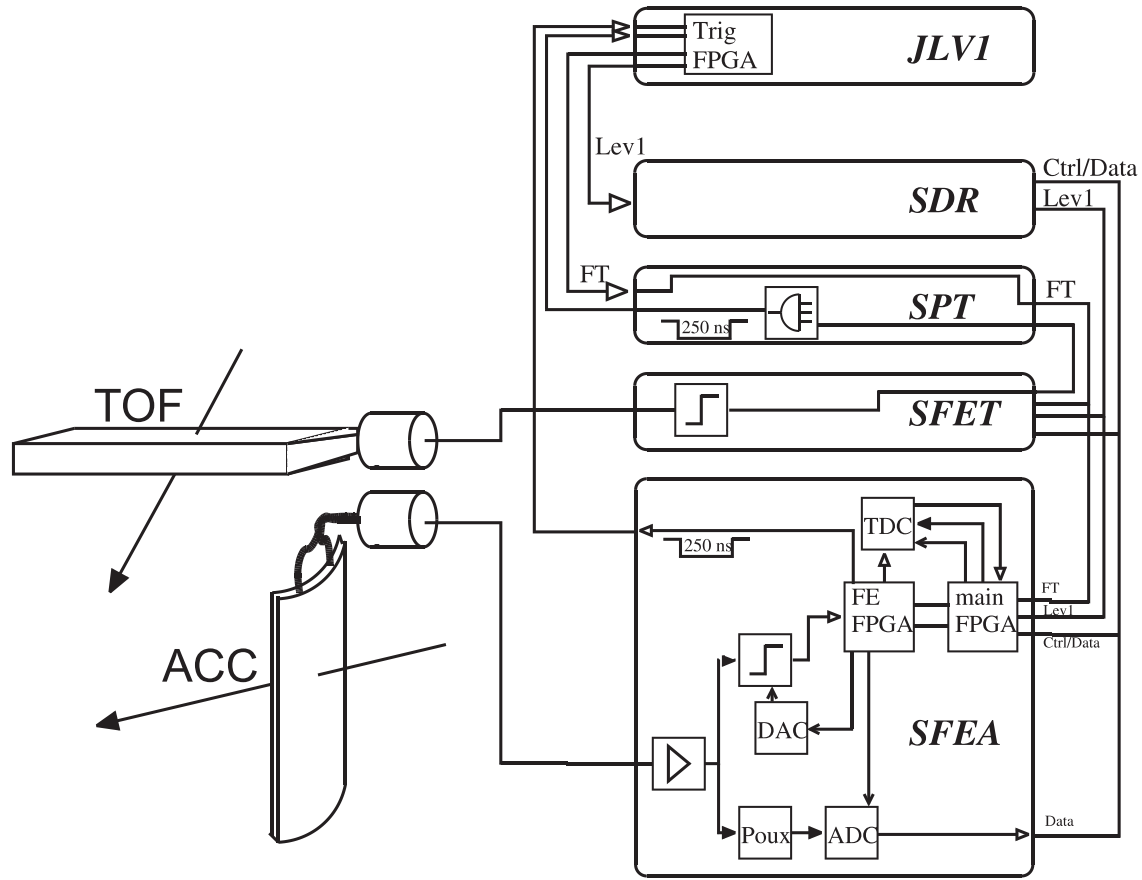,width=15cm}}\captionof{figure}{\label{f-acc_block}Signal processing for ACC and TOF \cite{schwering-2008}.}
\end{center}
\end{figure}

The readout of the different AMS-02 subdetectors and the data processing is carried out in different electronic subsystems called crates. The crates exist in three configurations. The first is the engineering module (EM) and is used for tests of the electronics layout. The second is the qualification module (QM) which is used for the space qualification  in a thermo vacuum test and on a vibration table and further performance tests. The third is the flight configuration (FM). The crates are mounted on heat removal radiators on the WAKE ($+y$) and RAM ($-y$) side of the AMS-02 detector. The ACC data acquisition and processing is done with the S-crates where also the data of the TOF are processed. The crate is structured in several units, the SFET2 (4$\times$), SFEA2 (1$\times$), SPT2 (1$\times$) and the SDR2 (1$\times$). There are four S-crates  in total mounted on top and bottom on each side. The level\,1 trigger decision for the readout of AMS-02 is formed by the JLV1 unit in the J-crate. The following describes the processing chain with the main emphasis to the ACC (Fig.~\ref{f-acc_block}):

The formation of the trigger for the AMS-02 detector is based in most cases on the signals of the TOF system. These signals of the TOF panels first enter SFET2 which includes three different discriminators. The high threshold discriminator (HT) produces a signal of 500\,ns width in case the amplitude exceeds the preset threshold. From the discriminated signals of 20 TOF panels which are processed by one S-crate the SPT2 generates two logical outputs by forming the OR for two groups of ten channels each. A group of ten channels correspond to one side of one TOF plane. The outputs of all SPT2s from the four S-crates are sent to the J-crate for the fast trigger (FT) decision in the JLV1. The FT goes back to the SPT2 to latch the input HTs and to generate the pre-trigger pattern which will be stored in the data. It also sets the reference time for all time measurements in the S-crate. The JLV1 does not only generate the FT but also the level\,1 trigger. This trigger is again formed on the basis of TOF inputs or also in combination with the ACC signals in order to veto bad events. To this purpose the ACC signals enter to the SFEA2 where they are amplified before further processing to gain higher sensitivity to small signals. The user sets discriminator thresholds in the SFEA2s for the ACC signals. These discriminated signals can be used in the JLV1 in the decision for the level\,1 trigger generation. The level\,1 trigger is sent via the SPT2s to the SFET2s and the SFEA2s and starts the readout process. The digitized data will be sent to SDR2. The next FT resets the integration process of the Pouxe chip.

Each of the SFEA2 units has an ADC for the charge measurement and a TDC for the timing measurements on two sides for reasons of redundancy. These sides are called A and B. After amplification the signal in SFEA2 is split to the discriminator and to the ADC branch. In addition to the discriminator decision, the TDC stores the time stamps for each signal transaction above the discriminator threshold. Each hit is recorded by the TDC with a precision of 55\,ps and causes a dead time of up to 30\,ns of the TDC clock. This happens in addition to the storage of the JLV1 decision in the J-crate. The TDC is programmed such that hits about 9\,\textmu s before and 7\,\textmu s after the level\,1 trigger can be stored.

\begin{figure}
\begin{center}
\centerline{\epsfig{file=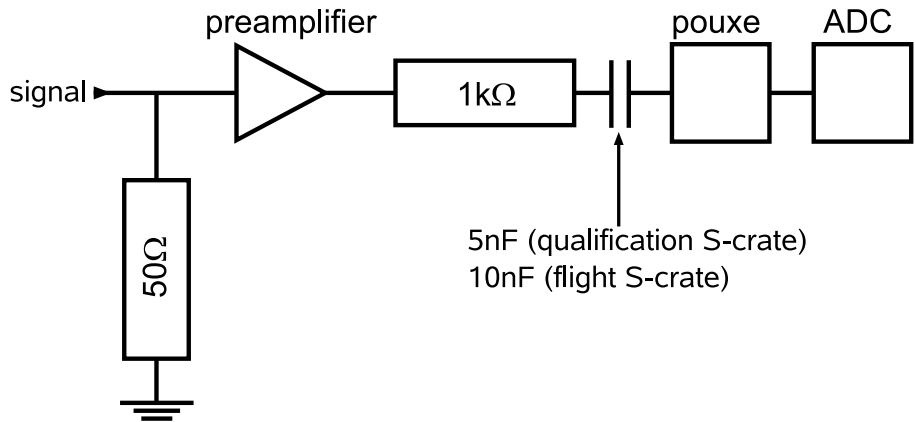,width=12cm}}\captionof{figure}{\label{f-electronics_inputstage}Schematics of the ADC branch for the ACC PMT readout.}
\end{center}
\end{figure}

The schematics of the ADC branch are shown in Fig.~\ref{f-electronics_inputstage}. Before the preamplified analog signal is measured in the ADC, the charge is amplified again in the so called Pouxe chip. The integrated charge is transformed to a pulse whose maximum is proportional to the input charge. The sample and hold time (s/h) defines the moment when the pulse charge is stored on a capacitor and is measured by the ADC. In the test setups used in the following the s/h time was optimized to measure the maximum of the input pulse after the level\,1 trigger. The difference between the qualification and flight modules is a change in the input capacitor of the ADC branch from 5\,nF to 10\,nF. This capacitor influences the optimal s/h time and the pulse shape because of different time constants in the two configurations. During flight there will be two types of events in the ACC. The first one is caused by additional external particles crossing the detector from the side and will not have any correlation between the TOF and ACC hits. In this case an optimization with respect to the level\,1 trigger caused by the TOF is not possible and also not needed as these events are not relevant for the analysis and can be simply vetoed by the discriminator branch. The second type of events in the ACC are close in time with the TOF signals and will therefore be measured in the s/h optimized ADC conditions. These particles arise from internal interactions in the detector or from external particles crossing the ACC at the same time as the particle which causes the TOF trigger.

\subsubsection{Calibration Measurements with the Qualification Electronics}

\begin{figure}
\begin{center}
\begin{minipage}[b]{.4\linewidth}
\centerline{\epsfig{file=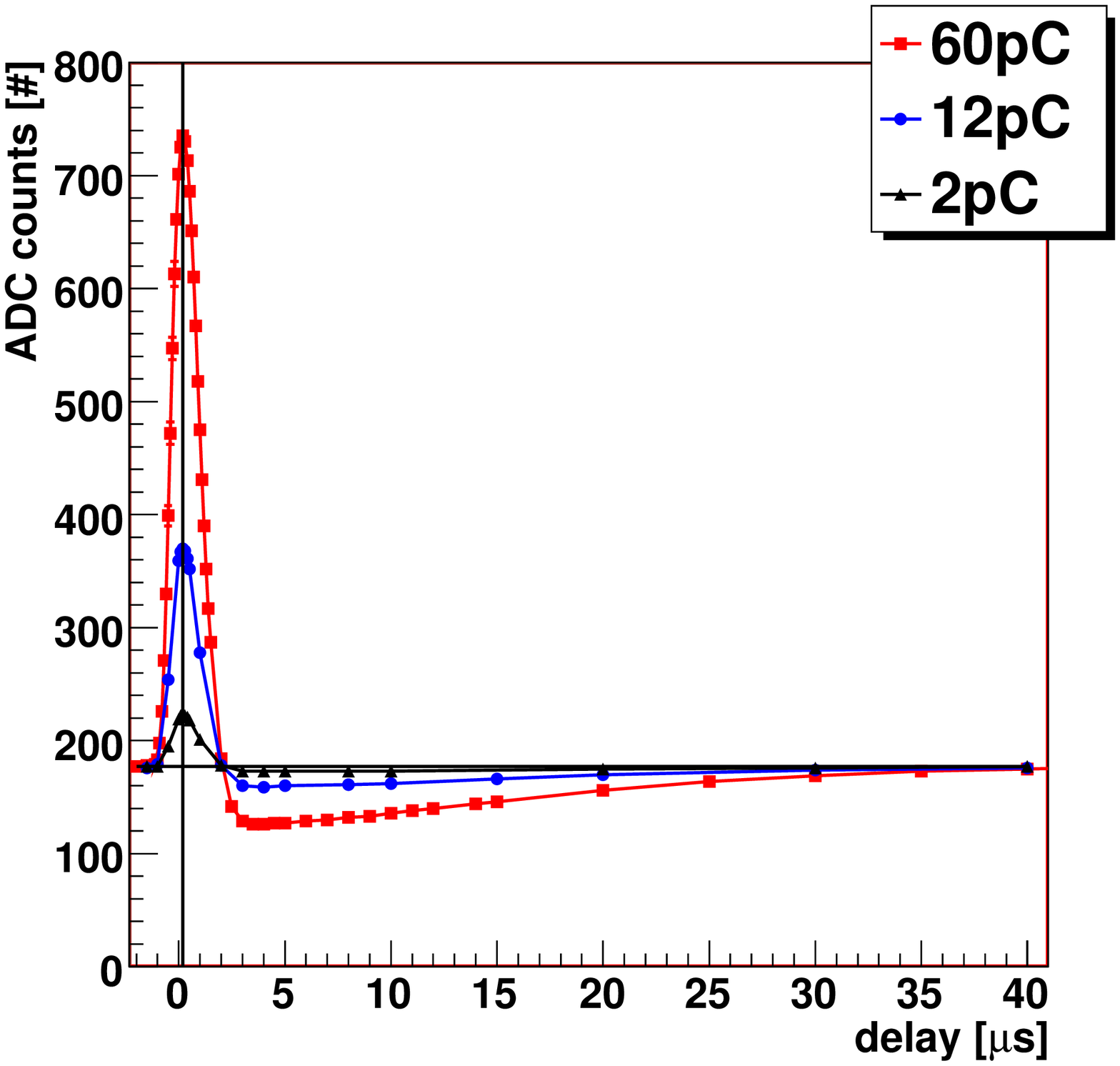,width=8cm}}\captionof{figure}{\label{f-delayscan_gain25_1.2k_5nF_ch2_sideA_QM}Delay scans of the SFEA2 ADC with test pulses. The black horizontal line indicates the pedestal position.}
\end{minipage}
\hspace{.1\linewidth}
\begin{minipage}[b]{.4\linewidth}
\centerline{\epsfig{file=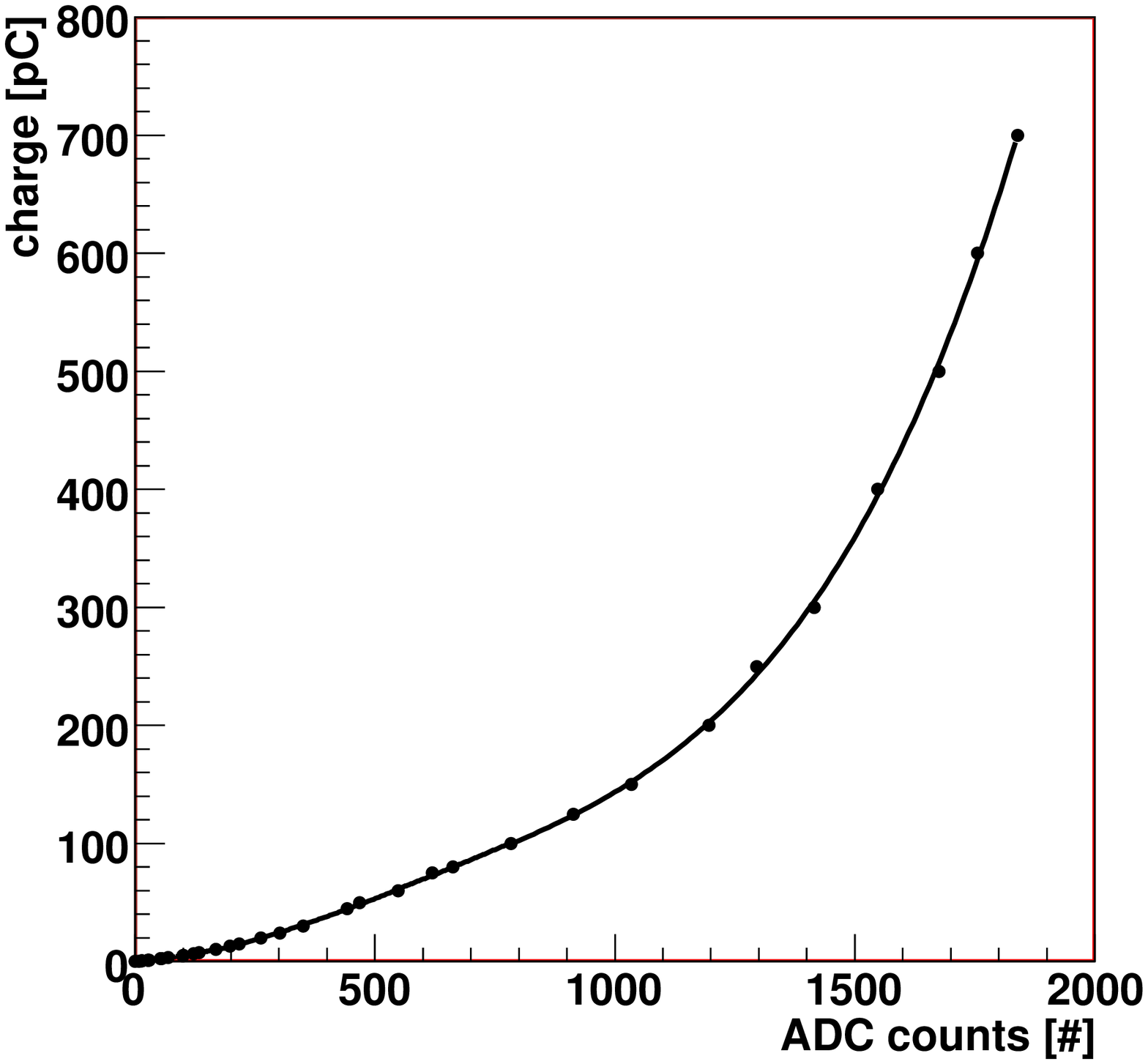,width=8cm}}\captionof{figure}{\label{f-lin_inv}Injected charge of a test pulse as a function of the SFEA2 ADC value.}
\end{minipage}
\end{center}
\end{figure}

Calibration measurements were performed before collecting data on atmospheric muons with the qualification S-crate. In addition to the delay of the complete readout chain, the sample and hold time of the Pouxe chip can be adjusted to get the maximum value for the measured charge. This delay scan was performed for three different charges by injecting a test pulse via a 100\,pF capacitor to the SFEA2 inputs and then shifting the signal in time to determine the optimal delay for a maximum pulseheight (Fig.~\ref{f-delayscan_gain25_1.2k_5nF_ch2_sideA_QM}). In the qualification module, with a 5\,nF capacitor input stage on the SFEA2 board, the maximum charge is found at 200\,ns s/h time for the three different injected charges of 2\,pC, 12 \,pC and 60\,pC. The MOP value of the ACC signal corresponds to about 20\,pC. The sampled charge of the 12\,pC test pulse has a maximum  undershoot of about 15\,ADC counts with a length of about 30\,\textmu s. As the choice of the input stage capacitor influences the s/h delay, the capacitor is changed to 10\,nF for the flight version because the TOF boards are also using 10\,nF capacitors and the s/h time, determined by the TOF system to be 600\,ns, can only be set to one value for all S-crates. The change from 5\,nF to 10\,nF will cause a slightly longer but on average smaller undershoot.

Here for the QM S-crate, the s/h time was set to 200\,ns and the linearity of the ADC was studied by injecting different charges into the SFEA2 inputs (Fig.~\ref{f-lin_inv}). A saturation at about 2000\,ADC counts is seen due to saturation of the input amplifier which has to deal with a wide range of charge values from $10^{-1}$ to $10^3$\,pC. As noted above, the ACC requirements emphasize the need to measure small charges with high resolution, so the lower resolution and saturation for large charges is not an issue. The knowledge of the non-linearity will be used later to calculate the number of photo-electrons from the measured ADC counts. The SFEA2 shows a higher resolution than the CAMAC electronics with 3.37\,ADC/pC. 

\begin{figure}
\begin{center}
\begin{minipage}[b]{.4\linewidth}
\centerline{\epsfig{file=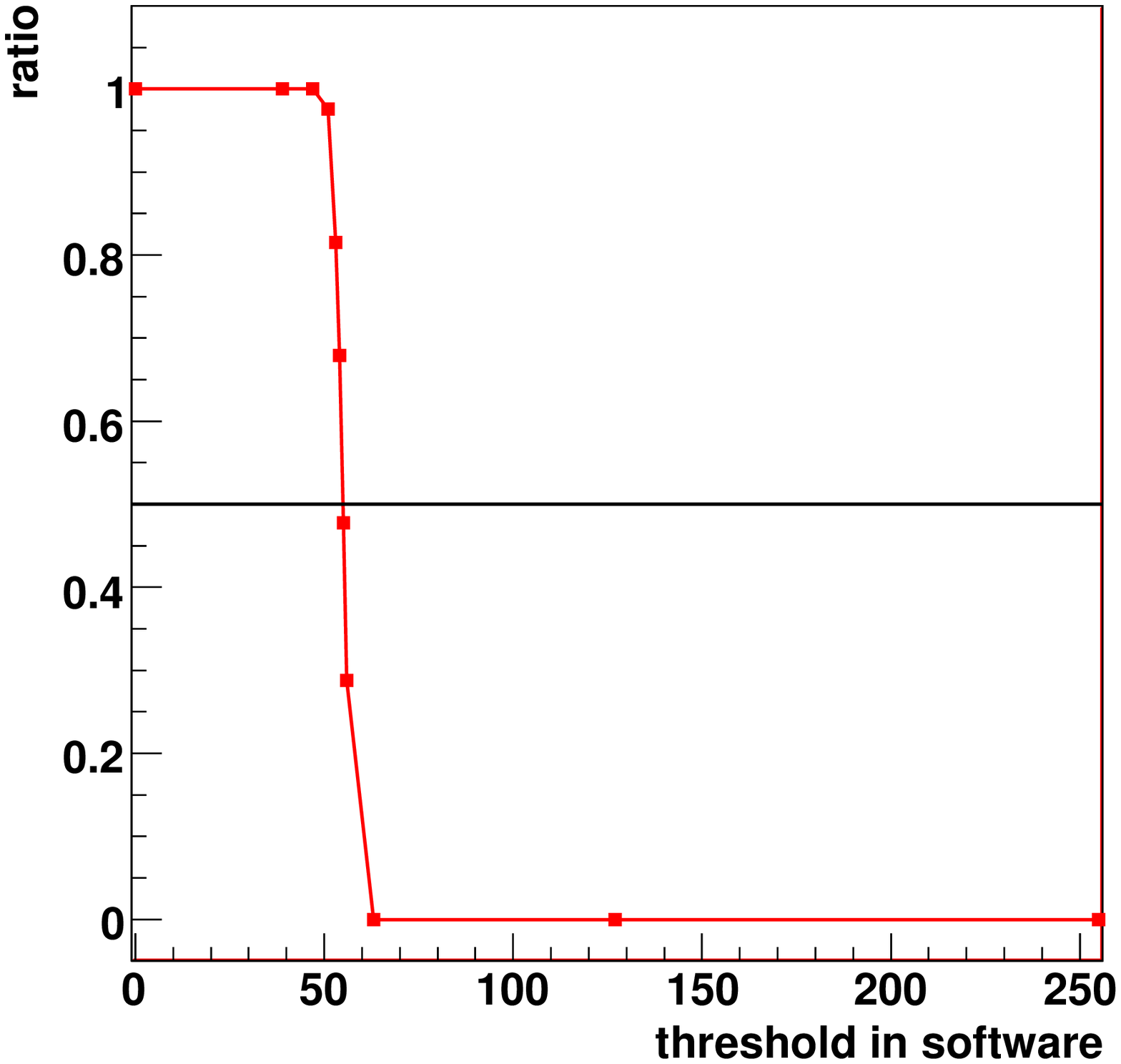,width=8cm}}\captionof{figure}{\label{f-thr100_gain25_1.2k_5nF_ch2_sideA_QM}Threshold scan of the TDC discriminator for the test pulse of 12.5\,mV amplitude. The ratio is defined as the number of hits above to those hits below the threshold. The threshold is set by the user in the control software.}
\end{minipage}
\hspace{.1\linewidth}
\begin{minipage}[b]{.4\linewidth}
\centerline{\epsfig{file=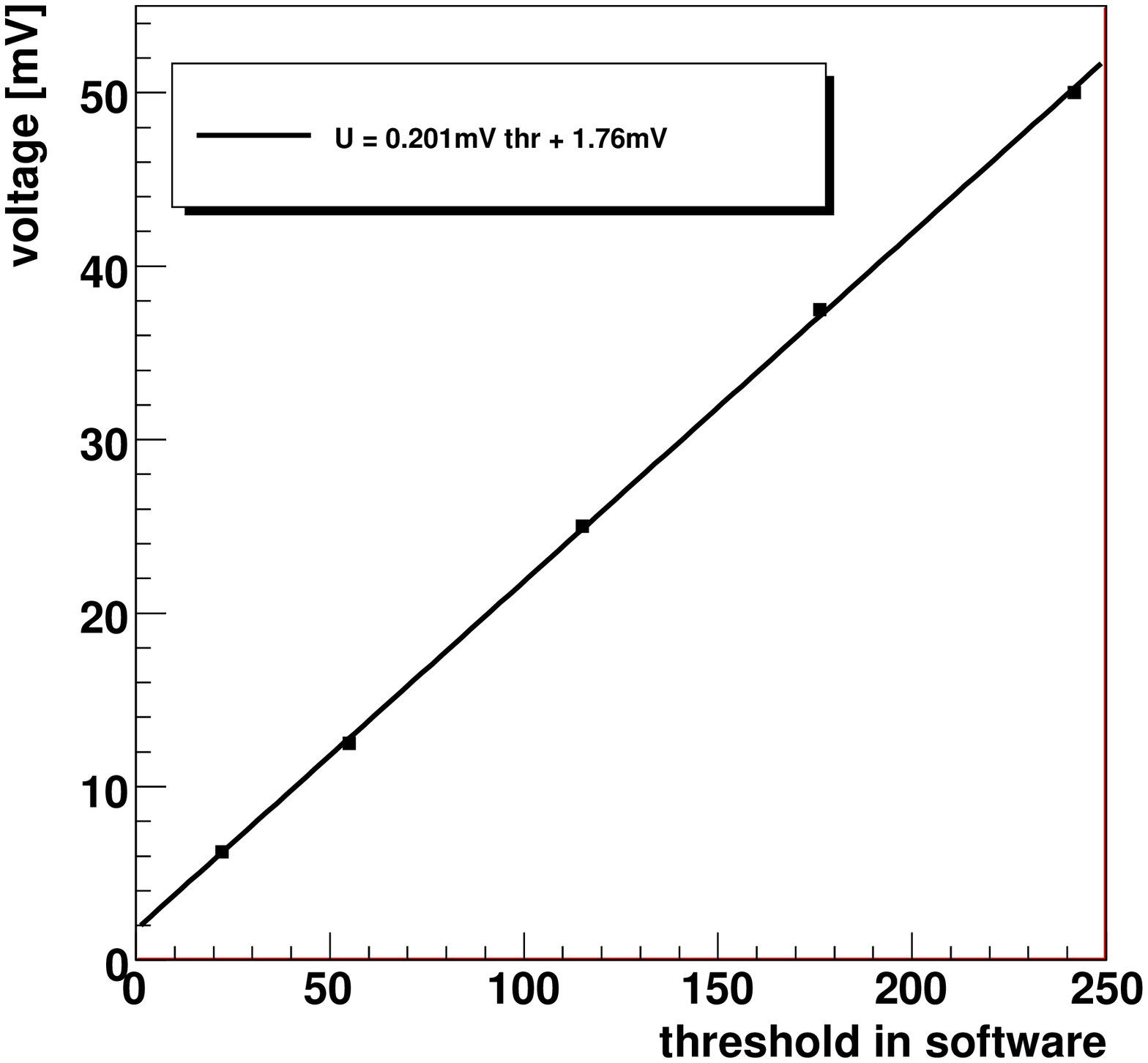,width=8cm}}\captionof{figure}{\label{f-mv_thr_gain25_1.2k_5nF_ch2_sideA_QM}Correlation of test pulse amplitudes and thresholds set by the user in the control software.}
\end{minipage}
\end{center}
\end{figure}

The next procedure concerns setting the discriminator threshold. The threshold is adjustable in 250 steps and adapted to accept events with small energy depositions in the scintillator which result in small pulse amplitudes of the PMTs. The discriminator threshold is sensitive to the pulse amplitude. The calibration is done with test pulses of 20\,ns length and different amplitudes. Varying the threshold for a given input pulse results in a sharp edge in the ratio between pulses above and below threshold (Fig.~\ref{f-thr100_gain25_1.2k_5nF_ch2_sideA_QM}). For a given pulse amplitude the threshold is defined at a ratio value of 0.5. These kind of measurements were performed for different thresholds and show a linear behavior (Fig.~\ref{f-mv_thr_gain25_1.2k_5nF_ch2_sideA_QM}):
\be U=(0.201\pm0.006)\,\text{mV}\cdot thr +(1.76\pm0.82)\,\text{mV}\label{e-thrmv}\ee where $thr$ is the user adjustable threshold in the control software.

\subsubsection{Atmospheric Muon Measurements with the Qualification Electronics}

\begin{figure}
\begin{center}
\begin{minipage}[b]{.4\linewidth}
\centerline{\epsfig{file=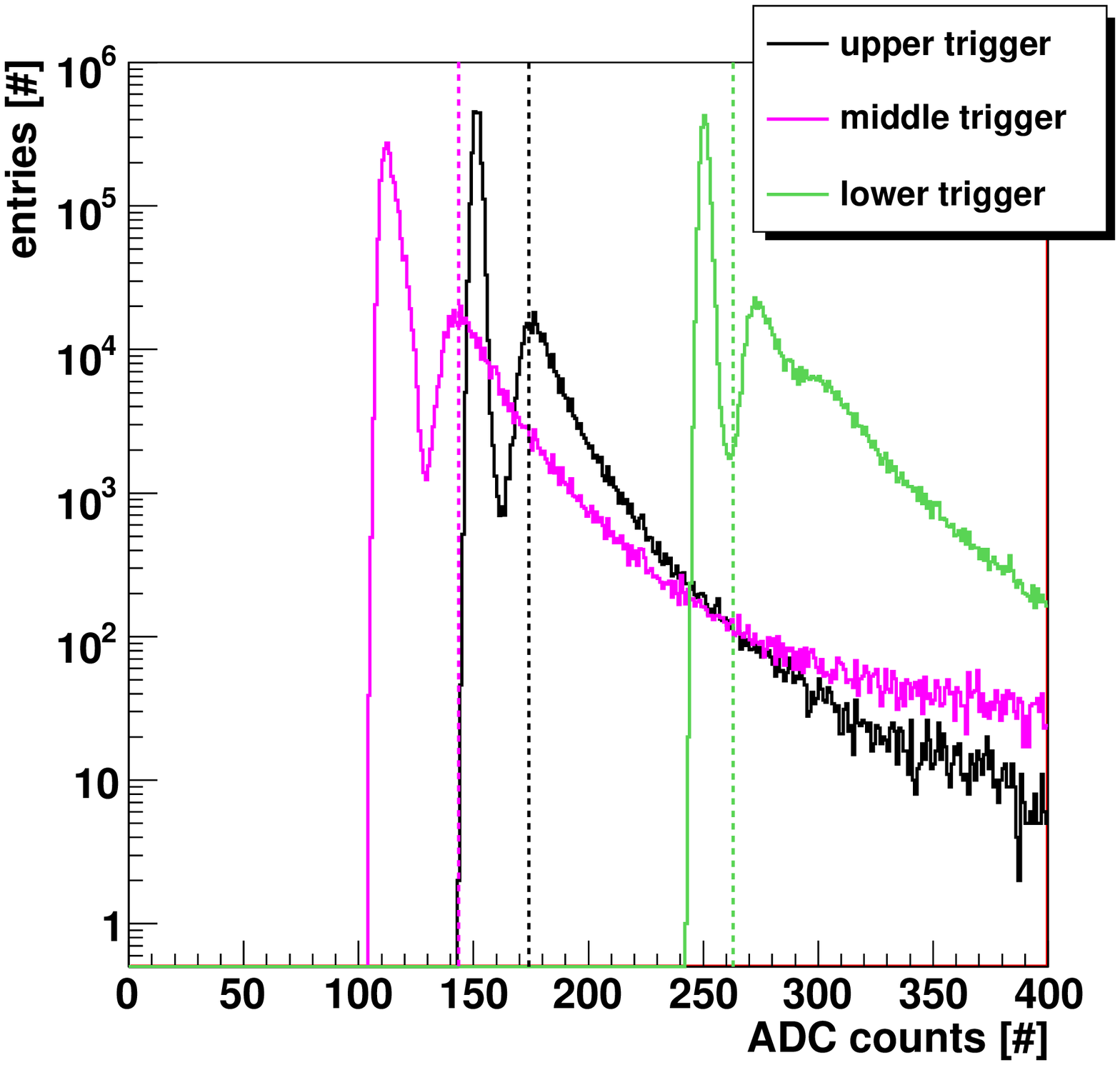,width=8cm}}\captionof{figure}{\label{f-total_HV32_2016V_HV33_2200V_sfet_no_cut}Data collected by the trigger counters, with cuts (dashed) used for the analysis.}
\end{minipage}
\hspace{.1\linewidth}
\begin{minipage}[b]{.4\linewidth}
\centerline{\epsfig{file=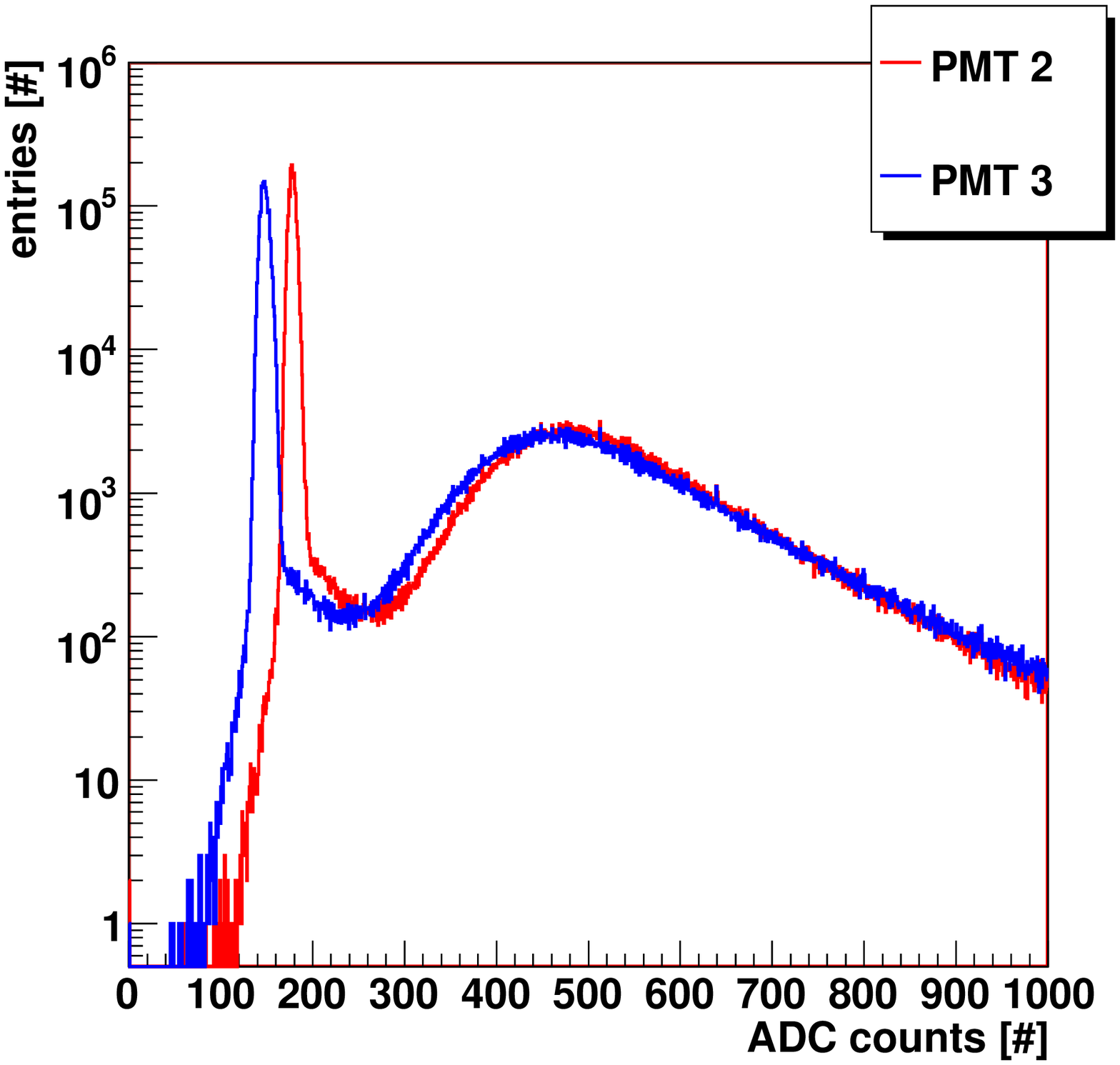,width=8cm}}\captionof{figure}{\label{f-total_HV32_2016V_HV33_2200V_acc_no_cut}All data collected by the ACC panel.}
\end{minipage}
\end{center}
\end{figure}
\begin{figure}
\begin{center}
\begin{minipage}[b]{.4\linewidth}
\centerline{\epsfig{file=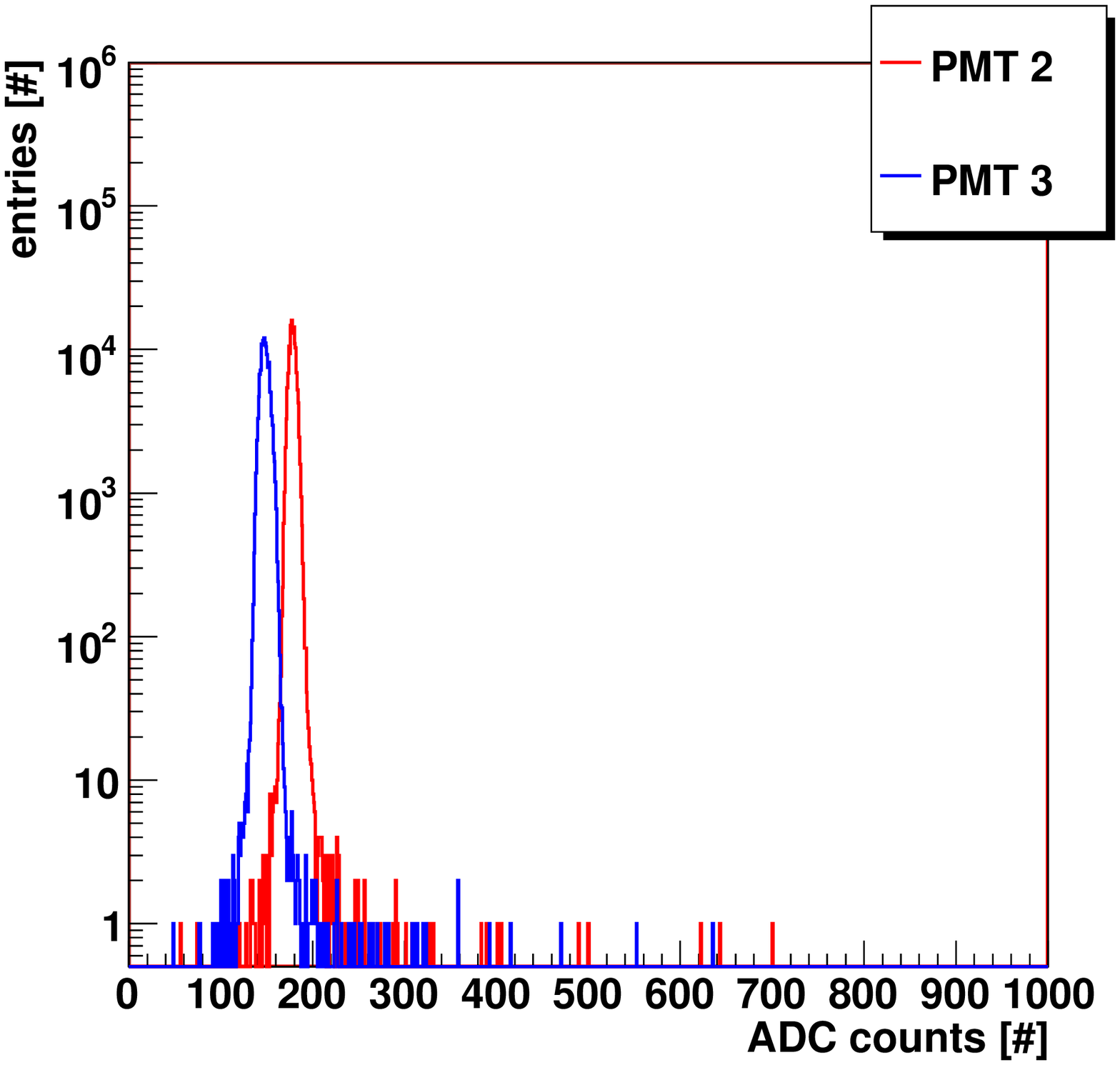,width=8cm}}\captionof{figure}{\label{f-total_HV32_2016V_HV33_2200V_ped}ADC pedestal spectra of the ACC PMTs.}
\end{minipage}
\hspace{.1\linewidth}
\begin{minipage}[b]{.4\linewidth}
\centerline{\epsfig{file=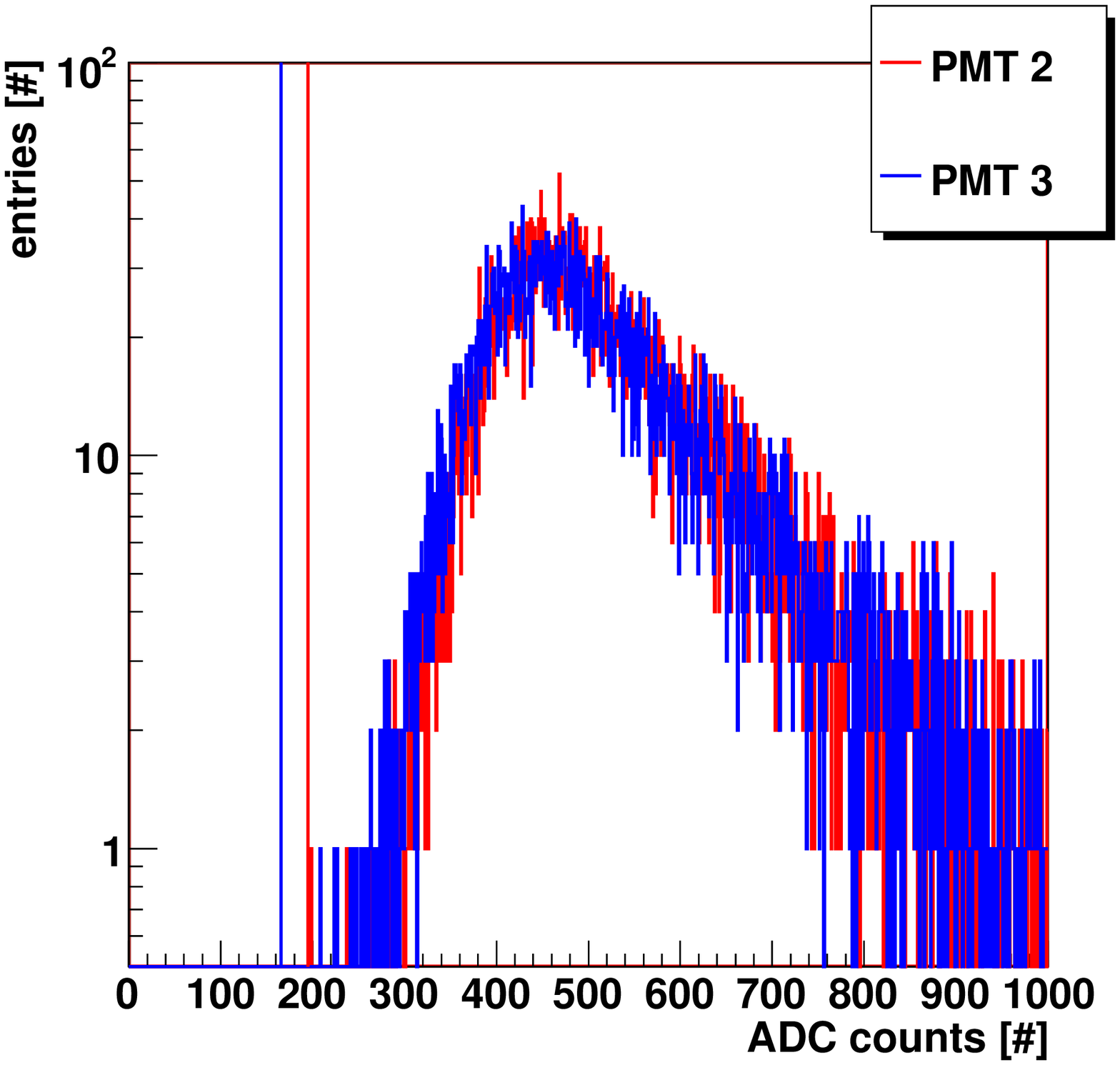,width=8cm}}\captionof{figure}{\label{f-total_HV32_2016V_HV33_2200V_acc}ADC spectra of ACC events with clean tracks. The vertical lines indicate the $3\sigma$ cut above the corresponding pedestals.}
\end{minipage}
\end{center}
\end{figure}

\begin{figure}
\begin{center}
\begin{minipage}[b]{.4\linewidth}
\centerline{\epsfig{file=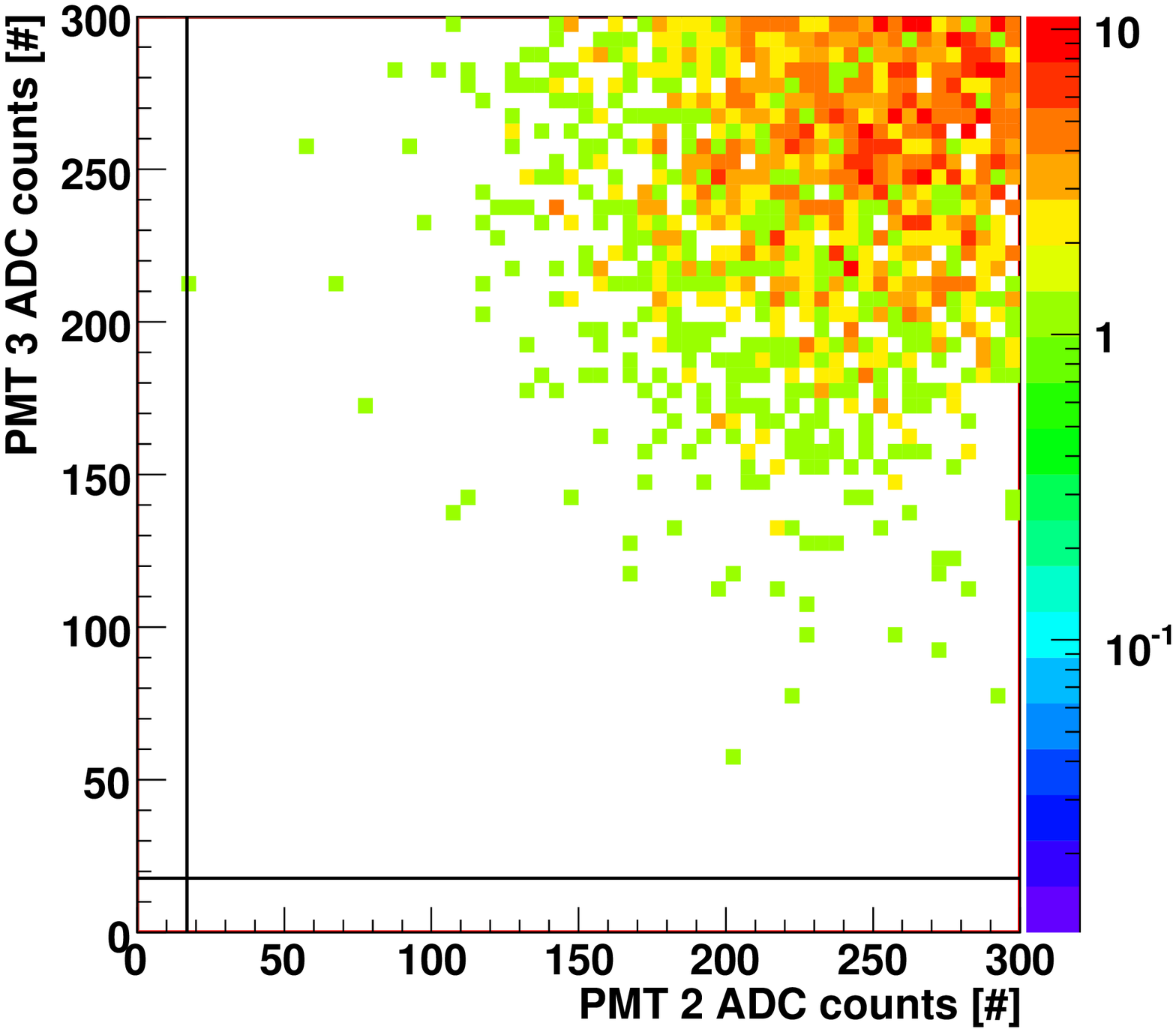,width=8cm}}\captionof{figure}{\label{f-total_HV32_2016V_HV33_2200V_acc0_acc1_adc}Correlation of pedestal corrected ADC values of the ACC PMTs. Cuts at $p_i+3\cdot \sigma_i$ are indicated. The color code on the right shows the number of entries.}
\end{minipage}
\hspace{.1\linewidth}
\begin{minipage}[b]{.4\linewidth}
\centerline{\epsfig{file=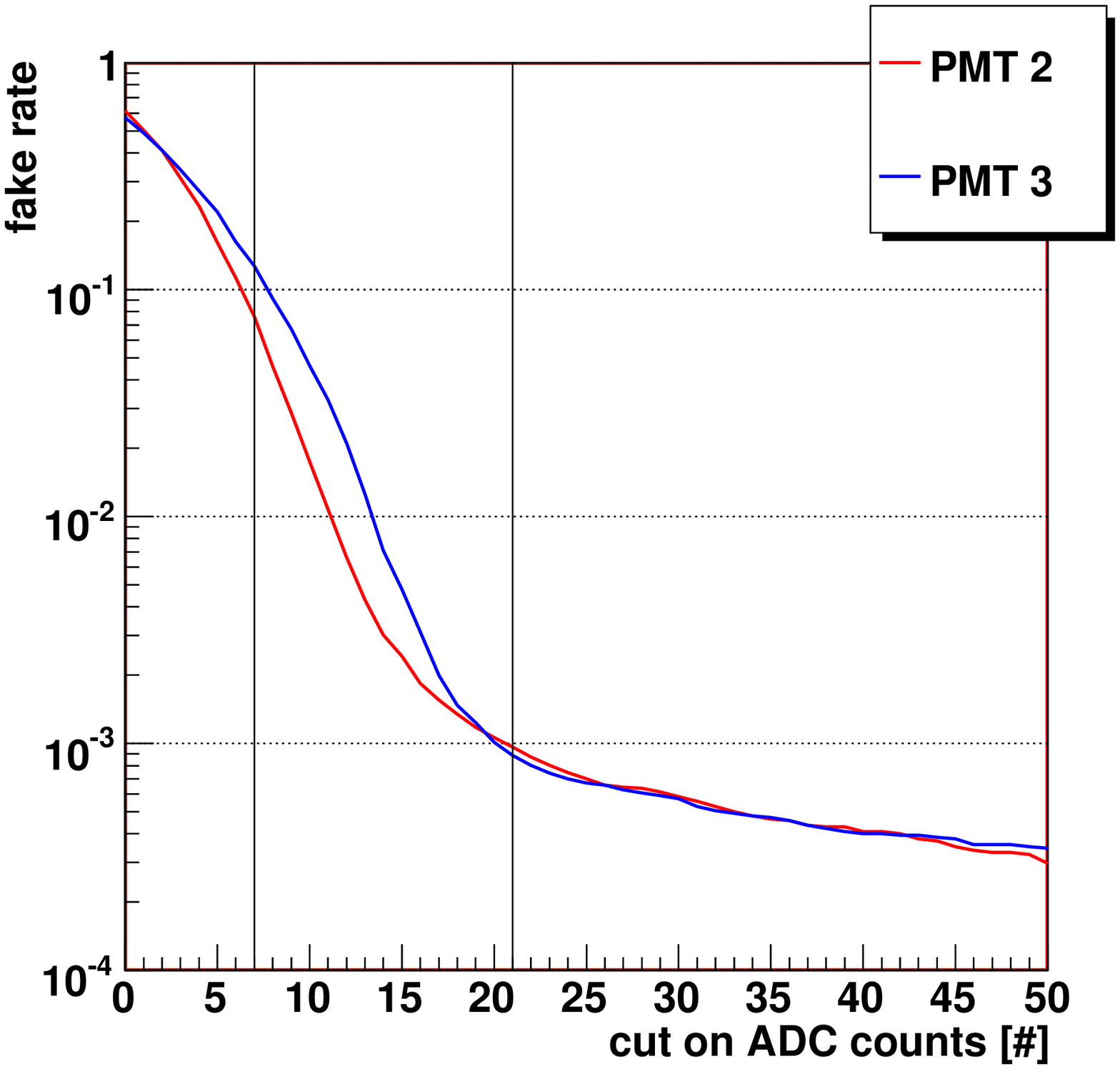,width=8cm}}\captionof{figure}{\label{f-total_HV32_2016V_HV33_2200V_eff_adc}Fake rate due to pedestal hits as a function of the cut used to define a good ACC event. The vertical lines indicate 1 and 3 RMS of the pedestal distribution, respectively.}
\end{minipage}
\end{center}
\end{figure}

The test with atmospheric muons was carried out with the setup described in Sec.~\ref {ss-perfdet} but using the qualification electronics. The trigger processing of the TOF was used and the level\,1 trigger was generated by the OR of the three trigger scintillator counters which started the readout. The scintillator counters were above and below the ACC panel (names: upper, middle, lower) and connected to the SFET2A. In addition, the internal trigger with a rate of 1\,Hz was used to collect pedestal entries. Fig.~\ref{f-total_HV32_2016V_HV33_2200V_sfet_no_cut} shows all data collected by the upper, middle and lower trigger counters with the respective cuts used for the analysis. Both PMTs of the ACC panel ran with the same gain at 2016\,V for PMT 2 and 2200\,V for PMT 3. The trigger decision on basis of the counters was made using the discriminated signals of the SFET2A for a threshold setting of 16. The raw data of the ACC panel collected with the SFEA2 board are shown in Fig.~\ref{f-total_HV32_2016V_HV33_2200V_acc_no_cut}. Fig.~\ref{f-total_HV32_2016V_HV33_2200V_ped} shows the pedestals collected using the internal random trigger for both ACC PMT channels. A track is now considered to be free of interactions if all trigger counters show pulseheights exceeding the respective cuts indicated in Fig.~\ref{f-total_HV32_2016V_HV33_2200V_sfet_no_cut}. The cuts $c_i$ used to define a clean ACC event are shown in Fig.~\ref{f-total_HV32_2016V_HV33_2200V_acc}. As in the previous analysis, good ACC events show at least one ADC value above the corresponding cut $c_i$:
\be c_i=p_i+3\cdot \sigma_i\ee where $p_i$ is the mean and $\sigma_i$ the RMS of the pedestal distribution. Only in one out of 8000 events the pulseheight of PMT\,2 is very close to the corresponding cut. Using the highest of the two PMT values significantly improves the performance of the ACC. For both PMTs the cuts can be increased to 140\,ADC counts before the first events missed by both channels occur. This is illustrated by the correlation of the ADC values of both PMTs (Fig.~\ref{f-total_HV32_2016V_HV33_2200V_acc0_acc1_adc}).

The fake rate is given by the percentage of events above the $c_i$ cut in the pedestal distributions (Fig.~\ref{f-total_HV32_2016V_HV33_2200V_ped}). It determines the fraction of events that will result in a veto without a particle crossing the ACC. As shown in Fig.~\ref{f-total_HV32_2016V_HV33_2200V_eff_adc}, the fake rate is $10^{-3}$ for the $3\sigma_i$ cut used here corresponding to about 20\,ADC counts.

\begin{figure}
\begin{center}
\begin{minipage}[b]{.4\linewidth}
\centerline{\epsfig{file=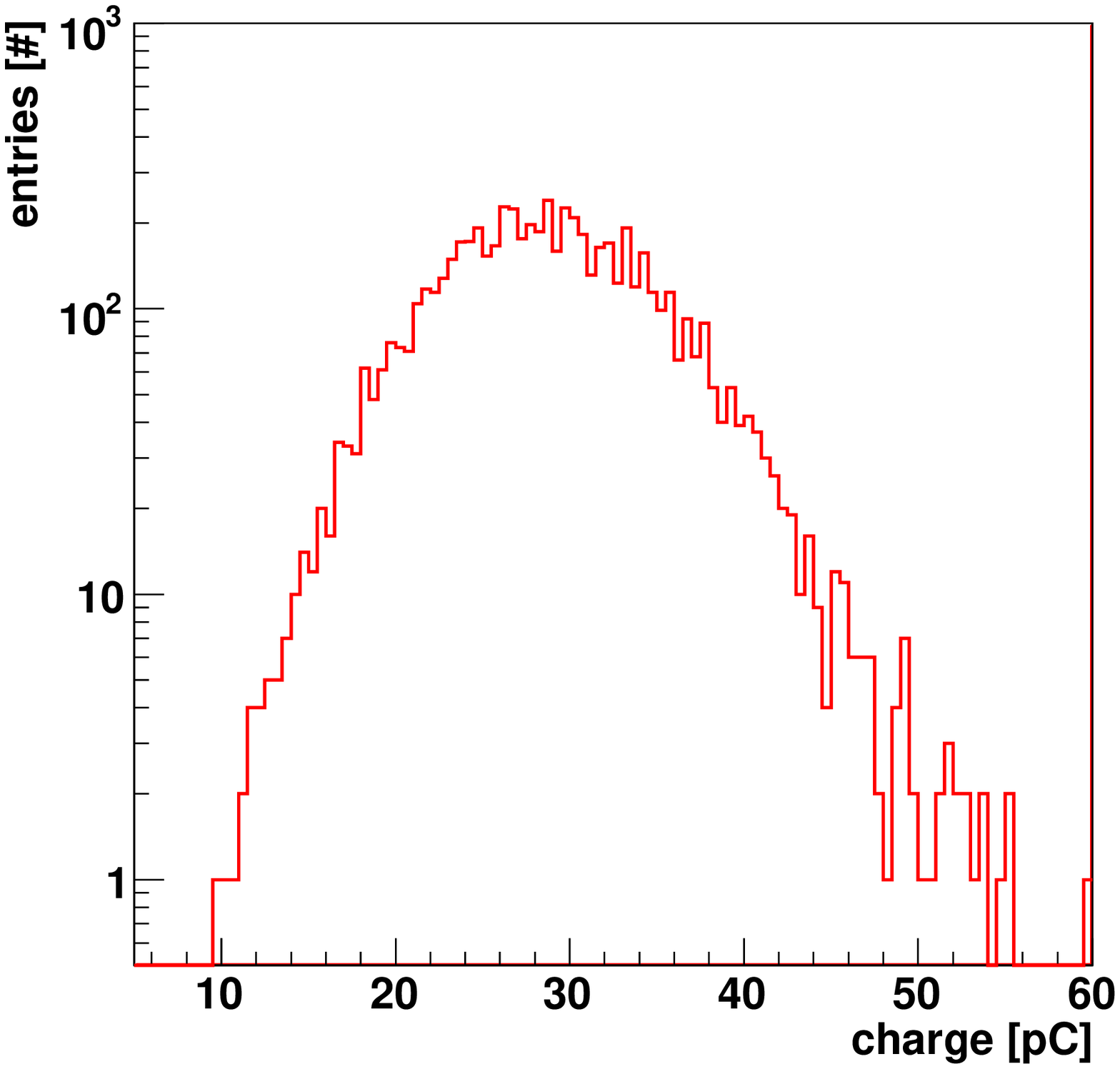,width=8cm}}\captionof{figure}{\label{f-080109_led_cal_HV32_2016V_HV33_2200V_ped.root_acc_sub_pc}Charge spectrum for PMT\,2 obtained by pulsing the integrated LED.}
\end{minipage}
\hspace{.1\linewidth}
\begin{minipage}[b]{.4\linewidth}
\centerline{\epsfig{file=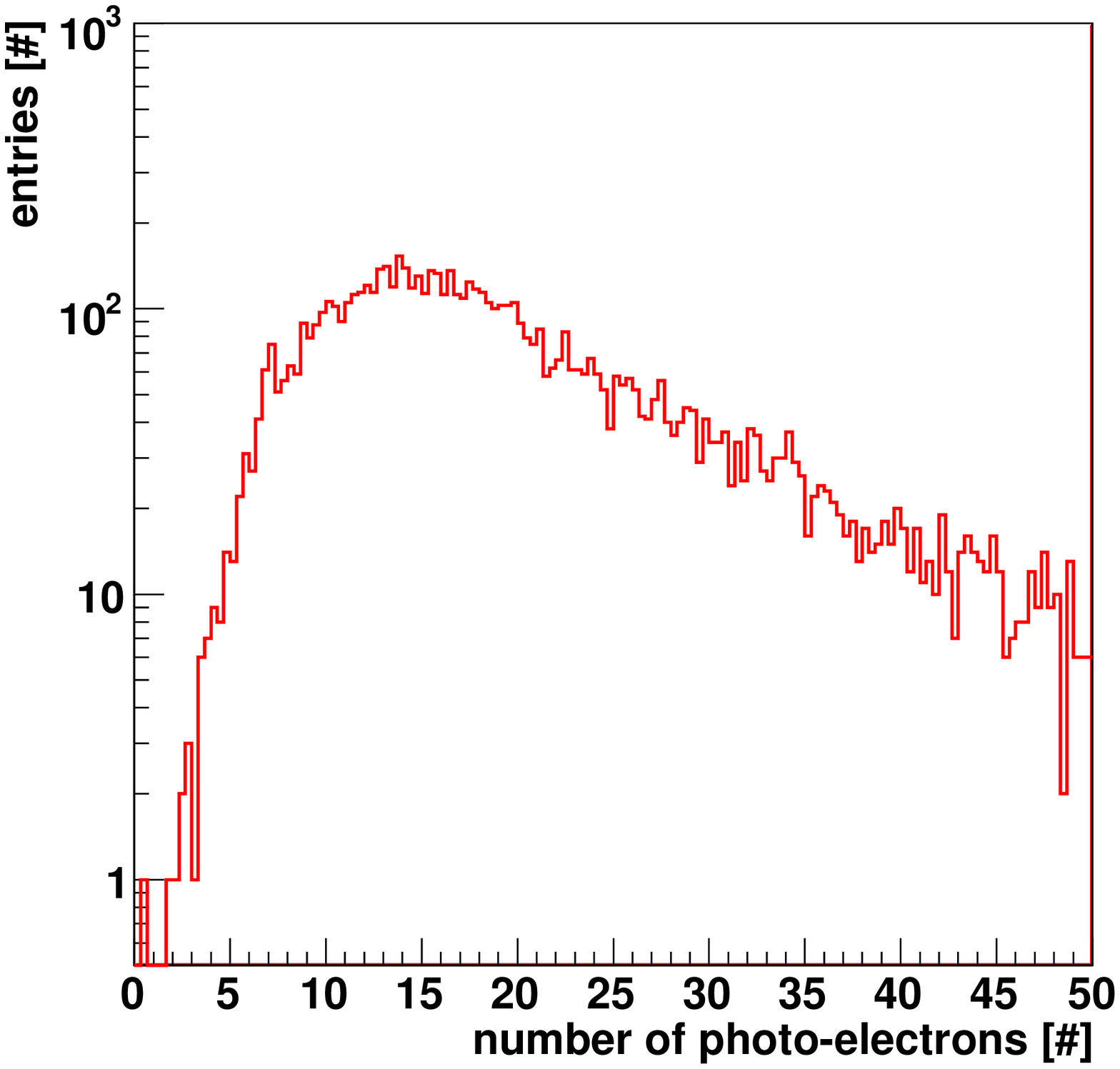,width=8cm}}\captionof{figure}{\label{f-total_HV32_2016V_HV33_2200V_acc_scint_pe}Number of photo-electrons (PMT\,2) for clean ACC events.}
\end{minipage}
\end{center}
\end{figure}

Furthermore it is now possible to use the charge calibration (Fig.~\ref{f-lin_inv}) to calculate the number of photo-electrons. Fig.~\ref{f-080109_led_cal_HV32_2016V_HV33_2200V_ped.root_acc_sub_pc} displays the charge spectrum obtained by pulsing one LED for PMT\,2 and applying the ADC to charge calibration. For clean ACC events, the number of photo-electrons for this channel is now calculated as in Sec.~\ref{ss-paneltest} and the distribution is shown in Fig.~\ref{f-total_HV32_2016V_HV33_2200V_acc_scint_pe}. The MOP value is at about 15 photo-electrons in agreement with the average of 16 photo-electrons measured for the whole system (Fig.~\ref{f-paneldis_systemtest}).

\begin{figure}
\begin{center}
\begin{minipage}[b]{.4\linewidth}
\centerline{\epsfig{file=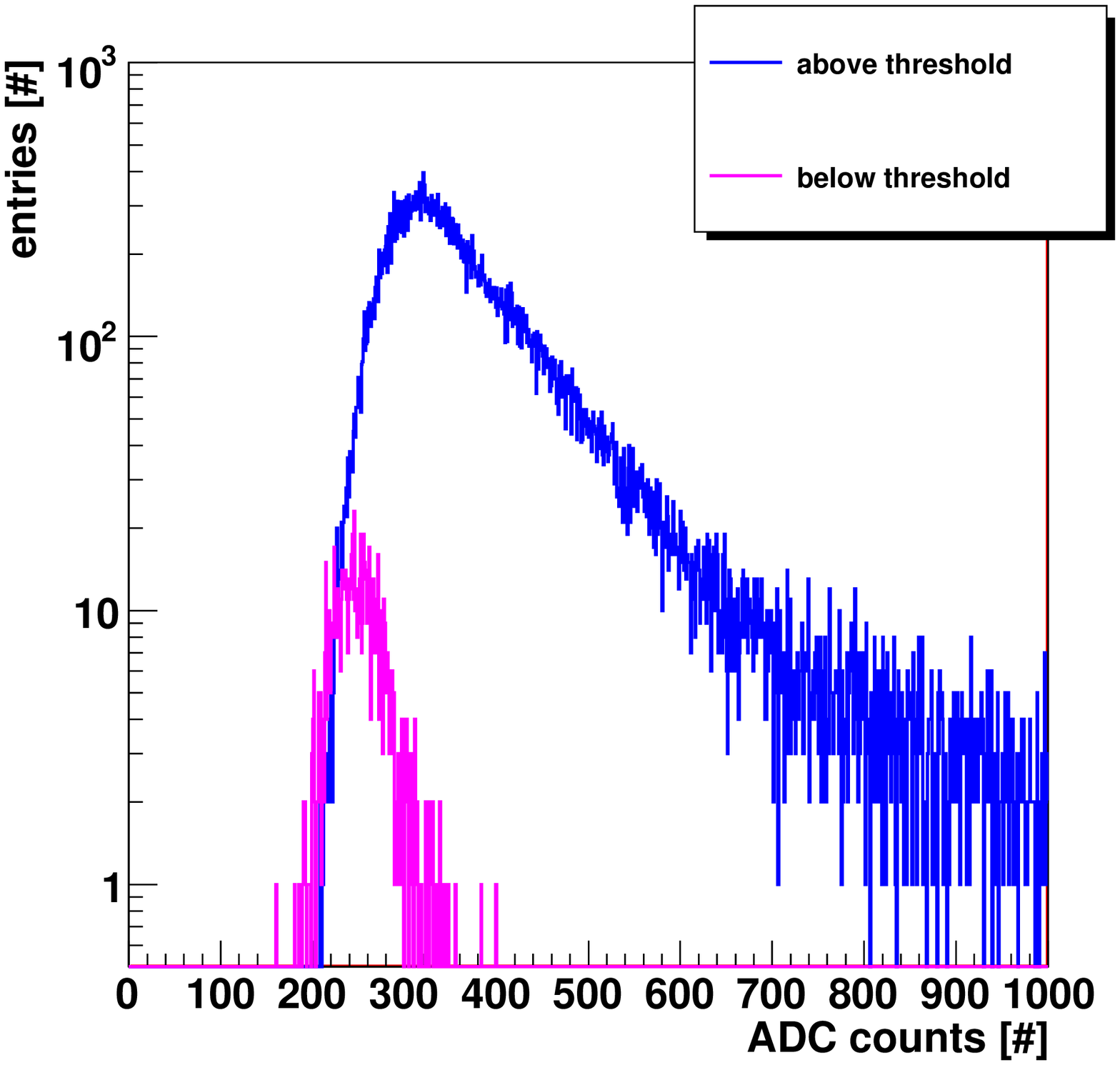,width=8cm}}\captionof{figure}{\label{f-total_on_off}PMT\,3: Discriminator decision for a threshold setting of 125.}
\end{minipage}
\hspace{.1\linewidth}
\begin{minipage}[b]{.4\linewidth}
\centerline{\epsfig{file=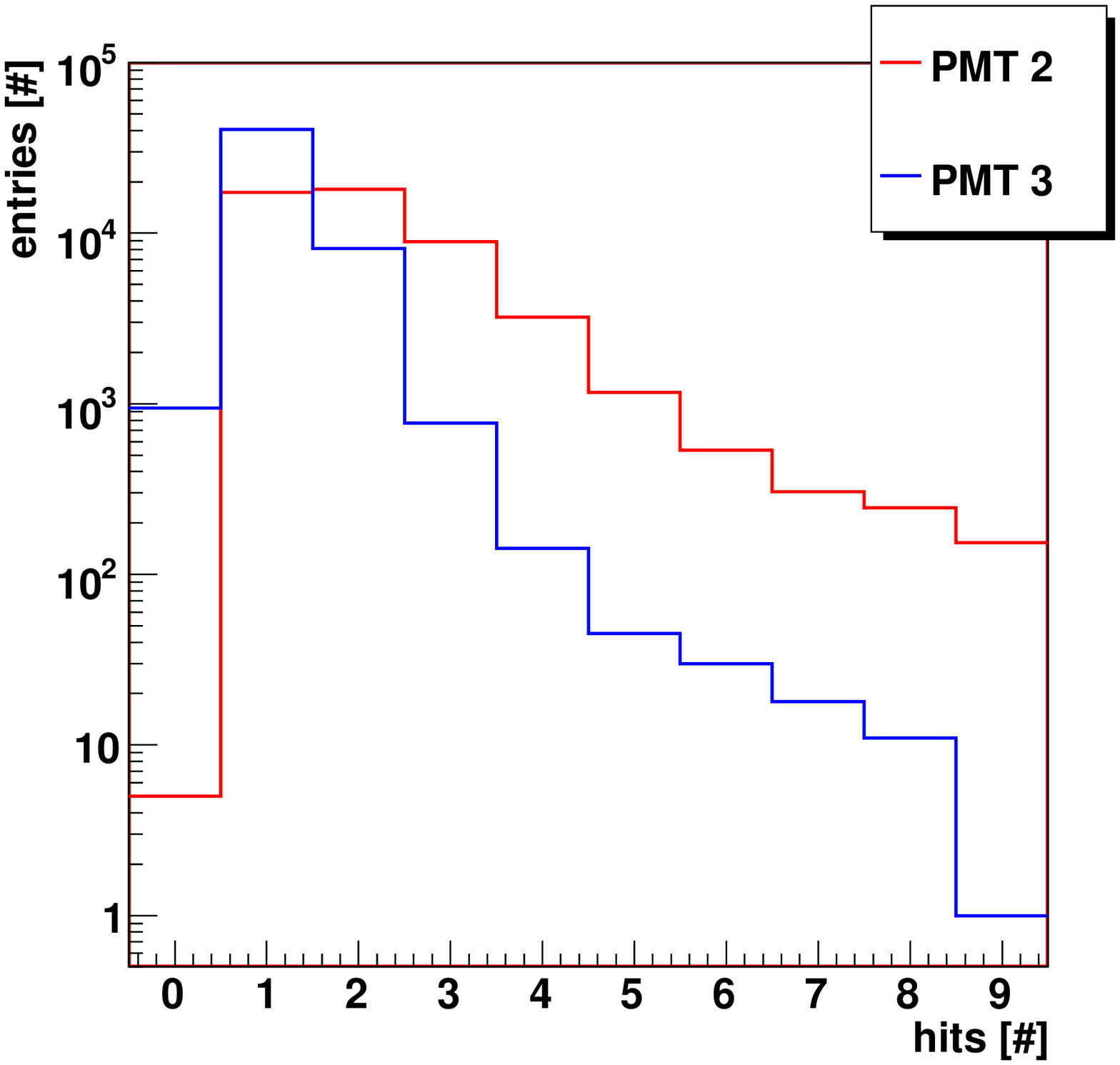,width=8cm}}\captionof{figure}{\label{f-total_sfea_hits}Number of hits in the TDC of the SFEA2 with PMT\,2 at higher gain than PMT\,3.}
\end{minipage}
\end{center}
\end{figure} 

Another issue of the ACC is to reduce the data rate during periods of very high fluxes, e.g. in the South Atlantic Anomaly. As mentioned above, the ACC can be used for the trigger decision by using the discriminator output in the JLV1. The discriminator output can be compared to the measured ADC values. The distribution of pulseheights with hits above and below threshold setting 125 ($\approx 26.9$\,mV) for PMT\,3 is shown in Fig.~\ref{f-total_on_off}. It is seen that the pulse amplitude and the total charge are not perfectly correlated. Since the discriminator reacts to the pulseheight and not to the charge, a clear separation cannot be achieved for signals from atmospheric muons with spikes after the main pulse. The signal amplitude may not exceed the threshold, although the total charge is close to the MOP value. Conversely, narrow pulses with large amplitude but small charge may fire the discriminator. This issue will also be further discussed in Sec.~\ref{ss-accpretest}.

Measurements with different thresholds showed that the trigger rate can be reduced by a factor $10^4$. Fig.~\ref{f-total_sfea_hits} shows the number of TDC time stamps for the case that PMT\,2 has a higher gain than PMT\,3 and a threshold setting of 125. As one would expect, the pulses of PMT\,2 are larger than the pulses of PMT\,3 resulting in more TDC hits. 

\begin{figure}
\begin{center}
\begin{minipage}[b]{.4\linewidth}
\centerline{\epsfig{file=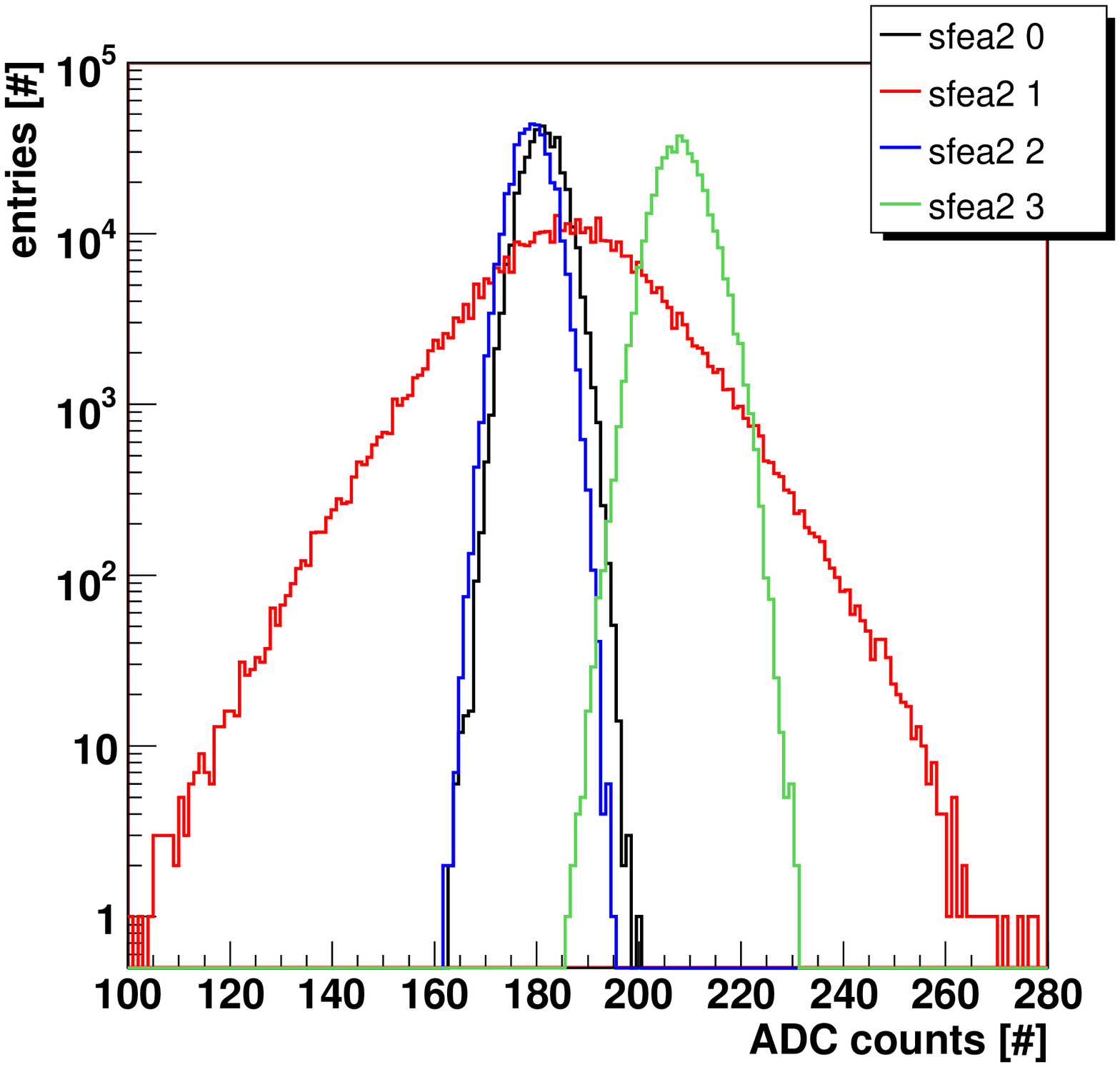,width=8cm}}\captionof{figure}{\label{f-run_1214394844.root_sfea2_adc_no_cut}ADC spectra of the SFEA2 board of the qualification S-crate during temperature variations in the range between -20°C and 60°C. A test pulse was injected in channel 1 only.}
\end{minipage}
\hspace{.1\linewidth}
\begin{minipage}[b]{.4\linewidth}
\centerline{\epsfig{file=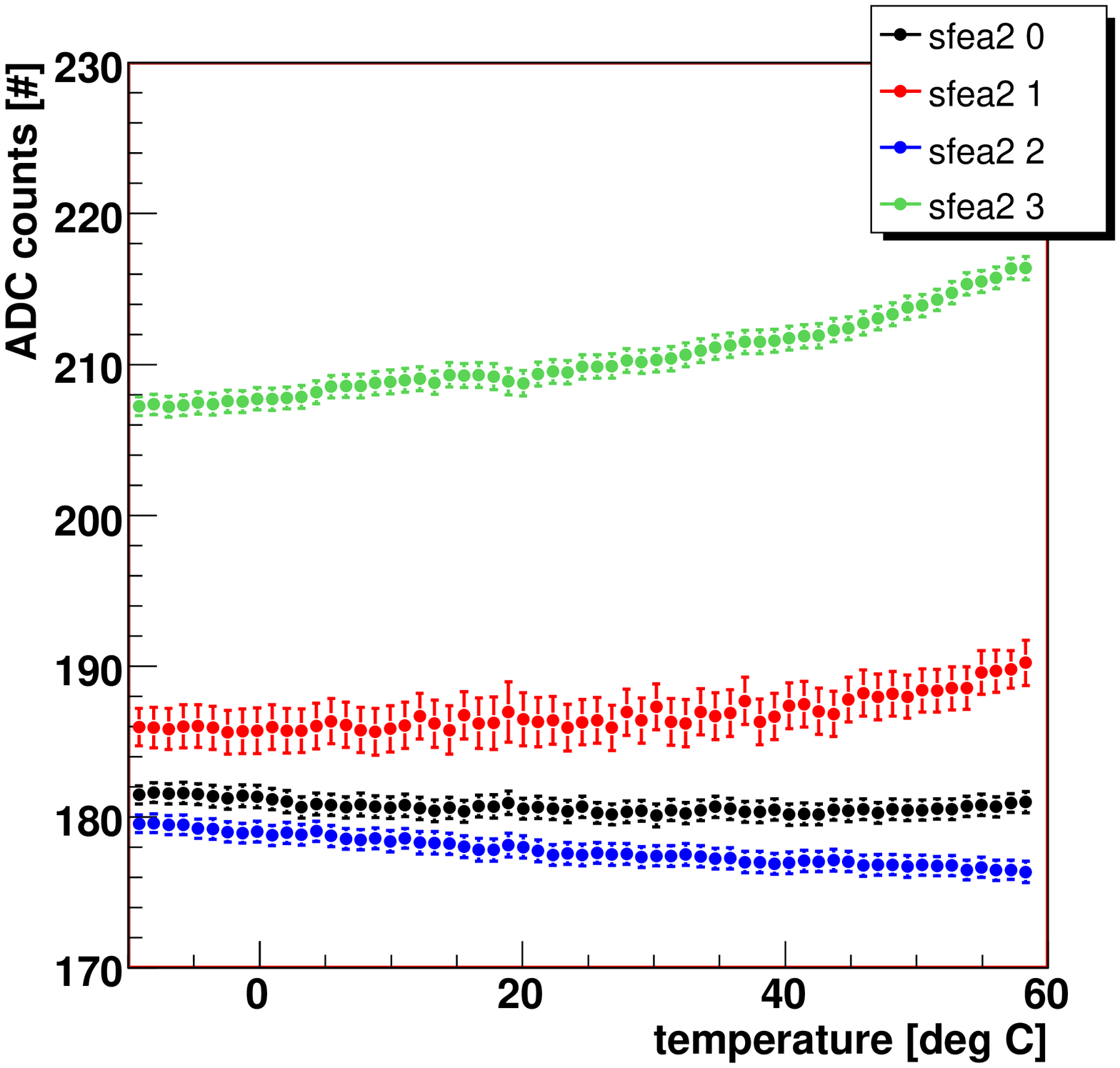,width=8cm}}\captionof{figure}{\label{f-run_1214394844.root_sfea2_temp_mean_adc}Mean ADC values of the SFEA2 board of the qualification S-crate as a function of temperature variation. A test pulse was injected in channel 1 only.}
\end{minipage}
\end{center}
\end{figure}

The ADC response as a function of the temperature was studied by placing the QM S-crate in a thermal chamber. Fig.~\ref{f-run_1214394844.root_sfea2_adc_no_cut} shows the ADC spectra for a temperature variation between -10°C and 60°C. A test pulse was injected in channel 1 while channels 0, 2 and 3 measured only pedestal. The mean ADC value for these channels as a function of the temperature shows that the temperature behavior depends on the individual channel. Channel 0 stays approximately constant while channel 2 shows a decrease by about 5\,ADC counts and channel 3 an increase by about 10\,ADC counts over the temperature range studied. This is also the reason for the increased width of the ADC spectra of channel 3 compared to channel 0 and 2. Channel 1 shows a weaker increase of about 5\,ADC counts and the spectrum is much wider than the others due to the test pulse. Because of the large temperature variations in space the individual behavior of the ADC channels must be taken into account for the pedestal correction. However, the variation is not too large such so a reliable operation and readout of the ACC will be possible.

In conclusion, the discriminator branch is able to reduce the trigger rate sufficiently while the ADC branch shows a high resolution and reliability in measuring even small charges.

\subsubsection{Calibration Measurements with the Flight Electronics}

\begin{figure}
\begin{center}
\begin{minipage}[b]{.4\linewidth}
\centerline{\epsfig{file=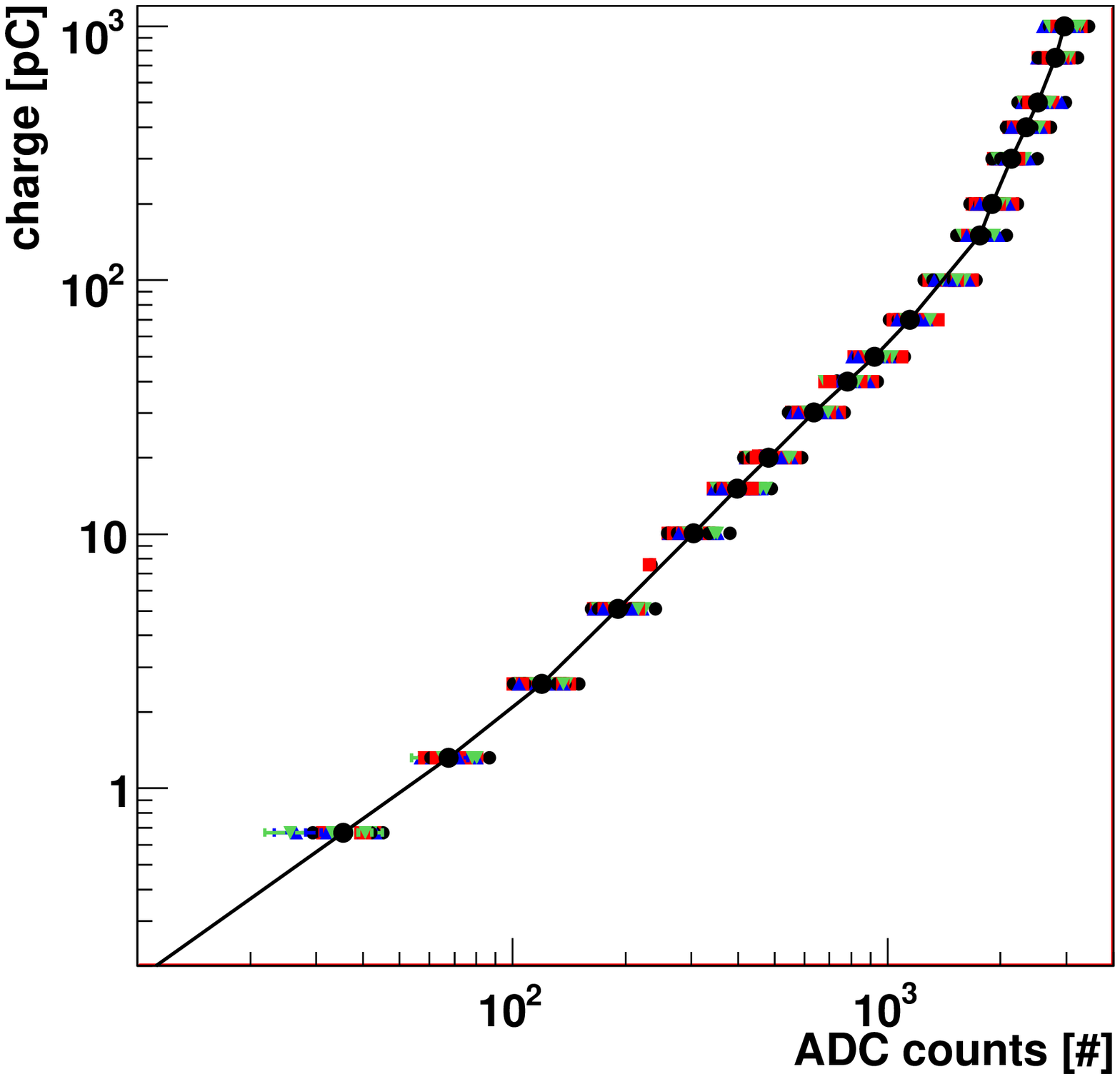,width=8cm}}\captionof{figure}{\label{f-all_cali}Non-linearity of the ADC for all flight S-crates and both sides A and B. The black dots are the mean values.}
\end{minipage}
\hspace{.1\linewidth}
\begin{minipage}[b]{.4\linewidth}
\centerline{\epsfig{file=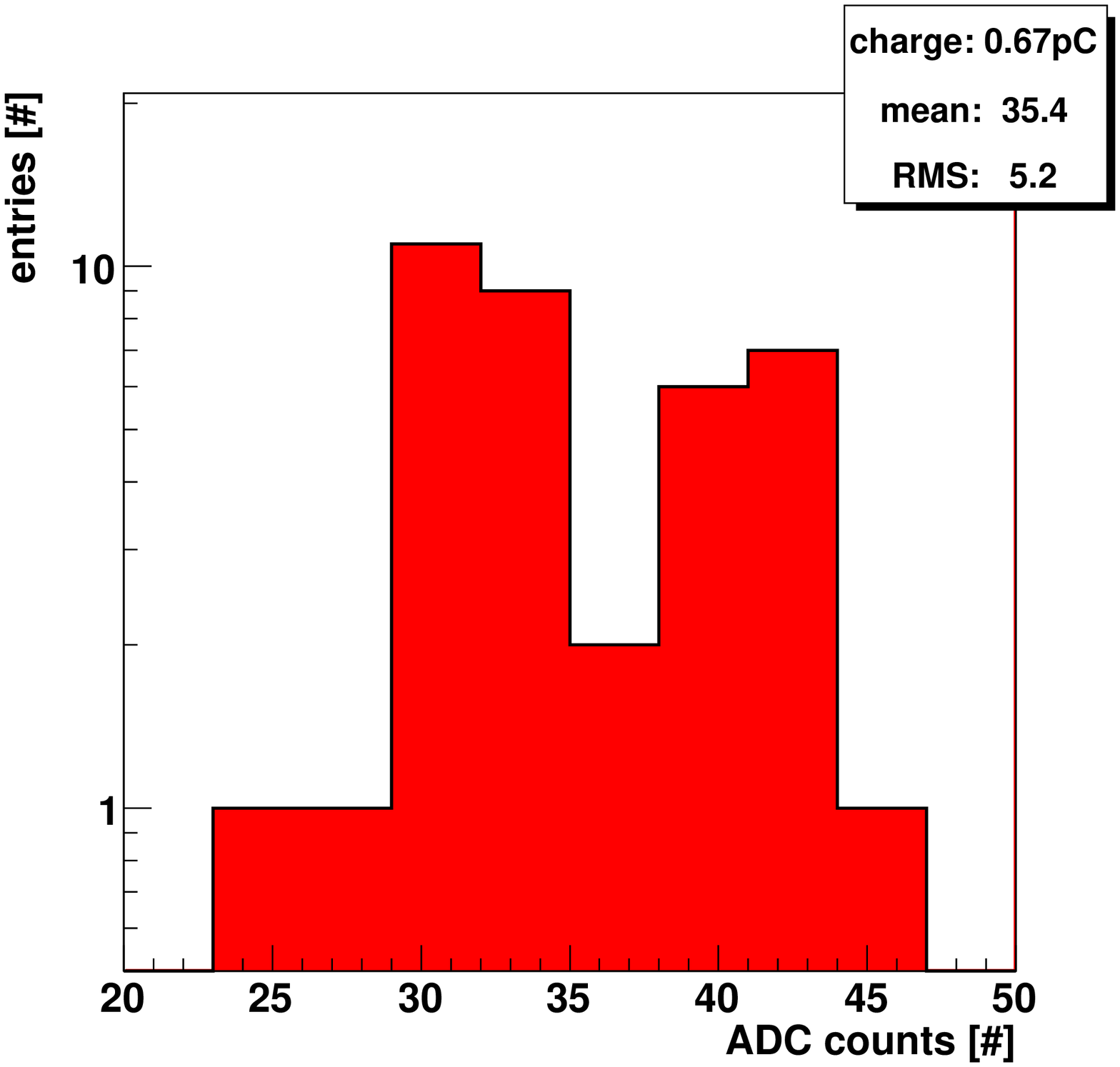,width=8cm}}\captionof{figure}{\label{f-res067}Variation of all SFEA2 ADC channels in the flight S-crates for an input test pulse of 0.67\,pC.}
\end{minipage}
\end{center}
\end{figure}

The ADCs on the SFEA2 boards of all four flight S-crates and the one spare crate must also be calibrated with test pulses. As mentioned above, each crate has one SFEA2 board with a double redundancy in the ADC. The different sides are called A and B. As for the qualification module calibration before (Fig.~\ref{f-lin_inv}), a test pulse of known charge was injected via a test capacitor of 100\,pC. The calibrations were done at a sample and hold time of 600\,ns which will be the setting during flight. The measurements of all crates on both sides are shown in Fig.~\ref{f-all_cali} together with their average values. As for the qualification S-crate, a saturation effect is also seen, but at a larger value of about 3000\,ADC counts due to the different input stage capacitor (10\,nF instead of 5\,nF). The variation in the response of the different ADCs to a small input charge of 0.7\,pC is 15\,\% (Fig.~\ref{f-res067}). It reduces to about 7\,\% for large input charge values of 1000\,pC. This necessitates a finer adjustment of the photomultiplier voltages during flight and will be discussed in Sec.~\ref{ss-accpretest}. 

\subsection{Tests after the Pre-Integration\label{ss-accpretest}}

\begin{figure}
\begin{center}
\centerline{\epsfig{file=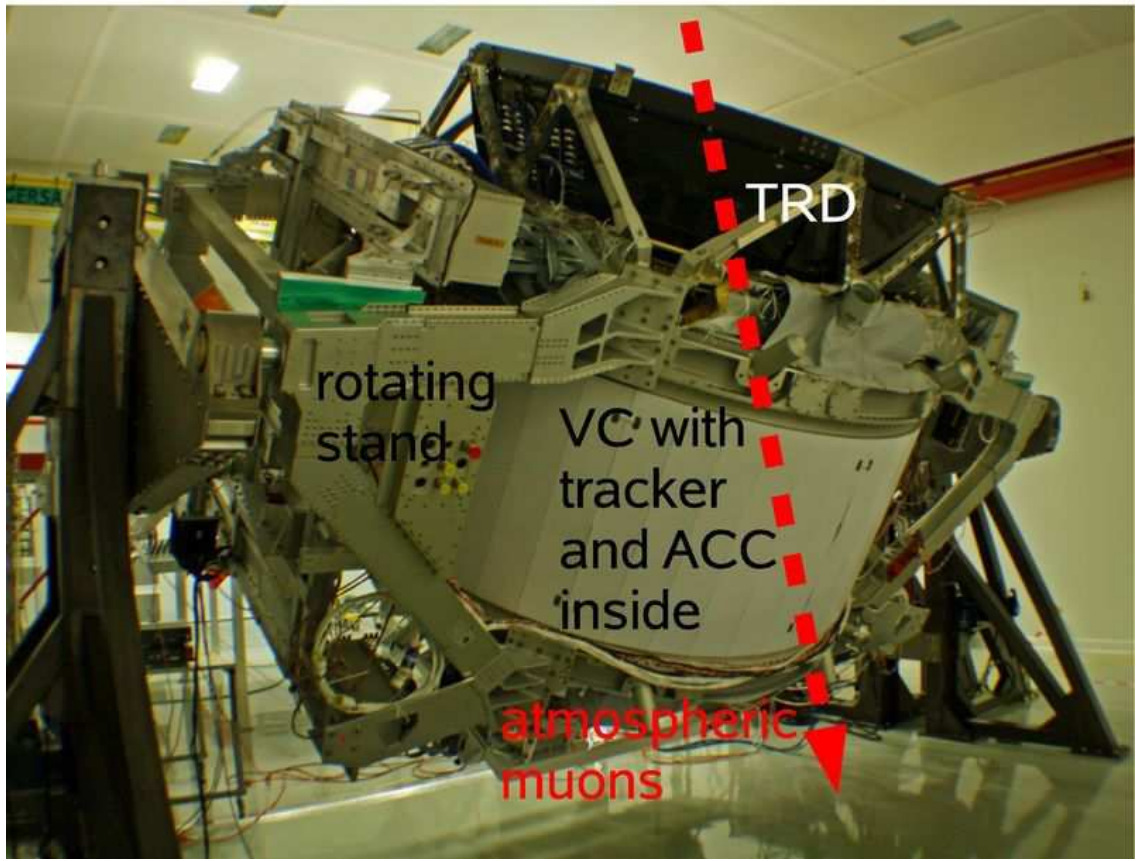,width=15cm}}\captionof{figure}{\label{f-cosmic_ams02}AMS-02 taking atmospheric muon data after the pre-integration. VC denotes the vacuum case for the helium tank.}
\end{center}
\end{figure}

A test with atmospheric muons was performed in Geneva after finishing the pre-integration of all components, but the magnet, into the AMS-02 detector. This period was used for intensive testing of the data acquisition chain and for calibration of the subdetectors. Fig.~\ref{f-cosmic_ams02} shows the experiment in the clean room mounted on a rotation stand at a 30° angle. In this section raw data and calibration of PMT voltages of the ACC system are presented and the analysis of the data reconstructed by extrapolating tracks from the TRD and the tracker to the ACC is discussed.
 
\subsubsection{ACC Raw Data}
 
\begin{figure}
\begin{center}
\begin{minipage}[b]{.4\linewidth}
\centerline{\epsfig{file=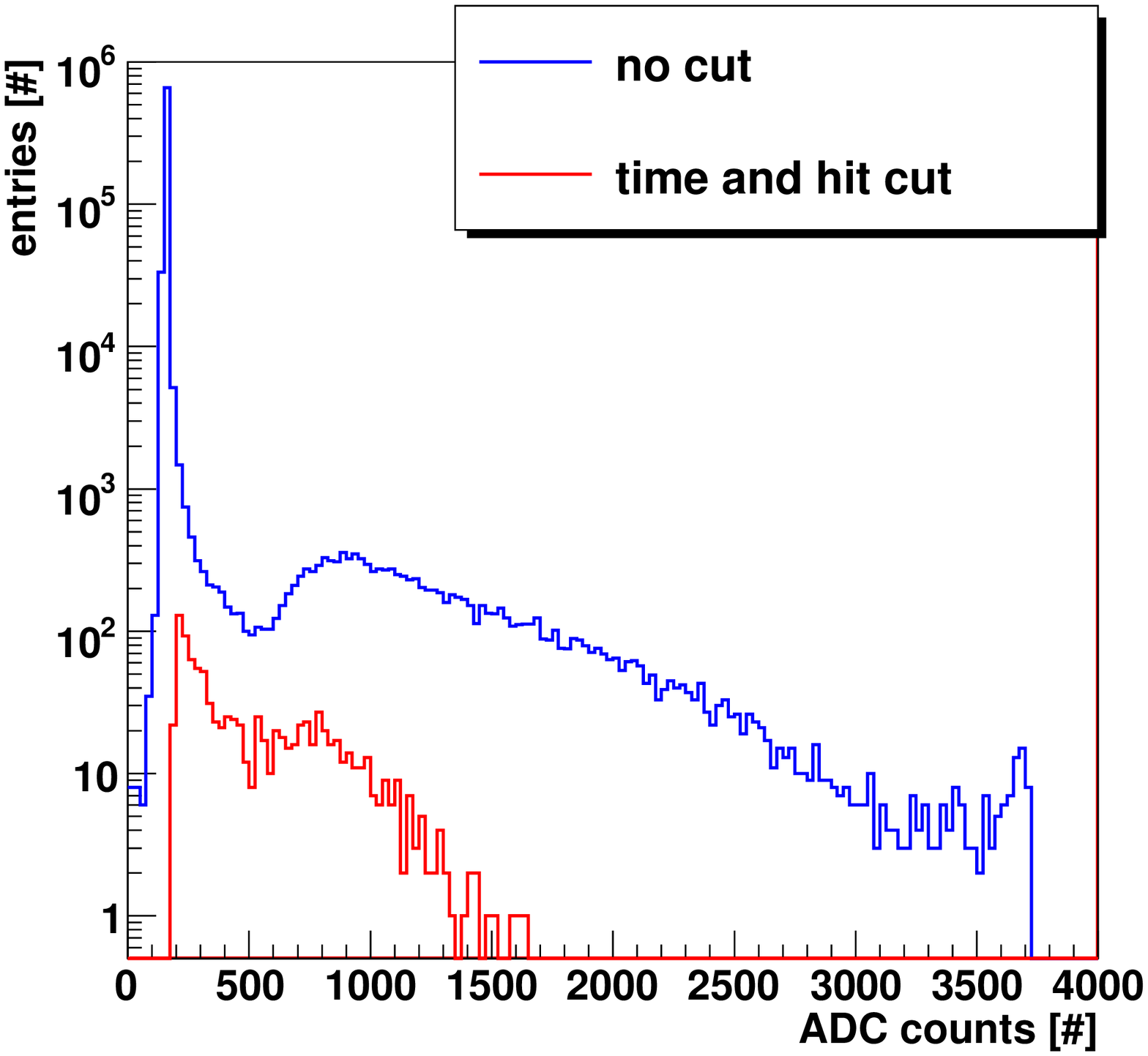,width=8cm}}\captionof{figure}{\label{f-1212576870_S3_sfea2_adc}Typical ADC spectra of PMTs without and with cuts (only one time stamp in the interval [8.76,8.78]\,\textmu s) connected to the SFEA2 cards in the S-crates.}
\end{minipage}
\hspace{.1\linewidth}
\begin{minipage}[b]{.4\linewidth}
\centerline{\epsfig{file=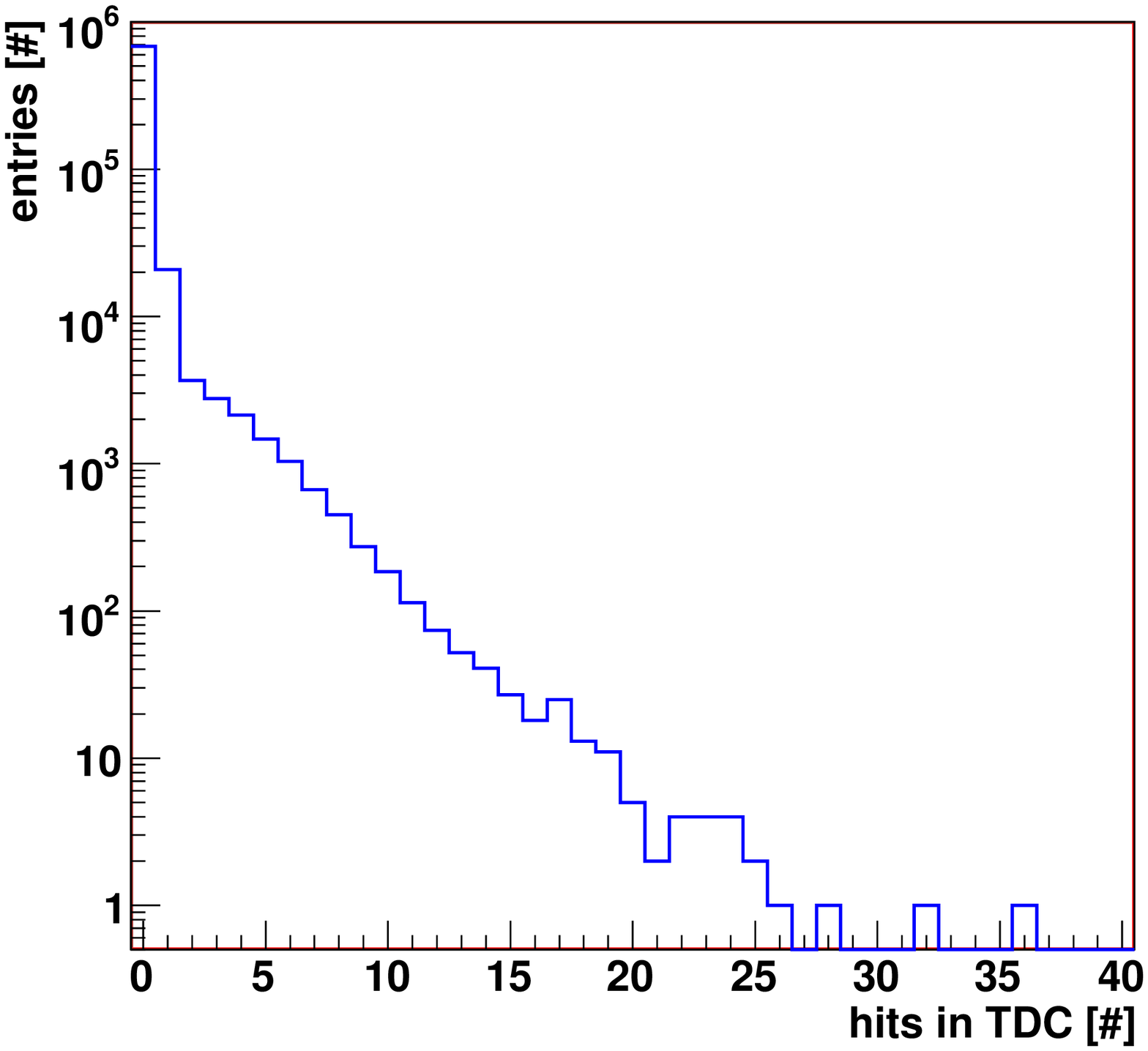,width=8cm}}\captionof{figure}{\label{1212576870_S3_sfea2_tdc_hits_no_cut}Typical number of time stamps in the TDC of PMTs connected to the SFEA2 card in the S-crates.}
\end{minipage}
\end{center}
\end{figure}

The collected raw data of all ACC PMTs is used to the calibrate the system. It is expected that the ACC PMTs collect pedestal entries most of the time because the muons have a distribution peaking at small zenith angles and the angular acceptance of the ACC in this direction is small \cite{biallass-2007}. A typical raw ADC spectrum of the PMTs connected to a flight module S-crate without any cuts is shown for one run in Fig.~\ref{f-1212576870_S3_sfea2_adc} (no cut). The raw data of all crates are used to adjust the voltages of the PMTs in order to get a homogeneous response of the detector with all MOP values at the same distance from the corresponding pedestal. The interpolated voltage values of the system test were taken as start values (Tab.~\ref{t-voltages}). The calibrated voltages for the ACC after pre-integration are shown in Tab.~\ref{t-cosmic_hv} together with the corresponding register settings for the control software. 

Fig.~\ref{1212576870_S3_sfea2_tdc_hits_no_cut} shows the number of TDC hits for a typical channel. As expected the PMT collects most of the time only pedestal such that the distribution peaks at 0. The second most common number of hits is 1 but also many events with more hits in the TDC exist. This number depends on the threshold which is here set to a value of 103 in the run control software and corresponds to 22\,mV (eq. \ref{e-thrmv}). A common PMT pulse has a maximum amplitude of 50 - 100\,mV. Hits above the threshold by PMT noise are rare because the threshold is about ten times larger than the pedestal width. After-pulses or additional pulses must be the explanation for the TDC hit multiplicity. Such pulses can also be seen on the oscilloscope where smaller spikes occur between the main pulses. 

\begin{table}
\begin{center}
\captionof{table}{\label{t-cosmic_hv}Voltage adjustment of the ACC PMTs and assignment to the S-crates and the corresponding SHV channels. The register setting is needed for the voltage adjustment in the user software.}
\begin{tabular}{c|c|c|c|c}
\hline
\hline
Crate	&PMT prod. no.	&SHV channel	&Register setting &Voltage [V]\\
\hline
S0&	8&	20&	858&	2098\\
S0&	12&	21&	851&	2082\\
S0&	10&	22&	863&	2111\\
S0&	1&	23&	838&	2051\\
\hline
S1&	18&	20&	797&	1950\\
S1&	13&	21&	904&	2211\\
S1&	15&	22&	890&	2177\\
S1&	19&	23&	762&	1863\\
\hline
S2&	6&	20&	833&	2036\\
S2&	4&	21&	894&	2185\\
S2&	9&	22&	881&	2155\\
S2&	17&	23&	811&	1983\\
\hline
S3&	11&	20&	821&	2007\\
S3&	21&	21&	918&	2245\\
S3&	14&	22&	871&	2130\\
S3&	7&	23&	808&	1976\\
\hline
\end{tabular}
\end{center}
\end{table}

\begin{figure}
\begin{center}
\begin{minipage}[b]{.4\linewidth}
\centerline{\epsfig{file=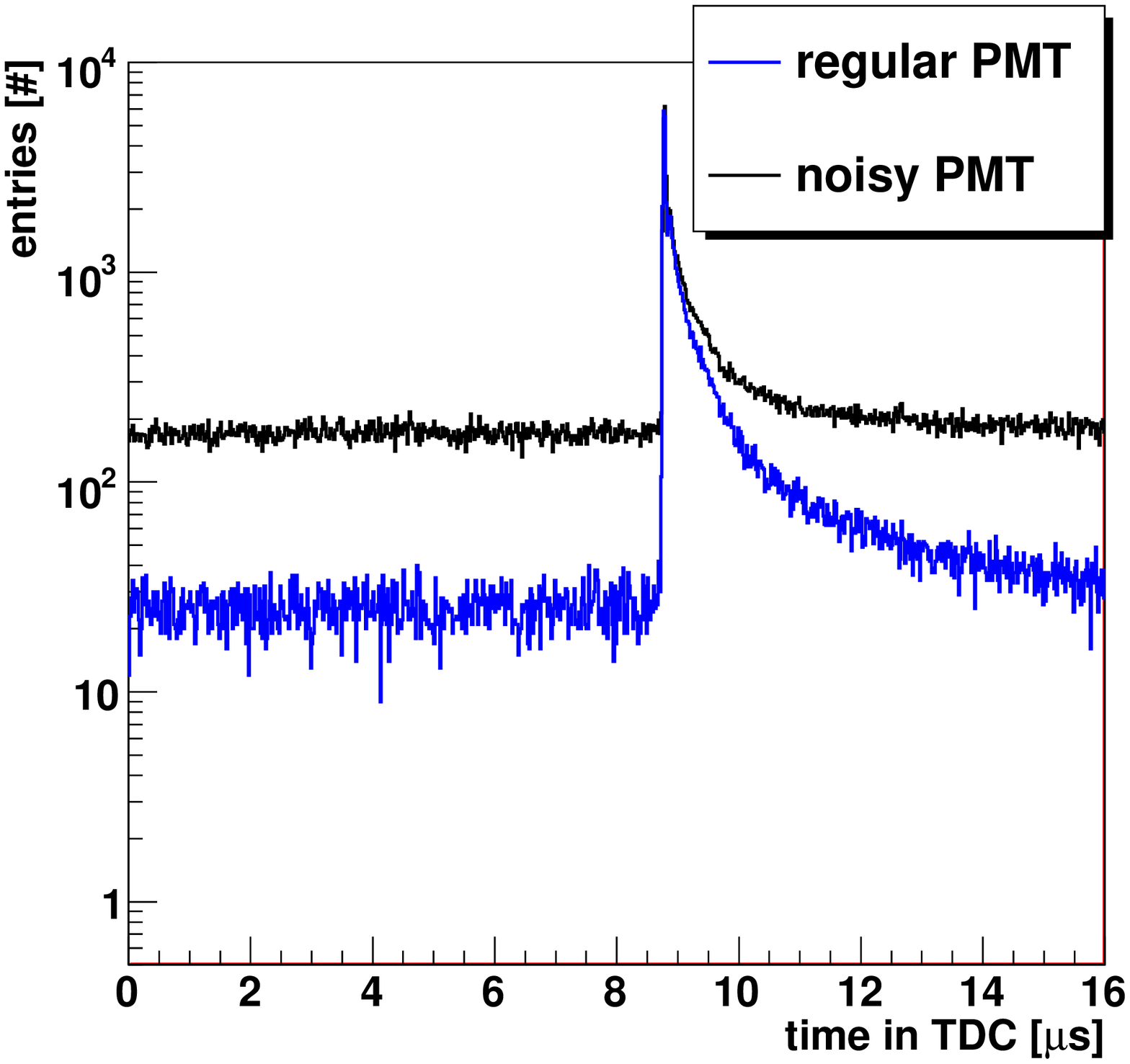,width=8cm}}\captionof{figure}{\label{f-1212576870_S3_sfea2_time_no_cut}A regular and a noisy distribution of the TDC time stamps of PMTs connected to the SFEA2 cards in the S-crates.}
\end{minipage}
\hspace{.1\linewidth}
\begin{minipage}[b]{.4\linewidth}
\centerline{\epsfig{file=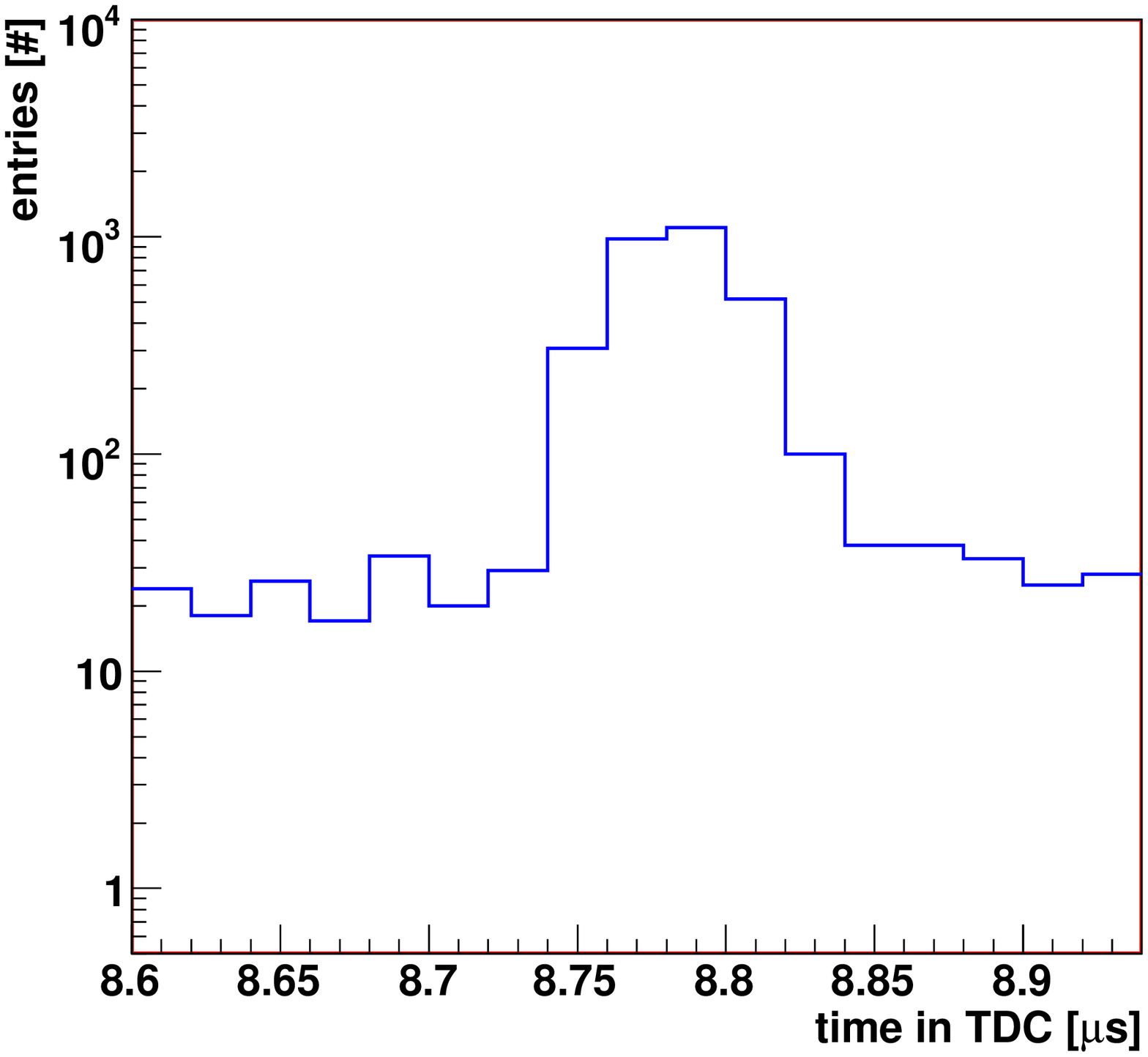,width=8cm}}\captionof{figure}{\label{f-1212576870_S3_sfea2_time_no_cut_1}Time in the TDC for events with only one time stamp of PMTs connected to the SFEA2 cards in the S-crates.}
\end{minipage}
\end{center}
\end{figure}

The after-pulses can also be seen in the time distribution by looking at events with more than 1 TDC hit (Fig.~\ref{f-1212576870_S3_sfea2_time_no_cut}, regular PMT). Before the level\,1 delay of about 8.7\,\textmu s no particles cross the ACC and again only an uniform pedestal distribution is seen. If a particle caused the TOF trigger and crossed the ACC, after-pulses of the ACC PMTs after the level\,1 trigger could occur resulting in multiple TDC hits. This is the cause of the falling edge in the distribution.

The TDC time distribution for events with exactly 1 hit in the TDC can be used to determine the trigger time window in which the main PMT pulse occurs (Fig.~\ref{f-1212576870_S3_sfea2_time_no_cut_1}). The level\,1 trigger is internally delayed by about 8.7\,\textmu s to be able to store information before and after the trigger. The distribution is uniform outside of the interval [8.7,8.84]\,ns where entries arise from noise in the PMTs with small ADC values. 

The saturation effect of the ADC which was observed with single test pulses before (Fig.~\ref{f-lin_inv} and \ref{f-all_cali}) is not visible for the case of multiple hits in the TDC (Fig.~\ref{f-1212576870_S3_sfea2_adc}, no cut). A distance in time between successive hits allows the preamplifier to recover partly and the Pouxe chip is able to collect a higher total charge from multiple hit events than for single pulses. When requiring only one TDC hit close to the peak of the TDC time distribution (Fig.~\ref{f-1212576870_S3_sfea2_time_no_cut_1}), the ADC distribution shows the saturation effect clearly (Fig.~\ref{f-1212576870_S3_sfea2_adc}, time and hit cut).

During the pre-integration data taking period the TDC time distribution for the S3-crate showed one channel with an increased number of entries in the complete time range (Fig.~\ref{f-1212576870_S3_sfea2_time_no_cut}, noisy PMT). This would strongly effect the level\,1 veto decision on the basis of this PMT (11) during flight such that it will be replaced by PMT\,2 with clear cable\,21 for the final integration.

\subsubsection{Reconstruction of TRD Tracks via the Tracker to the ACC}

\begin{figure}
\begin{center}
\centerline{\epsfig{file=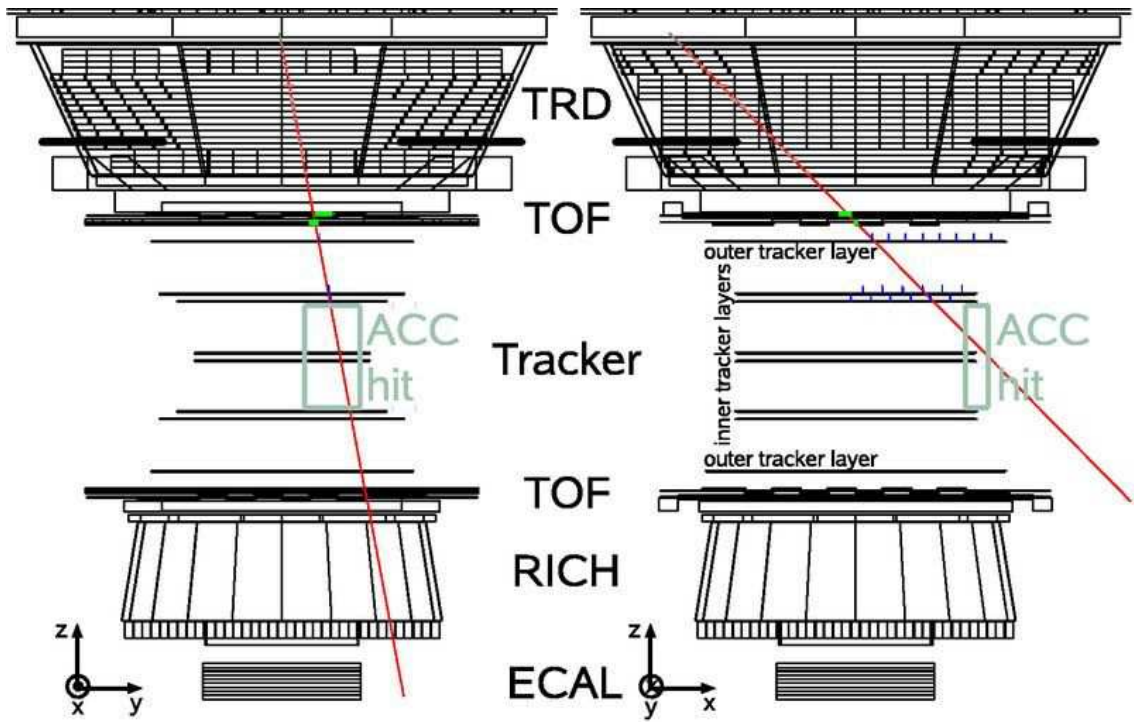,width=15cm}}\captionof{figure}{\label{f-eventdisplay_1209556604_33077}Clean event for the ACC analysis during the AMS-02 pre-integration runs with atmospheric muons. The outline of the ACC hit shows only a part of a panel.}
\end{center}
\end{figure}

\begin{figure}
\begin{center}
\captionof{table}{\label{t-cosmic_cuts}Event selection for AMS-02 runs with atmospheric muons after the pre-integration. $\Delta x$ and $\Delta y$ are the horizontal distances between tracks and hits. $\Delta\phi$ and $\Delta\theta$ are the differences between the direction angles of the TRD and the tracker track. Further explanations can be found in the text.}
\begin{tabular}{l|l|l}
\hline
\hline
Detector	& Criteria		& Cuts\\
\hline\hline
TOF		& trigger		& only both upper TOF layers have fired\\
\hline
TRD		& hits on track		& $\geq 3$ hits of the upper 4 layers within $\sqrt{\Delta x^2+\Delta y^2}<0.6$\,cm\\
		& 	   		& $\geq 10$ hits of the middle 12 layers within $\sqrt{\Delta x^2+\Delta y^2}<0.6$\,cm\\
		& 			& $\geq 3$ hits of the lower 4 layers within $\sqrt{\Delta x^2+\Delta y^2}<0.6$\,cm\\
		& noise			& ratio of total hits in TRD to hits on track $<1.5$\\
\hline
Tracker		& TRD road		& tracker hits within $\sqrt{\Delta x^2+\Delta y^2}<1.6$\,cm for tracker layer 1 \\
		&			& and within 4.6\,cm for layer 8 (linear interpolation in between)\\
		& fit			& exactly one rec. tracker hit in the first tracker layer\\
		&			& $\geq 3$  layers with exactly one reconstructed hit\\
		& goodness		& $\chi^2$ cut: $p\geq0.1$\\
		& TRD match		& direction angle matching: $|\Delta\phi|\leq47$\,mrad and $|\Delta\theta|\leq22$\,mrad\\		
		& position		& $-40.0\,\text{cm}<z<40.0\,\text{cm}$ on the ACC cylinder\\	
		& noise			& ratio of total rec. hits in tracker to hits on track $<1.3$\\
\hline
ACC		& detected event	& signal $\geq3\sigma$ (average pedestal RMS $\sigma = 7$\,ADC counts)\\
\hline
\end{tabular}
\end{center}
\end{figure}

\begin{figure}
\begin{center}
\begin{minipage}[b]{.4\linewidth}
\centerline{\epsfig{file=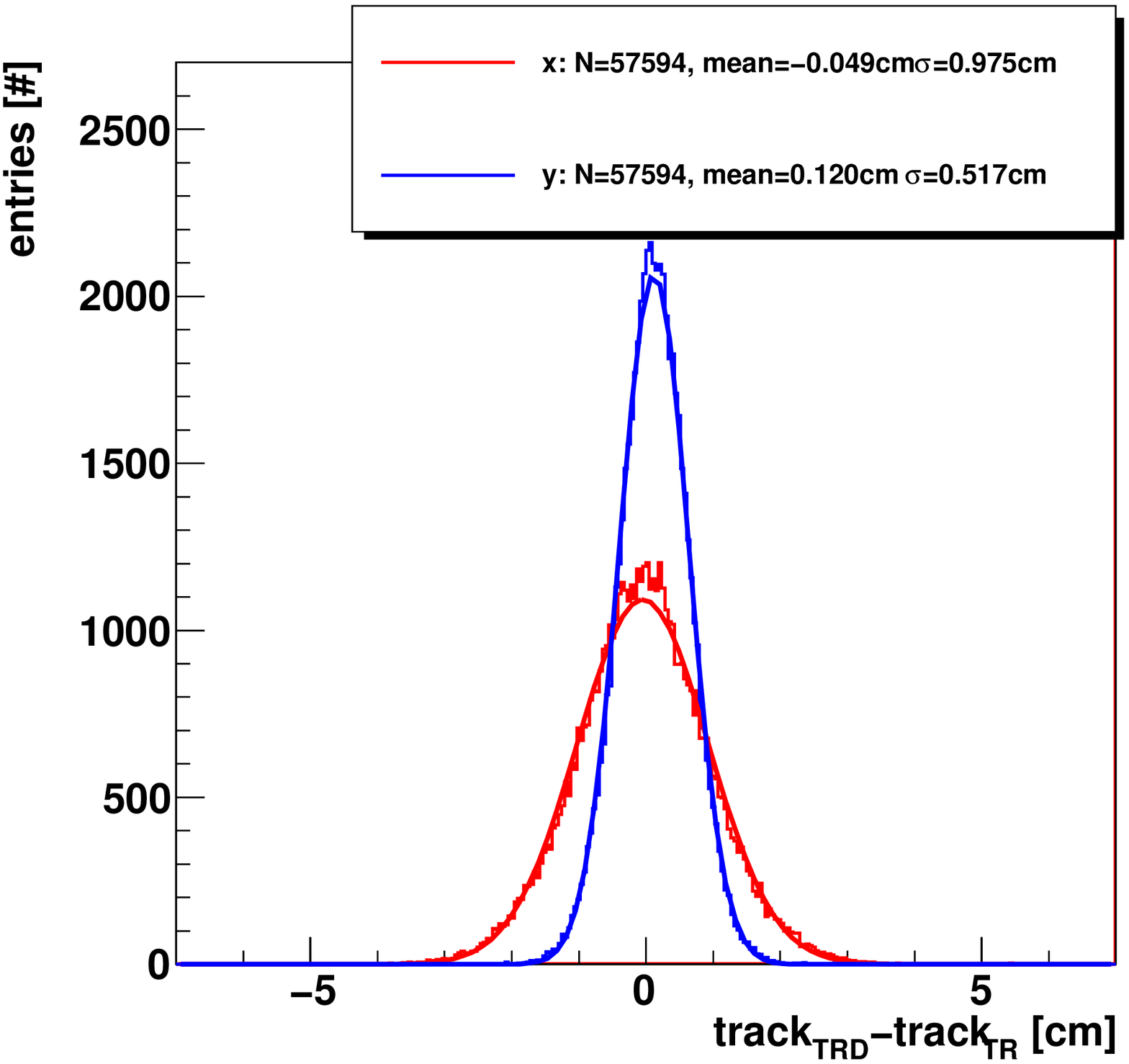,width=8cm}}\captionof{figure}{\label{f-090216_4_0_6_0_4_2_3_track_trd_tr_tr1}Distance between reconstructed tracker and TRD tracks at the $z$ position of tracker layer 1.}
\end{minipage}
\hspace{.1\linewidth}
\begin{minipage}[b]{.4\linewidth}
\centerline{\epsfig{file=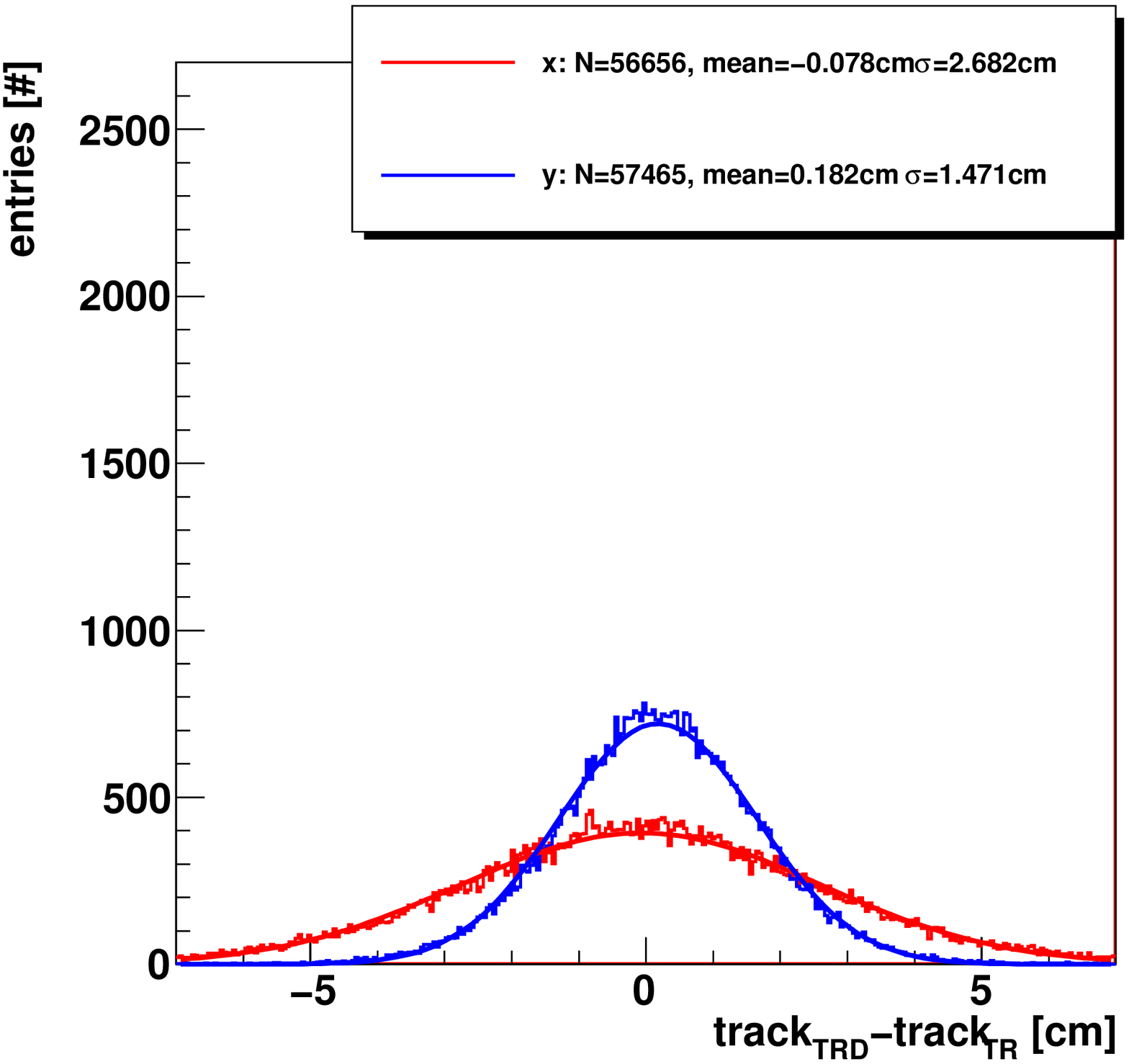,width=8cm}}\captionof{figure}{\label{f-090216_4_0_6_0_4_2_3_track_trd_tr_tr8}$x$, $y$ projections. Distance between reconstructed tracker and TRD tracks at the $z$ position of tracker layer 8.}
\end{minipage}
\end{center}
\end{figure}

\begin{figure}
\begin{center}
\begin{minipage}[b]{.4\linewidth}
\centerline{\epsfig{file=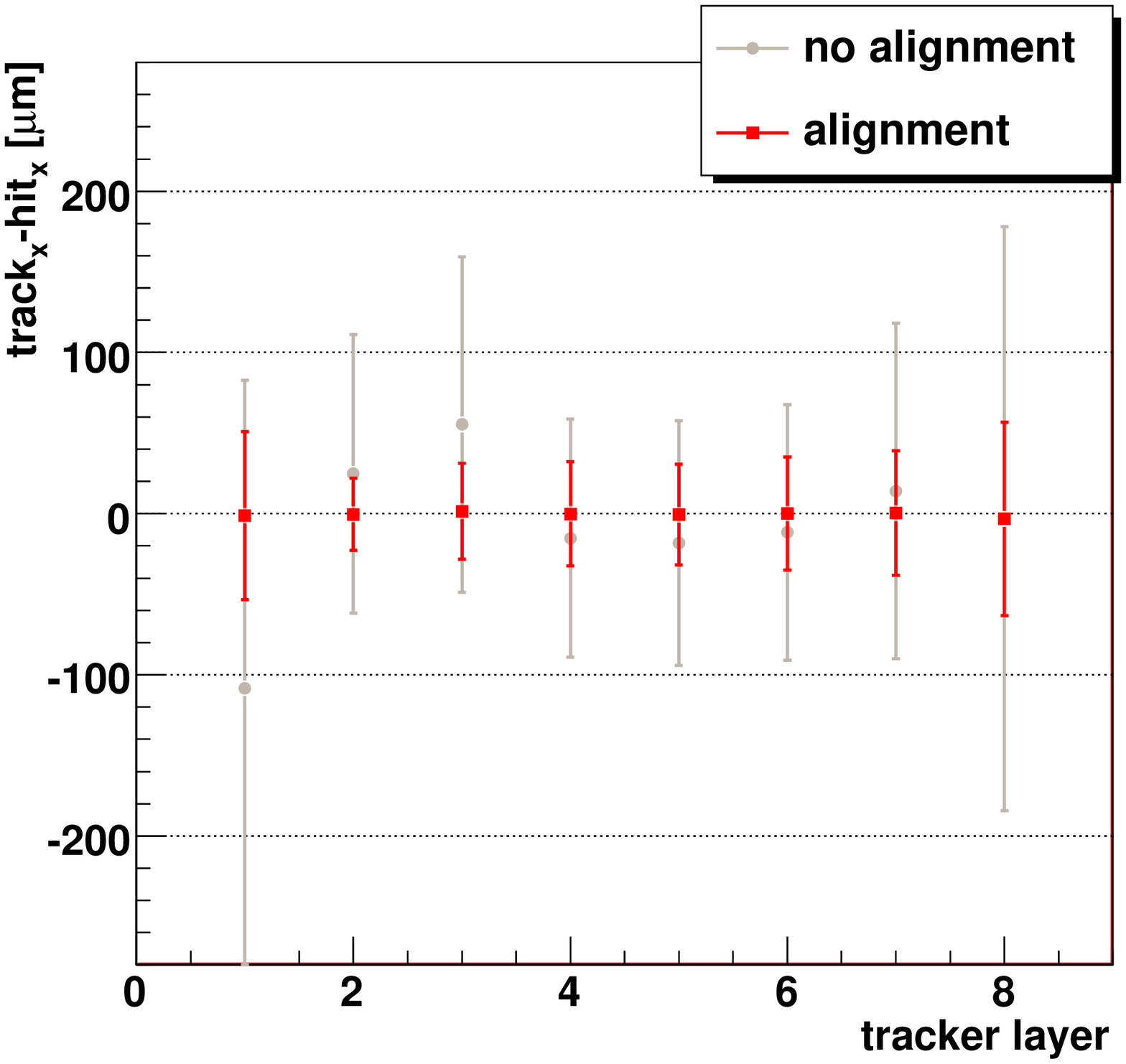,width=8cm}}\captionof{figure}{\label{f-track_acc_tr_mean_rms_x}Mean tracker residuals in $x$ direction before and after alignment with RMS as error bars.}
\end{minipage}
\hspace{.1\linewidth}
\begin{minipage}[b]{.4\linewidth}
\centerline{\epsfig{file=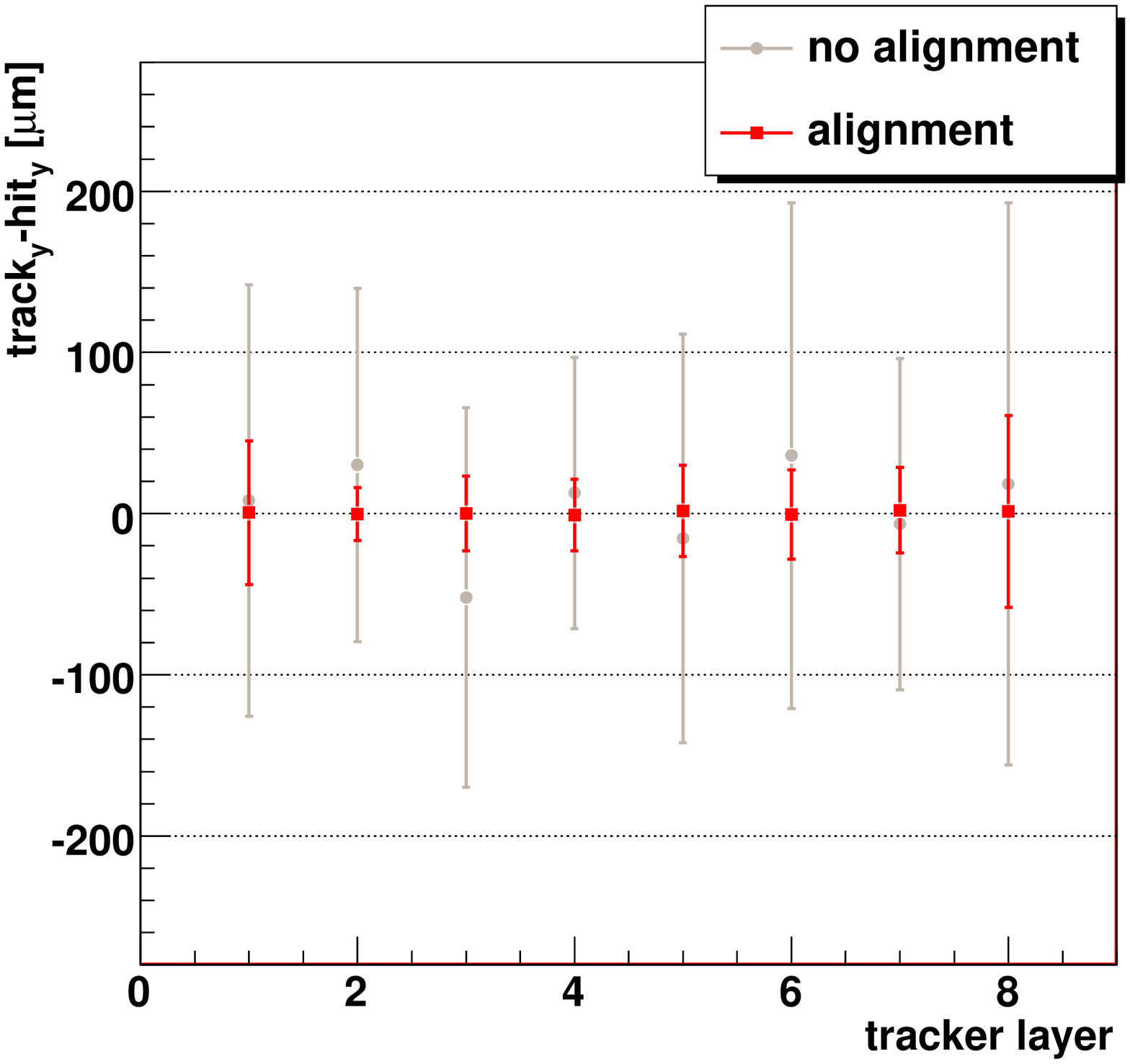,width=8cm}}\captionof{figure}{\label{f-track_acc_tr_mean_rms_y}Mean tracker residuals in $y$ direction before and after alignment with RMS as error bars.}
\end{minipage}
\end{center}
\end{figure}

In this analysis TRD and tracker tracks are extrapolated to the ACC in order to find the piercing point on the ACC to determine the ACC detection efficiency. From now on the reconstructed data is used for analysis. A new tracker fit was developed to find also tracks which do not cross the complete detector and point to the ACC because the track fit of the official AMS-02 software is optimized to find tracks crossing the complete tracker. The TRD track is used to set the seed for the new tracker fit for several reasons. The main purpose is to improve the spatial resolution of the extrapolated TRD track with the much higher tracker resolution for a reliable inefficiency study. Furthermore, the tracker readout in $x$ direction is multiplexed to save on power and payload weight and the requirement of a reasonable agreement between TRD and tracker track works as a momentum filter such that only particles without any significant interactions in the TOF or tracker are taken into account for analysis. The idea is now to extrapolate the TRD track to the tracker and define a road around it. The tracker hits within this road are used for a new track fit such so a reliable track with high spatial resolution can be achieved even with a small number of tracker hits.

Fig.~\ref{f-eventdisplay_1209556604_33077} shows a typical event used in the following analysis. In addition, the AMS-02 coordinate system is shown. The origin is located at the center of the tracker. The axis of symmetry of the ACC cylinder is the same as the $z$ axis of the AMS-02 coordinate system. Events selected for analysis satisfy the requirements listed in Tab.~\ref{t-cosmic_cuts} and are discussed below.

The selected runs are from the pre-integration data taking period. As mentioned above, the data taking was used for intensive testing of all subdetectors. The experimental conditions were sometimes changed from run to run and as a result not all runs are useful for the ACC analysis. Reasons to exclude runs in the following are e.g. trigger studies, data acquisition and setting problems. At the end of the data taking period AMS-02 was rotated. These runs are also included in this analysis.

\begin{figure}
\begin{center}
\begin{minipage}[b]{.24\linewidth}
\centerline{\epsfig{file=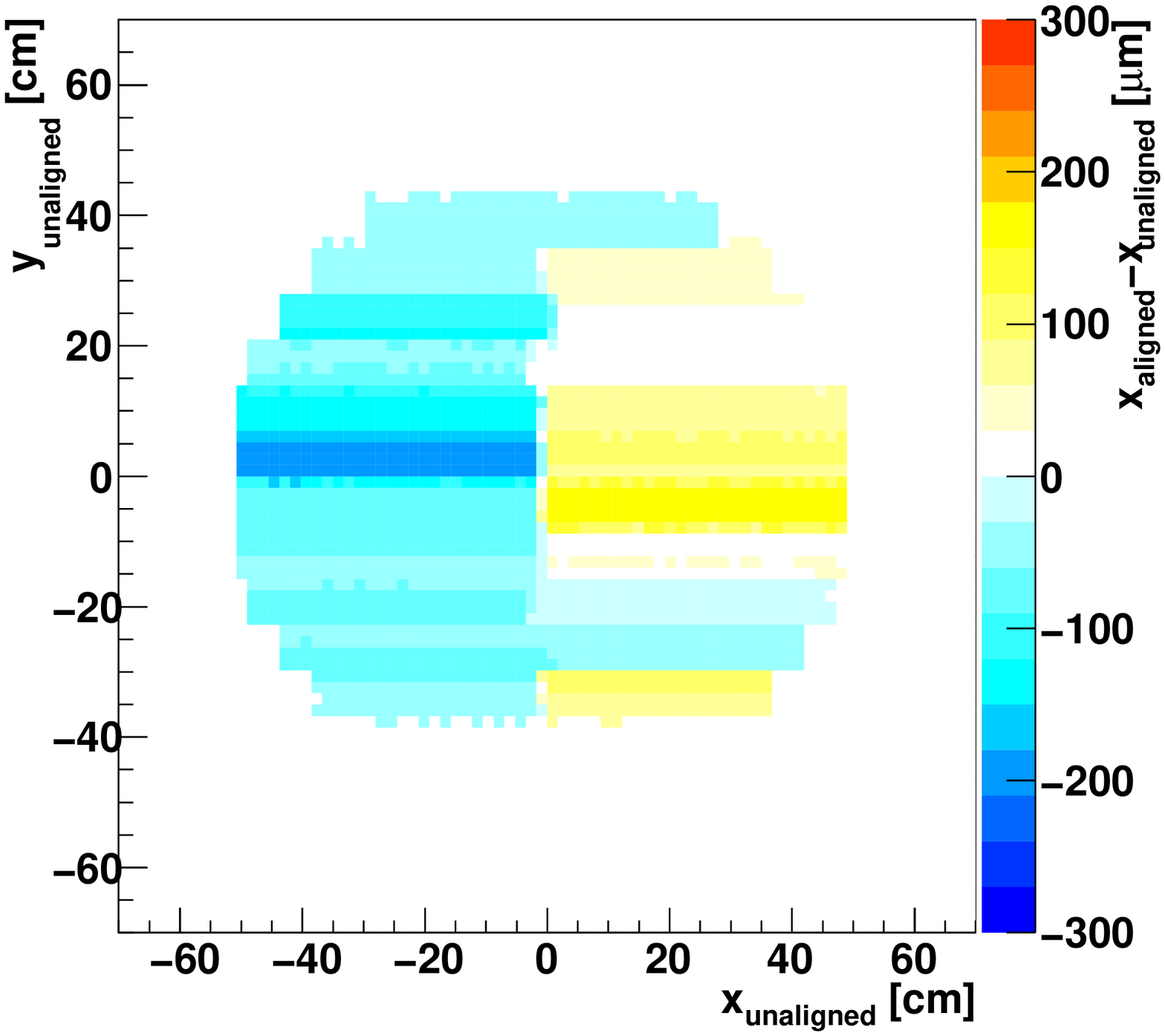,width=5cm}}\captionof{figure}{\label{f-shift_x_6}Alignment shift in $x$ direction for tracker hits in layer 6.}
\end{minipage}
\hspace{.1\linewidth}
\begin{minipage}[b]{.24\linewidth}
\centerline{\epsfig{file=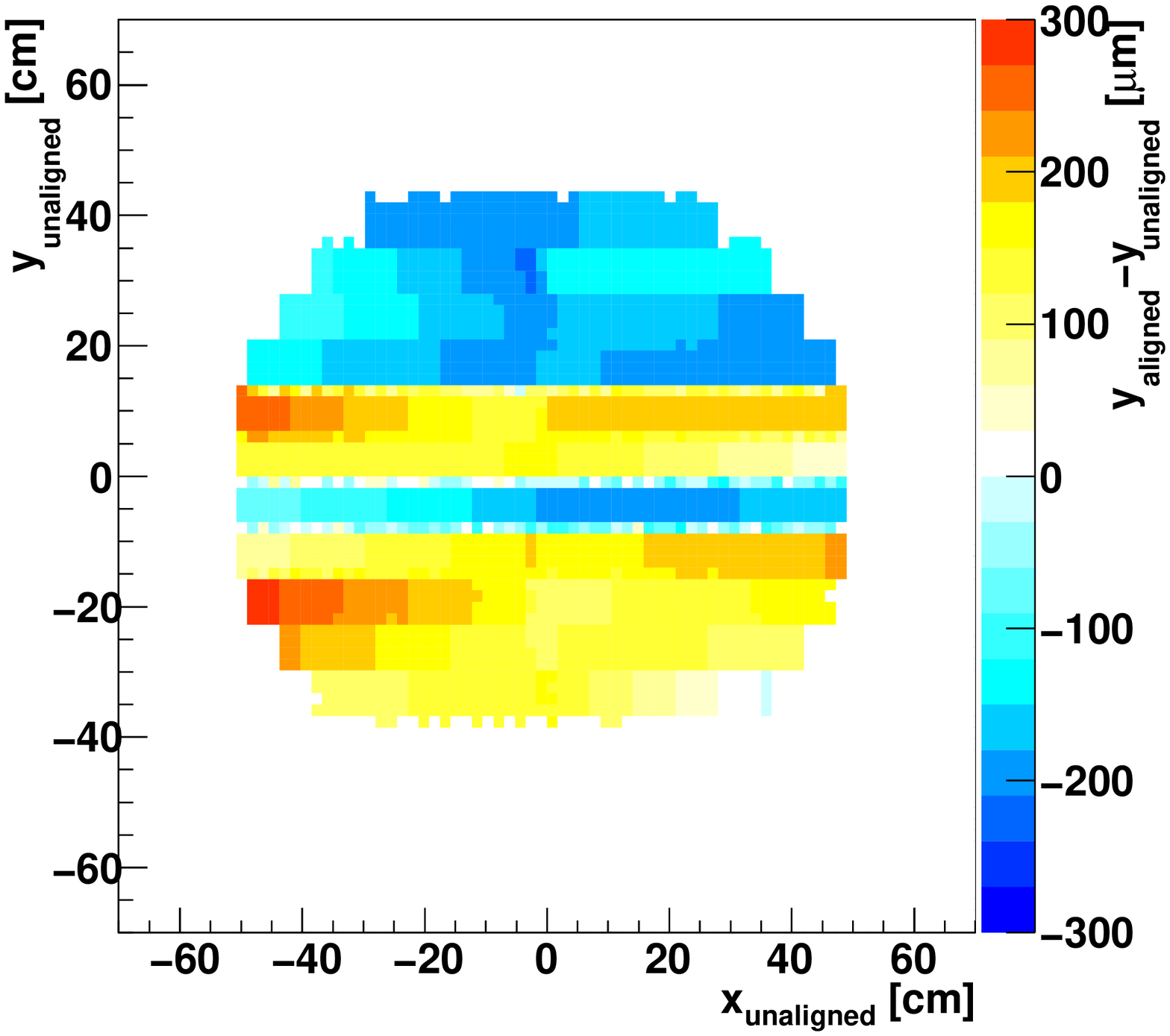,width=5cm}}\captionof{figure}{\label{f-shift_y_6}Alignment shift in $y$ direction for tracker hits in layer 6.}
\end{minipage}
\hspace{.1\linewidth}
\begin{minipage}[b]{.24\linewidth}
\centerline{\epsfig{file=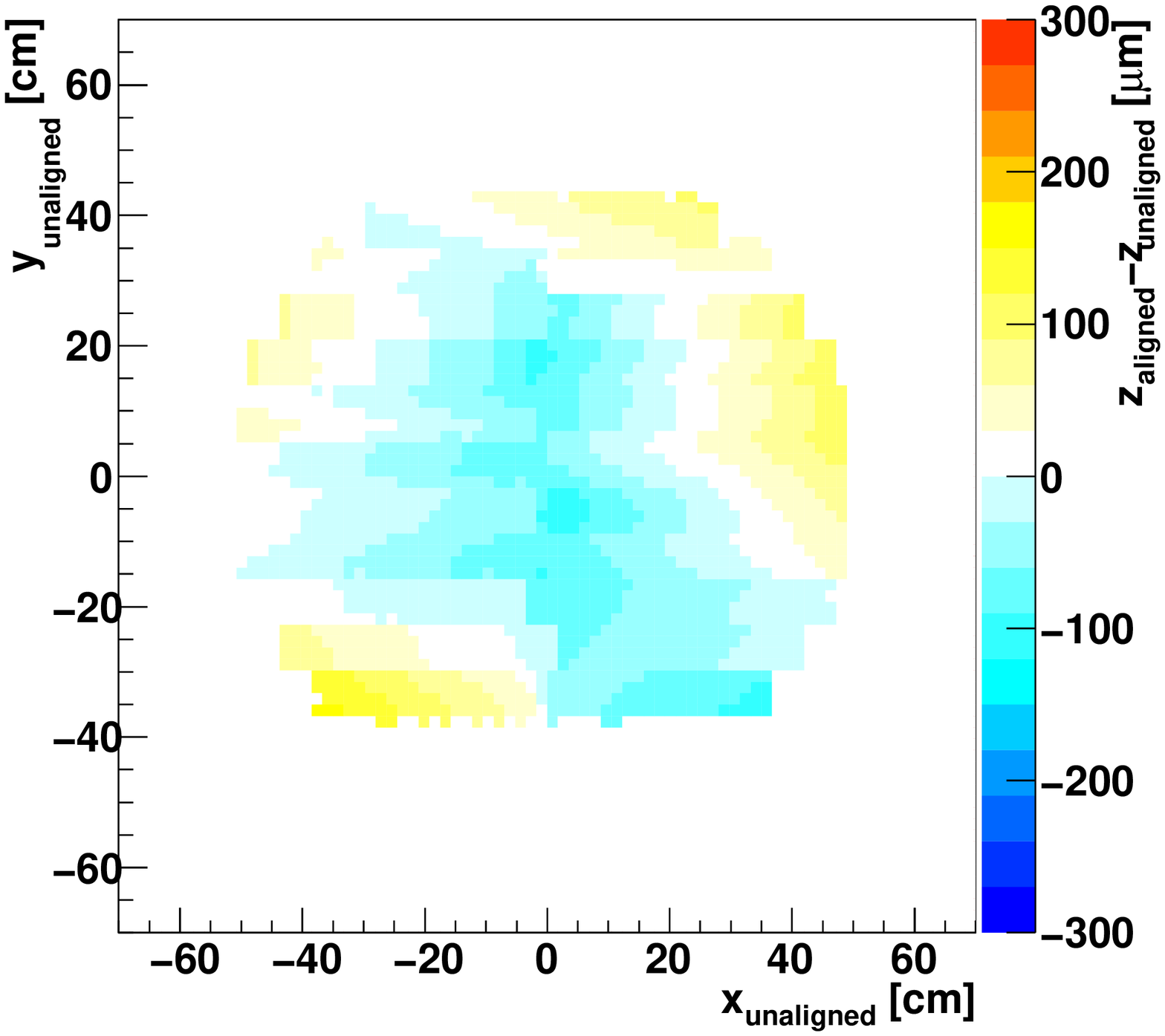,width=5cm}}\captionof{figure}{\label{f-shift_z_6}Alignment shift in $z$ direction for tracker hits in layer 6.}
\end{minipage}
\end{center}

\begin{center}
\begin{minipage}[b]{.4\linewidth}
\centerline{\epsfig{file=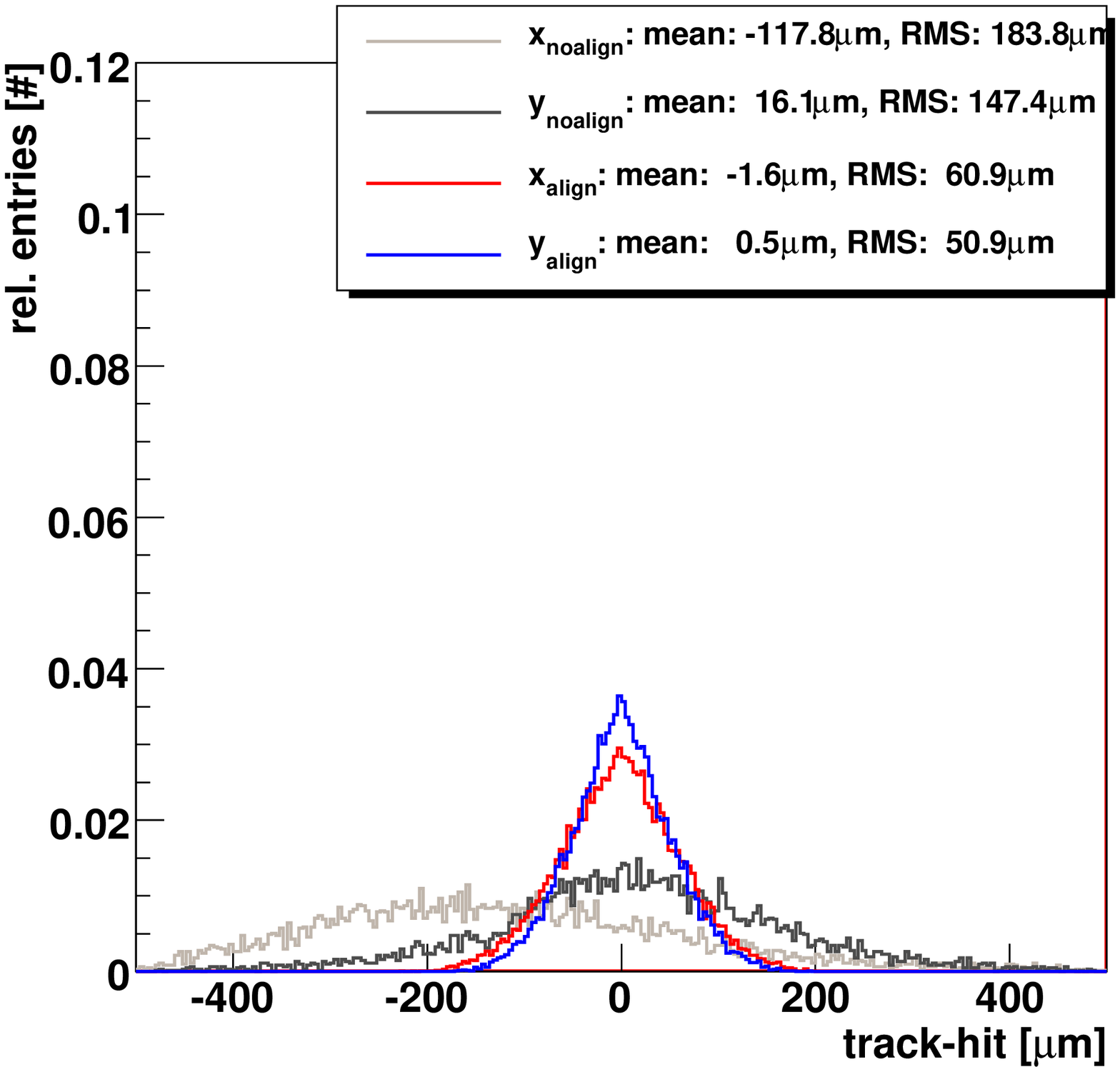,width=8cm}}\captionof{figure}{\label{f-track_straight_xy_compare_1_ctr_fit_8}Tracker residuals in $x$ and $y$ direction before and after alignment in tracker layer 1.}
\end{minipage}
\hspace{.1\linewidth}
\begin{minipage}[b]{.4\linewidth}
\centerline{\epsfig{file=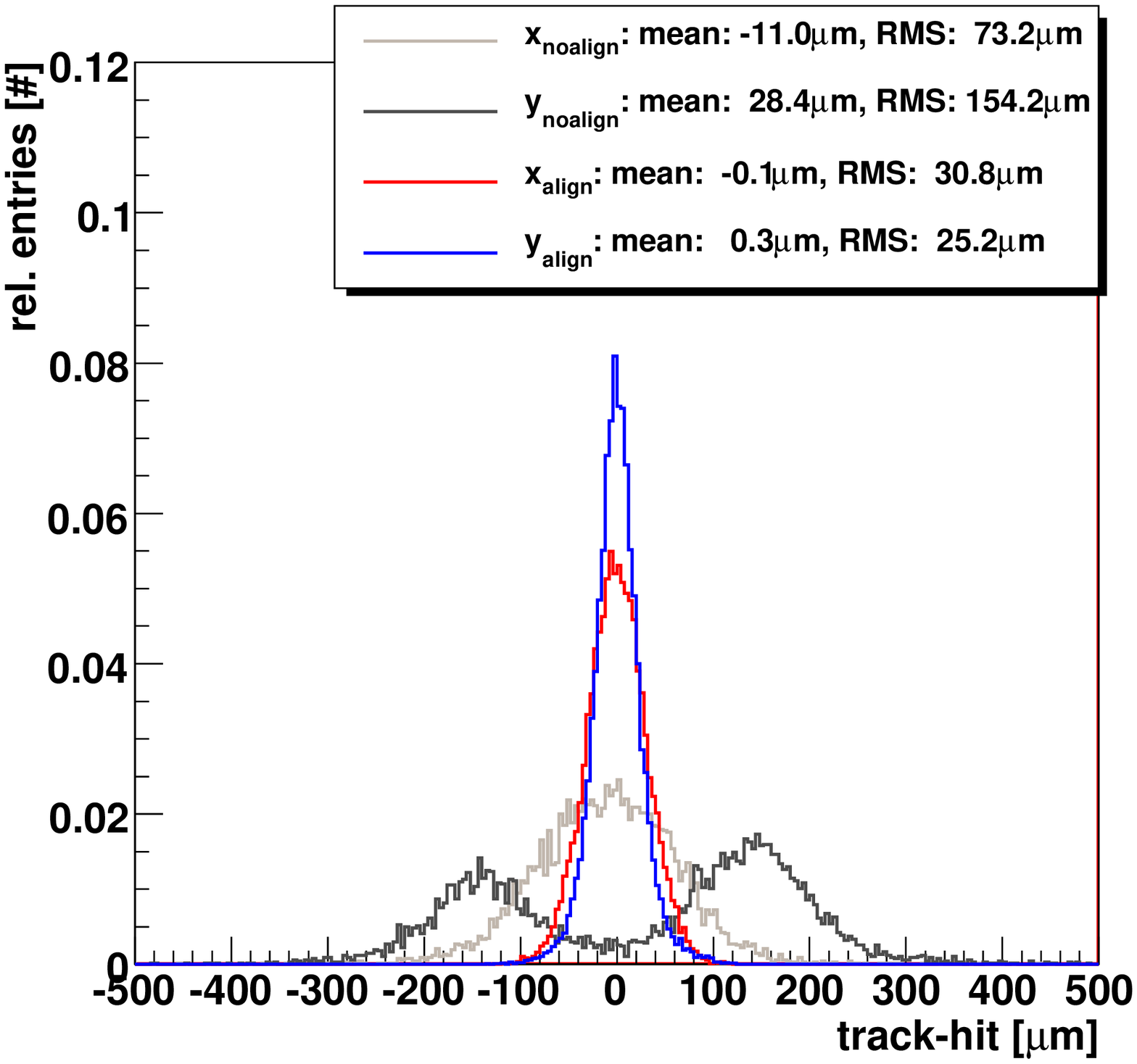,width=8cm}}\captionof{figure}{\label{f-track_straight_xy_compare_6_ctr_fit_8}Tracker residuals in $x$ and $y$ direction before and after alignment in tracker layer 6.}
\end{minipage}
\end{center}
\end{figure}

\begin{figure}
\begin{center}
\begin{minipage}[b]{.4\linewidth}
\centerline{\epsfig{file=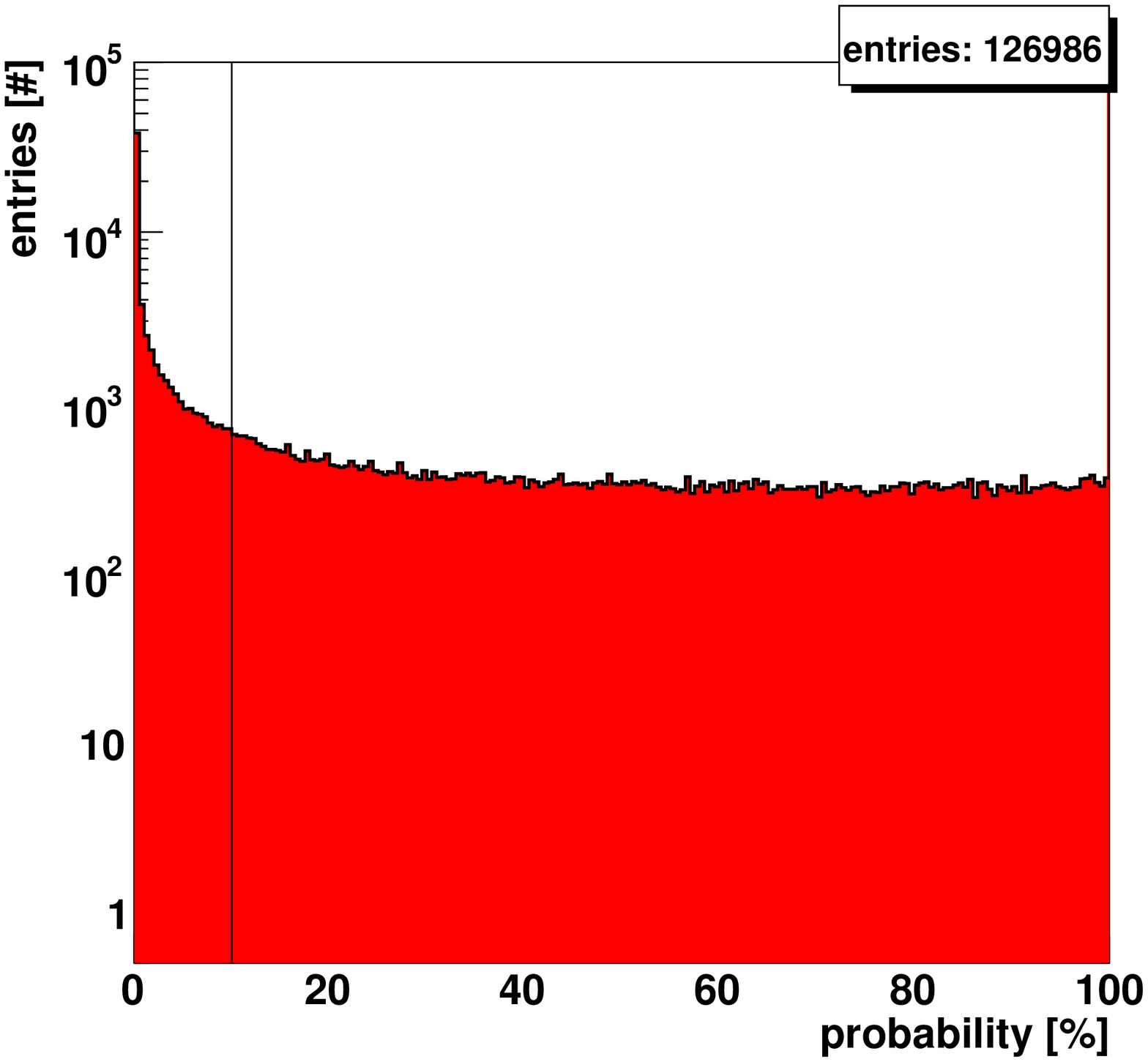,width=8cm}}\captionof{figure}{\label{f-090216_4_0_6_0_4_2_3_track_acc_cumchi2}Cumulative $\chi^2$ distribution. The vertical line indicates the applied cut.}
\end{minipage}
\hspace{.1\linewidth}
\begin{minipage}[b]{.4\linewidth}
\centerline{\epsfig{file=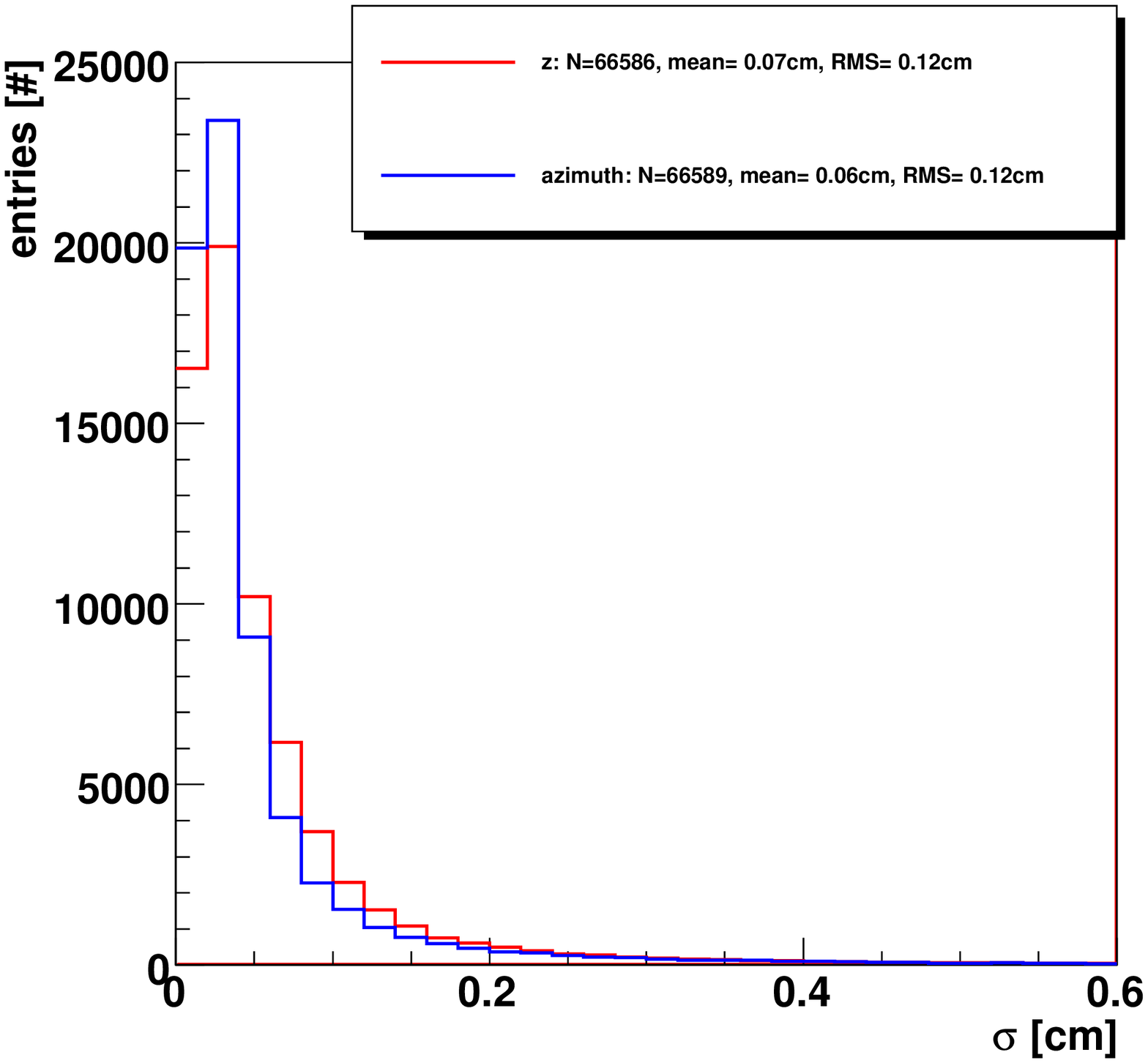,width=8cm}}\captionof{figure}{\label{f-090216_4_0_6_0_4_2_3_track_acc_error_prop}Distribution of propagated errors along and across an ACC panel.}
\end{minipage}
\end{center}
\end{figure}

In most runs a trigger is given by any two out of the four TOF planes. The trigger condition for events used in the analysis show hits in both upper TOF layers only to avoid e.g. events with multiple scattering of particles that crossed the vacuum tank before striking TOF or ACC which could result in an unclean event topology. The analysis starts with the TRD track \cite{siedenbu} which is defined by requiring that at least three out of the four upper layers of the TRD, at least ten of the twelve intermediate layers and again at least three of the four lower layers have hits on the track. A hit is considered to be on the track if its horizontal distance $\sqrt{\Delta x^2+\Delta y^2}$ from the track is less than 0.6\,cm (= straw tube diameter).

Before attempting to extrapolate the TRD track to the tracker, the alignment of the two subdetectors is checked by extrapolating the TRD track and the tracker track, reconstructed with the official AMS-02 software package \cite{choutko}, to the first and last tracker layer and examining their distance at these positions. The distribution of distances in $x$ and $y$ are shown in  Fig.~\ref{f-090216_4_0_6_0_4_2_3_track_trd_tr_tr1} and Fig.~\ref{f-090216_4_0_6_0_4_2_3_track_trd_tr_tr8}, respectively. As expected, the distributions are wider for the last layer than for the first layer due to the longer track. Now, the TRD road in the tracker is defined with the help of the mean standard deviations $\bar\sigma$ of the residual distributions \be\bar\sigma=\sqrt{\sigma_x^2+\sigma_y^2}.\ee The new tracker fit uses only reconstructed tracker hits within 1.5\,$\bar\sigma$ of the TRD track in the corresponding layer. In the first layer the tracks agree within 0.5 - 1\,mm. These small shifts of the TRD with respect to the tracker are not important for the following analysis.

Alignment of the tracker is the next step before carrying out the new track fit. The gray data points in Fig.~\ref{f-track_acc_tr_mean_rms_x} and \ref{f-track_acc_tr_mean_rms_y} show the mean residuals in $x$ and $y$ direction for each tracker layer for straight line track fits of particles crossing the upper and lower TOF planes without additional alignment. The error bars indicate the RMS of the residual distributions. Mean residuals of up to 100\,\textmu m and RMS up to 200\,\textmu m are seen. This is well compatible with the tracker integration precision of about 100\,\textmu m. The tracker consists of silicon wavers which are mounted with a precision of about 10\,\textmu m to ladders but the ladders are mounted to planes with about 100\,\textmu m precision such that the alignment of the tracker can be based on the alignment of the ladders on each plane. Here, it is extracted from the tracks made with the AMS-02 software. The mean differences between aligned and unaligned coordinates as a function of the unaligned coordinates $x$ and $y$ are shown exemplarily for tracker layer 6 in Fig.~\ref{f-shift_x_6} - \ref{f-shift_z_6}. The ladder structure of the layer is clearly seen. The corresponding alignment shifts for all layers are applied to all reconstructed tracker hits. This results in the red data points for a new straight line fit with aligned hits in Fig.~\ref{f-track_acc_tr_mean_rms_x} and \ref{f-track_acc_tr_mean_rms_y}. The mean values are corrected to nearly 0\,\textmu m and the RMS are drastically reduced. The mean RMS in $x$ direction is reduced from 112\,\textmu m to 36\,\textmu m and in $y$ direction from 126\,\textmu m to 31\,\textmu m. In addition, the residual distributions in Fig.~\ref{f-track_straight_xy_compare_1_ctr_fit_8} and \ref{f-track_straight_xy_compare_6_ctr_fit_8} show the effect of the alignment even clearer. The distributions are centered at 0 and the RMS are much smaller.

The new straight line fits are performed independently for the $(x,z)$ and $(y,z)$ plane and the track parameters are defined by:
\be z=m_xx+a_x\quad\text{and}\quad z=m_yy+a_y\ee
where $m_{x/y}$ are the slopes and $a_{x/y}$ are the $x/y$ intercepts with their corresponding errors $\sigma_{m,x/y}$ and $\sigma_{a,x/y}$ resulting from the fit. The fit uses the individual RMS for each layer of the aligned tracker as errors. Furthermore, the fit requires at least three different layers with only one reconstructed hit. Due to cooling problems the upper half of the tracker was not powered at all times during data taking, therefore it is important to require always exactly one hit in the first tracker layer for a reliable track fit. The track is considered to be good if the requirement $p\geq0.1$ is fulfilled. The $p$-value is defined as:\be p=\int_{\chi^2}^\infty f(t,n)\text{d}t\ee with Pearson's $\chi^2$ statistic, the number of degrees of freedom $n$ and the $\chi^2$ probability density function $f(\chi^2,n)$ \cite{cowan}. From purely statistics it is expected that the $p$-value distribution is uniform. A peak at $p=0$ corresponds to too many large $\chi^2$ values and is not in agreement with statistical fluctuations. For track quality reasons only the uniform part of the distribution is taken into account (Fig.~\ref{f-090216_4_0_6_0_4_2_3_track_acc_cumchi2}). Good tracks should also obviously satisfy $|z|<40$\,cm when extrapolated to the mean ACC radius of 54.95\,cm. 

\begin{figure}
\begin{center}
\begin{minipage}[b]{.4\linewidth}
\centerline{\epsfig{file=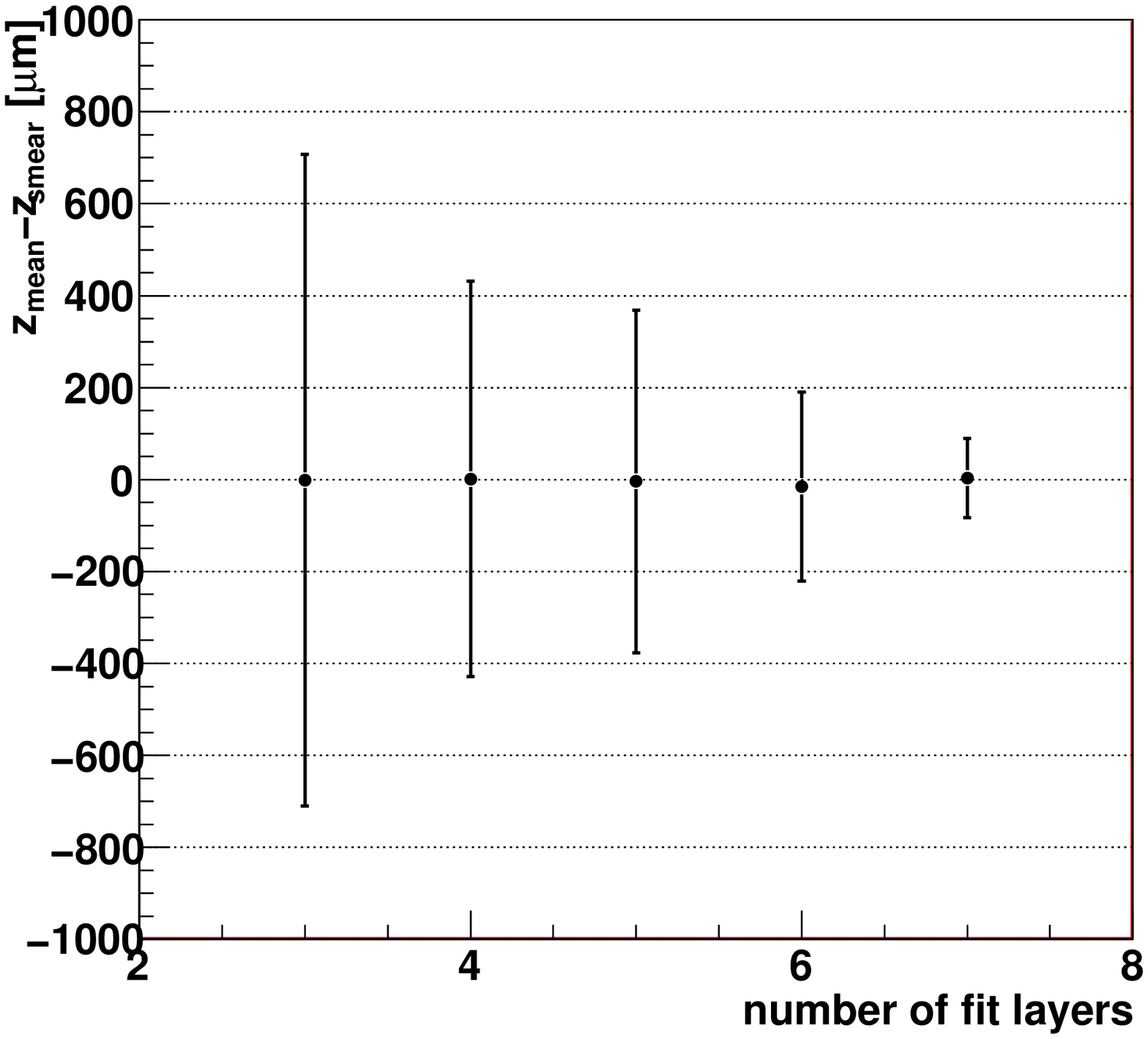,width=8cm}}\captionof{figure}{\label{f-090216_4_0_6_0_4_2_3_track_acc_z_ctr_fit}Difference between $z\sub{mean}$ and $z\sub{smear}$ extrapolated from the (smeared) track on the ACC cylinder.}
\end{minipage}
\hspace{.1\linewidth}
\begin{minipage}[b]{.4\linewidth}
\centerline{\epsfig{file=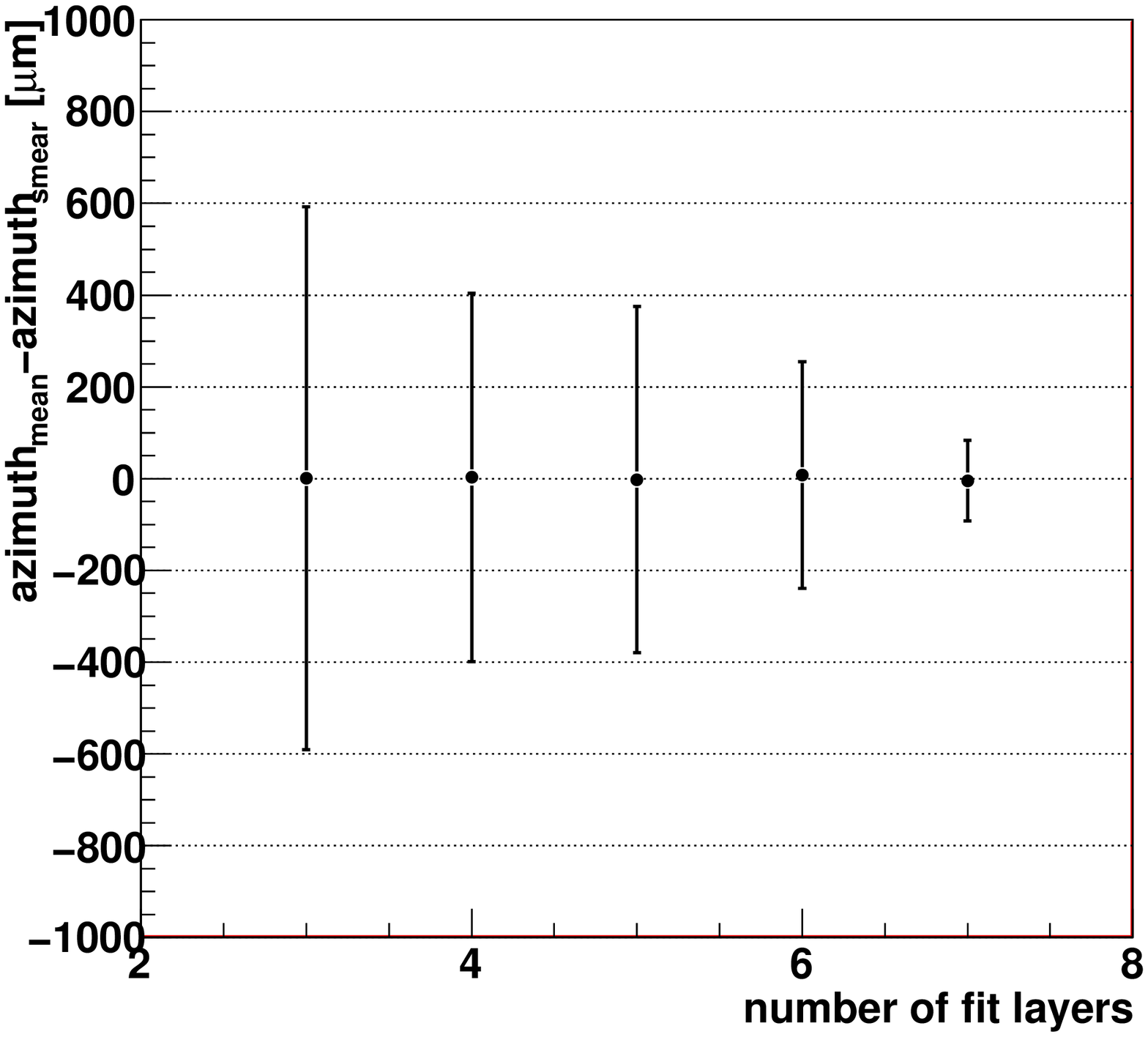,width=8cm}}\captionof{figure}{\label{f-090216_4_0_6_0_4_2_3_track_acc_phi_ctr_fit}Difference between mean azimuthal angle and azimuthal angle from the (smeared) track on the ACC cylinder.}
\end{minipage}
\end{center}
\end{figure}

\begin{figure}
\begin{center}
\begin{minipage}[b]{.4\linewidth}
\centerline{\epsfig{file=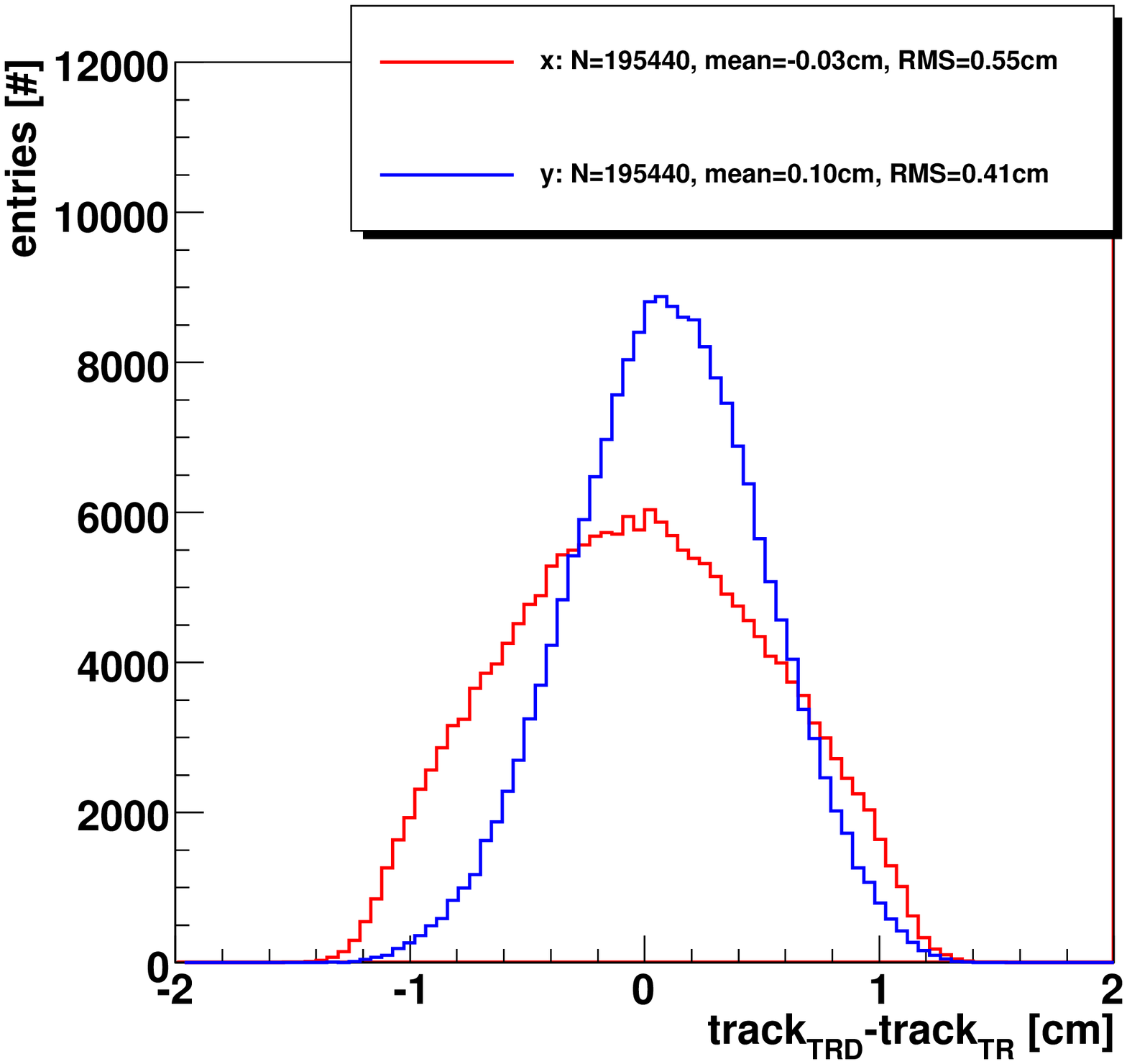,width=8cm}}\captionof{figure}{\label{f-090216_4_0_6_0_4_2_3_track_trd_tr_xy}Distance between TRD and new tracker tracks at the mean position of the upper TOF ($z=62$\,cm).}
\end{minipage}
\hspace{.1\linewidth}
\begin{minipage}[b]{.4\linewidth}
\centerline{\epsfig{file=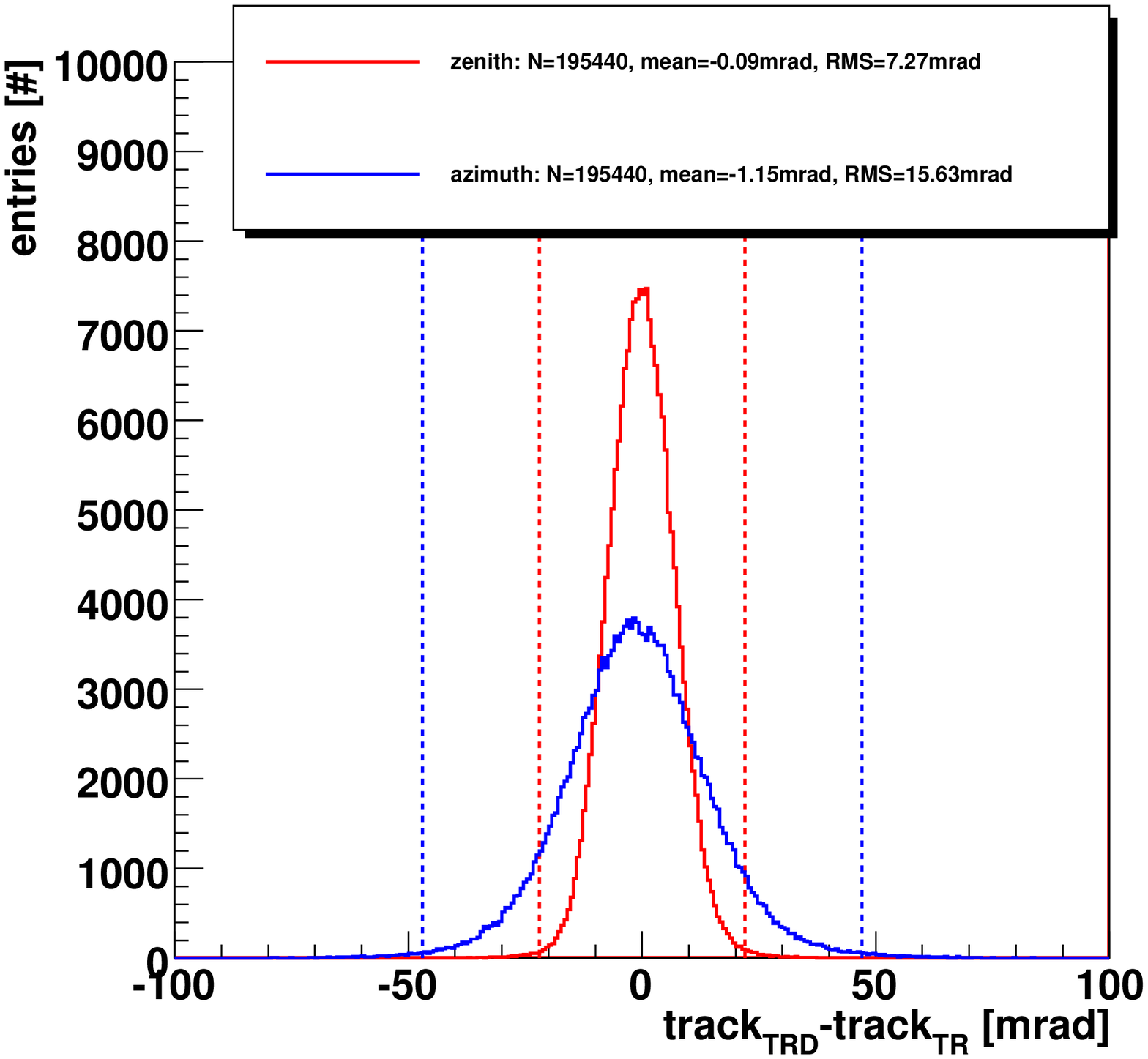,width=8cm}}\captionof{figure}{\label{f-track_trd_tr_thetaphi}Difference of direction angles between TRD and new tracker tracks at the mean position of the upper TOF ($z=62$\,cm). The lines indicate the event selection cuts.}
\end{minipage}
\end{center}
\end{figure}

The coordinates on the ACC cylinder ($z\sub{mean}$: position along the cylinder, $\phi\sub{mean}$: azimuthal angle) are extrapolated from the track. The distributions of propagated errors in $z$ and azimuth angle on the ACC cylinder show mean values of about 700\,\textmu m (Fig.~\ref{f-090216_4_0_6_0_4_2_3_track_acc_error_prop}). The position resolution on the ACC cylinder can also be extracted by randomly smearing the track parameters $m_{x/y}$ and $a_{x/y}$ with a Gaussian distribution using the corresponding errors. This results in a smeared track with the corresponding coordinates $z\sub{smear}$ and $\phi\sub{smear}$ on the ACC cylinder. Mean and RMS of the residual distributions $z\sub{mean}-z\sub{smear}$ and $\phi\sub{mean}-\phi\sub{smear}$ as a function of the number of fit layers are shown in Fig.~\ref{f-090216_4_0_6_0_4_2_3_track_acc_z_ctr_fit} and \ref{f-090216_4_0_6_0_4_2_3_track_acc_phi_ctr_fit}, respectively. As expected, the RMS gets smaller with increasing number of fit layers. Most particles traverse three tracker layers before hitting the ACC and the RMS for three tracker layers is compatible with the propagated errors on the ACC cylinder (600 - 800\,\textmu m). The residuals of the TRD track and the new tracker track in $x$ and $y$ direction at the mean upper TOF position are shown in Fig.~\ref{f-090216_4_0_6_0_4_2_3_track_trd_tr_xy}. The distributions are centered at $x=-0.03$\,cm and $y=0.10$\,cm. The differences in direction angles $\phi$ and $\theta$ are used to apply a quality cut at 3 RMS for both angles (47\,mrad and 22\,mrad, respectively) to reject particles with large direction angle changes due to interactions (Fig.~\ref{f-track_trd_tr_thetaphi}).

\begin{figure}
\begin{center}
\begin{minipage}[b]{.4\linewidth}
\centerline{\epsfig{file=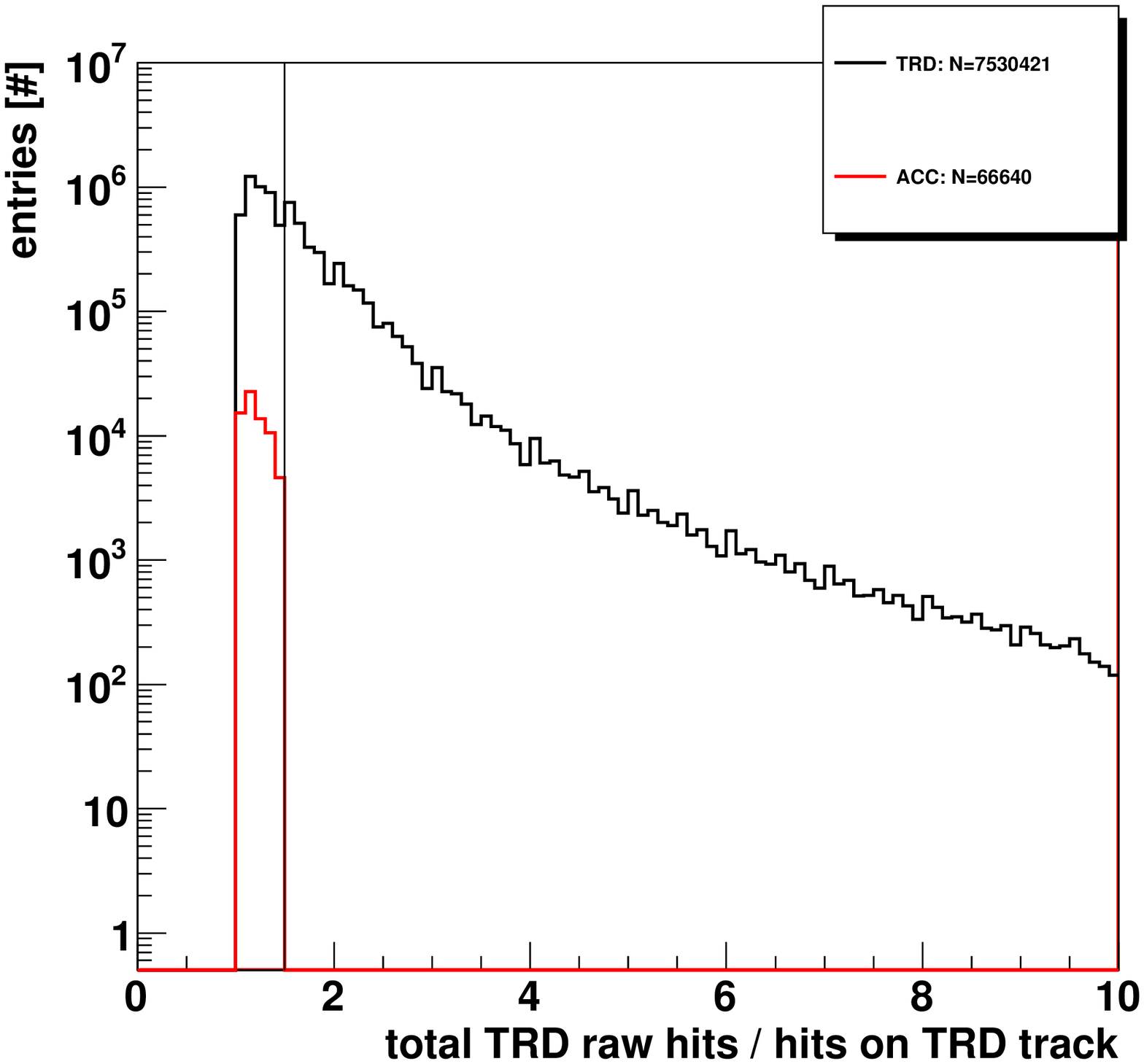,width=8cm}}\captionof{figure}{\label{f-090216_4_0_6_0_4_2_3_trd_hitratio} Ratio of raw TRD hits to hits on TRD track. The black distribution shows the ratio values for TRD tracks pointing to the ACC matching the run, TOF and TRD cuts in Tab.~\ref{t-cosmic_cuts} but not the TRD noise requirement. The red ratio distribution shows the values of events selected for the ACC analysis matching all requirements in Tab.~\ref{t-cosmic_cuts}. The vertical line indicates the cut for the event selection.}
\end{minipage}
\hspace{.1\linewidth}
\begin{minipage}[b]{.4\linewidth}
\centerline{\epsfig{file=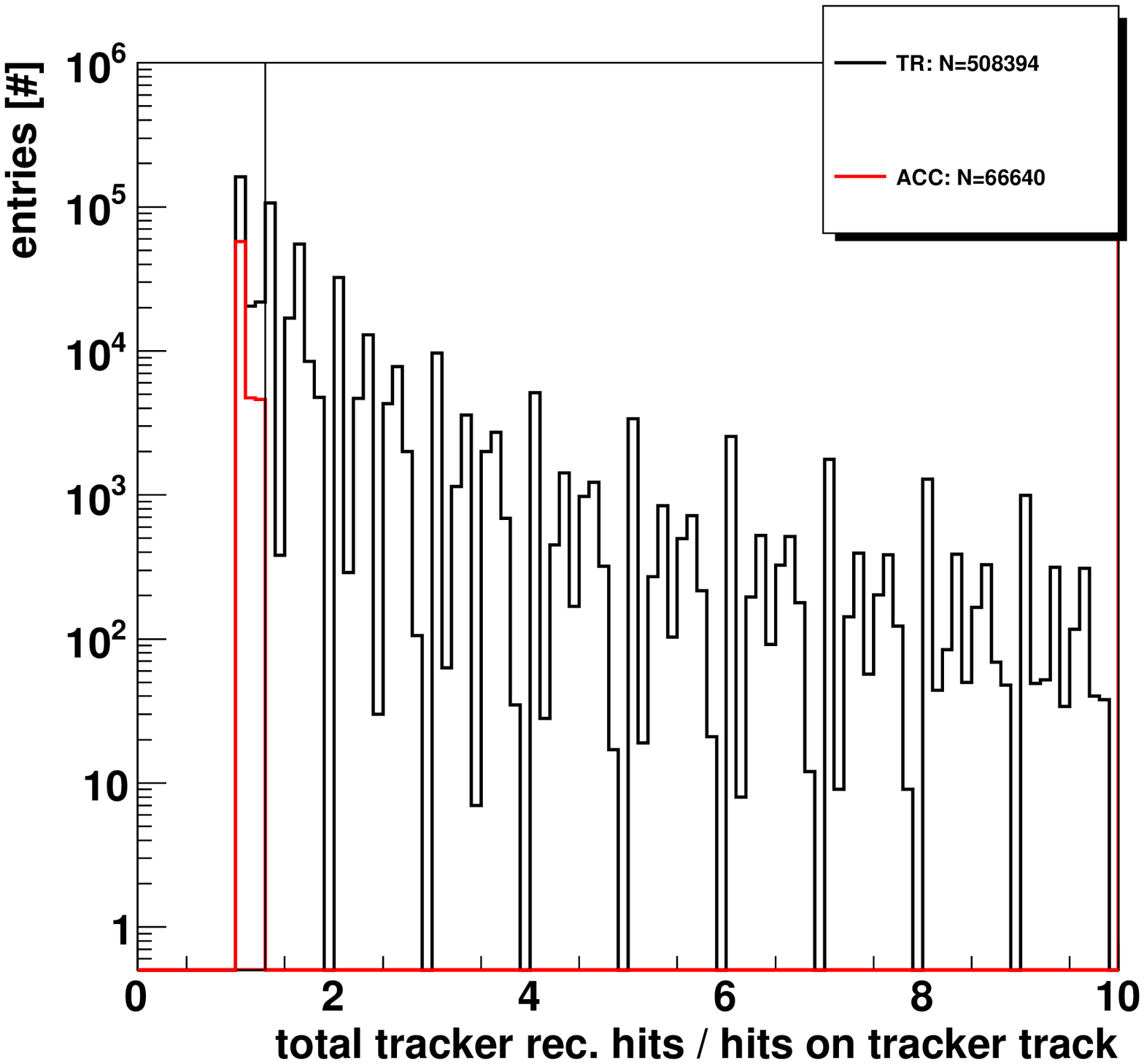,width=8cm}}\captionof{figure}{\label{f-090216_4_0_6_0_4_2_3_tr_hitratio}Ratio of reconstructed tracker hits to tracker hits on the fitted track. The black distribution shows the ratio values for tracker tracks pointing to the ACC matching the run, TOF, TRD and tracker cuts in Tab.~\ref{t-cosmic_cuts} but not the tracker noise requirement. The red ratio distribution shows the values of events selected for the ACC analysis matching all requirements in Tab.~\ref{t-cosmic_cuts}. The vertical line indicates the cut for the event selection.}
\end{minipage}
\end{center}
\end{figure}
\begin{figure}
\begin{center}
\begin{minipage}[b]{.4\linewidth}
\centerline{\epsfig{file=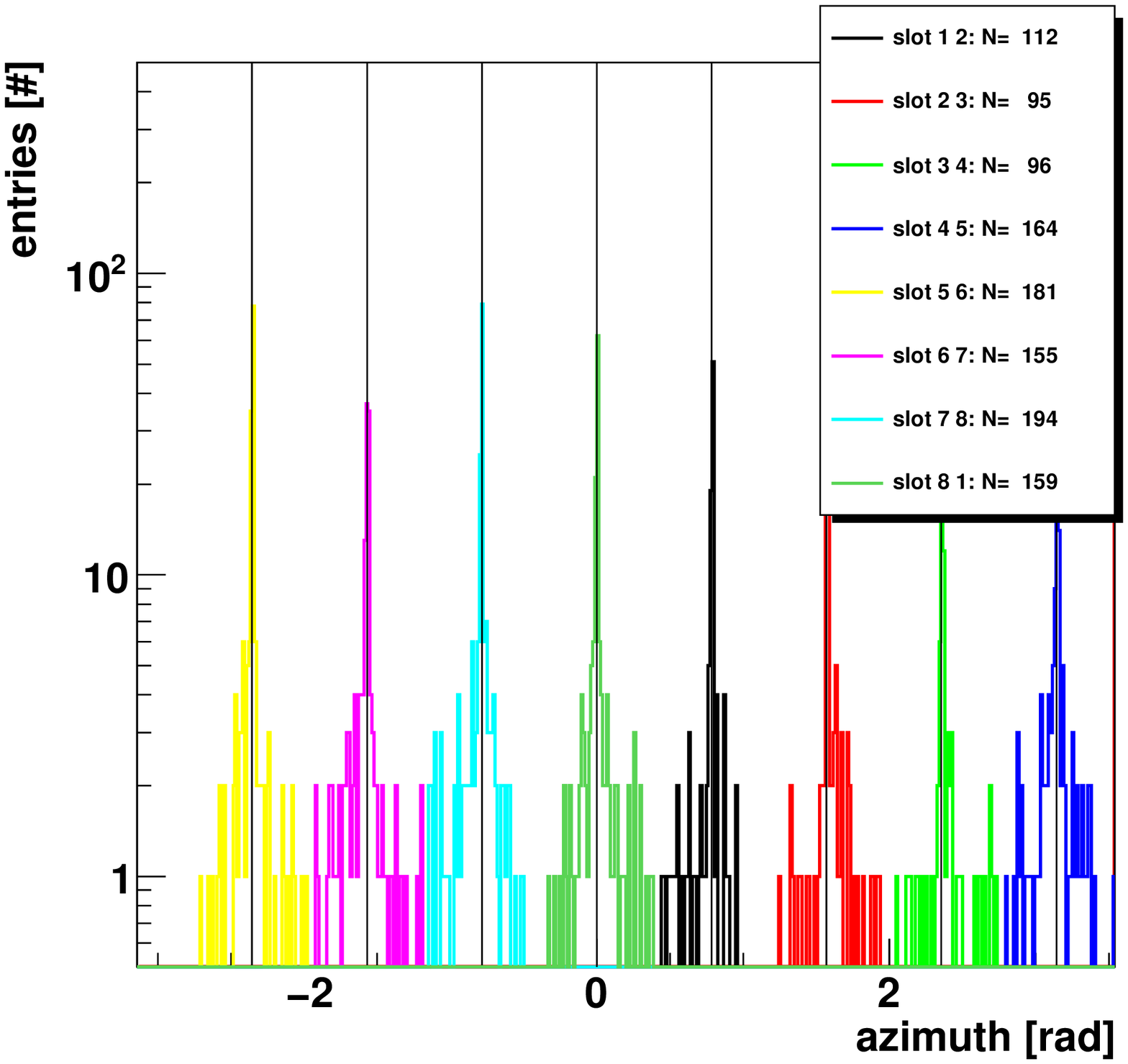,width=8cm}}\captionof{figure}{\label{f-090216_4_0_6_0_4_2_3_acc_sector}Positions of slot regions. The vertical lines indicate the positions used in the analysis ($n\cdot\pi/4, (n=-3,\dots,4)$).}
\end{minipage}
\hspace{.1\linewidth}
\begin{minipage}[b]{.4\linewidth}
\centerline{\epsfig{file=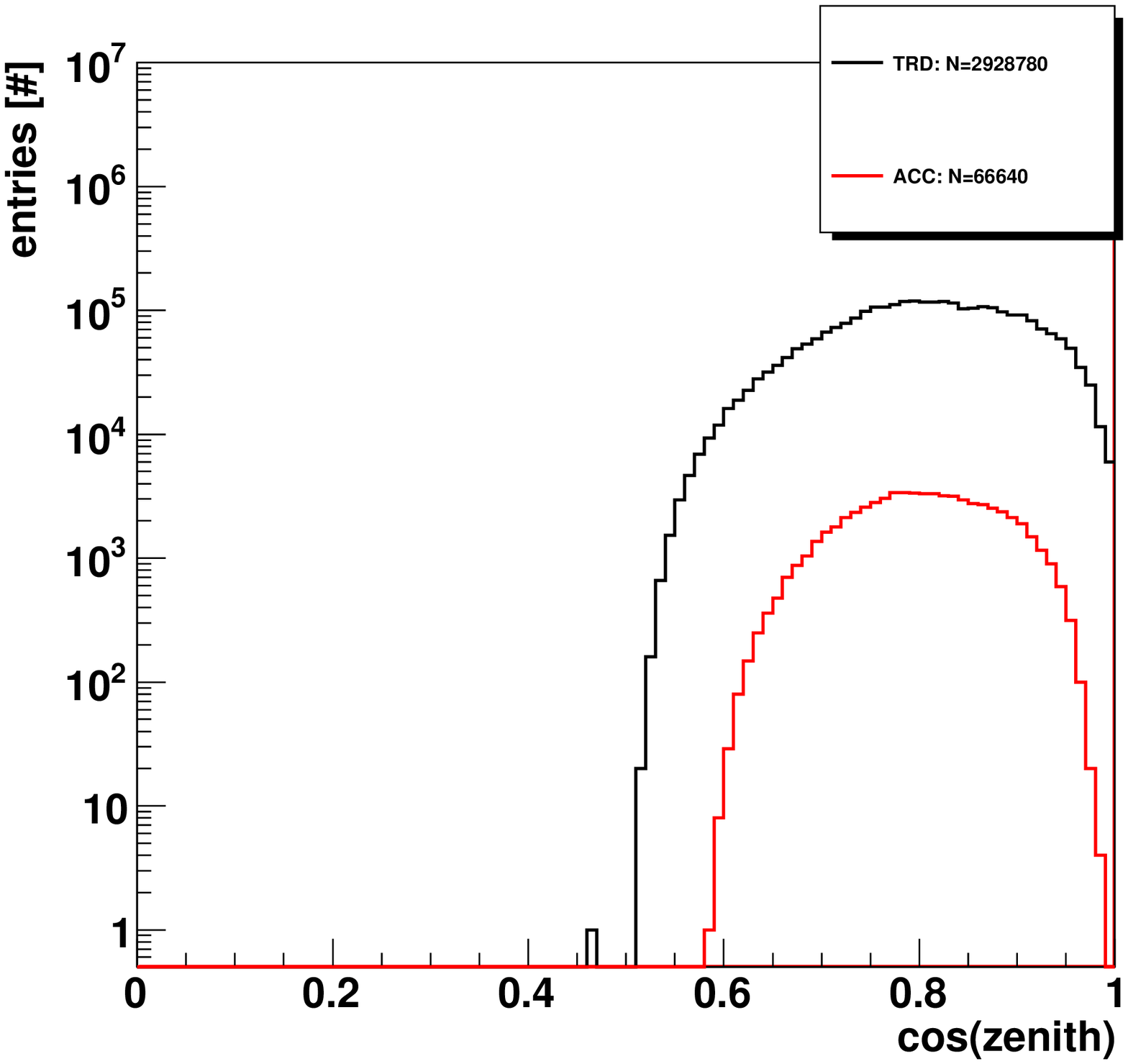,width=8cm}}\captionof{figure}{\label{f-090216_4_0_6_0_4_2_3_theta}Zenith angle distributions on the ACC using only TRD tracks and for new tracks in the tracker surviving all cuts.}
\end{minipage}
\end{center}
\end{figure}
\begin{figure}
\begin{center}
\begin{minipage}[b]{.4\linewidth}
\centerline{\epsfig{file=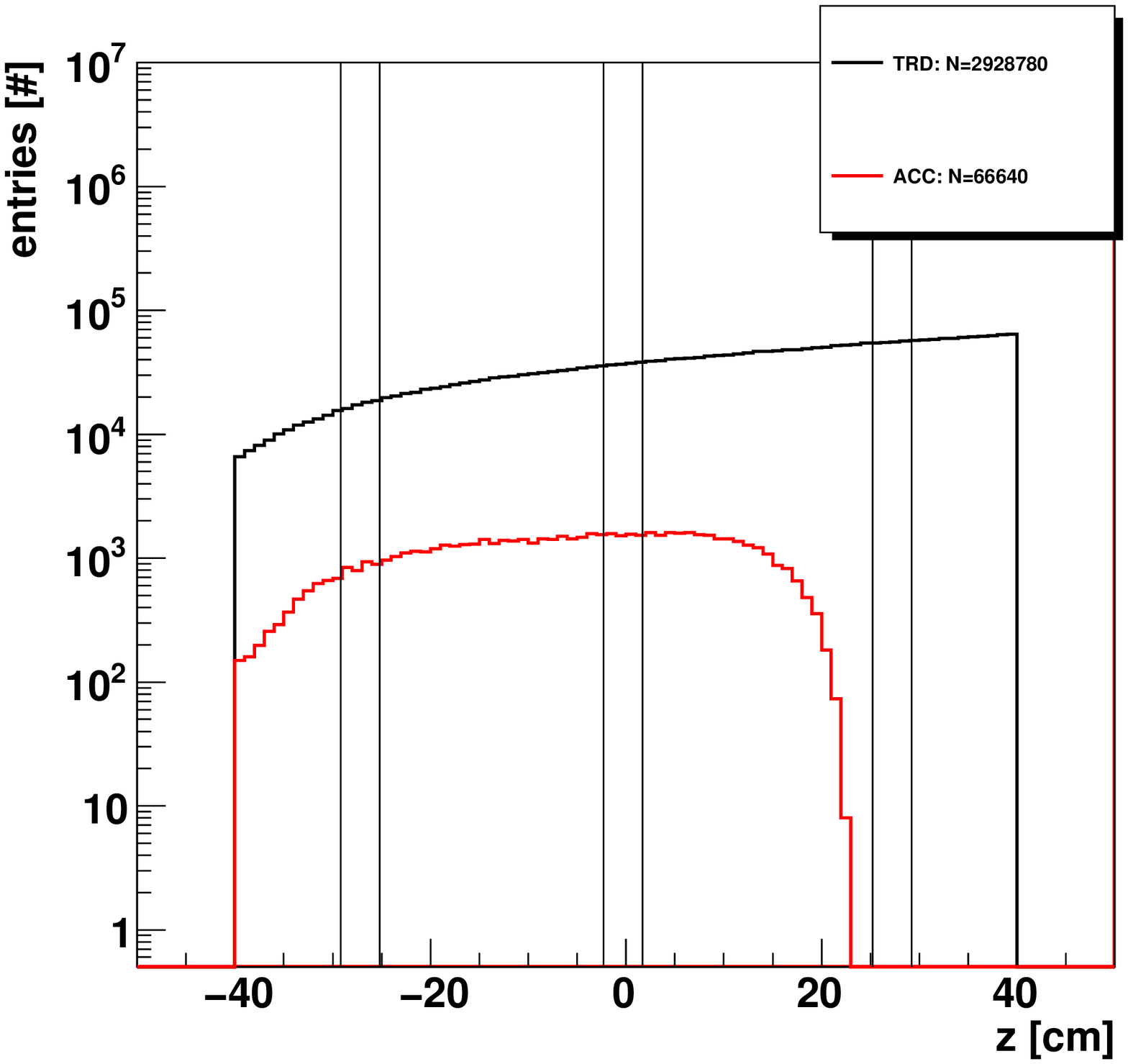,width=8cm}}\captionof{figure}{\label{f-090216_4_0_6_0_4_2_3_z}$z$ occupancy distributions on the ACC using only TRD tracks and for new tracks in the tracker surviving all cuts.}
\end{minipage}
\hspace{.1\linewidth}
\begin{minipage}[b]{.4\linewidth}
\centerline{\epsfig{file=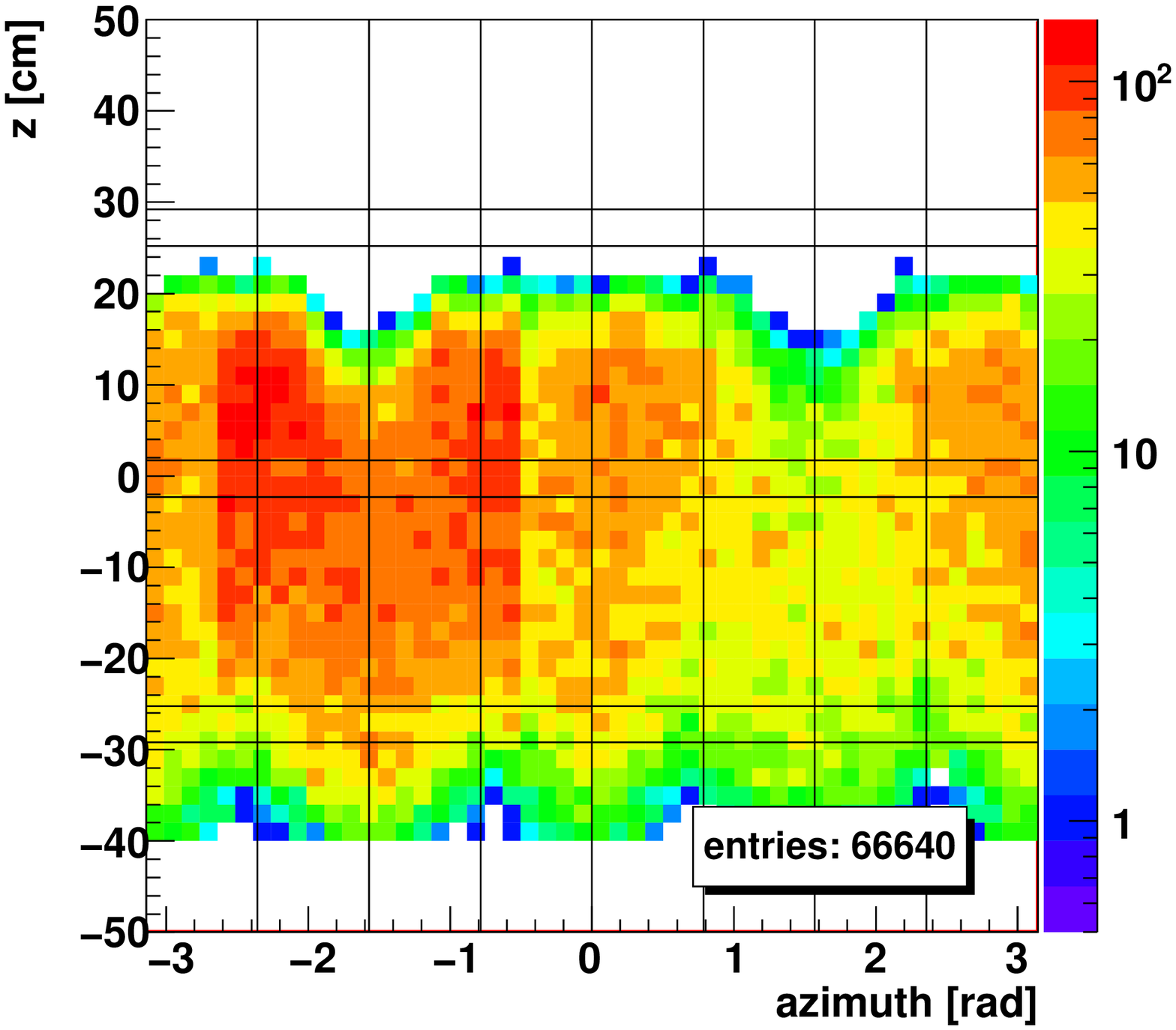,width=8cm}}\captionof{figure}{\label{f-090216_4_0_6_0_4_2_3_phi_z}Azimuthal and $z$ occupancy on the ACC of fitted tracks surviving all cuts. The vertical lines indicate the positions of the slot regions between two sectors and the horizontal lines the positions of the tracker layers. The color code on the right shows the number of entries.}
\end{minipage}
\end{center}
\end{figure}

For a very reliable track fit two additional cuts are applied to accept only very clean events for the analysis. The ratio of the total number of hits in the TRD to the number of hits on the TRD track is not allowed to be larger than 1.5 (Fig.~\ref{f-090216_4_0_6_0_4_2_3_trd_hitratio}). In a similar way, the ratio of the total number of reconstructed tracker hits to the number of hits on the fitted track must be smaller than 1.3 as shown in Fig.~\ref{f-090216_4_0_6_0_4_2_3_tr_hitratio}. The number of different total reconstructed tracker hits per event is extracted by counting the number of different $(y,z)$ hit coordinate pairs because the readout in $x$ direction is multiplexed by a factor of about seven depending on the tracker layer. These two ratio requirements lower the number of events having noise hits or additional hits resulting from interactions of the primary particle in the detector that could significantly spoil the track fit.

The average RMS of the pedestal distributions of the ACC PMTs is 7\,ADC counts. A good ACC event is defined to show at least in one PMT of the sector to which the fitted track is extrapolated an ADC value larger than 21\,ADC counts. ACC events not fulfilling this requirement are called 'missed'. A sector is defined as the two ACC panels sharing their PMTs. The position of the seven slots between the eight ACC sectors is determined by requiring all four PMTs of adjacent sectors to exceed the threshold of 21\,ADC counts. The frequency distribution of these events is shown as a function of the azimuth angle in Fig.~\ref{f-090216_4_0_6_0_4_2_3_acc_sector}. The new tracker fit is able to reproduce the geometry of the ACC because the peak positions are consistent with the expectation of $n\cdot\pi/4, (n=-3,\dots,4)$.

Also the zenith angle distributions of the atmospheric muons show the expected behavior. The angle of tracks hitting the ACC cannot be larger than the maximum acceptance angle of the tracker of about 55° and small zenith angles are suppressed for the ACC because the panels are perpendicular to the TOF planes. The distribution of all TRD tracks surviving the TRD selection criteria and pointing to the ACC is also shown. Here, track angles up to about 60° are possible. The ACC occupancy in $z$ direction is quite uniform (Fig.~\ref{f-090216_4_0_6_0_4_2_3_z}, black) but shows a drop for $z\approx20$\,cm when the first two inner tracker planes are reached because it was required that the first tracker plane and at least two further tracker show hits on the track. Additionally shown is the $z$ occupancy for TRD tracks. The track occupancy distribution in azimuth angle and $z$ on the ACC cylinder shows a structure due to the shape of the TRD, the TOF and the tracker planes (Fig.~\ref{f-090216_4_0_6_0_4_2_3_phi_z}). The number of entries in the azimuth range from $-\pi+1/2$\,rad to $-1/2$\,rad is increased because of a period where only the RAM side of the electronics was connected. Here, only tracks in this azimuth angle range were considered for the analysis to assure that the inefficiency is not determined by unconnected electronics.

\subsubsection{Comparison to atmospheric Muon Simulations}

\begin{figure}
\begin{center}
\centerline{\epsfig{file=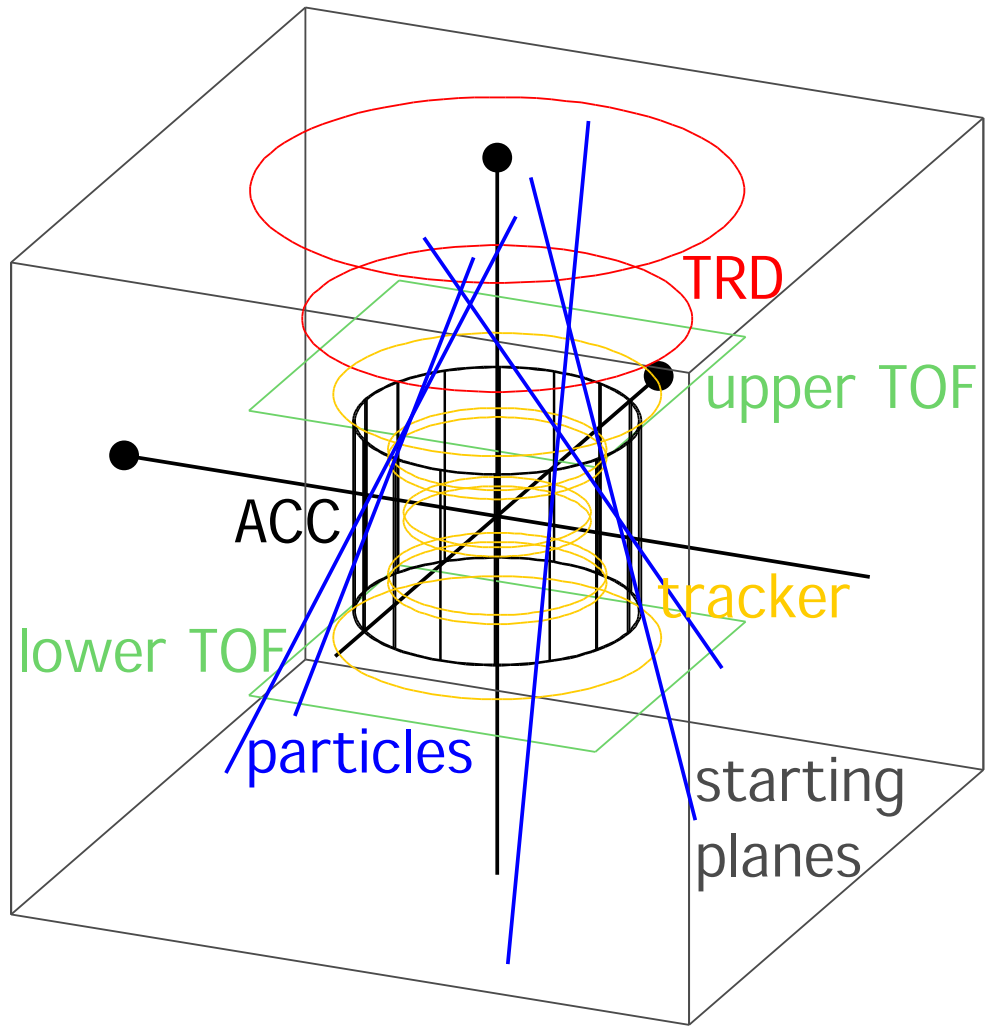,width=13cm}}\captionof{figure}{\label{f-0711_acceptance_acc}Model for simple AMS-02 simulations of flux, acceptance and pathlength in the scintillator material of the ACC. Calculations are made without magnetic field. The gray box indicates the starting planes for the simulation and the black lines the ACC cylinder. The blue lines illustrate particles hitting the TRD (red), the upper TOF (green), at least 3 tracker layers (orange) and the ACC.}
\end{center}
\end{figure}

\begin{figure}
\begin{center}
\begin{minipage}[b]{.4\linewidth}
\centerline{\epsfig{file=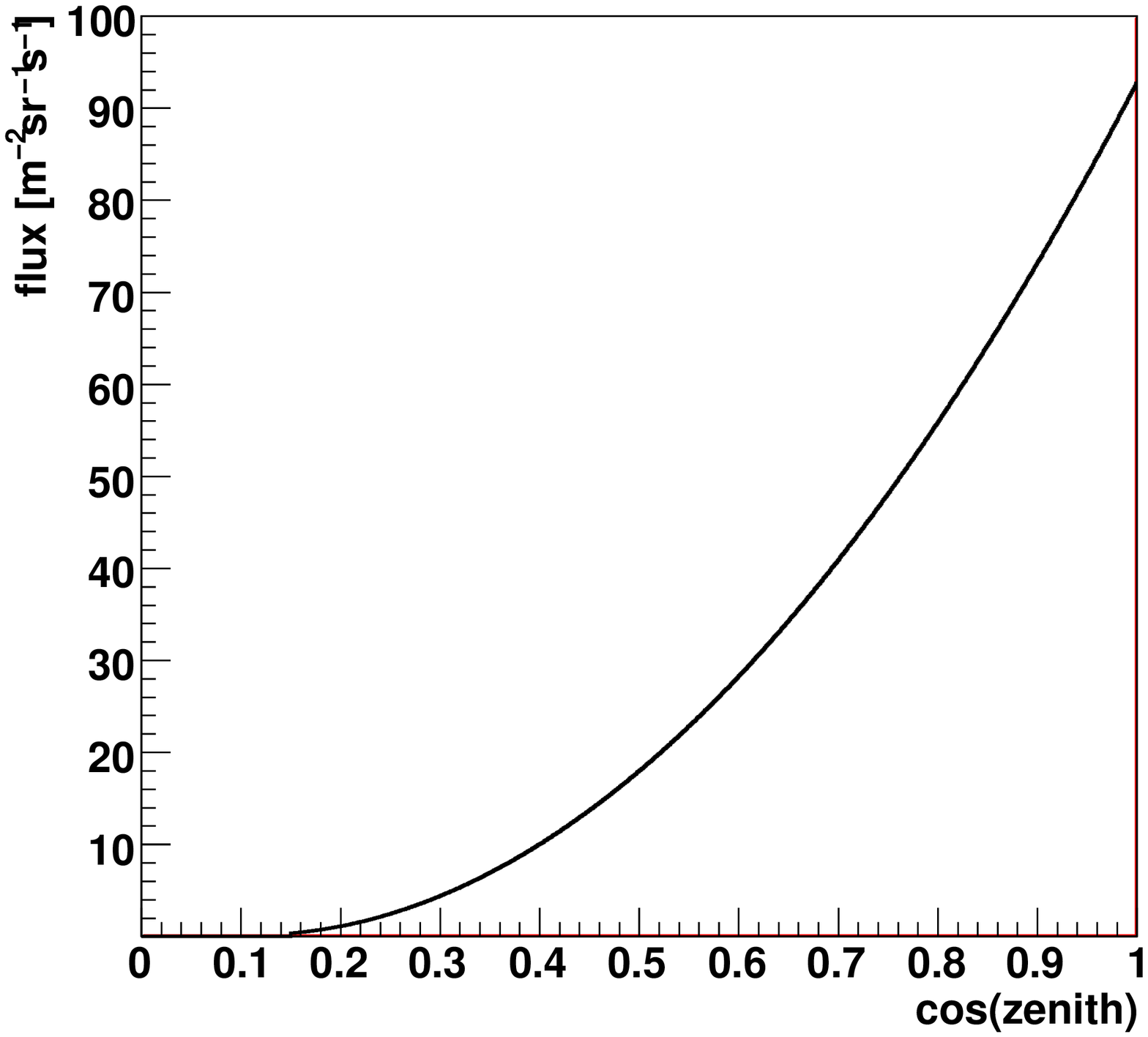,width=8cm}}\captionof{figure}{\label{f-Phi_c}Atmospheric muon flux as a function of the zenith angle in Geneva.}
\end{minipage}
\hspace{.1\linewidth}
\begin{minipage}[b]{.4\linewidth}
\centerline{\epsfig{file=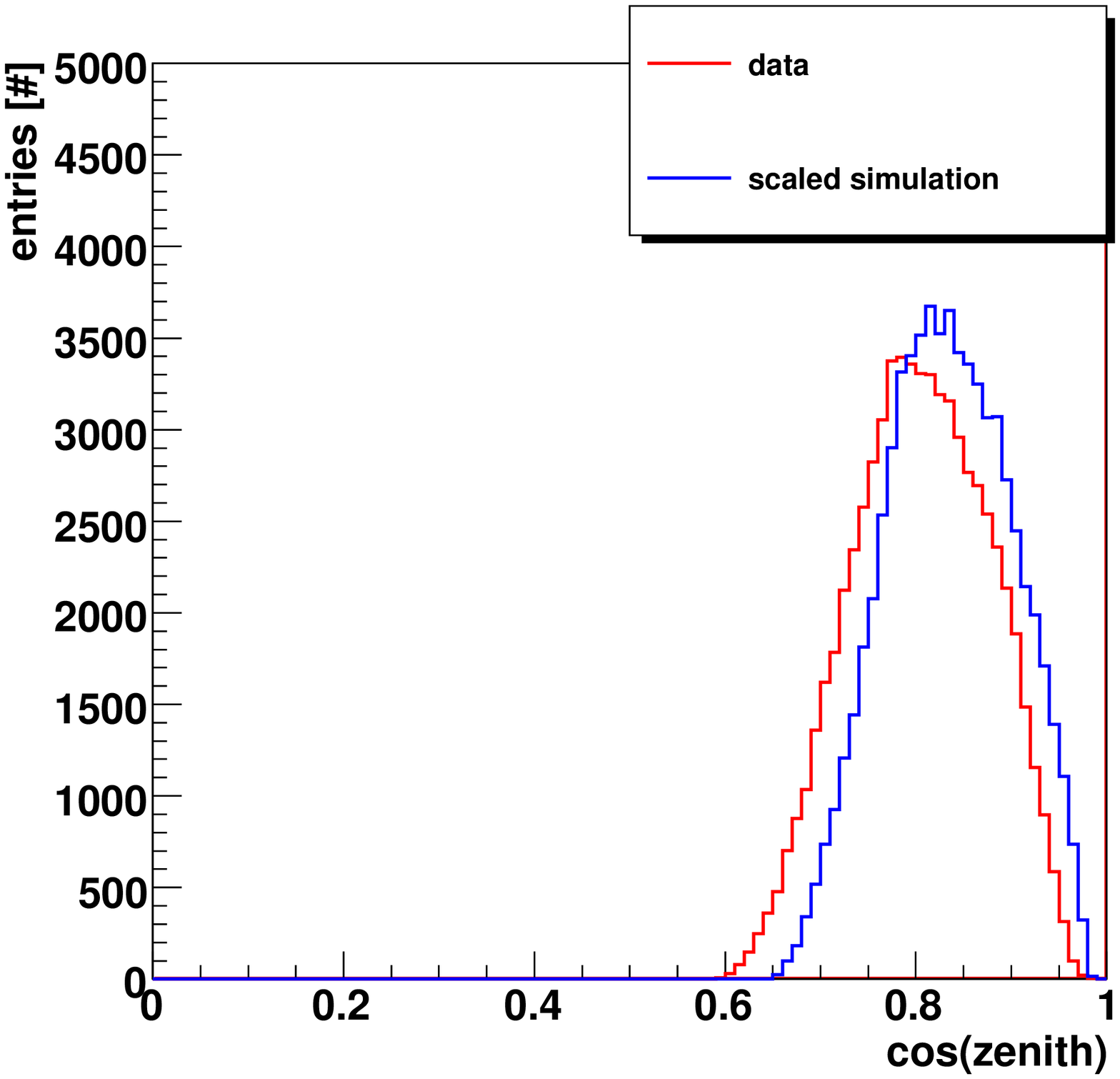,width=8cm}}\captionof{figure}{\label{f-theta_sim_data}Comparison between the measured zenith angle distribution of particles crossing the ACC and the simulated one. The simulation is scaled to the measured number of entries.}
\end{minipage}
\end{center}
\end{figure}

To validate the new track fit the measurements are compared to atmospheric muon simulations by placing a simplified AMS-02 detector in a cubic volume (Fig.~\ref{f-0711_acceptance_acc}). The starting positions were uniformly distributed on the walls of the cube and the particle tracks were simulated by straight lines. The framework allows to either generate an isotropic particle distribution or the distribution of atmospheric muons on Earth's surface in Geneva. The isotropic distribution is achieved by uniformly distributed azimuthal angles in the range $[0,2\pi]$ and by an uniform distribution in $\cos^2$ on the walls for the zenith angles in the interval $[0,\pi/2]$. The atmospheric muon distribution is started only from the upper cube plane and follows the parametrization of muon flux at ground level in Geneva weighted by an additional factor of $\cos(\text{zenith})$ for the starting plane (Fig.~\ref{f-Phi_c}) \cite{biallass-2007}. The octagonal shape of the TRD is simplified by two circular planes representing the upper and lower TRD planes. The TOF consists of two quadratic planes and the tracker of eight circular planes with different radii. The ACC is made of two cylinders. 

The simulations were analysed in a similar way as the data before. A selected event intersects the upper TOF, both TRD planes, at least three tracker planes and the ACC. The comparison between the simulated and measured zenith angle distributions is shown in Fig.~\ref{f-theta_sim_data}. For a better comparison only runs with AMS-02 in vertical position are respected for the data distribution. Both distributions show a very similar behavior. Only very few particles are selected at small zenith angles, the maximum is at about 40° and no particles are above about 50°. The differences arise from the fact that the simplified simulation uses approximated detector shapes for TOF, TRD and tracker but in general the ACC zenith angle distribution is well understood. A detailed analysis would require a more sophisticated simulation.

The same framework can also be used to study the trigger rate during the thermal vacuum test of the complete detector at ESTEC\footnote{European Space Research and Technology Centre} where AMS-02 will be rotated by 90°. For the trigger condition any two out of four TOF planes the trigger rate will drop from about 400\,Hz to 200\,Hz.

\subsubsection{ACC Inefficiency Determination for the Pre-Integration}

\begin{figure}
\begin{center}
\begin{minipage}[b]{.4\linewidth}
\centerline{\epsfig{file=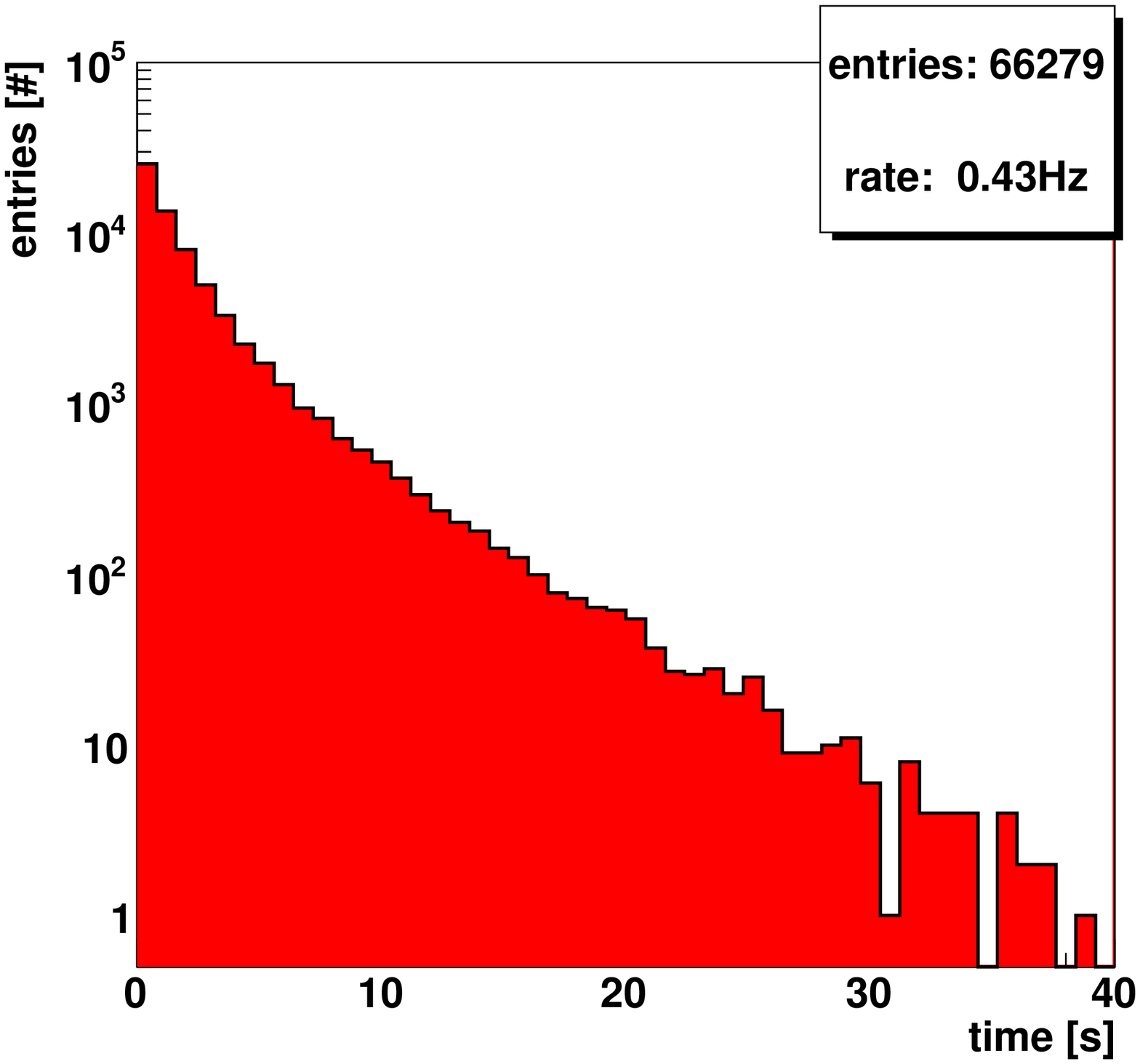,width=8cm}}\captionof{figure}{\label{f-090216_4_0_6_0_4_2_3_acc_jlv1_time}Time interval between two ACC events surviving all cuts.}
\end{minipage}
\hspace{.1\linewidth}
\begin{minipage}[b]{.4\linewidth}
\centerline{\epsfig{file=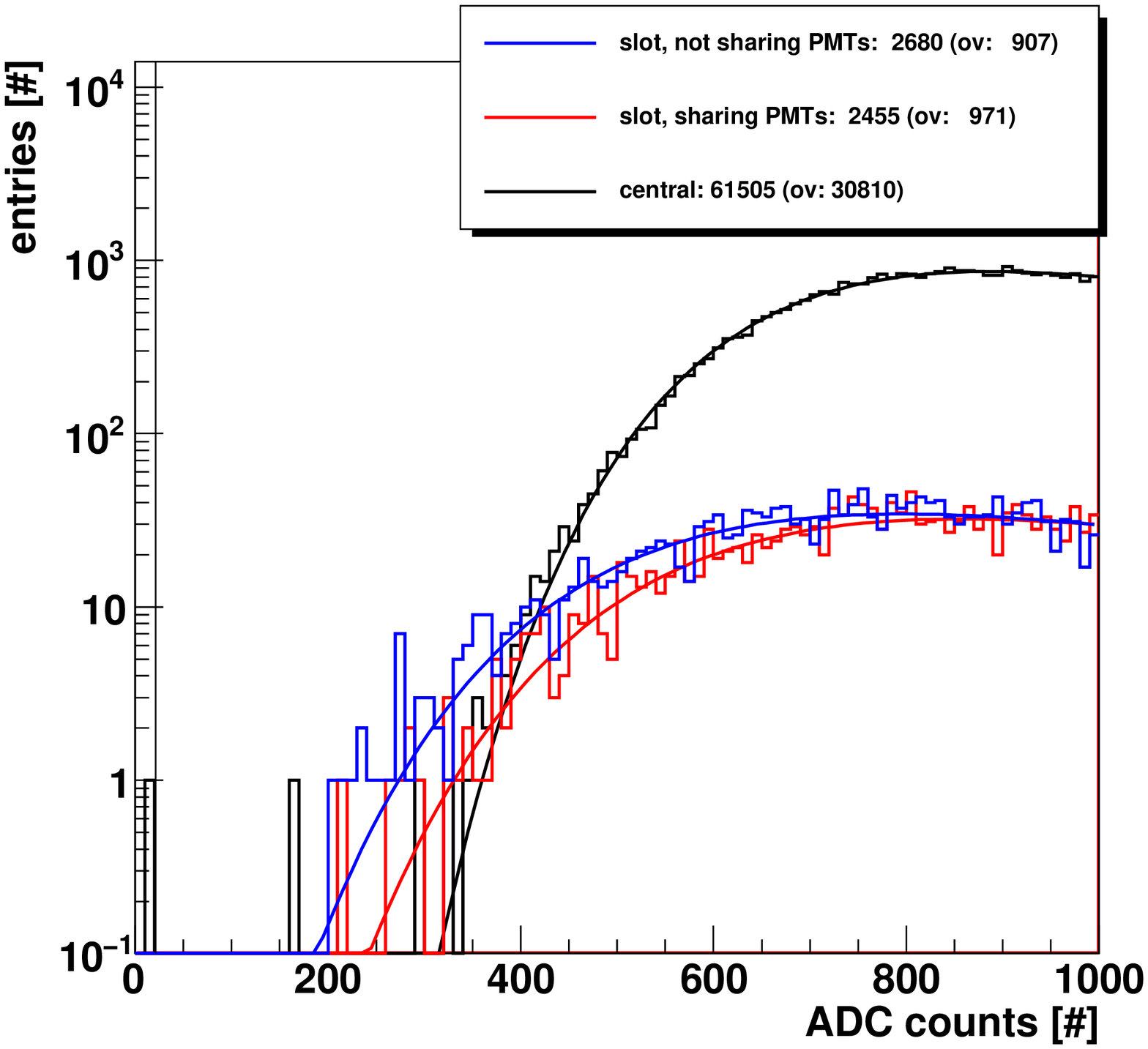,width=8cm}}\captionof{figure}{\label{f-090216_4_0_6_0_4_2_3_acc_adc_highest_slot_central}Highest ADC values of the complete ACC system for the central region of the ACC panels and the two different slot regions in an interval of 15\,mrad around the slots. The vertical line indicates the cut for the definition of a good event.}
\end{minipage}
\end{center}
\end{figure}

\begin{figure}
\begin{center}
\centerline{\epsfig{file=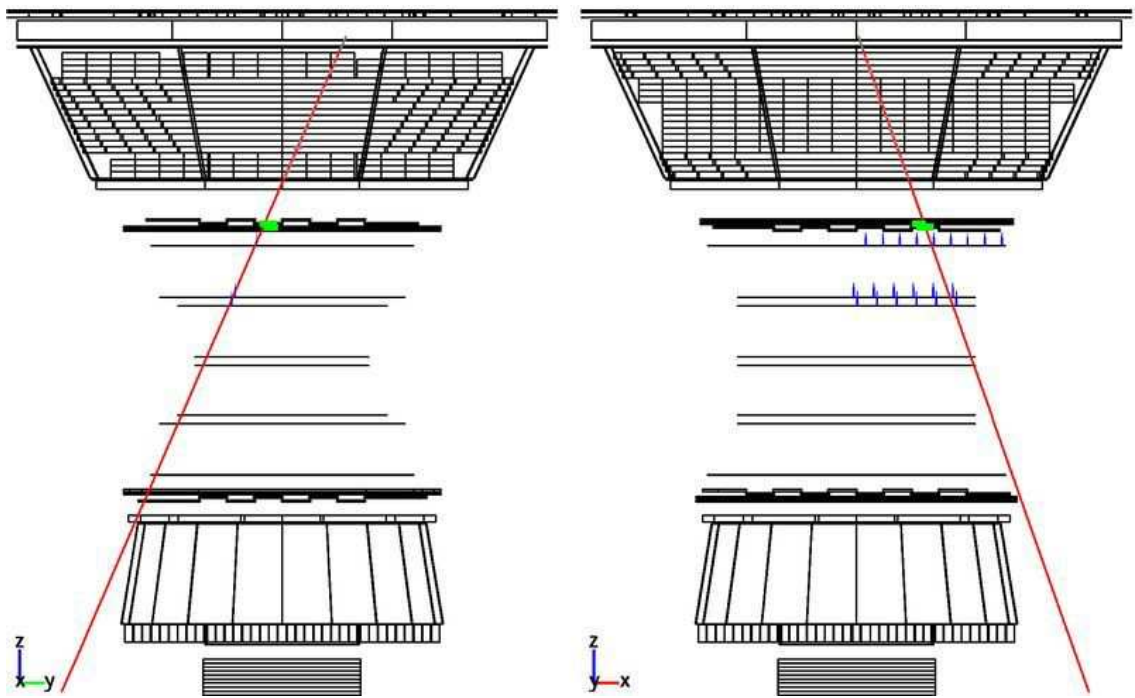,width=15cm}}\captionof{figure}{\label{f-1213347818_661137}June 13 11:56:16 2008: run: 1213347818, event: 661137.}
\end{center}
\end{figure}

The previous discussions showed that the new track fit is reliable and the inefficiency of the ACC can now be calculated. In total about 0.9\,\% of the good TRD events pointing to the ACC survive all subsequent requirements and are cleanly extrapolated to the ACC. The time interval distribution between good ACC events shows an approximately exponential behavior and has a mean event rate of 0.4\,Hz (Fig.~\ref{f-090216_4_0_6_0_4_2_3_acc_jlv1_time}). The highest ADC value distributions for the central part of a panel, the slot region between two sectors and the slot within a sector are shown in Fig.~\ref{f-090216_4_0_6_0_4_2_3_acc_adc_highest_slot_central}. As expected, the slot between two sectors shows the smallest MOP value. In addition, the spectrum of the central part shows one missed event below the cut of 21\,ADC counts which cannot be explained by statistical fluctuations of the fitted Landau distribution. The mean inefficiency for the pre-integration of the complete ACC system is calculated to be: \be\bar I=1.5^{+2.3}_{-1.1}\cdot 10^{-5} \qquad\text{or}\qquad \bar I<7.2\cdot10^{-5} \text{@ 95\,\% confidence level}.\ee This result is well beyond the requirement of $10^{-4}$. The missed event was measured while the complete experiment was rotated around the $y$ axis by -90° (Fig.~\ref{f-1213347818_661137}) such that the muon probably crossed the ACC from the outside before traversing tracker and TRD. In this case the probability for unclean events due to interactions, e.g. in the tank, is much higher. 

\begin{figure}
\begin{center}
\begin{minipage}[b]{.4\linewidth}
\centerline{\epsfig{file=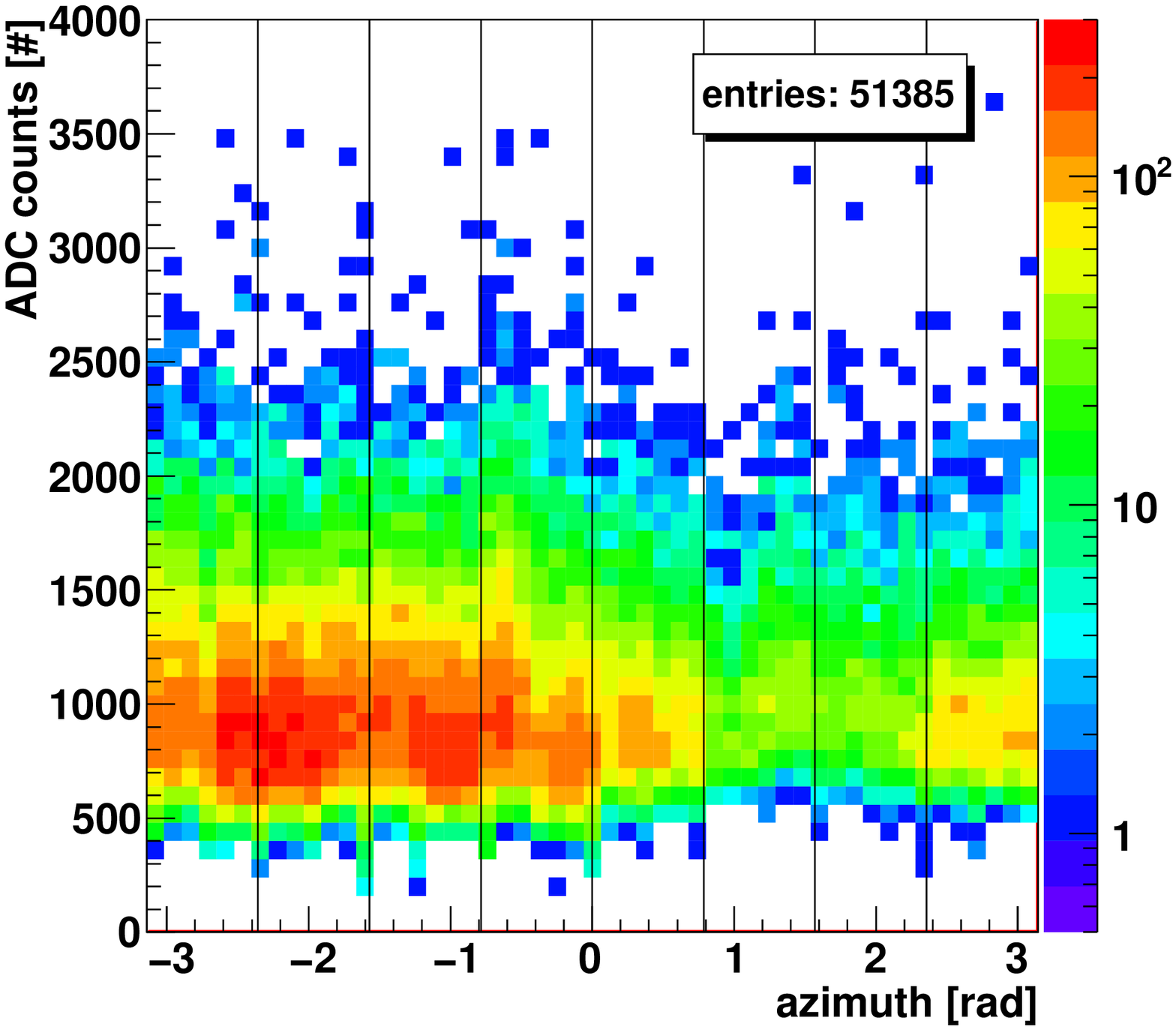,width=8cm}}\captionof{figure}{\label{f-090216_4_0_6_0_4_2_3_acc_phi_adc_highest_lower}ADC values of the $-z$ PMTs as a function of the azimuthal angle for the case that these PMTs have recorded the highest ADC value. The vertical lines indicate the slots between the ACC sectors. The color code on the right shows the number of entries.}
\end{minipage}
\hspace{.1\linewidth}
\begin{minipage}[b]{.4\linewidth}
\centerline{\epsfig{file=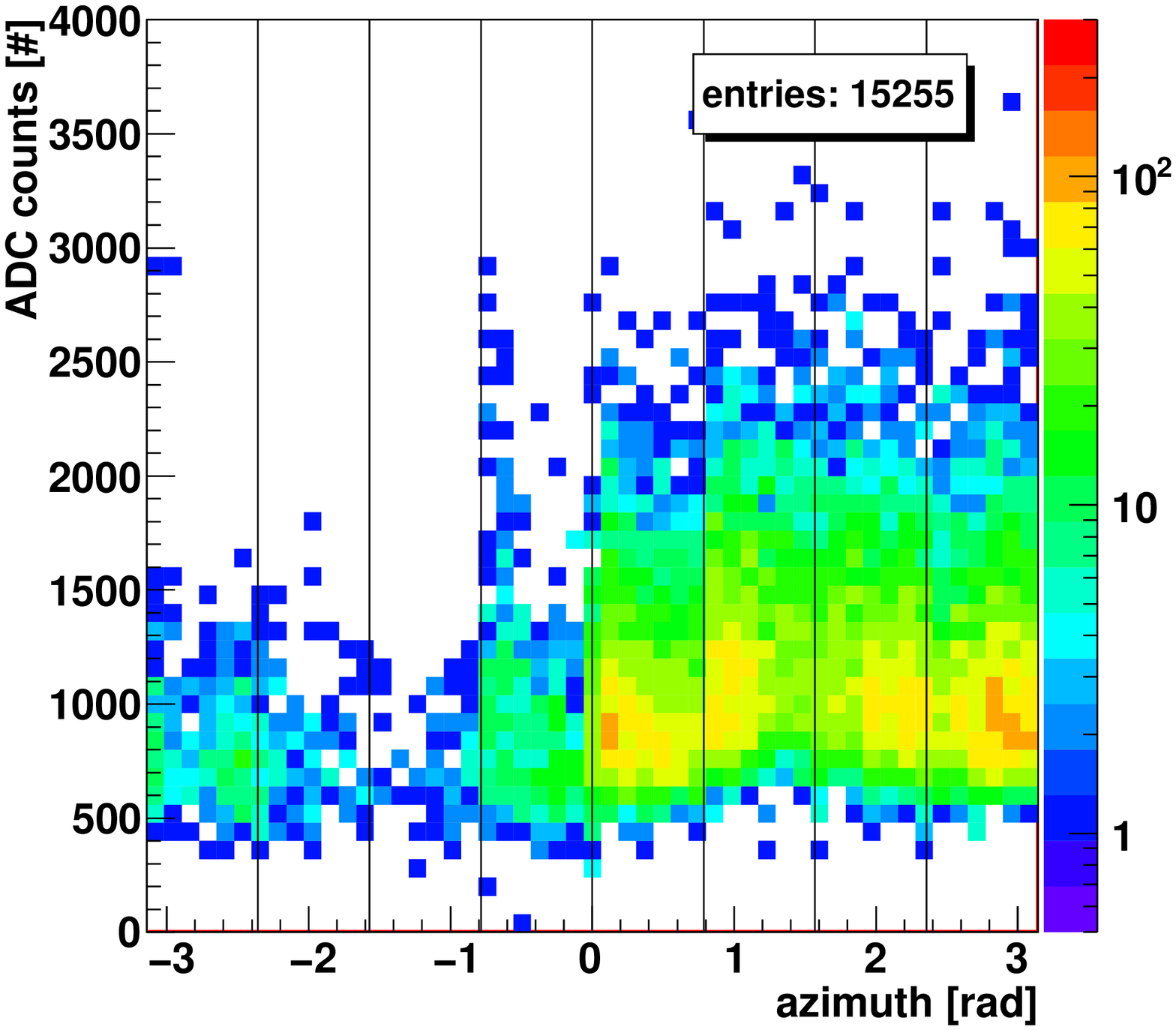,width=8cm}}\captionof{figure}{\label{f-090216_4_0_6_0_4_2_3_acc_phi_adc_highest_upper}ADC values of the $+z$ PMTs as a function of the azimuthal angle for the case that these PMTs have recorded the highest ADC value. The vertical lines indicate the slots between the ACC sectors. The color code on the right shows the number of entries.}
\end{minipage}
\end{center}
\end{figure}

An additional effect may also be relevant for the missed event and is discussed on the basis of Fig.~\ref{f-090216_4_0_6_0_4_2_3_acc_phi_adc_highest_lower} and \ref{f-090216_4_0_6_0_4_2_3_acc_phi_adc_highest_upper}. The measured ADC distributions for the highest of the ADC PMT measurements as a function of the azimuthal angle position show a homogeneous position of the pedestal corrected MOP value at about 900\,ADC counts for both the lower PMTs and the upper PMTs in the interval $[0,\pi]$. The deviation from this behavior for the upper PMTs between $[-\pi,0]$ arises from the fact that the qualification S-crate connected to this position was equipped with a smaller capacitor in the input stage (5\,nF) than foreseen for the flight version (10\,nF). This capacitor strongly influences the charge measurement with the ADC as already discussed in Sec.~\ref{ss-flightelec}. The missed event shows the highest ADC value (12\,ADC counts) in sector\,7 connected to the QM S-crate. The FM S-crate would maybe improve the measurement. The special conditions during this run can be used to argue that the derived inefficiency can be understood as an upper bound. Measurements in nominal conditions with a well calibrated detector would give a more reliable result.

\subsubsection{Further ACC Properties}

\begin{figure}
\begin{center}
\begin{minipage}[b]{.4\linewidth}
\centerline{\epsfig{file=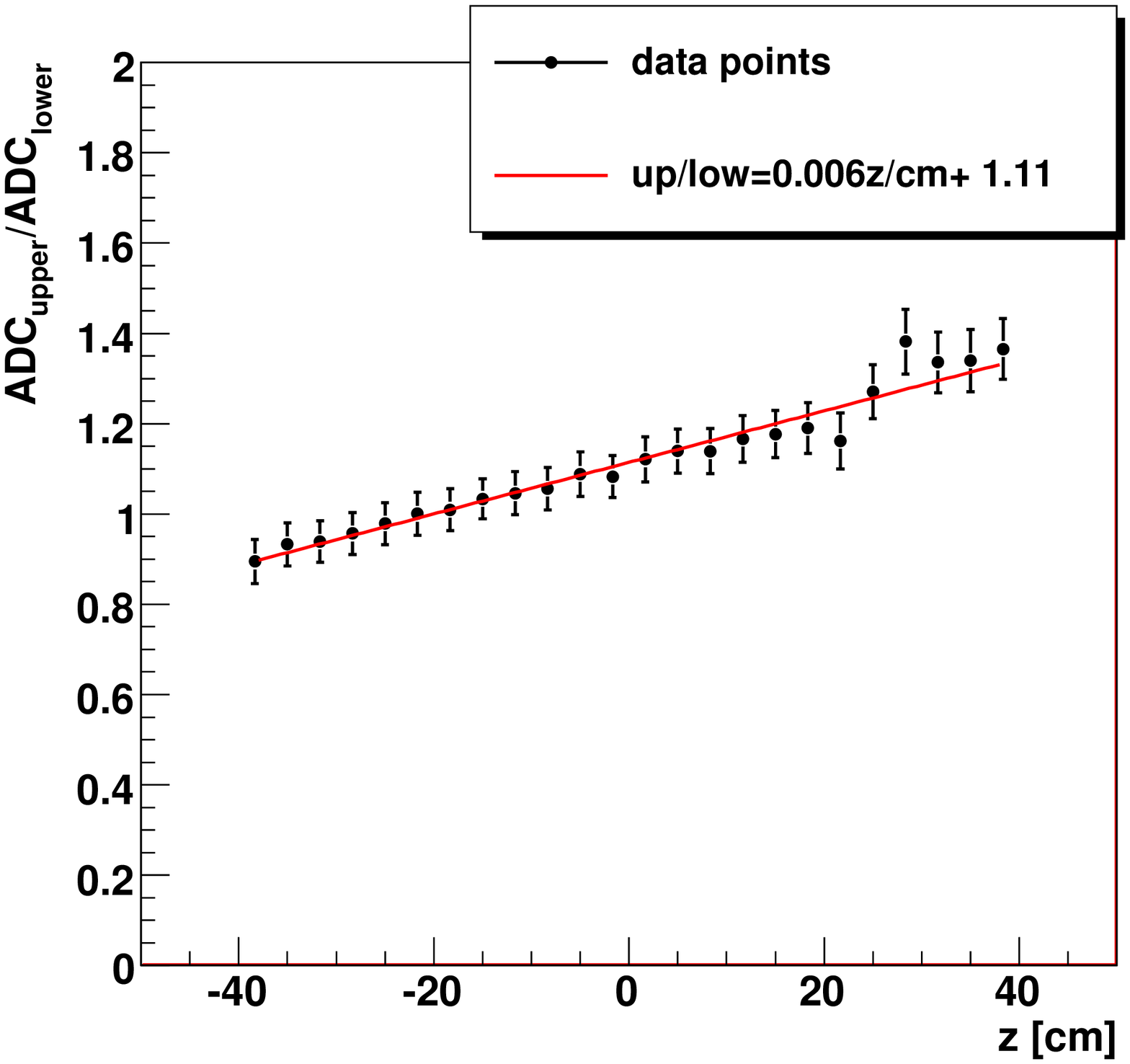,width=8cm}}\captionof{figure}{\label{f-081209_4_0_6_0_4_2_3_acc_z_frac_mean}Signal fraction of upper and lower PMTs as a function of the $z$ position.}
\end{minipage}
\hspace{.1\linewidth}
\begin{minipage}[b]{.4\linewidth}
\centerline{\epsfig{file=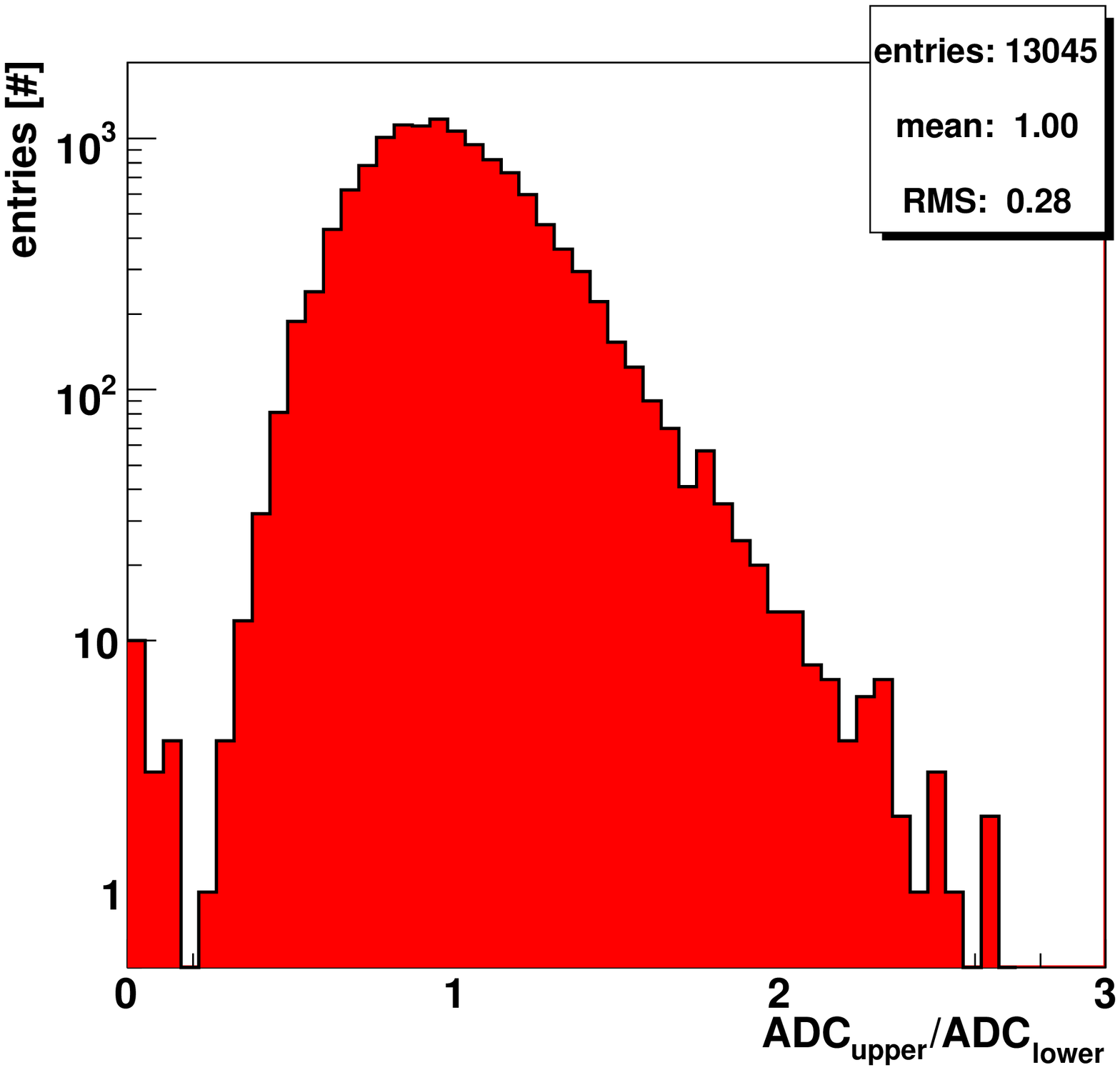,width=8cm}}\captionof{figure}{\label{f-081209_4_0_6_0_4_2_3_acc_z_frac_norm_center}Signal ratio of upper and lower PMTs in the central region $z=0$\,cm normalized to a mean of 1.}
\end{minipage}
\end{center}
\end{figure}

Besides the inefficiency study the data were used to extract further detector properties. In the following, the TOF requirement for the analysis was changed such that hits either in all four TOF planes or only in the upper two are required to study the signal behavior on the ACC panel in the complete $z$ range (-40\,cm - 40\,cm). As mentioned above, the ratio of the upper and lower PMT signals may provide a rough estimation of the hit position in $z$ direction (Fig.~\ref{f-inefficiency_several_ratio}). The upper PMTs connected to the qualification S-crate with the 5\,nF input capacitor are not taken into account for this analysis. The ratio $ADC\sub{upper}/ADC\sub{lower}$ increases with $z$ and a straight line fit gives (Fig.~\ref{f-081209_4_0_6_0_4_2_3_acc_z_frac_mean}):
\be R = (0.0057\pm0.0005)\,\text{cm}^{-1}\cdot z+(1.11\pm0.01).\label{e-pmtratio2}\ee
The value $R=(1.11\pm0.01)$ at $z=0$\,cm will be 1 for the flight configuration with the final high voltage calibration for the photomultipliers. As shown in Fig.~\ref{f-081209_4_0_6_0_4_2_3_acc_z_frac_norm_center} for the central region around $z=0$, the RMS of the signal ratio of upper and lower PMTs is about 30\,\%. In about 0.2\,\% of the events only one PMT shows a good entry and the ratio value is small and cannot be explained within statistical fluctuations. This emphasizes again the need of redundancy to improve the detection efficiency. Eq.~\ref{e-pmtratio2} is transformed in a position determination on the basis of the signal ratio $R$:
\be z = (175.4\pm15.4)\,\text{cm} \cdot ((1.00\pm0.28)\cdot R - (1.11\pm0.01)),\ee compatible with the previous measurements (Eq.~\ref{e-pmtratio_p}). A very similar result is achieved by comparing the $z$ position calculated from the PMT ratio $R$ with the $z$ position of the track which results in an RMS of the residual distribution of about 45\,cm. Therefore, the position determination with the PMT ratio is not very meaningful.

\begin{figure}
\begin{center}
\begin{minipage}[b]{.4\linewidth}
\centerline{\epsfig{file=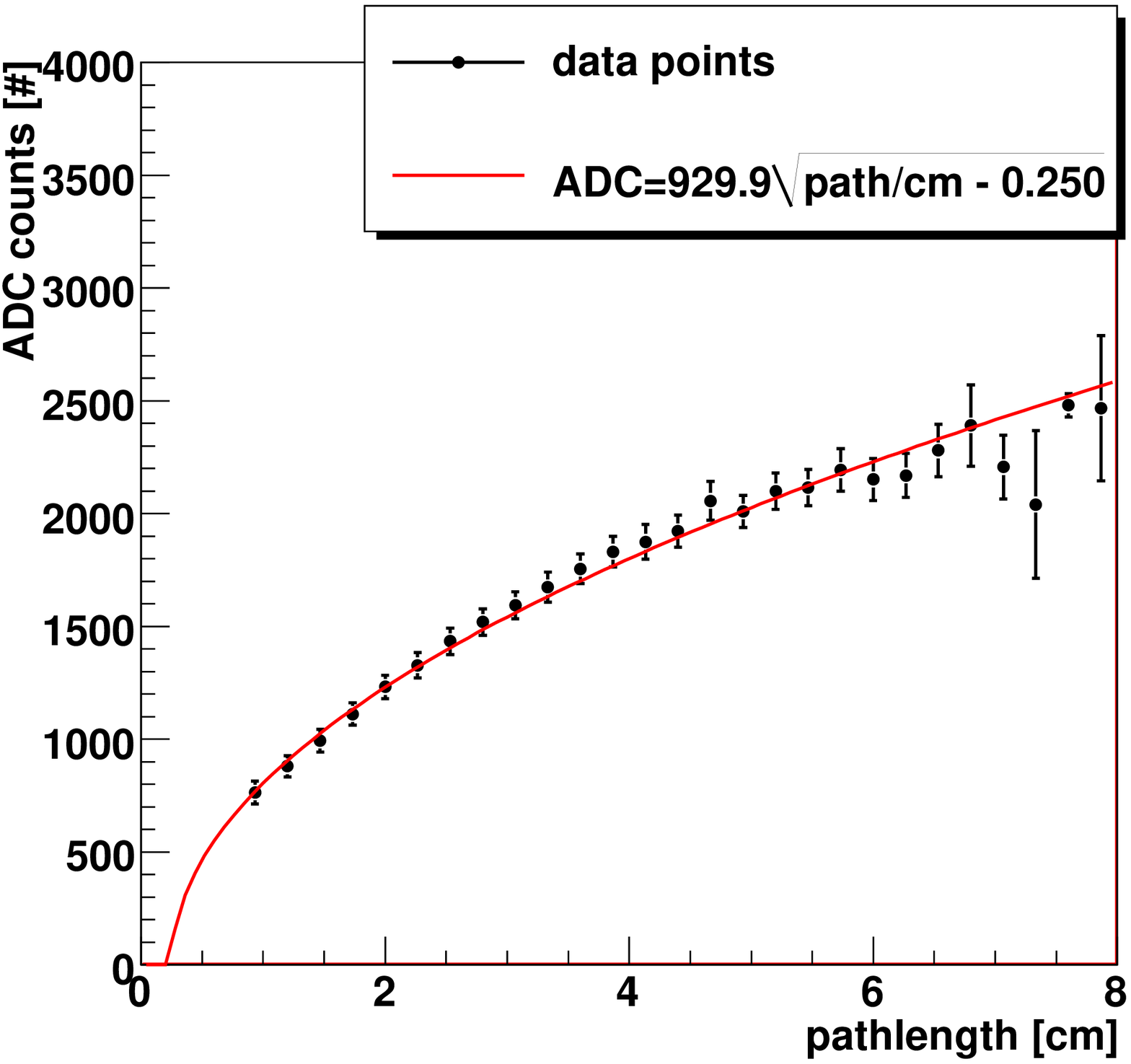,width=8cm}}\captionof{figure}{\label{f-081209_4_0_6_0_4_2_3_acc_path_adc_mean}Mean ADC value as a function of pathlength in the ACC material.}
\end{minipage}
\hspace{.1\linewidth}
\begin{minipage}[b]{.4\linewidth}
\centerline{\epsfig{file=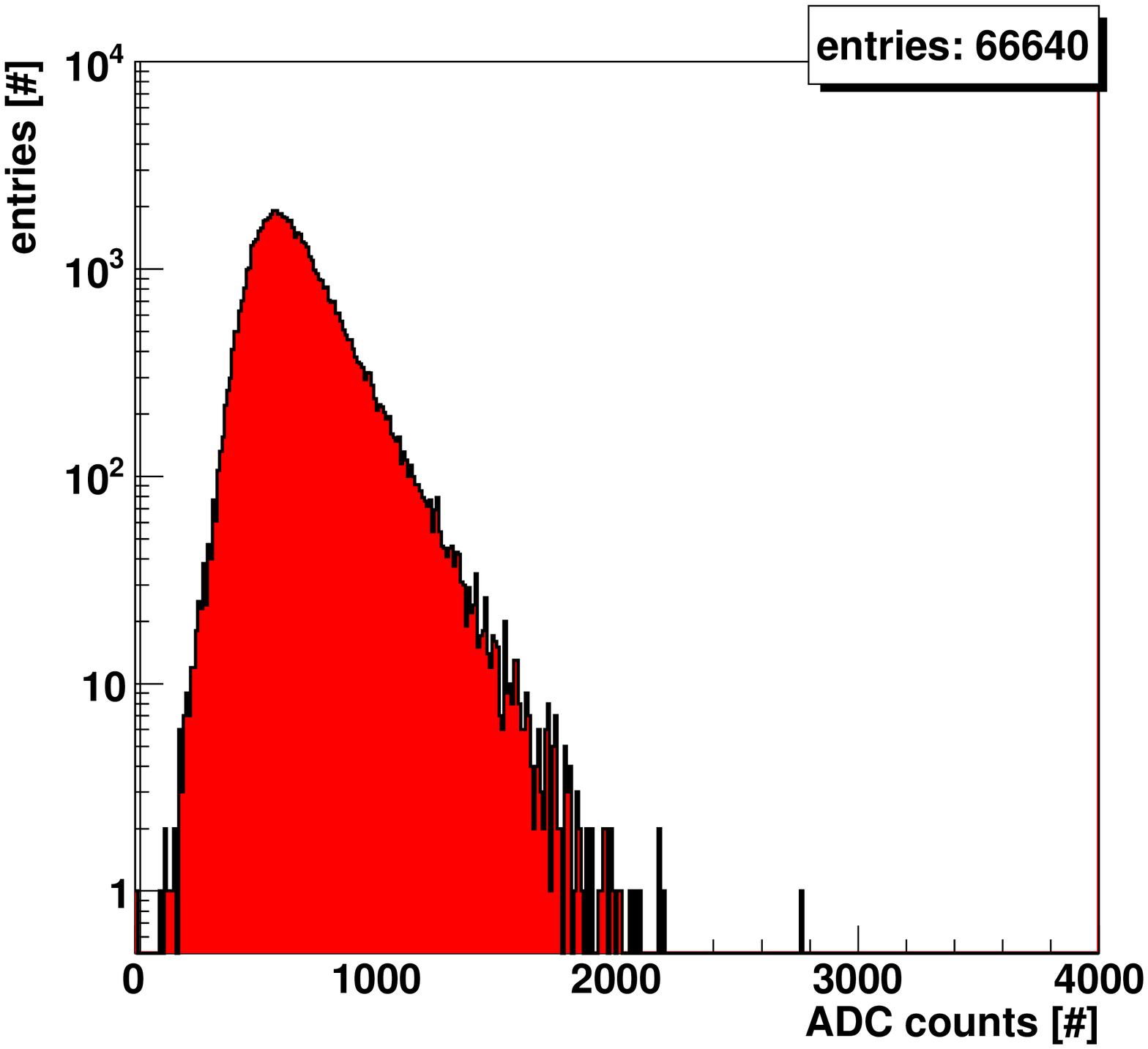,width=8cm}}\captionof{figure}{\label{f-090216_4_0_6_0_4_2_3_acc_adc_norm_highest}Highest ADC values normalized to the pathlength in the scintillator.}
\end{minipage}
\end{center}
\end{figure}

Knowing the direction and the piercing point of a particle with the ACC, the pathlength in the scintillator can be calculated and compared to the measured ADC value. Fig.~\ref{f-081209_4_0_6_0_4_2_3_acc_path_adc_mean} shows that the mean ADC value increases non-linearly with pathlength as already discussed for the ADC calibration of the flight electronics (Sec.~\ref{ss-flightelec}). A fit to the data points gives the mean ADC value as a function of the pathlength $l$ which can be parametrized by:
\be ADC(l)=(929.9\pm12.4)\cdot\sqrt{l/\text{cm}-(0.25\pm0.07)}.\ee
 This is now used to normalize all measured charges to the panel thickness $d=0.8$\,cm: \be ADC\sub{norm}(l)=ADC\sub{measure}\cdot \frac{ADC(d)}{ADC(l)}.\ee The distribution of the highest normalized ADC values is shown in Fig.~\ref{f-090216_4_0_6_0_4_2_3_acc_adc_norm_highest}. The normalized ADC values will also be used in the ACC modeling described in Sec.~\ref{ss-ineffdet}.

\begin{figure}
\begin{center}
\begin{minipage}[b]{.4\linewidth}
\centerline{\epsfig{file=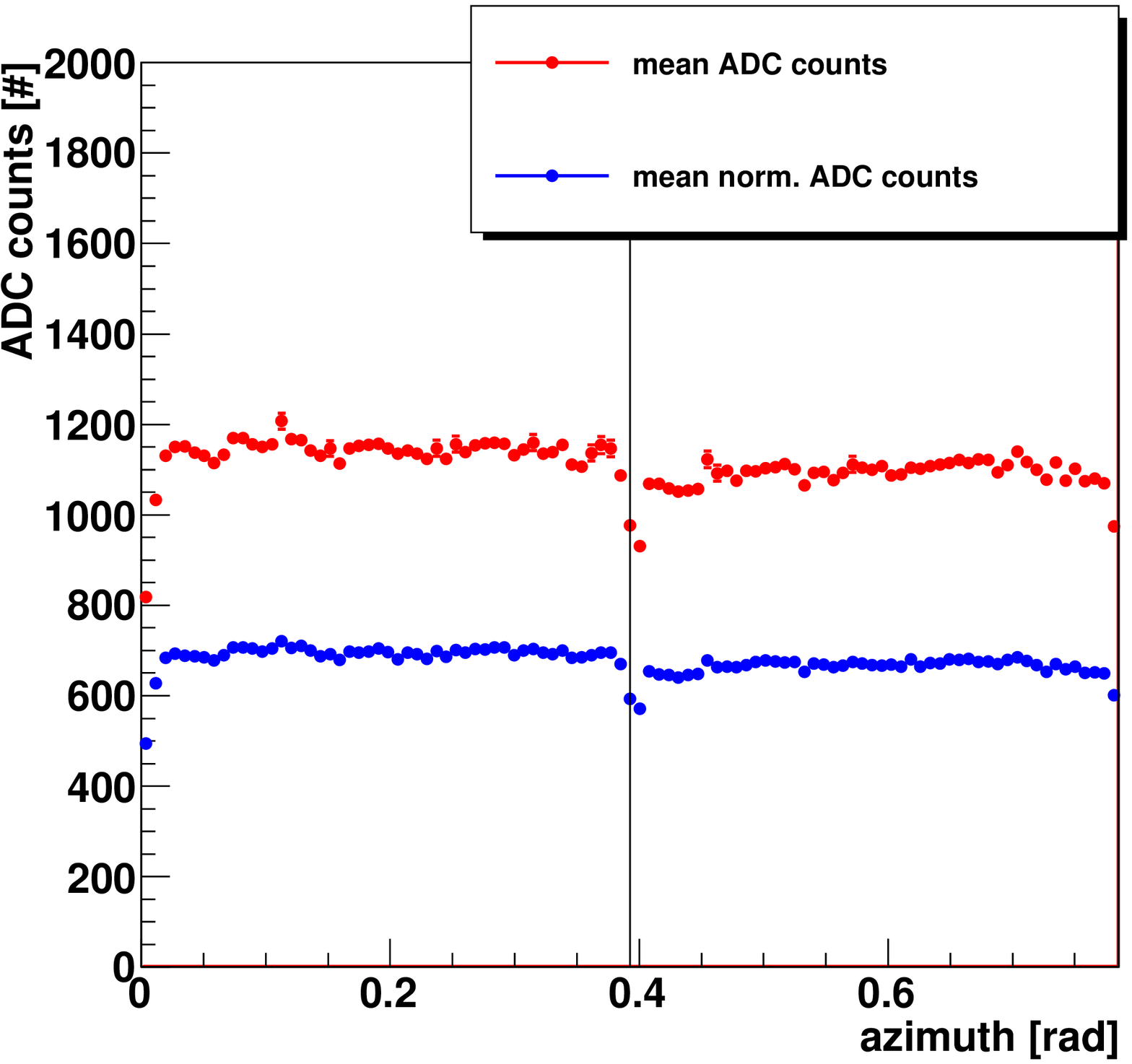,width=8cm}}\captionof{figure}{\label{f-acc_phi_mod_adc_highest_bothmean}Mean (normalized) ADC values vs. the azimuthal position in an ACC sector. The line indicates the slot between the two panels of the sector.}
\end{minipage}
\hspace{.1\linewidth}
\begin{minipage}[b]{.4\linewidth}
\centerline{\epsfig{file=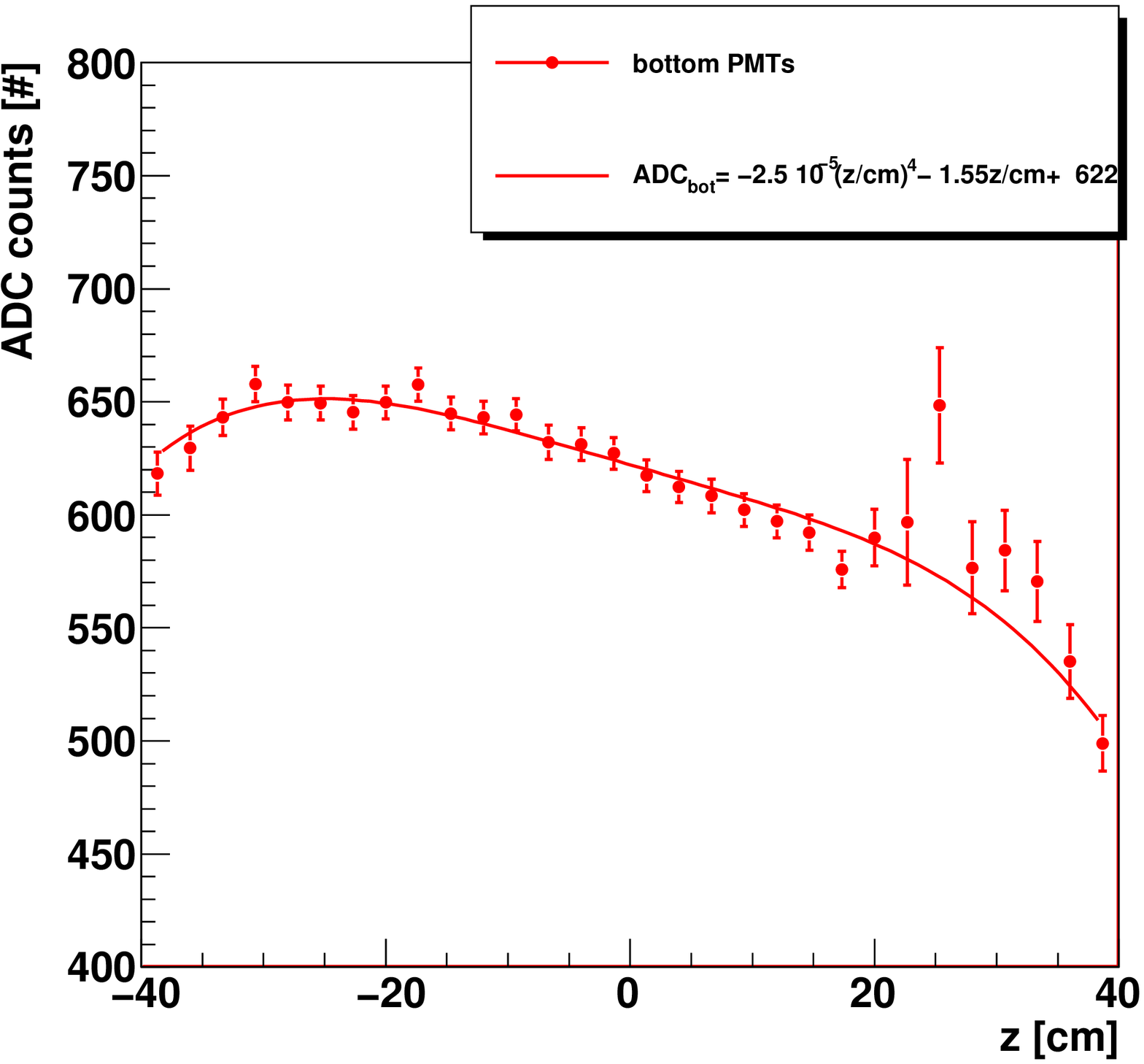,width=8cm}}\captionof{figure}{\label{f-081209_4_0_6_0_4_2_3_acc_z_adc_norm_highest_mean}Mean ADC values of the lower PMTs normalized to the pathlength in the scintillator as a function of $z$ position along the panel.}
\end{minipage}
\end{center}
\end{figure}

Within a sector the mean ADC values as a function of azimuth angle show clear drops at the slot positions (Fig.~\ref{f-acc_phi_mod_adc_highest_bothmean}). The mean drop ($\approx20$\,\%) for the slot region between panels sharing their PMTs is not as strong as for the slots between sectors ($\approx30$\,\%). The signal drop in the testbeam before was larger than that ($\approx50$\,\%, Fig.~\ref{f-0705_yadc24_beam_2acc_ODER_-151_-111_-151_-111} and \ref{f-inefficiency_several_ratio_width}). This is probably due to a much tighter installation during the pre-integration than for the testbeam. Tongues and grooves of the panels are compressed harder thus the gap without scintillator material at the slot regions is smaller.

In addition to attenuation as one moves away from the PMT, Fig.~\ref{f-081209_4_0_6_0_4_2_3_acc_z_adc_norm_highest_mean} shows that the signal drops at both ends of the panel. This is due to the method of embedding the WLS fibers. Within a region of about 5\,cm from each panel end the fraction of fiber to scintillator material is smaller compared to the rest of the panel (Fig.~\ref{f-prod_panel}). The excess with large error bars at about 30\,cm is due to the low statistics at this position because of the event selection requirements (first inner tracker layers). The signal reduction at both ends have not been observed in the measurements shown in Fig.~\ref{f-inefficiency_several_ratio_single} because the trigger counters had a width of 10\,cm and the effect was smeared out. Taking this difference into account a similar behavior in both experimental setups could be found. The signal decreases from the value at the center to the opposite site of the considered PMT in both cases by about 30\,\%. A polynomial fit to the signal behavior in Fig.~\ref{f-081209_4_0_6_0_4_2_3_acc_z_adc_norm_highest_mean} as a function of position $z$ gives:
\be ADC(z) = (-2.5\pm 0.3)\cdot10^{-5}\cdot(z/\text{cm})^4-(1.55\pm0.09)\cdot z/\text{cm}+(622\pm2)\ee
and can be used for the development of an ACC signal model in the next section.

\subsection{Determination of the ACC Inefficiency in Space\label{ss-ineffdet}}

The results of the previous qualification, testbeam, flight electronics and pre-integration measurements can be used to determine the inefficiency of the complete ACC system. In the testbeam only tracks with perpendicular incident on the ACC panel surface were measured. The distribution of particles in space will be isotropic and can result in longer pathlengths in the ACC scintillator material. The particle energy loss of about 1 - 2 \,MeV/cm in the material \cite{pdgbook} is negligible and so the measured charge scales linearly with the pathlength. This was also already found in Sec.~\ref{ss-perfdet}.

\begin{figure}
\begin{center}
\begin{minipage}[b]{.4\linewidth}
\centerline{\epsfig{file=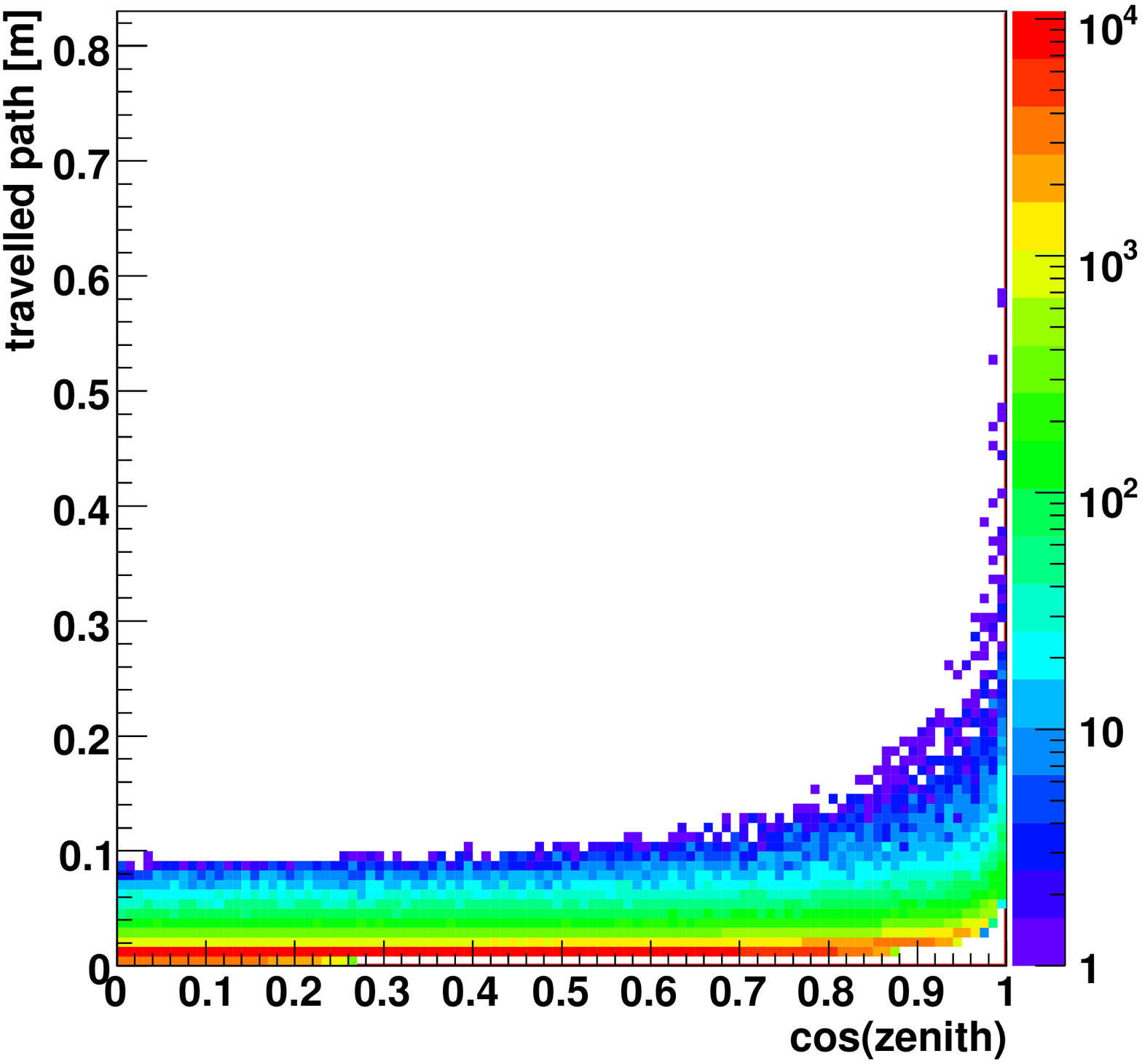,width=8cm}}
\captionof{figure}{\label{f-path_theta}Pathlength travelled in the ACC as a function of the zenith angle. The color code on the right shows the number of entries.}
\end{minipage}
\hspace{.1\linewidth}
\begin{minipage}[b]{.4\linewidth}
\centerline{\epsfig{file=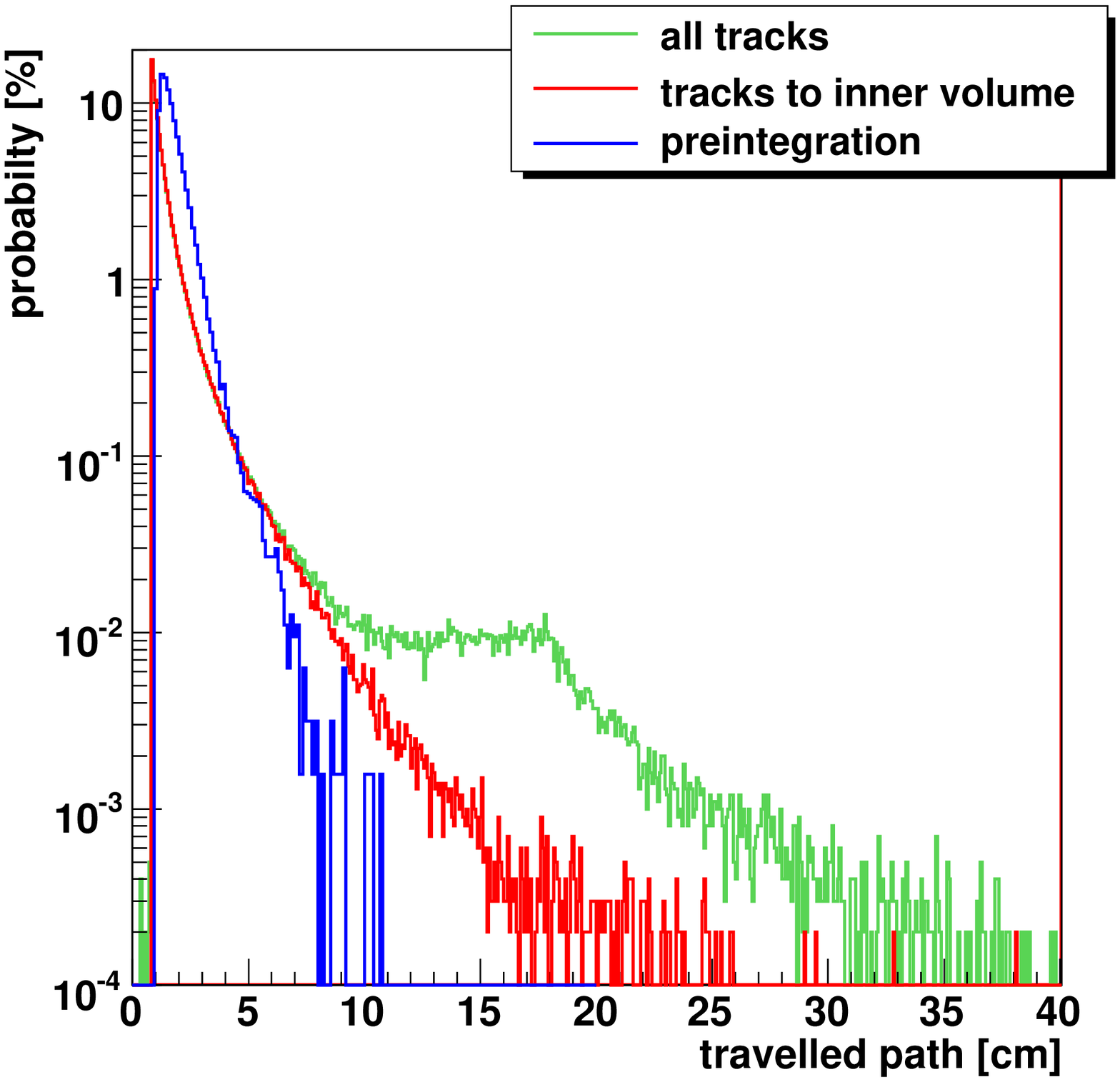,width=8cm}}
\captionof{figure}{\label{f-travelled_path}ACC pathlength probability distribution.}
\end{minipage}
\end{center}
\end{figure}

The same ACC simulation framework described in the last section (Fig.~\ref{f-0711_acceptance_acc}) was used to find the distribution of pathlengths. Particles were started on the walls of the cube with an isotropic distribution and the piercing points with the ACC cylinder were determined. The distribution of pathlengths in the scintillator material as a function of the zenith angle is shown in Fig.~\ref{f-path_theta}. As expected, small zenith angles with long pathlengths are suppressed by the angular acceptance of the ACC. The result for all particles crossing the ACC and reaching the inside of the cylinder is shown as the red histogram in Fig.~\ref{f-travelled_path}. The distribution shows a peak at the panel thickness of 0.8\,cm and has a tail up to much longer pathlengths. The distribution for all possible tracks is shown by the green histogram in the same figure. Starting at about 10\,cm it rises above the red histogram due to tracks which strike the ACC only tangentially and do not cross the inner ACC wall. In comparison, the distribution of pathlengths from pre-integration is shown. The mean of the pre-integration is nearly the same as for the isotropic distribution.

\begin{figure}
\begin{center}
\begin{minipage}[b]{.4\linewidth}
\centerline{\epsfig{file=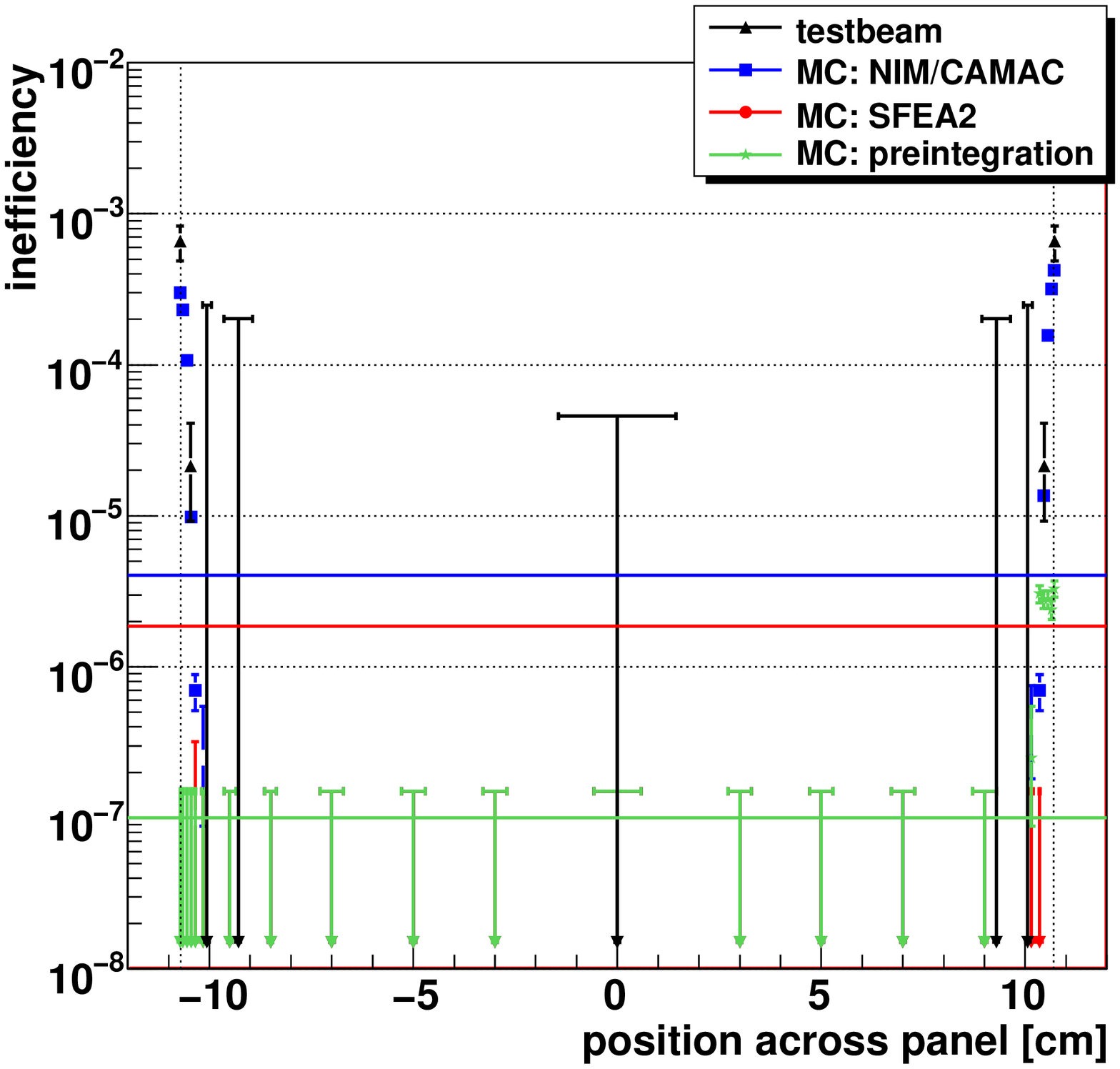,width=8cm}}\captionof{figure}{\label{f-eff_y_model}Expected inefficiencies across an ACC panel. The horizontal lines indicate the corresponding mean values. Points illustrating upper limits are at 95\,\% confidence level.}
\end{minipage}
\hspace{.1\linewidth}
\begin{minipage}[b]{.4\linewidth}
\centerline{\epsfig{file=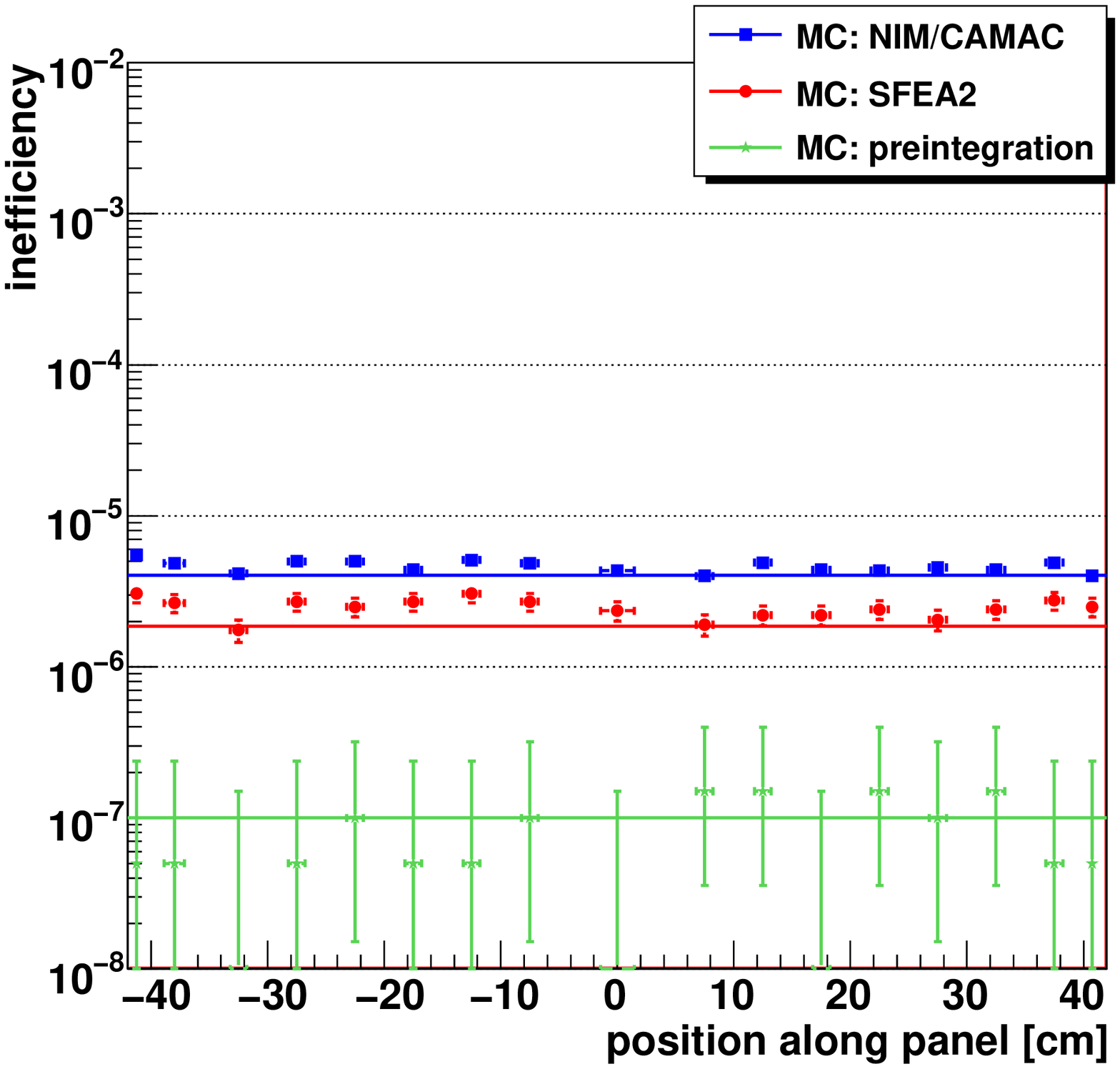,width=8cm}}\captionof{figure}{\label{f-eff_x_model}Expected inefficiencies along an ACC panel. The horizontal lines indicate the corresponding mean values.}
\end{minipage}
\end{center}
\end{figure}

To simulate the signal output of the complete ACC system in space the charges measured in the testbeam $Q\sub{testbeam}$ (Sec.~\ref{ss-testbeam}) must be modified. The first modification factor $f\sub{panel}$ results from the system test of the ACC in flight configuration (Sec.~\ref{ss-systemtest}) and is defined as the ratio of the number of photo-electrons measured in the system test to that measured in the testbeam for the central region ($N\sub{pe, central}=15$). The second factor $f\sub{path}$ takes into account the increase of the charge $Q$ with pathlength in the scintillator. In the simulation these factors are randomly chosen according to the distributions in Fig.~\ref{f-paneldis_systemtest} for $f\sub{panel}$ and for $f\sub{path}$ in Fig.~\ref{f-travelled_path}. The resulting charge $Q$:
\be
Q=Q\sub{testbeam}\cdot f\sub{panel}\cdot f\sub{path}\label{e-qmod}.
\ee
is used to calculate the inefficiency. This increases the MOP value by about 35\,\%. The detection inefficiency is of course dominated by the losses in the slot region between the panels where about 0.1\,\% of the events are missed. An upper inefficiency limit is obtained on the basis of the testbeam measurements for the mean inefficiency across the ACC panel: \be\bar I\sub{testbeam}<1.1\cdot10^{-4}\text{ @ 95\,\% confidence level.}\ee The inefficiency data points corrected for $f\sub{panel}$ and $f\sub{path}$ are shown in Fig.~\ref{f-eff_y_model}.

\begin{figure}
\begin{center}
\begin{minipage}[b]{.4\linewidth}
\centerline{\epsfig{file=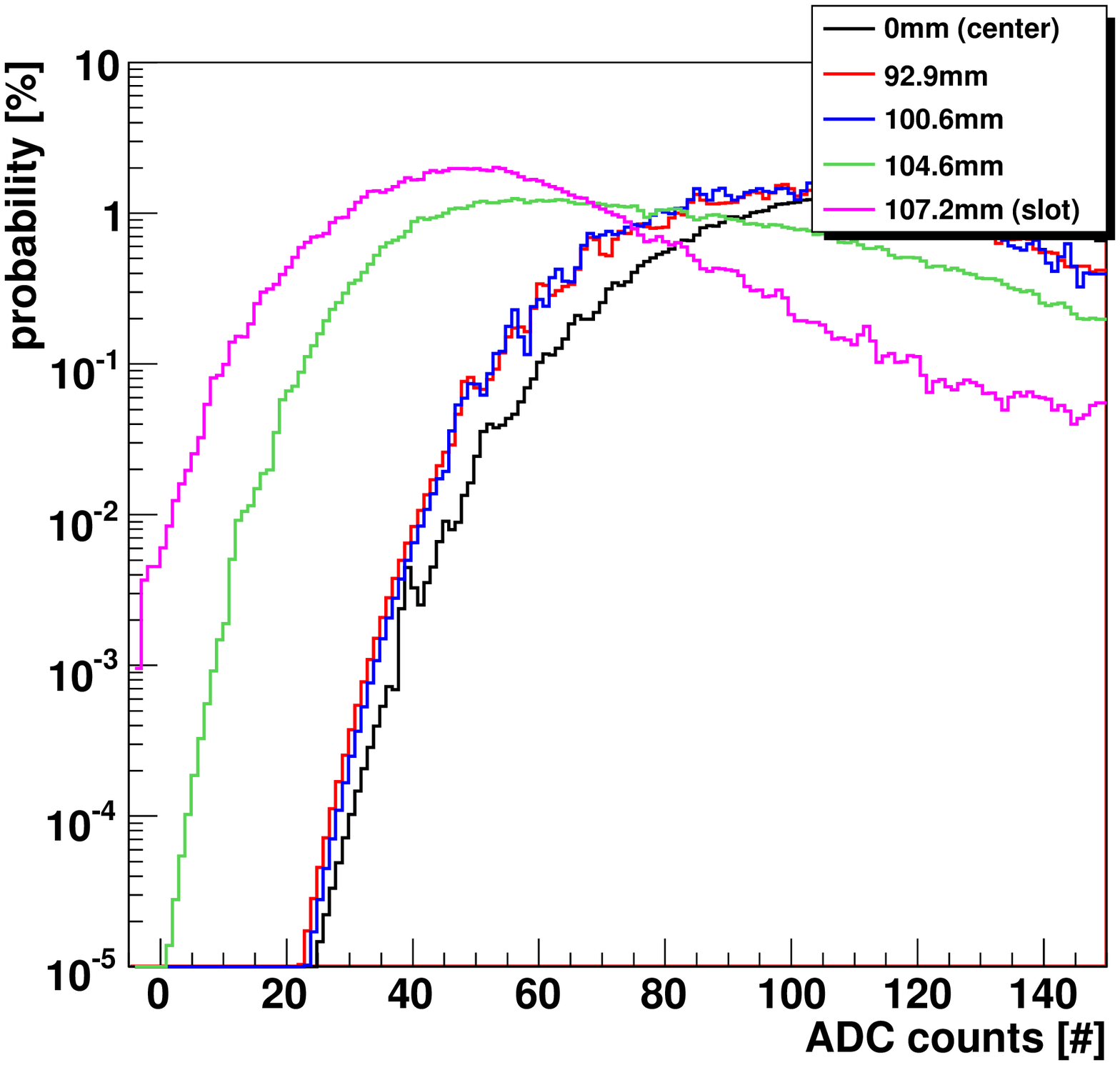,width=8cm}}\captionof{figure}{\label{f-testbeam_highest_norm_adc}Normalized and smoothed testbeam ADC spectra for the inefficiency calculation.}
\end{minipage}
\hspace{.1\linewidth}
\begin{minipage}[b]{.4\linewidth}
\centerline{\epsfig{file=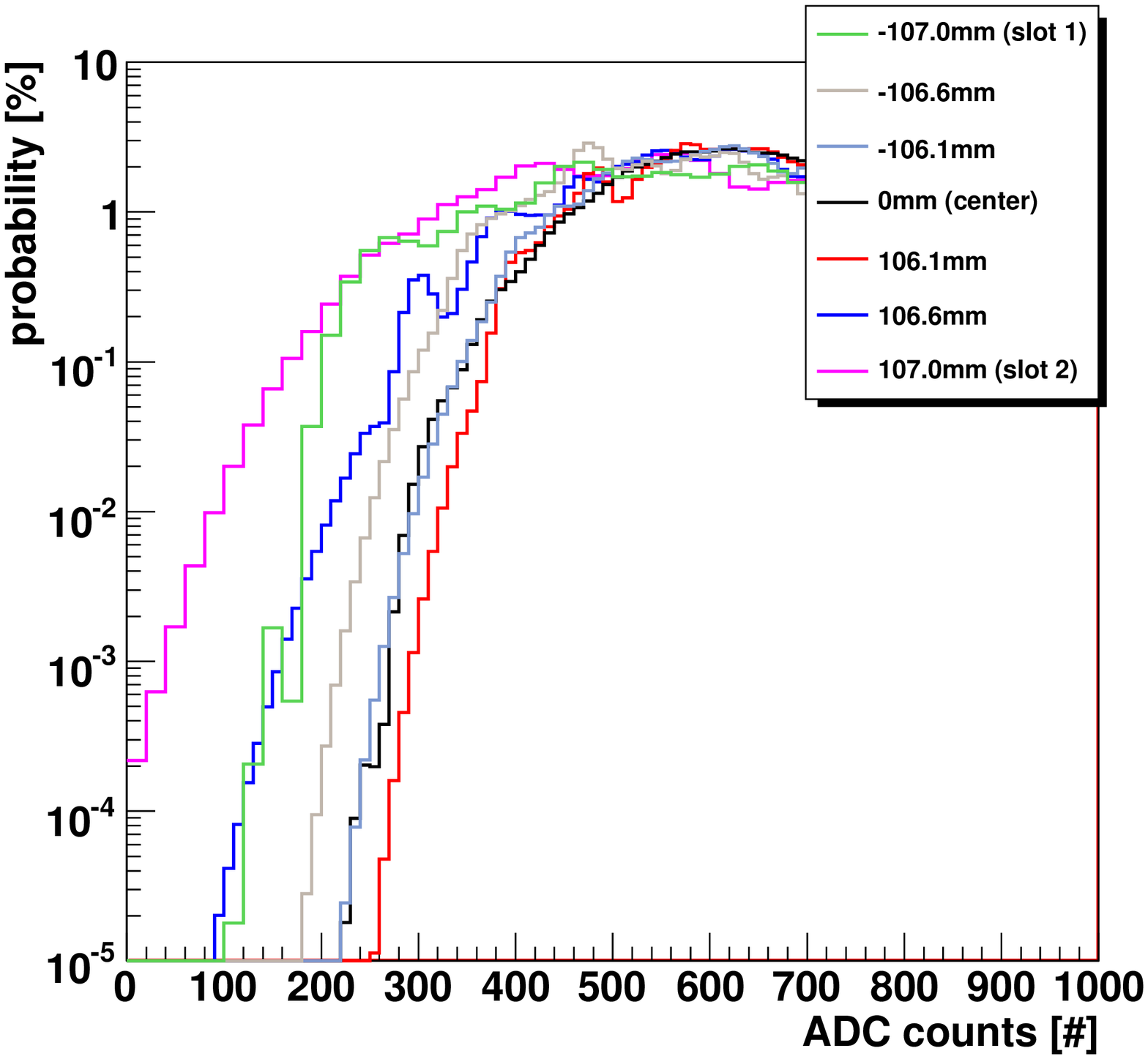,width=8cm}}\captionof{figure}{\label{f-pre-integration_highest_norm_adc}Normalized and smoothed pre-integration ADC spectra for the inefficiency calculation.}
\end{minipage}
\end{center}
\end{figure}

To estimate the inefficiency over the whole panel a simulation was performed assuming only statistical signal fluctuations and no additional effects (e.g. inefficiencies of the electronics):

\bi
\st The position $(x,y)$ on the panel is chosen randomly in the panel coordinate system introduced in Sec.~\ref{ss-testbeam}.
\st The testbeam spectra measured at different $y$ positions are smoothed by fitting them with a Landau function on the rising edge of the spectra (Fig.~\ref{f-testbeam_highest_norm_adc}). They are then used to obtain an interpolated spectrum $S_y$ for the chosen $y$ position according to: \be S_y = w\sub{below}\cdot S\sub{below} + w\sub{above}\cdot S\sub{above}\ee where $S\sub{below}$ and $S\sub{above}$ are the measured testbeam spectra for $y\sub{below}<y<y\sub{above}$ and weight factors are given by \be w\sub{below} = 1-\frac{b -y}{y\sub{above}-y\sub{below}}\quad\wedge\quad w\sub{above} = \frac{b -y}{y\sub{above}-y\sub{below}},\ee $b=10.72$\,cm being the half-panel width.
\st Two random ADC values are generated from these spectra to simulate the two PMTs.
\st The values are smeared according to signal attenuation during transport to the PMT (Fig.~\ref{f-inefficiency_several_acc0_acc1_ratio}) and by the two random factors of Eq.~\ref{e-qmod}.
\st The $x$ position along the panel influences the ratio of the two ADC values. They must be multiplied by factors according to the fit shown in Fig.~\ref{f-inefficiency_several_ratio_single}.
\st If the particle traverses the slot regions two more random ADC values are generated in the same way as above to take into account the adjacent panel. The two different types of slot regions (same sector and between two sectors) must be treated in different ways:

Two ADC values for each panel side are added in the case of panels within the same sector. Four individual values are kept if the particle traverses between two sectors.
\st The highest ADC value is determined and used for the next step.
\st In the last step the ADC values are corrected for the relative calibration of the testbeam electronics (CAMAC) and the flight electronics (SFEA2 board). Therefore, the charge in pC is calculated from the ADC values of the testbeam using the calibration (Fig.~\ref{f-0712_pedcorr_beam_1acc_ODER_-1000_0} and \ref{f-integrated_charge_26_28}). Then this charge is converted to the ADC value of the flight electronics according to Fig.~\ref{f-all_cali}.
\ei

A very similar simulation based on the pre-integration data was also carried out. Neglecting the missed event, the averaged spectra normalized to the panel thickness of all PMTs are also extrapolated to smaller ADC values (Fig.~\ref{f-pre-integration_highest_norm_adc}). The signals are smeared due to signal attenuation during transport according to Fig.~\ref{f-081209_4_0_6_0_4_2_3_acc_z_frac_norm_center} and the mean signal change as a function of pathlength is done following the behavior in Fig.~\ref{f-081209_4_0_6_0_4_2_3_acc_path_adc_mean}. The factor due to the position along the panel results from the fit in Fig.~\ref{f-081209_4_0_6_0_4_2_3_acc_z_adc_norm_highest_mean}.

Fig.~\ref{f-eff_y_model} and \ref{f-eff_x_model} show the inefficiencies across and along the panel. In Fig.~\ref{f-eff_y_model} the slot region between panels within the same sector is at -10.72\,cm and the slot region between two sectors is at 10.72\,cm. A good event is again defined by showing at least one pulseheight above 3 RMS of the corresponding pedestal distribution. As expected, the inefficiency increases at the slot positions and stays nearly constant along the panel. In addition, the mean inefficiencies for the NIM/CAMAC electronics, the SFEA2 electronics and the pre-integration configuration are shown. They are calculated from the average signal distribution of the complete panel. Based on the testbeam measurements the mean inefficiency is for NIM/CAMAC electronics:\be\bar I\sub{NIM/CAMAC}=(4.05\pm0.45)\cdot10^{-6}.\ee  The flight electronics can improve this value to: \be\bar I\sub{SFEA2}=(1.85\pm0.30)\cdot10^{-6}\ee due to a better resolution for small charges. The simulation based on the pre-integration data gives an even smaller mean inefficiency for the complete ACC system of \be\bar I\sub{pre}=1.0^{+0.9}_{-0.6}\cdot10^{-7}\qquad\text{or}\qquad \bar I\sub{pre}<3.2\cdot10^{-7} \text{@ 95\,\% confidence level}\ee due to the tighter compression of tongues and grooves of the panels. All these calculations show that the ACC system is able to detect charged particles very reliably.

\section{Projection of Antiparticle Measurements with AMS-02 \label{s-ams_proj}}

The following section discusses the ability of the AMS-02 detector to measure antihelium nuclei, antiprotons and positrons. First the role of the ACC is considered and then the search for antiparticles is discussed.

\subsection{Role of the ACC in the Measurements}

\begin{figure}
\begin{center}
\begin{minipage}[b]{.4\linewidth}
\centerline{\epsfig{file=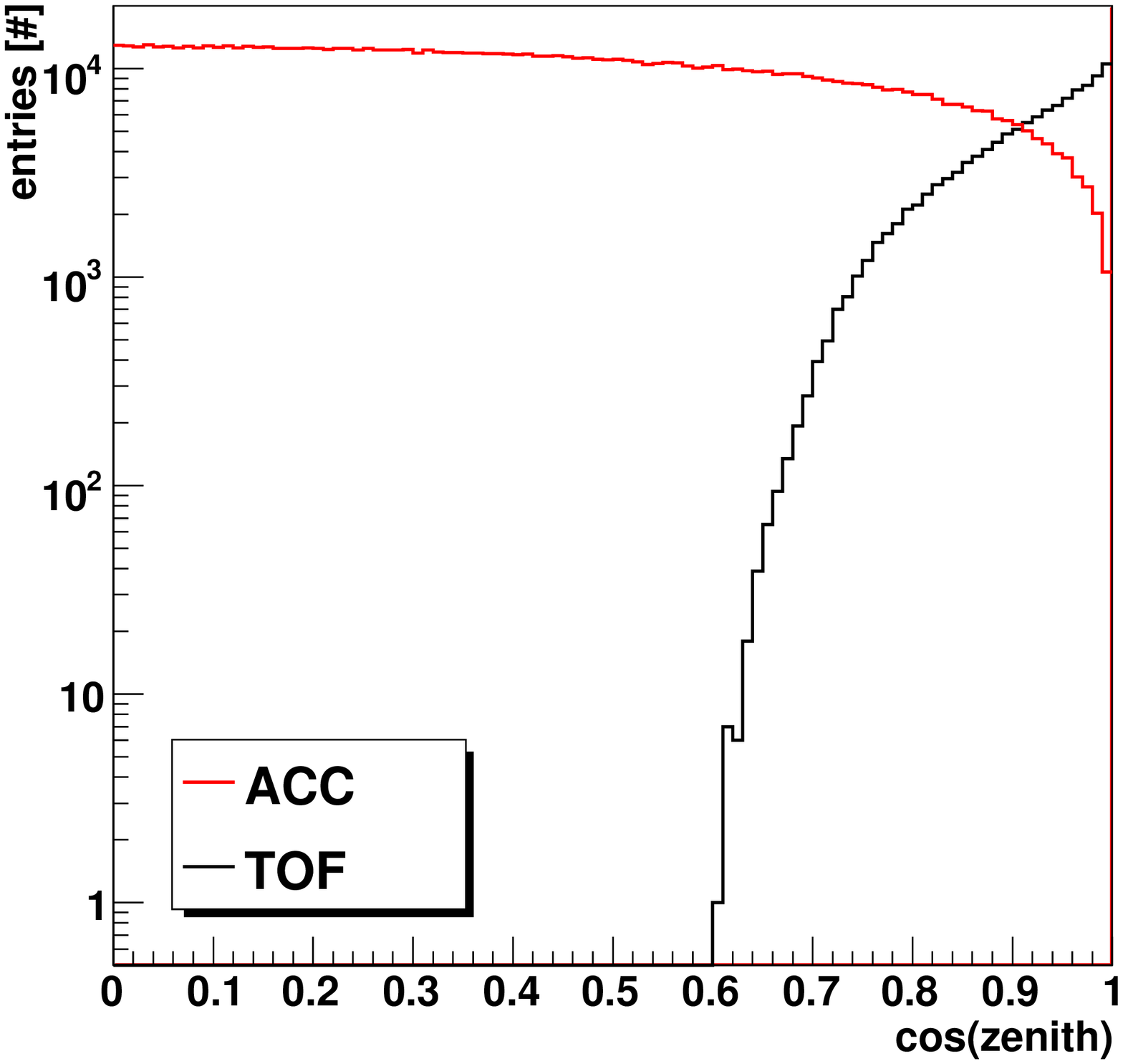,width=8cm}}\captionof{figure}{\label{f-c_theta}Zenith angle distribution of cosmic rays in TOF and ACC.}
\end{minipage}
\hspace{.1\linewidth}
\begin{minipage}[b]{.4\linewidth}
\centerline{\epsfig{file=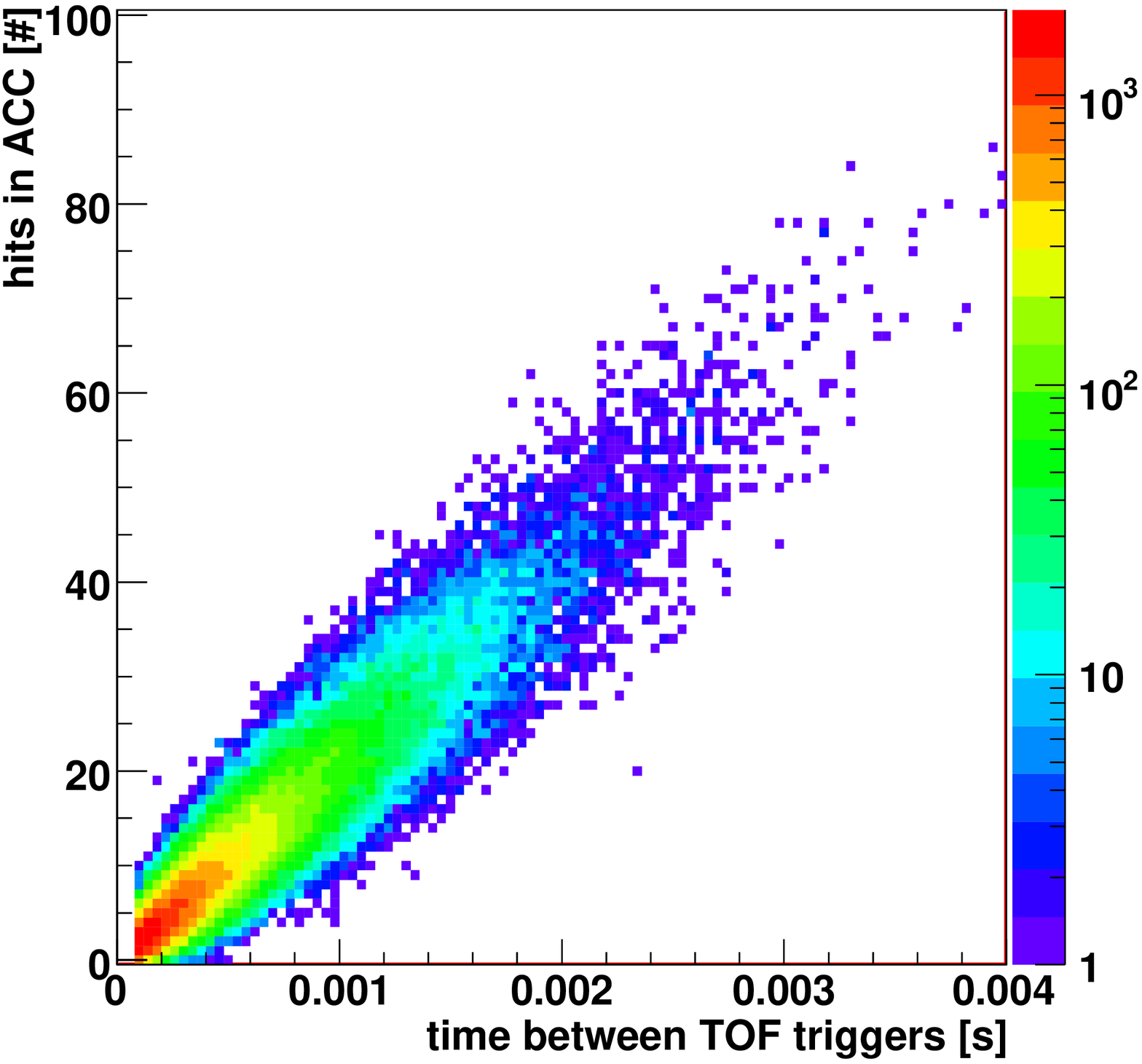,width=8cm}}\captionof{figure}{\label{f-c_time_hits}Correlation between number of hits in the ACC and time between TOF triggers at 3\,kHz rate.}
\end{minipage}
\end{center}
\end{figure}

The anticoincidence counter has been designed for two purposes. The first is to reduce the trigger rate during periods of high particle flux. The ACC Monte Carlo simulation framework described above is used to calculate the trigger reduction capability of the ACC. With reference to Fig.~\ref{f-0711_acceptance_acc} the acceptance $G\sub{ACC}$ of the ACC for particles reaching the inner ACC volume is given by \cite{sullivan-1971}:
\begin{eqnarray}
G\sub{ACC} &=& G\sub{cube}\cdot\frac{\displaystyle N\sub{hit}}{\displaystyle N\sub{total}},\\
G\sub{cube} &=& \pi\cdot 6l^2
\end{eqnarray}
where $l$ is the length of the cube edge, $N\sub{total}$ is the total number of trajectories generated and $N\sub{hit}$ the number of particles crossing the ACC panels. For $N\sub{hit} = 10^6$ and for $l=1.4$\,m the acceptance is $G\sub{ACC}=(8.877\pm 0.010)$\,m$^2$sr. The acceptance of the TOF is calculated in the same way and is modeled in a simplified way as two square planes of 1.3\,m length at a distance of 1.3\,m. A TOF hit is defined as a particle which strikes both the upper and lower planes. The resulting acceptance is $G\sub{TOF}=(1.060\pm0.003)$\,m$^2$sr. Thus on average 8.4 more particles hit the ACC cylinder than the number of TOF triggers generated. Fig.~\ref{f-c_theta} shows the zenith angle distributions for the TOF and the ACC. The angular acceptance of the ACC has a different shape since the walls of the ACC cylinder are perpendicular to the TOF planes. This is crucial as the reduction of the data rate during periods of high particle flux depends on vetoing events with a TOF trigger generated by one particle while a second particle enters the detector from the side. The maximum trigger rate for AMS-02 is limited by the readout and trigger decision of the ACC. The whole AMS-02 readout chain can be reliably operated up to a maximum event rate $R\sub{max}\approx3$\,kHz without using the ACC to reduce the trigger\cite{ams}. The corresponding maximum flux integrated over all energies $F\sub{int}$ without ACC reduction can be calculated to be:
\be F\sub{int} = \frac{R\sub{max}}{G\sub{TOF}}.\ee
The ACC becomes important for values of $F\sub{int}$ in the range $10^3$ - $10^4$\,m$^{-2}$sr$^{-1}$s$^{-1}$ which is of the order of the total proton flux. It was shown in Sec.~\ref{ss-flightelec} that the rate can be reduced with the ACC by about $10^4$. The maximum of the low-energetic proton flux at about 400\,km altitude in the South Atlantic Anomaly is about $7\cdot10^7$\,GeV$^{-1}$m$^{-2}$s$^{-1}$ \cite{buhler-2002}. The ACC allows reasonable measurements with AMS-02 up to total fluxes of about $10^5$\,m$^{-2}$sr$^{-1}$s$^{-1}$, respecting the trigger decision time of 1.4\,\textmu s\cite{kounine-2009} and the acceptance.

\begin{figure}
\begin{center}
\begin{minipage}[b]{.4\linewidth}
\centerline{\epsfig{file=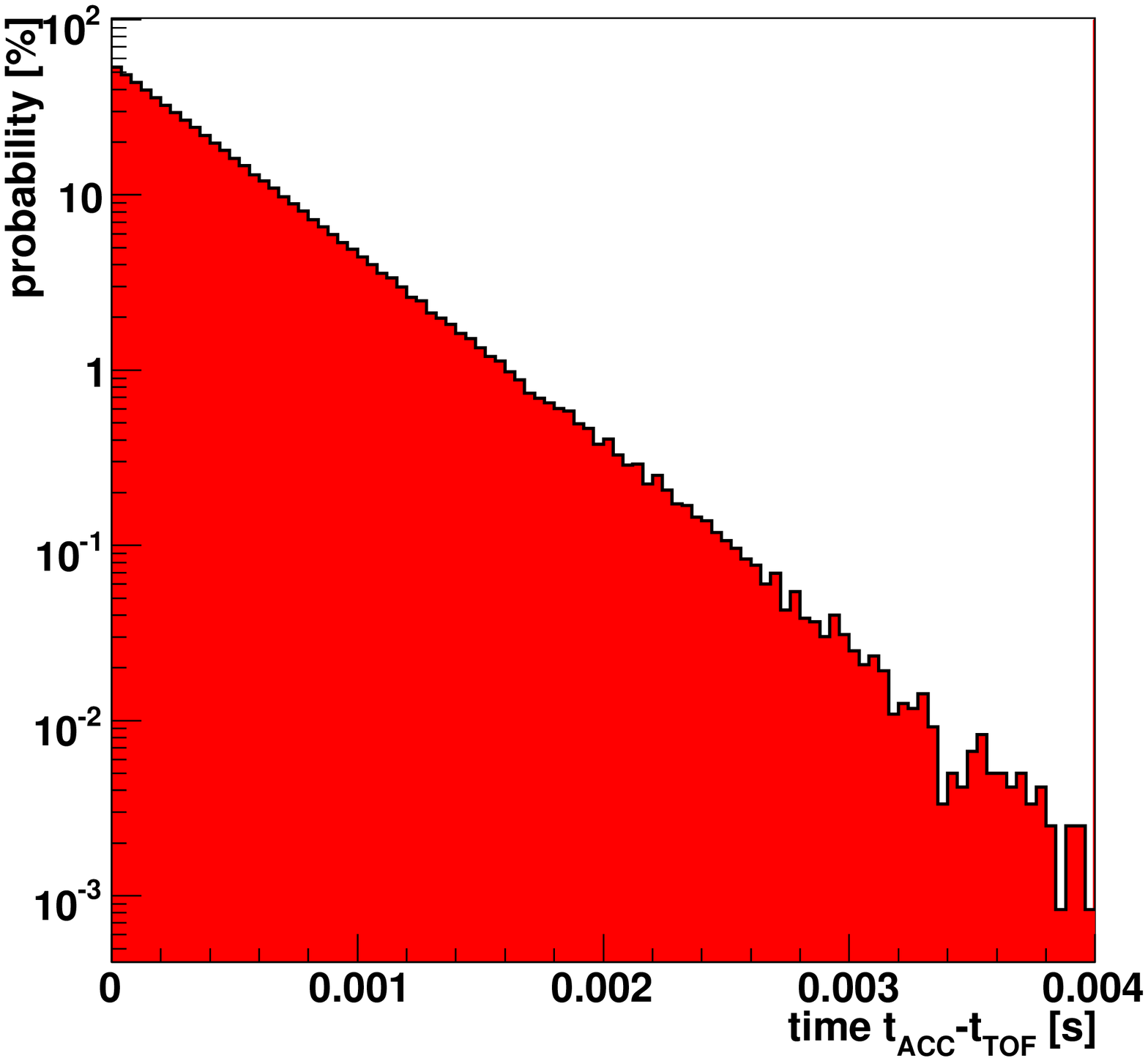,width=8cm}}\captionof{figure}{\label{f-c_time_tof_acc}Time between external events which hit at least one tracker plane and TOF trigger events normalized to the number of TOF triggers.}
\end{minipage}
\hspace{.1\linewidth}
\begin{minipage}[b]{.4\linewidth}
\centerline{\epsfig{file=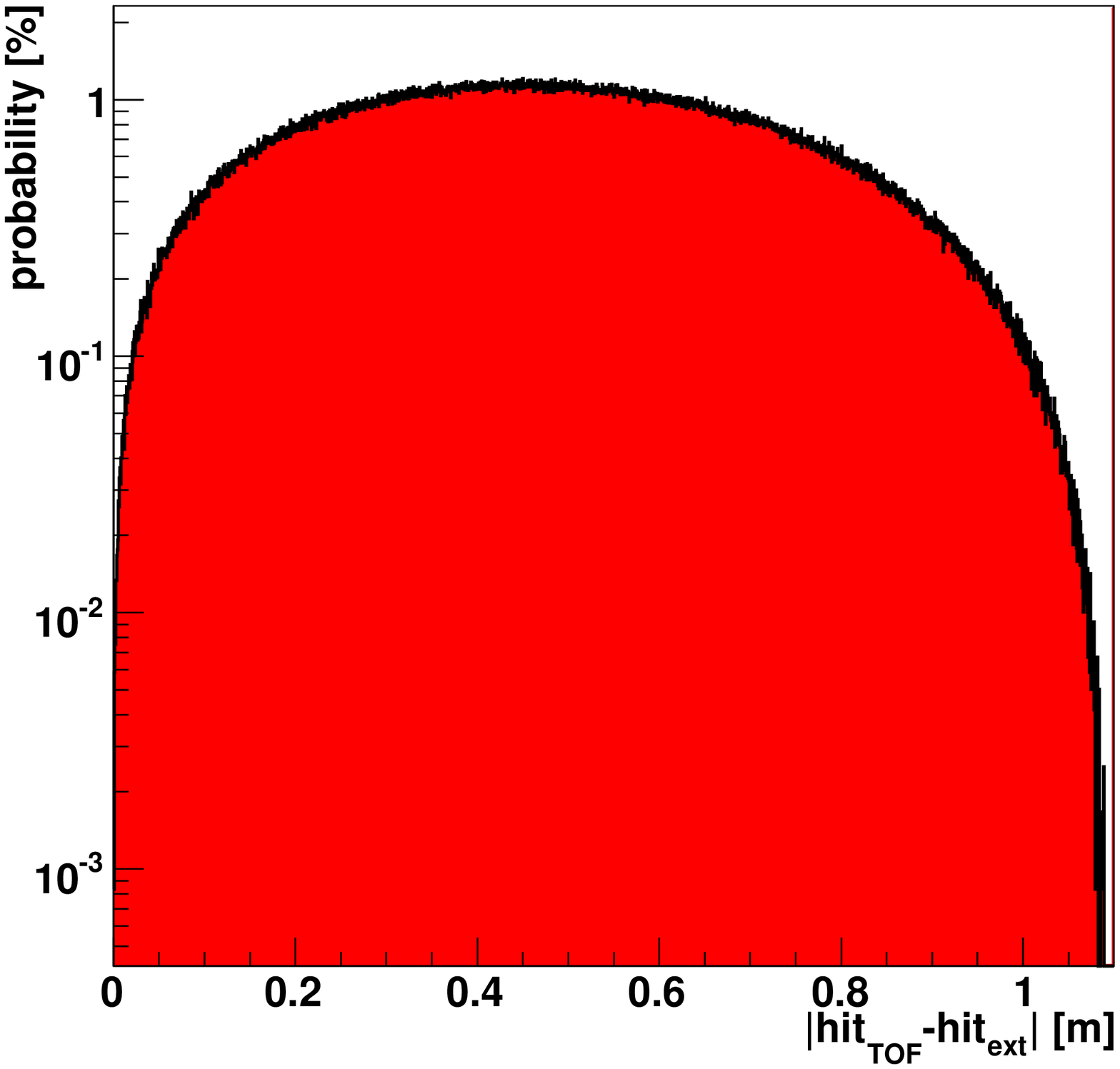,width=8cm}}\captionof{figure}{\label{f-c_diff_tof_acc_r}Probability for the distance between external events and TOF events in the tracker normalized to the number of TOF triggers.}
\end{minipage}
\end{center}
\end{figure}

The second task of the ACC is to assure a clean track reconstruction. It is important to reject external events crossing the tracker from the side between two TOF triggers and to detect secondary interactions in the detector (Fig.~\ref{f-ext_event_acc} and \ref{f-int_event_acc}). These two effects can spoil the charge determination and limit the precision of the measurement.

The probability that an external particle spoils the charge measurement depends on its proximity in time and space to one or more hits in the tracker resulting from particles satisfying the TOF trigger. At a constant flux the time interval between the detection of particles is described by a homogeneous Poisson process and the interval length follows an exponential distribution. In the simulation, the exponential time distribution between particles generated on the TOF planes is adjusted so that the maximum TOF trigger rate of 3\,kHz is achieved. As expected, the number of external events in the time interval between two TOF triggers increases linearly with the length of this interval (Fig.~\ref{f-c_time_hits}). From the tests with atmospheric muons it is known that the first ACC PMT pulse occurs within a 140\,ns interval around the TOF trigger (Fig.~\ref{f-1212576870_S3_sfea2_time_no_cut_1}). Fig.~\ref{f-c_time_tof_acc} shows the probability distribution for the interval between TOF triggers and external events in the ACC normalized to the total number of TOF triggers. The exponential distribution clearly favors hits that are close in time. Indeed about 50\,\% of the TOF trigger events are followed within 40\,\textmu s by external ACC events.

The distribution of spatial distances $r=\sqrt{\Delta x^2 + \Delta y^2}$ between two hits, one from an external event and one from a TOF triggered event, in the eight tracker planes is shown in Fig.~\ref{f-c_diff_tof_acc_r}. The distribution drops close to the maximum and minimum possible values. 

The overall probability for hits to be close both in time and space can be calculated from the convolution of these two distributions. The probability for external particles crossing the ACC and hitting the tracker within 1\,mm and 140\,ns in relation to a TOF trigger is determined to be about $\cal{O}$$(10^{-8})$. Taking into account the values for the measured ACC inefficiency, external events can cause a wrong determination in about 1 out of $10^{13}$ cases. However, in the two gaps of 114\,mm between the ACC and the highest and lowest tracker layers external particles can get into the detector without crossing the ACC which results in an inefficiency of $\cal{O}$$(10^{-10})$. It should be stressed though that these values are conservative upper limits since the tracker resolution is about $\cal{O}$(10\,\textmu m) rather than 1\,mm.

The rejection against particles arising from internal secondary interactions in the tracker, the electromagnetic calorimeter or elsewhere can only be obtained from a full detector simulation.

\subsection{Capability of Antiparticle Measurements with AMS-02\label{ss-antimeas}}

As mentioned above (Sec.~\ref{s-gev}), checking for deviations from the expected background for antiparticles like positrons or antiprotons could hint at the existence of new so far unknown sources of antiparticles. In the following, the AMS-02 detector capability to measure spectra and fractions of positrons and antiprotons with and without using the electromagnetic calorimeter will be investigated. The ECAL lowers the acceptance from 0.45\,m$^2$sr to 0.095\,m$^2$sr \cite{choutko2} due to its size and position and therefore reduces statistics by a factor of about 5.

\begin{figure}
\begin{center}
\begin{minipage}[b]{.4\linewidth}
\centerline{\epsfig{file=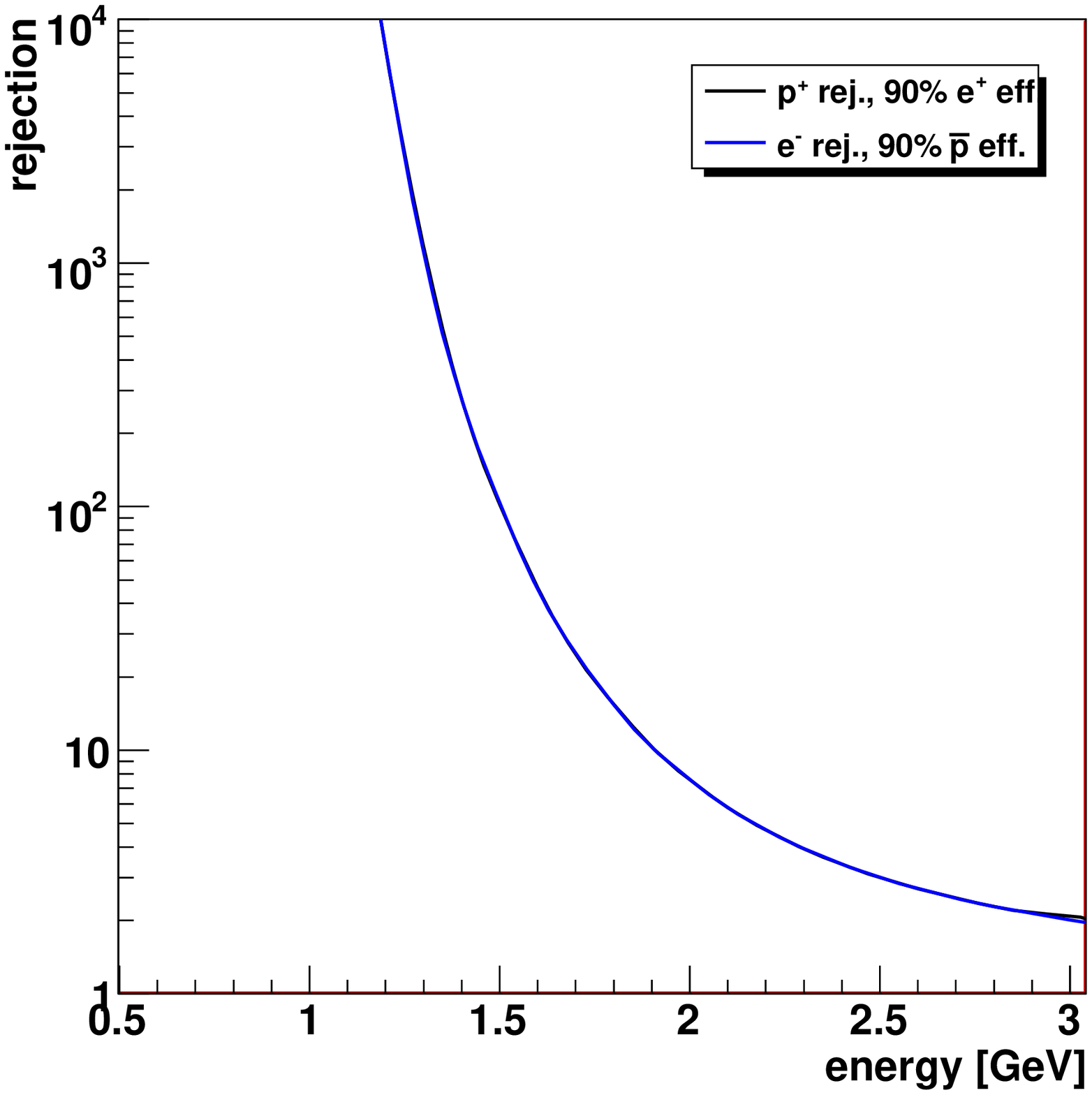,width=8cm}}\captionof{figure}{\label{f-tof_ams_rejection}Proton and electron rejection factors in the TOF. The two curves are on top of each other.}
\end{minipage}
\hspace{.1\linewidth}
\begin{minipage}[b]{.4\linewidth}
\centerline{\epsfig{file=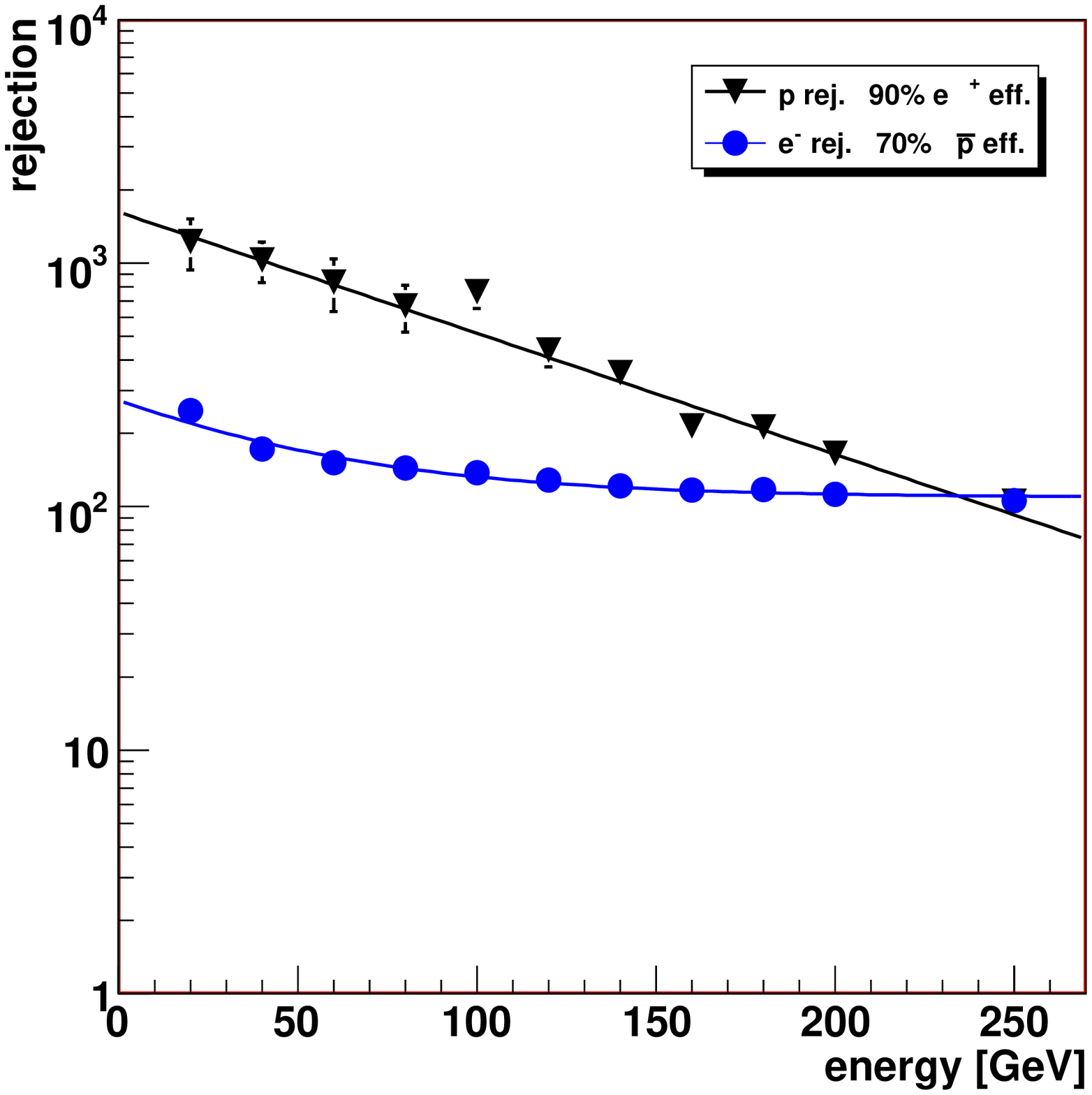,width=8cm}}\captionof{figure}{\label{f-e_p_rejec}Proton and electron rejection factors in the TRD \cite{doetinchem-2006-558}.}
\end{minipage}
\end{center}
\end{figure}

Discriminating positrons from protons and antiprotons from electrons is done by TOF and TRD. As for the PEBS analysis the rejection requirement depends on the respective flux ratios (Fig.~\ref{f-ratio_mod} and Sec.~\ref{ss-atmosim}). The calculation of the rejection by the TOF is done in the same way as for PEBS (Eq.~\ref{e-tof1} - \ref{e-tof2}). The distance between the TOF planes is 1.3\,m and the time resolution is $\sigma_t=100$\,ps. Fig.~\ref{f-tof_ams_rejection} shows the rejection factor for protons (electrons) as a function of particle energy, calculated for 90\,\% detection efficiency of positrons (antiprotons). The TRD rejection is known from testbeam measurements (Fig.~\ref{f-e_p_rejec}).

\begin{figure}
\begin{center}
\begin{minipage}[b]{.4\linewidth}
\centerline{\epsfig{file=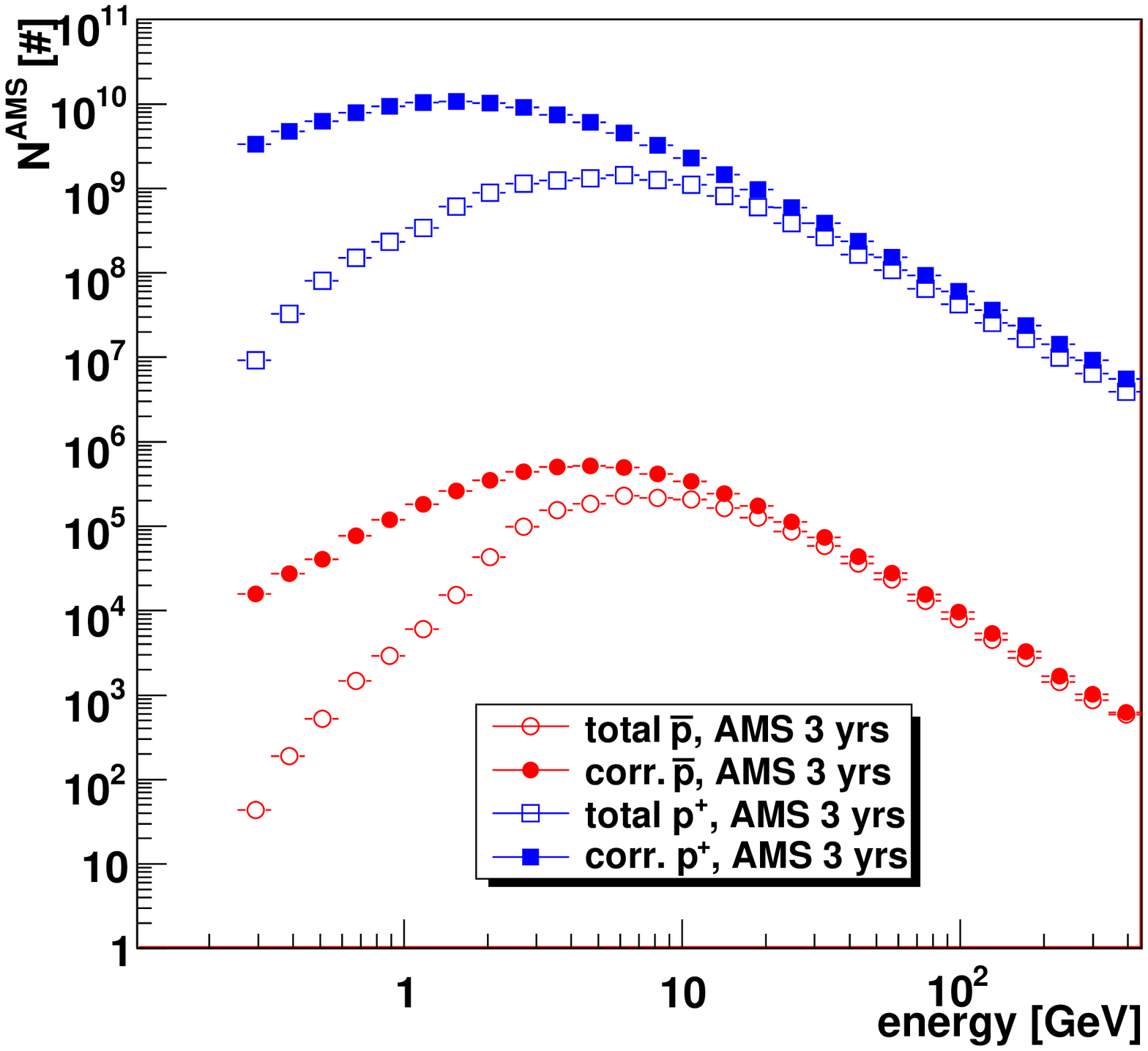,width=8cm}}\captionof{figure}{\label{f-pbar_p_nams_dm_hg_phi550_40000_m_southpole_1095_days}Projected number of total and corrected proton and antiproton events using the TOF, TRD and tracker subdetectors.}
\end{minipage}
\hspace{.1\linewidth}
\begin{minipage}[b]{.4\linewidth}
\centerline{\epsfig{file=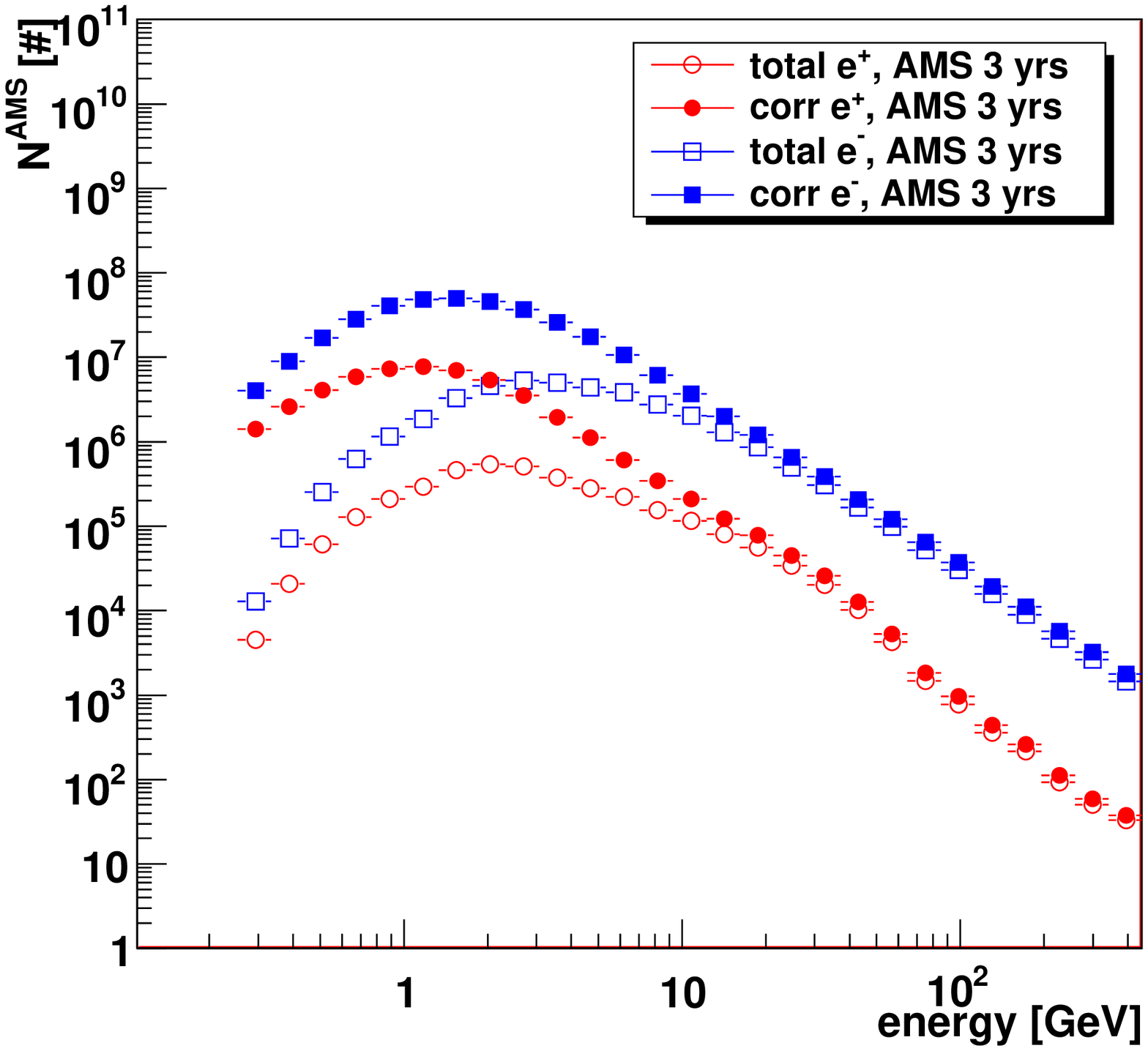,width=8cm}}\captionof{figure}{\label{f-e+_e-_nams_dm_hg_phi550_40000_m_southpole_1095_days}Projected number of total and corrected electron and positron events using the TOF, TRD, tracker and ECAL subdetectors.}
\end{minipage}
\end{center}
\end{figure}

\begin{figure}
\begin{center}
\begin{minipage}[b]{.4\linewidth}
\centerline{\epsfig{file=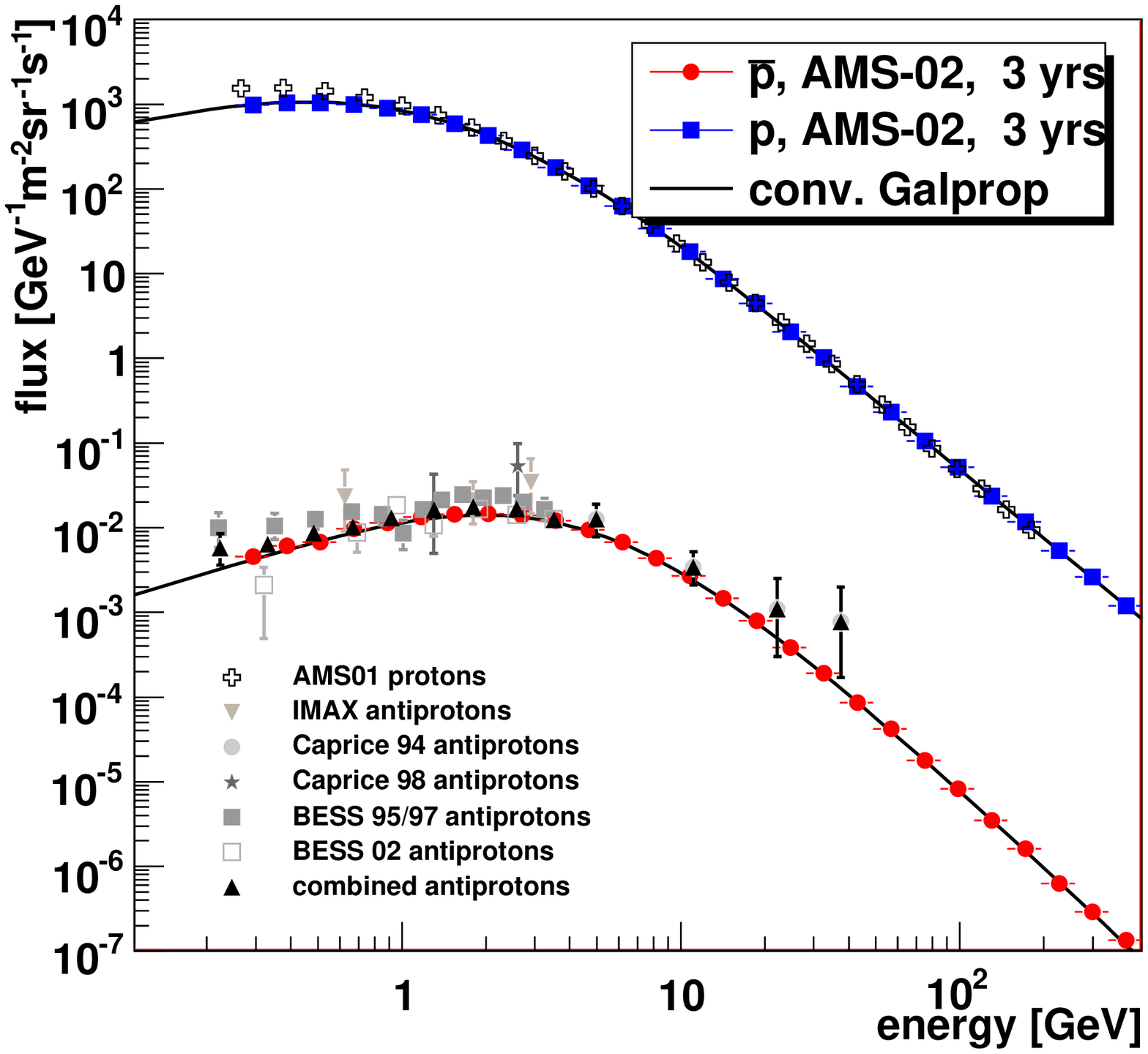,width=8cm}}\captionof{figure}{\label{f-pbar_p_fluxes_dm_hg_phi550_40000_m_southpole_1095_days}Projected proton and antiproton flux measurements by AMS-02 using the TOF, TRD and tracker subdetectors.}
\end{minipage}
\hspace{.1\linewidth}
\begin{minipage}[b]{.4\linewidth}
\centerline{\epsfig{file=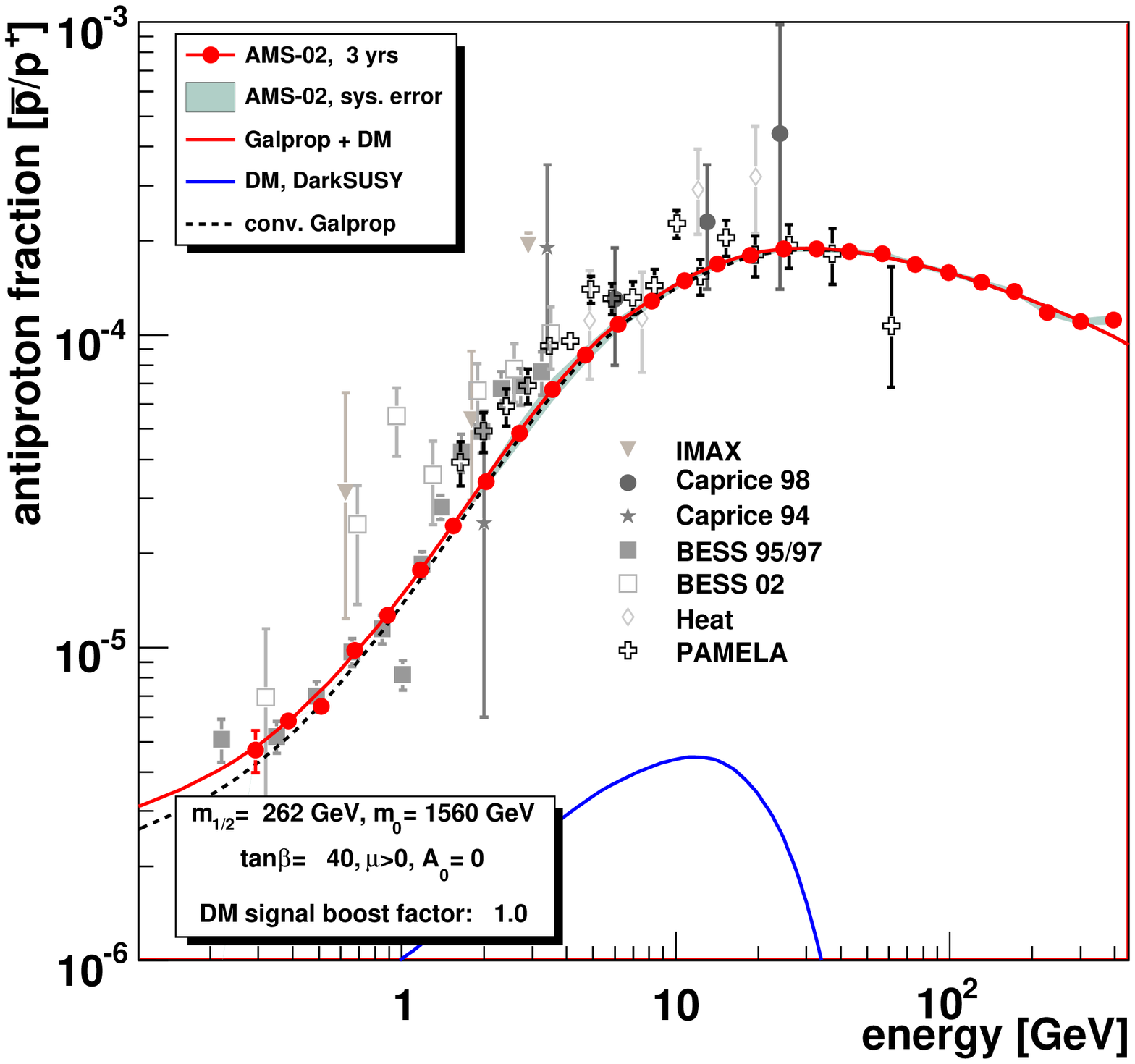,width=8cm}}\captionof{figure}{\label{f-pbar_p_fraction_dm_hg_phi550_40000_m_southpole_1095_days}Projected antiproton fraction measurements by AMS-02 with TOF, TRD and tracker subdetectors.}
\end{minipage}
\end{center}
\end{figure}

\begin{figure}
\begin{center}
\begin{minipage}[b]{.4\linewidth}
\centerline{\epsfig{file=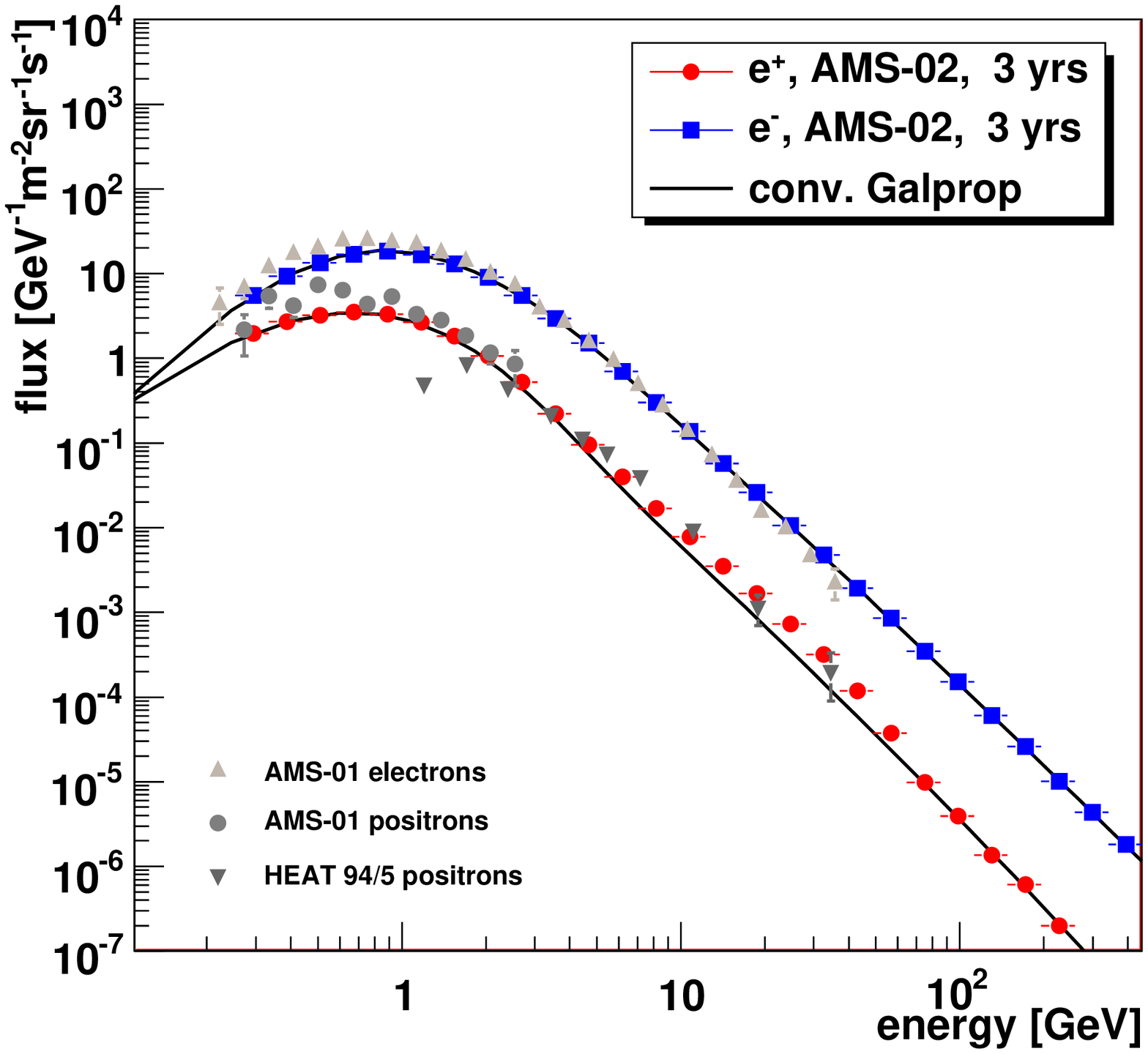,width=8cm}}\captionof{figure}{\label{f-e+_e-_fluxes_dm_hg_phi550_40000_m_southpole_1095_days}Projected electron and positron flux measurements by AMS-02 with TOF, TRD, tracker and ECAL subdetectors.}
\end{minipage}
\hspace{.1\linewidth}
\begin{minipage}[b]{.4\linewidth}
\centerline{\epsfig{file=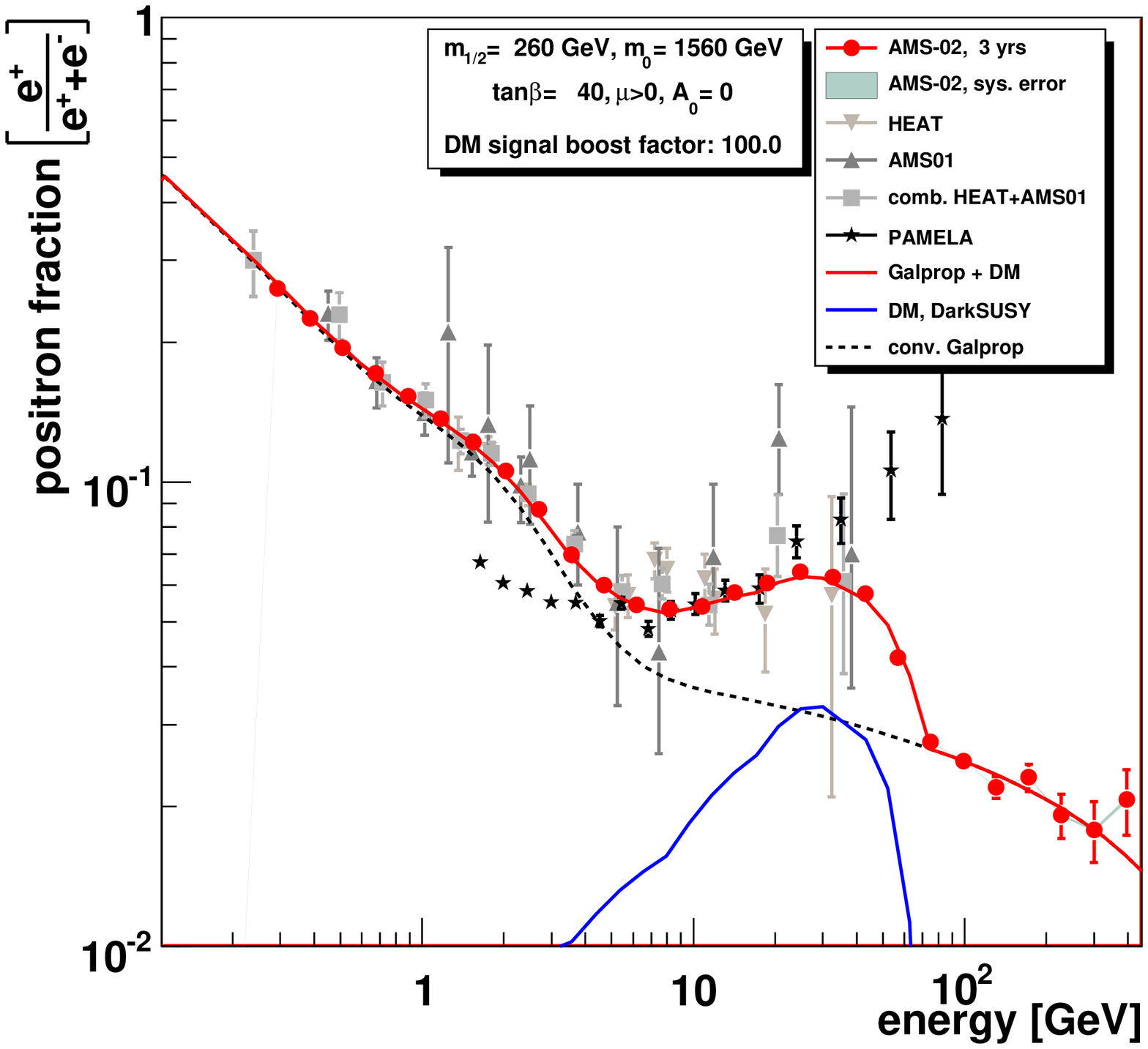,width=8cm}}\captionof{figure}{\label{f-e+_e-_fraction_dm_hg_phi550_40000_m_southpole_1095_days}Projected positron fraction measurements by AMS-02 with TOF, TRD, tracker and ECAL subdetectors.}
\end{minipage}
\end{center}
\end{figure}

The calculation assumes a measurement time of 3\,years, an acceptance of 0.45\,m$^2$sr without ECAL and 0.095\,m$^2$sr with ECAL and 10\,\% systematic error for the rejection factors and detection efficiencies. The probability for a particle to get through to the ISS as a function of rigidity is also respected and reduces the statistics at low energies (Fig.~\ref{f-cutoff_issorbit}). The momentum resolution of the tracker does not play an important role in the energy range up to about 500\,GeV (Eq.~\ref{e-tr}). Fig.~\ref{f-pbar_p_nams_dm_hg_phi550_40000_m_southpole_1095_days} and \ref{f-e+_e-_nams_dm_hg_phi550_40000_m_southpole_1095_days} show the difference between the total number of particles classified by the detector as a certain particle type and the corrected number of particles. The corrections in the low energy range are dominated by the geomagnetic cut-off effect. Starting from about 10\,GeV only small corrections due to detector efficiencies are needed. AMS-02 will be able to detect about $10^{9}$ - $10^{10}$ protons and $10^{6}$ antiprotons without the ECAL. Due to the shape of the proton to positron flux ratio, a reliable identification of positrons in the presence of the enormous proton background is possible up to energies of about 40\,GeV using TRD and TOF systems alone. For higher energies the ECAL must be used. Therefore, AMS-02 will be able to detect about $10^{8}$ electrons and $10^{6}$ positrons.

Fig.~\ref{f-pbar_p_fluxes_dm_hg_phi550_40000_m_southpole_1095_days} and \ref{f-pbar_p_fraction_dm_hg_phi550_40000_m_southpole_1095_days} show the results expected for the antiproton flux and fraction using only the TOF, TRD and tracker subdetectors. The systematic errors are not shown for the flux and are nearly not visible for the fraction. The antiproton fraction is also compared to the signal expected from a favored supersymmetric model for neutralino annihilations \cite{gast-2008}. The effect is small and emphasizes the need of good modeling of the solar and geomagnetic modulation in the low energy range for data analysis. Fig.~\ref{f-e+_e-_fluxes_dm_hg_phi550_40000_m_southpole_1095_days} and \ref{f-e+_e-_fraction_dm_hg_phi550_40000_m_southpole_1095_days} show the expected flux and positron fraction measurements for electrons and positrons using the ECAL with an rejection factor of $10^4$ for the complete energy range.

\newpage
\mbox{}
\newpage

\chapter{Conclusions}

Particle physics and astrophysics are merging. Experimental techniques and data are shared to build up one consistent theory for both subjects. The standard model of particle physics describes observations up to energies of 200\,GeV very well and is based on gauge symmetries with particles and antiparticles. It is known today that the Universe consists only of about 5\,\% of the standard model kind of matter. The dark matter makes up 23\,\% and the vast rest of about 70\,\% is some completely unknown type called dark energy. Furthermore, the asymmetry between matter and antimatter observed in the Universe cannot be explained by the standard model of particle physics. New theories and observations are needed to answer these fundamental questions. Therefore, new experiments in particle physics, astrophysics and astroparticle physics were designed and constructed.

The balloon-borne PEBS and the space-based AMS-02 missions are planned to use cosmic rays as messengers from space to constrain the properties of astrophysical objects, the nature of dark matter and theories for the baryon asymmetry. PEBS is planned to fly on Earth's poles and AMS-02 on the International Space Station. Only both experiments together are able to cover the complete sky. This might be important to reveal so far unknown cosmic ray sources. Cosmic rays travel through the galactic and interstellar medium and magnetic fields. Acceleration happens e.g. in supernovae remnants. Interactions with molecular clouds of primary cosmic rays like protons, helium nuclei and electrons can contribute additional antiparticles like positrons and antiprotons. These secondary particles have a smaller abundance. The cosmic rays are mostly composed of protons followed by helium and other nuclei, electrons, positrons, photons and antiprotons. There are so far no primary sources known for antiparticles. An observation of deviations from their small fluxes induced by secondary interactions could be an indicator for new effects. The current positron flux measurements show an excess starting at about 30\,GeV which could be explained by e.g. dark matter annihilations or nearby pulsars. Higher statistics and extended energy range are needed for a reliable analysis. Several theories like supersymmetry or Kaluza-Klein deliver viable candidates for dark matter. The mystery why no antimatter exists in our Universe has not been solved yet and no cosmic antinuclei were observed. It is believed that antinuclei like antihelium cannot form on time scales of the age of the Universe in a matter dominated environment. The measurement of one antinucleus would put tight constraints to antimatter theories. It is believed that small fractions of antimatter $<10^{-6}$ compared to matter could exist in antigalaxies about 10\,Mpc away from our galaxy. The observational challenge in the detection is to identify antiparticles or antinuclei against background particles and nuclei. The detectors described consist of subdetectors which are able to deliver a good separation up to particle energies of 100 - 1000\,GeV.

\begin{figure}
\begin{center}
\begin{minipage}[b]{.4\linewidth}
\centerline{\epsfig{file=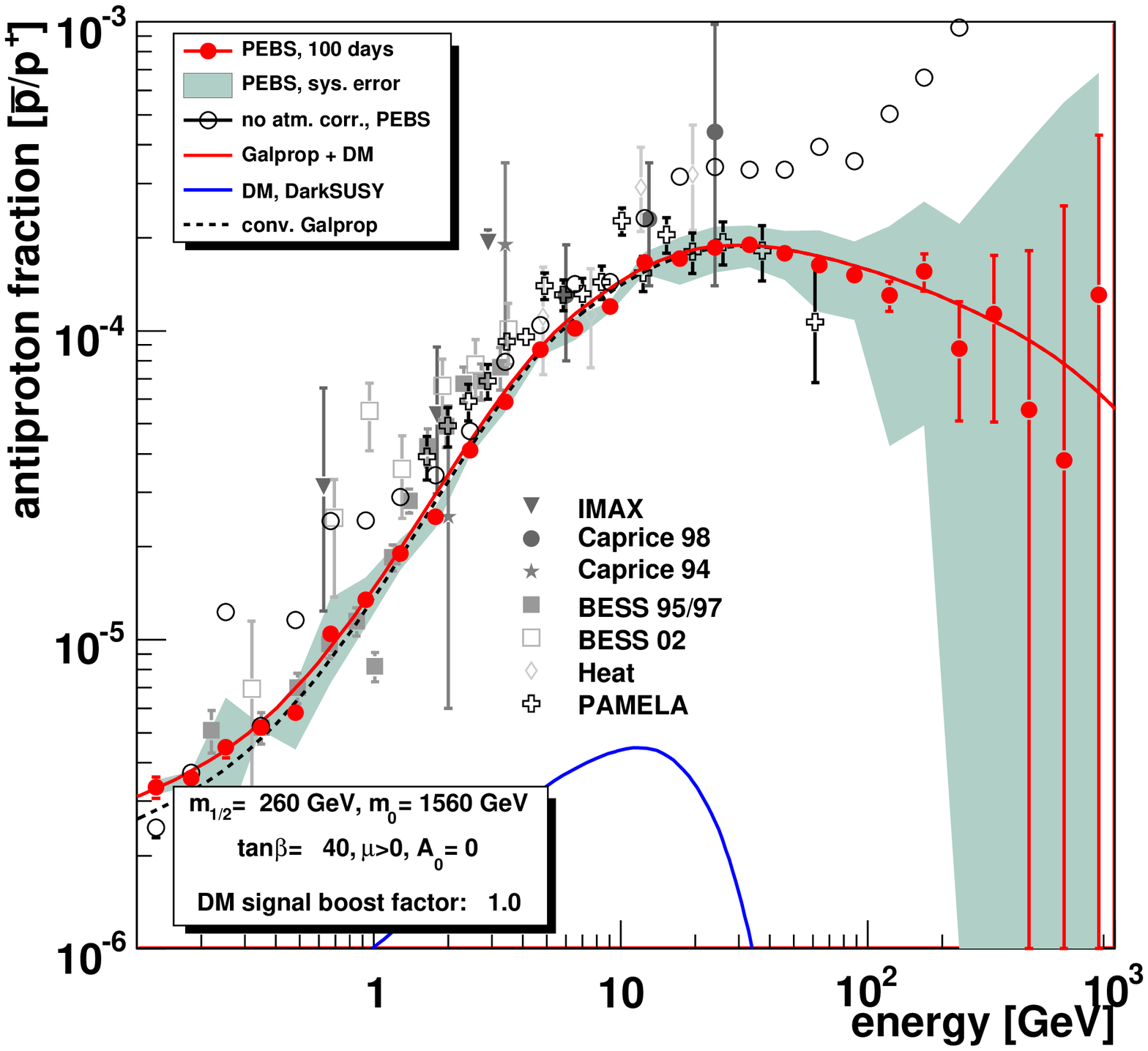,width=8cm}}
\captionof{figure}{\label{f-pbar_p_fraction_dm_hg_phi550_40000_m_southpole_100_days_conc}Projected antiproton fraction with a supersymmetric dark matter signal from neutralino annihilations.}
\end{minipage}
\hspace{.1\linewidth}
\begin{minipage}[b]{.4\linewidth}
\centerline{\epsfig{file=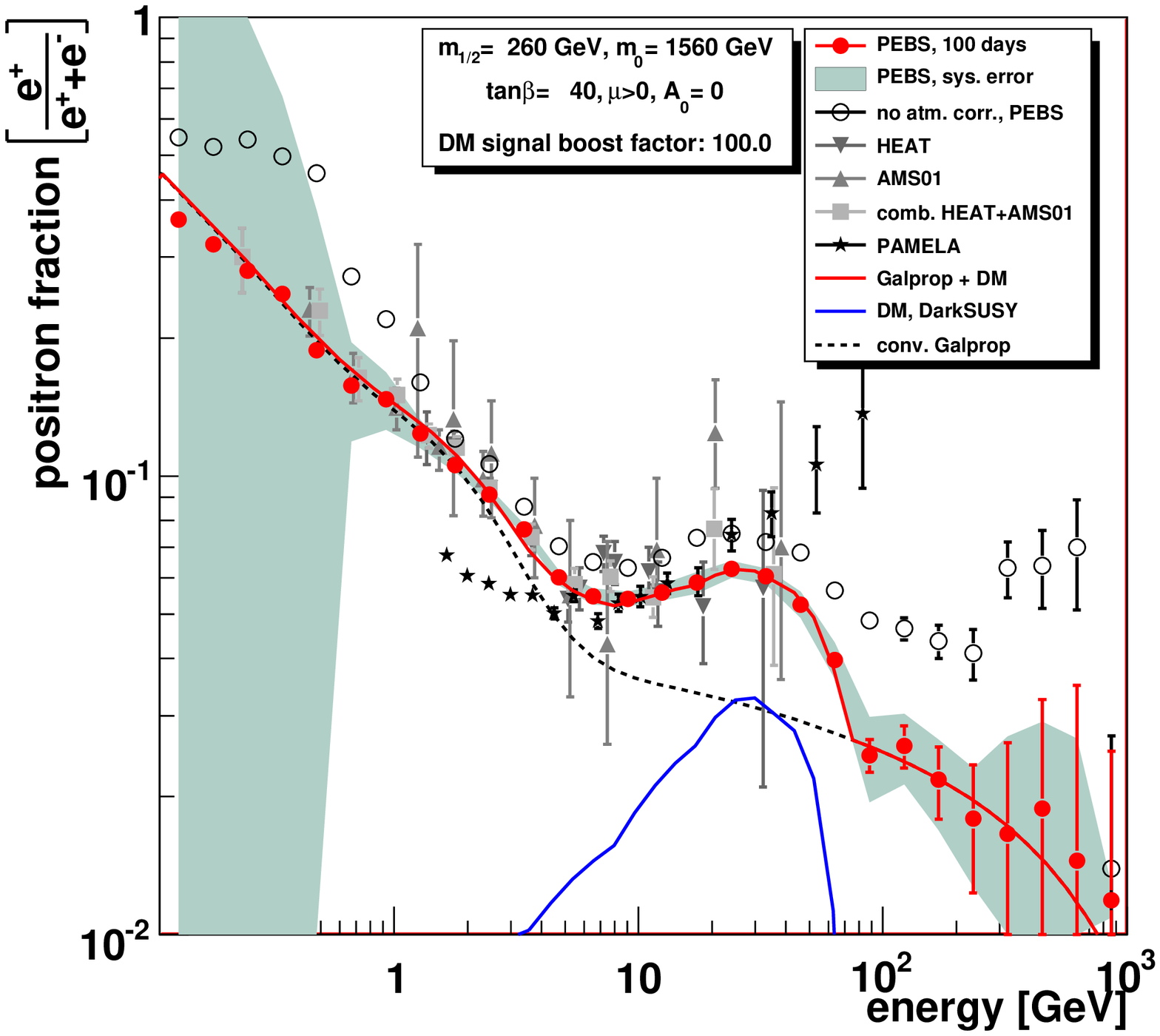,width=8cm}}
\captionof{figure}{\label{f-e+_e-_fraction_dm_hg_phi550_40000_m_southpole_100_days_conc}Projected positron fraction with a supersymmetric dark matter signal from neutralino annihilations.}
\end{minipage}
\end{center}
\end{figure}

One focus of this work was the calculation of atmospheric effects for the balloon-borne PEBS experiment which is planned to fly at Earth's poles at an altitude of 40\,km for 100\,days in total. The main goal is to improve the current measurements of cosmic positrons, electrons and antiprotons. The software package PLANETOCOSMICS based on GEANT4 was used to simulate the atmosphere and the magnetic field. The atmospheric effects cannot be neglected if one compares the atmospheric depth before 40\,km of 3.8\,g/cm$^2$ for perpendicular trajectories to the depth cosmic rays passed on their way to Earth of $6$ - $10$\,g/cm$^2$. Also important is the magnetic field which deflects particles depending on their rigidities and therefore geomagnetic cut-offs exist. A big advantage of cosmic-ray measurements at the poles compared to the AMS-02 orbit is the negligible cut-off. In this way it is also possible to study solar modulation effects down to low energies. As expected, the number of radiation lengths a particle has to cross before detection depends on the zenith angle. Therefore, large angles of cosmic rays are suppressed due to interactions while atmospheric secondary particles are nearly isotropic. This results in a reduced background for the measurement because atmospheric particles of large angles are not within the angular detector acceptance. The comparison of atmospheric fluxes measured by other experiments show good agreement to these simulations. The cosmic-ray antiprotons constitute about 60 - 70\,\% of the measured total flux classified as antiproton up to 100\,GeV. For energies higher than 100\,GeV the pion contribution becomes dominant. 80 - 90\,\% of the total number of particles classified as positrons at 40\,km are of cosmic origin in the energy range of 1 - 100\,GeV. The atmospheric contribution becomes huge in the energy range below 1\,GeV. Misidentified protons spoil the positron measurement starting from 700\,GeV. The fractions with statistical error bars and the systematic error contour are shown in Fig.~\ref{f-pbar_p_fraction_dm_hg_phi550_40000_m_southpole_100_days_conc} and \ref{f-e+_e-_fraction_dm_hg_phi550_40000_m_southpole_100_days_conc}. The measurements will have a higher accuracy starting from 1\,GeV than all experiments completed so far and will very well constrain propagation models and possible effects by new sources.

The AMS-02 detector will be installed on the International Space Station for three years in 2010 to measure cosmic rays without the interference of Earth's atmosphere. The charge sign and momentum measurements in the AMS-02 experiment are done with a silicon tracker in a high magnetic field of 0.8\,T created by a superconducting magnet. The anticoincidence counter (ACC) surrounds this cylindrical detector to assure a very clean track reconstruction. External particles crossing the detector from the side or particles from interactions inside the detector are able to generate hits in the tracker which could falsify the momentum and charge reconstruction. Events with hits in the ACC are going to be treated with special care and can be rejected from the analysis if needed. In addition, the ACC is needed to reduce the trigger rate of the experiment by vetoing the trigger decision during periods of high fluxes, e.g. in the South Atlantic Anomaly. Its acceptance is about eight times larger than for the rest of the experiment. This becomes important if the maximum data acquisition rate of 3\,kHz is exceeded at fluxes of $10^3$ - $10^4$\,m$^{-2}$sr$^{-1}$s$^{-1}$.

The development and test of the AMS-02 anticoincidence counter was the other focus of this thesis. The detector design concentrates on a reliable operation in a space environment and a high magnetic field after a flight in a Space Shuttle and installation on the International Space Station. The system needs a fast response and only a very small fraction of charged particles of $<10^{-4}$ may be missed. The ACC has a modular design consisting of 16 plastic scintillator panels made of Bicron BC-414 with a thickness of 8\,mm which form a cylinder around the tracker with a height of 830\,mm and a diameter of 1100\,mm. The scintillator light is guided by Kuraray Y-11(200)M wavelength shifting fibers (WLS) embedded in the panels via Toray PJU-FB1000 clear fibers to Hamamatsu R5946 fine mesh photomultiplier tubes (PMT). The signal has to be transported up to 2\,m away from the scintillator to the PMT because the photomultipliers are mounted at the position with the smallest possible magnetic stray field of 0.12\,T. The attenuation of the wavelength shifting fiber is too large for the complete transport length and a coupling to clear fibers can increase the final signal output. Therefore, it is important to match the angular acceptances of the fibers for a high transmission efficiency. The Toray PJU-FB1000 clear fiber has been chosen because it has a small attenuation at the green light of the WLS fiber and a large angular acceptance. The average total damping of the coupling and transportation is 2.1\,dB with an RMS of 0.1\,dB. The panels were tested after fabrication with a set of reference PMTs without clear fiber coupling and show an average of 19 photo-electrons at the photocathode with an RMS of 1 for the 16 flight panels. In addition, the PMTs have undergone space qualification tests with temperature cycles in the non-operational range of -35 - 50°C and the operational range of -30 - 45°C. The PMTs were tested with a reference panel and the selection for flight was based on the gain and the number of photo-electrons. The complete ACC system with the flight combination of panels, clear fiber cables and photomultipliers has an average output of 16 photo-electrons at the photocathode with an RMS of 1 and was installed into the complete AMS-02 detector with a reproducibility for the signal output of 99\,\% and an RMS of 8\,\%. 

\begin{figure}
\begin{center}
\begin{minipage}[b]{.4\linewidth}
\centerline{\epsfig{file=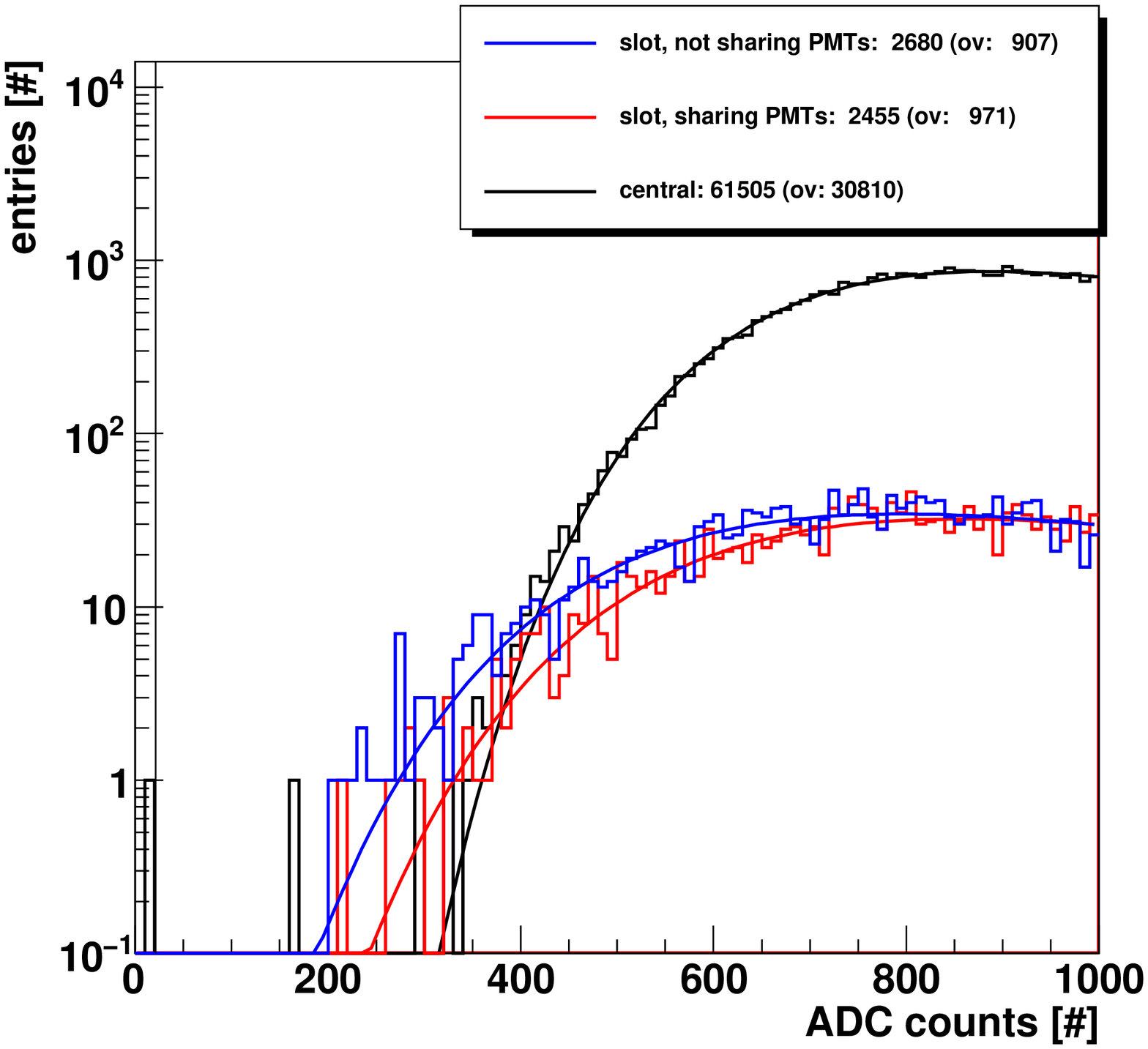,width=8cm}}\captionof{figure}{\label{f-080807_0_acc_adc_highest_conc}Highest ADC values out of all PMTs for each event during the atmospheric muon test with the complete AMS-02 experiment for tracks extrapolated from the TRD via the tracker to the ACC.}
\end{minipage}
\hspace{.1\linewidth}
\begin{minipage}[b]{.4\linewidth}
\centerline{\epsfig{file=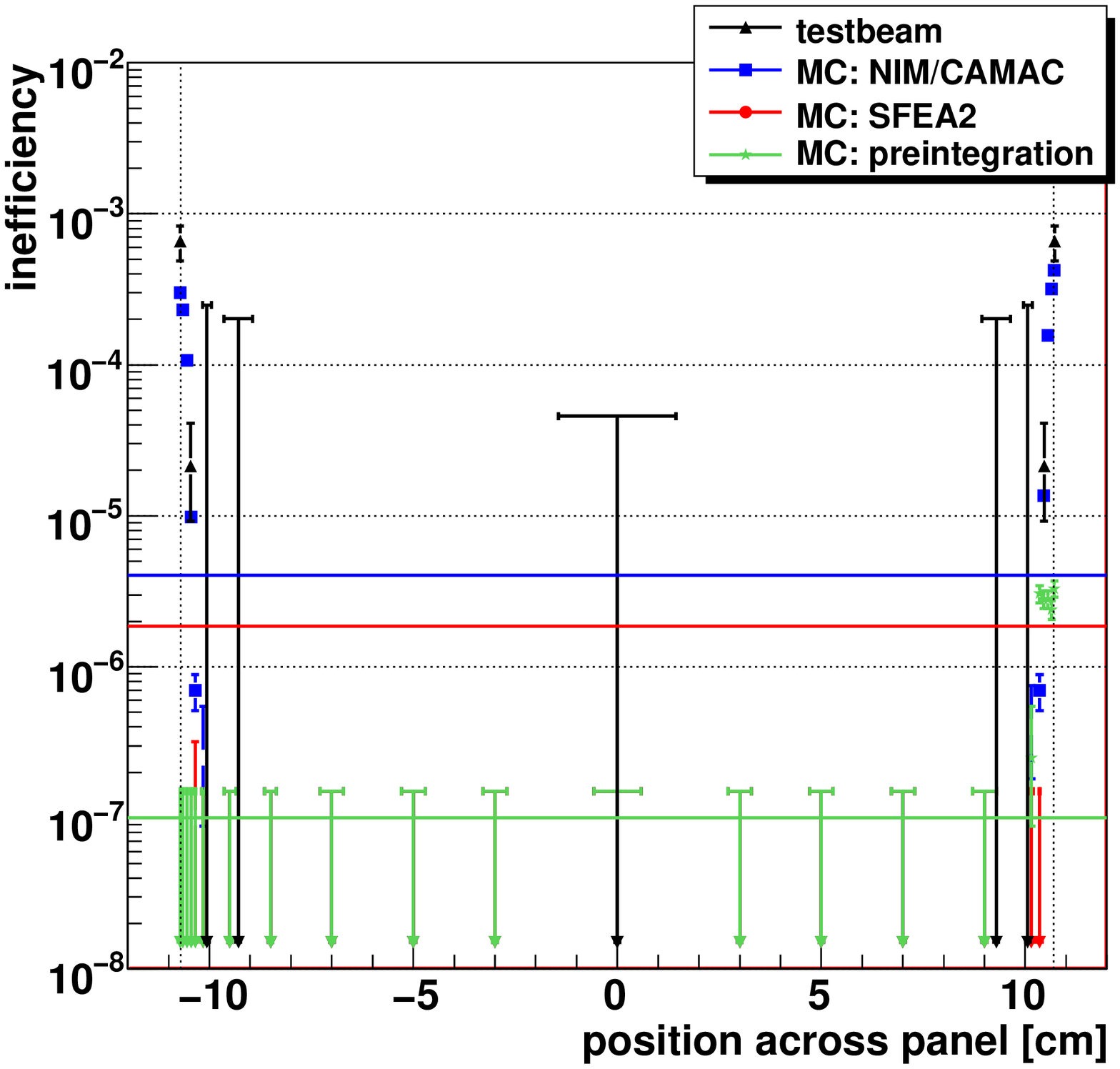,width=8cm}}\captionof{figure}{\label{f-eff_y_model_conc}Projected inefficiencies across the slot region between two ACC panels.}
\end{minipage}
\end{center}
\end{figure}

Special care had to be taken in the slot regions between two scintillator panels. This part dominates the determination of the mean ACC detection inefficiency. Testbeam measurements with conventional laboratory electronics show a sharp signal drop in these regions. The determination of the final inefficiency depends also on the final electronics and on the particle distribution in space which increases the average pathlength in the scintillator compared to the testbeam measurements by 35\,\%. The final readout is done in the S-crates where also the main trigger is processed. Tests with the electronics showed that the data rate can be reduced by $10^4$ and the resolution is good enough to resolve charges down to 0.7\,pC which corresponds to less than a photo-electron. The analysis of the atmospheric muons collected with the whole AMS-02 experiment allowed the extraction of further properties. The inefficiency was calculated by using tracks in the TRD and the tracker. The tracks in the transition radiation detector were extrapolated to the silicon tracker and hits in the tracker close to this track were used for a new fit with high resolution. Tracks pointing to an ACC panel were analysed and an detection inefficiency of about $10^{-5}$ could be derived for the complete detector (Fig.~\ref{f-080807_0_acc_adc_highest_conc}). Measurements with a well calibrated detector in nominal conditions may be able to improve this inefficiency as the data was taken during the first calibration period of the experiment. A mean inefficiency in the order of $10^{-7}$ - $10^{-6}$ was derived from simulations based on an ACC signal model considering only statistical fluctuations of the signal and neglecting further effects (Fig.~\ref{f-eff_y_model_conc}). 

This result enters the calculation of the expected capability to measure antimatter and to set upper bounds for a wrong charge and momentum reconstruction probability. Events could have a spoiled charge determination if external particles cross the ACC from the side and are close to a TOF triggered event in time and space. The upper bound for a wrong charge reconstruction probability for this type of event is $\cal O$$(10^{-13})$. The ACC does not close the tracker volume hermetically because of cable feed-throughs. External particles passing the gap can be excluded to be responsible for a wrong track reconstruction with an upper bound of $\cal O$$(10^{-10})$. The ACC plays also an important role to reject events with backsplash from the electromagnetic calorimeter or hard interactions in the tracker which change the inclination and the curvature of the track. These effects require further simulations.

The capability to measure antiparticle fluxes with AMS-02 without the electromagnetic calorimeter is interesting because it reduces the acceptance by a factor of about 5. Antiprotons can be well measured up to very high energies $\cal O$(500\,GeV) while for positrons this is only up to 40\,GeV the case. A reliable positron measurement at higher energies has to make use of the electromagnetic calorimeter.

The future cosmic-ray experiments discussed here for the energy range up to 500\,GeV will provide data with so far unseen precision and give strong constraints on the understanding of cosmic rays and their origin, on the nature of dark matter and on theories for antimatter in the Universe. AMS-02 is planned to be installed on the ISS in 2010 and the PEBS proposal foresees the first long duration flight in 2012. In the mean time the PAMELA and FERMI satellite-borne missions will take more data. The recent publications of PAMELA, ATIC-2, HESS and FERMI showed that very interesting theories are possible within the error bars. Just very recently HESS\cite{2009arXiv0905.0105H} and FERMI\cite{2009arXiv0905.0025F} published new electron data up to a few TeV which look very interesting but maybe suffer from systematic effects\cite{2009arXiv0905.0444S}. Thus, new experiments with high precision data are more than welcome.

\pagebreak

\clearpage
\phantomsection

\bibliographystyle{nature1}

\newpage
\mbox{}
\newpage

\chapter*{Acknowledgments}
\thispagestyle{empty}
First of all I would like to thank my supervisors Prof.~St.~Schael and Prof.~K.~Lübelsmeyer for giving me the opportunity to participate in this very interesting field of physics. I had the chance to learn a lot from them in many different fields over the last years. They always motivated me and always gave me good advice.

The fabrication, testing and installation of the anticoincidence counter would not have been possible without the collaboration with Th.~Kirn. We shared many fruitful hours together in offices, laboratories, clean rooms, cars to Geneva and last but not least on our bikes on the tortuous way to the Col de la Faucille. Thank you! I also want to thank of course all the others who made this project possible for all their small and large contributions. Particularly, besides many others: A.~Bachmann, A.~Basili, B.~Beischer, V.~Bindi, T.~Bruch, M.~Capell, E.~Choumilov, V.~Choutko, A.~Schukraft, A.~Schultz v.~Dratzig, A.~Kounine, A.~Lebedev, H.~Poisel, L.~Quadrani, G.~Schwering, Th.~Siedenburg and W.~Wallraff. Very special thanks go to Prof.~S.~Ting who is keeping the AMS-02 project alive and is doing everything he can to make it reality.

As a large part of my thesis is based on hardware there would be much less to write about without the workshops in Aachen. I would like to thank the mechanical workshop with B.~Debye, F.~Gillessen, F.~Müller, H.~Müller, G.~Kirchhoff and H.~Paprotney and the electronics workshop with M.~Dohmen, F.~Franzke, I.~Özen and St.~Schmitz for their very good and precise work. I want to thank in particular M.~Wlochal and W.~Karpinski for very helpful discussions and strong support.

For the work on the atmospheric simulations I have to thank L.~Desorgher for the development of the PLANETOCOSMICS code and H.~Gast for countless enlightening discussions.

I would also like to thank H.~Gast, Th.~Kirn and D.~Pandoulas for the revision of the manuscript.

I also really want to thank all my other colleagues for working with me, sharing the office together, the nice work climate, sharing a beer and especially for all the discussions not related to physics and the chats helping to keep an even balance: R.~Adolphi, T.~Bingler, R.~Brauer, C.H.~Chung, F.~Doemmecke, N.~Drießen, A.~Fischer, S.~Funk, F.~Giovacchini, J.~Hattenbach, K.~Klein, R.~Greim, Ch.~Kukulies, M.~Millinger, J.~Olzem, F.~Raupach, G.~Roper, M.~Schaefbauer, M.~Thomas and all the others I forgot.

Heartfelt thanks go to all my friends out there. You do not have to be blood to be family! Just to drop a few names: Rolf Berghammer, Philipp Biallaß, Martin Eckhardt, Tim Echtermeyer, Henning Gast, Patrik Grychtol, Erik Heigenhauser, Rebecca Johr, and Arno Kellermann. I really do not forget all the others I shared time with here and there, taught me lessons about life, went to school with, discussed with, partied together, survived the mosh pit with, tattooed me, had great days in the powder with or met while travelling. Thank you!

Sandra Spickermann saved my soul with her love and helped me to carry on while I was down. Thank you so very much!

The last but of course not least wholehearted thanks go to my family Gerlinde, Magnus, Verena and Moritz von Doetinchem and Amo Lorenz. They have always been there for me, supported me and helped me along in all phases of my life.

So let's see what the future brings and never forget:

\begin{center}
\begin{tabular}{l}
\textbf{\guillemotright[\dots] Because life's too short, so I do what I can to get by.}\\
\textbf{I'll decide where my time is spent and you can bet there'll be a smile on my face.}\\
\textbf{How about yours? How about yours!?\guillemotleft}\\
- Gorilla Biscuits, Time Flies
\end{tabular}
\end{center}

\end{document}